\documentclass[a4,11pt,twocolumn,twoside]{rsk-book}
\usepackage{epsf}
\usepackage{graphicx}
\usepackage{makeidx}
\usepackage{rotating}
\usepackage{multicol}
\usepackage[T1]{fontenc}
\usepackage{ae}
\usepackage{palatino}
\usepackage{mathpple}
\usepackage[round,authoryear]{natbib}

\renewcommand{\baselinestretch}{1.05}

\newcommand{\sig}{\:\lower0.6ex\hbox{$\stackrel{\textstyle >}{\sim}$}\:}
\newcommand{\sil}{\:\lower0.6ex\hbox{$\stackrel{\textstyle <}{\sim}$}\:}
\newcommand{\sigs}{\:\lower0.4ex\hbox{$\stackrel{\scriptstyle
      >}{\scriptstyle \sim}$}\,}
\newcommand{\sils}{\:\lower0.4ex\hbox{$\stackrel{\scriptstyle
      <}{\scriptstyle \sim}$}\,}
\newcounter{saveeqna}

\newcounter{saveeqnb}

\newcommand{\rsksection}{\chapter}
\newcommand{\rsksubsection}{\section}
\newcommand{\rsksubsubsection}{\subsection}
\newcommand{\rskparagraph}{\subsubsection}
\def\etal{{\em et {al.}}}%
\begin{document}

%%%%%%%%%%%%%%%%%%%%%%%%%%%%%%%%%%%%%%%%%%%%%%%%%%%%%%%%%%%%%%%%%%%%%%
\frontmatter
% ===========
% Titelseiten
% ===========
%
\onecolumn
\setcounter{page}{-9}

\newfont{\Giga}{cmssbx10 scaled 5200}
\newfont{\giga}{cmssbx10 scaled 4300}
\newfont{\Mega}{cmssbx10 scaled 3200}
\newfont{\mega}{cmssbx10 scaled 2500}
\newfont{\Kilo}{cmssbx10 scaled 2000}
\newfont{\kilo}{cmssbx10 scaled 1600}
\newfont{\Deca}{cmssbx10 scaled 1450}
\newfont{\deca}{cmssbx10 scaled 1200}

\thispagestyle{empty}
\vspace*{5mm}

%\vspace{1cm}
\begin{center}
{\giga                The Relation\\[0.5cm]  between \\[0.5cm]
Interstellar Turbulence\\[0.5cm] and\\[0.5cm] Star Formation } 
\end{center}

\vspace{2.5cm}
\begin{figure}[h]
\unitlength1.0cm
\begin{picture}(16,10)
\put( 2.2, 0.0) {\epsfxsize=13cm\epsfbox{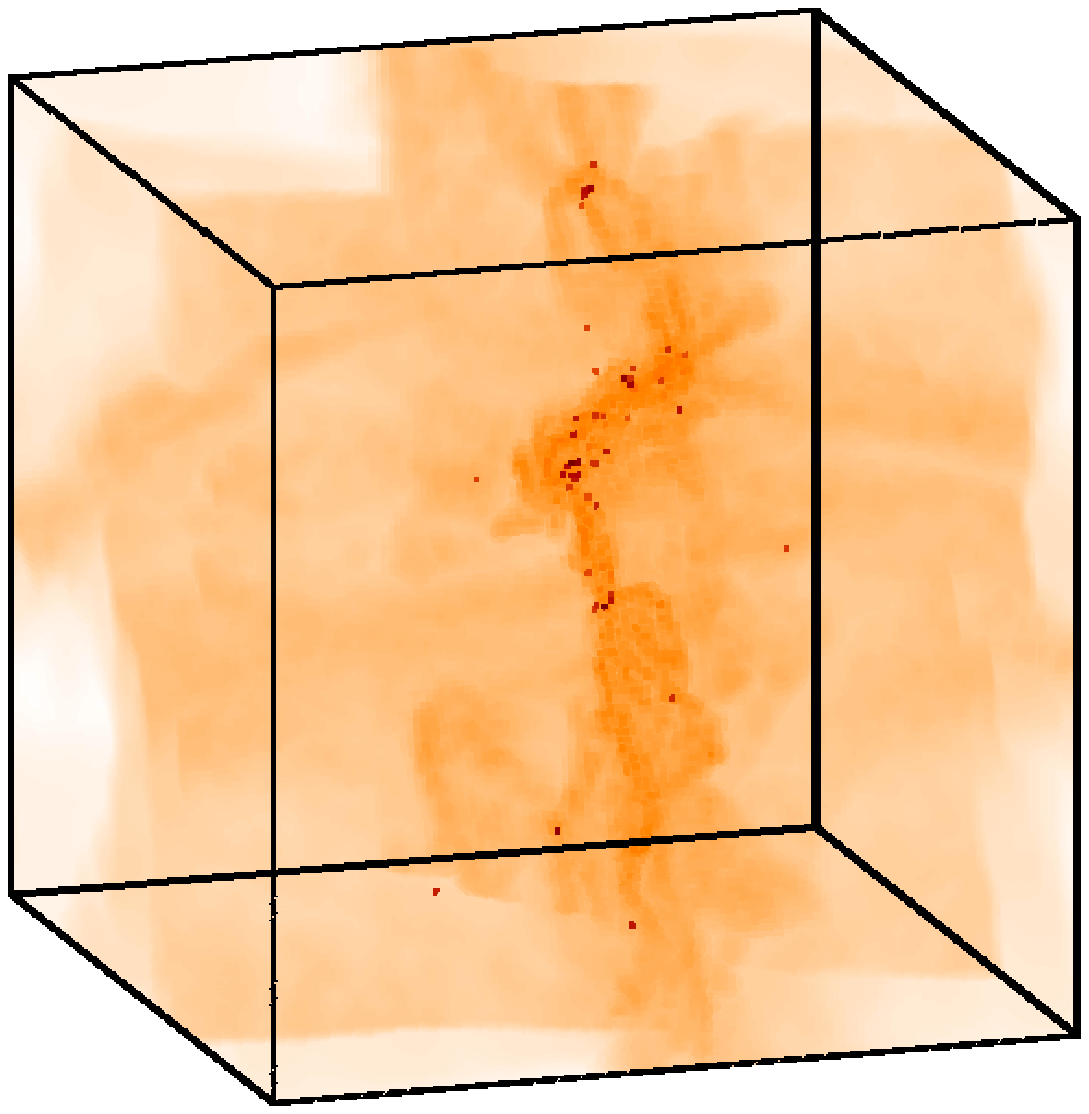}}
\end{picture}
\end{figure}

\vspace{0.5cm}
\begin{center}
{\Mega Ralf Klessen}\\
\end{center}

\newpage
\thispagestyle{empty}
\mbox{~}
\newpage
\thispagestyle{empty}

\begin{center}
{\Mega  The Relation between \\[0.4cm]
Interstellar Turbulence and \\[0.4cm]Star Formation  }\\
\vspace{3cm}
{\huge \sf Habilitationsschrift}\\
\vspace{1cm}
{\Large zur Erlangung} \\
\vspace{1ex}
{\Large der venia legendi} \\
\vspace{1ex}
{\Large f{\"u}r das Fach Astronomie} \\
\vspace{1ex}
{\Large an der Universit{\"a}t Potsdam}\\
\vfill
{\Large vorgelegt von}\\
\vspace{2ex}
{\huge \sf Ralf S.\ Klessen}\\
\vspace{2ex}
{\Large aus Landau a.~d.~Isar }\\
\vspace{3ex}
{\Large (im M{{\"a}}rz 2003)}
\end{center}

%%%%%%%%%%%%%%%%%%%%%%%%%%%%%%%%%%%%%%%%%%%%%%%%%%%%%%%%%%%%%%%%%%%%%%

\newpage\rule[0mm]{0mm}{0mm}
\thispagestyle{empty}
\newpage

%%%%%%%%%%%%%%%%%%%%%%%%%%%%%%%%%%%%%%%%%%%%%%%%%%%%%%%%%%%%%%%%%%%%%%

\newpage\rule[0mm]{0mm}{0mm}
\thispagestyle{empty}

{\large\bf\sf Deutsche Zusammenfassung:}

Eine der zentralen Fragestellungen der modernen Astrophysik ist es,
unser Verst{\"a}ndnis f{\"u}r die Bildung von Sternen und Sternhaufen in
unserer Milchstra{\ss}e zu erweitern und zu vertiefen. Sterne entstehen
in interstellaren Wolken aus molekularem Wasserstoffgas. In den
vergangenen zwanzig bis drei{\ss}ig Jahren ging man davon aus, da{\ss} der
Proze{\ss} der Sternentstehung vor allem durch das Wechselspiel von
gravitativer Anziehung und magnetischer Absto{\ss}ung bestimmt ist.
Neuere Erkenntnisse, sowohl von Seiten der Beobachtung als auch der
Theorie, deuten darauf hin, da{\ss} nicht Magnetfelder, sondern
{\"U}berschallturbulenz die Bildung von Sternen in galaktischen
Molek{\"u}lwolken bestimmt.

Diese Arbeit fa{\ss}t diese neuen {\"U}berlegungen zusammen, erweitert sie
und formuliert eine {\sl neue Theorie der Sternentstehung} die auf dem
komplexen Wechselspiel von Eigengravitation des Wolkengases und der
darin beobachteten {\"U}berschallturbulenz basiert.  Die kinetische
Energie des turbulenten Geschwindigkeitsfeldes ist typischerweise
ausreichend, um interstellare Gaswolken auf gro{\ss}en Skalen gegen
gravitative Kontraktion zu stabilisieren. Auf kleinen Skalen jedoch
f{\"u}hrt diese Turbulenz zu starken Dichtefluktuationen, wobei einige
davon die lokale kritische Masse und Dichte f{\"u}r gravitativen Kollaps
{\"u}berschreiten k{\"o}nnen. Diese Regionen schockkomprimierten Gases sind
es nun, aus denen sich die Sterne der Milchstra{\ss}e bilden. Die
Effizienz und die Zeitskala der Sternentstehung h{\"a}ngt somit
unmittelbar von den Eigenschaften der Turbulenz in interstellaren
Gaswolken ab. Sterne bilden sich langsam und in Isolation, wenn der
Widerstand des turbulenten Geschwindigkeitsfeldes gegen gravitativen
Kollaps sehr stark ist.  {\"U}berwiegt hingegen der Einflu{\ss} der
Eigengravitation, % auf gro{\ss}en Skalen, 
dann bilden sich Sternen in dichten Gruppen oder Haufen sehr rasch und
mit gro{\ss}er Effizienz.

Die Vorhersagungen dieser Theorie werden sowohl auf Skalen einzelner
Sternentstehungsgebiete als auch auf Skalen der Scheibe unserer
Milchstra{\ss}e als ganzes untersucht. Es zu erwarten, da{\ss} protostellare
Kerne, d.h.\ die direkten Vorl{\"a}ufer von Sternen oder
Doppelsternsystemen, eine hochgradig dynamische Zeitentwicklung
aufweisen, und keineswegs quasi-statische Objekte sind, wie es in der
Theorie der magnetisch moderierten Sternentstehung vorausgesetzt wird.
So mu{\ss} etwa die Massenanwachsrate junger Sterne starken zeitlichen
Schwankungen unterworfen sein, was wiederum wichtige Konsequenzen f{\"u}r
die statistische Verteilung der resultierenden Sternmassen hat. Auch
auf galaktischen Skalen scheint die Wechselwirkung von Turbulenz und
Gravitation ma{\ss}geblich. Der Proze{\ss} wird hier allerdings noch
zus{\"a}tzlich moduliert durch chemische Prozesse, die die Heizung und
K{\"u}hlung des Gases bestimmen, und durch die differenzielle Rotation
der galaktischen Scheibe. Als wichtigster Mechanismus zur Erzeugung
der interstellaren Turbulenz l{\"a}{\ss}t sich die {\"U}berlagerung vieler
Supernova-Explosionen identifizieren, die das Sterben massiver Sterne
begleiten und gro{\ss}e Mengen an Energie und Impuls freisetzen.
Insgesamt unterst{\"u}tzen die Beobachtungsbefunde auf allen Skalen das
Bild der turbulenten, dynamischen Sternentstehung, so wie es in dieser
Arbeit ge\-{\mbox{zeich}}\-net wird.

\newpage\rule[0mm]{0mm}{0mm}
\thispagestyle{empty}
%%%%%%%%%%%%%%%%%%%%%%%%%%%%%%%%%%%%%%%%%%%%%%%%%%%%%%%%%%%%%%%%%%%%%%
\newpage\rule[0mm]{0mm}{0mm}
\thispagestyle{empty}

{\large\bf\sf Summary:}

  Understanding the formation of stars in galaxies is central to much
  of modern astrophysics. For several decades it has been thought that
  the star formation process is primarily controlled by the interplay
  between gravity and magnetostatic support, modulated by neutral-ion
  drift.  Recently, however, both observational and numerical work has
  begun to suggest that supersonic interstellar turbulence rather than
  magnetic fields controls star formation.
  
  This review begins with a historical overview of the successes and problems
  of both the classical dynamical theory of star formation, and the standard
  theory of magnetostatic support from both observational and theoretical
  perspectives.  We then present the outline of a new paradigm of star
  formation based on the interplay between supersonic turbulence and
  self-gravity.  Supersonic turbulence can provide support against
  gravitational collapse on global scales, while at the same time it produces
  localized density enhancements that allow for collapse on small scales. The
  efficiency and timescale of stellar birth in Galactic gas clouds
  strongly depend on the properties of the interstellar turbulent velocity
  field, with slow, inefficient, isolated star formation being a hallmark of
  turbulent support, and fast, efficient, clustered star formation occurring
  in its absence.
  
  After discussing in detail various theoretical aspects of supersonic
  turbulence in compressible self-gravitating gaseous media relevant for star
  forming interstellar  clouds, we explore the consequences of the new
  theory for both local star formation and galactic scale star formation.  The
  theory predicts that individual star-forming cores are likely not
  quasi-static objects, but dynamically evolving.  Accretion onto
  these objects will vary with time and depend on the properties of the
  surrounding turbulent flow. This has important consequences for the
  resulting stellar mass function. Star formation on scales of galaxies as a
  whole is expected to be controlled by the balance between gravity and
  turbulence, just like star formation on scales of individual interstellar
  gas clouds, but may be modulated by additional effects like cooling and
  differential rotation.  The dominant mechanism for driving interstellar
  turbulence in star-forming regions of galactic disks appears to be
  supernovae explosions. In the outer disk of our Milky Way or in low-surface
  brightness galaxies the coupling of rotation to the gas through magnetic fields
  or gravity may become important.

\newpage\rule[0mm]{0mm}{0mm}
\thispagestyle{empty}
%%%%%%%%%%%%%%%%%%%%%%%%%%%%%%%%%%%%%%%%%%%%%%%%%%%%%%%%%%%%%%%%%%%%%%

 \newpage\rule[0mm]{0mm}{0mm}
 \thispagestyle{empty}
 
 \newcommand{\rsksf}{\bf}

 The scientific content of this {\em habilitation thesis} is based on
 the following refereed publications. The corresponding sections of the
 {\em thesis} are indicated in parentheses:

  \renewcommand{\labelitemi}{$\circ$}
 \begin{itemize}
 \item Heitsch, F., M.-M.\ Mac~Low, and {\rsksf R.\ S.\ Klessen},
   2001, {\em The Astrophysical Journal}, {\bf 547}, 280 -- 291:
   ``Gravitational Collapse in Turbulent Molecular Clouds. II.
   Magnetohydrodynamical Turbulence''\hfill\mbox{\sf (parts of \S\S2.1 -- 2.6)}
 \item {\rsksf Klessen, R.~S.}, 2003, {\em Reviews in Modern Astronomy
     16}, in press (Ludwig Biermann Lecture, astro-ph/0301381, 33
   pages): ``Star Formation from Interstellar Clouds'' \\\mbox{~~~~~~~~~~~}\hfill \mbox{\sf
     (parts of \S1, \S\S2.3 -- 2.6, parts of \S6)}
 \item {\rsksf Klessen, R.~S.}, 2001, {\em The Astrophysical Journal},
   {\bf 556}, 837 -- 846: ``The Formation of Stellar Clusters: Mass
   Spectra from Turbulent Fragmentation'' \hfill \mbox{\sf (\S4.4, \S4.7)}
 \item {\rsksf Klessen, R.~S.}, 2001, {\em The Astrophysical Journal},
   {\bf 550}, L77 -- L80: ``The Formation of Stellar Clusters: Time
   Varying Protostellar Accretion Rates''\hfill \mbox{\sf (\S4.4, \S4.5)}
 \item {\rsksf Klessen, R.~S.}, 2000, {\em The Astrophysical Journal},
   {\bf 535}, 869 -- 886: ``One-Point Probability Distribution
   Functions of Supersonic, Turbulent Flows in Self-Gravitating Media''
 \hfill {\sf (\S3.2)}
 \item {\rsksf Klessen, R.~S.}, and A.\ Burkert, 2000, {\em The
     Astrophysical Journal Supplement Series}, {\bf 128}, 287 -- 319:
   ``The Formation of Stellar Clusters: Gaussian Cloud Conditions I''\hfill
   \mbox{\sf(parts of \S\S4.1 -- 4.4)} 
 \item {\rsksf Klessen, R.\ S.}, and A.\ Burkert, 2001, {\em The
     Astrophysical Journal}, {\bf 549}, 386 -- 401: ``The Formation of
   Stellar Clusters: Gaussian Initial Conditions II'' \hfill
   \mbox{\sf(parts of \S\S4.1 -- 4.4)} 
 \item {\rsksf Klessen, R.\ S.}, and D.\ N.\ C.\ Lin, 2003, {\em
     Physical Review E}, in press: ``Diffusion in Supersonic,
   Turbulent, Compressible Flows''\hfill \mbox{\sf (\S3.1)}
 \item {\rsksf Klessen, R.~S.}, F.\ Heitsch, and M.-M.\ Mac~Low, 2000,
   {\em The Astrophysical Journal}, {\bf 535}, 887 -- 906:
   ``Gravitational Collapse in Turbulent Molecular Clouds: I.
   Gasdynamical Turbulence'' \\ \mbox{~}\hfill\mbox{\sf (parts of \S\S2.1 -- 2.6, \S3.3)}
 \item Mac~Low, M.-M., and {\rsksf R.\ S.\ Klessen}, 2003, {\em Reviews
     of Modern Physics}, submitted (astro-ph/0301093, 86 pages): ``The
   Control of Star Formation by Supersonic Turbulence'' \\ \mbox{~}\hfill\mbox{\sf
     (parts of \S\S1, 2, 4 -- 6)}
 \item Mac~Low, M.-M., {\rsksf R.\ S.\ Klessen}, A.\ Burkert, M.\ D.\ 
   Smith, 1998, {\em Physical Review Letters}, {\bf 80}, 2754 -- 2757:
   ``Kinetic Energy Decay Rates of Supersonic and Super-Alfv\'enic
   Turbulence in Star-Forming Clouds''\hfill \mbox{\sf (\S2.5.1)}
 \item Ossenkopf, V., {\rsksf R.\ S.\ Klessen}, and F.\ Heitsch, 2001,
   {\em Astronomy \& Astrophysics}, {\bf 379}, 1005 -- 1016: ``On the
   Structure of Turbulent Self-Gravitating Molecular Clouds''\hfill \mbox{\sf (\S3.4)}
 \item Wuchterl, G., and {\rsksf R.\ S.\ Klessen}, 2001, {\em The
     Astrophysical Journal}, {\bf 560}, L185 -- L188: ``The First
   Million Years of the Sun''\hfill \mbox{\sf (\S4.5)}\hfill 
 \end{itemize}

 \newpage\rule[0mm]{0mm}{0mm}
 \thispagestyle{empty}

%%%%%%%%%%%%%%%%%%%%%%%%%%%%%%%%%%%%%%%%%%%%%%%%%%%%%%%%%%%%%%%%%%%%%%

\newpage\rule[0mm]{0mm}{0mm}
\thispagestyle{empty}
\hfill {\em \begin{minipage}[t]{11.0cm}\begin{tabular}{p{0.7cm}p{9.9cm}}
 6.44 & Nicht `wie' die Welt ist, ist das Mystische,
  sondern `da{\ss}' sie ist.\\
 6.52 & Wir f{\"u}hlen, da{\ss} selbst, wenn alle `m{\"o}glichen'
 wissenschaftlichen Fragen beantwortet sind, unsere Lebensprobleme
 noch gar nicht ber{\"u}hrt sind. Freilich bleibt dann eben keine
 Frage mehr; und eben dies ist die Antwort.\\[0.3cm]
& (Ludwig Wittgenstein, Logisch-Philosophische Abhandlung)
\end{tabular}
\end{minipage}
} 
\vfill
\hfill {\Large \em meiner Familie gewidmet}

\newpage\rule[0mm]{0mm}{0mm}
\thispagestyle{empty}

\twocolumn

\onecolumn
\thispagestyle{empty}
{\Huge{\sf CONTENTS}}\vskip1pc
{\sf {\large \contentsline {chapter}{\numberline {1}INTRODUCTION}{1}}
\contentsline {section}{\numberline {1.1}Overview}{1}
\contentsline {section}{\numberline {1.2}Turbulence}{3}
\contentsline {section}{\numberline {1.3}Outline}{5}
{\large \contentsline {chapter}{\numberline {2}TOWARDS A NEW PARADIGM}{7}}
\contentsline {section}{\numberline {2.1}Classical Dynamical Theory}{9}
\contentsline {section}{\numberline {2.2}Problems with Classical Theory}{13}
\contentsline {section}{\numberline {2.3}Standard Theory of Isolated Star Formation}{14}
\contentsline {section}{\numberline {2.4}Problems with Standard Theory}{19}
\contentsline {subsection}{\numberline {2.4.1}Singular Isothermal Spheres}{19}
\contentsline {subsection}{\numberline {2.4.2}Observations of Clouds and Cores}{21}
\contentsline {subsubsection}{Magnetic Support}{21}
\contentsline {subsubsection}{Infall Motions}{23}
\contentsline {subsubsection}{Density Profiles}{24}
\contentsline {subsubsection}{Chemical Ages}{24}
\contentsline {subsection}{\numberline {2.4.3}Observations of Protostars and Young Stars}{25}
\contentsline {subsubsection}{Accretion Rates}{25}
\contentsline {subsubsection}{Embedded Objects}{26}
\contentsline {subsubsection}{Stellar Ages}{27}
\contentsline {section}{\numberline {2.5}Beyond the Standard Theory}{28}
\contentsline {subsection}{\numberline {2.5.1}Maintenance of Supersonic Motions}{28}
\contentsline {subsection}{\numberline {2.5.2}Turbulence in Self-Gravi\discretionary {-}{}{}ta\discretionary {-}{}{}ting Gas}{28}
\contentsline {subsection}{\numberline {2.5.3}A Numerical Approach}{30}
\contentsline {subsection}{\numberline {2.5.4}Global Collapse}{30}
\contentsline {subsection}{\numberline {2.5.5}Local Collapse in Globally Stable Regions}{31}
\contentsline {subsection}{\numberline {2.5.6}Effects of Magnetic Fields}{33}
\contentsline {subsection}{\numberline {2.5.7}Promotion and Prevention of Local Collapse}{35}
\contentsline {subsection}{\numberline {2.5.8}The Timescales of Star Formation}{37}
\contentsline {subsection}{\numberline {2.5.9}Scales of Interstellar Turbulence}{37}
\contentsline {subsection}{\numberline {2.5.10}Efficiency of Star Formation}{39}
\contentsline {subsection}{\numberline {2.5.11}Termination of Local Star Formation}{39}
\contentsline {section}{\numberline {2.6}Outline of a New Theory of Star Formation}{41}
{\large \contentsline {chapter}{\numberline {3}PROPERTIES OF SUPERSONIC TURBULENCE}{43}}
\contentsline {section}{\numberline {3.1}Transport Properties}{43}
\contentsline {subsection}{\numberline {3.1.1}Introduction}{43}
\contentsline {subsection}{\numberline {3.1.2}A Statistical Description of Turbulent Diffusion}{44}
\contentsline {subsection}{\numberline {3.1.3}Numerical Method}{45}
\contentsline {subsection}{\numberline {3.1.4}Flow Properties}{47}
\contentsline {subsection}{\numberline {3.1.5}Transport Properties in an Absolute Reference Frame}{48}
\contentsline {subsection}{\numberline {3.1.6}Transport Properties in Flow Coordinates}{49}
\contentsline {subsection}{\numberline {3.1.7}A Mixing Length Description}{50}
\contentsline {subsection}{\numberline {3.1.8}Summary}{53}
\contentsline {section}{\numberline {3.2}One-Point Probability Distribution Function}{54}
\contentsline {subsection}{\numberline {3.2.1}Introduction}{54}
\contentsline {subsection}{\numberline {3.2.2}PDF's and Their Interpretation}{55}
\contentsline {subsubsection}{Turbulence and PDF's}{55}
\contentsline {subsubsection}{PDF's of Observable Quantities}{55}
\contentsline {subsubsection}{Statistical Definitions}{57}
\contentsline {subsection}{\numberline {3.2.3}PDF's from Gaussian Velocity Fluctuations}{57}
\contentsline {subsection}{\numberline {3.2.4}Analysis of Decaying Su\discretionary {-}{}{}per\discretionary {-}{}{}sonic Turbulence without Self-Gravity}{60}
\contentsline {subsection}{\numberline {3.2.5}Analysis of Decaying Turbulence with Self-Gravity}{63}
\contentsline {subsection}{\numberline {3.2.6}Analysis of Driven Tur\discretionary {-}{}{}bu\discretionary {-}{}{}len\discretionary {-}{}{}ce with Self-Gravity}{68}
\contentsline {subsection}{\numberline {3.2.7}Summary}{70}
\contentsline {section}{\numberline {3.3}Fourier Analysis}{72}
\contentsline {subsection}{\numberline {3.3.1}Fourier Spectra as Function of Driving Wavelength}{73}
\contentsline {subsection}{\numberline {3.3.2}Fourier Spectra During Collapse}{73}
\contentsline {subsection}{\numberline {3.3.3}Summary}{74}
\contentsline {section}{\numberline {3.4}$\Delta $-Variance}{76}
\contentsline {subsection}{\numberline {3.4.1}Introduction}{76}
\contentsline {subsection}{\numberline {3.4.2}Turbulence Models}{76}
\contentsline {subsection}{\numberline {3.4.3}Density Structure}{78}
\contentsline {subsubsection}{$\Delta $-Variance Analysis}{78}
\contentsline {subsubsection}{Collapse of a Gaussian Density Field}{78}
\contentsline {subsubsection}{Interaction Between Gravity and Turbulence}{79}
\contentsline {subsubsection}{Influence of the Numerical Model}{82}
\contentsline {subsubsection}{Magnetic Fields}{82}
\contentsline {subsection}{\numberline {3.4.4}Velocity Structure}{83}
\contentsline {subsection}{\numberline {3.4.5}Comparison with Observations}{84}
\contentsline {subsubsection}{Dust Observations}{84}
\contentsline {subsubsection}{Molecular Line Observations}{85}
\contentsline {subsection}{\numberline {3.4.6}Summary}{87}
{\large \contentsline {chapter}{\numberline {4}LOCAL STAR FORMATION}{89}}
\contentsline {section}{\numberline {4.1}Molecular Clouds}{89}
\contentsline {subsection}{\numberline {4.1.1}Composition of Molecular Clouds}{89}
\contentsline {subsection}{\numberline {4.1.2}Density and Velocity Structure of Molecular Clouds}{89}
\contentsline {subsection}{\numberline {4.1.3}Support of Molecular Clouds}{93}
\contentsline {subsection}{\numberline {4.1.4}Scaling Relations for Molecular Clouds}{94}
\contentsline {section}{\numberline {4.2}Star Formation in Mo\discretionary {-}{}{}le\discretionary {-}{}{}cu\discretionary {-}{}{}lar Clouds}{95}
\contentsline {section}{\numberline {4.3}Properties of Protostellar Cores}{97}
\contentsline {section}{\numberline {4.4}Dynamical Interactions in Clusters}{102}
\contentsline {section}{\numberline {4.5}Accretion Rates}{104}
\contentsline {section}{\numberline {4.6}Protostellar Evolutionary Tracks}{106}
\contentsline {subsection}{\numberline {4.6.1}Dynamical PMS Calculations}{108}
\contentsline {subsection}{\numberline {4.6.2}Formation of a $1\,$M$_{\odot }$-Star}{109}
\contentsline {subsection}{\numberline {4.6.3}Implications}{111}
\contentsline {section}{\numberline {4.7}Initial Mass Function}{112}
\contentsline {subsection}{\numberline {4.7.1}The Observed IMF}{112}
\contentsline {subsection}{\numberline {4.7.2}Models of the IMF}{114}
\contentsline {subsection}{\numberline {4.7.3}Mass Spectra from Turbulent Fragmentation}{115}
{\large \contentsline {chapter}{\numberline {5}GALACTIC SCALE STAR FORMATION}{117}}
\contentsline {section}{\numberline {5.1}When is Star Formation Efficient?}{117}
\contentsline {subsection}{\numberline {5.1.1}Overview}{117}
\contentsline {subsection}{\numberline {5.1.2}Gravitational Instabilities in Galactic Disks}{118}
\contentsline {subsection}{\numberline {5.1.3}Thermal Instability}{121}
\contentsline {section}{\numberline {5.2}Formation and Lifetime of Molecular Clouds}{122}
\contentsline {section}{\numberline {5.3}Driving Mechanisms}{126}
\contentsline {subsection}{\numberline {5.3.1}Magnetorotational Instabilities}{126}
\contentsline {subsection}{\numberline {5.3.2}Gravitational Instabilities}{127}
\contentsline {subsection}{\numberline {5.3.3}Protostellar Outflows}{127}
\contentsline {subsection}{\numberline {5.3.4}Massive Stars}{128}
\contentsline {subsubsection}{Stellar Winds}{128}
\contentsline {subsubsection}{{\rm H}{\sc ii} Region Expansion}{129}
\contentsline {subsubsection}{Supernovae}{129}
\contentsline {section}{\numberline {5.4}Applications}{130}
\contentsline {subsection}{\numberline {5.4.1}Low Surface Brightness Galaxies}{130}
\contentsline {subsection}{\numberline {5.4.2}Galactic Disks}{131}
\contentsline {subsection}{\numberline {5.4.3}Globular Clusters}{131}
\contentsline {subsection}{\numberline {5.4.4}Galactic Nuclei}{132}
\contentsline {subsection}{\numberline {5.4.5}Primordial Dwarfs}{132}
\contentsline {subsection}{\numberline {5.4.6}Starburst Galaxies}{132}
{\large \contentsline {chapter}{\numberline {6}CONCLUSIONS}{135}}
\contentsline {section}{\numberline {6.1}Summary}{135}
\contentsline {section}{\numberline {6.2}Future Research Problems}{137}
{\large \contentsline {chapter}{BIBLIOGRAPHY}{139}}
{\large \contentsline {chapter}{THANKS...}{153}}
}
\twocolumn

%\tableofcontents

%%%%%%%%%%%%%%%%%%%%%%%%%%%%%%%%%%%%%%%%%%%%%%%%%%%%%%%%%%%%%%%%%%%%%%
\renewcommand{\S}{{Section}}%
%%%%%%%%%%%%%%%%%%%%%%%%%%%%%%%%%%%%%%%%%%%%%%%%%%%%%%%%%%%%%%%%%%%%%%
\mainmatter
\rsksection{INTRODUCTION}
\label{sec:introduction}
%%%
%%%
%%%
%%%\section{INTRODUCTION}
%%%\label{sec:introduction}

\rsksubsection{Overview}

Stars are important. They are the primary source of radiation, with
competition only from the 3K black body radiation of the cosmic microwave
background and from accretion processes onto black holes in active
galactic nuclei, which themselves are likely to have formed from
stars. And stars have produced the bulk of all chemical elements
heavier than H and He that made up the primordial gas.  The Earth
itself consists primarily of these heavier elements, called metals in
astronomical terminology.  These metals are made by nuclear fusion processes in
the interior of stars, with the heaviest elements originating from the
passage of the final supernova shockwave through the most massive
stars.  To reach the chemical abundances observed today in our solar
system, the material had to go through many cycles of stellar birth
and death.  In a literal sense, we are star dust.

Stars are also our primary source of astronomical information and, hence, are
essential for our understanding of the universe and the physical processes
that govern its evolution. At optical wavelengths almost all natural light we
observe in the sky originates from stars. 
During day this is more than obvious, but it is also true at
night. The Moon, the second brightest object in the sky, reflects
light from our Sun, as do the planets, while virtually every other
extraterrestrial source of visible light is a star or collection of
stars.
%(except maybe for Jupiter who/which is slowly contracting and thus
%radiates more energy away than receiving from the Sun, however, does
%so mostly at infra-red wavelengths --- see e.g.\ Hubbard, Burrows, \&
%Lunine 2002). The other light sources visible by naked eye on a clear
%dark night are, again, to the overwhelming majority associated with
%stars, either nearby single stars or star clusters in the Milky Way,
%or other galaxies (like Andromeda on the northern hemisphere, or the
%two Magellanic Clouds on the southern sky) where again we see the
%light emitted from their stars. 
Throughout the millenia, these objects have been the observational targets of
traditional astronomy, and define the celestial landscape, the constellations.
When we look at a dark night sky, we can also note dark patches of obscuration
along the band of the Milky Way.  These are clouds of dust and gas that block
the light from stars further away.

Since about a century ago we know that these clouds are associated with the
birth of stars. The advent of new observational instruments and techniques
gave access to astronomical information at wavelengths far shorter and longer
that visible light.  It is now possible to observe astronomical objects at
wavelengths ranging high-energy $\gamma$-rays down to radio frequencies.
Especially useful for studying these dark clouds are radio and sub-mm
wavelengths, at which they are transparent. Observations now show that {\em
all} star formation occurring in the Milky Way is associated with
these dark clouds of molecular hydrogen and dust.

Stars are common. The mass of the Galactic disk plus bulge is about $6\times
10^{10}\,$M$_{\odot}$ (e.g.\ Dehnen \& Binney 1998), where $1\,$M$_{\odot} =
2\times10^{33}\,$g is the mass of our Sun. Thus, there are of order $10^{12}$
stars in the Milky Way, assuming standard values for the stellar mass spectrum
(e.g.\ Kroupa 2002). Stars are constantly forming. Roughly 10\% of the disk
mass of the Milky Way is in the form of gas, which is forming stars at a rate
of about $1\,$M$_{\odot}\,$yr$^{-1}$.  Although stars dominate the baryonic
mass in the Galaxy, it is dark matter that determines the overall mass budget,
invisible material that indicates its presence only via its contribution to
the gravitational potential. The dark matter halo of our Galaxy is about 10
times more massive than gas and stars together.  At larger scales this
inbalance is even more pronouced. Stars are estimated to make up only 0.4\% of
the total mass of the Universe (Lanzetta, Yahil, \& Fernandez-Soto 1996), and
about 17\% of the total baryonic mass (Walker \etal\ 1991).

Mass is the most important parameter determining the evolution of individual
stars. Massive stars with high pressures at their centers have strong nuclear
fusion there, making them short-lived but very luminous, while low-mass stars
are long-lived but extremely faint.  For example, a star with $5\,$M$_{\odot}$
lives only for $2.5\times 10^7\,$yr, while a star with $0.2\,$M$_{\odot}$
survives for $1.2\times 10^{13}\,$yr, that is longer than the current age of
the universe. For comparison the Sun with an age of $4.5\times 10^9\,$yr has
reached approximately half of its life span. The mass luminosity relation is
quite steep with roughly $L\propto M^{3.2}$ (Kippenhahn \& Weigert 1990).
During its short life the $5\,$M$_{\odot}$ star will then emit a luminosity of
$1.5\times 10^4\,$L$_{\odot}$, while the brightness of the $0.2\,$M$_{\odot}$
star is only $\sim10^{-3}\,$L$_{\odot}$. For reference, the luminosity of the
Sun is $1\,$L$_{\odot} = 3.85\times 10^{33}$erg$\,$s$^{-1}$.
%  }

\begin{figure}[t]
\begin{center}
\unitlength1cm
\begin{picture}( 8.00, 9.00)
%\put( 0.00, 0.50){\epsfxsize=7.4cm \epsfbox{mk03_fig1.ps}}
\put( 0.20, 0.00){\epsfxsize=7.3cm \epsfbox{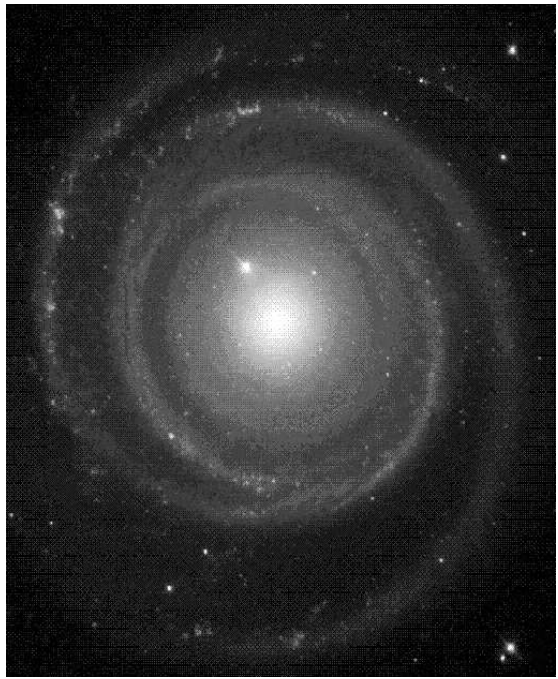}}
\end{picture}
\end{center}
\caption{\label{fig:NGC4622} Optical image of the
  spiral galaxy NGC$\,$4622 observed with the Hubble Space Telescope.
  (Courtesy of NASA and The Hubble Heritage Team --- STScI/AURA) }
\end{figure}

The light from star-forming external galaxies in the visible and blue
wavebands is dominated by young, massive stars. This is the reason why we
observe beautiful spiral patterns in many disk galaxies, like NGC$\,$4622
shown in Figure \ref{fig:NGC4622}, as spiral density density waves lead to gas
compression and subsequent star formation at the wave locations. As optical
emission from external galaxies is dominated by massive stars, and these
massive star are always young, they do not have sufficient time to disperse in
the galactic disk, but still trace the characteristics of the instability that
triggered their formation.  Hence, understanding dynamical properties of
galaxies requires an understanding of how, where, and under which conditions
stars form.

In a simple approach, galaxies can be seen as gravitational potential
wells containing gas that has been able to radiatively cool in less
than the current age of the universe.  In the absence of any
hindrance, the gas would then collapse gravitationally to form stars
on a free-fall time (Jeans 1902)
\begin{equation}
\label{eqn:free-fall-time}
\tau_{\rm ff} = \left(\frac{3\pi}{32 G \rho}\right)^{1/2} = 150 \mbox{ Myr}
\left(\frac{n}{0.1\,{\rm cm}^{-3}}\right)^{-1/2},
\end{equation}
where $n$ is the number density of gas molecules scaled to typical Galactic
values. Interstellar gas consists of one part He for every ten parts H. Then
$\rho = \mu n$ is the mass density with $\mu = 2.11 \times 10^{-24}\,$g, and
$G$ is the gravitational constant. The free-fall time $\tau_{\rm ff}$ is very
short compared to the age of the Milky Way, which is about $10^{10}\,$yr.
However, there is still gas left in the Galaxy and stars continue to form from
this gas that presumably has already been cool for many billions years.  What
physical processes regulate the rate at which gas turns into stars, or
differently speaking, what prevented that Galactic gas from forming stars at
high rate immediately after it first cooled?

Observations of the star formation history of the universe demonstrate
that stars did indeed form more vigorously in the past than today
(e.g.\ Lilly \etal\ 1996, Madau \etal\ 1996, Baldry \etal\ 2002,
Lanzetta \etal\ 2002), with as much as 80\% of star formation being
complete by redshift $z=1$, more than 6~Gyr before the present. What
mechanisms allowed rapid star formation in the past, but reduce its
rate today?

The clouds in which stars form are dense enough, and well enough protected from
dissociating UV radiation by  self-shielding and dust scattering in their
surface layers  for hydrogen to be mostly in molecular form in their interior.
The density and velocity structure of these molecular clouds is extremely
complex and follows hierarchical scaling relations that appear to be
determined by supersonic turbulent motions (e.g.\ Blitz \& Williams 1999).
Molecular clouds are large, and their masses exceed the threshold for
gravitational collapse by far when taking only thermal pressure into account.
Just like galaxies as a whole, naively speaking, they should be contracting
rapidly and form stars at very high rate. This is generally not observed.
The star formation efficiency of molecular clouds in the solar neighborhood is
estimated to be of order of a few percent (e.g.\ Elmegreen 1991, McKee 1999).

For many years it was thought that support by magnetic pressure
against gravitational collapse offered the best explanation for the
slow rate of star formation.  In this theory, developed by Shu (1977;
and see Shu, Adams, \& Lizano 1987), Mouschovias (1976; and see
Mouschovias 1991b,c), Nakano (1976), and others, interstellar magnetic
fields prevent the collapse of gas clumps with insufficient mass to flux ratio,
%(see Section \ref{sub:classicalprobs})
leaving dense cores in magnetohydrostatic equilibrium.  The magnetic
field couples only to electrically charged ions in the gas, though, so
neutral atoms can only be supported by the field if they collide
frequently with ions.  The diffuse interstellar medium (ISM) with
number densities $n$ of order unity remains ionized highly enough so that
neutral-ion collisional coupling is very efficient (as we discuss
below in \S~\ref{sub:standard}).  In dense cores, where $n >
10^5$~cm$^{-3}$, ionization fractions drop below parts per million.
Neutral-ion collisions no longer couple the neutrals tightly to the
magnetic field, so the neutrals can diffuse through the field in a
process known in astrophysics as ambipolar diffusion.  (The same term
is used by plasma physicists to describe ion-electron diffusion.)
This ambipolar diffusion allows gravitational collapse to proceed in
the face of magnetostatic support, but on a timescale as much as an
order of magnitude longer than the free-fall time, drawing out the
star formation process.

We review a body of work that suggests that
magnetohydrostatic support modulated by ambipolar diffusion fails to
explain the star formation rate, and indeed appears inconsistent with
observations of star-forming regions.  Instead, this work suggests
that support by supersonic turbulence is both sufficient to explain
star formation rates, and more consistent with observations.  In this
picture, most gravitational collapse is prevented by turbulent
motions, and any gravitational collapse that does occur does so
quickly, with no passage through hydrostatic states.

\rsksubsection{Turbulence}

At this point, we should briefly discuss the concept of turbulence, and the
differences between supersonic, compressible (and magnetized) turbulence, and
the more commonly studied incompressible turbulence.  We mean by turbulence,
in the end, nothing more than the gas flow resulting from random motions at
many scales.  We furthermore will use in our discussion only the very general
properties and scaling relations of turbulent flows, focusing mainly on
effects of compressibility. Some additional theoretical aspects of supersonic
turbulent self-gravitating flows relevant for star-forming interstellar gas
clouds are introduced in Section \ref{sec:properties}, however, for a more
detailed and fundamental discussion of the complex statistical characteristics
of turbulence, we refer the reader to the book by Lesieur (1991).
% Mordecai: I modified the above. I sounded strange to me. Please check,
% whether you like or not.
% 
% We make no particular
% assumptions about the statistical properties of the flow, although
% some such properties are sufficiently general that we may indeed learn
% something from them, as we will discuss. However, turbulence in the
% ISM may be driven in extremely non-uniform ways by processes such as
% supernova explosions, and may reflect that driving in its properties.

Most studies of turbulence treat incompressible turbulence,
characteristic of most terrestrial applications.
% (see Lesieur 1991).
Root-mean-square (rms) velocities are subsonic, and density remains
almost constant.  Dissipation of energy occurs entirely on the scales
of the smallest vortices, where the dynamical scale $\ell$ is shorter than
the length on which viscosity acts $\ell_{\rm visc}$.  Kolmogorov (1941a)
described a heuristic theory based on dimensional analysis that
captures the basic behavior of incompressible turbulence surprisingly
well, although subsequent work has refined the details substantially.
He assumed turbulence driven on a large scale $L$, forming eddies at
that scale.  These eddies interact to from slightly smaller eddies,
transferring some of their energy to the smaller scale.  The smaller
eddies in turn form even smaller ones, until energy has cascaded all
the way down to the dissipation scale $\ell_{\rm visc}$.  

In order to maintain a steady state, equal amounts of energy must be
transferred from each scale in the cascade to the next, and eventually
dissipated, at a rate
\begin{equation}
\dot{E} = \eta v^3/L,
\end{equation}
where $\eta$ is a constant determined empirically. This leads to a
power-law distribution of kinetic energy $E\propto v^2 \propto
k^{-10/3}$, where $k = 2\pi/\ell$ is the wavenumber, and density does
not enter because of the assumption of incompressibility.  Most of the
energy remains near the driving scale, while energy drops off steeply
below $\ell_{\rm visc}$.  Because of the local nature of the cascade in
wavenumber space, the viscosity only determines the behavior of the
energy distribution at the bottom of the cascade below $\ell_{\rm visc}$,
while the driving only determines the behavior near the top of the
cascade at and above $L$.  The region in between is known as the
inertial range, in which energy transfers from one scale to the next
without influence from driving or viscosity.  The behavior of the flow
in the inertial range can be studied regardless of the actual scale at
which $L$ and $\ell_{\rm visc}$ lie, so long as they are well separated.
The behavior of higher order structure functions $S_p(\vec{r}) =
\langle \{v(\vec{x}) - v(\vec{x}+\vec{r})\}^p \rangle$ in
incompressible turbulence has been successfully modeled by She \&
Leveque (1994) by assuming that dissipation occurs in the filamentary
centers of vortex tubes.

Gas flows in the ISM vary from this idealized picture in a number of
important ways.  Most significantly, they are highly compressible, with
Mach numbers ${\cal M}$ ranging from order unity in the warm, diffuse
ISM, up to as high as 50 in cold, dense molecular clouds.
Furthermore, the equation of state of the gas is very soft due to
radiative cooling, so that pressure $P\propto \rho^{\gamma}$ with the
polytropic index falling in the range $0.4 < \gamma < 1.2$ (e.g.\ 
Spaans \& Silk 2000, Ballesteros-Paredes, V{\'a}zquez-Semadeni, \&
Scalo 1999b, Scalo \etal\ 1998). Supersonic flows in highly
compressible gas create strong density perturbations.  Early attempts
to understand turbulence in the ISM (von Weizs\"acker 1943, 1951,
Chandrasekhar 1949) were based on insights drawn from incompressible
turbulence.  Although the importance of compressibility was already
understood, how to incorporate it into the theory remained unclear.
Furthermore, compressible turbulence is only one physical process that may
cause the strong density inhomogeneities observed in the ISM. Others are
thermal phase transitions (Field, Goldsmith, \& Habing 1969, McKee \& Ostriker
1977, Wolfire \etal\ 1995) or gravitational collapse (e.g.\ Wada \& Norman
1999).

In supersonic turbulence, shock waves offer additional possibilities
for dissipation.  Shock waves can transfer energy between widely
separated scales, removing the local nature of the turbulent cascade
typical of incompressible turbulence.  The spectrum may shift only
slightly, however, as the Fourier transform of a step function
representative of a perfect shock wave is $k^{-2}$, so the associated
energy spectrum should be close to $\rho v^2 \propto k^{-4}$, as was
indeed found by Porter\& Woodward (1992) and Porter, Pouquet, \&
Woodward (1992, 1994).  However, even in hypersonic turbulence, the
shock waves do not dissipate all the energy, as rotational motions
continue to contain a substantial fraction of the kinetic energy,
which is then dissipated in small vortices.   Boldyrev (2002)
has proposed a theory of structure function scaling based on the work
of She \& Leveque (1994) using the assumption that dissipation in
supersonic turbulence primarily occurs in sheet-like shocks, rather
than linear filaments.  First comparisons to numerical models show
good agreement with this model (Boldyrev, Nordlund, \& Padoan 2002a),
and it has been extended to the density structure functions by
Boldyrev, Nordlund, \& Padoan (2002b).

The driving of interstellar turbulence is neither uniform nor
homogeneous.  Controversy still reigns over the most important energy
sources at different scales, as described in \S~\ref{sub:driving}, but
it appears likely that isolated and correlated supernovae 
dominate.  However, it is not yet understood at what scales expanding,
interacting blast waves contribute to turbulence.  Analytic estimates
have been made based on the radii of the blast waves at late times
(Norman \& Ferrara 1996), but never confirmed with numerical models
(much less experiment).  Indeed, the thickness of the blast waves may
be more important 

Finally, the interstellar gas is magnetized.  Although magnetic field
strengths are difficult to measure, with Zeeman line splitting being the best
quantitative method, it appears that fields within an order of magnitude of
equipartition with thermal pressure and turbulent motions are pervasive in the
diffuse ISM, most likely maintained by a dynamo driven by the motions of the
interstellar gas.  A model for the distribution of energy and the scaling
behavior of strongly magnetized, incompressible turbulence based on the
interaction of shear Alfv\'en waves is given by Goldreich \& Sridhar (1995,
1997) and Ng \& Bhattacharjee (1996).  The scaling properties of the structure
functions of such turbulence was derived from the work of She \& Leveque
(1994) by M\"uller \& Biskamp (2000; also see Biskamp \& M\"uller 2000) by
assuming that dissipation occurs in current sheets.  A theory of very weakly
compressible turbulence has been derived by using the Mach number ${\cal M}
\ll 1$ as a perturbation parameter (Lithwick \& Goldreich 2001), but no
further progress has been made towards analytic models of strongly
compressible magnetohydrodynamic (MHD) turbulence with ${\cal M} \gg
1$. See also Cho \& Lazarian (2003), Cho \etal\ (2002).

With the above in mind, we propose that stellar birth is regulated by
interstellar turbulence and its interplay with gravity.  Turbulence,
even if strong enough to counterbalance gravity on global scales, will
usually provoke local collapse on small  scales.  Supersonic turbulence
establishes a complex network of interacting shocks, where converging
flows generate regions of high density. This density enhancement can
be sufficient for gravitational instability. Collapse sets in.
However, the random flow that creates local density enhancements also
may disperse them again.  For local collapse to actually result in the
formation of stars, collapse must be sufficiently fast for the region
to `decouple' from the flow, i.e.\ it must be shorter than the typical
time interval between two successive shock passages.  The shorter this
interval, the less likely a contracting region is to survive. Hence,
the efficiency of star formation depends strongly on the properties of
the underlying turbulent velocity field, on its lengthscale and
strength relative to gravitational attraction.  This principle holds
for star formation throughout all scales considered in this review,
ranging from small local star forming regions in the solar
neighborhood up to galaxies as a whole. For example, we predict in
star burst galaxies self-gravity to completely overwhelm any turbulent
support, whereas in the other extreme, in low surface brightness
galaxies we argue that turbulence is strong enough to essentially
quench any noticeable star formation activity.

\rsksubsection{Outline}

To lay out this new picture of star formation in more detail, in
\S~\ref{sec:paradigm} we first critically discuss the historical
development of star formation theory, and then argue that star
formation is controlled by the interplay between gravity and
supersonic turbulence.  We begin this section by describing the
classical dynamic theory, and then move on to what has been until
recently the standard theory, where the star formation process is
controlled by magnetic fields.  After describing the theoretical and
observational problems that both approaches have, we present work that 
% We challenge both approaches to star
%formation theory and present recent findings that are more successful
%in describing the birth of stars in our Galaxy.  
leads us to an outline of the new theory of star formation. Then we introduce
some further properties of supersonic turbulence in self-gravitating gaseous
media relevant to star-forming interstellar gas clouds in \S\ 
\ref{sec:properties}. We consider the transport properties of supersonic
turbulence, discuss energy spectra in Fourier space, and quantify the
structural evolution of gravitational collapse in turbulent flows by means of
one-point probability distribution functions of density and velocity and by
calculating the $\Delta$-variance. In \S~\ref{sec:local} we then apply the new
theory of turbulent star formation, first to local star forming regions in the
Milky Way in. We discuss the properties of molecular clouds, stellar clusters,
and protostellar cores (the direct progenitors of individual stars), and we
investigate the implications of the new theory on protostellar mass accretion,
and on the subsequent distribution of stellar masses. In
\S~\ref{sec:galactic}, we discuss the control of star formation by supersonic
turbulence on galactic scales. We ask when is star formation efficient, and
how are molecular clouds formed and destroyed.  We review the possible
mechanisms that generate and maintain supersonic turbulence in the
interstellar medium, and come to the conclusion that supernova explosions
accompanying the death of massive stars are the most likely
agents. %To connect to star formation on local scales, we then turn to
%the question of molecular cloud formation. And after that, 
Then we apply the theory to various types of galaxies, ranging from
low surface brightness galaxies to massive star bursts.  Finally, in
\S~\ref{sec:conclusions} we summarize, and describe unsolved problems
open for future research.

%%% Local Variables: 
%%% mode: latex
%%% TeX-master: "habil"
%%% End: 

\rsksection{TOWARDS A NEW PARADIGM}
\label{sec:paradigm}
%%%
%%% RMP-Module for the "Classical Theory of SF"
%%%
%%% checked out by RSK: 26.06.01
%%%
%%%\rsksubsection{Classical dynamical theory}
%%%\label{sub:classical}
%%%

% In the classical dynamical theory it is the interplay between
% self-gravity on the one side and thermal pressure plus possible
% additional contributions from microscopic turbulence on the other that
% determines how and when stars form. In order for molecular cloud
% material to collapse and form stars the gravitational attraction must
% overwhelm the combined action of all dispersive forces.
 
Stars form from gravitational contraction of molecular cloud
material. A crude estimate of the stability of such a system
against gravitational collapse can be made by simply considering its
energy balance. To become unstable gravitational attraction must outweigh the
combined action of all dispersive or resistive forces. In the most
simplistic case, the absolute value of the potential energy of a system
in virial equilibrium is exactly twice the total kinetic energy, $E_{\rm
pot} + 2\,E_{\rm kin} = 0$. If $E_{\rm pot} + 2\,E_{\rm kin}<0$ the
system collapses, while for $E_{\rm pot} + 2\,E_{\rm kin}>0$ it
expands. This estimate can easily be extended by including surface
terms and additional physical forces. In particular taking magnetic
field effects into account may become important for describing
interstellar clouds (Chandrasekhar, 1953; see also McKee \etal, 1993, for a
more recent discussion). In the presence of turbulence
the total kinetic energy not only includes the internal energy but
also the contribution from turbulent gas motions. Simple energy
considerations can in general already provide qualitative insight into
the dynamical behavior of a system (Bonazzola \etal, 1987).

A thorough investigation, however, requires a linear stability
analysis.  For the case of an non-magnetic, isothermal, infinite,
homogeneous, self-gravitating medium at rest (i.e.\ without turbulent
motions) Jeans (1902) derived a relation between the oscillation
frequency $\omega$ and the wavenumber $k$ of small perturbations,
\begin{equation}
\omega^2 - c_{\rm s}^2 k^2 + 4\pi G\,\rho_0 = 0\;,
\label{eqn:jeans-dispersion-rel}
\end{equation}
where $c_{\rm s}$ is the isothermal sound speed, $G$ the gravitational
constant, and $\rho_0$ the initial mass density. The derivation
neglects viscous effects and assumes that the
linearized version of the Poisson equation describes only the relation
between the perturbed potential and the perturbed density (neglecting
the potential of the homogeneous solution, the so-called `Jeans
swindle', see e.g.\ Binney and Tremaine, 1997).  The third term
in Equation (\ref{eqn:jeans-dispersion-rel}) is responsible for the
existence of decaying and growing modes, as pure sound waves stem from
the dispersion relation $\omega^2 - c_{\rm s}^2 k^2 =0$. Perturbations
are unstable against gravitational contraction if their wavenumber is
below a critical value, the Jeans wavenumber $k_{\rm J}$, i.e.~if
\begin{equation}
k^2 < k_{\rm J}^2 \equiv \frac{4 \pi G \rho_0}{c_{\rm s}^2}\;, 
\label{eqn:jeans-wave-number}
\end{equation}
or equivalently if the wavelength of the perturbation exceeds a
critical size given by $\lambda_{\rm J} \equiv 2 \pi k_{\rm J}^{-1}$.
Assuming the perturbation is spherical with diameter  $\lambda_{\rm
J}$, this directly translates into a mass limit
\begin{equation}
M_{\rm J} \equiv  \frac{4\pi}{3}\rho_0 \left(\frac{\lambda_{\rm J}}{2}\right)^3 =  \frac{\pi}{6}\left( \frac{\pi}{G}
\right)^{3/2} \rho_0^{-1/2} {c_{\rm s}^3}.
\label{eqn:jeans-mass}
\end{equation}
All perturbations exceeding the Jeans mass $M_{\rm J}$ will collapse
under their own weight.  For isothermal gas $c_{\rm s}^2 \propto T$
and subsequently  $M_{\rm J}\propto \rho_0^{-1/2} T^{3/2}$. The
critical mass $M_{\rm
J}$ decreases when the density $\rho_0$ grows or when the
temperature $T$ sinks.

The Jeans instability has a simple physical interpretation in terms of
the energy budget. The energy density of a sound wave is
positive. However, its gravitational energy is negative, because the
enhanced attraction in the compressed regions outweighs the reduced
attraction in the dilated regions. The instability sets in at the
wavelength $\lambda_{\rm J}$ where the net energy density becomes
negative. The perturbation will grow allowing the energy to decrease
even further. For a fundamental derivation of this instability from
the canonical ensemble in statistical physics see de~Vega and
S{\'a}nchez (2001). In isothermal gas, there is no mechanism that
prevents complete collapse. In reality, however, during the collapse
of molecular gas clumps, the opacity increases and at densities of
$n({\rm H}_2) \approx 10^{10}\:$cm$^{-3}$ the equation of state
becomes adiabatic. Then collapse proceeds slower. Finally at very high
central densities ($\rho \approx 1\:$g$\,$cm$^{-3}$) fusion processes
set in. This energy source leads to a new equilibrium (e.g.\ Tohline
1982): a new star is born.

Attempts to include the effect of turbulent motions into the star
formation process were already being made in the middle of the
$20^{\rm th}$ century by von Weizs\"acker (1943, 1951) based on
Heisenberg's (1948a,b) concept of turbulence. He also considered the
production of interstellar clouds from the shocks and density
fluctuations in compressible turbulence. A more quantitative theory
was proposed by Chandrasekhar (1951a,b), who
investigated the effect of microturbulence in the subsonic regime. In
this approach the scales of interest, e.g. for gravitational collapse,
greatly exceed the outer scale of turbulence.  If turbulence is
isotropic (and more or less incompressible), it simply contributes to
the pressure on large scales, and Chandrasekhar derived a dispersion
relation similar to Equation (\ref{eqn:jeans-dispersion-rel}) by
introducing an effective sound speed 
\begin{equation}
c_{{\rm s},e\!\!\!\;f\!\!f}^2 = c_{\rm s}^2 + 1/3 \, \langle v^2 \rangle\,,
\label{eqn:eff-sound-speed}
\end{equation}
 where $\langle v^2 \rangle$ is the rms velocity dispersion due to
turbulent motions.  

%The developments through the mid-eighties are reviewed by Scalo (1985).
%Particularly noteworthy is the work of Sasao (1973), who may have been
%the first to quantitatively show that the generation of density
%enhancements by turbulence, which Chandrasekhar (1951) neglected,
%might be as important as turbulent support.  
In reality, however, the outer scales of turbulence typically exceed
or are at least comparable to the size of the system (e.g.\ Ossenkopf
and Mac~Low, 2001), and the assumption of microturbulence is invalid.
In a more recent analysis, Bonazzola \etal\ (1987) therefore
suggested a wavelength-dependent effective sound speed $c_{{\rm
s},e\!\!\!\;f\!\!f}^2(k) = c_{\rm s}^2 + 1/3 \, v^2(k)$ for Equation
(\ref{eqn:jeans-dispersion-rel}).
%, leading to a dispersion
%relation
%\begin{equation}
%\omega^2 - \left (c_{\rm s}^2 + \frac{1}{3}v^2(k)\right) k^2 + 4\pi G\,\rho_0 = 0\;.
%\label{eqn:bonazzola-dispersion-rel}
%\end{equation}
In this description, the stability of the system depends not only on
the total amount of energy, but also on the wavelength distribution
of the energy, since $v^2(k)$ depends on the turbulent power spectrum.
A similar  approach was also adopted by V{\'a}zquez-Semadeni and Gazol
(1995), who added Larson's (1981) empirical scaling relations to the analysis.

A most elaborate investigation of the stability of turbulent,
self-gravitating gas was undertaken by Bonazzola \etal\ (1992), who used
renormalization group theory to derive a dispersion relation with a
generalized, wavenumber-dependent, effective sound speed and an
effective kinetic viscosity that together account for turbulence at
all wavelengths shorter than the one in question.  According to their
analysis, turbulence with a power spectrum steeper than $P(k)\propto
1/k^3$ can support a region against collapse at large scales, and
below the thermal Jeans scale, but not in between.  On the other hand,
they claim that turbulence with a shallower slope, as is expected for
incompressible turbulence (Kolmogorov 1941a,b), Burgers turbulence
(Lesieur 1997), or shock dominated flows (Passot, Pouquet
\& Woodward 1988), cannot support clouds against collapse at scales
larger than the thermal Jeans wavelength.

Analytic attempts to characterize turbulence  have a fundamental limitation,
so far they  are all restricted to incompressible flows. However, molecular cloud
observations clearly show extremely non-uniform structure.  It may
even be possible to describe the equilibrium state as an inherently
inhomogeneous thermodynamic critical point (de~Vega, S{\'a}nchez and
Combes, 1996a,b; de~Vega and S{\'a}nchez, 1999). This may render all
applications of incompressible turbulence to the theory of star
formation meaningless. In fact, it is the main goal of this review to
introduce and stress the importance of compressional effects in
supersonic turbulence for determining the outcome of star formation.
%
%As a consequence of
%the assumption of homogeneity, the further assumption of
%microturbulence must then be made.  Interstellar turbulence, however,
%does appear to have an outer scale that exceeds or at least is
%comparable to the size of the molecular cloud complex (Ossenkopf and
%Mac~Low 2001) such a cut-off in the power spectrum, but rather extends
%over {\em all} spatial scales present in the system.  A further
%corollary of the assumption of homogeneity is that the turbulent
%dynamical time scale is much shorter than the collapse time scale
%$\tau_{\rm ff}$, which is only justified if the assumption of
%microturbulence holds.
%
%We will come back to these points and introduce the properties of
%compressible supersonic turbulence relevant for astrophysical
%processes in \S~\ref{sub:beyond}, and apply these insights to derive what
%we bravely call the new paradigm of turbulent dynamical star formation in 
%\S~\ref{sub:new}. 

In order to do that, we need to recapitulate the development of our
understanding of the star formation process over the last few decades.
We begin with the classical dynamical theory (\S~\ref{sub:classical})
and describe the problems that it encounters in its original form
(\S~\ref{sub:classicalprobs}). In particular the timescale problem lead
astrophysicists think about the influence of magnetic fields. This
line of reasoning resulted in the construction of the paradigm of
magnetically mediated star formation, which we discuss in
\S~\ref{sub:standard}.  However, it became clear that this so called
``standard theory'' has a variety of very serious shortcomings
(\S~\ref{sub:beyond}). They lead us to rejuvenate the earlier dynamical
concepts of star formation and to reconsider them in the modern
framework of compressible supersonic turbulence
(\S~\ref{sub:new}). Consequently, we propose in \S~\ref{sub:new} a new
paradigm of dynamical  turbulent star formation.

\rsksubsection{Classical Dynamical Theory}
\label{sub:classical}
%\input{classical.tex} % RSK[out:25.06./in:05.07.]
%%%
%%% RMP-Module for the "Classical Theory of SF"
%%%
%%% checked out by RSK: 26.06.01
%%%
%%%\rsksubsection{Classical dynamical theory}
%%%\label{sub:classical}
%%%
%%% RSK: references checked -- 07.09.02
%%%
%%%
The classical dynamical theory focuses on the
interplay between self-gravity on the one side and pressure gradients
on the other. Turbulence can be taken into account, but only on
microscopic scales significantly smaller than the collapse scales. In
this microturbulent regime random gas motions yield an isotropic
pressure which can be absorbed into the equations of
motion by defining an effective sound speed as in Equation
(\ref{eqn:eff-sound-speed}). The dynamical behavior of the system
remains unchanged, and we will not distinguish between
effective and thermal sound speed $c_{\rm s}$ in this and the
following two sections.

\begin{figure*}[ht]
\begin{center}
\unitlength1cm
\begin{picture}( 16.00, 7.80)
%\put( 0.00, 0.50){\epsfxsize=16cm \epsfbox{figure-classical-03.eps}}
\put( 0.00, 0.0){\epsfxsize=16cm \epsfbox{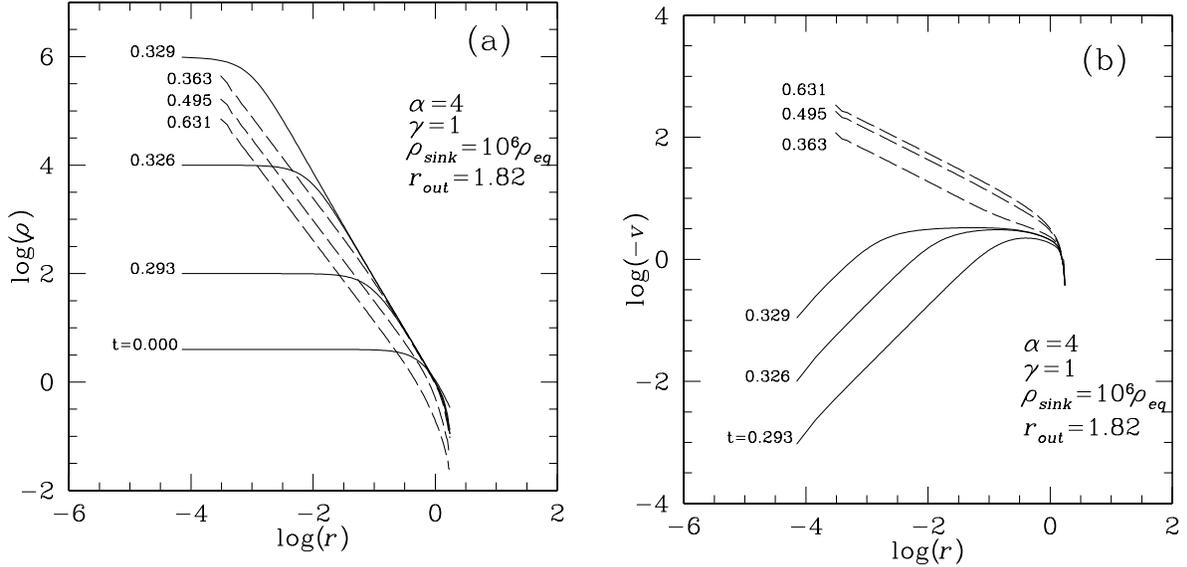}}
\end{picture}
\end{center}
\caption{\label{fig:larson-density-evolution} Radial density profile
(a) and infall velocity profile (b) depicted at various stages of
dynamical collapse. All quantities are given in normalized units. The
initial configuration at $t=0$ corresponds to a critical isothermal
($\gamma=1$) Bonnor-Ebert sphere with outer radius $r_{\rm out} =
1.82$. It carries $\alpha=4$ times more mass than allowed by
hydrostatic equilibrium, and therefore begins to contract. The numbers
on the left denote the evolutionary time and illustrate the `runaway'
nature of collapse. Since the relevant collapse timescale, the
free-fall time $\tau_{\rm ff}$, scales with density as $\tau_{\rm
ff}\propto \rho^{-1/2}$ central collapse speeds up as $\rho$
increases. When density contrast reaches a value of $10^6$ a ``sink''
cell is created in the center, which subsequently accretes all
incoming matter.  This time roughly corresponds to the formation of
the central protostar, and allows for following its subsequent
accretion behavior. The profiles before the formation of the central
point mass indicated by solid lines, and for later times by dashed
lines. The figure is from Ogino \etal\ (1999).  }
\end{figure*}

Because of the importance of gravitational instability for stellar
birth, Jeans' (1902) pioneering work has triggered numerous attempts to
derive solutions to the collapse problem, rigorous analytical as well
as numerical ones. Particularly noteworthy are the studies by Bonnor
(1956) and Ebert (1957) who independently of each other derived
analytical solutions for the equilibrium structure of spherical
density perturbations in self-gravitating isothermal ideal gases and a
criterion for gravitational collapse (see Lombardi and Bertin, 2001,
for a recent analysis; and studies by Schmitz, 1983, 1984, 1986, 1988, and
Schmitz \& Ebert, 1986, 1987, for the treatment of generalized
polytropic equations-of-state and/or rotation). It has been shown
recently that this may be a good description for the density
distribution in quiescent molecular cloud cores just before they begin
to collapse and form stars (Bacman \etal\ 2000, Alves, Lada, and Lada
2001). The first numerical calculations of protostellar collapse
became possible in the late 1960's (e.g.\ Bodenheimer \& Sweigart,
1968; Larson, 1969; Penston, 1969a,b) and
showed that gravitational contraction proceeds in a highly
nonhomologous manner, contrary to what has previously been assumed
(Hayashi 1966). This is illustrated in Figure
\ref{fig:larson-density-evolution}, which shows the radial density
distribution of a protostellar core at various stages of the
isothermal collapse phase. The gas sphere initially follows a
Bonnor-Ebert critical density profile but carries a four times larger
mass than allowed by the equilibrium conditions. Therefore it is
gravitationally unstable and begins to collapse. As the inner part has
no pressure gradient it contracts in free fall. As matter falls
inwards, the density in the interior grows and decreases in the outer
parts. This builds up pressure gradients in the outer parts, where
contraction becomes significantly retarded from free fall. In the
interior, however, the collapse remains approximately free
falling. This means it actually speeds up, because the free-fall
timescale $\tau_{\rm ff}$ scales with density as $\tau_{\rm ff}\propto
\rho^{-1/2}$. Changes in the density structure occur in a smaller and
smaller region near the center and on shorter and shorter timescale,
while practically nothing happens in the outer parts.  As a result the
overall matter distribution becomes strongly centrally peaked with
time, and approaches $\rho \propto r^{-2}$. This the well known
density profile of isothermal spheres. The establishment of a central
singularity corresponds to the formation of the protostar which grows
in mass by accreting the remaining envelope until the reservoir of gas
is exhausted.

In reality, however, the isothermal collapse phase ends when the central
density reaches 
densities of $n({\rm H}_2) \approx 10^{10}\,$cm$^{-3}$. Then gas
becomes optically thick and the heat generated by the collapse is no
longer radiated away freely. The central region begins to heat up and
contraction comes to a first halt. But as the temperature reaches
$T\approx 2000\,$K molecular hydrogen begins to dissociate. The core
becomes unstable again and collapse sets in anew. Most of the released
gravitational energy goes into the dissociation of ${\rm
H}_2$-molecules so that the temperature rises only slowly. This
situation is similar to the first isothermal collapse phase. When all
molecules in the core are dissociated, the temperature rises sharply
and pressure gradients again become able to halt the collapse. The
second hydrostatic core has formed. This is the first occurrence of the
protostar which subsequently grows in mass by the accretion of the
still infalling material from the outer parts of the original cloud
fragment. As this matter is still in free fall, most of the luminosity
of the protostar at that stage is generated in a strongly supersonic
accretion shock. Consistent dynamical calculations of all phases of
protostellar collapse are presented by Masunaga, Miyama, \& Inutsuka
(1998), Masunaga \& Inutsuka (2000a,b), Wuchterl \& Klessen (2001), and
Wuchterl \& Tscharnuter (2002).

{%\narrowtext
\begin{table}
\caption{\label{tab:larson-penston}
Properties of the Larson-Penston solution of isothermal
  collapse.} 
\begin{center}
\begin{tabular}{@{\extracolsep{\fill}}p{1.2cm}p{3.0cm}p{3.0cm}}
%\begin{tabular*}{8.6cm}[top]{@{\extracolsep{\fill}}p{2.3cm}p{3cm}p{3cm}}
\hline \hline
 & before core formation  & after core formation\\
 & $(t<0)$                  & $(t>0)$               \\
\hline
%\tableline
density   & $\rho \propto (r^2+r_0^2)^{-1}$    & $\rho
\propto r^{-3/2}$, \hfill$r\rightarrow 0$\\ 
profile   & ($r_0 \!\rightarrow\!  0$ as $t\!\rightarrow \!0_{-}$)  & $\rho
\propto r^{-2},$ \hfill$r\rightarrow \infty$\\ 
%& {\em \mbox{flattened isotherm.} sphere} & \\
& {\em \mbox{isotherm. sphere}} & \\
& {\em \mbox{with flat core}} & \\
velocity   & $v\propto r/t$ as $t \!\rightarrow \!0_{-}$  &
$v \propto r^{-1/2}$, \hfill$r\rightarrow 0$\\
profile & $v\approx -3.3\,c_{\rm s}$ & $v\approx -3.3\,c_{\rm s}$\\
 &  \hfill as $r\rightarrow \infty$ & \hfill as $r\rightarrow \infty$\\
accretion  & &  $\dot{M}= 47\,c^3_{\rm s}/G$ \\
rate  & &  \\
\hline \hline
%\end{tabular*}
\end{tabular}
\end{center}
\end{table}
%\widetext
}

\begin{figure}
\begin{center}
\unitlength1cm
%\begin{picture}(13.00,10.00)
%\put( 1.40, 0.50){\epsfxsize=9.0cm \epsfbox{mk03_fig3.ps}}
%\end{picture}
\includegraphics[width=0.45\textwidth]{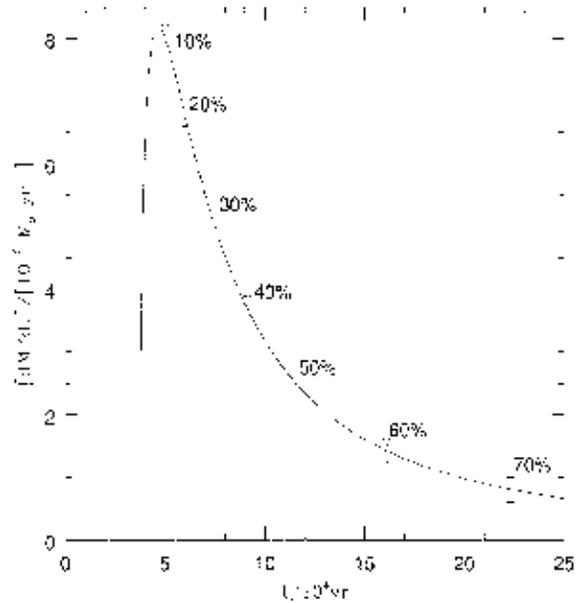}
\end{center}
\caption{\label{fig:larson-accretion-rate}
  Time evolution of the protostellar accretion rate for the collapse
  of a gas clump with Plummer-type density distribution similar to
  observed protostellar cores. For details see Whitworth \&
  Ward-Thompson (2001) 
}
\end{figure}

It was Larson (1969) who realized that the dynamical evolution in the
initial isothermal collapse phase can be described by an analytical
similarity solution. This was independently discovered also by Penston
(1969b), and later extended by Hunter (1977) into the regime after the
protostar has formed. This so called Larson-Penston solution describes
the isothermal collapse of homogeneous ideal gas spheres initially at
rest. Its properties are summarized in 
%Table~I
Table \ref{tab:larson-penston}.
Two predictions are most relevant for the astrophysical context. The
first is the occurrence of supersonic infall velocities that extend
over the entire protostellar cloud. Before the formation of the
central protostar the infall velocity tends towards -3.3 times the
sound speed $c_{\rm s}$, and afterwards approaches free fall collapse
in the center with $v\propto r^{-1/2}$ while still maintaining
$v\approx 3.3 c_{\rm c}$ in the outer envelope for some time (Hunter
1977). Second, the Larson-Penston solution predicts protostellar
accretion rates which are constant and of order $\dot{M} \approx 30
c^3_{\rm s}/G$. It is important to note that the dynamical models
conceptually allow for time-varying protostellar mass accretion rates,
if the gradient of the density profile of a collapsing cloud core
varies with radius. Most relevant in the astrophysical context, if the
core has a flat inner region and decreasing density outwards (as it is
observed in low-mass cores, see \S~\ref{sub:standardprobs}), then
$\dot{M}$ has a high initial peak, when the flat core gets accreted,
and later declines as the lower-density outer-envelope material is
falling in (e.g.\ Ogino \etal\ 1999). For the collapse of a
Plummer-type sphere
%(which have flat inner density profile followed by an outer
%power-law decline) 
with specifications such as to fit the protostellar
core L1544, the time evolution of $\dot{M}$ is illustrated in Figure
\ref{fig:larson-accretion-rate} (see Whitworth \& Ward-Thompson
2001).
Plummer-type spheres have flat inner density profile followed
by an outer power-law decline, and thus similar basic
properties as the Larson-Penston spheres in mid collapse. 
%Note that the original Larson-Penston similarity solution (i.e.\ for the
%collapse of initially homogeneous isolated isothermal spheres)
%predicts constant mass inflow after core formation with  $\dot{M}
%\approx 30  c^3_{\rm s}/G$ -- time variations of $\dot{M}$ require
%radial variations of the slope of the density profile. 
%
The dynamical properties of the Larson-Penston solution set it
clearly apart from the inside-out collapse model (Shu 1977) derived
for magnetically mediated star formation
(\S~\ref{sub:standard}). One-dimensional numerical simulations of the
dynamical collapse of homogeneous isothermal spheres typically
demonstrate global convergence to the Larson-Penston solution, but
also show that certain deviations occur, e.g.\ in the time evolution $\dot{M}$, due to
pressure effects (Bodenheimer \& Sweigart 1968; Larson 1969, Hunter
1977; Foster \& Chevalier 1993; Tomisaka 1996b; Basu 1997; Hanawa
\& Nakayama, 1997; Ogino \etal\ 1999).

With the rapid advances in computer technology, both two-dimensional
and three-dimensional computations became possible. Some of the first
two dimensional calculations are reported by Larson (1972),
Tscharnuter (1975), Black \& Bodenheimer (1976), Fricke, Moellenhoff,
\& Tscharnuter (1976), Nakazawa, Hayashi, \& Takahara (1976),
Bodenheimer \& Tscharnuter (1979), Boss (1980a), and Norman, Wilson,
\& Barton (1980). Two-dimensional dynamical modeling has the advantage
to be fast compared to three-dimensional simulations, and therefore
allows for including a larger number of physical processes while
reaching higher spatial resolution. The obvious disadvantage is that
only axissymetric perturbations can be studied.  Initial attempts to
study collapse in three dimensions are reported by Cook \& Harlow
(1978), Bodenheimer \& Boss (1979), Boss (1980b), Rozyczka \etal\ 
(1980), or Tohline (1980). Since these early studies, numerical
simulations of the collapse of isolated isothermal objects have been
extended, for example, to include highly oblate cores (Boss, 1996) or
elongated filamentary cloud cores (e.g.\ Bastien \etal\ 1991; Inutsuka
\& Miyama, 1997), differential rotation (Boss \& Myhill, 1995), and
different density distributions for the initial spherical cloud
configuration with or without bar-like perturbations (Burkert \&
Bodenheimer 1993; Klapp, Sigalotti, \& de~Felice 1993; Burkert \&
Bodenheimer 1996; Bate \& Burkert 1997; Burkert, Bate, \& Bodenheimer
1997; Truelove \etal\ 1997, 1978; Tsuribe \& Inutsuka 1999a; Klein
1999; Boss \etal\ 2000). The inclusion of magnetic fields into the
treatment will be discussed in \S~\ref{sub:standardprobs}.

Whereas spherical collapse models can only treat the formation of
single stars, the two- and three-dimensional calculations show that
the formation of binary and higher-order multiple stellar systems can
well be described in terms of the classical dynamical theory and is a
likely outcome of protostellar collapse and molecular cloud
fragmentation (for a
comprehensive overview see Bodenheimer \etal\ 2000). Observationally, the fraction of binary and multiple
stars relative to single stars is about 50\% for the field star
population in the solar neighborhood. This has been determined for all
known F7-G9 dwarf stars within 22 pc from the Sun by Duquennoy \&
Mayor (1991) and for M dwarfs out to similar distances by Fischer \&
Marcy (1992; also Leinert \etal\ 1997). The binary fraction for
pre-main sequence stars appears to be at least equally high (see e.g.\ 
Table 1 in Mathieu \etal\ 2000). These findings put strong constraints
on the theory of star formation, as {\em any} reasonable model needs
to explain the observed high number of binary and multiple stellar
systems. It has long been suggested that sub-fragmentation and
multiple star formation is a natural outcome of isothermal collapse
(Hoyle 1954), however, stability analyses show that the growth time of
small perturbations in the isothermal phase is typically small
compared to the collapse timescale itself (e.g.\ Silk \& Suto 1988;
Hanawa \& Nakayama 1997). Hence, in order to form multiple stellar
systems, either perturbations to the collapsing core must be external
and strong, or subfragmentation occurs at later non-isothermal phase
of collapse after a protostellar disk has formed. This disk may become
gravitationally unstable if the surface density exceeds a critical
value given by the epicyclic frequency and the sound speed (Safranov,
1960; Toomre 1964) and fragment into multiple objects (as summarized
by Bodenheimer \etal\ 2000). This naturally leads to two distinct
modes of multiple star formation.

Contracting gas clumps with strong external perturbation occur
naturally in turbulent molecular clouds or when stars form in
clusters.  While collapsing to form or feed protostars, clumps may
loose or gain matter from interaction with the ambient turbulent flow
(Klessen \etal\ 2000). In a dense cluster environment, collapsing
clumps may merge to form larger clumps containing multiple
protostellar cores, which subsequently compete with each other for
accretion form the common gas environment (Murray \& Lin, 1996;
Bonnell \etal\ 1997, Klessen \& Burkert, 2000, 2001). Strong external
perturbations and capture through clump merger leads to {\em wide}
binaries or multiple stellar systems. Stellar aggregates with more
that two stars are dynamically unstable, hence, some protostars may
become ejected again from the gas rich environment they accrete from.
This not only terminates their mass growth, but leaves the remaining
stars behind more strongly bound. These dynamical effects may
transform the original wide binaries into close binaries (see also
Kroupa 1995a,b,c). Binary stars that form through disk fragmentation
are close binaries right from the beginning, as typical sizes of
protostellar disks are of order of a few 100$\,$AU\footnote{One
astronomical unit is the mean radius of Earths orbit around the Sun,
$1\,{\rm AU}=1.5\times10^{13}\,{\rm cm}$.}.

The formation of clusters of stars (as opposed to binary or small multiple
stellar systems) is easily accounted for in the classical dynamical theory by
simply considering larger and more massive molecular cloud regions. The
proto-cluster cloud will fragment and build up a cluster of stars if it has
highly inhomogeneous density structure similar to the observed clouds (Keto,
Lattanzio, \& Monaghan 1991; Inutsuka \& Miyama 1997, Klessen \& Burkert 2000,
2001) or, equivalently, if it is subject to strong external perturbations,
e.g.\ from cloud-cloud collisions (Whithworth \etal\ 1995; Turner \etal\ 
1995), or is highly turbulent (see \S \ref{sub:beyond} and
\S~\ref{sub:new}).

\rsksubsection{Problems with Classical Theory}
\label{sub:classicalprobs}
%\input{classicalprobs.tex} 

%%%
%%% RMP-Module for the "Problems with the Classical Theory"
%%%
%%% 
%%%
%\rsksubsection{Problems with classical theory}
%\label{sub:classicalprobs}

The classical theory of gravitational collapse balanced by pressure
and microturbulence did not take into account the conservation of
angular momentum and magnetic flux during collapse.  It became clear
from observations of polarized starlight (Hiltner 1949, 1951) that
substantial magnetic fields thread the interstellar medium
(Chandrasekhar \& Fermi 1953a), forcing the magnetic flux problem to
be addressed, but also raising the possibility that the solution to
the angular momentum problem might be found in the action of magnetic
fields.  The typical strength of the magnetic field in the diffuse ISM
was not known to an order of magnitude, though, with estimates ranging
as high as 30 $\mu$G from polarization (Chandrasekhar \& Fermi 1953a)
and synchrotron emission (e.g.\ Davies \& Shuter 1963).  Lower values
from Zeeman measurements of H{\sc i} (Troland \& Heiles 1986) and
from measurements of pulsar rotation and dispersion measures (Rand \&
Kulkarni 1989, Rand \& Lyne 1994) comparable to the modern value of
around 3~$\mu$G only gradually became accepted over the next two
decades.

The presence of a field, especially one as strong as was then
considered possible, formed a major problem for the classical theory
of star formation.  To see why, let us consider the behavior of a
field in an isothermal region of gravitational collapse (Mestel \&
Spitzer 1956, Spitzer 1968).  If we neglect all surface terms except
thermal pressure $P_0$ (a questionable assumption as shown by
Ballesteros-Paredes \etal\ 1999a, but the usual one at the time), and
assume that the field, with magnitude $B$ is uniform within a region of average
density $\rho$ and effective spherical radius R, we can write the
virial equation as
\begin{equation} \label{virial}
4 \pi R^3 P_0 = 3 \frac{Mk_BT}{\mu} - \frac{1}{R} 
   \left(\frac35 GM^2 - \frac13 R^4 B^2\right),
\end{equation}
where $M = 4/3 \pi R^3 \rho$ is the mass of the region, $k_B$ is
Boltzmann's constant, $T$ is the temperature of the region, and $\mu$
is the mean mass per particle.  So long as the ionization is
sufficiently high for the field to be frozen to the matter, the flux
through the cloud $\Phi = \pi R^2 B$ must remain constant.  Therefore,
the opposition to collapse due to magnetic energy given by the last
term on the right hand side of equation~(\ref{virial}) will remain
constant during collapse.  If it is insufficient to prevent collapse
at the beginning, it remains insufficient as the field is compressed.

If we write the radius $R$ in terms of the mass and density of the
region, we can rewrite the two terms in parentheses on the right hand
side of equation~(\ref{virial}) to show that gravitational attraction
can only overwhelm magnetic repulsion if
% \begin{equation} \label{eqn:crit-rho}
% M > M_{\rm cr} \equiv \frac{5^{3/2}}{48\pi^2} \frac{B^3}{G^{3/2} \rho^2} = 
% (4 \times 10^6 \mbox{M}_{\odot})\left(\frac{n}{1\,{\rm
% cm}^{-3}}\right)^2 \left(\frac{B}{\mbox{3 $\mu$G}}\right)^3, 
% \end{equation}
\begin{eqnarray} \label{eqn:crit-rho}
\lefteqn{M > M_{\rm cr} \equiv \frac{5^{3/2}}{48\pi^2} \frac{B^3}{G^{3/2} \rho^2} =} \hspace{0.0cm}\nonumber\\ 
\!\!\!&&\!\!\!(4 \times 10^6 \mbox{M}_{\odot})\left(\frac{n}{1\,{\rm cm}^{-3}}\right)^2 \left(\frac{B}{\mbox{3 $\mu$G}}\right)^3,\nonumber\\[-0.3cm] 
\end{eqnarray}
where the numerical constant is correct for a uniform sphere, and the
number density $n$ is computed with mean mass per particle $\mu = 2.11
\times 10^{-24}$~g~cm$^{-3}$.  Mouschovias \& Spitzer (1976) noted
that the critical mass can also be written in terms of a critical
mass-to-flux ratio
\begin{equation} \label{eqn:crit-phi}
\left(\frac{M}{\Phi}\right)_{\rm cr} = \frac{\zeta}{3\pi}
\left(\frac{5}{G}\right)^{1/2} = 490 \:\mbox{g}\,{\rm G}^{-1}\,\mbox{cm}^{-2},
\end{equation}
where the constant $\zeta = 0.53$ for uniform spheres (or flattened
systems, as shown by Strittmatter 1966) is used in the final
equality. (Assuming a constant mass-to-flux ratio in a region results
in $\zeta = 0.3$ [Nakano \& Nakamura 1978]).  For a typical
interstellar field of 3~$\mu$G, the critical surface density for
collapse is $7 \mbox{M}_{\odot}$~pc$^{-2}$, corresponding to a number
density of 230~cm$^{-2}$ in a layer of thickness 1~pc.  A cloud is
termed {\em subcritical} if it is magnetostatically stable and {\em
supercritical} if it is not.

The very large value for the magnetic critical mass in the diffuse ISM
given by equation~\ref{eqn:crit-rho} forms a crucial objection to the
classical theory of star formation.  Even if such a large mass could
be assembled, how could it fragment into objects with stellar masses
of 0.01--100~M$_{\odot}$, when the critical mass should remain invariant
under uniform spherical gravitational collapse?

Two further objections to the classical theory were also prominent.
First was the embarrassingly high rate of star formation predicted by
a model governed by gravitational instability, in which objects should
collapse on roughly the free-fall timescale, Equation (\ref{eqn:free-fall-time}), 
%
%\begin{equation}
%\tau_{\rm ff} = (3 \pi / 32 G \rho)^{1/2} = (46 \mbox{ Myr}) n^{-1/2},
%\end{equation}
%
orders of magnitude shorter than the ages of typical galaxies.  

Second was the gap between the angular momentum contained in a parcel
of gas participating in rotation in a galactic disk and the much
smaller angular momentum contained in any star rotating slower than
breakup (Spitzer 1968).  The disk of the Milky Way rotates with
angular velocity $\Omega \simeq 10^{-15}$~s$^{-1}$.  A uniformly
collapsing cloud with initial radius $R_0$ formed from material with
density $\rho_0 = 2 \times 10^{-24}$~g~cm$^{-3}$ rotating with the
disk will find its angular velocity increasing as $(R_0/R)^2$, or as
$(\rho/\rho_0)^{2/3}$.  By the time it reaches a typical stellar
density of $\rho = 1$~g~cm$^{-3}$, its angular velocity has increased
by a factor of $6 \times 10^{15}$, giving a rotation period of well
under a second.  The centrifugal force $\Omega^2 R$ exceeds the
gravitational force by eight orders of magnitude for solar
parameters. A detailed discussion including a demonstration that
binary formation does not solve this problem can be found in
Mouschovias (1991b).

The observational discovery of bipolar outflows from young stars
(Snell, Loren \& Plambeck 1980) was a surprise that was unanticipated
by the classical model of star formation.  It has become clear that
the driving of these outflows is one part of the solution of the
angular momentum problem, and that magnetic fields transfer the
angular momentum from infalling to outflowing gas (e.g. K{\"o}nigl \&
Pudritz 2001).

Finally, mm-wave observations of emission lines from dense molecular
gas revealed a further puzzle: extremely superthermal linewidths
indicating that the gas was moving randomly at hypersonic
velocities. (Zuckerman \& Palmer 1974). Such motions in unmagnetized
gas generate shocks that would dissipate the energy of the motions
within a crossing time because of shock formation (e.g. Field 1978).
Attempts were made using clump models of turbulence to show that the
decay time might be longer (Scalo \& Pumphrey 1982, Elmegreen 1985).
In hindsight, isolated spherical clumps turn out not to be a good
model for turbulence however, so these models failed to accurately
predict its behavior (Mac Low \etal\ 1998).

\rsksubsection{Standard Theory of Isolated Star Formation}
\label{sub:standard}
%\input{standard.tex}
%%%
%%% RMP-Module for "Standard theory of isolated star formation''
%%%
%%% 
%%%
%\rsksubsection{Standard theory of isolated star formation}
%\label{sub:standard}

The problems outlined in the preceeding subsection were addressed in
what we will call the ``standard theory'' of star formation that has
formed the base of most work in the field for the past two decades.
Mestel \& Spitzer (1956) first noted that the problem of magnetic
support against fragmentation could be resolved if mass could move
across field lines, and proposed that this could occur in mostly
neutral gas through the process of ion-neutral drift, usually known as
ambipolar diffusion in the astrophysical community.\footnote{In plasma
physics, the term ambipolar diffusion is applied to ions and electrons held together
electrostatically rather than magnetically while drifting together out
of neutral gas.}  The other problems outlined then appeared solvable
by the presence of strong magnetic fields, as we now describe.

Ambipolar diffusion can solve the question of how magnetically
supported gas can fragment if it allows neutral gas to gravitationally
condense across field lines.  The local density can then increase
without also increasing the magnetic field, thus decreasing the critical
mass for gravitational collapse $M_c$ given by
Equation (\ref{eqn:crit-rho}).  This can also be interpreted as increasing
the local mass-to-flux ratio, approaching the critical value given by
Equation (\ref{eqn:crit-phi}).  

The timescale  $\tau_{\rm AD}$ on which this occurs can be derived by
considering the relative drift velocity of neutrals and ions
$\vec{v_{\rm D}} =\vec{v}_i - \vec{v}_n$ under the influence of the magnetic
field $\vec{B}$ (Spitzer 1968).  So long as the ionization fraction is
small and we do not care about instabilities (e.g.\ Wardle 1990), the
inertia and pressure of the ions may be neglected.  The ion momentum
equation then reduces in the steady-state to a balance between Lorentz
forces and ion-neutral drag,
\begin{equation} \label{eqn:drift}
\frac{1}{4\pi} (\nabla \times \vec{B}) \times \vec{B} = \alpha \rho_i
\rho_n (\vec{v}_i -  \vec{v}_n),
\end{equation}
where the coupling coefficient $\alpha = \langle \sigma v \rangle /(m_i + m_n)$, with
$m_i$ and $m_n$ the mean mass per particle for the ions and neutrals,
and $\rho_i$ and $\rho_n$ the ion and neutral densities.  Typical
values in molecular clouds are $m_i = 10 \,m_{\rm H}$, $m_n = (7/3) m_{\rm H}$, and
$\alpha = 9.2 \times 10^{13}$.  This is roughly independent of the
mean velocity, as the cross-section $\sigma$ scales linearly with
velocity in the regime of interest (Osterbrock 1961, Draine 1980).  To
estimate the typical timescale, consider drift occurring across a
cylindrical region of radius $R$, with a typical bend in the field
also of order $R$ so the Lorentz force can be estimated as roughly
$B^2/4\pi R$.  Then the ambipolar diffusion timescale can be derived by
solving for $v_{\rm D}$ in Equation (\ref{eqn:drift}) to be
% \begin{equation}
% \label{eqn:AD}
% \tau_{\rm AD}  =  \left.\frac{R}{v_{\rm D}}\right. = \left(\frac{4\pi \alpha \rho_i \rho_n R}{(\nabla \times \vec{B}) \times
% \vec{B}}\right) \approx \frac{4\pi \alpha
% \rho_i \rho_n R^2} {B^2}
%            = (25\,{\rm{Myr}}) 
%                 \left(\frac{B}{3\,\mu\mbox{G}       }\right)^{-2}
%                 \left(\frac{n_n}{10^2\,\mbox{cm}^{-3}}\right)^{2}
%                 \left(\frac{R}{1\,\mbox{pc}        }\right)^{2}
%                 \left(\frac{x}{10^{-6}               }\right)^{},
% \end{equation}
\begin{eqnarray}
\label{eqn:AD}
\lefteqn{\tau_{\rm AD}  =  \left.\frac{R}{v_{\rm D}}\right. = \left(\frac{4\pi \alpha \rho_i \rho_n R}{(\nabla \times \vec{B}) \times
\vec{B}}\right) \approx}\mbox{\hspace{0.3cm}}\nonumber \\ 
&&\frac{4\pi \alpha
\rho_i \rho_n R^2} {B^2}
           = (25\,{\rm{Myr}}) 
                \left(\frac{B}{3\,\mu\mbox{G}
                    }\right)^{-2}\nonumber \\
&&                \times\left(\frac{n_n}{10^2\,\mbox{cm}^{-3}}\right)^{2}
                \left(\frac{R}{1\,\mbox{pc}        }\right)^{2}
                \left(\frac{x}{10^{-6}               }\right)^{},\nonumber\\[-0.3cm]
\end{eqnarray}

For ambipolar diffusion to solve the magnetic flux problem on an
astrophysically relevant timescale, the ionization fraction $x$ must be extremely
small.  With the direct observation of dense molecular gas (Palmer \&
Zuckerman 1967, Zuckerman \& Palmer 1974) more than a decade after the
original proposal by Mestel \& Spitzer (1956), such low ionization
fractions came to seem plausible.  Nakano (1976, 1979) and Elmegreen
(1979) computed the detailed ionization balance of molecular clouds
for reasonable cosmic ray ionization rates, showing that at densities
greater than $10^4\,$cm$^{-3}$, 
the ionization fraction was roughly
\begin{equation}
x \simeq (5 \times 10^{-8}) \left(\frac{n}{10^5 \mbox{ cm}^{-3}}\right)^{1/2}
\end{equation}
(Elmegreen 1979), becoming constant at densities higher than $10^7\,$cm$^{-3}$
or so.  Below densities of $10^4\,$cm$^{-3}$, the ionization is controlled by
the external UV radiation field, and the gas is tightly coupled to the
magnetic field.

With typical molecular cloud parameters $\tau_{\rm AD}$ is of order
$10^7\,$yr (Equation \ref{eqn:AD}). 
%This holds for molecular cloud cores as well as for molecular
%clouds as a whole. 
The  ambipolar diffusion timescale $\tau_{\rm AD}$
is thus about $10-20$ times longer than the corresponding dynamical timescale
$\tau_{\rm ff}$ of the system (e.g.\ McKee \etal\ 1993).  The delay induced by
waiting for ambipolar diffusion to occur has the not incidental benefit of
explaining why star formation is not occurring in a free-fall time, i.e.\ at
rates far higher than observed in normal galaxies. On the other hand, the
timescale is short enough to apparently explain why magnetic fields in the
standard model do not completely shut off any star formation at all by fully
preventing the collapse and fragmentation of molecular clouds.  Altogether the
ambipolar diffusion timescale appeared to be consistent with molecular cloud
lifetimes, which in the 1980's were thought to be about 30-100$\,$Myr (Solomon
\etal\ 1987, Blitz \& Shu 1980; see however Ballesteros-Paredes \etal\ 1999, and
Elmegreen 2000, who argue for much shorter cloud lifetimes).

These considerations lead scientists to investigate star formation
models that are based on magnetic diffusion as dominant physical
process rather than rely on simple hydrodynamical collapse. In
particular Shu (1977) proposed the self-similar collapse of initially
quasi-static singular isothermal spheres as the most likely
description of the star formation process.  He assumed that ambipolar
diffusion in a magnetically subcritical isothermal cloud core would
lead to the build-up of a quasi-static $1/r^2$-density structure which
contracts on timescales of order of $\tau_{\rm AD}$.  This
evolutionary phase is denoted quasi-static because $\tau_{\rm
AD}\gg\tau_{\rm ff}$. Ambipolar diffusion is supposed to eventually
lead to the formation of a singularity in central density, at that
stage the system becomes unstable and undergoes inside-out collapse.
During collapse the model assumes that magnetic fields are no longer
dynamically important and they are subsequently ignored in the
original formulation of the theory. A rarefaction wave moves outward
with the speed of sound with the cloud material behind the wave
free-falling onto the core and matter ahead still being at rest.  The
Shu (1977) model predicts constant mass accretion onto the central
protostar at a rate $\dot{M}=0.975\,c_{\rm s}^3/G$.  This is
significantly below the values derived for Larson-Penston collapse. In
the latter case the entire system is collapsing dynamically and
delivers mass to the center very efficiently, in the former case
inward mass transport is comparatively inefficient as the cloud
envelope remains at rest until reached by the rarefaction wave.  The
density structure of the inside-out collapse, however, is essentially
indistinguishable from the predictions of dynamical collapse. To
observationally differentiate between the two models one 
needs to obtain kinematical data and determine magnitude and spatial
extent of infall motions with high accuracy. The basic predictions of
inside-out collapse are summarized in 
%Table~II.
Table \ref{tab:shu}.  
As singular isothermal spheres per definition carry infinite mass, the
growth of the central protostar is thought to come to a halt when
feedback processes (like bipolar outflows, stellar winds, etc.) become
important and terminate further infall.
{%\narrowtext
\begin{table}[t]
\caption{\label{tab:shu}
Properties of the Shu solution of isothermal
  collapse.} 
\begin{center}
\begin{tabular}{@{\extracolsep{\fill}}p{1.5cm}p{2.8cm}p{2.9cm}}
%\begin{tabular*}{8.6cm}[top]{@{\extracolsep{\fill}}p{2.3cm}p{3cm}p{3cm}}
\hline \hline
& before core formation  & after core formation\\
 & $(t<0)$                  & $(t>0)$               \\
\hline
%\tableline
density    & $\rho \propto r^{-2}$, $\forall$  $r$    & $\rho
\propto r^{-3/2}$,\\ 
 profile   &  {\em\mbox{sing.\ isothermal}}   &  \hfill$\forall$$r\le c_{\rm s}t$\\ 
& {\em sphere} &  $\rho
\propto r^{-2}$,\\
& & \hfill$\forall$$r> c_{\rm s}t$\\
velocity  & $v\equiv 0$, $\forall$  $r$ &
$v \propto r^{-1/2}$, \\
profile  &  & \hfill$\forall$$r\le c_{\rm s}t$\\
& & $v\equiv 0$, \hfill$\forall$$r>c_{\rm s}t$\\
accretion  & &  $\dot{M}
=0.975\,c^3_{\rm s}/G$ \\
 rate & &  {\em (const.)}\\
\hline \hline
%\end{tabular*}
\end{tabular}
\end{center}
\end{table}
%\widetext
}

\begin{figure*}[ht]
\begin{center}
%\unitlength1cm
%\begin{picture}( 12.00, 10.00)
%\put( 0.00, 0.50){\epsfxsize=12cm \epsfbox{core-evolution.eps}}
%\end{picture}
\includegraphics[width=12cm]{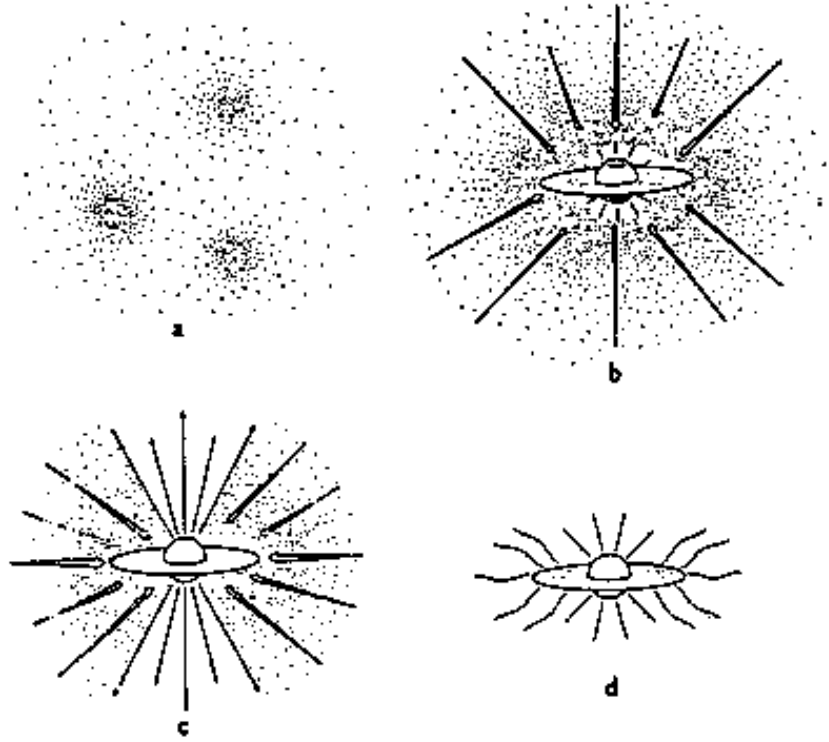}
\end{center}
\vspace*{-0.5cm}
\caption{\label{fig:4-stages} The main stages of star formation. (a)
{\em Prestellar phase.}  Protostellar cores form within molecular
clouds in areas where self-gravity overwhelms non-thermal support
mechanisms (i.e.\ magnetic flux loss via ambipolar diffusion in the
`standard scenario', or localized decay of supersonic turbulence in
the new paradigm). (b) {\em Class 0 phase.} A protostellar object
surrounded by an accretion disk has formed and grows in mass at the
center of the infalling cloud. The object is visible in sub-mm and
infra-red wavelength only because the central star is deeply embedded and its
visible light is completely obscured by the massive cloud
envelope. (c) {\em Class I phase.} Stellar winds and radiation break
out along the rotational axis of the system, create a bipolar outflow
and dissolve the envelope. The object becomes observable in the
optical. (d) {\em Class II phase.}  The envelope has become almost
fully accreted or has been removed by stellar feedback processes and
infall terminates. The young star and its (proto) planetary disk is
fully revealed. This picture is still valid when star formation is
controlled by supersonic turbulence. --- Adopted from Shu \etal\ (1987).
}
\end{figure*}

The overall picture of magnetic mediation and collapse of the singular
isothermal sphere has become known as the so called `standard theory'
of star formation (as best summarized in the review by Shu \etal\
1987). The process of stellar birth can be subdivided into four stages:
as visualized in Figure \ref{fig:4-stages}: 
(a) The {\em prestellar
phase} describes the evolution of molecular cloud cores before the
formation of a central protostar. Subcritical clumps contract slowly
due to leakage of magnetic support by ambipolar diffusion, those cores
form single stars. Supercritical cores evolve rapidly and may
fragment to form multiple stellar systems.  (b) Once the central
density has reached a singular state, i.e.\ the protostar has formed,
the system goes into inside-out collapse and feeds the protostar at
constant rate $\dot{M}=0.975c_{\rm s}^3/G$. In this evolutionary phase
the central protostar and its disk are deeply embedded in the
protostellar envelope of dust and gas. The mass of the envelope
$M_{\rm env}$ largely exceeds the combined mass $M_{\star}$ of star
and disk. The main contribution to the total luminosity is accretion,
and the system is best observable at sub-mm and infra-red
wavelengths. In the astronomical classification scheme it is called
`class 0' object. (c) At later times, powerful protostellar outflows
develop which clear out the envelope along the rotational axis.  This
is the `class I' stage at which the system is observable in infra-red
and optical wavebands  and for which $M_{\rm env}\ll
M_{\star}$. The central protostar is directly visible when looking
along the outflow direction.  (d) During the `class II' phase, the
outflow eventually removes the envelope completely. This terminates
further mass accretion and the protostar enters the classical pre-main
sequence contraction phase. It still is surrounded by a very low-mass
disk of gas and dust which adds infrared `excess' to the spectral
energy distribution of the system (which is already dominated by the
stellar Planck spectrum at visible wavelength, see e.g.\ Beckwith
1999).
This is the stage during which planets are believed to form
(e.g.\ Lissauer 1993, Ruden 1999). Protostellar systems at that stage
are commonly called T Tauri stars (Bertout 1989). As time evolves
further the disk becomes more and more depleted until only a tenuous
dusty debris disk remains that is long-lived and lasts (i.e.\
continuously reforms) into and throughout the stellar main-sequence
phase (Zuckermann 2001).

Largely within the framework of the `standard theory', numerous
(analytical) extensions to the simplistic original inside-out collapse
model have been proposed. The stability of isothermal gas clouds with
rotation for example has been investigated by Schmitz (1983, 1984,
1986), Tereby, Shu, \& Cassen (1984), Schmitz \& Ebert (1986, 1987),
Inutsuka \& Miyama (1992), Nakamura, Hanawa, \& Nakano (1995),
and Tsuribe \& Inutsuka (1999b).
The effects of magnetic fields on the equilibrium structure of clouds
and later during the collapse phase (where they have been neglected in
the original inside-out scenario) are considered by Schmitz (1987),
Baureis, Ebert, \& Schmitz (1989), Tomisaka, Ikeuchi, \& Nakamura
(1988a,b, 1989a,b, 1990), Tomisaka (1991, 1995, 1996a,b), Galli \& Shu
(1993a,b), Li \& 
Shu (1996, 1997), Galli \etal\ (1999, 2001), and Shu \etal\ (2000). The
proposed picture is that ambipolar diffusion of initially subcritical
protostellar cores that are threaded by uniform magnetic fields will
lead to the build-up of disk-like structures with constant
mass-to-flux ratio. These disks are called `isopedic'. The
mass-to-flux ratio increases steadily with time. As it exceeds the
maximum value consistent with magnetostatic equilibrium the entire
core becomes supercritical and begins to collapse from the inside out
with the mass-to-flux ratio assumed to remain approximately constant.
It can be shown (Shu \& Li 1997), that for isopedic disks the forces
due to magnetic tension are just a scaled version of the disks
self-gravity with opposite sign (i.e.\ obstructing gravitational
collapse), and that the magnetic pressure scales as the gas pressure
(although the proportionality factor in general is spatially varying
except in special cases). These findings allow to apply many results
derived for unmagnetized disks to the magnetized regime with only
little modification to the equations. One application of this result
is that for isopedic disks the derived mass accretion rate is just a
scaled version of the original Shu (1977) rate, i.e.\ $\dot{M}\approx
(1+{H}_0)\,c_{\rm s}^3/G$, with the dimensionless parameter $H_0$
depending on the effective mass-to-flux ratio.  Note, however, that
the basic assumption of constant mass-to-flux ratio during the
collapse phase appears inconsistent with detailed numerical
calculations of ambipolar diffusion processes (see 
\S~\ref{subsub:SIS}). In these computations the
mass-to-flux ratio in the central region increases more rapidly than
in the outer parts of the cloud. This leads to a separation into a
dynamically collapsing inner core with $(M/\Phi)_{\rm n} > 1$, and an
outer envelope with $(M/\Phi)_{\rm n} < 1$ that is still held up by
the magnetic field. The parameter $(M/\Phi)_{\rm n}$ is the dimensionless
mass-to-flux ratio normalized to the critical value as given in
Equation (\ref{eqn:crit-phi}). The isopedic description may therefore
only be valid in the central region with $(M/\Phi)_{\rm n} > 1$.

Note also that the `standard theory' introduces a somehow artificial
dichotomy to the star formation process, in the sense that low-mass
stars are thought to form from low-mass magnetically subcritical
cores, whereas high-mass stars (or entire stellar clusters) form from
magnetically supercritical cloud cores (e.g.\ Lizano \& Shu
1989). This distinction became necessary as it became clear that the
formation of very massive stars or stellar clusters cannot be
regulated by magnetic fields and ambipolar diffusion processes (see
\S~\ref{subsub:obs-stars}). We will argue in the next
section that this probably is true for low-mass stars also, and
therefore that star formation is {\em not} regulated by magnetic
mediation on {\em any} scale, but instead is mediated by interstellar
turbulence (\S~\ref{sub:new}). The new theory constitutes a
unifying scheme for both low-mass {\em and} high-mass star formation,
thus removing the undesired artificial dichotomy introduced by the
`standard theory'.

Finally let us remark, that the presence of strong magnetic fields was
suggested as a way to explain the universally observed (Zuckerman \&
Palmer 1974) presence of hypersonic random motions in molecular clouds
by Arons \& Max (1975). They noted that linear Alfv\'en waves have no
dissipation associated with them, as they are purely transverse. In a
cloud with Alfv\'en speed $v_{\rm A} = B/(4\pi\rho)^{1/2}$ much greater than
the sound speed $c_{\rm s}$, such Alfv\'en waves could produce the observed
motions without necessarily forming strong shocks.  This was
generally, though incorrectly, interpreted to mean that these waves
could therefore survive from the formation of the cloud, explaining
the observations without reference to further energy input into the
cloud.  The actual work acknowledged that ambipolar diffusion would
still dissipate these waves (Kulsrud \& Pearce 1969, Zweibel \&
Josafatsson 1983) at a rate substantial enough to require energy input
from a driving source to maintain the observed motions.

Strong magnetic fields furthermore provided a mechanism to reduce the angular
momentum in collapsing molecular clouds through magnetic braking.
Initially this was treated assuming that clouds were rigid rotating
spheres (Ebert, von Hoerner, \& Temesv\'ary 1960), but was accurately
calculated by Mouschovias \& Paleologou (1979, 1980) for both
perpendicular and parallel cases.  They showed that the criterion for
braking to be effective was essentially that the outgoing helical
Alfv\'en waves from the rotating cloud had to couple to a mass of gas
equal to the mass in the cloud.  Mouschovias \& Paleologou (1980)  show
that this leads to a characteristic deceleration time for a parallel
rotator of density $\rho$ and thickness $H$ embedded in a medium of
density $\rho_0$ and Alfv\'en velocity $v_{\rm A} = B/(4\pi\rho_0)^{1/2}$ of
\begin{equation}
\tau_{\parallel} = (\rho/\rho_0) (H/2v_{\rm A}),
\end{equation}
and a characteristic time for a perpendicular rotator with radius $R$,
\begin{equation}
\tau_{\perp} = \frac12 \left[\left(1 +
\frac{\rho}{\rho_0}\right)^{1/2} - 1\right] \frac{R}{v_{\rm A}}.
\end{equation}
For typical molecular cloud parameters, these times can be less than
the free-fall time, leading to efficient transfer of angular momentum
away from collapsing cores. This may help to resolve the so called
angular momentum problem in star formation (e.g.\ Bodenheimer 1995)
which describes the puzzle that single stars have considerably smaller
specific angular momenta $j$ compared to the observed molecular cloud
cores they supposedly form from. The angular momentum problem occurs
essentially on all scales, as values of $j$ in molecular cloud cores
again are smaller than in the average molecular cloud material which
they condense out of, and on the largest scales molecular cloud
complexes as a whole seem again to have smaller specific angular
momenta than the global interstellar medium in differentially rotating
galactic disks.

\rsksubsection{Problems with Standard Theory} % RSK[summer 2001]
\label{sub:standardprobs}
%\input{standardprobs.tex}

%%%
%%% RMP-Module for the "Problems with the Standard Theory"
%%%
%%% 
%%%
%\rsksubsection{Problems with standard theory}
%\label{sub:standardprobs}

During the 1980's the theory of magnetically mediated star formation
discussed in the previous section was widely advocated and generally
accepted as the standard theory of low-mass star formation, almost
completely replacing the earlier dynamical models. However, despite
its success and intellectual beauty, the picture of magnetically
mediated star formation suffers from a series of severe
shortcomings. It may not actually bear much relevance for the
astrophysical problem of how stars form and grow in mass -- although
for its simplicity and elegance it clearly is a pedagogically
important model of the star formation process. The prediction that
low-mass stars, and hence the vast majority of all stars, form from
molecular cloud cores that closely resemble quasi-static, singular,
isothermal spheres which built up via ambipolar diffusion processes
from magnetically supported gas on timescales of several tens of the
free-fall timescale deserves criticism from several sides -- for
theoretical as well as for observational reasons. This became obvious
in the 1990's with improved numerical simulations and the advent of
powerful new observational techniques, especially in the sub-mm and
infrared wavelength regime. Critical summaries are given by Whitworth
\etal\ (1996) and Nakano (1998). Note also that the theory
traditionally was applied to the formation of low-mass stars, it was
never seriously held accountable for describing the birth of very
high-mass stars and stellar clusters (Shu \etal\ 1987). This lead to
speculations about two distinct modes of star formation, low-mass
stars forming from magnetically supported cloud cores and high-mass
stars forming from magnetically supercritical cloud material. We shall
discuss in Sections \ref{sub:beyond} and \ref{sub:new} that replacing
magnetic fields as the pivotal physical mechanism of the theory and
introducing instead interstellar turbulence as the central mediating
agent of star formation naturally leads to a unified scheme for all
mass and length scales and removes this disturbing dichotomy.

Before we introduce the new dynamical theory of star formation based
on interstellar turbulence, we need to analyze in detail the
properties and shortcomings of the theory we seek to replace. The
following therefore is a list of critical remarks and summarizes  the
inconsistencies of models of magnetically mediated star formation. We
begin with the {\em theoretical} considerations that make the
inside-out collapse of quasistatic singular isothermal spheres a very
unlikely description of stellar birth, which constitutes the essence
of the `standard theory'. We then discuss the {\em observational}
inconsistencies of the theory.

\rsksubsubsection{Singular Isothermal Spheres}
\label{subsub:SIS}
The collapse of singular isothermal spheres is the astrophysically
most unlikely and unstable member of a large family of self-similar
solutions to the 1-dimensional collapse problem. Ever since the
seminal studies by Bonnor (1956) and Ebert (1957) and by Larson (1969)
and Penston (1969a) much attention in the star formation community has
been focused on finding astrophysically relevant analytic asymptotic
solutions to the 1-dimensional collapse problem. The standard
solution was derived by Shu (1977) considering the evolution of
initially singular isothermal spheres as they leave equilibrium. His
findings subsequently were extended by Hunter (1977, 1986), and
Whitworth \& Summers (1985) demonstrated that all solutions to the
isothermal collapse problem are members of a two-parameter family with
the Larson-Penston-type solutions (collapse of spheres with
uniform central density) and the Shu-type solutions
(expansion-wave collapse of singular spheres) populating extreme
ends of parameter space. The problem set has been extended to include
a polytropic equation of state (Suto \& Silk 1988), shocks (Tsai and
Hsu 1995), and/or cylinder and disk-like geometries (Inutsuka and
Miyama 1992; Nakamura, Hanawa, \& Nakano 1995). In addition,
mathematical generalization using a Lagrangian formulation has been
proposed by Hendriksen (1989, see also Hendriksen, Andr{\'e}, and
Bontemps 1997).

Of all proposed initial configurations for protostellar collapse, 
quasi-static, singular, isothermal spheres seem to be the most difficult
to realize in nature. Stable equilibria for self-gravitating spherical
isothermal gas clouds embedded in an external medium of given pressure
are only possible up to a density contrast of $\rho_{\rm c}/\rho_{\rm s}\approx
14$ between the cloud center and surface. Clouds that are more
centrally concentrated than that critical value can only find unstable
equilibrium states. Hence, all evolutionary paths that could yield a
central singularity lead through instability, and collapse is likely
to set in long before a $1/r^2$ density profile is established at
small radii $r$ (Whitworth \etal\ 1996; also Silk \& Suto 1988, and
Hanawa \& Nakayama 1997). External perturbations tend to break
spherical symmetry in the innermost region and degrade the overall
density profile at small radii. It will become less strongly
peaked. The resulting behavior in the central region then more closely
resembles the Larson-Penston description of collapse.  Similar
behavior is found if outward propagating shocks are considered (Tsai
and Hsu 1995).  As a consequence, the existence of physical processes
that are able to produce singular isothermal equilibrium spheres in
nature is highly questionable. Furthermore, the original proposal of
ambipolar diffusion processes in magnetostatically supported gas does
not yield the desired result either.
%, as discussed in the next section.
%but instead results in a
%Larson-Penston-type collapse in the inner part where magnetic support
%is lost. In the outer part, indeed a $1/r^2$ envelope is established,
%but matter is hold up primarily by the field and not in equilibrium
%between pressure and self-gravity only.

%
%\rskparagraph{
%\noindent {\em (b)}
Ambipolar diffusion in initially magnetically supported gas clouds
results in dynamical Larson-Penston-type collapse of the central
region where magnetic support is lost, while the outer part is still
hold up primarily by the field (and develops a $1/r^2$ density
profile). Mass is fed to the center not due to an outward moving
expansion wave, but due to ambipolar diffusion in the outer envelope.
The proposal that singular isothermal spheres may form through
ambipolar diffusion processes in magnetically subcritical cores has
been extensively studied by Mouschovias and collaborators in a series
of numerical simulations with ever increasing accuracy and
astrophysical detail (Mouschovias 1991, Mouschovias \& Morton 1991,
1992a,b, Fiedler \& Mouschovias 1992, 1993, Morton \etal\ 1994,
Ciolek \& Mouschovias 1993, 1994, 1995, 1996, 1998, Basu and
Mouschovias 1994, 1995a,b, Desch \& Mouschovias 2001; see however
also Nakano 1979, 1982, 1983, Lizano \& Shu 1989, or Safier \etal\
1997). The numerical results indicate that the decoupling between
matter and magnetic fields occurs over several orders of magnitude in
density becoming important at densities $n({\rm H}_2) >
10^{10}\,$cm$^{-3}$ (i.e.\ there is no single critical density below
which matter is fully coupled to the field and above which it is not),
and that ambipolar diffusion indeed is the dominant physical
decoupling process (e.g.\ Desch \& Mouschovias 2001). As a
consequence of ambipolar diffusion, initially subcritical protostellar
gas clumps separate into a central nucleus, which becomes both
thermally and magnetically supercritical, and an extended envelope
that is still held up by static magnetic fields. The central region
goes into rapid collapse sweeping up much of its residual magnetic
flux with it. The dynamical behavior of this core more closely
resembles the Larson-Penston description of collapse than the Shu
solution (Basu 1997).  The low-density envelope contains most of the
mass, roughly has a $1/r^2$ density profile, and contracts only
slowly. It feeds matter into the central collapse region on the
ambipolar diffusion timescale.

%\noindent {\em (c)} 
Star formation from singular isothermal spheres is biased against
binary formation. The collapse of rotating singular isothermal spheres
very likely will result in the formation of single stars, as the
central protostellar object forms very early and rapidly increases in
mass with respect to a simultaneously forming and growing rotationally
supported protostellar disk (e.g.\ Tsuribe \& Inutsuka 1999a,b).  By
contrast, the collapse of cloud cores with flat inner density profiles
will deliver a much smaller fraction of mass directly into the central
protostar within a free-fall time. More matter will go first into a
rotationally supported disk-like structure. These disks tend to be
more massive with respect to the central protostar in a
Larson-Penston-type collapse scenario compared to collapsing singular
isothermal spheres, and they are more likely become unstable against
subfragmentation resulting in the formation of binary or higher-order
stellar systems (see the review by Bodenheimer \etal\ 2000).  Since
the majority of stars seems to form as part of binary or higher-order
system (e.g.\ Mathieu \etal\ 2000), star formation in nature appears
incompatible with collapse from strongly centrally peaked initial
conditions (Whitworth \etal\ 1996).

\rsksubsubsection{Observations of Clouds and Cores}
\label{subsub:obs-clouds}
Before we consider the observational evidence against the `standard
theory' of magnetically mediated star formation, let us recapitulate
its basic predictions as introduced in Section \ref{sub:standard}. The
theory predicts (a) constant accretion rates and (b) infall motions
that are confined to regions that have been passed by a rarefaction
wave that moves outwards with the speed of sound, while the parts of a
core that lie further out are assumed to remain static.  The theory
furthermore (c) relies on the presence of magnetic field strong enough
to hold up the protostellar gas from collapsing, i.e.\ it predicts
that protostellar cores should be magnetically subcritical during the
largest fraction of their lifetime. In the following we demonstrate
{\em all} these predictions are contradicted by observations.

\rskparagraph{Magnetic Support}
%\noindent {\em (d)} 
Most (if not all) protostellar cores with magnetic field measurements
are supercritical.  When the theory was promoted in the late 1970's
and 1980's accurate measurements of magnetic field strength in
molecular clouds and cloud cores did not exist or were highly
uncertain. Consequently, magnetic fields in molecular cloud were
essentially {\em assumed} to have the desired properties necessary for
the theory to work and to circumvent the observational problems
associated with the classical dynamical theory (Section
\ref{sub:classicalprobs}). In particular the field was thought to have
a strong fluctuating component associated with magnetohydrodynamic
waves that give rise to the super-thermal linewidths ubiquitously
observed in molecular cloud material and non-thermal support against
self-gravity.  Even today, accurate determinations of magnetic field
strengths in molecular cloud cores are rare. Most field estimates rely
on measuring the Zeeman splitting in molecular lines, typically OH,
which is very difficult and observationally challenging (e.g.\ Heiles
1993). Hence, the number of clouds in which the Zeeman effect has been
detected above the $3\sigma$ significance level at present is less
than 20, whereas the number of non-detections or upper limits is
considerably larger. For a compilation of field strengths in low-mass
cores see Crutcher (1999), or more recent work by Bourke \etal\
(2001). Results from studying masers around high-mass stars are
excluded in these analyses as in that environment the gas density and
velocity dispersion are very uncertain, and the presence of the
high-mass star is likely to significantly alter the local density and
magnetic field structure with respect to the initial cloud properties
at the onset of star formation.

\begin{figure}
\begin{center}
\unitlength1cm
\begin{picture}( 8.00, 10.00)
%\put( 0.00, 0.50){\epsfxsize=8cm \epsfbox{figure-classicalprobs-01.eps}}
\put(-0.30,-0.50){\epsfxsize=8.2cm \epsfbox{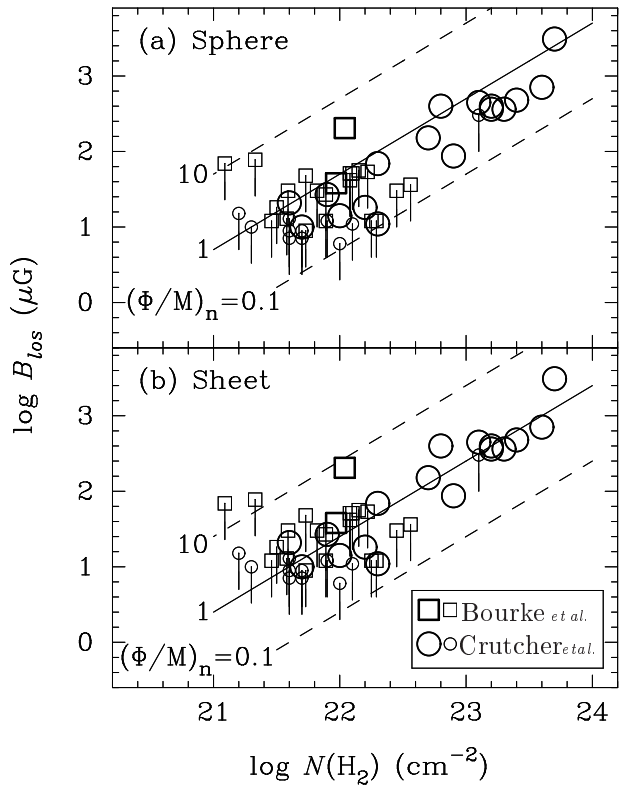}}
\end{picture}
\end{center}
\caption{\label{fig:B-vs-N} Line-of-sight magnetic field strength
$B_{\rm los}$ versus column density $N({\rm H_s}$ for various
molecular cloud cores. Squares indicated values determined by Bourke
\etal\ (2001) and circles denote observations summarized by Crutcher
(1990), Sarma \etal\ (2000), and Crutcher \& Troland (2000). The
large symbols represent clear detections of the Zeeman effect, whereas
the small denote $3\sigma$ upper limits to the field strength. The
lines drawn for the upper limits illustrate the shift from the
$3\sigma$ to the $1\sigma$ level of detection. To guide your eye,
lines of constant flux-to-mass ratio $(\Phi/M)_{\rm n}$ are indicated
normalized to the critical value, i.e.\ to the inverse of equation
(\ref{eqn:crit-phi}). The observed line-of-sight component $B_{\rm los}$
of the field is being (statistically) deprojected to obtain the
absolute value of the field $B$. The upper panel (a) hereby assumes
spherical core geometry, while the flux-to-mass ratios in the lower
panel (b) are computed for sheetlike structures. Values $(\Phi/M)_{\rm
n}<1$ indicate magnetic field strengths insufficient of supporting
against gravitational contraction, i.e.\ cores are magnetically
supercritical. On the other hand, $(\Phi/M)_{\rm n}>1$ indicates
magnetic support as required by the `standard theory'. Note that
almost all cores are magnetically supercritical. This is evident when
assuming spherical symmetry, but also in the case of sheetlike
protostellar cores the average ratio is $\langle (\Phi/M)_{\rm n}
\rangle \approx 0.4$ when considering the $1\sigma$ upper limits. This
is significantly lower than the critical value. (From Bourke \etal\ 2001.)
%The one clear
%exception is RCW57 which however exhibits two velocity components,
%hence, the Zeeman value has to be considered with some doubt. For
%further discussion see Bourke \etal\ 2001, the figure presented here
%is from their study.
}
\end{figure}
%%%
In his critical review of the `standard theory' of star formation Nakano
(1998) pointed out that {\em no} convincingly magnetically subcritical
core had been found up to that time. Similar conclusions still hold
today, {\em all} magnetic field measurements are consistent with
protostellar cores typically being magnetically supercritical or at most
marginally critical. This is indicated in Figure \ref{fig:B-vs-N},
which plots the observed line-of-sight magnetic field $B_{\rm los}$
against the column density $N({\rm H_2})$ determined from CO
measurements. In particular when including non-detections and upper
limits into the analysis, the general conclusion is that magnetic
fields are too weak to prevent or considerably postpone the
gravitational collapse of protostellar cores.  The basic assumption of
the `standard theory' of magnetically mediated star formation therefore
seems in  contradiction to the observational facts.
% The measurements also suggest that low column density
% (and therefore low mass) cores are more strongly supercritical than
% their high column-density counterparts, but this may be an
% observational bias due to beam dilution effects. 
Partial reconciliation, however, between the theory and the
observations may be achieved when taking the extreme geometrical
assumption that all protostellar cores are highly flattened, i.e.\
essentially sheetlike objects (Shu \etal\ 1999). Only then,
flux-to-mass ratios can be derived that come close to the critical
value of equilibrium between magnetic pressure and gravity, but even
for sheetlike `cores' the average flux-to-mass ratio lies by a factor
of 2 to 3 below the critical value when taking all measurements into
account including the upper limits at the $1\sigma$ level (Bourke
\etal\ 2001). In addition, highly flattened morphologies appears to be
inconsistent with the observed density structure of protostellar
cores. They typically appear as `roundish' objects (like the dark
globule B68 which almost perfectly resembles a Bonnor-Ebert sphere,
see Alves \etal\ 2001) and more likely are moderately prolate (with
axis ratios of about 2:1) than highly oblate (with axis ratios $\sim
6$:1) when statistically deprojected (Myers \etal\ 1991, Ryden 1996;
however, some authors prefer the oblate interpretation, see Li \& Shu
1996, Jones, Basu, \& Dubinski 2001).  Altogether the observational
evidence suggests that stars form from magnetically supercritical cloud
core rather than from magnetically supported structures.

%{\bf if GMCs are gravitationally bound, are they collapsing?}
%\noindent {\em (e)}
Molecular clouds as a whole are magnetically highly supercritical and
therefore subject to dynamical collapse.  It is generally believed
that molecular clouds and giant molecular cloud complexes as a whole
are dynamical objects that are magnetically supercritical and bound by
self-gravity, rather than being magnetically subcritical and prevented
from expansion by external pressure or large-scale galactic magnetic
fields. This follows from a careful analysis of the virial balance
equations (McKee 1989, McKee \etal\ 1993, Williams \etal\ 2000). This
analysis can be extended to the substructure within molecular clouds,
to clumps and protostellar cores. Bertoldi \& McKee (1992) already
argued that very massive clumps that form stellar clusters need to be
magnetically supercritical.

This conclusion was extended to low-mass cloud cores by Nakano (1998)
on the basis of the following argument: Clumps and cores in molecular
clouds are generally observed as regions of significantly larger
column density compared to the cloud as a whole (e.g.\ Benson and
Myers 1989, Tatematsu \etal\ 1993). If a protostellar core were
magnetically subcritical it needs to be confined by the mean cloud
pressure or mean magnetic field in the cloud, otherwise it would
quickly expand and disappear. Calculations of the collapse of strongly
subcritical cores such as those by Ciolek \& Mouschovias (1994) fix
the magnetic field at the outer boundaries, artificially confining the
cloud. Under the assumption of virial equilibrium, typical values for
the mean pressure and mean magnetic field in molecular clouds demand
column densities in cores that are comparable to those in the ambient
molecular cloud material. This contradicts the observed large column
density enhancements in cloud cores, and is additional evidence that
also low-mass cores are magnetically supercritical and collapsing.

\rskparagraph{Infall Motions}
%\noindent{\em (f) }
Protostellar infall motions are detected on scales larger and with
velocities greater than predicted by the `standard theory'. One of the
basic assumptions of the standard theory
%of magnetically mediated star formation 
is the existence of a long-lasting quasi-static phase in protostellar
evolution. Before the formation of the central young stellar object
(YSO), molecular cloud cores are held up by strong magnetic fields and
evolve slowly as matter filters through the field lines by ambipolar
diffusion on timescales exceeding the free-fall time by a factor of
ten or more. Once the central singularity is established, the system
undergoes collapse from the inside out, set in motion by an expansion
wave that moves outwards with the speed of sound. Gas inside the
expansion wave approaches free-fall and feeds the central protostar at
a constant accretion rate, while gas further out remains at
rest. Therefore the theory predicts that prestellar cores (i.e.\ cloud
cores without central YSO, see e.g.\ Andr{\'e} \etal\ 2000 for a
discussion) should show no signatures of infall motions, and that
protostellar cores at later stages of the evolution should exhibit
collapse motions confined to their central regions. This can be tested
by mapping molecular cloud cores at the same time in optically thick
as well as thin lines. Inward motions can in principle be inferred
from the asymmetry of optically thick lines, however the signal is
convolved with signatures of rotation and possible outflows (when
looking at protostellar cores that already contain embedded
YSO's). The optically thin line is needed to determine the zero point
of the velocity frame (see e.g.\ Myers \etal\ 1996).

One of the best studied examples is the `starless' core L1544 which
exhibits infall asymmetries (implying velocities up to
$0.1\,$km$\,$s$^{-1}$) that are too extended ($\sim 0.1\,$pc) to be
consistent with inside-out collapse (Tafalla \etal\ 1998, Williams
\etal\ 1999). Similar conclusions can be derived for a variety of
other sources (see the review of Myers, Evans, \& Ohashi 2000; or the
extended survey for infall motions in prestellar cores by Lee, Myers,
and Tafalla 1999, 2001). Typical contraction velocities in the
prestellar phase are between $0.05$ and $0.1\,$km$\,$s$^{-1}$,
corresponding to mass infall rates ranging from a few $10^{-6}$ to a
few $10^{-5}\,$M$_{\odot}$yr$^{-1}$. The sizes of the infalling regions
(e.g.\ as measured in CS) typically exceed the sizes of the
corresponding cores as measured in high-density tracers like N$_2$H$^+$
by a factor of 2--3. 
Even the dark globule B335, which was
considered ``a theorists dream'' (Myers \etal\ 2000) and which was
thought to match `standard theory' very well (Zhou \etal\ 1993) is
also consistent with Larson-Penston collapse (e.g.\ Masunaga and
Inutsuka 2001) when using improved radiation transfer techniques but
relying on single dish data only. The core however
exhibits considerable sub-structure and complexity (clumps, outflows,
etc.)  when observed with high spatial and spectral resolution using
interferometry (Wilner \etal\ 2001). This raises questions about the
applicability of {\em any} 1D isothermal collapse model to objects
where high resolution data are available.  

Extended
inward motions are a common feature in prestellar cores, and appear a
necessary ingredient for the formation of stars as predicted dynamical
theories (Sections \ref{sub:classical} and \ref{sub:new}).

\begin{figure}[ht]
\begin{center}
\unitlength1cm
\begin{picture}( 8.00, 7.20)
%\put( 0.00, 0.50){\epsfxsize=9cm \epsfbox{figure-classicalprobs-02.eps}}
\put(-0.20, 0.00){\epsfxsize=9cm \epsfbox{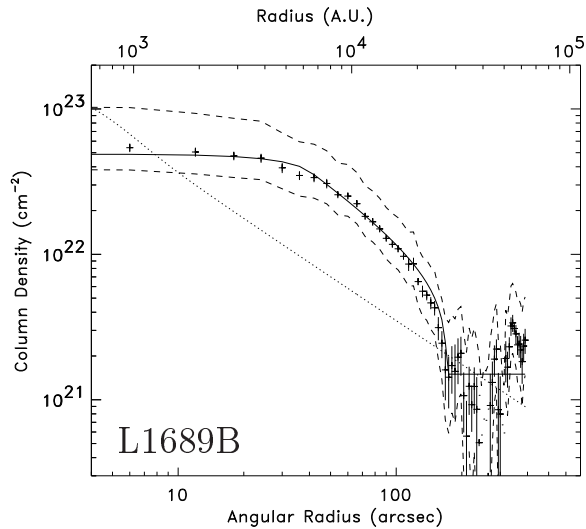}}
\end{picture}
\end{center}
\caption{\label{fig:L1689B} Radial column density profile of the
prestellar core L1689B derived from combined infrared absorption and
$1.3\,$mm continuum emission maps. Crosses show the observed values
with the corresponding statistical errors, while the total
uncertainties in the method are indicated by the dashed lines. For
comparison, the solid line denotes the best-fitting Bonnor-Ebert
sphere and the dotted line the column density profile of an singular
isothermal sphere. The observed profile is well reproduced by an
unstable Bonnor-Ebert sphere with a density contrast of $\sim 50$, see
Bacmann \etal\ 2000 for a further details.}
\end{figure}
\begin{figure*}[ht]
\unitlength1cm
\begin{picture}(16.00, 8.50)
%\put( 0.00, 0.50){\epsfxsize=16cm \epsfbox{figure-classicalprobs-03.eps}}
\put( 0.50, 0.50){\epsfxsize=16cm \epsfbox{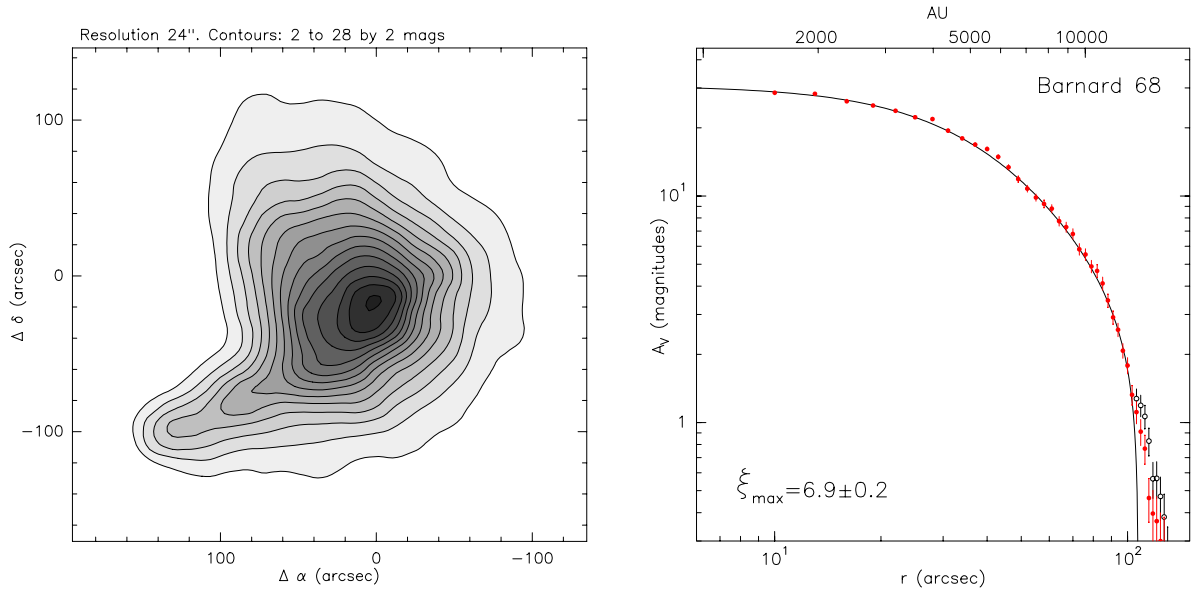}}
\end{picture}
\caption{\label{fig:B68} Optical extinction map (left image) and
resulting azimuthally averaged radial surface density profile (right
image) of the dark globule B68 (from Alves \etal\ 2001). The
resolution of the map is $24''$ and the contour levels indicate steps
of two magnitudes of extinction. The density profile is remarkably
well fitted by a marginally supercritical Bonnor-Ebert sphere with
concentration parameter $\xi_{\rm max} \approx 6.9$ (where the critical
value is $\xi_{\rm max} = 6.5$). Filled red symbols give the observed
radial profile and corresponding errors when the elongated tail at the
lower left is dismissed, open symbols denote the result from the
complete map. Spherical symmetry is an excellent approximation for the
inner parts of B68, noticeable deviations occur at the surface layers
at the 10\%-level of the central density. Data are from Alves \etal\ (2001).}
\end{figure*}

%%%
\rskparagraph{Density Profiles}
%\noindent {\em (g) }
Prestellar cores have flat inner density profiles.  The basis of the
Shu (1977) model is the singular isothermal sphere, i.e.\ the theory
assumes radial density profiles $\rho \propto 1/r^2$ at all radii $r$
as starting conditions of protostellar collapse. The advent of a new
generation of infrared detectors and powerful receivers in the radio
and sub-mm waveband in the late 1990's made it possible to directly
test this hypothesis and determine the radial (column) density profile
of prestellar cores with high sensitivity and resolution (e.g.\
Ward-Thompson \etal\ 1994, Andr{\'e} \etal\ 1996, Motte, Andr{\'e},
and Neri 1998, Ward-Thompson \etal\ 1999, Bacmann \etal\ 2001, Motte
and Andr{\'e} 2001). These studies show that starless cores typically
have flat inner density profiles out to radii of a few $10^{-2}\,$pc
followed by a radial decline of roughly $\rho\propto 1/r^2$ and
possibly a sharp outer edge at radii $0.05$ -- $0.3\,$pc (see the
review of Andr{\'e} \etal\ 2000).
% This is in sharp contrast to protostellar cores that
% already bear a YSO in there interior, these cores exhibit a density
% structure $\rho\propto 1/r^2$ at all radii $r$ (e.g.\ Ladd \etal\ 
% 1991, Motte \& Andr{\'e} 1999).  
This is illustrated in Figure \ref{fig:L1689B} which shows the
observed column density of the starless core L1689B derived from
combining mid-infrared absorption maps with 1.3 mm dust continuum
emission maps (Bacmann \etal\ 2001).  The density structure in the
prestellar phase appears consistent with pressure-bounded Bonnor-Ebert
spheres (Bonnor 1956, Ebert 1955) with sufficient density contrast to
imply instability against gravitational collapse. This impression is
strengthened further by recent findings from multi-wavelength stellar
extinction studies of dark globules.  For example, the density
distribution of the dark globule B68 is in nearly perfect agreement
with being a marginally supercritical Bonnor-Ebert sphere, as
illustrated in Figure \ref{fig:B68} (Alves, Lada, \& Lada
2001). Altogether high-resolution mapping of prestellar cores provides
the most direct evidence against singular isothermal spheres as
initial conditions of protostellar collapse. 
%%%

\rskparagraph{Chemical Ages}
%\noindent {\em (k) }
The chemical age of substructure in molecular cloud is much smaller
than the ambipolar diffusion time. This poses a timescale argument against
magnetically regulated star formation, and  comes from the investigation of
the chemical status of density fluctuations in molecular clouds. The
comparison of multi-molecule observations of cloud cores with
time-dependent chemical models indicates typical ages of about
$10^5\,$years (e.g.\ Bergin \& Langer 1997, Pratap \etal\ 1997,
Aikawa \etal\ 2001; or see the reviews by van~Dishoeck \etal\ 1993,
van~Dishoeck \& Black 1998, and  Langer \etal\ 2000). This is
significantly shorter than the timescale required for ambipolar
diffusion to become important as required in the standard model.  The
inferred chemical ages of molecular cloud cores appear only compatible
with supersonic, and super-Alv{\'e}nic turbulence as being the main
agent that determines molecular cloud structure and regulates the star
formation process (see Sections \ref{sub:beyond} and \ref{sub:new}).
%However, the strong model-dependence of this
%conclusion makes this argument somehow less conclusive than others in
%this Section.

\rsksubsubsection{Observations of Protostars and Young Stars}
\label{subsub:obs-stars}
%\noindent {\em (h) 
\rskparagraph{Accretion Rates}
Protostellar accretion rates decline with time. As
an immediate consequence of the assumed singular $1/r^2$ initial
density profile, the Shu (1977) model predicts constant protostellar
accretion rates $\dot{M}_{\star}= 0.975 c_{\rm s}^3/G$, with sound
speed $c_{\rm s}$ and gravitational constant $G$. As matter falls onto
the central protostar it goes through a shock and releases energy that
is radiated away giving rise to a luminosity $L_{\rm acc} \approx
GM_{\star}\dot{M}_{\star}/R_{\star}$ (Shu \etal\ 1987, 1993). The fact
that most of the matter first falls onto a protostellar disk, where it
gets transported inwards on a viscous timescale before it is able to
accrete onto the star does not alter the expected overall luminosity
by much (see e.g.\ Hartmann 1998).

\begin{figure}[ht]
\begin{center}
\unitlength1cm
\begin{picture}( 8.00, 6.00)
%\put( 0.00, 0.50){\epsfxsize=9cm \epsfbox{Andre-PPIV-fig6b.eps}}
\put(-0.20,-0.10){\epsfxsize=9cm \epsfbox{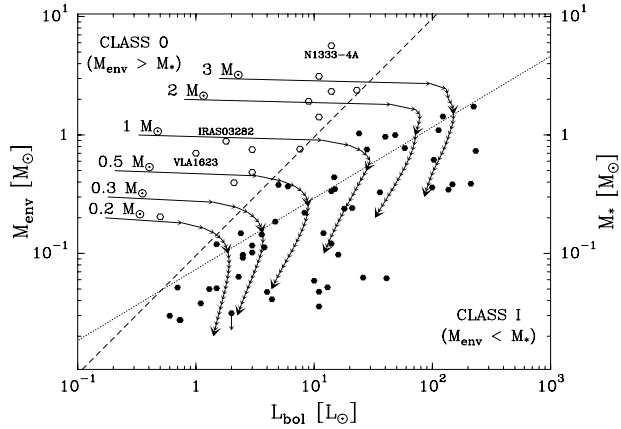}}
\end{picture}
\end{center}
\caption{\label{fig:L_bol-M_env} $M_{\rm env} - L_{\rm bol}$ diagram
for a sample of protostellar cores in the main accretion phase from
Andr{\'e} \& Montmerle (1994) and Saraceno \etal\ (1996). Open
circles indicate objects for which the envelope mass exceeds the mass
of the central protostar ($M_{\rm env} > M_{\rm \star}$) and filled
circles denote the later evolutionary stage where $M_{\rm env} <
M_{\rm \star}$.  Overlayed are evolutionary tracks that assume bound
initial configuration of finite mass and that have $L_{\rm bol} =
GM_{\star} \dot{M_{\star}}/R_{\star} + L_{\star}$ with $L_{\star}$
from Stahler (1988) and with both, $M_{\rm env}$ and
$\dot{M_{\star}}=M_{\rm env}/\tau$ ($\tau \approx 10^5\,$yr),
declining exponentially with time (Bontemps \etal\ 1996, also Myers
\etal\ 1998). Exponentially declining $\dot{M_{\star}}$ show better
agreement with the data than do constant accretion rates. Small arrows
are plotted on the tracks every $10^4\,$yr, and large arrows when 50\%
and 90\% of the total mass is accreted onto the central YSO. The
dashed and dotted lines indicate the transition from $M_{\rm env}>
M_{\star}$ to $M_{\rm env}< M_{\star}$ using two different relations,
$M_{\star}\propto L_{\rm bol}$ and $M_{\star}\propto L_{\rm
bol}^{0.6}$, respectively, indicating the range  proposed in the literature (e.g.\
Andr{\'e} \& Montmerle 1994, or Bontemps \etal\ 1996). The latter
relation is suggested by the accretion scenario adopted in the
tracks. The figure is adapted from Andr{\'e} \etal\ (2000).}
\end{figure}

During the early phases of
protostellar collapse, i.e.\ as long as the mass $M_{\rm env}$ of the
infalling envelope exceed the mass $M_{\star}$ of the central
protostar, the accretion luminosity $L_{\rm acc}$ by far exceeds the
intrinsic luminosity $L_{\star}$ of the young star. Hence the observed
bolometric luminosity $L_{\rm bol}$ of the object is a direct measure
of the accretion rate as long as reasonable estimates of $M_{\star}$
and $R_{\star}$ can be obtained. Determinations of bolometric
temperature $T_{\rm bol}$ and luminosity $L_{\rm bol}$ therefore
should provide a fair estimate of the evolutionary stage of a
protostellar core (e.g.\ Chen \etal\ 1995, Myers \etal\
1998). Scenarios in which the accretion rate decreases with time and
increases with total mass of the collapsing cloud fragment yield
qualitatively better agreement with the observations than do models
with constant accretion rate (Andr{\'e} \etal\ 2000, see however
Jayawardhana, Hartmann, \& Calvet 2001, for an alternative
interpretation based on environmental conditions).  A comparison of
observational data with theoretical models where $\dot{M}_{\star}$
decreases exponentially with time is shown in Figure
\ref{fig:L_bol-M_env}.  
%%%

\begin{figure}[ht]
\begin{center}
\unitlength1cm
\begin{picture}( 8.00, 7.40)
%\put( 0.00, 0.50){\epsfxsize=9cm \epsfbox{Andre-PPIV-fig7.eps}}
\put(-0.40,-0.30){\epsfxsize=8.5cm \epsfbox{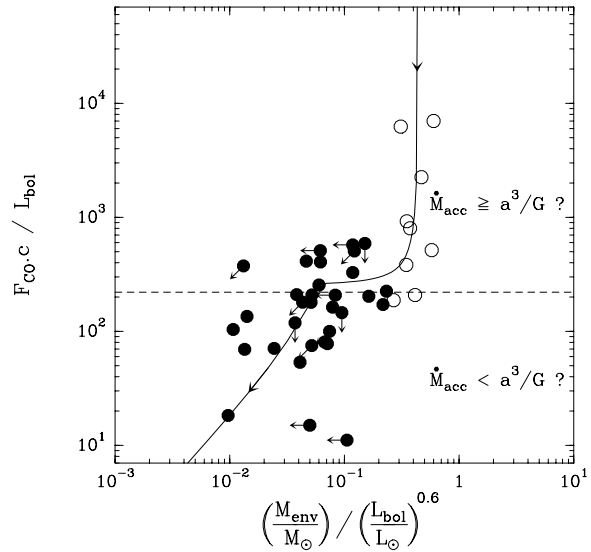}}
\end{picture}
\end{center}
\caption{\label{fig:decreasing-dM/dt}
Outflow momentum flux $F_{\rm CO}$ 
plotted against envelope mass
$M_{\rm env}$ normalized to the bolometric luminosity $L_{\rm bol}$
using the relations $M_{\rm env}\propto L_{\rm bol}^{0.6}$ and $F_{\rm
CO}c\propto L_{\rm bol}$. Protostellar cores with $M_{\rm env}>
M_{\star}$ are indicated by open circles, and $M_{\rm env}> M_{\star}$
by filled circles (data from Bontemps \etal\ 1996). $F_{\rm
CO}c/L_{\rm bol}$ is an empirical tracer for the accretion rate, and the
speed of light $c$ is invoked in order to obtain a dimensionless quantity. 
$M_{\rm env}/L_{\rm bol}^{0.6}$ is an evolutionary indicator that
decreases with time. The abscissa therefore corresponds to a time axis
with early times at the right and late evolutionary stages at the
left. Overlayed on the data is a evolutionary model that assumes a
flat inner density profile (for details see Hendriksen
\etal\ 1997, where the figure was published originally).
}
\end{figure}

A closely related method to estimate the
accretion rate $\dot{M}_{\star}$ is by determining protostellar
outflow strengths (Bontemps \etal\ 1996). It is known that most
embedded young protostars are associated with powerful molecular
outflows (Richer \etal\ 2000) and that the outflow activity decreases
towards later evolutionary stages. For protostars at the end of their
main accretion phase there exists a clear correlation between the
outflow momentum flux $F_{\rm CO}$ and the bolometric luminosity
$L_{\rm bol}$ (e.g.\ Cabrit \& Bertout 1992). Furthermore, $F_{\rm
CO}$ is well correlated with $M_{\rm env}$ for all protostellar cloud
cores (Bontemps \etal\ 1996, Hoherheijde \etal\ 1998, Henning and
Launhardt 1998). This result is independent of the $F_{\rm CO}-L_{\rm
bol}$ relation and most likely results from a progressive decrease of
outflow power with time during the main accretion phase. With the
linear correlation between outflow mass loss and protostellar
accretion rate (Hartigan \etal\ 1995) these observations therefore
suggest stellar accretion rates $\dot{M}_{\star}$ that decrease with
time. This is illustrated in Figure \ref{fig:decreasing-dM/dt} which
compares the observed values of the normalized outflow flux and the
normalized envelope mass for a sample of $\sim40$ protostellar cores
with a simplified dynamical collapse model with decreasing accretion
rate $\dot{M}_{\star}$ (Hendriksen \etal\ 1997). The model describes
the data relatively well, as opposed to models of constant
$\dot{M}_{\star}$.
%%%

\rskparagraph{Embedded Objects}
%\noindent {\em (i) }
The fraction of protostellar cores with embedded protostellar objects
is very high. Further indication that the standard theory may need to
be modified comes from estimates of the time spent by protostellar
cores during various stages of their evolution. For a sample of
protostars, the relative numbers of objects in distinct evolutionary
phases roughly correspond to the relative time spent in each
phase. Beichman \etal\ (1986) used the ratio of numbers of starless
cores to the numbers of cores with embedded objects detected with the
{\em Infrared Astronomical Satellite} (IRAS) and estimated that the
duration of the prestellar phase is about equal to the time needed for
a young stellar object to completely accrete the protostellar envelope
it is embedded in. As the standard model assumes cloud cores in the
prestellar phase evolve on ambipolar diffusion timescales, which are
an order of magnitude longer than the dynamical timescales of the
later accretion phase, one would expect a significantly larger number
of starless cores than cores with embedded protostars.  Millimeter
continuum mapping of pre-stellar cores gives similar results
(Ward-Thompson \etal\ 1994, 1999), leading Andr{\'e} \etal\ (2000) to
argue that the timespan of cores to increase their central density
$n({\rm H}_2)$ from $\sim10^4$ to $\sim10^5\,$cm$^{-3}$ is about the
same as to go from $n{\rm H}_2)\approx 10^5\,$cm$^{-3}$ to the
formation of the central protostar. This clearly disagrees with
standard ambipolar diffusion models (e.g.\ Ciolek \& Mouschovias
1994) which predict a six times longer duration.  Ciolek \& Basu
(2000) were indeed able to accurately model infall in L1544 using an
ambipolar diffusion model, but they did so by using initial conditions
that were already almost supercritical, so that very little ambipolar
diffusion had to occur before dynamical collapse would set in.  Ciolek
\& Basu (2001) quantify the central density required to match the
observations, and conclude that observed pre-stellar cores are either
already supercritical or just about to be. Altogether these
considerations suggest that already in the prestellar phase the
timescales of core contraction are determined by fast dynamical
processes rather than by slow ambipolar diffusion.

\rskparagraph{Stellar Ages}
%\noindent {\em (j) }
Stellar age spreads in young clusters are small. If the contraction
time of individual cloud cores in the prestellar phase is determined
by ambipolar diffusion rather than by dynamical collapse (or turbulent
dissipation, see Sections \ref{sub:beyond} and \ref{sub:new}), then
the age spread in a stellar population (say in a young cluster) should
considerably exceed the relevant dynamical timescales. Within a
star-forming region high-density protostellar cores will evolve and
form a central YSO faster than their low-density counterparts, and the
age distribution is roughly determined by the evolution time of the
lowest-density condensation. Note, that ambipolar diffusion time
$\tau_{\rm AD}$ and free-fall time $\tau_{\rm ff}$ both are inversely
proportional to square root of the density, $\tau_{\rm AD} \propto
\tau_{\rm ff} \propto \rho^{-1/2}$, and that $\tau_{\rm AD} \approx 10
\tau_{\rm ff}$ under typical conditions (e.g.\ McKee \etal\ 1993).

However, the age spread in star clusters is very short. For example in
the Orion Trapezium cluster it is less than $10^6\,$years (Prosser
\etal\ 1994, Hillenbrand 1997, Hillenbrand \& Hartmann 1998), and the
same holds for L1641 (Hodapp \& Deane 1993). The age spread is
comparable to the dynamical time in these clusters. Similar
conclusions can be obtained for Taurus (Hartmann 2001), NGC$\,$1333
(Bally \etal\ 1996, Lada \etal\ 1996), NGC$\,$6531 (Forbes 1996), and
a variety of other clusters (see Elmegreen \etal\ 2000 for a review,
also Palla \& Stahler 1999, Hartmann 2001). There is a relation
between the duration of star formation and the size of the star
forming region. Larger regions form stars for a longer timespan. This
correlation is comparable to the linewidth-size relation, or the
crossing time-size relation, respectively, found for molecular gas,
and suggests that typical star formation times correspond to about 2
to 3 turbulent crossing times in that region (Efremov \& Elmegreen
1998).  This is very fast compared to the ambipolar diffusion
timescale, which is about 10 crossing times in a uniform medium with
cosmic ray ionization (Shu \etal\ 1987) and is even longer if stellar
UV sources contribute to the ionization (Myers \& Khersonsky 1995) or
if the cloud is very clumpy (Elmegreen \& Combes 1992). Magnetic
fields, therefore, cannot regulate star formation on scales of stellar
clusters.\\

% The timescale argument alone, however, cannot rule out that
% magnetic fields determine the formation of individual stars, this is
% only possible in combination with the other caveats listed here.

%{\bf [Mordecai: I am not sure that the argument is really that
%convincing, as the considered densities are different. The global
%free-fall time is determined for the average cloud density, which is
%probaly 1 or 2 orders of magnitude lower than the mean densities of
%individual protostellar cores, hence, $\tau_{\rm AD}(core) \approx
%\tau_{\rm ff}(cluster)$. With typical age spread being of the order of
%$\tau_{\rm ff}(cluster)$ I am not really fully convinced that the age
%spread issue is very convincing. What do you think?]}

%I think that $tau_{AD}(cluster)$ is what is proposed for the star
%formation time scale, and that therefore age spreads of order
%$\tau_{\rm AD}(core)$ are indeed a counter argument.}

%%%

\rsksubsection{Beyond the Standard Theory}
\label{sub:beyond}
%\input{beyond.tex}
%%%
%%% RMP-Module
%%%
%%%
%%%\rsksubsection{Beyond the standard theory}
%%%\label{sub:beyond}
%%%

New understanding of the behavior of turbulence has suggested that it
may be more important than previously thought in the support of
molecular clouds against gravitational collapse.  Indeed, it may take
many of the roles traditionally assigned to magnetic fields in the
standard model.

\begin{figure*}[tb]
\begin{center}
\vspace*{0.5cm}
\includegraphics[width=.8\textwidth]{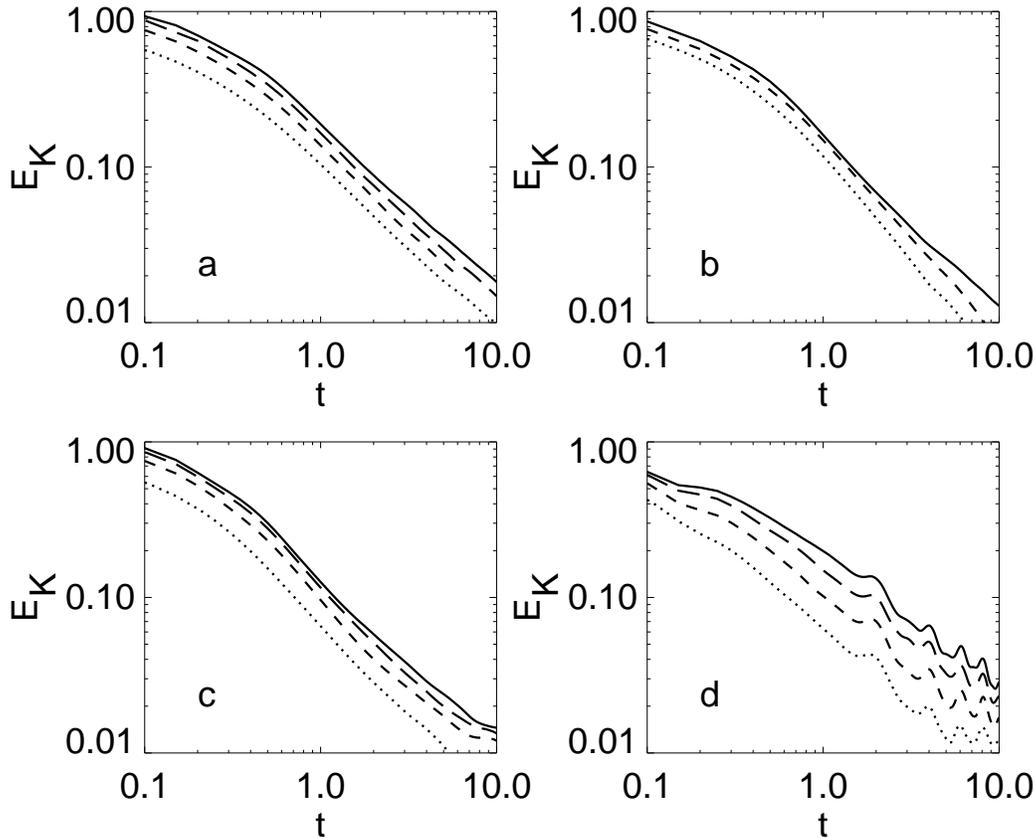}
\end{center}
\caption[Kinetic energy vs.\ time in decaying
turbulence]{Three-dimensional resolution studies of the decay of
supersonic turbulence for initial Mach number $M=5$, isothermal
models.  ZEUS models have $32^3$ ({\em dotted}), $64^3$ ({\em short
dashed}), $128^3$ ({\em long dashed}), or $256^3$ ({\em solid}) zones,
while the SPH models have 7000 ({\em dotted}), 50,000 ({\em short
dashed}), or 350,000 ({\em solid}) particles. Panels show {\em a)}
hydro runs with ZEUS, {\em b)} hydro runs with SPH, {\em c)} $A=5$ MHD
runs with ZEUS, and {\em d)} $A=1$ MHD runs with ZEUS (From Mac Low et
al.\ (1998)).}
\label{fig:prlres}
\end{figure*}

\rsksubsubsection{Maintenance of Supersonic Motions}
\label{subsub:motions}
We first consider the question of how to maintain the observed
supersonic motions.  As described above, magnetohydrodynamic waves
were generally thought to provide the means to prevent the dissipation
of interstellar turbulence.  However, numerical models have now shown
that they probably do not.  One-dimensional simulations of decaying,
compressible, isothermal, magnetized turbulence by Gammie \& Ostriker
(1996) showed quick decay of kinetic energy $K$ in the absence of
driving, but found that the quantitative decay rate depended strongly
on initial and boundary conditions because of the low dimensionality.
Mac Low \etal\ (1998), Stone, Ostriker \& Gammie (1998), and Padoan
\& Nordlund (1999) measured the decay rate in direct numerical
simulations in three dimensions, using a number of different numerical
methods.  They uniformly found rather faster decay, with Mac Low
\etal\ (1998) characterizing it as $E_{\rm kin} \propto t^{-\eta}$, with $0.85 <
\eta < 1.1$. A resolution and algorithm study is shown in
Figure~\ref{fig:prlres}.  Magnetic fields with strengths ranging up to
equipartition with the turbulent motions (ratio of thermal to magnetic
pressures as low as $\beta = 0.025$) do indeed reduce $\eta$ to the
lower end of this range, but not below that, while unmagnetized
supersonic turbulence shows values of $\eta \simeq$~1--1.1.

Stone \etal\ (1998) and Mac Low (1999) showed that supersonic
turbulence decays in less than a free-fall time under molecular cloud
conditions, regardless of whether it is magnetized or unmagnetized.
The hydrodynamical result agrees with the high-resolution,
transsonic, decaying models of Porter \& Woodward (1992, 1994).  Mac
Low (1999) showed that the formal dissipation time $\tau_{\rm d} = K/\dot{K}$
scaled in units of the free fall time $t_{\rm ff}$ is
% \begin{equation} \label{eqn:decay}
% \tau_{\rm d}/\tau_{\rm ff} = \frac{1}{4 \pi \xi} \left(\frac{32}{3}\right)^{1/2}
% \frac{\kappa}{{\cal M}_{\rm rms}} \simeq \,3.9 \,\frac{\kappa}{{\cal
% M}_{\rm rms}},
% \end{equation}
\begin{equation} \label{eqn:decay}
\tau_{\rm d}/\tau_{\rm ff} = \frac{1}{4 \pi \xi} \left(\frac{32}{3}\right)^{1/2}
\frac{\kappa}{{\cal M}_{\rm rms}} \simeq \,3.9 \,\frac{\kappa}{{\cal
M}_{\rm rms}},
\end{equation}
where $\xi = 0.21/\pi$ is the energy-dissipation coefficient, ${\cal M}_{\rm
rms} = v_{\rm rms}/c_{\rm s}$ is the rms Mach number of the turbulence, and
$\kappa$ is the ratio of the driving wavelength to the Jeans
wavelength $\lambda_{\rm J}$.  In molecular clouds, ${\cal M}_{\rm rms}$ is
typically observed to be of order 10 or higher.  If the ratio $\kappa
< 1$, as is probably required to maintain gravitational support
(L\'eorat \etal\ 1990), then even strongly magnetized turbulence will
decay long before the cloud collapses and not markedly retard the
collapse.

Either observed supersonic motions must be continually driven, or molecular
clouds must be less than a single free-fall time old.  As we discuss in
\S~\ref{sub:clouds}, the observational evidence does suggest that clouds are a
few free-fall times old, on average, though perhaps not more than two or
three, so there is likely some continuing energy input into the clouds.
% Mordecai- redundant: Magnetic fields do not notably reduce the
% energy decay rate. 

\rsksubsubsection{Turbulence in Self-Gravi\-ta\-ting Gas}
\label{subsub:self-grav}

This leads to the question of what effects supersonic turbulence will have on
self-gravitating clouds.  Can turbulence alone delay gravitational
collapse beyond a free-fall time?  In \S~\ref{sec:introduction}
and \S~\ref{sub:classical}, we summarized analytic approaches to this question,
and pointed out that they were all based on the assumption that the turbulent
flow is close to incompressible.  Analytic attempts to statistically
characterize highly compressible 
turbulence such as that we actually see in molecular clouds have usually been
based on heuristic models (Elmegreen 1993), although recently they have been
able to start making some progress in recovering the velocity structure
(Boldyrev 2002, Boldyrev, Nordlund, \& Padoan 2002).

Numerical models of highly compressible, self-gravitating turbulence
have shown the importance of density fluctuations generated by the
turbulence to understanding support against gravity.  Early models
were done by Bonazzola \etal\ (1987), who used low resolution ($32
\times 32$ collocation points) calculations with a two-dimensional
spectral code to support their analytical results.  The hydrodynamical
studies by Passot \etal\ (1988), L{\'e}orat, Passot \& Pouquet (1990),
V{\'a}zquez-Semadeni, Passot, \& Pouquet (1995) and
Ballesteros-Paredes, V{\'a}zquez-Semadeni \& Scalo~(1999), were also
restricted to two dimensions, and were focused on the interstellar
medium at kiloparsec scales rather than molecular clouds, although
they were performed with far higher resolution (up to $800 \times 800$
points).  Magnetic fields were introduced in these models by Passot,
V\'azquez-Semadeni, \& Pouquet~(1995), and extended to three
dimensions with self-gravity (though at only $64^3$ resolution) by
V\'azquez-Semadeni, Passot, \& Pouquet~(1996).  One-dimensional
computations focused on molecular clouds, including both MHD and
self-gravity, were presented by Gammie \& Ostriker~(1996) and Balsara,
Crutcher \& Pouquet (1999).  Ostriker, Gammie, \& Stone (1999)
extended their work to 2.5 dimensions more recently.

These early models at low resolution, low dimension, or both, suggested
several important conclusions. First, gravitational collapse, even in the
presence of magnetic fields, does not generate sufficient turbulence to
markedly slow continuing collapse. Second, turbulent support against
gravitational collapse may act at some scales, but not others.  More recent
three-dimensional, high-resolution computations by Klessen \etal\ (1998, 2000)
Klessen (2000), Klessen \& Burkert (2000, 2001), and Heitsch \etal\ (2001a)
have now confirmed both of these results. In the following subsections, we
give a brief description of the numerical methods used, give more details on
these results, and draw consequences for the theory of star formation.

\rsksubsubsection{A Numerical Approach}
\label{subsub:numerics}

%Most analytic approaches to molecular cloud stability (e.g.\ as reviewed in
%Klessen \etal\ 2000) make the strong assumption of a homogeneous equilibrium
%state with constant density $\rho_0$. This substantially limits their
%reliability, as observations clearly show that molecular clouds are extremely
%non-uniform (see \S~\ref{sub:regions}).  The restrictions
%of a purely analytical approach can be circumvented with numerical modeling.  A
%detailed numerical analysis of self-gravitating turbulent clouds performed by
%Klessen \etal\ (2000) and Heitsch \etal\ (2001) forms the basis for the next
%sections.
%
Klessen \etal\ (2000) and Heitsch \etal\ (2001a) used two different numerical
methods: ZEUS-3D (Stone \& Norman 1992ab, Hawley \& Stone 1995), an Eulerian
MHD code; and an implementation of smoothed particle hydrodynamics (SPH; Benz
1990, Monaghan 1992), a Lagrangian hydrodynamics method using particles as an
unstructured grid, while Klessen \etal\ (1998), Klessen (2000), and Klessen \&
Burkert (2000, 2001) used only SPH computations.  Both codes were used to
examine the gravitational stability of three-dimensional hydrodynamical
turbulence at high resolution.  The use of both Lagrangian and Eulerian
methods to solve the equations of self-gravitating hydrodynamics in three
dimensions (3D) allowed them to attempt to bracket reality by taking advantage
of the strengths of each approach.  This also gave them some protection
against interpreting numerical artifacts as physical effects.

SPH can resolve very high density contrasts because it increases the particle
concentration, and thus the effective spatial resolution, in regions of high
density, making it well suited for computing collapse problems.  By the same
token, though, it resolves low-density regions poorly. Shock structures tend
to be broadened by the averaging kernel in the absence of adaptive techniques.
The correct numerical treatment of gravitational collapse requires the
resolution of the local Jeans mass at every stage of the collapse (Bate \&
Burkert 1997).  In the current code, once an object with density beyond the
resolution limit of the code has formed in the center of a collapsing gas
clump it is replaced by a `sink' particle (Bate, Bonnell, \& Price 1995).
Adequately replacing high-density cores and keeping track of their further
evolution in a consistent way prevents the time step from becoming
prohibitively small. This allows modeling of the collapse of a large number
of cores until the overall gas reservoir becomes exhausted.

ZEUS-3D, conversely, gives equal resolution in all regions, and allows
good resolution shocks everywhere, as well as allowing the inclusion
of magnetic fields (see \S~\ref{subsub:MHD}).  On the other hand,
collapsing regions cannot be followed to scales less than one or two
grid zones.  The numerical resolution required to follow gravitational
collapse must be considered.  For a grid-based simulation, the
criterion given by Truelove \etal\ (1997) holds.  Equivalent to the
SPH resolution criterion, the mass contained across two or three grid
zones has to be rather smaller than the local Jeans mass throughout
the computation.

The computations presented here are done on periodic cubes, with an isothermal
equation of state, using up to $256^3$ zones (with one model at $512^3$ zones)
or $80^3$ SPH particles. To generate turbulent flows Gaussian velocity
fluctuations are introduced with power only in a narrow interval $k-1 \leq
|\vec{k}| \leq k$, where $k = L/\lambda_{\rm d}$ counts the number of driving
wavelengths $\lambda_{\rm d}$ in the box (Mac Low~\etal\ 1998).  This offers a
simple approximation to driving by mechanisms that act on that scale. To drive
the turbulence, this fixed pattern is normalized to maintain constant kinetic
energy input rate $\dot{E}_{\rm in} = \Delta E / \Delta t$ (Mac~Low 1999).
Self-gravity is turned on only after a state of dynamical equilibrium has been
reached. In Table \ref{tab:models} we summarize the numerical models used in
the subsequent analysis and give a list of their basic properties. 

\begin{table*}[tbh]
%{\large \bf Tables:}
\begin{center}
\begin{tabular}{cccccccc}
\hline \hline
Name &  Method  & Resolution    & $k_{\rm drv}$ & $\dot{E}_{\rm in}$ & 
$E_{\rm kin}^{eq}$ & $\langle M_{\rm J}\rangle_{\rm turb}$ & $t_{5\%}$  \\
%%% Times have been normalized to the free-fall time!
\hline
${\cal A}1$  &  SPH  & $200\,000$ & $1-2$ & $0.1$ & $0.15$ & $0.6$  & $0.5$  \\
${\cal A}2$  &  SPH  & $200\,000$ & $3-4$ & $0.2$ & $0.15$ & $0.6$  & $0.7$  \\
${\cal A}3$  &  SPH  & $200\,000$ & $7-8$ & $0.4$ & $0.15$ & $0.6$  & $2.2$  \\
\hline                                                                     
${\cal B}1$  &  SPH  & $50\,000$ &   $1-2$ & $0.5$ &  $0.5$ & $3.2$  &  $0.5$ \\
~$\,{\cal B}1h$ &  SPH  & $200\,000$&$1-2$ & $0.5$ &  $0.5$ & $3.2$  &  $0.4$ \\
%~${\cal B}2\ell$ &  SPH  & $20\,000$&$3-4$ & $1.0$ &  $0.5$ & $3.2$  &  $1.6$ \\
${\cal B}2$  &  SPH  & $50\,000$ &   $3-4$ & $1.0$ &  $0.5$ & $3.2$  &  $1.5$ \\
~$\,{\cal B}2h$ &  SPH  & $200\,000$&$3-4$ & $1.0$ &  $0.5$ & $3.2$  &  $1.4$ \\
${\cal B}3$  &  SPH  & $50\,000$ &   $7-8$ & $2.4$ &  $0.5$ & $3.2$  &  $6.0$ \\
${\cal B}4$  &  SPH  & $50\,000$ & $15-16$ & $5.0$ &  $0.5$ & $3.2$  & $8.0$  \\
${\cal B}5$  &  SPH  & $50\,000$ & $[39-40]$ & $[5.9]$ &  $[0.3]$ & $[1.7]$  &  ---  \\ 
\hline                                                                     
${\cal C}2$  &  SPH  & $50\,000$ &   $3-4$ & $7.5$ &  $2.0$ & $18.2$  &  $6.0$ \\
\hline                                                                     
%~${\cal D}1\ell$ & ZEUS  &  $64^3$ & $1-2$   & $0.4$ &  $0.5$ & $3.2$  &  *   \\
%~${\cal D}2\ell$ & ZEUS  &  $64^3$ & $3-4$   & $0.8$ &  $0.5$ & $3.2$  &  *   \\
%~${\cal D}3\ell$ & ZEUS  &  $64^3$ & $7-8$   & $1.6$ &  $0.5$ & $3.2$  &  *   \\
${\cal D}1$      & ZEUS  & $128^3$ & $1-2$   & $0.4$ &  $0.5$ & $3.2$  &$0.4$ \\
${\cal D}2$      & ZEUS  & $128^3$ & $3-4$   & $0.8$ &  $0.5$ & $3.2$  &$1.2$ \\
${\cal D}3$      & ZEUS  & $128^3$ & $7-8$   & $1.6$ &  $0.5$ & $3.2$  &$2.4$ \\

~${\cal D}1h$    & ZEUS  & $256^3$ & $1-2$   & $0.4$ &  $0.5$ & $3.2$  &$0.4$ \\
~${\cal D}2h$    & ZEUS  & $256^3$ & $3-4$   & $0.8$ &  $0.5$ & $3.2$  &$1.2$ \\
~${\cal D}3h$    & ZEUS  & $256^3$ & $7-8$   & $1.6$ &  $0.5$ & $3.2$  &$3.1$ \\
\hline \hline
\end{tabular}
\end{center}
\caption{\label{tab:models}
Overview of the models.  The columns give model name,
numerical method, resolution, driving wave lengths $k$, energy input
rate $\dot{E}_{\rm in}$, equilibrium value of kinetic energy without
self-gravity $E_{\rm kin}^{eq}$, turbulent Jeans mass $\langle M_{\rm
J}\rangle_{\rm turb}$, and the time required to reach a core mass
fraction $M_* = 5$\%.  The resolution is given for SPH as particle
number and for ZEUS as number of grid cells. Dashes in the last column
indicate that no sign of local collapse was observed within
$20\tau_{\rm ff}$, while stars indicate that the numerical resolution
was insufficient for unambiguous identification of collapsed
cores. The total mass in the system is $M=1$. Model ${\cal B}5$ focuses on a
subvolume with mass $M=0.25$ and 
decreased sound speed $c_{\rm s} = 0.05$. When scaled up to the standard cube
this 
corresponds to the {\em effective} values given in square brackets.
 Adapted from Klessen \etal\ (2000).}
\end{table*}

\rsksubsubsection{Global Collapse}
\label{subsub:global}

First we examine the question of whether gravitational collapse can
generate enough turbulence to prevent further collapse. Hydrodynamical
SPH models initialized at rest with Gaussian density perturbations
show fast collapse, with the first collapsed objects forming in a
single free-fall time (Klessen, Burkert, \& Bate 1998; Klessen \&
Burkert 2000, 2001). Models set up with a freely decaying turbulent
velocity field behaved similarly (Klessen 2000).  Further accretion of
gas onto collapsed objects then occurs over the next free-fall time,
defining the predicted spread of stellar ages in a freely-collapsing
system. The turbulence generated by the collapse (or virialization)
does not prevent further collapse as suggested by many people (e.g.\ 
Elmegreen 1993). Such a mechanism only works for thermal pressure
support in systems such as galaxy cluster halos when dissipation is
ineffective, while the dissipation for turbulence is quite effective
(Equation~\ref{eqn:decay}).

Models of freely collapsing, magnetized gas remain to be done, but models of
self-gravitating, decaying, magnetized turbulence by Balsara, Ward-Thompson,
\& Crutcher (2001) using an MHD code incorporating a Riemann solver suggest
that the presence of magnetic fields is unlikely to markedly extend collapse
timescales.  They further show that accretion down filaments aligned with
magnetic field lines onto cores occurs readily.  This allows high mass-to-flux
ratios to be maintained even at small scales, which is necessary for
supercritical collapse to continue after fragmentation occurs.

\begin{figure*}[ht]
\unitlength1.0cm
\begin{picture}(16,14.3)
%\put(0.0,-3.0){\epsfxsize=14.5cm \epsfbox{ms50734-fig01.ps}}
\put(3.0,0.0){\epsfxsize=10cm \epsfbox{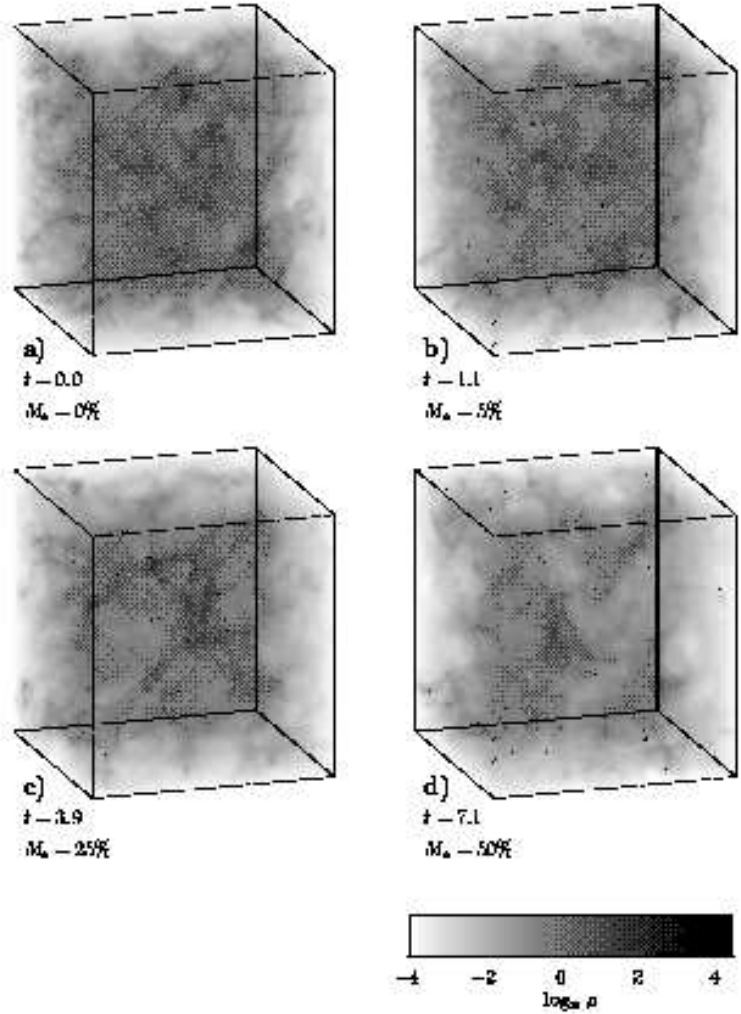}}
\end{picture}
\caption{\label{fig:3D-cubes}
  Density cubes for model ${\cal B}2h$ from Klessen \etal\ (2000),
  which is driven in the at intermediate wavelength, shown (a) at the
  time when gravity is turned on, (b) when the first collapsed cores are
  formed and have accreted $M_* = 5$\% of the mass, (c) when the mass
  in dense cores is $M_* = 25$\%, and (d) when $M_* = 50$\%. Time is
  measured in units of the global system free-fall time scale
  $\tau_{\rm ff}$, dark dots indicate the location of the collapsed cores. }
\end{figure*}

\rsksubsubsection{Local Collapse in Globally Stable Regions}
\label{subsub:local}
Second, we examine whether continuously driven turbulence can support against
gravitational collapse.  The models of driven, self-gravitating turbulence by
Klessen \etal\ (2000) and Heitsch \etal\ (2001a) described above
(\S~\ref{subsub:numerics}) show that {\em local} collapse occurs even when the
turbulent velocity field carries enough energy to counterbalance gravitational
contraction on global scales.  This confirms the results of two-dimensional
(2D) and low-resolution ($64^3$) 3D computations with and without magnetic
fields by V\'azquez-Semadeni \etal\ (1996).  An example of local collapse in a
globally supported cloud is given in Figure~\ref{fig:3D-cubes}. A hallmark of
global turbulent support is isolated, inefficient, local collapse.

Local collapse in a globally stabilized cloud is not predicted by any
of the analytic models (as discussed in Klessen \etal\ 2000). The
resolution to this apparent paradox lies in the requirement that any
substantial turbulent support must come from supersonic flows, as
otherwise pressure support would be at least equally important.
Supersonic flows compress the gas in shocks; in isothermal gas with
density $\rho$ the postshock gas has density $\rho' = {\cal M}^2
\rho$, where ${\cal M}$ is the Mach number of the shock.  The
turbulent Jeans length $\lambda_{\rm J} \propto \rho'^{-1/2}$ in these
density enhancements, so it drops by a factor of ${\cal M}$ in
isothermal shocks.

\begin{figure}[h]
\begin{center}
\unitlength1.0cm
\begin{picture}(8,5.3)
%\put(-4.0,-8.0){\epsfxsize=16cm \epsfbox{ms50734-fig04.ps}}
\put(-3.4,-10.0){\epsfxsize=17.5cm \epsfbox{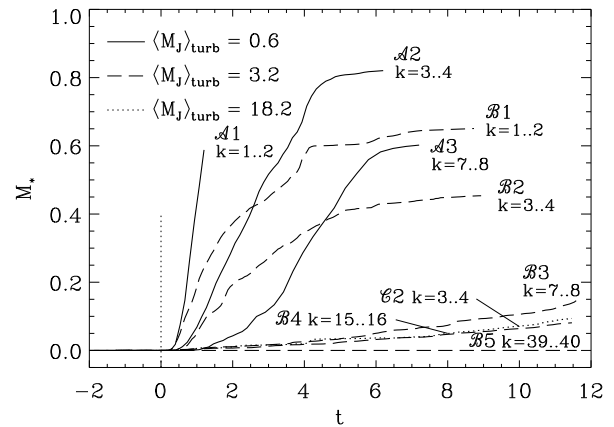}}
\end{picture}
\end{center}
\caption{\label{fig:accretion-history} Fraction $M_*$ of mass accreted
in dense cores as function of time for different models of
self-gravitating supersonic turbulence. The models differ by driving
strength and driving wavenumber, as indicated in the figure. The mass
in the box is initially unity, so the solid curves are formally
unsupported, while the others are formally supported.  The figure
shows how the efficiency of local collapse depends on the scale and
strength of turbulent driving.  Time is measured in units of the
global system free-fall time scale $\tau_{\rm ff}$. Only a model
driven strongly at scales smaller than the Jeans wavelength
$\lambda_J$ in shock-compressed regions shows no collapse at
all. (From Klessen \etal\ 2000.) }
\end{figure}

Klessen \etal\ (2000) demonstrated that turbulent support can
completely prevent collapse only when it can support not just the
average density, but also these high-density shocked regions, as shown
in Figure~\ref{fig:accretion-history}. (This basic point was appreciated
by Elmegreen [1993] and V\'azquez-Semadeni \etal\ [1995].)  Two
criteria must be fulfilled: the rms velocity must be sufficiently high
for the turbulent Jeans criterion to be met in these regions, and the
driving wavelength $\lambda_{\rm d} < \lambda_{\rm J}(\rho')$.  If these two
criteria are not fulfilled, the high-density regions collapse,
although the surrounding flow remains turbulently supported.  The
efficiency of collapse depends on the properties of the supporting
turbulence.  Sufficiently strong driving on short enough scales can
prevent local collapse for arbitrarily long periods of time, but such
strong driving may be rather difficult to arrange in a real molecular
cloud.  If we assume that stellar driving sources have an effective
wavelength close to their separation, then the condition that driving
acts on scales smaller then the Jeans wavelength in `typical' shock
generated gas clumps requires the presence of an extraordinarily large
number of stars evenly distributed throughout the cloud, with typical
separation 0.1 pc in Taurus, or only 350 AU in Orion.  This is not
observed. Very small driving scales seem to be at odds with the
observed large-scale velocity fields at least in some molecular clouds
(e.g.\ Ossenkopf \& Mac~Low 2002).

The first collapsed cores form in small groups randomly dispersed
throughout the volume.  Their velocities directly reflect the
turbulent velocity field of the gas from which they formed and
continue to accrete.  However, as more and more mass accumulates in
protostars, their mutual gravitational interaction becomes
increasingly important, beginning to determine the dynamical state of
the system. It behaves more and more like a collisional $N$-body
system, where close encounters occur frequently (see
\S~\ref{sub:clusters}).

\begin{figure}[ht]
\begin{center}
\unitlength1.0cm
\begin{picture}(10.0,15.0)
%\put(-1.0,-1.0){\epsfxsize=13cm \epsfbox{HMK2001-fig03-magstatsup.ps}}
\put(-2.0,-2.5){\epsfxsize=14.5cm \epsfbox{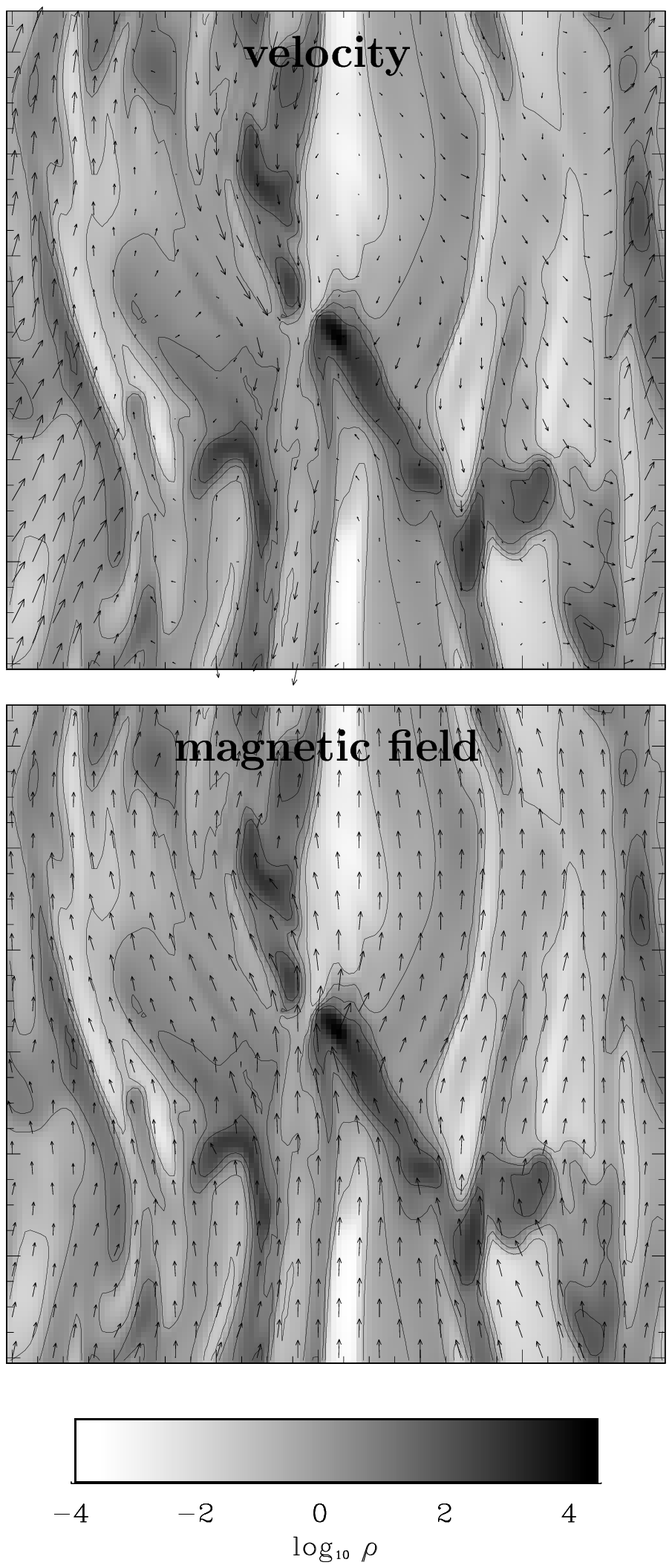}}
\end{picture}
\end{center}
\caption{\label{fig:magstatsup}
%fig03-magstatsup.ps
  Two dimensional slice through a cube of magnetostatically supported
  large-scale driven, self-gravitating turbulence from Heitsch \etal\ (2001a).
  The upper panel shows the velocity field vectors and the lower panel the
  magnetic field vectors.  The initial magnetic field is along the
  $z$-direction, i.e.\ vertically oriented in all plots presented.  The field
  is strong enough in this case not only to prevent the cloud from collapsing
  perpendicular to the field lines, but even suppress the turbulent motions in
  the cloud. The turbulence only scarcely affects the mean field. The picture
  is taken at $t=5.5t_{\rm ff}$. (From Heitsch \etal\ 2001a.)}
\end{figure}

\rsksubsubsection{Effects of Magnetic Fields}
\label{subsub:MHD}

So far, we have concentrated on the effects of purely hydrodynamic
turbulence. How does the picture discussed here change if we consider
the presence of magnetic fields?  Magnetic fields have been suggested
to support molecular clouds well enough to prevent gravitationally
unstable regions from collapsing (McKee 1999), either magnetostatically
or dynamically through MHD waves.

Assuming ideal MHD, a self-gravitating cloud of mass $M$ permeated by
a uniform flux $\Phi$ is stable unless the mass-to-flux ratio exceeds
the value given by equation~(\ref{eqn:crit-phi}).  Without any other
mechanism of support, such as turbulence acting along the field lines,
a magnetostatically supported cloud collapses to a sheet which is then
supported against further collapse. Fiege \& Pudritz (1999) found an
equilibrium configuration of helical field that could support a
filament, rather than a sheet, from fragmenting and collapsing, but
realizing this configuration in highly turbulent molecular
clouds appears difficult.

\begin{figure*}[ht]
\unitlength1.0cm
\begin{picture}(16,11.3)
%\put( 0.0, -6.5){\epsfxsize=15cm \epsfbox{HMK2001-fig05-2Dslices.ps}}
%\put( 0.4, -6.6){\epsfxsize=15cm \epsfbox{mk03_fig13.ps}}
\put( 1.4, -0.3){\epsfxsize=14.5cm \epsfbox{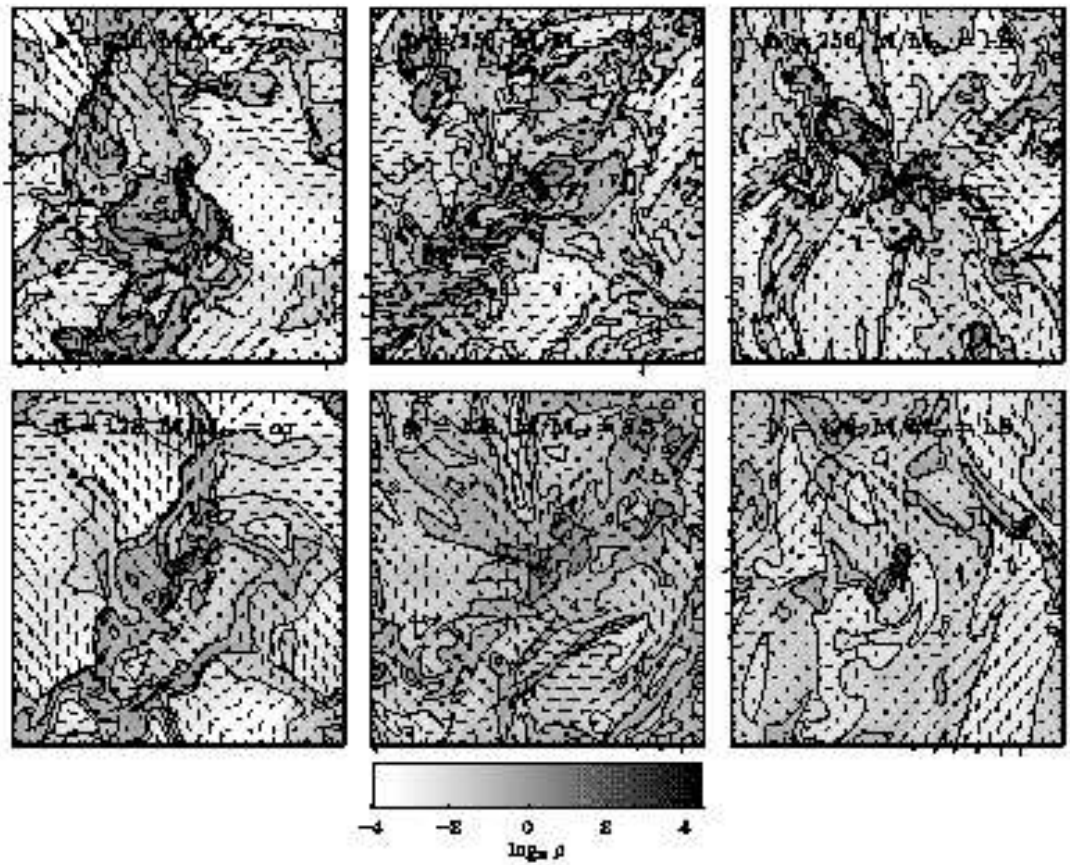}}
\end{picture}
\caption{\label{fig:2Dslices}Two-dimensional slices of $256^3$ models
from Heitsch \etal\ (2001a) driven at large scales with wavenumbers
$k=1-2$ hard enough that the mass in the box represents only 1/15
$\langle M_{\rm J}\rangle_{\rm turb}$, and with initially vertical
magnetic fields strong enough to give critical mass fractions as
shown.  The slices are taken at the location of the zone with the
highest density at the time when $10$\% of the total mass has been
accreted onto dense cores. The plot is centered on this zone.  Arrows
denote velocities in the plane. The length of the largest arrows
corresponds to a velocity of $v \sim 20 c_s$. The density greyscale is
given in the colorbar. As fields become stronger, they influence the
flow more, producing anisotropic structure. (From Heitsch \etal\ 2001b.)
%fig05-2Dslices.ps
%boundig box 45 420 595 620
}
\end{figure*}
Investigation of support by MHD waves concentrates mostly on the
effect of Alfv\'{e}n waves, as they (1) are not as subject to damping
as magnetosonic waves and (2) can exert a force {\em along} the mean
field, as shown by Dewar (1970) and Shu \etal\ (1987). This is because
Alfv\'{e}n waves are {\em transverse} waves, so they cause
perturbations $\delta \vec{B}$ perpendicular to the mean magnetic
field $\vec{B}$. McKee \& Zweibel (1994) argue that Alfv\'{e}n waves
can even lead to an isotropic pressure, assuming that the waves are
neither damped nor driven. However, in order to support a region
against self-gravity, the waves would have to propagate outwardly,
because inwardly propagating waves would only further compress the
cloud. Thus, this mechanism requires a negative radial gradient in
wave sources in the cloud (Shu \etal\ 1987).

It can be demonstrated (e.g.\ Heitsch \etal\ 2001a) that supersonic
turbulence does not cause a magnetostatically supported region to
collapse, and vice versa, that in the absence of magnetostatic
support, MHD waves cannot completely prevent collapse, although they
can retard it to some degree.  The case of a subcritical region with
$M < M_{cr}$ is illustrated in Figure \ref{fig:magstatsup}.
%, where an initially uniform magnetic field runs parallel to the
%$z$-axis.  The cloud is expected to collapse to a sheet, which in
%turn should be stable.  
Indeed, sheets form, though 
% the field is strong enough to force significant
%anisotropy in the flow, although the dense sheets that form do not
always perpendicular to the field lines.  This is because the
turbulent driving can shift the sheets along the field lines without
changing the mass-to-flux ratio. The sheets do not collapse further,
because the shock waves cannot sweep gas across field lines and the
entire region is initially supported magnetostatically.

\begin{figure}[ht]
\begin{center}
\unitlength1.0cm
\begin{picture}(8.0, 7.0)
%\put( 0.0, 0.0){\epsfxsize=8cm \epsfbox{HMK2001-fig07-variance.ps}}
\put(00.00,-0.8){\epsfxsize=7.6cm \epsfbox{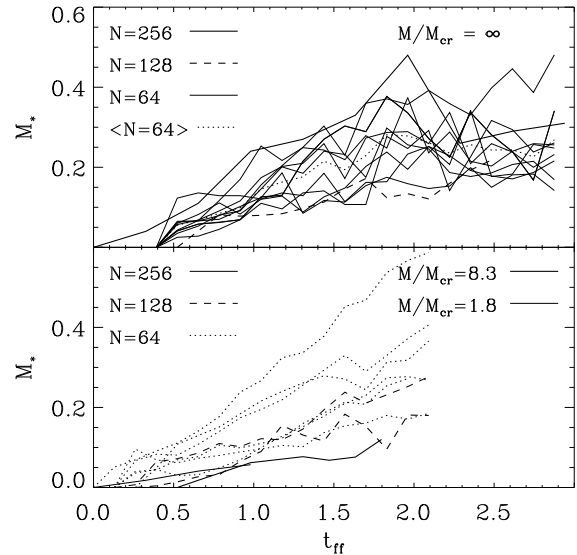}}
\end{picture}
\end{center}
\caption{\label{fig:variance}
%fig07-variance.ps
  {\em Upper panel:} Core-mass accretion rates for $10$ different
  low-resolution models ($N=64^3$ cells) of purely hydrodynamic
  turbulence with equal parameter set but different realizations of
  the turbulent velocity field. The thick line shows a ``mean
  accretion rate'', calculated from averaging over the sample. For
  comparison, higher-resolution runs with identical parameters but
  $N=128^3$ and $N=256^3$ are shown as well.  The latter one can be
  regarded as an envelope for the low resolution models.  {\em Lower
    panel:} Mass accretion rates for various models with different
  magnetic field strength and resolution.  Common to all models is the
  occurrence of local collapse and star formation regardless of the
  detailed choice of parameters, as long as the system is
  magnetostatically supercritically.  (For further details see Heitsch
  \etal\ 2001a.)}
\end{figure}

A supercritical cloud with $M > M_{cr}$
% is not magnetostatically supported, and 
could only be stabilized by MHD wave pressure.
%, assuming ideal MHD.
%However, as is the case for purely hydrodynamic turbulence, 
This is insufficient to completely prevent gravitational collapse, as
shown
%  Concerning the morphology of the cloud, the effects of magnetic
%  fields are weak as long at the cloud remains supercritical. This is
%  illustrated  
in Figure \ref{fig:2Dslices}.
%, which compares the morphology of
%hydrodynamical (left), weakly magnetized (middle), and strongly
%magnetized (right) supercritical models driven at large scales at a
%resolution of $256^3$ zones. The figure presents two dimensional
%slices through the three dimensional simulation volume centered on the
%locations of the most massive clumps. In order to compare the models
%at similar stages of their evolution, the snapshots are taken at a
%time when roughly 10\% of the total mass has been accreted onto
%collapsing clumps.  All three runs show well developed turbulence,
%rarefied regions, shocked regions, and at least one clump.  However,
%the weakly magnetized model, with $M/M_{cr} = 8.3$, seems to contain
%more power on small scales than the pure hydro run (with $M/M_{cr} =
%\infty$).  In the strongly magnetized model ($M/M_{cr}=1.8$), the
%vertically oriented mean field (in the plane of the Figure) starts
%producing some anisotropy.  This model represents a morphological
%transition from the pure hydrodynamical model with completely randomly
%oriented motions to the magnetostatically supported model presented in
%Figure \ref{fig:magstatsup}, with its ordered strucures.
Collapse occurs in all models of unmagnetized and magnetized
turbulence regardless of the numerical resolution and magnetic field
strength as long as the system is magnetically supercritical. This is
shown quantitatively in Figure~\ref{fig:variance}. Increasing the
resolution makes itself felt in different ways in hydrodynamical and
MHD models.  In the hydrodynamical case, higher resolution results in
thinner shocks and thus higher peak densities.  These higher density
peaks form cores with deeper potential wells that accrete more mass
and are more stable against disruption.  Higher resolution in the MHD
models, on the other hand, better resolves short-wavelength MHD waves,
which apparently can delay collapse, but not prevent it.  This result
extends to models with $512^3$ zones (Heitsch \etal\ 2001b, Li \etal\
2001).

The delay of local collapse seen in our magnetized simulations is
caused mainly by weakly magnetized turbulence acting to decrease
density enhancements due to shock interactions. Although a simple
additional pressure term will model this effect for small field
strength, this approximation cannot be used to follow the subsequent
collapse of the cores, as done by Boss (2000, 2002), as it entirely
neglects the effects of magnetic tension on angular momentum.  As a
result, magnetic braking of rotating cores (\S~\ref{sub:standard}) is
neglected, allowing binary formation to proceed where it would
otherwise not occur.

\rsksubsubsection{Promotion and Prevention of Local Collapse}

Highly compressible turbulence both promotes and prevents collapse.
Its net effect is to inhibit collapse globally, while perhaps
promoting it locally.  This can be seen by examining the dependence of
the Jeans mass  $M_{\rm J} \propto \rho^{-1/2} c_{\rm s}^3$, Equation
(\ref{eqn:jeans-mass}), 
on the rms turbulent velocity $v_{\rm rms}$.  If we follow the
classical picture that treats turbulence as an additional pressure,
then we define $c_{{\rm s},e\!\!\!\;f\!\!f}^2 = c_{\rm s}^2 + v_{\rm rms}^2/3$.
However, compressible turbulence in an isothermal medium causes local
density enhancements that increase the density by ${\cal M}^2 
\propto v_{\rm rms}^2$.  Combining these two effects, we find that 
\begin{equation}
M_{\rm J} \propto v_{\rm rms}^2
\end{equation}
for $v_{\rm rms} \gg c_{\rm s}$, so that ultimately turbulence does inhibit
collapse.  However, there is a broad intermediate region, especially
for long wavelength driving, where local collapse occurs despite
global support.

\begin{figure*}[th]
\unitlength1.0cm
\vspace{0.2in}
\begin{picture}(16,7.6)
%\put(-2.0,-5.0){\epsfxsize=14.5cm \epsfbox{ms50734-fig03.ps}}
%\put(-0.5,-5.8){\epsfxsize=14.5cm \epsfbox{mk03_fig15.ps}}
\put(2.5,-0.1){\epsfxsize=10.5cm \epsfbox{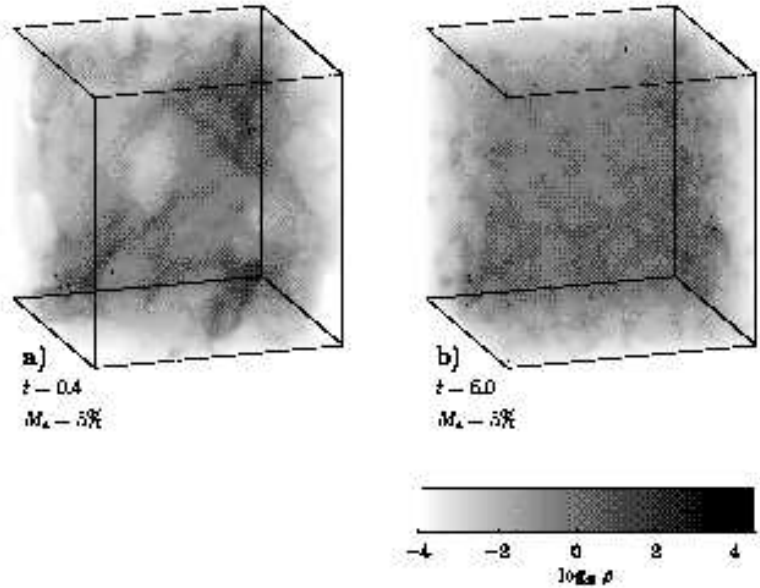}}
\end{picture}
\caption{\label{fig:3D-cubes-1..2+7..8} Density cubes for (a) a model
of large-scale driven turbulence (${\cal B}1h$) and (b) a model of
small-scale driven turbulence (${\cal B}3$) at dynamical stages where
the core mass fraction is $M_* = 5$\%. Compare with
Figure~\ref{fig:3D-cubes}b. Together they show the influence of
different driving wavelengths for otherwise identical physical
parameters. Larger-scale driving results in collapsed cores in more
organized structure, while smaller-scale driving results in more
randomly distributed cores. Note the different times at which $M_*
= 5\%$ is reached. (From Klessen \etal\ 2000.)}
\end{figure*}
The total mass and lifetime of a fluctuation determine whether it will
actually collapse.  Roughly speaking, the lifetime of a clump is
determined by the interval between two successive passing shocks: 
the first creates it, while the second one, if strong enough, may disrupt
the clump again   (Klein, McKee \&
Colella 1994, Mac Low \etal\ 1994).  If the timeinterval between two
shocks is sufficiently long, however,  a
Jeans unstable clump can contract to  high densities to
effectively decouple from the ambient gas flow and  becomes able
to survive the encounter with further shock fronts (e.g.\ Krebs \&
Hillebrandt 1983). Then it continues to accrete from the surrounding gas,
forming a dense core.  The weaker the passing shocks, and the greater
the separation between them, the more likely that collapse will occur.
Therefore, weak driving and long driving wavelengths enhance
collapse.  The influence of the driving wavelength is enhanced
because individual shocks sweep up more mass when the typical wavelength is longer, so density enhancements resulting from the
interaction of shocked layers have larger masses, and so are more
likely to exceed their local Jeans limit.  Turbulent driving
mechanisms that act on large scales produce large coherent structures
(filaments of compressed gas with embedded dense cores) on relatively
short timescales compared to small-scale driving even if the total
kinetic energy in the system is the same
%We demonstrate the effect of the driving wavelength in
(Figure~\ref{fig:3D-cubes-1..2+7..8},
% which compares a model with
%large-scale driving (${\cal B}1h$) to a  model with small-scale
%driving (${\cal B}3$) at a time when 5\% of the gas mass is accreted
%onto protostellar cores. These density cubes 
can be directly compared
with Figure~\ref{fig:3D-cubes}b). 
%, which shows the intermediate-scale
%turbulence at the same evolutionary stage.)  Note the difference in
%the morphology of the density structures.
%Figure~\ref{fig:3D-cubes-1..2+7..8}a is dominated by {\em one} large
%shock front that traverses the volume, which is the sole site of core
%formation. On the other hand, the density structure in model ${\cal
%  B}3$ (figure \ref{fig:3D-cubes-1..2+7..8}b) is far more homogeneous,
%without any large-scale structure.  Cores form alone, randomly
%dispersed throughout the volume.

A more detailed understanding of how local collapse proceeds comes
from examining the full time history of accretion for different models
(Figure \ref{fig:accretion-history}). 
% shows the accretion history for
%three sets of SPH models.  For each set of models, the driving
%strength is held constant while the effective driving wavelength is
%varied, showing the pronounced effect of the wavelength at equal
%driving strength.  At the extreme, if the driving is at wavelengths
%below the Jeans wavelength of the shocked layers local collapse does
%not occur (model ${\cal B}5$).  The ${\cal A}$ models have lower
%driving strength than the ${\cal B}$ and ${\cal C}$ models,
%demonstrating the effect of driving strength at each driving wave
%length (full details of these models are given in Table 1 of Klessen
%\etal\ 2000).
%
The cessation of strong accretion onto cores occurs long before all
gas has been accreted, with the mass fraction at which this occurs
depending on the properties of the turbulence.  This is because the
time that dense cores spend in shock-compressed, high-density regions
decreases with increasing driving wave number and increasing driving
strength.  In the case of long wavelength driving, cores form
coherently in high-density regions associated with one or two large
shock fronts that can accumulate a considerable fraction of the total
mass of the system, while
%The overall accretion rate is high and cores spend
%suffient time in this environment to accrete a large fraction of the
%total mass in the region.  Any further mass growth has to occur from
%chance encounters with other dense regions.  
in the case of short wavelength driving, the network of shocks is tightly knit, and cores form in
smaller clumps and remain in them for shorter times. 
%hock generated clumps of small masses because individual shocks are
%not able to sweep up much matter. Furthermore, in this rapidly
%changing environment the time interval between the formation of clumps
%and their destruction is short.  The period during which individual
%cores are located in high-density regions where they are able to
%accrete at high rate is short as well.  Global
%accretion rates are small and begin to saturate at lower values of
%$M_*$ as the driving wavelength is decreased.

\begin{figure*}[th]
\unitlength1.0cm
\begin{picture}(16,10.5)
%\put( 0.0,-3.40){\epsfxsize=14cm \epsfbox{ms50734-fig13.ps}}
%\put(-1.0,-4.40){\epsfxsize=17.0cm \epsfbox{mk03_fig16.ps}}
\put(1.5,-0.40){\epsfxsize=13.0cm \epsfbox{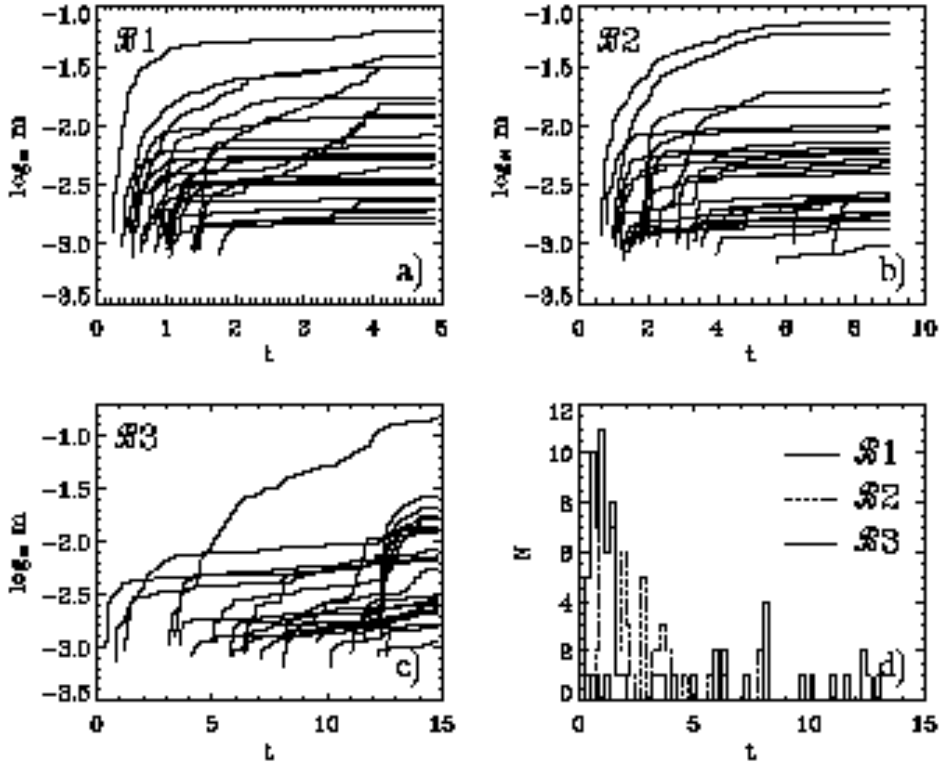}}
\end{picture}
\caption{\label{fig:core-formation-histogram} 
  Masses of individual protostars as function of time in SPH models
  (a) ${\cal B}1$ driven at large scales with $k=1-2$ driving, (b)
  ${\cal B}2$ with $k=3-4$ driving, i.e.\ at intermediate scales, and
  (c) ${\cal B}3$ with $k=7-8$ small-scale driving.  The curves
  represent the formation and accretion histories of individual
  protostars. For the sake of clarity, only every other core is shown
  in (a) and (b), whereas in (c) the evolution of every single core is
  plotted.  Time is given in units of the global free-fall time
  $\tau_{\rm ff}$. Note the different time scale in each plot. In the
  depicted time interval models ${\cal B}1$ and ${\cal B}2$ reach a
  core mass fraction $M_* =70$\%, and both form roughly 50 cores.
  Model ${\cal B}3$ reaches $M_* =35$\% and forms only 25 cores.
  Figure (d) compares the distributions of formation times. The age
  spread increases with decreasing driving scale showing that
  clustered core formation should lead to a coeval stellar population,
  whereas a distributed stellar population should exhibit considerable
  age spread. (From Klessen \etal\ 2000.)  }
\end{figure*}

\rsksubsubsection{The Timescales of Star Formation}
Turbulent control of star formation predicts that stellar clusters
form predominantly in regions that are insufficiently supported by
turbulence or where only large-scale driving is active.  In the
absence of driving, molecular cloud turbulence decays more quickly
than the free-fall timescale $\tau_{\rm ff}$ (Equation~\ref{eqn:decay}),
so dense stellar clusters will form on the free-fall time scale. 
% is thus the
%typical timescale on which dense stellar clusters will form in the
%absence of support or in the presence of decaying turbulence.  
Even in the presence of support from large-scale driving, substantial
collapse still occurs within a few free-fall times 
(Figures~\ref{fig:accretion-history}
and~\ref{fig:core-formation-histogram}a).  If the dense cores followed
in these models continue to collapse on a short timescale to build up
stellar objects in their centers, then this directly implies the star
formation time.  Therefore the age distribution will be roughly
$\tau_{\rm ff}$ for stellar clusters that form coherently with high
star formation efficiency.  When scaled to low densities, say $n({\rm
H}_2) \approx 10^2\,{\rm cm}^{-3}$ and $T\approx10\,$K, the global
free-fall timescale in the models is $\tau_{\rm ff} = 3.3 \times
10^6\,$years.  If star forming clouds such as Taurus indeed have ages
of order $\tau_{\rm ff}$, as suggested by Ballesteros-Paredes \etal\
(1999), then the long star formation timescale computed here is quite
consistent with the very low star formation efficiencies seen in
Taurus (e.g.\ Leisawitz \etal\ 1989, Palla \& Stahler 2000, Hartmann
2001), as the cloud simply has not had time to form many stars.  In
the case of high-density regions, $n({\rm H}_2) \approx 10^5\,{\rm
cm}^{-3}$ and $T\approx10\,$K, the dynamical evolution proceeds much
faster and the corresponding free-fall timescale drops to $\tau_{\rm
ff} = 10^5\,$years.  These values are indeed supported by
observational data such as the formation time  of the Orion
Trapezium cluster.  It is inferred to stem from gas of density $n({\rm
H}_2) \sil 10^5\,{\rm cm}^{-3}$, and is estimated to be less than
$10^6$ years old (Hillenbrand \& Hartmann 1998).  The age spread in
the models increases with increasing driving wave number $k$ and
increasing $\langle M_{\rm J} \rangle _{\rm turb}$, as shown in
Figure~\ref{fig:core-formation-histogram}. Long periods of core
formation for globally supported clouds appear consistent with the low
efficiencies of star-formation in regions of isolated star formation,
such as Taurus, even if they are rather young objects with ages of
order $\tau_{\rm ff}$.

\rsksubsubsection{Scales of Interstellar Turbulence}
\label{subsub:scales}

Turbulence only has self-similar properties on scales between the
driving and dissipation scales.  What are these scales for
interstellar turbulence?
% The upper end typically is associated with the global extent
%of the system considered or with the scale at which energy is
%inserted. The lower scale is the energy dissipation scale of the
%system where turbulent kinetic energy is converted into random motion,
%into heat. 
In a purely hydrodynamic system the dissipation scale is the scale
where molecular viscosity becomes important.  In interstellar clouds
the situation may be different.  Zweibel \& Josafatsson (1983) showed
that ambipolar diffusion (ion-neutral drift) is the most important
dissipation mechanism in typical molecular clouds with very low
ionization fractions $x = \rho_i/\rho_n$, where $\rho_i$ is the
density of ions, $\rho_n$ is the density of neutrals, and the total
density $\rho = \rho_i + \rho_n$.  An ambipolar diffusion strength can
be defined as
\begin{equation}
\lambda_{\rm AD} = v_{\rm A}^2 / \nu_{ni},
\end{equation}
where $v_{\rm A}^2 = B^2/4\pi\rho_n$ approximates the effective Alfv\'en
speed for the coupled neutrals and ions if $\rho_n \gg \rho_i$, and
$\nu_{ni} = \gamma \rho_i$ is the rate at which each neutral is hit by
ions.  The coupling constant depends on the cross-section for
ion-neutral interaction, and for typical molecular cloud conditions
has a value of $\gamma \approx 9.2 \times
10^{13}$~cm$^3\,$s$^{-1}$g$^{-1}$ (e.g.\ Smith \& Mac Low 1997).
Zweibel \& Brandenburg (1997) define an ambipolar diffusion Reynolds
number as
 \begin{equation}
 R_{\rm AD} = \tilde{L}\tilde{V} / \lambda_{\rm AD} = {\cal M}_{\rm A} \tilde{L} \nu_{ni}/v_{\rm A},
 \end{equation}
which must fall below unity for ambipolar diffusion to be important
(also see Balsara 1996), where $\tilde{L}$ and $\tilde{V}$ are the
characteristic length and velocity scales, and ${\cal M}_{\rm A} =
\tilde{V}/v_{\rm A}$ is the characteristic Alfv\'en Mach number. In
our situation we again can take the rms velocity as typical value for
$\tilde{V}$.  By setting $R_{\rm AD} = 1$, we can derive a critical
lengthscale below which ambipolar diffusion is important
% \begin{equation} \label{lcrit1}
% \tilde{L}_{\rm cr} = \frac{v_{\rm A}}{{\cal M}_{\rm A} \nu_{ni}} 
%          \approx (0.041 \mbox{ pc})\left(\frac{B}{10\,\mu{\rm G}}\right) {\cal M_{\rm
%          A}}^{-1} \left(\frac{x}{10^{-6}}\right)^{-1}
%          \left(\frac{n_n}{10^3\,{\rm cm}^{-3}}\right)^{-3/2},
% \end{equation}
\begin{eqnarray} \label{lcrit1}
\lefteqn{\!\!\!\!\!\!\!\!\!\!\!\!\!\!\!\tilde{L}_{\rm cr} = \frac{v_{\rm A}}{{\cal M}_{\rm A} \nu_{ni}} 
         \approx (0.041 \mbox{ pc})\left(\frac{B}{10\,\mu{\rm G}}\right) {\cal M_{\rm
         A}}^{-1}}\nonumber \\&& \!\!\!\!\!\!\!\!\!\!\!\times\left(\frac{x}{10^{-6}}\right)^{-1}
         \left(\frac{n_n}{10^3\,{\rm cm}^{-3}}\right)^{-3/2}\!\!\!,
\end{eqnarray}
with the magnetic field strength $B$, the ionization fraction $x$, the
neutral number density $n_n$, and where we have taken $\rho_n = \mu
n_n$, with $\mu = 2.36\,m_{\rm H}$.  This is consistent with typical
sizes of protostellar cores (e.g.\ Bacmann \etal\ 2000), if we assume
that ionization and magnetic field both depend on the density of the
region and follow the empirical laws $n_i = 3 \times 10^{-3}\,{\rm
cm}^{-3}\,(n_n / 10^5\,{\rm cm}^{-3})^{1/2}$ (e.g.\ Mouschovias 1991b)
and $B \approx 30\,\mu{\rm G}\, (n_n/10^3\,{\rm cm}^{-3})^{1/2}$
(e.g.\ Crutcher 1999).  Balsara (1996) notes that there are wave
families that can survive below $L_{\rm cr}$ that resemble hydrodynamical
sound waves.  This means that this scale may determine where the
magnetic field becomes uniform, but not necessarily where the
hydrodynamical turbulent cascade cuts off.

On large scales, a maximum upper limit to the turbulent cascade in the
Milky Way is given by the extent and thickness of the Galactic
disk. If indeed molecular clouds are created at least in part by
converging large-scale flows generated by the collective influence of
recurring supernovae explosions in the gaseous disk of our Galaxy, as
we argue in \S~\ref{sub:driving}, then the extent of the Galactic disk
is indeed the true upper scale of turbulence in the Milky Way. For
individual molecular clouds this means that turbulent energy is fed in
at scales well above the size of the cloud itself.

The initial compression that assembles the cloud may generate the bulk
of a clouds turbulent energy content (see \S~\ref{sub:clouds}).  If
the surrounding flow is not strong enough to continue to drive the
cloud, the turbulence will quickly dissipate, resulting in collapse
and active star formation.  If the same compressional motions that
created the cloud in the first place can also act as a continuing
source of kinetic energy, they may be strong enough to again destroy
the cloud after several crossing times (e.g.\ Hartmann \etal\ 2001).
Either way, the result is short cloud life times, as argued by
Ballesteros-Paredes \etal\ (1999) and Elmegreen (2000).  This picture
of molecular cloud turbulence being driven by large-scale, external
sources is supported by the observation that density and velocity
structure shows power-law scaling extending up to the largest scales
observed in all clouds that have been analyzed (Ossenkopf \& Mac Low
2002).

\rsksubsubsection{Efficiency of Star Formation}

The {\em global} star formation efficiency in normal molecular clouds
is usually estimated to be of the order of a few per cent. Their life
times may be on the order of a few crossing times, i.e.\ a few $10^6$
years (e.g.\ Ballesteros-Paredes \etal\ 1999, Fukui \etal\ 1999, Elmegreen 2000).  In
this case nearly all models of interstellar turbulence discussed below
are consistent with the observed overall efficiencies. If molecular
clouds survive for several tens of their free-fall time $\tau_{\rm
ff}$ (i.e.\ a few $10^7$ years as proposed by Blitz \& Shu 1980),
turbulence models are more strongly constrained. However, even in this
case models with parameters reasonable for Galactic molecular clouds
can maintain global efficiencies below $M_* = 5$\% for $~10\,\tau_{\rm
ff}$ (Klessen \etal\ 2000).  Furthermore, it needs to be noted that
the {\em local} star formation efficiency in molecular clouds can
reach very high values. For example, the Trapezium star cluster in
Orion is likely to have formed with an efficiency of about 50\%
(Hillenbrand \& Hartmann 1998).
%
%compared to a value of 5\% proposed for Taurus-Aurigae ({\bf is that
%  true}).

\rsksubsubsection{Termination of Local Star Formation}
\label{subsub:termination}
It remains quite unclear what terminates stellar birth on scales of
individual star forming regions, and even whether these processes are
the primary factor determining the overall efficiency of star
formation in a molecular cloud.  Three main possibilities exist.
First, feedback from the stars themselves in the form of ionizing
radiation and stellar outflows may heat and stir surrounding gas up
sufficiently to prevent further collapse and accretion.  Second,
accretion might peter out either when all the high density,
gravitationally unstable gas in the region has been accreted in
individual stars, or after a more dynamical period of competitive
accretion, leaving any remaining gas to be dispersed by the background
turbulent flow.  Third, background flows may sweep through, destroying
the cloud, perhaps in the same way that it was created. Most likely
the astrophysical truth lies in some combination of all three
possibilities. 

If a stellar cluster formed in a molecular cloud contains OB stars,
then the radiation field and stellar wind from these high-mass stars
strongly influence the surrounding cloud material. The UV flux ionizes
gas out beyond the local star forming region. Ionization heats the
gas, raising its Jeans mass, and possibly preventing further
protostellar mass growth or new star formation.  The termination of
accretion by stellar feedback has been suggested at least since the
calculations of ionization by Oort \& Spitzer (1955). Whitworth (1979)
and Yorke \etal\ (1989) computed the destructive effects of individual
blister H{\sc ii} regions on molecular clouds, while in series of
papers, Franco \etal\ (1994), Rodriguez-Gaspar \etal\ (1995), and
Diaz-Miller \etal\ (1998) concluded that indeed the ionization from
massive stars may limit the overall star forming capacity of molecular
clouds to about 5\%.  Matzner (2002) analytically modeled the effects
of ionization on molecular clouds, concluding as well that turbulence
driven by H{\sc ii} regions could support and eventually destroy
molecular clouds.  The key question facing these models is whether
H{\sc ii} region expansion couples efficiently to clumpy,
inhomogeneous molecular clouds, a question probably best addressed
with numerical simulations.

Bipolar outflows are a different manifestation of protostellar
feedback, and may also strongly modify the properties of star forming
regions (Norman \& Silk 1980, Lada \& Gautier 1982, Adams \& Fatuzzo
1996).  Recently Matzner \& McKee (2000) modeled the ability of
bipolar outflows to terminate low-mass star formation, finding that
they can limit star formation efficiencies to 30--50\%, although they
are ineffective in more massive regions.  How important these
processes are compared to simple exhaustion of available reservoirs of
dense gas (Klessen \etal\ 2000, V{\'a}zquez-Semadeni \etal\ 2003)
remains an important question.

The models relying on exhaustion of the reservoir of dense gas argue
that only dense gas will actually collapse, and that only a small
fraction of the total available gas reaches sufficiently high
densities, due to cooling (Schaye 2002), gravitational collapse and
turbulent triggering (Elmegreen 2002), or both (Wada, Meurer, \&
Norman 2002). This of course pushes the question of local star
formation efficiency up to larger scales, which may indeed be the
correct place to ask it.

Other models focus on competitive accretion in local star formation,
showing that the distribution of masses in a single group or cluster
can be well explained by assuming that star formation is fairly
efficient in the dense core, but that stars that randomly start out
slightly heavier tend to fall towards the center of the core and
accrete disproportionately more gas (Bonnell \etal\ 1997; 2001a).
These models have recently been called into question by the
observation that the stars in lower density young groups in Serpens
simply have not had the time to engage in competitive accretion, but
still have a normal IMF (Olmi \& Testi 2002).

Finally, star formation in dense clouds created by turbulent flows may
be terminated by the same flows that created them.  Ballesteros-Paredes
\etal\ (1999) suggested that the coordination of star formation over
large molecular clouds, and the lack of post-T Tauri stars with ages
greater than about 10$\,$Myr tightly associated with those clouds, could
be explained by their formation in a larger-scale turbulent flow.
Hartmann \etal\ (2001) make the detailed argument that these flows may
disrupt the clouds after a relatively short time, limiting their star
formation efficiency that way.  Below, in \S~\ref{sub:driving} we will
argue that field supernovae are the most likely driver for this
background turbulence, at least in the star-forming regions of galaxies.

\rsksubsection{Outline of a New Theory of Star Formation}
\label{sub:new}
%\input{new.tex}
%%% 
%%% RMP-Module 
%%% 
%%% 
%%%\rsksubsection{Outline of A New Theory of Star Formation} 
%%%\label{sub:newx} 
%%% 

The support of star-forming clouds by supersonic turbulence can explain many
of the same observations successfully explained by the standard theory, while
also addressing the inconsistencies between observation and the standard
theory described in the previous section.  The key point that is new in our
argument is that supersonic turbulence produces strong density fluctuations in
the interstellar gas (Padoan \& Nordlund 1999), sweeping gas up from large
regions into dense sheets and filaments (V\'azquez-Semadeni, Passot, \&
Pouquet 1996, Klessen \etal\ 2000), even in the presence of magnetic fields
(Passot, V\'azquez-Semadeni, \& Pouquet 1995, Heitsch \etal\ 2001a,b). Supersonic
turbulence decays quickly (Mac Low \etal\ 1998, Stone \etal\ 1998, Mac Low
1999), but so long as it is maintained by input of energy from some driver
(\S~\ref{sub:driving}), it can support regions against gravitational collapse.

Such support comes at a cost, however.  The very turbulent flows that
support the region produce density enhancements in which the Jeans
mass determining gravitational collapse drops as $M_{\rm J} \propto
\rho^{-1/2}$, (Equation \ref{eqn:jeans-mass}), and the magnetic critical
mass above which magnetic fields can no longer support against that
collapse drops even faster, as $M_{\rm cr} \propto \rho^{-2}$ in Equation
(\ref{eqn:crit-rho}).  For local collapse to actually result in the
formation of stars, Jeans-unstable, shock-generated, density
fluctuations must collapse to sufficiently high densities on time
scales shorter than the typical time interval between two successive
shock passages.  Only then can they decouple from the ambient flow and
survive subsequent shock interactions.  The shorter the time between
shock passages, the less likely these fluctuations are to
survive. Hence, the timescale and efficiency of protostellar core
formation depend strongly on the wavelength and strength of the
driving source (\S~\ref{sub:beyond}), and the accretion histories of
individual protostars are strongly time varying
(\S~\ref{sub:accretion}).  Global support by supersonic turbulence
thus tends to produce local collapse and low rate star formation
(Klessen \etal\ 2000, Heitsch \etal\ 2001a,b), exactly as seen in
low-mass star formation regions characteristic of the disks of spiral
galaxies.  Conversely, lack of turbulent support results in regions
that collapse freely.  In hydrodynamic simulations (Wada \& Norman
1999, Klessen \& Burkert 2000), freely collapsing gas forms a web of
density enhancements in which star formation can proceed efficiently,
as seen in regions of massive star formation and starbursts.

The regulation of the star formation rate then occurs not just at the
scale of individual star-forming cores through ambipolar diffusion
balancing magnetostatic support, but rather at all scales (Elmegreen
2002), via the dynamical processes that determine whether regions of
gas become unstable to prompt gravitational collapse.  Efficient star
formation occurs in collapsing regions; apparent inefficiency occurs
when a region is turbulently supported and only small subregions get
compressed sufficiently to collapse.  The star formation rate is
determined by the balance between turbulent support and local density,
and is a continuous function of the strength of turbulent support for
any given region.   Fast and efficient star formation is the natural
behavior of gas lacking sufficient turbulent support for its local
density.

Regions that are gravitationally unstable in this picture collapse
quickly, on the free-fall time scale.  They never pass through a
quasi-equilibrium state as envisioned by the standard model.
Large-scale density enhancements such as molecular clouds could be
caused either by gravitational collapse, or by ram pressure from
turbulence (Ballesteros-Paredes \etal\ 1999).  If collapse does not
succeed, the same large-scale turbulence that formed molecular clouds
can destroy them again (\S~\ref{sub:clouds}).

\rsksection[PROPERTIES OF SUPERSONIC TURBULENCE]{SOME FURTHER PROPERTIES OF SUPERSONIC TURBULENCE}
\label{sec:properties}

%%% Local Variables: 
%%% mode: latex
%%% TeX-master: "habil"
%%% End: 

\newcommand{\pdf}{%
PDF%
}

In the current review we argue that it is the subtle interplay between
self-gravity and supersonic turbulence in interstellar gas clouds that
determines where, when, and with which overall efficiency stars will form. We
claim that stars form in shock-generated molecular cloud clumps where density
and mass exceed the threshold for gravitational collapse to set in.  In the
previous Section we therefore have concentrated on discussing the effects of
compressibility. However, turbulence is a highly complex physical phenomenon
with further statistical properties that are relevant for the star formation
process. We begin this Section by studying the transport properties of
supersonic turbulence which is important for understanding element mixing in
interstellar gas clouds and the distribution of stellar chemical abundances
(Section \ref{subsec:mixing}). We also investigate of the one-point
probability distribution functions (\pdf) of density and velocity in turbulent
compressible flows (Section \ref{subsec:pdf}). Then, we analyze the Fourier
spectrum of turbulent velocity fields (\ref{subsec:fourier}) and use the
$\Delta$-variance to quantify the statistical properties of the density
distribution on star-forming clouds (Section \ref{subsec:delta}).

%%%%%%%%%%%%%%%%%%%%%%%%%%%%%%%%%%%%%%%%%%%%%%%%%%%%%%%%%%%%%%%%%%%%%%%%%%%%%%%
% [KL03] Klessen & Lin, 2003, PRE
%%%%%%%%%%%%%%%%%%%%%%%%%%%%%%%%%%%%%%%%%%%%%%%%%%%%%%%%%%%%%%%%%%%%%%%%%%%%%%%

%\rsksubsection{Transport Properties of Supersonic Turbulence}
\rsksubsection{Transport Properties}
\label{subsec:mixing}

\rsksubsubsection{Introduction}
\label{subsubsec:introduction}
Laboratory and terrestrial gases and liquids are usually well
described by incompressible flows.  In
contrast, the dynamical behavior of typical astrophysical gases, are
characterized by poorly understood highly compressible supersonic
turbulent motion.  For example, the large
observed linewidths in large molecular clouds show direct evidence for
the presence of chaotically oriented velocity fields with magnitudes
in excess of the sound speed. This random motion carries enough
kinetic energy to counterbalance and sometimes overcompensate the
effects of self-gravity of these clouds (\S\ \ref{sub:regions}).
The intricate interplay between supersonic turbulence and self-gravity
determines the overall dynamical evolution of these clouds and their
observable features such as their density structure, the star formation rate
within them, and their lifetimes.  Thus, it is importance for the description
of many astrophysical systems to understand in detail the momentum and heat
transfer properties of compressible turbulent gases.

Some important clues on the nature and efficiency of mixing associated with
the clouds' supersonic turbulence can be constrained by the observed
metallicity distribution of the stars formed within them.  In the Pleiades
cluster, stars which emerged from the same molecular cloud have nearly
identical metal abundance (Wilden \etal\ 2002).  This astronomical context
therefore imposes a strong motivation for a general analysis of the transport
and mixing processes in compressible supersonically turbulent media.

Analytical and numerical studies of diffusion processes are typically
restricted to certain families of statistical processes, like random walk
(Metzler \& Klaafter 2000) or remapping models or certain Hamiltonian systems
(Isichenko 1992). The direct numerical modeling of turbulent physical flows
mostly concentrates on incompressible media (Domolevo \& Sainsaulieu 1997;
Moser \etal\ 1999; Ossia \& Lesieur 2001), but some studies have been extended
into the weakly compressible regime (Coleman \etal\ 1995; Huang \etal\ 1995;
Porter \etal\ 1992; Porter \etal\ 1999).  Although highly compressible
supersonic turbulent flows have been studied in several specific astrophysical
contexts (see V{\'a}zquez-Semadeni \etal\ 2000 for a review)\footnote{See also
  Ballesteros-Paredes \& Mac Low (2002), Ballesteros-Paredes \etal\ (1999b),
  Balsara \& Pouquet (1999), Balsara \etal\ (2001), Boldyrev \etal\ (2002a),
  Gomez \etal\ (2001), Heitsch \etal\ (2001a,b), Klessen \etal\ (2000),
  Klessen (2001a,b), Mac Low (1999), Mac Low \etal\ (1998, 2001), Ostriker
  \etal\ (1999, 2001), Padoan \& Nordlund (1999), Padoan \etal\ (2000), Passot
  \etal\ (1995), Passot \& V{\'a}zquez-Semadeni (1998), Porter \& Woodward
  (2000), Smith \etal\ (2000), Stone \etal\ (1998), Sytine \etal\ (2000), or
  V{\'a}zquez-Semadeni \etal\ (1995)}, the diffusion properties of such flows
have not been investigated in detail.

It is the goal of this Section to analyze transport phenomena in supersonic
compressible turbulent flows and to demonstrate that -- analogous to the
incompressible case -- a simple mixing length description can be found even
for strongly supersonic and highly compressible turbulence. We first briefly
recapitulate in Section \ref{subsubsec:formalism} the Taylor formalism for
describing the efficiency of turbulent diffusion in subsonic flows.  In \S\ 
\ref{subsubsec:num-meth} we describe the numerical method which we use to
integrate the Navier-Stokes equation. In \S\ \ref{subsubsec:transport} we
report the diffusion coefficient obtained in our numerical models, and in \S\ 
\ref{subsubsec:mixing} we introduce an extension of the well known mixing
length approach to diffusion into the supersonic compressible regime. Finally,
in \S\ \ref{subsubsec:summary} we summarize our results.

\rsksubsubsection{A Statistical Description of Turbulent Diffusion}
\label{subsubsec:formalism}

Transport properties in fluids and gases can be characterized by
studying the time evolution of the second central moment of some
representative fluid-elements' displacement in the medium,
\begin{equation}
\label{eqn:def-sigma}
\xi^2_{\vec{r}}(t-t') =   \langle [\vec{r}_i(t)-\vec{r}_i(t')]^2
\rangle_i \;,
\end{equation}
where the average $\langle \cdot \rangle_i$ is taken over an ensemble
of passively advected tracer particles $i$ (e.g. dye in a fluid, or
smoke in air) that are placed in the medium at a time $t'$ at
positions $\vec{r}_i(t')$; or where the average is taken over the
fluid molecules themselves (or equivalently, over sufficiently small
and distinguishable fluid elements). The dispersion in one spatial
direction, say along the $x$-coordinate, is $\xi^2_{x}(t-t') = \langle
[x_i(t)-x_i(t')]^2\rangle_i$. For isotropic turbulence it follows that
$\xi^2_{x} = \xi^2_{y} = \xi^2_{z} = 1/3 \;\xi^2_{\vec{r}}$.  For
fully-developed stationary turbulence, the initial time $t'$ can be
chosen at random and for simplicity is set to zero in what follows.

The quantity $\xi_{\vec{r}}(t)$ can be associated with the diffusion
coefficient $D$ as derived for the classical diffusion equation,
\begin{equation}
\label{eqn:diff}
\frac{\partial n}{\partial t} = D \vec{\nabla}^2 n\;,
\end{equation}
where $n(\vec{r}_i,t)$ is the probability distribution function (\pdf)
for finding a particle $i$ at position $\vec{r}_i(t)$ at time $t$ when
it initially was at a location $\vec{r}_i(0)$.  This holds if the
particle position is a random variable with a Gaussian distribution
(Batchelor 1949). In the classical sense, $n(\vec{r},t)$ may correspond to
the contaminant density in the medium. Equation (\ref{eqn:diff}) holds
for normal diffusion processes and for time scales larger than the
typical particles' correlation time scale $\tau$.

In general, however, the Lagrangian diffusion coefficient is time
dependent and can be defined as 
\begin{equation}
\label{eqn:D(t)-1}
D(t) =  \frac{d\xi^2_{\vec{r}}(t)}{dt}= 2 \langle
\vec{r}_i(t) \cdot \vec{v}_i(t) \rangle_i \;,
\end{equation}
where $\vec{v}_i(t)=d\vec{r}_i(t)/dt$ is the Lagrangian velocity of
the particle.  The diffusion coefficient along one spatial direction,
say along the $x$-coordinate, follows accordingly as $D_x =
d\xi^2_{x}(t)/dt = 2 \langle x_i(t) v_{x.i}(t) \rangle_i$. Equation
(\ref{eqn:D(t)-1}) holds for homogeneous turbulence with zero mean
velocity. From $\vec{r}_i(t) = \vec{r}_i(0) + \int_0^t \vec{v}_i(t')
dt'$ it follows that
\begin{eqnarray}
  \label{eqn:D(t)-2}
  D(t) &=& 2 \left \langle \left[ \vec{r}_i(0) + \int_0^t \vec{v}_i(t') dt'
  \right] \cdot \vec{v}_i(t)\right \rangle_i \nonumber\\
        &=& 2 \int_0^t \langle
\vec{v}_i(t')\cdot\vec{v}_i(t) \rangle_i dt'
\;.
\end{eqnarray}
% \begin{equation}
%   \label{eqn:D(t)-2}
%   D(t) = 2 \left \langle \left[ \vec{r}_i(0) + \int_0^t \vec{v}_i(t') dt'
%   \right] \cdot \vec{v}_i(t)\right \rangle_i
%        = 2 \int_0^t \langle
% \vec{v}_i(t')\cdot\vec{v}_i(t) \rangle_i dt'
% \;.
% \end{equation}
The above expression allows us to related $D(t)$ to the trace of the
Lagrangian velocity autocorrelation tensor ${\rm{tr}}\,{\cal C}(t-t') =
\langle \vec{v}_i(t') \cdot \vec{v}_i(t) \rangle_i$ as
\begin{equation}
  \label{eqn:correlation-tensor}
  D(t) = 2 \int_0^t {\rm{tr}}\,{\cal C}(t-t') dt' = 2 \int_0^t {\rm{tr}}\,{\cal C}(t') dt' \;,
\end{equation}
a result which was already derived by Taylor (1921). This formulation has the
advantage that it is fully general and that it allows us to study anomalous
diffusion processes. Note, that strictly speaking any transport process with
$\xi_{\vec{r}}(t)$ not growing linearly in time is called anomalous diffusion.
This is always the case for time intervals shorter than the correlation time
$\tau$, but sometimes anomalous diffusion can also occur for $t\gg\tau$.  If
$\xi_{\vec{r}}(t) \propto t^{\alpha}$ and if $\alpha < 1$ transport processes
are called {\em subdiffusive}, if $\alpha >1$ they are called {\em
  superdiffusive} (Lesieur 1997; Isichenko 1992; Castiglione \etal\ 1999;
Lillo \& Mantegna 2000).  Studying transport processes directly in terms of
the particle displacement, i.e.\ Equation (\ref{eqn:def-sigma}), is useful when
attempting to find simple approximations to the diffusion coefficient $D(t)$
for example in a mixing length approach.

\begin{table*}[th]
{\caption{Model properties}
\label{tab:models-3}
}
% \tiny
\scriptsize
%\footnotesize
\begin{center}
\setlength{\tabcolsep}{0.17cm}
\begin{tabular}[t]{ccccccccccccccc}\hline\hline
\footnotesize{(1)}  & \footnotesize{(2)}  & \footnotesize{(3)} &
\footnotesize{(4)}  & \footnotesize{(5)}  & \footnotesize{(6)} &
\footnotesize{(7)}  & \footnotesize{(8)}  & \footnotesize{(9)} &
\footnotesize{(10)} & \footnotesize{(11)} & \footnotesize{(12)}&
\footnotesize{(13)} \\
   {model}     & {$k$}     & {${\cal M}$}  & {$t_{\rm cross}$}   & {$\bar{\sigma}_x$} 
 & {$\bar{\sigma}_y$}   & {$\bar{\sigma}_z$}   & $D_x(\infty)$   & $D_y(\infty)$   & $D_z(\infty)$
 & $2\bar{\sigma}_x/k$  & $2\bar{\sigma}_y/k$  & $2\bar{\sigma}_z/k$\\
%                                 
%
%   { }              & {$\cal M$}     & {$k$}             & {$\tilde{v}$}     
% & {$\tilde{\ell}$} & {$\tau$}       & {references}      \\
% %                                 
%   {model}          & {Mach}         & {driving}         & {typical}       
% & {typical}        & {correlation}  & { }               \\
% %                                 
%   {identifier}     & {number$^a$}   & {wavenumber}      & {velocity}        
% & {lengthscale}    & {time}         & { }               \\
% %
\hline \hline
{\em 0}$\ell$ & $1..2$ &   0.6 &  35.3 & 0.030 & 0.028 & 0.027 & 0.027
& 0.021 & 0.019 & 0.030 -- 0.060 & 0.028 -- 0.057 & 0.027 -- 0.054 \\
{\em 0i} & $3..4$ &   0.5 &  39.1 & 0.026 & 0.026 & 0.025 & 0.010 &
0.010 & 0.009 & 0.013 -- 0.017 & 0.013 -- 0.017 & 0.013 -- 0.017 \\
{\em 0s} & $7..8$ &   0.4 &  46.2 & 0.021 & 0.022 & 0.022 & 0.005 &
0.005 & 0.005 & 0.005 -- 0.006 & 0.005 -- 0.006 & 0.005 -- 0.006 \\
\hline
{\em 1}$\ell$ & $1..2$ &   1.9 &  10.4 & 0.106 & 0.084 & 0.098 & 0.140
& 0.069 & 0.111 & 0.106 -- 0.213 & 0.084 -- 0.167 & 0.098 -- 0.196 \\ 
{\em 1i} & $3..4$ &   1.9 &  10.6 & 0.097 & 0.096 & 0.092 & 0.042 &
0.047 & 0.038 & 0.048 -- 0.065 & 0.048 -- 0.064 & 0.046 -- 0.061 \\     
{\em 1s} & $7..8$ &   1.7 &  11.5 & 0.086 & 0.089 & 0.087 & 0.025 &
0.026 & 0.024 & 0.021 -- 0.024 & 0.022 -- 0.025 & 0.022 -- 0.025 \\
\hline 
{\em 2}$\ell$ & $1..2$ &   3.1 &   6.5 & 0.173 & 0.129 & 0.158 & 0.223
& 0.103 & 0.169 & 0.173 -- 0.346 & 0.129 -- 0.257 & 0.158 -- 0.315 \\
{\em 2i} & $3..4$ &   3.1 &   6.4 & 0.167 & 0.155 & 0.151 & 0.084 &
0.071 & 0.063 & 0.083 -- 0.111 & 0.077 -- 0.103 & 0.075 -- 0.100 \\
{\em 2s} & $7..8$ &   3.2 &   6.3 & 0.154 & 0.163 & 0.157 & 0.044 &
0.054 & 0.047 & 0.038 -- 0.044 & 0.041 -- 0.046 & 0.039 -- 0.045 \\
\hline 
{\em 3}$\ell$ & $1..2$ &   5.2 &   3.8 & 0.301 & 0.252 & 0.227 & 0.314
& 0.245 & 0.169 & 0.301 -- 0.603 & 0.252 -- 0.505 & 0.227 -- 0.454 \\ 
{\em 3i} & $3..4$ &   5.8 &   3.5 & 0.261 & 0.287 & 0.316 & 0.131 &
0.189 & 0.233 & 0.130 -- 0.174 & 0.143 -- 0.191 & 0.158 -- 0.211 \\
{\em 3s} & $7..8$ &   5.8 &   3.4 & 0.297 & 0.288 & 0.289 & 0.106 &
0.092 & 0.091 & 0.074 -- 0.085 & 0.072 -- 0.082 & 0.072 -- 0.083 \\
\hline 
{\em 4}$\ell$ & $1..2$ &   8.2 &   2.4 & 0.467 & 0.318 & 0.444 & 0.693
& 0.241 & 0.558 & 0.467 -- 0.933 & 0.318 -- 0.635 & 0.444 -- 0.887 \\
{\em 4i} & $3..4$ &   9.7 &   2.1 & 0.451 & 0.478 & 0.520 & 0.248 &
0.323 & 0.349 & 0.225 -- 0.301 & 0.239 -- 0.319 & 0.260 -- 0.347 \\ 
{\em 4s} & $7..8$ &  10.4 &   1.9 & 0.532 & 0.513 & 0.519 & 0.194 &
0.167 & 0.170 & 0.133 -- 0.152 & 0.128 -- 0.147 & 0.130 -- 0.148 \\ 
\hline\hline\end{tabular}
\end{center}
\vspace*{-0.3cm}
{%
{\em 1.\ column:} Model identifier,  with the letters  $\ell$, $i$, and $s$ standing for
 large-scale, intermediate-wavelength, and short-wavelength turbulence, respectively.\\
{\em 2.\ column:} Driving wavelength interval.\\
{\em 3.\ column:} Mean Mach number,  defined  
 as ratio between the time-averaged one-dimensional  velocity dispersion
 $\bar{\sigma}_v = 3^{-1/2} (\bar{\sigma}^{\,2}_x + \bar{\sigma}^{\,2}_y  + \bar{\sigma}^{\,2}_y)^{1/2}$ 
  and the isothermal sound speed $c_{\rm s}$, ${\cal M}=
 \bar{\sigma}_v/c_{\rm s}$. The values for the
 different velocity components $x$, $y$, and $z$ may differ
 considerably, especially for large-wavelength turbulence. Please
 recall from Section \ref{subsubsec:num-meth} that the speed of sound is
 $c_{\rm s} = 0.05$, and thus the sound crossing time $t_{\rm sound} =
 20$.\\
{\em 4.\ column:} Average shock crossing time through the computational volume.\\
{\em 5.\ to 7.\ column:} Time averaged  velocity dispersion along the
 three principal axes $x$, $y$, and $z$, e.g.\ for the $x$-component
 $\bar{\sigma}_x^{\,2} = \int_0^t \langle (v_{x.i}(t') - \langle
 v_{x.i}(t')  \rangle_i )^2\rangle_i dt' / t$. \\
{\em 8.\ to 10.\ column:} Mean-motion corrected diffusion coefficients along
 the  three principal axes computed from Equation (\ref{eqn:D(t)-1}) for
 time intervals $t\gg\tau$.\\
{\em 11.\ to 13.\ column:} Predicted values of the mean motion corrected
 diffusion coefficients  $D^\prime_x$, $D^\prime_y$, and $D^\prime_z$
 from  extending mixing length theory into the 
 supersonic regime (Section \ref{subsubsec:mixing}).\\
}
\end{table*}

\rsksubsubsection{Numerical Method}
\label{subsubsec:num-meth}
In order to utilize the above formalism, we carry out a series of
numerical simulation of supersonic turbulent flows.  A variety of
numerical schemes can be used to describe the time evolution of gases
and fluids. By far the most widely-used and thoroughly-studied class
of methods is based on the finite difference representations of the
equations of hydrodynamics (e.g.\ Potter 1977). In the most simple
implementation, the fluid properties are calculated on equidistant
spatially fixed grid points in a Cartesian coordinate system. Finite
difference schemes have well defined mathematical convergence
properties, and can be generalized to very complex, time varying,
non-equidistant meshes with arbitrary geometrical properties.
However, it is very difficult to obtain a Lagrangian description,
which is essential when dealing with compressible supersonic
turbulence with a high degree of vorticity.  Methods that do not rely
on any kind of mesh representation at all are therefore highly
desirable.

For the current investigation we use smoothed particle hydrodynamics (SPH),
which is a fully Lagrangian, particle-based method to solve the equations of
hydrodynamics. The fluid is represented by an ensemble of particles, where
flow properties and thermodynamic observables are obtained as local averages
from a kernel smoothing procedure (typically based on cubic spline functions)
(Benz 1990; Monaghan 1992). Each particle $i$ is characterized by mass $m_i$,
velocity $\vec{v}_i$ and position $\vec{r}_i$ and carries in addition density
$\rho_i$, internal energy $\epsilon_i$ or temperature $T_i$, and pressure
$p_i$.  The SPH method is commonly used in the astrophysics community because
it can resolve large density contrasts simply by increasing the particle
concentration in regions where it is needed.  This versatility is important
for handling compressible turbulent flows where density fluctuations will
occur at random places and random times. The same scheme that allows for high
spatial resolution in high-density regions, however, delivers only limited
spatial resolution in low-density regions. There, the number density of SPH
particles is small and thus the volume necessary to obtain a meaningful local
average tends to be large. Furthermore, SPH requires the introduction of a
von~Neumann Richtmyer artificial viscosity to prevent interparticle
penetration, shock fronts are thus smeared out over two to three local
smoothing lengths. Altogether, the performance and convergence properties of
the method are well understood and tested against analytic models and other
numerical schemes, for example in the context of turbulent supersonic
astrophysical flows (Mac~Low \etal\ 1998; Klessen \& Burkert 2000, 2001;
Klessen \etal\ 2000), and its intrinsic diffusivity is sufficiently low to
allow for the current investigation of turbulent diffusion phenomena (Lombardi
\etal\ 1999).

To simplify the analysis we assume the medium is infinite and isotropic on
large scales, and consider a cubic volume which is subject to periodic
boundary conditions. The medium is described as an ideal gas with an
isothermal equation of state, i.e.\ pressure $p$ relates to the density $\rho$
as $p=c^2_{\rm s}\rho$ with $c_{\rm s}$ being the speed of sound.  Throughout
this paper we adopt normalized units, where all physical constants (like the
gas constant), total mass $M$, mean density $\langle \rho \rangle$, and the
linear size $L$ of the cube all are set to unity. The speed of sound is
$c_{\rm s}=0.05$, hence, the sound crossing time through the cube follows as
$t_{\rm sound} = 20$.  In all models discussed here, the fluid is represented
by an ensemble of $205\,379$ SPH particles which gives sufficient resolution
for the purpose of the current analysis.

\begin{figure*}[th]
% \begin{picture}(16,6)
% \put( -4.0, 0.0){\epsfbox{./figure1.ps}}
% \end{picture}
%\includegraphics[width=15cm]{../rsk-fig02-2D-slices.ps}
\vspace{0.5cm}
\begin{center}
\includegraphics[width=15cm]{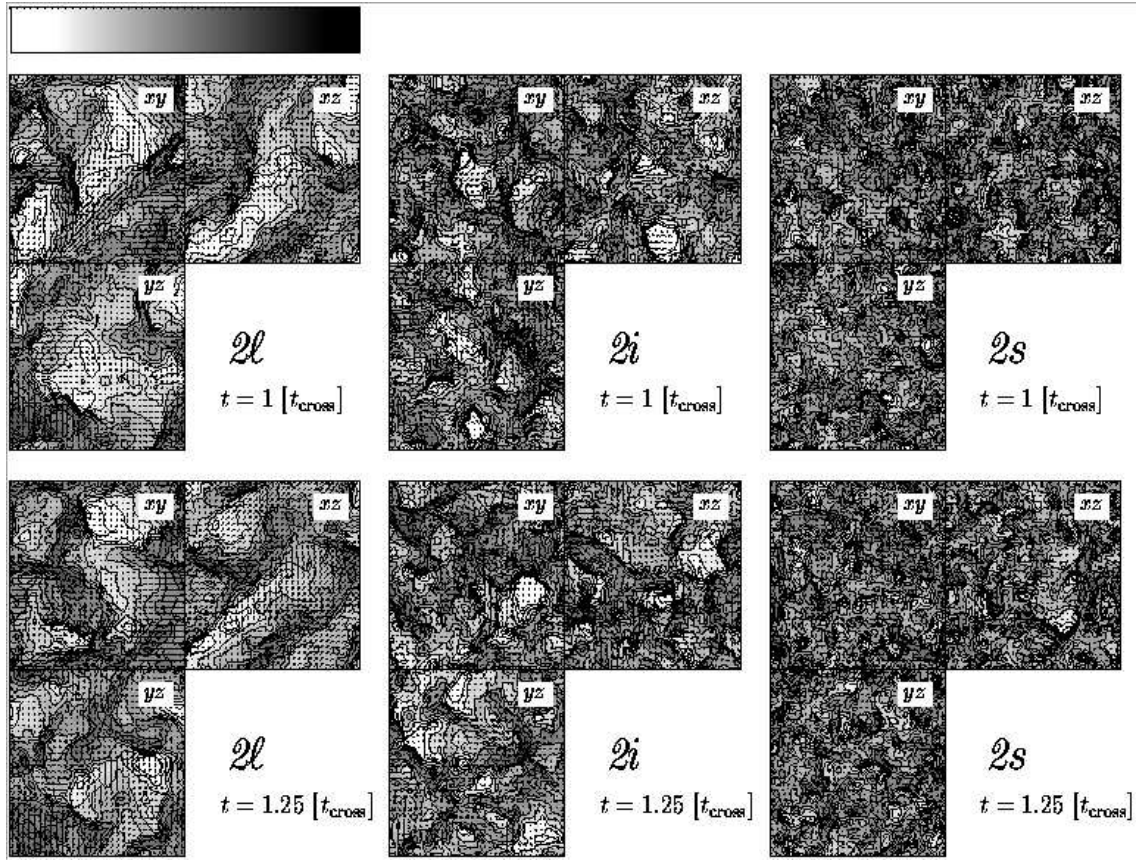}
\end{center}
\vspace{0.2cm}
\caption{\label{fig:3Dplot-cut}%
  Density and velocity structure of models {\em 2}$\ell$, {\em 2i},
  and {\em 2s} (from left to right). The panels show cuts through the
  center of the computational volume normal to the three principal
  axes of the system, after one shock crossing time $t_{\rm cross} =
  L/\sigma_v \approx 6.5$ and 1/4 $t_{\rm cross}$ later. Density is
  scaled logarithmically as indicated in the greyscale key at the
  upper left side. The maximum density is $\sim100$,
  while the mean density is one in the normalized units used.  Vectors
  indicate the velocity field in the plane. The rms Mach number is
  ${\cal M} \approx 3.1$. Large-scale turbulence ({\em 2}$\ell$) is
  dominated by large coherent density and velocity gradients leading a
  large degree of anisotropy, whereas small-scale turbulence ({\em
    2s}) exhibits noticeable structure only on small scales with the
  overall density structure being relatively homogeneous and
  isotropic. (From Klessen \& Lin 2003)}
\end{figure*}

Supersonic turbulence is known to decay rapidly (Mac~Low \etal\ 1998; Stone
\etal\ 1998; Padoan \& Nordlund 1999; Biskamp \& M{\"u}ller 2000; M{\"u}ller
\& Biskamp 2000). Stationary turbulence in the interstellar medium therefore
requires a continuous energy input. To generate and maintain the turbulent
flow we introduce random Gaussian forcing fields in a narrow range of
wavenumbers such that the total kinetic energy contained in the system remains
approximately constant. We generate the forcing field for each direction
separately and simply add up the three contributions. Thus, we excite both,
solenoidal as well as compressible modes at the same time. The typical ratio
between the solenoidal and compressible energy component is between 2:1 and
3:1 in the resulting turbulent flow (see e.g.\ Figure 8 in Klessen \etal\ 
2000). We keep the forcing field fixed in space, but adjust its amplitude in
order to maintain a constant energy input rate into the system compensating
for the energy loss due to dissipation (for further details on the method
see Mac~Low 1999 or Klessen \etal\ 2000). This non-local driving scheme allows
us to exactly control 
the (spatial) scale which carries the peak of the turbulent kinetic energy. It
is this property that motivated our choice of random Gaussian fields as
driving source. In reality the forcing of turbulence in the interstellar
medium is likely to be a multi-scale phenomenon with appreciable contributions
from differential rotation (i.e.\ shear) in the Galactic disk and energy input
from supernovae explosions ending the lives of massive stars.  Comparable to
the values observed in interstellar gas, we study flows with Mach numbers in
the range 0.5 to 10, where we define the Mach number from the {\em
  one-dimensional} rms velocity dispersion $\sigma_v$ as ${\cal M} =
\sigma_v/c_{\rm s}$. For each value of the Mach number we consider three
different cases, one case where turbulence is driven on large scales only
(i.e.\ with wavenumbers $k$ in the interval $1\le k \le 2$),
intermediate-wavelength turbulence ($3\le k \le 4$), and small-scale
turbulence ($7 \le k \le 8$), as summarized in Table \ref{tab:models-3}. Note
that our models are not subject to global shear because of the adopted
periodic boundary conditions. We call turbulence "large scale" when the
Fourier decomposition of the velocity field is dominated by the largest scales
possible for the considered volume $L^3$, i.e.\ the system becomes isotropic
and homogeneous only on scales larger than $L$. On scales below $L$ it may
exhibit a considerable degree of anisotropy. This is most noticeable in the
case $1\le k \le 2$, because wavenumber space is very poorly sampled and
variance effects become significant. The system is dominated by one or two
large shock fronts that cross through the medium. In the interval $7\le k \le
8$ the number of Fourier modes contribution to the velocity field is large,
and the system appears more isotropic and homogeneous already on distances
smaller than $L$. This trend is clearly visible in Figure
\ref{fig:3Dplot-cut}.

Similar to any other numerical calculations, the models discussed here fall
short of describing real gases in comprehensive details as they cannot include
all physical processes that may act on the medium.  In interstellar gas
clouds, transport properties and chemical mixing will not only be determined
by the compressible turbulence alone, but the density and velocity structure
is also influenced by magnetic fields, chemical reactions, and radiation
transfer processes.  Furthermore, all numerical models are resolution limited.
The turbulent inertial range in our large-scale turbulence simulations spans
over about 1.5 decades in wavenumber. This range is considerably less than
what is observed in interstellar gas clouds.  The same limitation holds for
the Reynolds numbers achieved in the models, they fall short of the values in
real gas clouds by several orders of magnitude.  Nevertheless, despite these
obvious shortcomings, the results derived here do characterize global
transport properties in interstellar gas clouds and in other supersonically
turbulent compressible flows.

\begin{figure*}[ht]
%
% \begin{picture}(16,6)
% \put( -4.0, 0.0){\epsfbox{../figure01-mean-vel.ps}}
% \end{picture}
%
%\includegraphics[bb=0cm 0cm 15cm 10cm,width=15cm]{../figure01-mean-vel.ps}
%\vspace{0.5cm}
\begin{center}
\includegraphics[width=14cm]{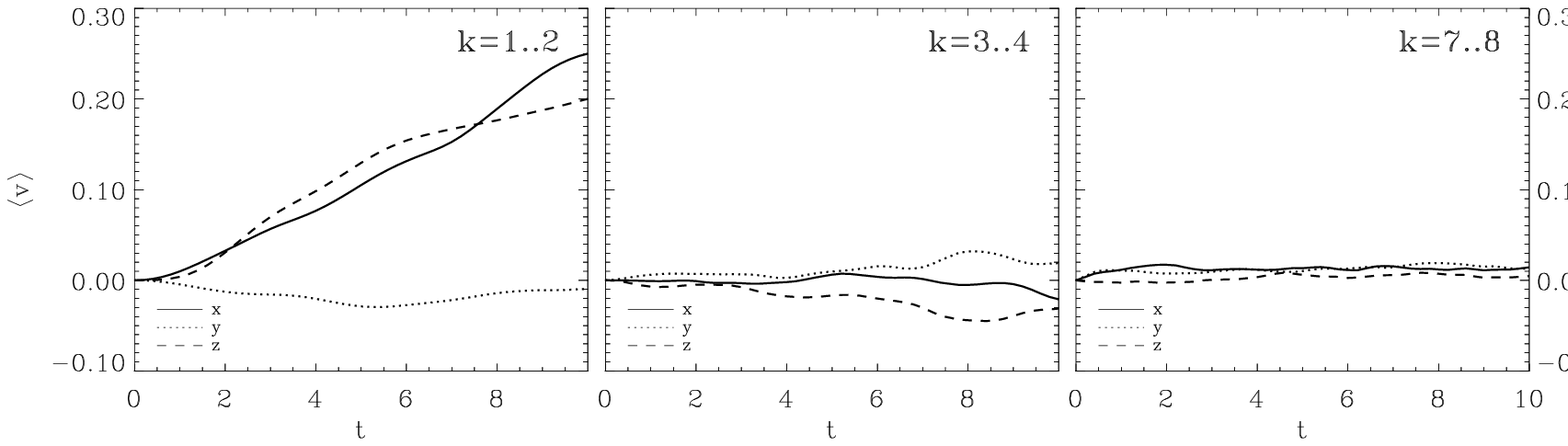}
\end{center}
\vspace{0.0cm}
\caption{\label{fig:v-mean}%
Time evolution of the mean flow velocity $\langle \vec{v}_i(t)
\rangle_i$ in models {\em 2}$\ell$,  {\em 2i},  and {\em 2s}. Time $t$
and velocity $v$ are given in normalized units. (From Klessen \& Lin 2003)}
\end{figure*}
\rsksubsubsection{Flow Properties}
\label{subsubsec:transport}
Supersonic turbulence in compressible media establishes a complex
network of interacting shocks.  Converging shock fronts locally
generate large density enhancements, diverging flows create rarefied
voids of low gas density.  The fluctuations in turbulent velocity
fields are highly transient, as the random flow that creates local
density enhancements can disperse them again. The life time of
individual shock-generated clumps corresponds to the time interval for
two successive shocks to pass through the same location in space,
which in turn depends on the length scale of turbulence and on the
Mach number of the flow.

The velocity field of turbulence that is driven at large wavelengths
is found to be dominated by large-scale shocks  which are very
efficient in sweeping up material, thus creating massive coherent
density structures. The shock passing time is rather long, and shock-generated
clumps can travel quite some distance before begin 
disrupted. %One expects efficient transport of material.  
On the contrary, when energy is inserted mainly on small scales, the network
of interacting shocks is very tightly knit.  Clumps have low masses and the
time interval between two shock fronts passing through the same location is
small, hence, swept-up gas cannot travel far before being
dispersed again. %One expects less efficient diffusion.

The density and velocity structure of three models with large-,
inter\-mediate-, and small-wavelength turbulence is visualized in
Figure~\ref{fig:3Dplot-cut}. It shows cuts through the centers of the
simulated volume. As turbulence is stationary, all times are
equivalent, and the snapshot in the upper panel is taken at some
arbitrary time. The lower panel depicts the system some time interval
later corresponding to $1/4$ shock crossing time through the cube. One
clearly notices markable differences in the density and velocity field
between the three models.

\rsksubsubsection{Transport Properties in an Absolute Reference Frame}
\label{subsubsec:Euler}
In order to drive supersonic turbulence and to maintain a given rms
Mach number in the flow, we use a random Gaussian velocity field with
zero mean to `agitate' the fluid elements at each timestep. However,
despite the fact that the driving scheme has zero mean, the system is
likely to experience a net acceleration and develop an appreciable
drift velocity, because of the compressibility of the medium. This
evolutionary trend is well illustrated in Figure \ref{fig:v-mean}
which plots the time evolution of the three components of the mean
velocity for models {\em 2}$\ell$, {\em 2i}, and {\em 2s}, with rms
Mach numbers ${\cal M} \approx 3.1$, where turbulence is driven on
{\em (a)} large (i.e.\ with small wavenumbers $1\le k \le2$), {\em
(b)} intermediate ($3\le k \le 4$), and {\em (c)} small scales (with
$7\le k \le 8$). The net acceleration is most pronounced when
turbulent energy is inserted on the global scales, as in this case
larger and more coherent velocity gradients can build up across the
volume compared to small-scale turbulence.

The tendency for the zero-mean Gaussian driving mechanisms to induce
significant center-of-mass drift velocities in highly compressible
media can be understood as follows. Suppose the gas is perturbed by
one single mode in form of a sine wave. If the medium is homogeneous
and incompressible, equal amounts of mass would be accelerated in the
forward as well as in the backward direction. But, if the medium is
inhomogeneous, there would be an imbalance between the two directions
and the result would be a net acceleration of the system.  If the
density distribution remains fixed, this acceleration would be
compensated by an equal amount of deceleration after half a period,
and the center of mass would simply oscillate.  However, if the system
is highly compressible and the driving field is a superposition of
plane waves, the density distribution would change continuously (and
randomly).  Any net acceleration at one instance in time would not be
completely compensated after some finite time interval later. This
will only occur for $t\rightarrow \infty$ assuming ergodicity of the
flow. Subsequently, the system is expected to develop a net flow
velocity in some random direction for $t<\infty$. This effect is most
clearly noticeable for long-wavelength turbulence, where density and
velocity structure is dominated by the coherent large-scale structure.
But the effect is small for turbulence that is excited on small
scales, because in this limit, there is a large number of accelerated
`cells' which in turn compensate for another's acceleration.

\begin{figure*}[tp]
\vspace{0.1cm}
\begin{center}
\includegraphics[width=14.5cm]{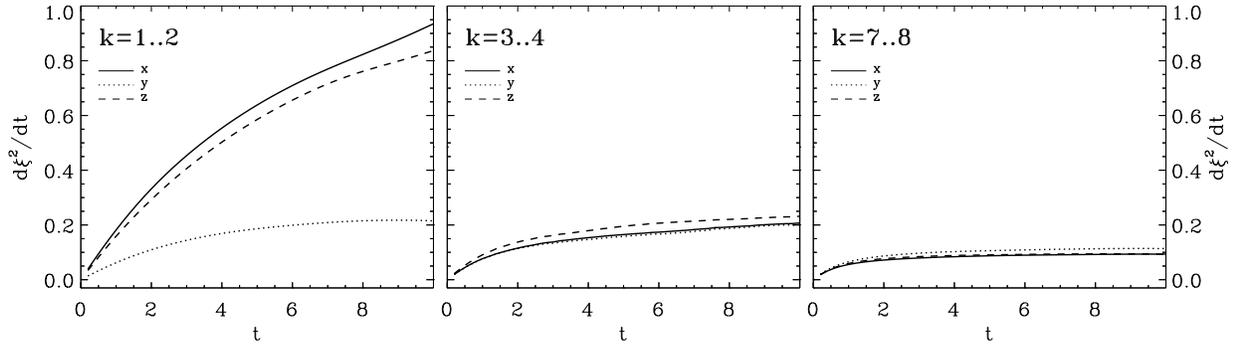}
\end{center}
\vspace{0.0cm}
    \caption{\label{fig:Euler-D}%
      Time evolution of the diffusion coefficient $D(t)=d\langle
      [\vec{r}_i(t)-\vec{r}_i(0)]^2 \rangle_i / dt$ for models {\em
        2}$\ell$, {\em 2i}, and {\em 2s} computed in an absolute
      reference frame. All units are normalized as described in
      \S \ref{subsubsec:num-meth}. (From Klessen \& Lin 2003) 
}
\end{figure*}
The property that the compressible turbulent flows are likely to
pick up average drift velocities, even when driven by Gaussian fields
with zero mean, has implications for the transport coefficients.
Figure \ref{fig:Euler-D} shows the time evolution of the absolute
(Eulerian) diffusion coefficients $D_x$, $D_y$, and $D_z$ in each
spatial direction computed from Equation (\ref{eqn:def-sigma}).
%
% Time $t$ is normalized to the average shock
% crossing time through the computational volume, $t_{\rm cross} =
% 1/\bar{\sigma}_{v}$. The value of $\bar{\sigma}_{v}$ may differ for
% different spatial directions...
% with $\bar{\sigma}_{v}$ beings the
% one-dimensional rms velocity dispersion of the flow $\sigma_v =
% 3^{-1/2}\langle (\vec{v}_i - \langle \vec{v}_i
% \rangle)^{\,2}\rangle^{1/2}$ averaged over all times. Recall that the
% total extent of the simulated cube is one. The diffusion coefficient
% $d(t)$ is scaled by multiplying with $T_{\rm corss}$.  
Note that for stationary turbulence, only time differences are
relevant and one is free to chose the initial time.  In order to
improve the statistical significance of the analysis, we obtain $D(t)$
and $\xi_{\vec{r}}(t)$ by further averaging over all time intervals
$t$ that `fit into' the full timespan of the simulation.
%
% Equations \ref{eqn:D(t)-1} and \ref{eqn:D(t)-2}, but instead of using
% the advected (Lagrangian) velocity, now the absolute (Eulerian)
% velocity is used. 
%

% Due to the (continuous) net acceleration experienced by the system,
% the quantity $\xi_{\vec{r}}^2(t)$ grows faster than linearly with
% time, even for asymptotically large times,  i.e.\ as $t
% \rightarrow \infty$.  The system resides in a superdiffusive regime,
% where $D(t)$ never saturates. Instead, $D(t)$ grows continuously with
% time, which is most evident in model {\em 2}$\ell$ of large-scale
% turbulence. The ever increasing drift velocity $\langle \vec{v}_i(t)
% \rangle_i$ causes strong velocity correlations leading to the
% divergence of the velocity autocorrelation tensor, i.e.\ $\int_0^{t}
% {\rm tr}\,{\cal C}(t')dt' \rightarrow \infty$ for $t \rightarrow
% \infty$.  This net motion, however, can be corrected for, allowing us
% to study the dispersion of particles in a reference frame that moves
% along with the average flow velocity of the system.

Due to the (continuous) net acceleration experienced by the system,
the quantity $\xi_{\vec{r}}^2(t)$ grows faster than linearly with
time, even for intervals much larger than the correlation time $\tau$,
i.e.\ for $\tau \ll t < \infty$.  The system resides in a
superdiffusive regime, where $D(t)$ does not saturate. Instead, $D(t)$
grows continuously with time, which is most evident in model {\em
  2}$\ell$ of large-scale turbulence. The ever increasing drift
velocity $\langle \vec{v}_i(t) \rangle_i$ causes strong velocity
correlations leading continuous growth of the velocity autocorrelation
tensor $\int_0^{t} {\rm tr}\,{\cal C}(t')dt'$.  This net motion,
however, can be corrected for, allowing us to study the dispersion of
particles in a reference frame that moves along with the average flow
velocity of the system.

\begin{figure*}[tp]
%
%\begin{picture}(16,28)
%\put(  -4.0, 0.0){\epsfbox{../figure-01-diffusion-coeff-D.ps}}
%\end{picture}
%
\begin{center}
\vspace*{0.2cm}
\includegraphics[width=12.0cm]{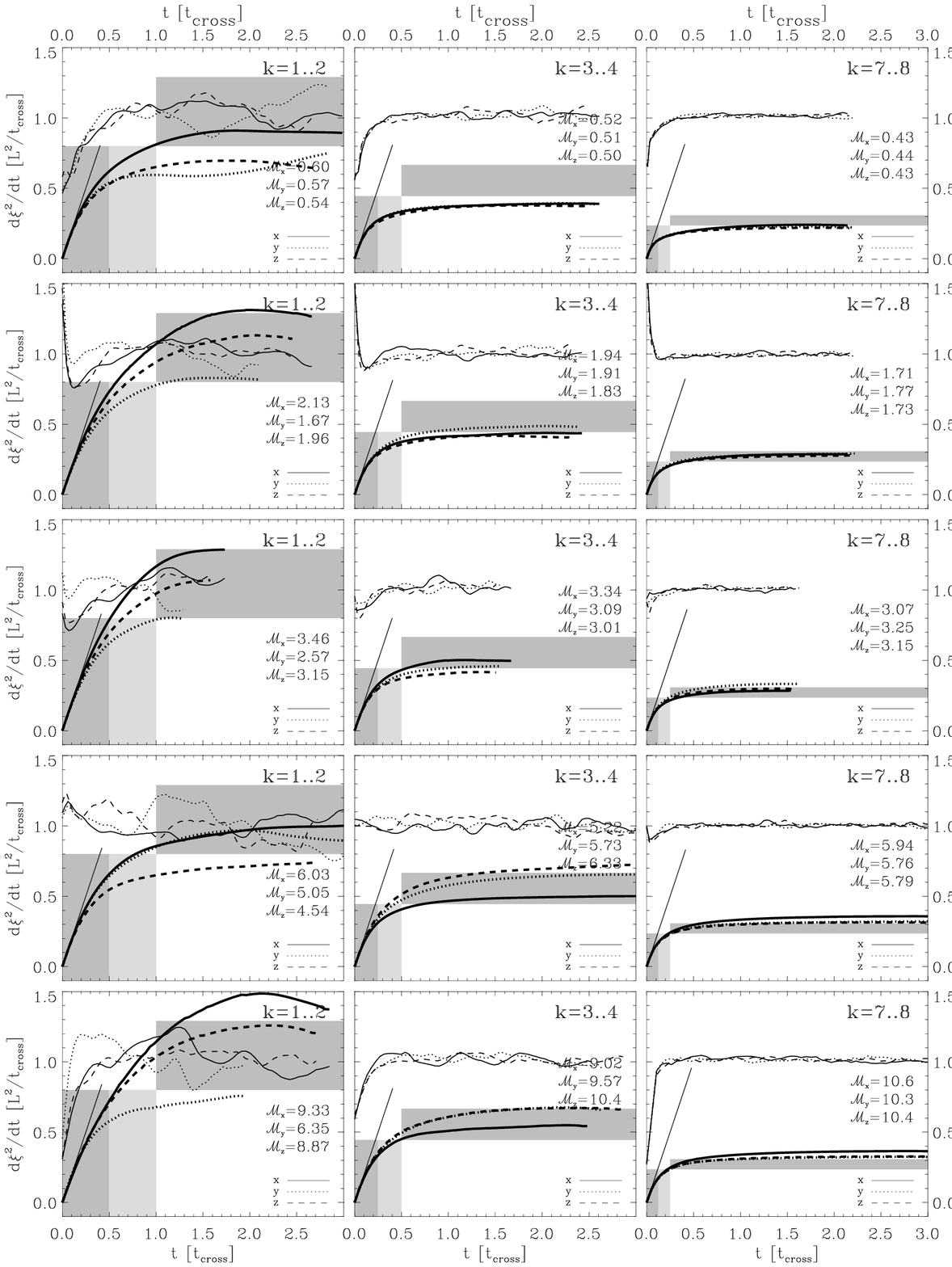}
\end{center}
\vspace{0.3cm}
    \caption{\label{fig:Lagrange-D}%
      Time evolution of the diffusion coefficient $D^\prime(t)=d\xi^2_i / dt$
      computed in a reference frame that follows the average flow velocity
      ({\em thick line}, axis scaling on the left ordinate), i.e.\ is centered
      on $\langle \vec{r}_i(t) \rangle_i = \int_0^t\langle \vec{v}_i(t')
      \rangle_i dt'$.  Velocity dispersions along the three major axes $x$,
      $y$, and $z$ are each normalized to unity using the time-averaged
      one-dimensional Mach number $\cal M$ (as indicated in each plot)
      together with the given value of the sound speed $c_{\rm s}$ ({\em thin
        lines}, axis scaling on the right ordinate). Times are rescaled to
      the rms shock crossing time through the simulated cube $t_{\rm cross} =
      L/\sigma_v = L/({\cal M}c_{\rm s})$. Details for each model are given in
      Table \ref{tab:models-3}. The horizontal gray shaded area indicates the
      mixing length prediction for $t\rightarrow \infty$, and the vertical
      gray and light gray shaded areas show a time interval of
      $\tau=L/(k{\cal M}c_{\rm s})$ and $2\tau$, respectively. For $t\ll\tau$
      diffusion should be anomalous and coherent, with $D^\prime(t)$ growing
      linearly with time. The expected behavior from mixing length theory in
      this regime is indicated by the straight line originating at $t=0$. It
      indeed gives a good fit. Note that all models driven on large scales
      ($1\le k \le 2$) exhibit a considerable degree of anisotropy, manifested by
      different rms Mach numbers $\cal M$ along the three principal axes and
      different values $D^\prime(t)$. For the models with intermediate-scale
      and small-scale driving anisotropy effects are increasingly less
      important. (From Klessen \& Lin 2003)}
\end{figure*}

\rsksubsubsection{Transport Properties in Flow Coordinates}
\label{subsubsec:Lagrange}

In order to gain further insight into the transport properties of
compressible supersonic turbulent flows, we study the time evolution
of the relative (Lagrangian) diffusion coefficient. In this
prescription, 
\begin{equation}
D^\prime (t)= \frac{d \xi^2_{\vec{r}}(t)}{dt} \nonumber
\end{equation} 
is obtained relative to a frame of reference which comoves with the
mean motion of the system $\langle \vec{v}_i(t) \rangle_i$ on the
trajectory $\langle \vec{r}_i(t) \rangle_i = \int_0^t\langle
\vec{v}_i(t') \rangle_i dt'$. Then, $\xi^2_{\vec{r}}(t-t')= \langle
[(\vec{r}_i(t)-\langle \vec{r}_i(t) \rangle_i)-(\vec{r}_i(t')- \langle
\vec{r}_i(t') \rangle_i)]^2 \rangle_i$ (see Equation
\ref{eqn:def-sigma}).
%}

In Figure \ref{fig:Lagrange-D}, we show the evolution of $D^\prime
(t)$ for each coordinate direction for the complete suite of models.
The rms Mach numbers range from about 0.5 to 10, each considered for
three cases where turbulence is driven on large, intermediate, and
small scales, respectively. The plots are rescaled such that the
time-averaged one-dimensional rms velocity dispersion $\bar{\sigma}_{v}$ is
normalized to unity (for each direction separately). We also rescale
the time $t$ with respect to the average shock crossing time
scale through the computational volume, $t_{\rm cross} =
L/\bar{\sigma}_{v}$.  Recall that $L=1$, and note that $\bar{\sigma}
_{v}$ usually differs between the three spatial directions because of
the variance effects, especially in models of large-scale turbulence.

In Figure \ref{fig:Lagrange-D}, we demonstrate that the magnitude of
$D^\prime (t)$ saturates for large time intervals $t> \tau$ in all
directions. In a reference frame that follows the mean motion of the
flow, diffusion in compressible supersonically turbulent media indeed
behaves in a normal manner.  For small time intervals $t< \tau$,
however, the system still exhibits an anomalous diffusion even with
the mean-motion correction. In this regime $D^\prime(t)$ grows roughly
linearly with time. For $t>\tau$ the diffusion coefficient $D^\prime
(t)$ reaches an asymptotic limit. This result holds for the entire
range of Mach numbers studied and for turbulence that is maintained by
energy input on very different spatial scales.

From Figure \ref{fig:Lagrange-D}, we find that diffusion in
compressible supersonic turbulent flows follows a universal law.  It
can be obtained by using the rms Mach number (together with the sound
speed $c_{\rm s}$) as characterizing parameter for rescaling the
velocity dispersion $\sigma_v$, and the rms shock crossing time scale
through the volume $t_{\rm cross} = L/({\cal M}c_{\rm s})$ for
rescaling the time.  The normalized diffusion coefficient $D^\prime(t)$
exhibits a universal slope of two at times $t<\tau$ (i.e.\ in the
superdiffusive regime), and approaches a constant value that depends
only on the length scale but not on the strength (i.e.\ the resulting
Mach number) of the mechanism that drives the turbulence.  Even for
highly compressible supersonic turbulent flows it is possible to find
simple scaling relations to characterize the transport properties ---
analogous to the mixing length description of diffusive processes in
incompressible subsonic turbulent flows.

\rsksubsubsection{A Mixing Length Description}
\label{subsubsec:mixing}
Incompressible turbulence is often described in terms of a hierarchy
of turbulent eddies, where each eddy contains multiple eddies of
smaller size on the lower levels of the hierarchy, while itself being
part of turbulent eddy at larger scales (Richardson 1922; Kolmogorov 1941;
Obukhov 1941). At each
level of the hierarchy, an eddy is characterized by a typical
lengthscale $\tilde{\ell}$ and a typical velocity $\tilde{v}$. The
typical lifetime of an eddy is its `turn-over' time $\tau=
\tilde{\ell} / \tilde{v}$.  This mixing length prescription is an
attempt to characterize the flow properties in terms of the typical
scales $\tilde{\ell}$ and $\tilde{v}$. For example, this classical
picture defines an effective `eddy' viscosity $\mu = \rho
\tilde{\ell}\tilde{v}$, where $\rho$ is the density. The mixing length
$\tilde{\ell}$ is interpreted to be the turbulent analogue of the mean
free path of molecules in the kinetic theory of gases, with
$\tilde{v}$ being the characteristic velocity of the turbulent
fluctuation. 

In such a model, the velocities of gas molecules within an eddy are
strongly correlated within a time interval $t<\tau$. They all follow
the eddy rotation; the diffusion process is coherent.  However, for
$t\gg\tau$ the velocities of gas molecules become uncorrelated, as the
eddy has long been destroyed and dispersed. Hence, the velocity
autocorrelation function vanishes for large time intervals, ${\cal
C}(t)\rightarrow 0$ for $t\rightarrow \infty$. Diffusion becomes
incoherent as in Brownian motion or the random walk.  The diffusion
coefficient in the mixing length approach simply is $D(t) \approx 2
\tilde{v}^2 t$ in the regime $t<\tau$, which follows from replacing
$\vec{r}(t)$ by $\tilde{v}t$ and $\vec{v}(t)$ by $\tilde{v}$ in
Equation (\ref{eqn:D(t)-1}). As the largest correlation length is the
eddy size, $\vec{r}(t)$ is substituted by $\tilde{\ell}=\tilde{v}\tau$
for times $t\gg\tau$, and the classical mixing length theory yields
$D(t) \approx 2 \tilde{\ell} \tilde{v} = 2 \tilde{v}^2 \tau = {\rm
constant}$.

Compressible, supersonic, turbulent flows rapidly build up a network
of interacting shocks with highly transient density and velocity
structure. Density fluctuations are generated by locally converging
flows, and their lifetimes are determined by the time $\tau$ between
two successive shock passages. This time interval is determined by the
typical shock velocity, which is roughly the rms velocity of the flow,
i.e.\ the Mach number times the sound speed, $\sigma_v = {\cal M}
c_{\rm s}$. It also depends on the length scale at which energy is
inserted into the system to maintain the turbulence, which in our case
is $L/k$ with $k$ being the driving wavenumber and $L$ being the size
of the considered region (recall that in our models $L$ is
unity). This length scale is also the typical traveling distance
before two shocks interact with each other.  As basic ingredients for
a supersonic compressible mixing length description we can thus
identify:
\begin{eqnarray}
\label{eqn:mixing-length}
\!\!\!\!\!\!\!\!\!\!{\rm shock~travel~length:}&\!\!\!\!\!\!\!&\tilde{\ell}  \approx  L/k , \\
\!\!\!\!\!\!\!\!\!\!{\rm rms~velocity:} &\!\!\!\!\!\!\!&\tilde{v}  \approx  \sigma_v = {\cal M} c_{\rm s} .
\end{eqnarray}
The Lagrangian  velocity correlation time scale, $\tau$, is
analogous to the time interval during which shock-generated density
fluctuation remains unperturbed and moves coherently before it is
being dispersed by the interaction with a new shock front. This time
interval is equivalent to the time scale a shock travels along its
`mean free path' $\tilde{\ell}$ with an rms velocity $\tilde{v}$. This
crossing time is $\tau = \tilde{\ell}\tilde{v} \approx \sigma_v\,L/k$.
For $t<\tau$ gas molecules can travel coherently within individual
shock generated density fluctuations, and the diffusion coefficient in
the mixing length prescription follows as
\begin{equation}
  \label{eqn:D(t)-small-t}
  D^\prime(t) \approx 2\tilde{v}^2t \approx 2\sigma_v^2 t\;.
\end{equation}
$D^\prime(t)$ grows linearly with time with slope $2\sigma_v^2$.
For large times, $t\gg\tau$, $D^\prime(t)$ approaches a constant value,
\begin{equation}
  \label{eqn:D(t)-large-t}
  D^\prime (t) \approx 2\tilde{v}^2\tau \approx 2\sigma_v\,L/k\;.
\end{equation}

This mixing length approach (Equations \ref{eqn:D(t)-small-t} and
\ref{eqn:D(t)-large-t}) suggests a unique scaling dependence of the
diffusion coefficients in supersonic compressible flows on the {\em
Mach number} $\cal M$ and on the {\em length scale} $\tilde{\ell}$ of
the most energy containing modes {\em with respect to the total size
$L$ of the system considered}.

We can use $\cal M$ (together with the given value of the sound speed)
to normalize the rms velocity: $\sigma_v= {\cal M} c_{\rm s} \mapsto
\sigma_v'=1$. And we can also rescale the time with respect to the rms
shock crossing time scale through the total volume, which is $t_{\rm
cross} = L/\sigma_v = L/({\cal M} c_{\rm s}) = t_{\rm sound}/{\cal M}$
with $t_{\rm sound} = L/c_{\rm s}$ being the sound crossing time, so that
$t \mapsto t'=t/t_{\rm cross}$. From this normalization procedure, we
get $D^\prime (t)\mapsto D''(t') = D^\prime (t) \,{\cal M} c_{\rm s}L$
and obtain the following universal profile for the diffusion
coefficient,
\begin{eqnarray}
\label{eqn:scaled-D1}
\!\!\!\!\!\!\!D^{\prime\prime}(t')=2t' & {\rm ~for~}\; & t' \ll 1/k \\
\label{eqn:scaled-D2} 
\!\!\!\!\!\!\!D^{\prime\prime}(t')=2/k & {\rm ~for~}\; & t' \gg 1/k \;.  
\end{eqnarray}

Note that this result holds for each velocity component separately, as
the results in Figure \ref{fig:Lagrange-D} indicate. In this case
$\sigma_v$ stands for $\sigma_x$, $\sigma_y$, or $\sigma_z$ in
Equations (\ref{eqn:D(t)-small-t}) and (\ref{eqn:D(t)-large-t}), and it
hold for the total diffusion coefficient, when using $\sigma_v =
({\sigma_x^2 + \sigma_y^2 + \sigma_z^2})^{1/2}$ instead.

\begin{figure*}[ht]
%
%\begin{picture}(16,6)
%\put(  4.0, 0.0){\epsfbox{../figure-mixing-length-vs-model.ps}}
%\end{picture}
\begin{center}
\vspace*{0.5cm}\hspace*{1.1cm}
\includegraphics[width=15cm]{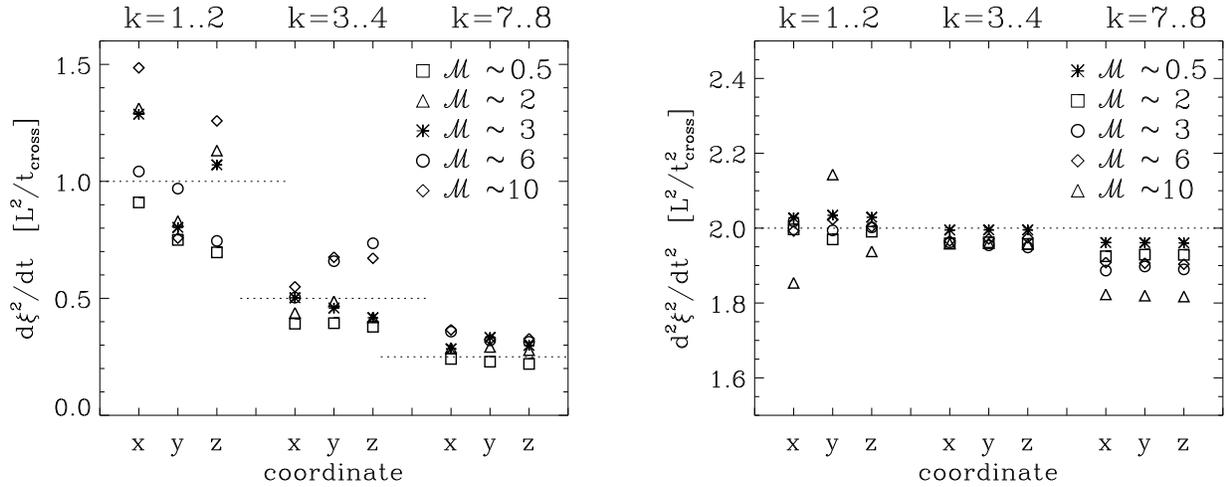}
\end{center}
\vspace{0.3cm}
    \caption{\label{fig:comparison}%
      Comparison between mixing length predictions and numerical
      models.  At the left we plot he normalized, mean-motion
      corrected diffusion coefficient $D''(t')$ for $t\rightarrow
      \infty$, and at the right its slope $dD''(t')/dt'$ for $t'\ll
      1/k$. For each suite of models, large-scale, intermediate-scale,
      and small-scale turbulence, respectively (as indicated by the
      forcing wavenumber $k$ at the top of each plot), we separately
      show the three velocity components (as indicated at the bottom).
      The different Mach numbers in each model suite are denoted by
      different symbols (as identified at the right-hand side of each
      plot). The dotted lines give the corresponding prediction of the
      mixing-length theory, $D''(t')=2/k$ and $dD''(t')/dt'=2$,
      respectively, where we take $k$ to be the maximum wavenumber of the
      forcing scheme indicated at the top of each plot. (From Klessen \& Lin
      2003)} 
\end{figure*}
The validity of the mixing length approximation is quantified in
Figure \ref{fig:comparison} which plots the mixing length predictions
against the values obtained from the numerical models. For large-scale
and intermediate scale turbulence, the mixing length approach gives
very satisfying results, only for small-scale turbulence it
underestimates the diffusion strength. This disparity probably has to
do with the numerical resolution of the code, in the sense that
driving wavenumbers of $k\approx 8$ come close to the dissipation
scale of the method and hence the inertial range of turbulence is
limited (Klessen \etal\ 2000). That limitation leads to an effective driving
for the turbulent motion on somewhat larger scales than $1/8$.
Consequently, it leads to a stronger diffusion, i.e.\ somewhat larger
diffusion coefficients than those predicted by Equations
(\ref{eqn:scaled-D1}) and (\ref{eqn:scaled-D2}). The same numerical effects also account for the
slightly shallower slope of $D'(t)$ for $t\ll\tau$ for models $7\le k\le 8$.

Figures \ref{fig:Lagrange-D} and \ref{fig:comparison} indicate that
the classical mixing length theory can be extended from incompressible
(subsonic) turbulence into the regime of supersonic turbulence of
highly compressible media. In this case, driving length $\tilde{\ell}$
and rms velocity dispersion $\sigma_v= {\cal M}c_{\rm s}$ act as
characteristic length and velocity scales in the mixing length
approach. Note, that this only applies to mean-motion corrected
transport. In general (i.e.\ in an absolute reference frame),
supersonic turbulence in compressible media leads to superdiffusion as
visualized in Figure \ref{fig:Euler-D}.

\rsksubsubsection{Summary}
\label{subsubsec:summary}
Supersonic turbulence in compressible media establishes a complex
network of interacting shocks.  Converging shock fronts locally
generate  large density enhancements, diverging flows create voids of
low gas density.  The fluctuations in turbulent velocity
fields are highly transient, as the random flow that creates local
density enhancements can disperse them again.

Due to compressibility, supersonically turbulent flows will usually
develop noticeable drift velocities, especially when turbulence is
driven on large scales, even when it is excited with Gaussian fields
with zero mean.  This tendency has consequences for the transport
properties in an absolute reference frame. The flow exhibits
super-diffusive behavior  (see also Balk 2001). However, when the
diffusion process is analyzed in a comoving coordinate system, i.e.\ 
when the induced bulk motion is being corrected, the system exhibits
normal behavior. The diffusion coefficient $D(t)$ saturates for large
time intervals, $t\rightarrow \infty$.

By extending classical mixing length theory into the supersonic regime
we propose a simple description for the diffusion coefficient based on
the rms velocity $\tilde{v}$ of the flow and the typical shock travel
distance $\tilde{\ell}$,
\begin{eqnarray}
D^{\prime}(t) = 2 \tilde{v}^2t & {\mbox{for}} & t\ll
\tilde{\ell}/\tilde{v} \nonumber\,,\\ 
D^{\prime} (t) = 2
\tilde{v}\tilde{\ell} & {\mbox{for}} & t\gg \tilde{\ell}/\tilde{v}
\nonumber\,.
\end{eqnarray}

This functional form may be used in those numerical models where knowledge of
the mixing properties of turbulent supersonic flows is required, but where
these flows cannot be adequately resolved. This is the case, for example, in
astrophysical simulations of galaxy formation and evolution, where the
chemical enrichment of the interstellar gas and the distribution and spreading
of heavy elements produced from massive stars throughout galactic disks needs
to be treated without being able to follow the turbulent motion of
interstellar gas on small enough scales relevant to star formation (Rana 1991;
Bertschinger 1998). Our results furthermore are directly relevant for
understanding the properties of individual star-forming interstellar gas
clouds within the disk of our Milky Way. These are dominated by supersonic
turbulent motions which can provide support against gravitational collapse on
global scales, while at the same time produce localized density enhancements
that allow for collapse, and thus stellar birth, on small scales. The
efficiency and timescale of star formation in galactic gas clouds depend on
the intricate interplay between their internal gravitational attraction and
their turbulent energy content.  The same is true for the statistical
properties of the resulting star clusters. For example, the element abundances
in young stellar clusters are found to be very homogeneous (Wilden \etal\ 2002),
implying that the gas out of which these stars formed must be have been
chemically well mixed initially.  On the basis of the results discussed here,
this observation can be used to constrain astrophysical models of interstellar
turbulence in star-forming regions. Understanding transport processes and
element mixing in supersonic turbulent flows thus is a prerequisite for
gaining deeper insight into the star formation phenomenon in our Galaxy.
 
%%%%%%%%%%%%%%%%%%%%%%%%%%%%%%%%%%%%%%%%%%%%%%%%%%%%%%%%%%%%%%%%%%%%%%%%%%%%%%%
% [K00] \pdf paper (Klessen 2000)
%%%%%%%%%%%%%%%%%%%%%%%%%%%%%%%%%%%%%%%%%%%%%%%%%%%%%%%%%%%%%%%%%%%%%%%%%%%%%%%

\rsksubsection{One-Point Probability Distribution Function}
\label{subsec:pdf}

\rsksubsubsection{Introduction}
\label{sec:intro}

Correlation and distribution functions of dynamical variables are
frequently deployed for characterizing the kinematical properties of
turbulent molecular clouds.  Besides using 2-point statistics
(e.g.~Scalo 1984, Kleiner \& Dickman 1987, Kitamura \etal\ 1993,
Miesch \& Bally 1994, LaRosa, Shore \& Magnani 1999), many studies
have hereby concentrated on 1-point statistics, namely on analyzing
the probability distribution function (\pdf) of the (column) density
and of dynamical observables, e.g.~of the centroid velocities of
molecular lines and their increments.  The density \pdf\ has been used
to characterize numerical simulations of the interstellar medium by
V{\'a}zquez-Semadeni (1994), Padoan, Nordlund, \& Jones (1997),
Passot, \& V{\'a}zquez-Semadeni (1998) and Scalo \etal\ (1998). Velocity \pdf's for several star-forming molecular clouds
have been determined by Miesch \& Scalo (1995) and Miesch, Scalo \&
Bally (1998). Lis \etal\ (1996, 1998) analyzed snapshots of a
numerical simulation of mildly supersonic, decaying turbulence
(without self-gravity) by Porter \etal\ (1994) and
applied the method to observations of the $\rho$-Ophiuchus
cloud. Altogether, the observed \pdf's exhibit strong non-Gaussian
features, they are often nearly exponential with possible evidence for
power-law tails in the outer parts. This disagrees with the nearly
Gaussian behavior typically found in experimental measurements and
numerical models of incompressible turbulence. The observed centroid
velocity {\em increment} \pdf's are more strongly peaked and show
stronger deviations from Gaussianity than numerical models of
incompressible turbulence predict.  Furthermore, the spatial
distribution of the largest centroid velocity differences (determining
the tail of the distribution) appears `spotty' across the face of the
clouds; there is no convincing evidence for filamentary
structure. Miesch \etal\ (1998) conclude that turbulence in molecular
clouds involves physical processes that are not adequately described
by incompressible turbulence or mildly supersonic decay simulations
(see also Mac~Low \& Ossenkopf 2000).

Based on Klessen (2000), we extend in this Section previous determinations of
\pdf's from numerical models into a regime more applicable for interstellar
turbulence by (1) by calculating fully supersonic flows, (2) by including
self-gravity, and (3) by incorporating a (simple analytic) description of
turbulent energy input.  To do this, we use numerical models introduced and
discussed in Section \ref{subsub:numerics}, and compare with existing
molecular cloud observations in the literature.  The \pdf's for the density,
for the line centroid velocity and for their increments are derived as
function of time and evolutionary state of the turbulent model.

% The structure of this paper is as follows: Section~\ref{sec:remarks}
% introduces and defines the statistical tools applied in the current
% study. It is followed in Sec.~\ref{sec:model} by a description of the
% numerical scheme used to compute the time evolution of the turbulent
% flows. Sec.~\ref{sec:init} shows that already simple variance effects
% in random Gaussian fields are able to introduce strong {\em
% non}-Gaussian distortions to the \pdf's which makes a clear-cut
% interpretation difficult. Section \ref{sec:decay-non-grav} contains
% the analysis of decaying, initially highly supersonic turbulence
% without self-gravity. This effect is then added to the simulations
% presented in Sec.~\ref{sec:decay-with-gravity}. The model most
% relevant for molecular cloud dynamics is discussed in
% Sec.~\ref{sec:driven-with-gravity}. It includes a simple driving term
% to replenish the turbulent cascade. Finally, in Sec.~\ref{sec:summary}
% all results are summarized.

\rsksubsubsection{PDF's and Their Interpretation}
\label{sec:remarks}
\rskparagraph{Turbulence and PDF's}
\label{subsec:turbulence}

The Kolmogorov (1941) approach to incompressible turbulence is a
purely phenomenological one and assumes the existence of a stationary
turbulent cascade. Energy is injected into the system at large scales
and cascades down in a self-similar way. At the smallest scales it
gets converted into heat by molecular viscosity. The flow at large
scales is essentially inviscid, hence for small wave numbers the
equation of motion is dominated by the advection term.  If the
stationary state of fully developed turbulence results from random
external forcing then one na\"{\i}vely expects the velocity
distribution in the fluid to be Gaussian on time scales larger than
the correlation time of the forcing, irrespectively of the statistics
of the forcing term which follows from the central limit theorem.
However, the situation is more complex (e.g.\ Frisch 1995, Lesieur
1997). One of the most striking (and least understood) features of
turbulence is its intermittent spatial and temporal behavior. The
structures that arise in a turbulent flow manifest themselves as high
peaks at random places and at random times. This is reflected in the
\pdf's of dynamical variables or passively advected scalars.  They are
sensitive measures of deviations from Gaussian statistics. Rare strong
fluctuations are responsible for extended tails, whereas the much
larger regions of low intensity contribute to the peak of the \pdf\ near
zero (for an analytical approach see e.g. Forster, Nelson \& Stephens
1977, Falkovich \& Lebedev 1997, Chertkov, Kolokolov \& Vergassola
1997, Balkovsky \etal\ 1997, Balkovsky \& Falkovich 1998). For
incompressible turbulence the theory predicts velocity \pdf's which are
mainly Gaussian with only minor enhancement at the far ends of the
tails. The distribution of velocity {\em differences} (between
locations in the system separated by a given shift vector $\Delta
\vec{r}$) is expected to deviate considerably from being normal and is
likely to resemble an exponential. This finding is supported by a
variety of experimental and numerical determinations (e.g.~Kida \&
Murakami 1989, Vincent \& Meneguzzi 1991, Jayesh \& Warhaft 1991, She
1991, She, Jackson \& Orszag 1991, Cao, Chen, \& She 1996, Vainshtein
1997, Lamballais, Lesieur, \& M{\'e}tais 1997, Machiels \& Deville
1998). Compressible turbulence has remained to be too complex for a
satisfying mathematical analysis.

\rskparagraph{PDF's of Observable Quantities}
It is not clear how to relate the analytical work on incompressible
turbulence to molecular clouds. In addition to the fact that
interstellar turbulence is highly supersonic and self-gravitating,
there are also observational limitations. Unlike the analytical
approach or numerical simulations, molecular cloud observations allow
access only to dimensionally reduced information.  Velocity
measurements are possible only along the line-of-sight, and the
spatial structure of a cloud is only seen in projection onto the plane
of the sky, i.e.~as variations of the column density. Although some
methods can yield information about the 3-dimensional spatial
structure of the cloud (see Stutzki \& G{\"u}sten 1990, Williams,
De~Geus, \& Blitz 1994), the result is always model dependent and
equivocal (see also Ballesteros-Paredes, V{\'a}zquez-Semadeni, \&
Scalo 1999).

A common way of obtaining knowledge about the velocity structure of
molecular clouds is to study individual line profiles at a large
number of various positions across the cloud. In the optical thin case
line shapes are in fact histograms of the radial velocities of gas
sampled along the telescope beam.  Falgarone \& Phillips (1990) and
Falgarone \etal\ (1994) showed that line profiles constructed from
high-sensitivity CO maps exhibit non-Gaussian wings and attributed
this to turbulent intermittency (see also Falgarone \etal\ 1998 on
results from the IRAM-key project).  Dubinski, Narayan, \& Phillips
(1995) demonstrated that non-Gaussian line profiles can be produced
from {\em any} Gaussian random velocity field if variance effects
become important (which is always the case for very steep or truncated
power spectra). They concluded that non-Gaussian line profiles do not
provide clear evidence for intermittency.

Another method of inferring properties of the velocity distribution in
molecular clouds is to analyze the \pdf\ of line centroid velocities
obtained from a large number of individual measurements scanning the
entire projected surface area of a cloud (Miesch \& Scalo 1995, Lis \etal\ 1998, Miesch \etal\ 1998). Each line profile (i.e. the \pdf\ {\em
along} the line-of-sight) is collapsed into one single number, the
centroid velocity, and then sampled {\em perpendicular} to the
line-of-sight. Hence, the two functions differ in the direction of the
sampling and in the quantity that is considered. A related statistical
measure is the \pdf\ of centroid velocity increments, it samples the
velocity differences between the centroids for line measurements which
are offset by a given separation.  The observational advantage of
using centroid and increment \pdf's is, that the line measurements can
typically be taken with lower sensitivity as only the centroid has to
be determined instead of the detailed line shape. These measures are
also less dependent on large-scale systematic motions of the cloud and
they are less effected by line broadening due to the possible presence
of warm dilute gas. However, to allow for a meaningful analysis of the
\pdf's especially in the tails, the number of measurements needs to be
very large and should not be less than about 1000. In order to sample
the entire volume of interstellar clouds, the molecular lines used to
obtain the \pdf's are optically thin. We follow this approach in the
present investigation and use a mass-weighted velocity sampling along
the line-of-sight to determine the line centroid. This zero-opacity
approximation does not require any explicit treatment of the radiation
transfer process.

The observed \pdf's are obtained from {\em averaged} quantities (from
column densities or line centroids). To relate these observational
measures to quantities relevant for turbulence theory, i.e.~to the
full 3-dimensional \pdf, numerical simulations are necessary as only
they allow unlimited access to all variables in phase space. A first
attempt to do this was presented by Lis \etal\ (1996, 1998) who
analyzed a simulation of mildly supersonic decaying hydrodynamic
turbulence by Porter \etal\ (1994). Since their model did neither
include self-gravity nor consider flows at high Mach number or
mechanisms to replenish turbulence, the applicability to the
interstellar medium remained limited.  This fact prompts the current
investigation which extends the previous ones by calculating {\em
highly supersonic flows}, and by including {\em self-gravity} and a
{\em turbulent driving scheme}.  The current study does not consider
magnetic fields. Their influence on the \pdf's needs to be addressed
separately.  However, the overall importance of magnetic fields and
MHD waves on the dynamical structure of molecular clouds may not be
large. The energy associated with the observed fields is of the order
of the (turbulent) kinetic energy content of molecular clouds
(Crutcher 1999).  Magnetic fields cannot prevent the decay of
turbulence (e.g.\ Mac Low \etal\ 1998) which implies the presence of
external driving mechanisms. These energy sources replenish the
turbulent cascade and may excite MHD waves explaining the inferred
equipartition between turbulent and magnetic energies.

\rskparagraph{Statistical Definitions}
\label{subsec:definitions} 
 The one-point probability distribution function $f(x)$ of a variable
$x$ is defined such that $f(x)dx$ measures the probability for the
variable to be found in the interval $[x,x+dx]$. The {\em density \pdf}
($\rho$-\pdf) discussed in this paper is obtained from the local
density associated with each SPH particle. It is basically the
normalized histogram summed over all particles in the simulation,
i.e.\ a mass-weighted sampling procedure is applied.  The {\em
line-of-sight velocity centroid \pdf} ($v$-\pdf) is more complicated to
compute.  The face of the simulated cube is divided into $64^2$
equal-sized cells. For each cell, the line profile is computed by
sampling the normal (line-of-sight) velocity component of all gas
particles that are projected into that cell. The line centroid is
determined as the abscissa value of the peak of the distribution.
This procedure corresponds to the formation of optically thin lines in
molecular clouds, where all molecules within a certain column through
the clouds contribute equally to the shape and intensity of the line.
To reduce the sampling uncertainties, this procedure is repeated with
the location of the cells shifted by half a cell size in each
direction. Altogether about 20$\,$000 lines contribute to the
\pdf. This is procedure is repeated for line-of-sights along all three
system axes to identify projection effects. The {\em line centroid
increment \pdf} ($\Delta v$-\pdf) is obtained in a similar
fashion. However, the sampled quantity is now the velocity {\em
difference} between line centroids obtained at two distinct locations
separated across the face of the cloud by a fixed shift vector $\Delta
\vec{r}$.  The $\Delta v$-\pdf\ for a spatial lag $\Delta r$ is obtained
as azimuthal average, i.e.~as superposition of all individual \pdf's
with shift vectors of length $\Delta r$.  

Also statistical moments of the distribution can be used to quantify
the spread and shape of \pdf's. For the current analysis we use the
first four moments.  Mean value $\mu$ and standard deviation $\sigma$
(the 1.\ and 2.\ moments) quantify the location and the width of the
\pdf\ and are given in units of the measured quantity. The third and
fourth moments, skewness $\theta$ and kurtosis $\kappa$, are
dimensionless quantities characterizing the shape of the
distribution. The skewness $\theta$ describes the degree of asymmetry
of a distribution around its mean. The kurtosis $\kappa$ measures the
relative peakedness or flatness of the distribution. We use a
definition where $\kappa=3$ corresponds to a normal
distribution. Smaller values indicate existence of a flat peak
compared to a Gaussian, larger values point towards a stronger peak or
equivalently towards the existence of prominent tails in the
distribution. A pure exponential results in $\kappa=6$.  Gaussian
random fields are statistically fully determined by their mean value
and the 2-point correlation function, i.e.~by their first two moments,
$\mu$ and $\sigma$. All higher moments can be derived from those. The
2-point correlation function is equivalent to the power spectrum in
Fourier space (e.g.\ Bronstein \& Semendjajew 1979).

Besides using moments there are other possibilities of characterizing
a distribution. Van den Marel \& Franx (1993) and Dubinski \etal\ (1995) applied Gauss-Hermite expansion series to quantify
non-normal contributions in line profiles.  A more general approach
has been suggested by Vio \etal\ (1994), who discuss alternatives to
the histogram representation of \pdf's. However, as astrophysical data
sets typically {\em are} histograms of various types and as histograms
are the most commonly used method to describe \pdf's, this approach is
also adopted here.

\begin{figure*}[th]
\unitlength1.0cm
\begin{picture}(18,13.5)
\put( -1.75,-13.25){\epsfbox{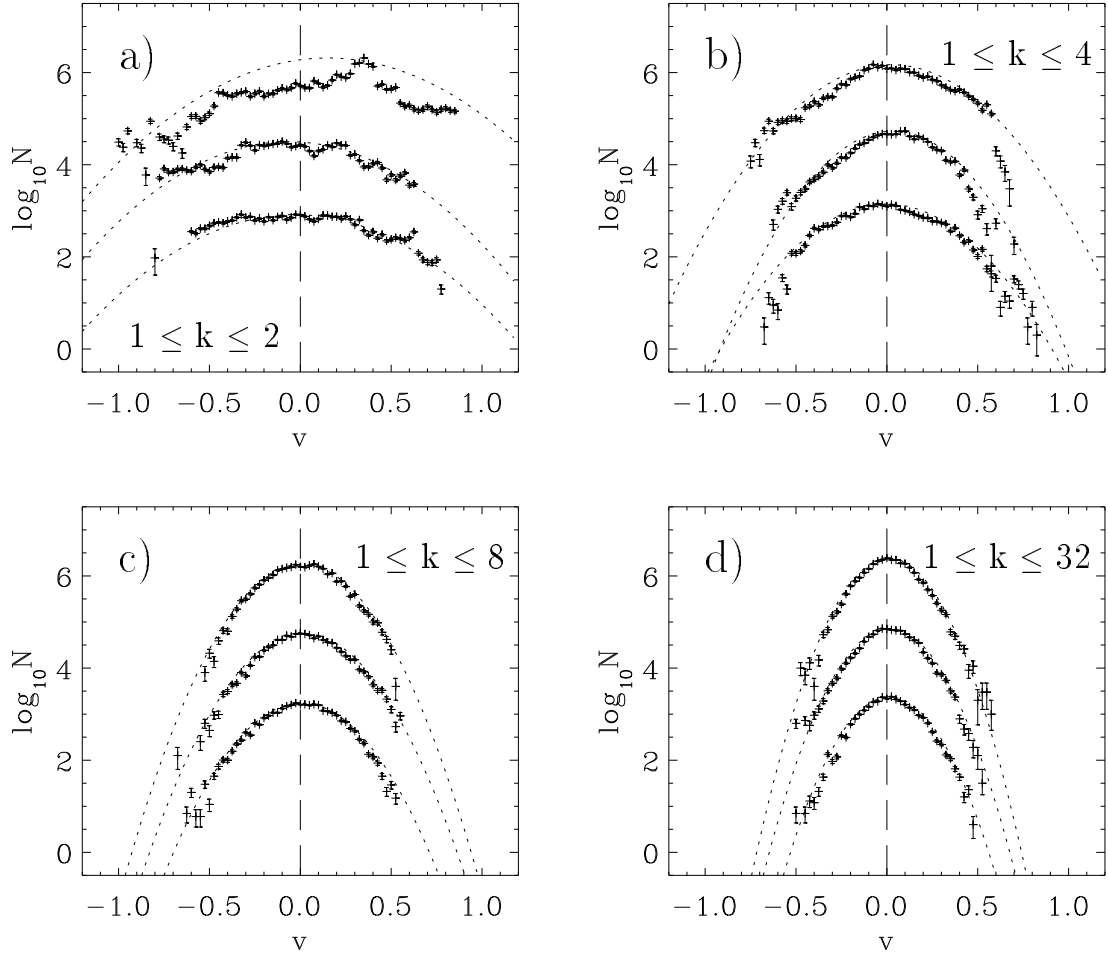}}
\end{picture}
%\begin{center}
%\includegraphics[width=10cm]{ms39926-fig01.ps}
%\end{center}
\caption{\label{fig:v-pdf-init} PDF's of line centroids for a
homogeneous gaseous medium with Gaussian velocity field. The power
spectrum is $P(k) = {\rm const.}$ with wave numbers in the intervals
(a) $1\leq k\leq2$, (b) $1\leq k\leq 4$, (c) $1\leq k\leq 8$, to (d)
$1\leq k\leq 32$. All other modes are suppressed. Each figure plots
\pdf's of the $x$-, $y$-, and $z$-component of the velocity offset by
$\Delta \log_{10} N = 1.5$ (lowest, middle, and upper distribution,
respectively). The length of the error bars
is determined by the square root of the numbers of entries per
velocity bin. The Gaussian fit from the first two moments is shown
with dotted lines. (From Klessen 2000)}
\end{figure*}

\rsksubsubsection{PDF's from Gaussian Velocity Fluctuations}
\label{sec:init}

Variance effects in poorly sampled Gaussian velocity fields can lead
to considerable {\em non}-normal contributions to the $v$- and $\Delta
v$-\pdf's.  If a random process is the result of sequence of
independent events (or variables), then in the limit of large numbers,
its distribution function will be a Gaussian around some mean value.
However, only the properties of a large {\em ensemble} of Gaussian
fields are determined in a statistical sense. Individual realizations
may exhibit considerable deviations from the mean. The effect is
strongest when only few (spatial) modes contribute to the field or,
almost equivalently, when the power spectrum falls off very
steeply. In this case, most kinetic energy is in large-scale motions.

This is visualized in Figure \ref{fig:v-pdf-init}, it shows $v$-\pdf's
for homogeneous gas (sampled by $64^3$ SPH particles placed on a
regular grid) with Gaussian velocity fields with power spectra $P(k) =
{\rm const.}$ which are truncated at different wave numbers $k_{\rm
max}$ ranging from (a) $k_{\rm max}=2$ to (d) $k_{\rm max}=32$. Each
realization is scaled such that the rms velocity dispersion is
$\sigma_v = 0.5$.  The figure displays the \pdf's for the $x$-, $y$-,
and $z$-component of the velocity. The \pdf's of the strongly truncated
spectrum (Figure \ref{fig:v-pdf-init}a) do not at all resemble normal
distributions. The Gaussian statistics of the field is very badly
sampled with only very few modes.  Note that the \pdf's of the same
field may vary considerably for different velocity {\em components},
i.e.~for different {\em projections}.  With the inclusion of larger
number of Fourier modes this situation improves, and in
Figure \ref{fig:v-pdf-init}d the \pdf's of all projections sample the
expected Gaussian distribution very well.

\begin{figure*}[th]
\unitlength1.0cm
\begin{picture}(18,13.4)
\put( -1.55,-13.25){\epsfbox{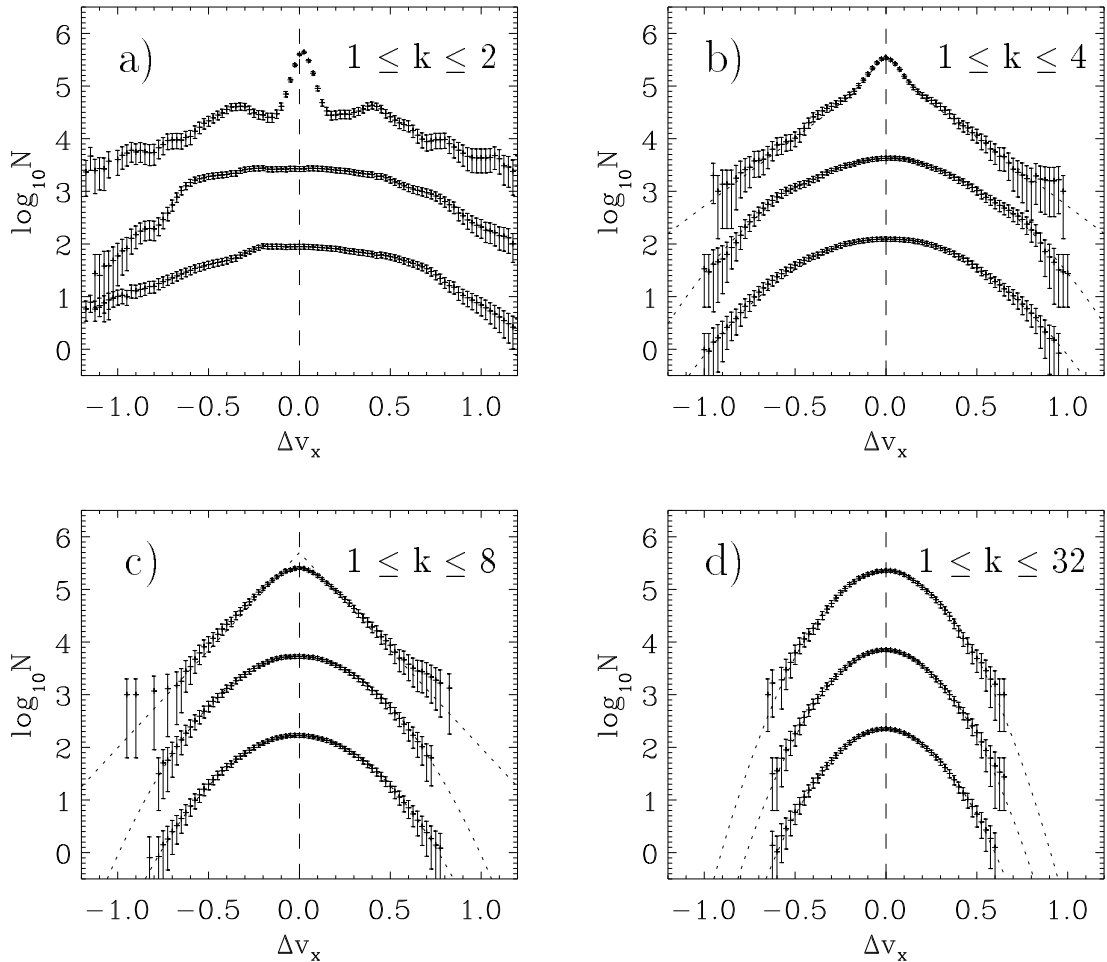}}
\end{picture}
\caption{\footnotesize{\label{fig:dv-pdf-init} PDF's of line centroid
    increments for the same systems as in Figure \ref{fig:v-pdf-init}: (a)
    $1\leq k\leq2$, (b) $1\leq k\leq 4$, (c) $1\leq k\leq 8$, to (d) $1\leq
    k\leq 32$. Each plot shows the distribution of centroid velocity
    differences between locations separated by the distance $\Delta r$ ---
    upper curve: $\Delta r = 1/32$, middle curve: $\Delta r = 10/32$, and
    lower curve: $\Delta r = 30/32$. Only the velocity component for the
    line-of-sight parallel to the $x$-axis is considered. Again, the dotted
    lines represent the best fit Gaussian, except for the upper curve in (b)
    and (c) where the best exponential fit is shown.  (From Klessen 2000)}  
}\end{figure*}

A similar conclusion can be derived for $\Delta v$-\pdf. This measure
is even more sensitive to deviations from Gaussian statistics.
Figure~\ref{fig:dv-pdf-init} plots the $\Delta v$-\pdf's for the same
sequence of velocity fields. For brevity, only the line-of-sight
component parallel to the $x$-axis is considered. Furthermore, from
the sequence of possible $\Delta v$-\pdf's (defined by the spatial lag
$\Delta r$) only three are shown, at small ($\Delta r=1/32$, upper
curve), medium ($\Delta r=10/32$, middle curve), and large spatial
lags ($\Delta r=30/32$, upper curve). Sampling the Gaussian field with
only two modes (Figure \ref{fig:dv-pdf-init}a) is again insufficient to
yield increment \pdf's of normal shape. The velocity field is very
smooth, and the line centroid velocity difference between neighboring
cells is very small. Hence, for $\Delta r=1/32$ the \pdf\ is dominated
by a distinct central peak at $\Delta v = 0$.  The tails of the
distribution are quite irregularly shaped. The situation becomes
`better' when sampling increasing distances, as regions of the fluid
separated by larger $\Delta r$ are less strongly correlated in
velocity. For $\Delta r=10/32$ and $\Delta r=30/32$ the \pdf's follow
the Gaussian distribution more closely although irregularities in the
shapes are still present. In Figures \ref{fig:dv-pdf-init}b and c the
$\Delta v$-\pdf's for medium to large lags are very well fit by
Gaussians. Deviations occur only at small $\Delta r$, the \pdf's are
exponential (and the distribution for $k_{\rm max} = 4$ is still a bit
cuspy). Finally, Figure \ref{fig:dv-pdf-init}d shows the three $\Delta
v$-\pdf's for the case where all available spatial modes contribute to
the velocity field ($1\leq k \leq 32$). The \pdf's follow a Gaussian
for all spatial lags.

\begin{figure*}[th]
\unitlength1.0cm
\begin{picture}(18,6.35)
\put( -1.75,-18.40){\epsfbox{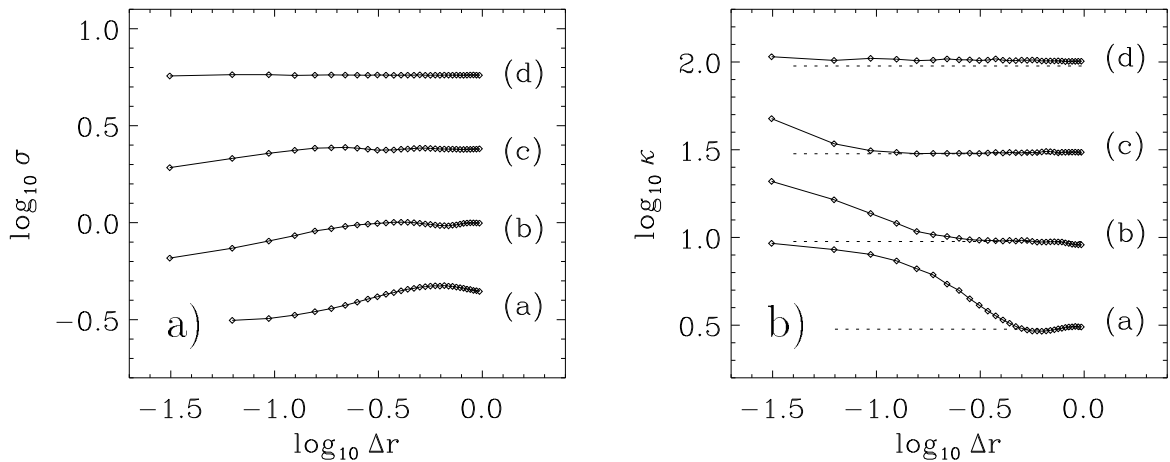}}
\end{picture}
\caption{\footnotesize{\label{fig:mom-init} (a) The second, dispersion $\sigma$, and
(b) the fourth moment, kurtosis $\kappa$, of the distribution of
velocity increments displayed in Figure \ref{fig:dv-pdf-init} as
function of spatial lag $\Delta r$. The letters on the right-hand
sight indicate the corresponding time with (a) $t=0$, (b) $t=0.5$, and
so forth.  Each plot is
offset by $\Delta \log_{10} \sigma=0.5$ and $\Delta \log_{10} \sigma =
0.5$, and in (b) the horizontal dotted line indicates the value for a
Gaussian $\kappa = 3$ ($\log_{10} \kappa = 0.48$). (From Klessen 2000) }
}\end{figure*}

This behavior is also seen in the variation of the moments of the
distribution as function of the spatial lag $\Delta r$.  Applied to
the above sequence of Gaussian velocity fields,
Figure \ref{fig:mom-init} displays the dispersion $\sigma$ and the
kurtosis $\kappa$ of the distribution. The corresponding models are
indicated at the right hand side of each plot. The width of the
distribution, as indicated by the dispersion $\sigma$
(Figure \ref{fig:mom-init}a), typically grows with increasing $\Delta
r$, reflecting the relative peakedness of the distribution at small
lags. For example, the distribution (a) yields a slope of ~0.3 in the
range $-0.6\leq \log_{10} \Delta r \leq -0.4$, and (b) leads to a
value of 0.2 in relatively large interval $-1.5 \leq \log_{10} \Delta
r \leq -0.5$. The effect disappears for the better sampled
fields. Typical values for that slope in observed molecular clouds are
$-0.3$ to $-0.5$ (Miesch \etal\ 1998).\footnote{Note, that Miesch \etal\ (1998) are plotting the function $\sigma^2$ versus the spatial lag
$\Delta r$. For a comparison with the present study, their numbers
have to be divided by a factor of two. Furthermore, they use a
relatively narrow range of $\Delta r$-values to compute the slope of
the function; larger intervals would on average tend to decrease these
values (see their Figure 14). In addition, Miesch \etal\ (1998) applied
spatial filtering to remove large-scale velocity gradients in the
clouds. These would lead to steeper slopes.  The fact that in the
present study the functions $\sigma$ and $\kappa$ level out for large
spatial lags $\Delta r$ is a consequence of the periodic boundary
conditions which do not allow for large-scale
gradients.\label{comment}} A direct measure of the peakedness of the
distribution is its fourth moment, the kurtosis $\kappa$
(Figure \ref{fig:mom-init}b).  At small lags $\Delta r$, clearly the
\pdf's of model (a) are more strongly peaked than exponential ($\kappa
=6$). Comparing the entire sequence reveals again the tendency of the
\pdf's to become Gaussian at decreasing $\Delta r$ with increasing
number of modes considered in the construction of the velocity field.

Taking all together, it is advisable to consider conclusions about
interstellar turbulence derived from solely analyzing one-point
probability distribution functions from molecular clouds with
caution. Similar to what has been shown by Dubinski \etal\ (1995) for
molecular line profiles, deviations from the regular Gaussian shape
found in $v$- and $\Delta v$-\pdf's need not be the signpost of
turbulent intermittency.  Gaussian velocity fields which are dominated
by only a small number of modes (either because the power spectrum
falls off steeply towards larger wave numbers, or because small wave
length distortions are cut away completely) will lead to very similar
distortions. In addition, the properties of the \pdf\ may vary
considerably between different projections. The same velocity field
may lead to smooth and Gaussian \pdf's for one velocity component,
whereas another projection may result in strong non-Gaussian wings
(see also Figure \ref{fig:v-pdf-xyz-decay-with-gravity}).

\rsksubsubsection{Analysis of Decaying Su\-per\-sonic Turbulence without Self-Gravity}
\label{sec:decay-non-grav}

In this section the \pdf's of freely decaying initially highly
supersonic turbulence without self-gravity are discussed.  They are
calculated from an SPH simulation with $350\,000$ particles (Mac~Low
\etal\ 1998, model G). Initially the system is homogeneous with a
Gaussian velocity distribution with $P(k) = {\rm const.}$ in the
interval $1 \leq k \leq 8$. The rms Mach number of the flow is ${\cal M}=5$.

After the onset of the hydrodynamic evolution the flow quickly becomes
fully turbulent resulting in rapid dissipation of kinetic energy. The
energy decay is found to follow a power law $t^{-\eta}$ with exponent
$\eta = 1.1 \pm 0.004$.  The overall evolution can be subdivided into
several phases. The first phase is very short and is defined by the
transition of the initially Gaussian velocity field into fully
developed supersonic turbulence. It is determined by the formation of
the first shocks which begin to interact with each other and build up
a complex network of intersecting shock fronts. Energy gets transfered
from large to small scales and the turbulent cascade builds up. The
second phase is given by the subsequent self-similar evolution of the
network of shocks. Even though individual features are transient, the
overall properties of this network change only slowly. In this phase
of highly supersonic turbulence the loss of kinetic energy is
dominated by dissipation in shocked regions. In the transsonic regime,
i.e.~the transition from highly supersonic to fully subsonic flow,
energy dissipation in vortices generated by shock interactions becomes
more and more important. Only the strongest shocks remain in this
phase. Surprisingly, the energy decay law does not change during this
transition. It continues to follow a power law with exponent $\eta
\approx 1$.  In the subsonic phase the flow closely resembles
incompressible turbulence.  Its properties are similar to those
reported from numerous experiments and simulations (e.g.~Porter \etal\ 1994, Lesieur 1997, Boratav \etal\ 1997). The simulation is
stopped at $t=20.0$ when the flow has decayed to a rms Mach number of
${\cal M}=0.3$.  Since the energy loss rate follows a power law, the duration
of each successive phase grows.

\begin{figure*}[th]
\unitlength1.0cm
\begin{picture}(18,6.2)
\put( -1.75,-18.40){\epsfbox{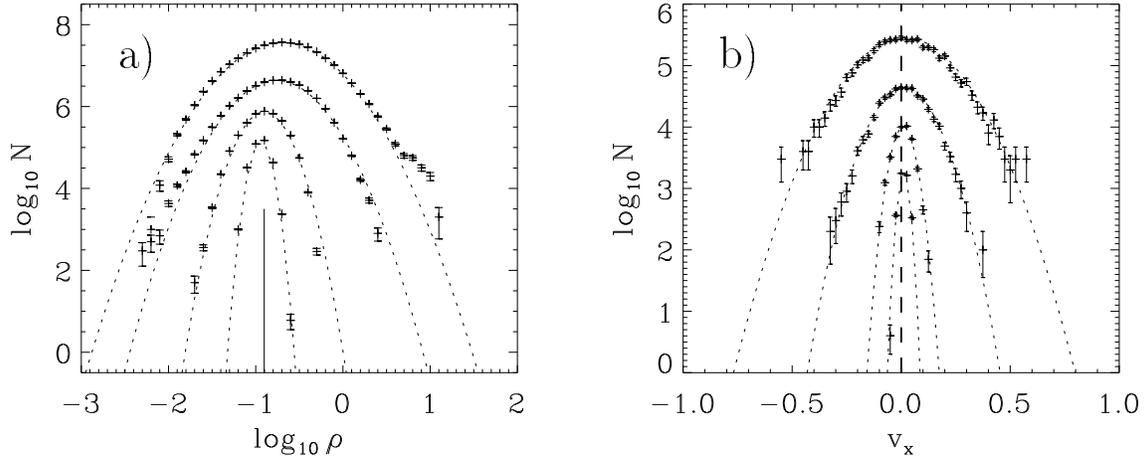}}
\end{picture}
\caption{\footnotesize{\label{fig:rho-v-pdf-decay-no-gravity}
    PDF's of (a) density and of (b) centroid velocities for the line-of-sight
    being parallel to the $x$-axis of the system. The \pdf's are obtained at
    four different phases of the dynamical evolution of the system (see the
    main text), at $t=0.2$ (upper curves), at $t=0.6$ (second curve from the
    top), at $t=3.5$ (third curve), and at $t=20.0$ (lowest curve). These
    times correspond to Mach numbers ${\cal M}=5.0$, ${\cal M}=2.5$, ${\cal
      M}=1.0$, and ${\cal M}=0.3$, respectively. For each distribution, the
    best-fit Gaussian is indicated using dotted lines. (From Klessen 2000) }
  }\end{figure*}

This sequence of evolutionary stages is seen in the \pdf's of the
system. One noticeable effect is the decreasing width of the
distribution functions as time progresses. As the kinetic energy
decays the available range of velocities shrinks. This not only leads
to `smaller' $v$- and $\Delta v$-\pdf's, but also to a smaller
$\rho$-\pdf\ since compressible motions lose influence and the system
becomes more homogeneous. This is indicated in
Figure \ref{fig:rho-v-pdf-decay-no-gravity}, it displays (a) the
$\rho$-\pdf\ and (b) $v$-\pdf\ at the following stages of the dynamical
evolution (from top to bottom): Shortly after the start, at $t=0.2$
when the first shocks occur, then at $t=0.6$ when the network of
interacting shocks is established and supersonic turbulence is fully
developed, during the transsonic transition at $t=3.5$, and finally at
$t=20.0$ when the flow has progressed into the subsonic regime. The
rms Mach numbers at these stages are ${\cal M}=5.0$, ${\cal M}=2.5$, ${\cal M}=1.0$, and
${\cal M}=0.3$, respectively. The density \pdf\ always closely follows a
log-normal distribution, i.e.~it is Gaussian in the {\em logarithm} of
the density. Also the distribution of line centroids at the four
different evolutionary stages of the system is best described by a
Gaussian with only minor deviations at the far ends of the velocity
spectrum.

\begin{figure*}[ht]
\unitlength1.0cm
\begin{picture}(16,12.5)
\put( -1.75,-13.55){\epsfbox{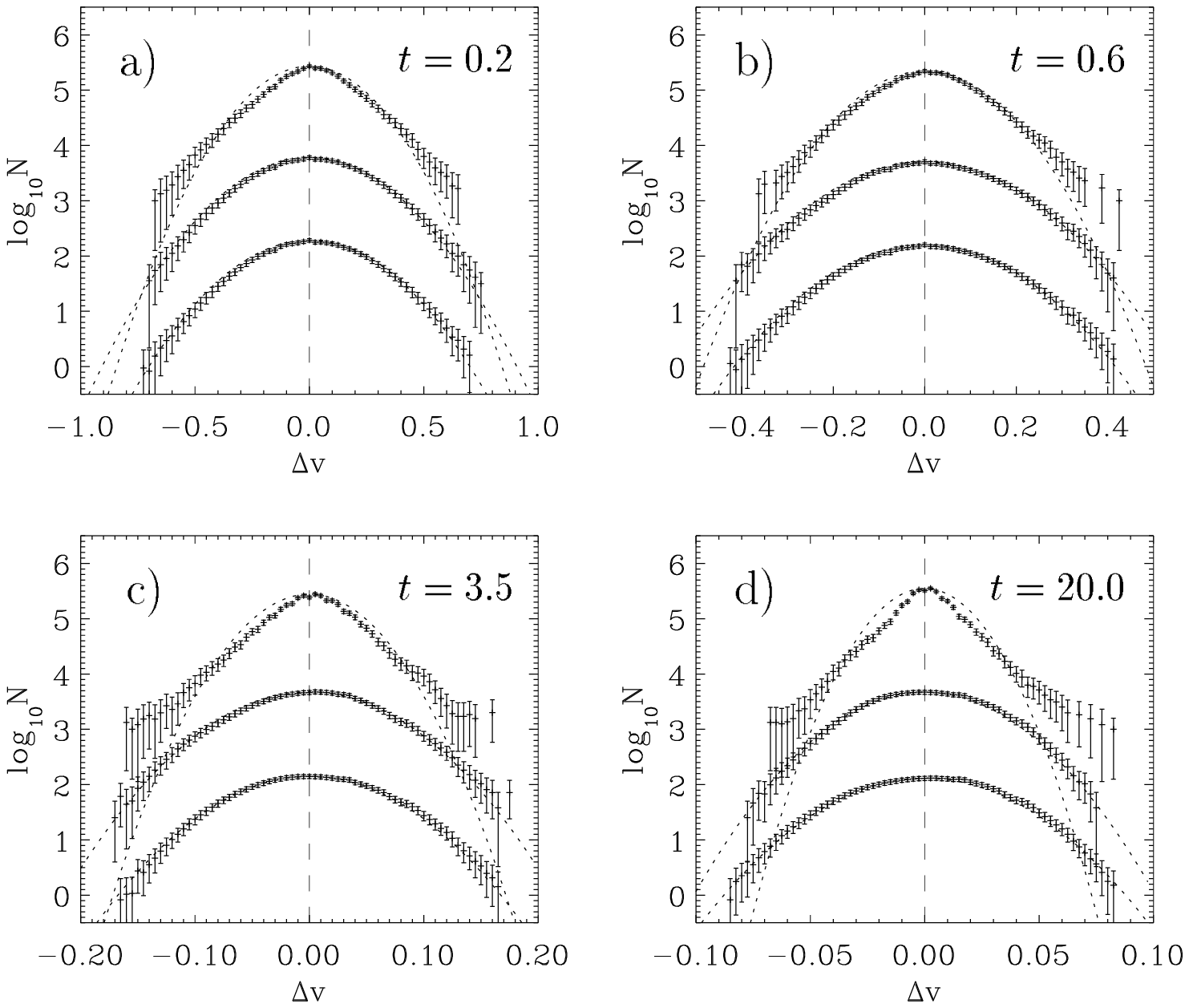}}
\end{picture}
\caption{\footnotesize{\label{fig:dv-pdf-decay-no-gravity} PDF's of the
$x$-component of the centroid velocity increments for three spatial
lags: upper curve -- $\Delta r = 1/32$, middle curve -- $\Delta r =
10/32$, and lower curve -- $\Delta r = 30/32$. As in
Figure \ref{fig:rho-v-pdf-decay-no-gravity}, the \pdf's are obtained at
(a) $t=0.2$ , (b) $t=0.6$, (c) $t=3.5$, and (d) $t=20.0$. The Gaussian
fits are again indicated by dotted lines. (From Klessen 2000) }
}\end{figure*}

For the same points in time, Figure \ref{fig:dv-pdf-decay-no-gravity}
shows the $\Delta v$-\pdf's for $x$-component of the velocity. The
displayed spatial lags are selected in analogy to
Figure \ref{fig:dv-pdf-init}.  Note the different velocity scaling in
each plot reflecting the decay of turbulent energy as the system
evolves in time.  Throughout the entire sequence, spatial lags larger
than about 10\% of the system size always lead to $\Delta v$-\pdf's
very close to Gaussian shape (the middle and lower curves).
Considerable deviations occur only at small spatial lags (the upper
curves). For those, the increment \pdf's exhibit exponential wings
during all stages of the evolution. When scaling the \pdf's to the same
width, the distribution in the subsonic regime (d) appears to be more
strongly peaked than during the supersonic or transsonic phase (a --
c). There, the central parts of the \pdf's are still reasonably well
described by the Gaussian obtained from the first two moments, whereas
in (d) the peak is considerably narrower, or vice versa, the tails of
the distribution are more pronounced.

These results can be compared with the findings by Lis \etal\ (1998). They report increment \pdf's for three snapshots of a
high-resolution hydrodynamic simulation of decaying mildly super-sonic
turbulence performed by Porter \etal\ (1994). They analyze the system
at three different times corresponding to rms Mach numbers of
${\cal M}\approx 0.96$, ${\cal M}\approx 0.88$, and ${\cal M}\approx 0.52$. Their first two
data sets thus trace the transition from supersonic to subsonic flow
and are comparable to phase (c) of the current model; their last data
set corresponds to to phase (d).  In the transsonic regime both
studies agree: Lis \etal\ (1998) report enhanced tails in the
increment \pdf's for the smallest spatial lags which they considered
and near Gaussian distributions for larger lags (however, the largest
separation they study is about 6\% of the linear extent of the
system). In the subsonic regime, Lis \etal\ (1998) find near Gaussian
\pdf's for very small spatial lags ($<1$\%), but extended wings in the
\pdf's for lags of 3\% and 6\% of the system size. They associate this
with the `disappearance' of large-scale structure. Indeed, their
Figure 7 exhibits a high degree of fluctuations on small scales which
they argue become averaged away when considering small spatial lags in
the $\Delta v$-\pdf. Comparing the \pdf\ with spatial lags of 3\% (upper
curves in Figure \ref{fig:dv-pdf-decay-no-gravity}, compared to the
\pdf's labeled with $\Delta = 15$ in Lis \etal\ 1998) both studies come
to the same result. At these scales the $\Delta v$-\pdf's tend to
exhibit more pronounced wings in the subsonic regime as in the
supersonic regime. The SPH calculations reported here do not allow for
a meaningful construction of $\delta v$-\pdf's for $\Delta r <
3$\%. The Gaussian behavior of \pdf's for very small spatial lags
reported by Lis \etal\ (1998) therefore cannot be examined.  However,
neither of the purely hydrodynamic simulations lead to \pdf's that
are in good agreement with the observations. Observed \pdf's typically
are much more centrally peaked at small spatial separation (see e.g.\
Figure 4 in Lis \etal\ 1998 and Miesch \etal\ 1998).

\begin{figure*}[th]
\unitlength1.0cm
\begin{picture}(16,15.3)
\put( -0.0,-7.4){\epsfxsize=17cm \epsfbox{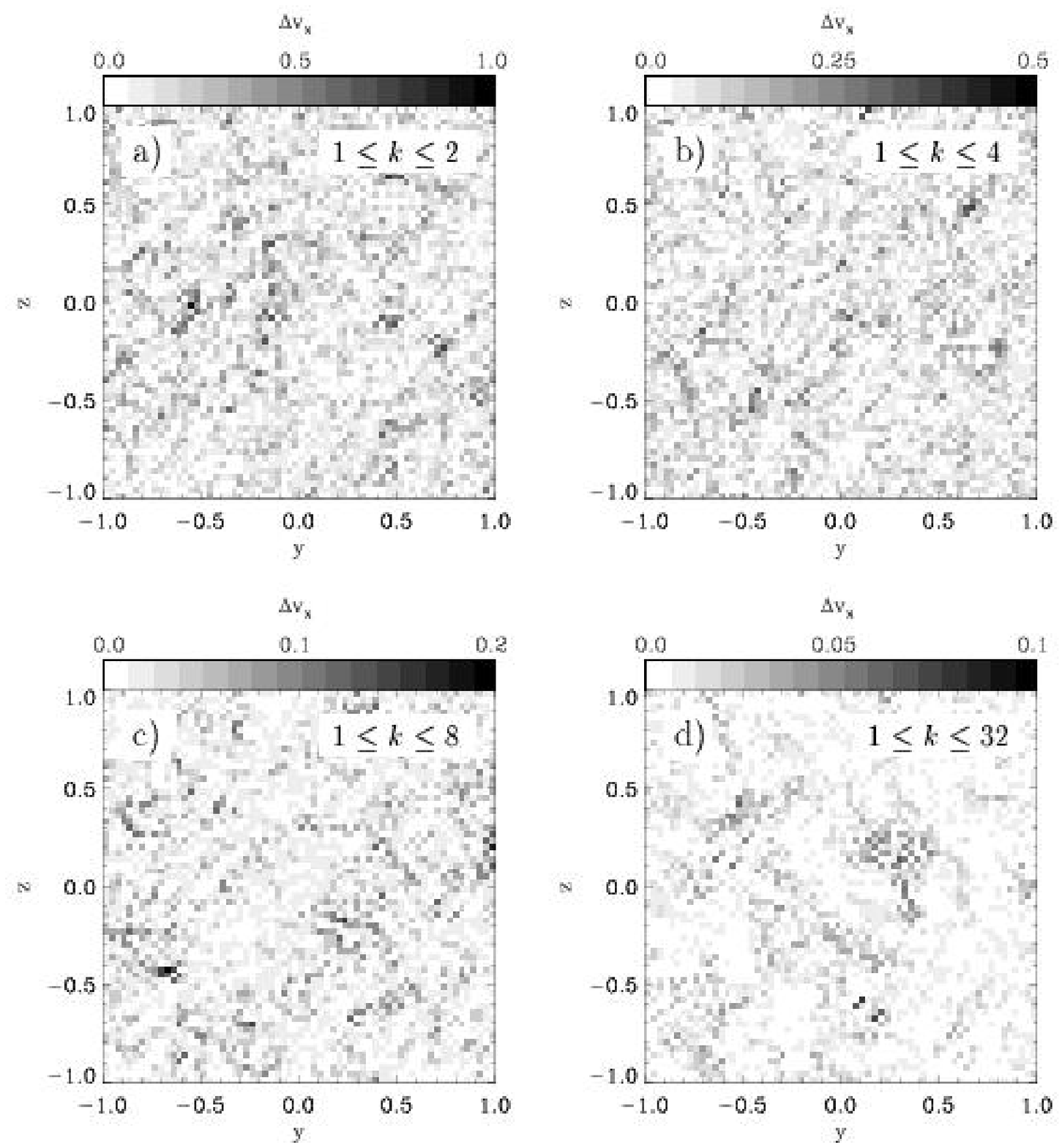}}
\end{picture}
\caption{\footnotesize{\label{fig:dv-array-decay-no-gravity}
2-dimensional distribution (in the $yz$-plane) of centroid
increments for velocity profiles along the $x$-axis of the system
between locations separated by a vector lag $\Delta \vec{r} =
(1/32,1/32)$. Analog to the previous figures, the data are displayed
for times (a)  $t=0.2$, (b) $t=0.6$, (c) $t=3.5$, and (d) $t=20.0$.
The magnitude of the velocity increment $\Delta v_{x}$ is indicated at
the top of each plot;  note the different scaling. (From Klessen 2000)
}
}\end{figure*}
Figure \ref{fig:dv-array-decay-no-gravity} shows the spatial
distribution of centroid velocity differences between cells separated
by a vector lag of $\Delta \vec{r} = (1/32,1/32)$ (i.e.~between
neighboring cells along the diagonal). Data are obtained at the same
times as above.  Each figure displays the array of the absolute values
of the velocity increments $\Delta v_x$ in linear scaling as indicated
at the top. Note the decreasing velocity range reflecting the decay of
turbulent energy.  The distribution of $\Delta v_x$ appears random,
there is no clear indication for coherent structures.  This is
corresponds to most observations. Miesch \etal\ (1998) find for their
sample of molecular clouds that high-amplitude velocity differences
for very small spatial lags typically are well distributed resulting
in a `spotty' appearance. Note, however, that using azimuthal
averaging Lis \etal\ 1998 report the finding of filamentary structures
for the $\rho$-Ophiuchus cloud. Altogether, filamentary structure is
difficult to define and a mathematical thorough analysis is seldomly
performed (for an astrophysical approach see Adams \& Wiseman 1994,
for a discussion of the filamentary vortex structure in incompressible
turbulence consult Frisch 1995 or Lesieur 1997). The visual inspection
of maps is often misleading and influenced by the parameters used to
display the image.  Larger velocity bins for instance tend to produce
a more `filamentary' structure than very fine sampling of the velocity
structure. Further uncertainty may be introduced by the fact that
molecular clouds are only seen in one projection as the signatures of
the dynamical state of the system can strongly depend on the viewing
angle.

\rsksubsubsection{Analysis of Decaying Turbulence with Self-Gravity}
\label{sec:decay-with-gravity}
In this section,  we concentrate on the properties of decaying, initially
supersonic turbulence in a self-gravitating medium. 
%
% Figure
% \ref{fig:3D-plot-decay-with-gravity} displays an SPH simulation with
% $200\,000$ particles at six different times of its dynamical
% evolution. Since the model is subject to periodic boundary conditions,
% every figure has to be considered infinitely replicated in each
% direction. 
%
We analyze an SPH simulation with  
 $205\,379$ particles which  is initially
homogeneous and its velocity field is generated with $P(k) = {\rm
const.}$ using modes with wave numbers $1\leq k \leq 8$. From the
choice $\alpha = 0.01$ it follows that the system contains 120 {\em
thermal} Jeans masses.  The initial rms velocity dispersion is
$\sigma_v = 0.5$ and with the sound speed $c_{\rm s} = 0.082$ the rms
Mach number follows as ${\cal M}=6$.  These values imply that the initial
turbulent velocity field contains sufficient energy to globally {\em
stabilize} the system against gravitational collapse.  Scaled to
physical units using a density $n({\rm H_2}) = 10^5\,$cm$^{-3}$, which
is typical for massively star-forming regions (e.g.\ Williams \etal\
2000), the system corresponds to a volume of $[0.32\,{\rm pc}]^3$ and
contains a gas mass of 200$\,$M$_{\odot}$.  As the simulation starts,
the system quickly becomes fully turbulent and loses kinetic energy.
Like in the case without self-gravity a network of intersecting shocks
develops leading to density fluctuations on all scales.  During the early evolution, there is enough
kinetic energy to prevent global collapse and the
properties of the system are similar to those of pure hydrodynamic
turbulence discussed in the previous section. However, as time progresses and turbulent energy decays
the effective Jeans mass decreases, and after sufficient time  also
large-scale collapse becomes possible.  Gas clumps follow the global
flow pattern towards a common center of gravity where they may merge
or sub-fragment. Gradually a cluster of dense cores is built up. In
the isothermal model this process continues until all available gas is
accreted onto the `protostellar' cluster (see the discussion in Section
 \ref{sec:paradigm}).

\begin{figure*}[th]
\unitlength1.0cm
\begin{picture}(16, 9.4)
\put( -1.75,-16.4){\epsfbox{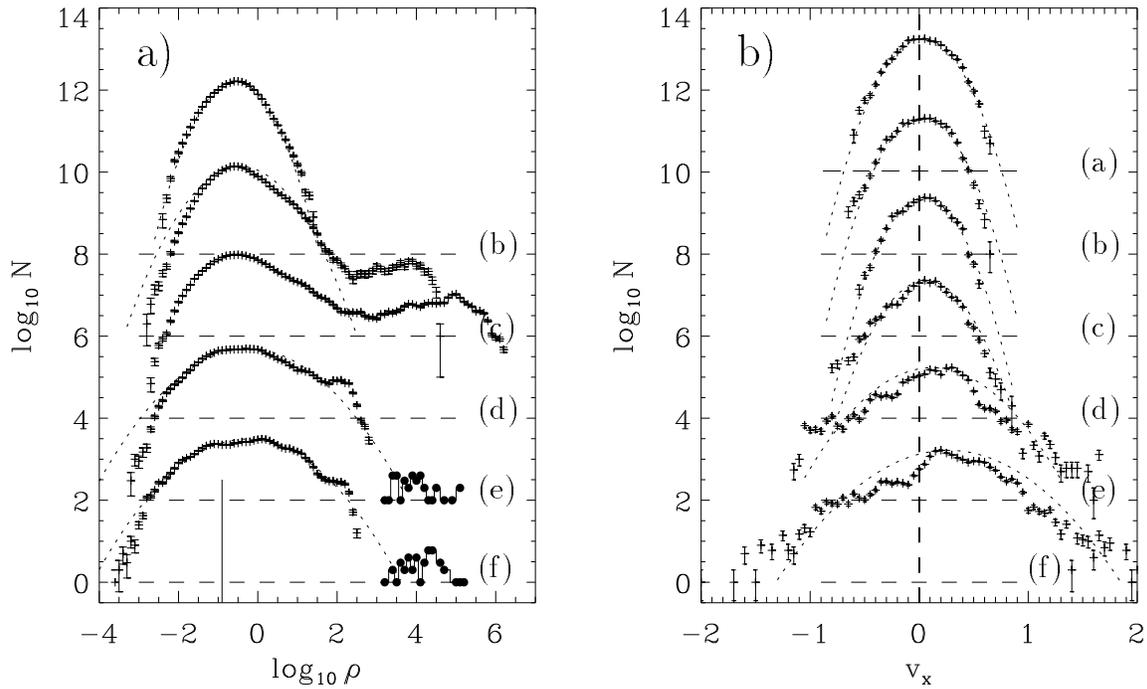}}
\end{picture}
\caption{\footnotesize{\label{fig:rho-v-pdf-decay-with-gravity} PDF's of (a) the
density and (b) the $x$-component of line centroids for the simulation
of initially supersonic, decaying turbulence in self-gravitating
gas. The time sequence again is (a) $t=0.0$, (b) $t=0.5$, and so forth. In the
left panel, the initial density is indicated by the vertical line at
$\rho = 1/8$. The density contributions from collapsed cores forming
in the late stages of the evolution are indicated by solid dots. The
core density corresponds to a mean value computed from the core mass
divided by its accretion volume.  In both figures, each \pdf\ is offset
by $\Delta \log_{10} N = 2.0$ with the base $\log_{10} N = 0.0$
indicated by horizontal dashed lines. The best-fit Gaussian curves are
shown as dotted lines.  (From Klessen 2000)}
}\end{figure*}

The \pdf's of (a) the density and of (b) the $x$-component of the line
centroid velocities for the above six model snapshots are displayed in
Figure \ref{fig:rho-v-pdf-decay-with-gravity}. The corresponding time is
indicated by the letters at the right side of each panel.  During the
dynamical evolution of the system the density distribution develops a
high density tail. This is the imprint of local collapse. The
densities of compact cores are indicated by solid dots (at $t=2.0$ and
$t=2.5$). Virtually all particles in the high density tails at earlier
times (at $t=1.0$ and more so at $t=1.5$) are accreted onto these
cores. The bulk of matter roughly follows a log-normal density
distribution as indicated by the dotted parabola. The $v$-\pdf's are
nearly Gaussian as long as the dynamical state of the system is
dominated by turbulence. Also the width of the \pdf\ remains roughly
constant during this phase. This implies that the decay of turbulent
kinetic energy is in balance with the gain of kinetic energy due to
gravitational (`quasi-static') contraction on large scales. The time
scale for this process is determined by the energy dissipation in
shocks and turbulent eddies.  However, once {\em localized} collapse
is able to set in, accelerations on small scales increase dramatically
and the evolution `speeds up'.  For times $t>2.0$ the centroid \pdf's
become wider and exhibit significant deviations from the original
Gaussian shape.  The properties of the \pdf's are similar to those
observed in star-forming regions (Miesch \& Scalo 1995, Lis \etal\ 1998, Miesch \etal\ 1998).  This is expected since gravitational
collapse is a necessary ingredient for forming stars.

Gravity creates {\em non}-isotropic density and velocity structure
structures. When analyzing $v$- and $\Delta v$-\pdf's, their appearance
and properties will strongly depend on the viewing angle.  This is a
serious point of caution when interpreting observational data, as
molecular cloud are seen only in {\em one} projection.  As
illustration, Figure \ref{fig:v-pdf-xyz-decay-with-gravity} plots the
centroid \pdf\ at the time $t=2.0$ for the line-of-sight projection
along all three axes of the system. Whereas the \pdf's for the $x$- and
the $y$-component of the velocity centroid are highly structured
(upper and middle curve -- the latter one is even double peaked), the
distribution of the $z$-component (lowest curve) is smooth and much
smaller in width, comparable to the `average' \pdf\ at {\em earlier}
stages of the evolution.  As the variations between different viewing
angles or equivalently different velocity components can be very
large, statements about the 3-dimensional velocity structure from only
observing one projection can be misleading.

\begin{figure}[th]
\unitlength1.0cm
\begin{picture}(18,6.2)
\put(-2.50,-18.4){\epsfbox{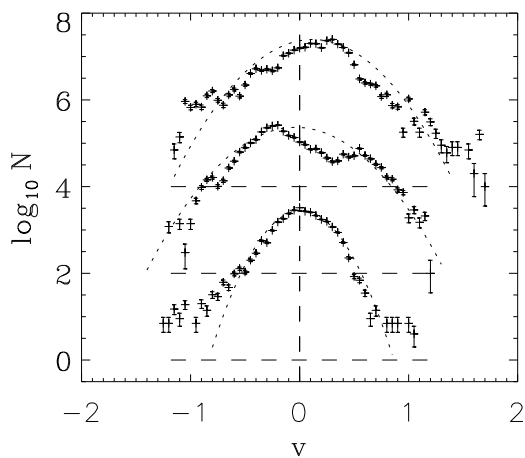}}
\end{picture}
\caption{\footnotesize{\label{fig:v-pdf-xyz-decay-with-gravity} Centroid velocity
PDF's for the simulation of initially supersonic, decaying turbulence
in self-gravitating gas at $t=2.0$ for the line-of-sight being along
the $x$-axis (upper curve -- it is identical to the fifth \pdf\ in
Figure \ref{fig:rho-v-pdf-decay-with-gravity}b), along the $y$-axis
(middle), and along the $z$-axis of the system (bottom).  Each
distribution is offset by $\Delta \log_{10} N= 2.0$ with the
horizontal lines indicating the base $\log_{10} N= 0.0$. The \pdf's of
various projections and velocity components can differ considerably. (From
Klessen 2000) 
}
}\end{figure}

\begin{figure*}[th]
\unitlength1.0cm
\begin{picture}(18,12.8)
\put( -1.75,-13.45){\epsfbox{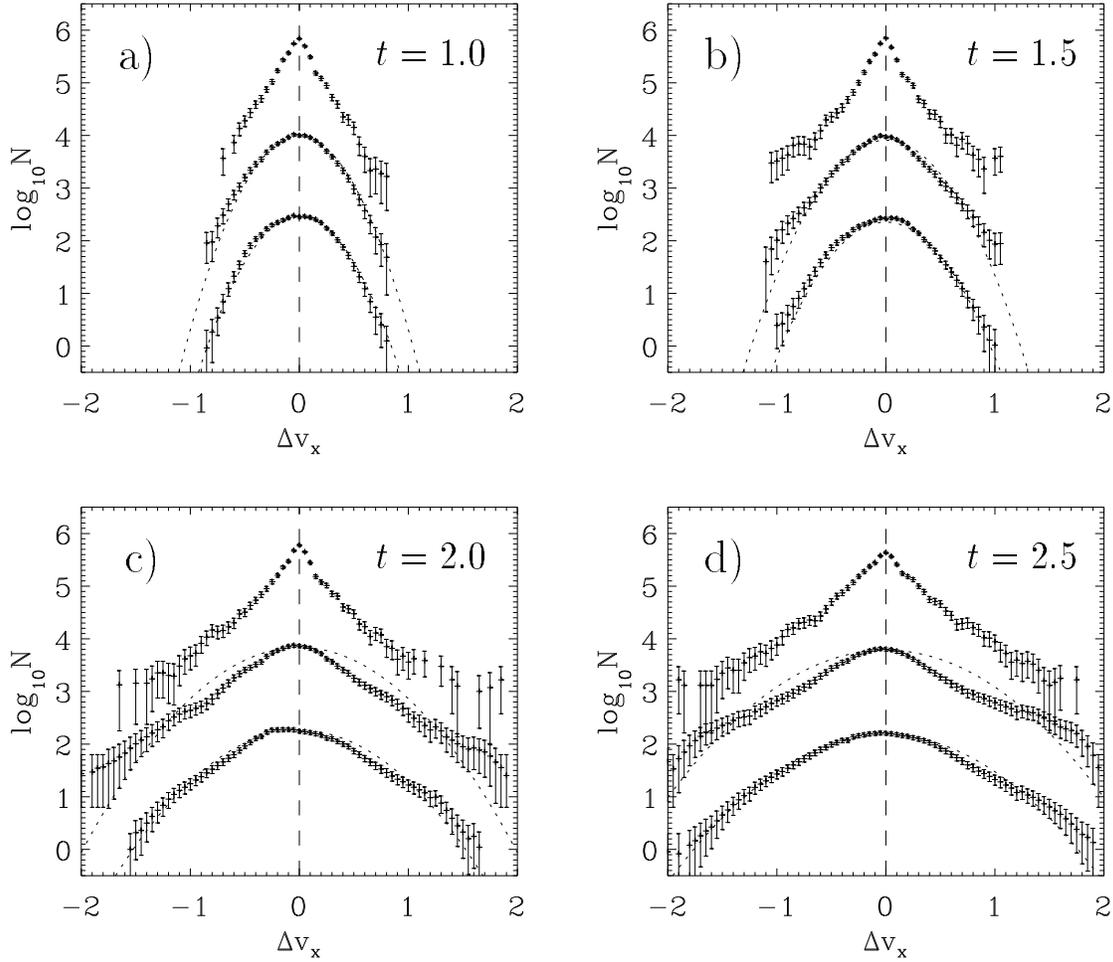}}
\end{picture}
\caption{\footnotesize{\label{fig:dv-pdf-decay-with-gravity} PDF's of the
$x$-component of the centroid velocity increments for three spatial
lags: upper curve -- $\Delta r = 1/32$, middle curve -- $\Delta r =
10/32$, and lower curve -- $\Delta r = 30/32$. The functions are
computed from the simulation of initially supersonic, decaying
turbulence in self-gravitating gas at (a) $t=1.0$, (b) $t=1.5$, (c)
$t=2.0$, and (d) $t=2.5$.  Where appropriate, the Gaussian curves
obtained from the first two moments of the distribution are indicated
by dotted lines. During the early phases of the evolution, the flow is
similar to pure hydrodynamic turbulence (the \pdf's are close to the
ones in Figure \ref{fig:dv-pdf-decay-no-gravity}). As turbulent energy
decays self-gravity gains influence and the late stages of the
evolution are dominated by gravitational contraction. Consequently the
\pdf's in the sequence (a) to (d) become more and more non-Gaussian
with the progression of time. This  concerns the \pdf's for
small to intermediate  lags $\Delta r$. (From Klessen 2000) }
}\end{figure*}

Gravity effects the $\Delta v$-\pdf. Figure
\ref{fig:dv-pdf-decay-with-gravity} displays the increment \pdf's at
small, intermediate and large spatial lags, analog to Figures
\ref{fig:dv-pdf-init} and \ref{fig:dv-pdf-decay-no-gravity}. Time
ranges from (a) $t=1.0$ to (d) $t=2.5$. The \pdf's for $t=0.0$ and
$t=0.5$ are not shown since at these stages supersonic turbulence
dominates the dynamic of the system and the \pdf's are comparable to
the ones without gravity
(Figure \ref{fig:dv-pdf-decay-no-gravity}). This still holds for
$t=1.0$. The increment \pdf's for medium to large spatial lags appear
Gaussian, however, the \pdf\ for the smallest lag follows a perfect
exponential all the way inwards to $\Delta v = 0$. Unlike in the case
without gravity, the peak of the distribution is not `round', i.e.~is
not Gaussian in the innermost parts (when scaled to the same
width). {\em It is a sign of self-gravitating systems that the
increment \pdf\ at smallest lags is very strongly peaked and remains
exponential over the entire range of measured velocity
increments}. This behavior is also seen Figures
\ref{fig:dv-pdf-decay-with-gravity}b--d. At these later stages of the
evolution in addition non-Gaussian behavior is also found at medium
lags. This results from the existence of large-scale filaments and
streaming motions. The same behavior is found for the increment \pdf's
from observed molecular clouds (for $\rho$-Ophiuchus see Lis et
al~1998; for Orion, Mon R2, L1228, L1551, and HH83 see Miesch \etal\ 1998). In each case, the distribution for the smallest lag (one
pixel size) is very strongly peaked at $\Delta v = 0$, in some cases
even more than exponential. The deviations from the Gaussian shape
remain for larger lags but are not so pronounced.  The inclusion of
self-gravity into models of interstellar turbulence leads to good
agreement with the observed increment \pdf's. However, this result may
{\em not} be unique as in molecular clouds additional processes are
likely to be present that could also lead to strong deviations from
Gaussianity. 

\begin{figure*}[t]
\unitlength1.0cm
\begin{picture}(16,7.5)
\put( -1.75,-16.15){\epsfbox{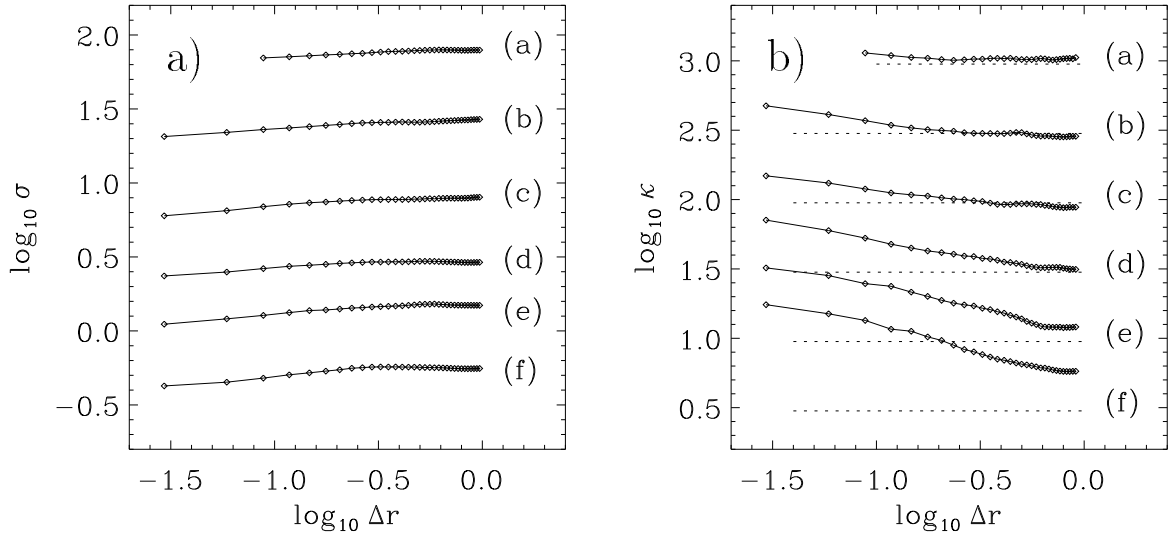}}
\end{picture}
\caption{\footnotesize{\label{fig:mom-decay-with-gravity} (a) The second, dispersion
$\sigma$, and (b) the fourth moment, kurtosis $\kappa$, as function of
spatial lag $\Delta r$ for the distribution of velocity increments in
the simulation of self-gravitating, decaying, supersonic
turbulence. The letters on the right-hand side indicate the time at
which the increment \pdf's are computed ranging from $t=0.0$ at the top
down to $t=2.5$ at the bottom 
(see Figure
%\ref{fig:3D-plot-decay-with-gravity} or
\ref{fig:rho-v-pdf-decay-with-gravity}). Each \pdf\ is offset by $\Delta
\log_{10} \sigma=0.5$ and $\Delta \log_{10} \kappa = 0.5$,
respectively. (From Klessen 2000)}
}\end{figure*}

The time evolution of the statistical moments of the $\Delta v$-\pdf\ 
for various spatial lags is presented in
Figure \ref{fig:mom-decay-with-gravity}. It plots (a) the dispersion
$\sigma$, and (b) the kurtosis $\kappa$. 
%The letters on the right-hand
%side indicate the corresponding time in
%Figure \ref{fig:3D-plot-decay-with-gravity}. 
At $t=0.0$ the width
$\sigma$ of the \pdf\ is approximately constant for all $\Delta r$ and
the kurtosis $\kappa$ is close to normal value of three.  Both
indicate that Gaussian statistics very well describes the initial
velocity field. As turbulent energy decays, gravitational collapse
sets in. Because of the gravitational acceleration, the amplitudes of
centroid velocity differences between separate regions in the cloud
grow larger, the width $\sigma$ of the $\Delta v$-\pdf's
increases. This becomes more important when sampling velocity
differences on larger spatial scales, hence $\sigma$ also increases
with $\Delta r$. The slope is $d \log_{10} \sigma/d\log_{10}\Delta r
{\:\lower0.6ex\hbox{$\stackrel{\textstyle <}{\sim}$}\:} 0.2$. For
$\log_{10}\Delta r > -0.4$ it levels out, which is a result of the
adopted periodic boundary conditions. They do not allow for
large-scale velocity gradients. The increasing `peakedness' of $\Delta
v$-\pdf\ is reflected in the large values of the kurtosis $\kappa$ at
the later stages of the evolution. For small spatial lags the \pdf's
are more centrally concentrated than exponential (i.e.~$\kappa > 6$),
and even at large spatial separations they are still more strongly
peaked than Gaussian ($\kappa > 3$). The slope at $t=2.5$ is $d
\log_{10} \kappa/d\log_{10}\Delta r \approx -0.4$ which is indeed
comparable to what is found in observed star-forming regions (Miesch
\etal\ 1998).

\begin{figure*}[th]
\unitlength1.0cm
\begin{picture}(16,15.2)
\put( -0.0,-7.5){\epsfxsize=17cm \epsfbox{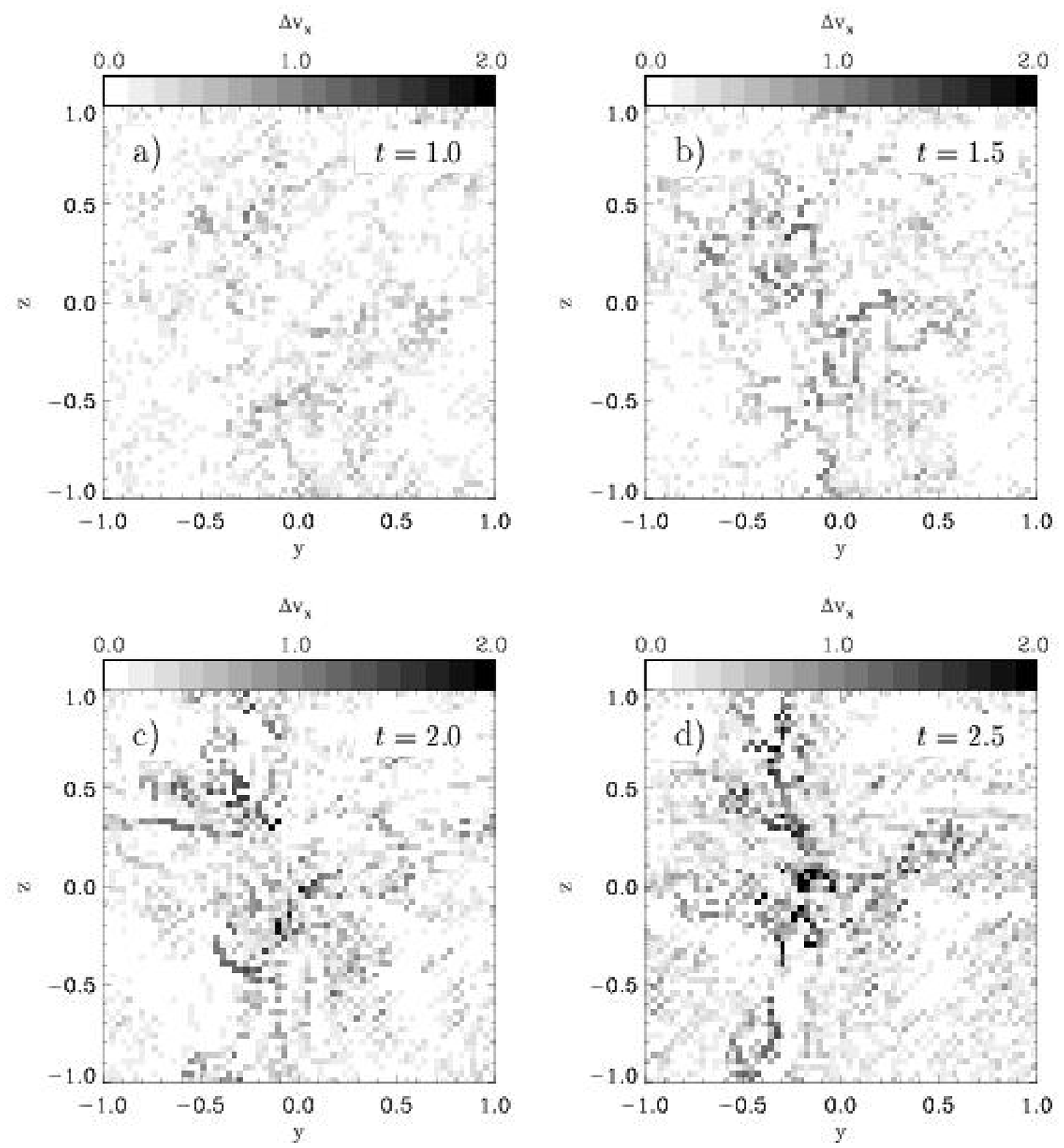}}
\end{picture}
\caption{\footnotesize{\label{fig:dv-array-decay-with-gravity} 2-dimensional
distribution (in the $yz$-plane) of centroid increments for velocity
profiles along the $x$-axis of the system between locations separated
by a vector lag $\Delta \vec{r} = (1/32,1/32)$ for the simulation of
self-gravitating, decaying, supersonic turbulence. Analog to
Figure \ref{fig:dv-pdf-decay-with-gravity}, the data are displayed for times
(a) $t=1.0$, (b) $t=1.5$, (c) $t=2.0$, and (d) $t=2.5$.  The magnitude
of the velocity increment $\Delta v_{x}$ is indicated at the top of
each plot. (From Klessen 2000) }
}\end{figure*}

For the above simulation of self-gravitating, decaying, supersonic
turbulence, Figure \ref{fig:dv-array-decay-with-gravity} plots the
2-dimensional distribution of centroid increments for a vector lag
$\Delta \vec{r} = (1/32,1/32)$. The velocity profiles are sampled
along the $x$-axis of the system.  The magnitude of the velocity
increment $\Delta v_{x}$ is indicated at the top of each plot. The
spatial distribution of velocity increments during the initial phases
appears random. Later on, gravity gains influence over the flow and
creates a network of intersecting filaments where gas streams onto and
flows along towards local potential minima. At that stage, the
velocity increments with the highest amplitudes tend to trace the
large-scale filamentary structure. This is the sign of the anisotropic
nature of gravitational collapse motions.

% \begin{figure*}[p]
% \unitlength1.0cm
% \begin{picture}(16,14.1)
% \put( -0.30,-10.5){\epsfysize=27cm\epsfbox{ms39926-fig13.ps}}
% \end{picture}
% \caption{\footnotesize{\label{fig:3D-plot-driven-with-gravity} The 3-dimensional gas
% distribution in the simulation of constantly driven turbulence in
% self-gravitating gas. Once the turbulent kinetic energy reaches the
% equilibrium level, gravity is turned on. This stage is displayed in
% (a).  The next three snapshots of the system are taken at times (b)
% $t=1.8$, when 20\% of the gas mass is in dense collapsed cores (as
% indicated by black dots -- cf. with
% Figure \ref{fig:3D-plot-decay-with-gravity}), at (c) $t=3.2$, when the
% mass in cores is 40\% of the total mass, and at (d) $t=4.8$, when the
% cluster of cores contains 60\% of the system mass. Time is given in
% units of the free-fall time, but unlike in the previous cases it is
% counted from the point gravity is turned on. (From Klessen 2000) }
% }\end{figure*}

\rsksubsubsection{Analysis of Driven Tur\-bu\-len\-ce with Self-Gravity}
\label{sec:driven-with-gravity}

For the analysis of continuously driven turbulence in self-gravitating gaseous
media, we take model $2{\cal B}h$ discussed in Section \ref{sub:beyond} (Table
\ref{tab:models};  for full details consult Klessen \etal\ 2000).   

\begin{figure*}[ht]
\unitlength1.0cm
\begin{picture}(18,6.6)
\put( -1.75,-18.0){\epsfbox{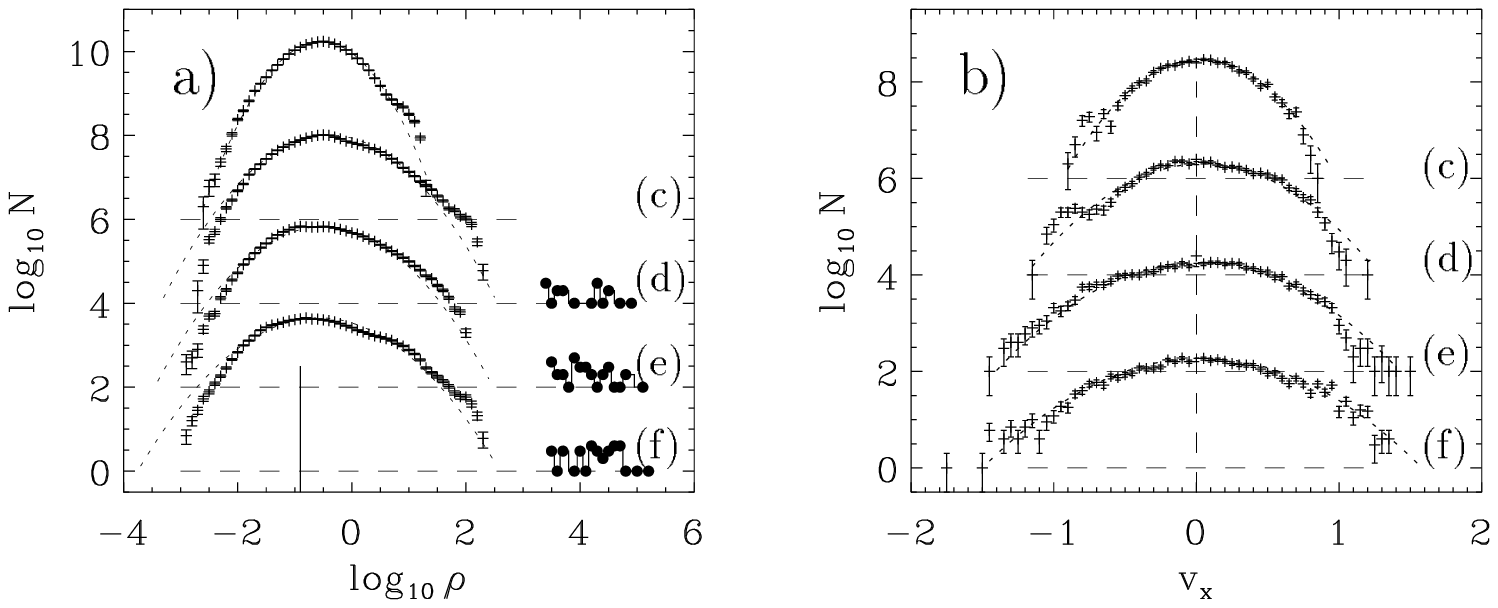}}
\end{picture}
\caption{\footnotesize{\label{fig:rho-v-pdf-driven-with-gravity} PDF's of (a) the
density and (b) the $x$-component of line centroids for the simulation
of driven turbulence in self-gravitating gas. The time sequence is the
same as in the previous figures as indicated by the letters to the
right. Each \pdf\ is offset by $\Delta \log_{10} N = 2.0$ with the base
$\log_{10} N = 0.0$ indicated by horizontal dashed lines. The best-fit
Gaussian curves are shown as dotted lines.  The density contributions
in (a) coming from collapsed cores are indicated by solid dots. (From Klessen 2000) }
}\end{figure*}

The \pdf's of (a) the density and (b) the $x$-component of the line
centroid velocities corresponding to the above four snapshots are
displayed in Figure \ref{fig:rho-v-pdf-driven-with-gravity}.  As in the
previous model, the bulk of gas particles that are not accreted onto
cores build up an approximately log-normal $\rho$-\pdf\ (indicated by
the dotted lines). Also the $v$-\pdf\ remains close to the Gaussian
value. This is different from the case of purely decaying
self-gravitating turbulence, where at some stage global collapse
motions set in and lead to very wide and distorted centroid
\pdf's. This is not possible in the simulation of driven turbulence, as
it is stabilized on the largest scales by turbulence. Collapse occurs
only locally which leaves the width of the \pdf's relatively unaffected
and only mildly alters their shape.

\begin{figure*}[th]
\unitlength1.0cm
\begin{picture}(18,6.2)
\put( -1.75,-17.2){\epsfbox{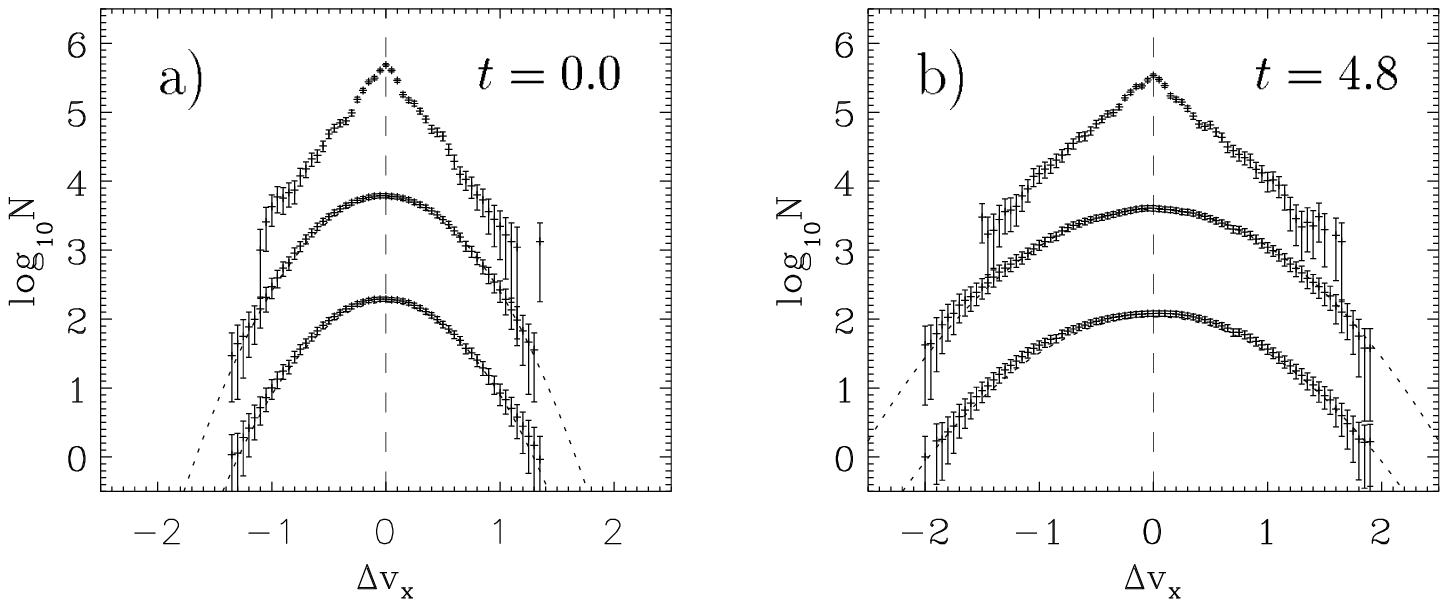}}
\end{picture}
\caption{\footnotesize{\label{fig:dv-pdf-driven-with-gravity} PDF's of the
$x$-component of the centroid velocity increments for three spatial
lags: upper curve -- $\Delta r = 1/32$, middle curve -- $\Delta r =
10/32$, and lower curve -- $\Delta r = 30/32$. The functions are
computed form the simulation of driven, self-gravitating, supersonic
turbulence at (a) $t=0.0$ and (b) $t=4.8$. As in the previous models the increment \pdf's
for small spatial lags are approximately exponential, however, the
\pdf's for larger separations remain close to Gaussian throughout the
evolution. (From Klessen 2000) }
}\end{figure*}

\begin{figure*}[ht]
\unitlength1.0cm
\begin{picture}(18,6.7)
\put( -1.75,-16.2){\epsfbox{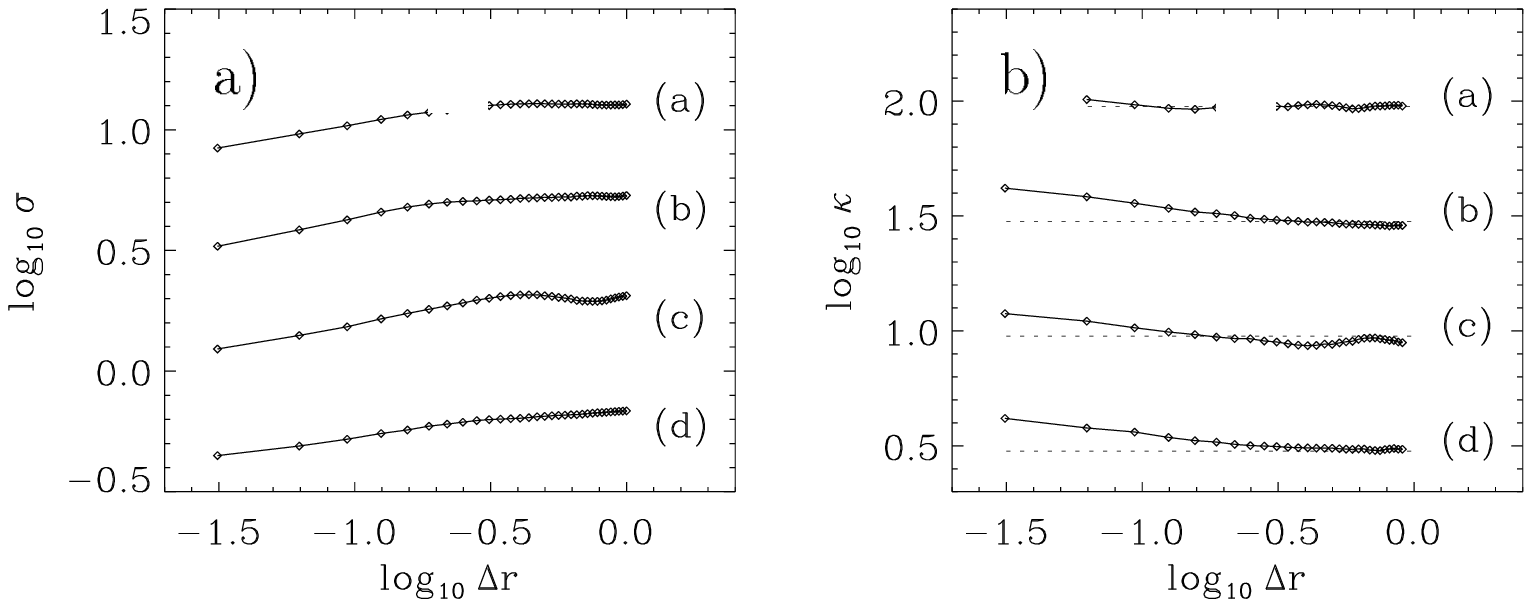}}
\end{picture}
\caption{\footnotesize{\label{fig:mom-driven-with-gravity} (a) The
second, dispersion $\sigma$, and (b) the fourth moment, kurtosis
$\kappa$, as function of spatial lag $\Delta r$ for the distribution
of velocity increments in the simulation of driven, self-gravitating,
supersonic turbulence. The letters on the right-hand side indicate
again the corresponding times.  Each \pdf\ is offset by
$\Delta \log_{10} \sigma=0.5$ and $\Delta \log_{10} \kappa = 0.5$,
respectively.} (From Klessen 2000) }\end{figure*}

Also the $\Delta v$-\pdf's show no obvious sign of evolution. For the
$x$-component of the velocity these functions are displayed in
Figure \ref{fig:dv-pdf-driven-with-gravity}, again for three different
spatial lags. The chosen times correspond (a) to the equilibrium state
at $t=0.0$, and (b) to $t=4.8$ which is the final state of the
simulation. The \pdf's only marginally grow in width. At every
evolutionary stage, the \pdf\ for the smallest spatial lag is
exponential, whereas the \pdf's for medium and large shift vectors
closely follow the Gaussian curve defined by the first two moments of
the distribution (dotted lines). The functions are similar to the ones
in the previous model before the large scale collapse motions set in
(Figure \ref{fig:dv-pdf-decay-with-gravity}a, b). Only overall
contraction will affect $\Delta v$-\pdf\ at medium to large lags. This
behavior also follows from comparing the statistical moments. Figure
\ref{fig:mom-driven-with-gravity} plots (a) the dispersion $\sigma$
and (b) the kurtosis $\kappa$ as function of the spatial lag $\Delta
r$.  Figures \ref{fig:mom-decay-with-gravity}a and
\ref{fig:mom-driven-with-gravity}a are very similar, as soon as
turbulence is established the width $\sigma$ of the \pdf\ increases with
$\Delta r$ with a slope of $d \log_{10} \sigma/d\log_{10}\Delta r
{\:\lower0.6ex\hbox{$\stackrel{\textstyle <}{\sim}$}\:} 0.2$ for small
to medium lags and levels out for larger ones. However, when comparing
the `peakedness' of the \pdf\ as indicated by $\kappa$ (Figures
\ref{fig:mom-decay-with-gravity}b and
\ref{fig:mom-driven-with-gravity}b) the model of decaying
self-gravitating turbulence yields much higher values since the \pdf's
are more strongly peaked due to the presence of large-scale collapse
motions.

\begin{figure*}[t]
\unitlength1.0cm
\begin{picture}(16,8.0)
\put( -0.0,-12.8){\epsfxsize=17cm \epsfbox{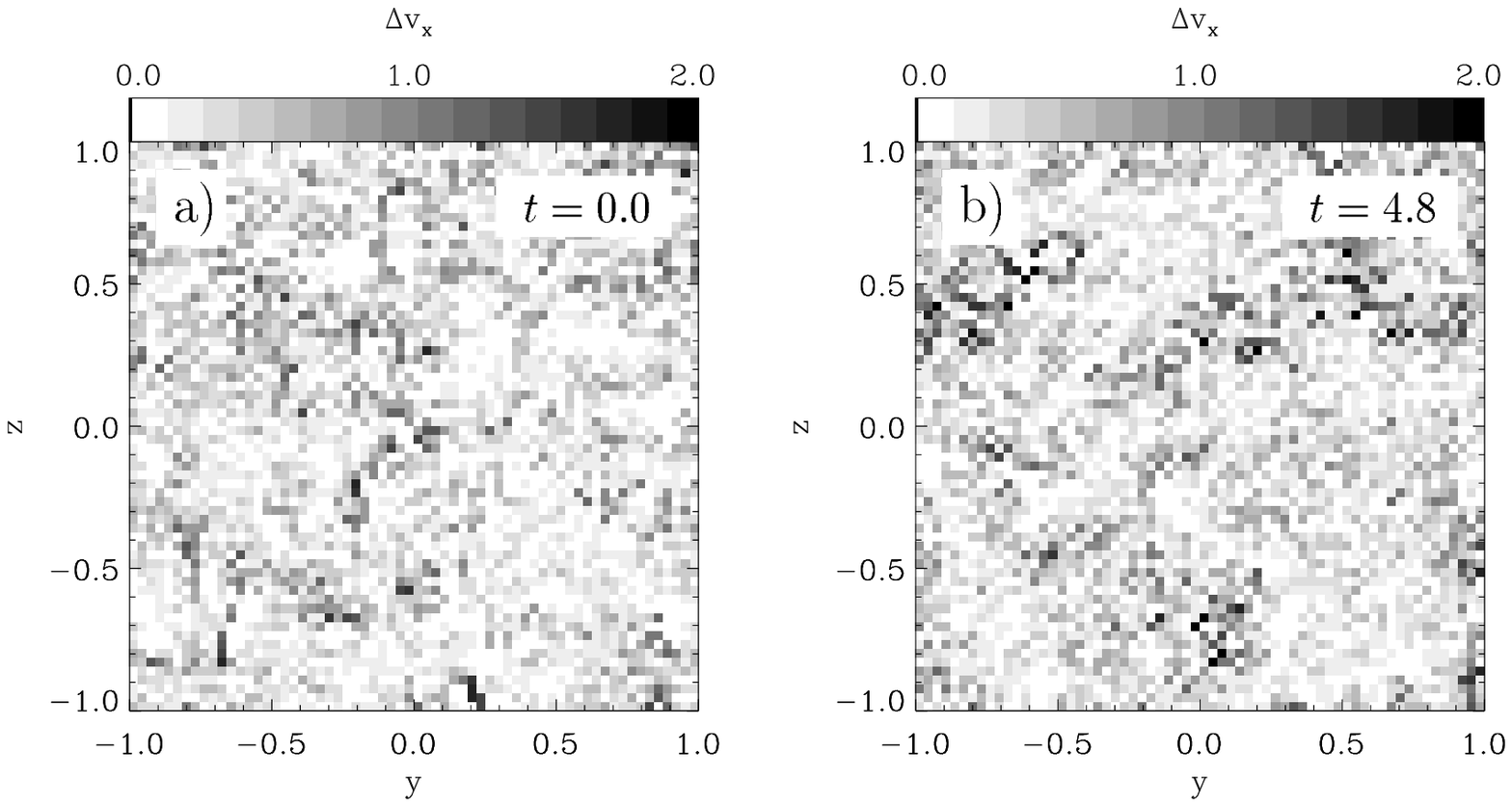}}
\end{picture}
\caption{\footnotesize{\label{fig:dv-array-driven-with-gravity} 2-dimensional
distribution (in the $yz$-plane) of the absolute value of the
$x$-component of centroid velocity increments between locations
separated by a vector lag $\Delta \vec{r} = (1/32,1/32)$ for the
simulation of driven self-gravitating supersonic turbulence. The data
are displayed at times (a) $t=0.0$, and (b) $t=4.8$.  The scaling is
indicated at the top of each figure. (From Klessen 2000) }
}\end{figure*}

Figure \ref{fig:dv-array-driven-with-gravity} finally shows the
spatial distribution of the $x$-component of the line centroid
increments for a vector lag $\Delta \vec{r} = (1/32,1/31)$. Since the
increment maps at different evolutionary times are statistically
indistinguishable, only times (a) $t=0.0$ and (b) $t=4.8$ are
displayed in the figure. As in the case of supersonic, purely
hydrodynamic turbulence the spatial distribution of velocity
increments appears  random and uncorrelated.

The adopted driving mechanism prevents global collapse. The bulk
properties of the system therefore resemble hydrodynamic, {\em
non}-self-gravitating turbulence. However, local collapse motions do
exist and are responsible for noticeable distortions away from the
Gaussian statistics. As the non-local driving scheme adopted here
introduces a bias towards Gaussian velocity fields, these distortions
are not very large. There is a need to introduce other, more realistic
driving agents into this analysis. These could lead to much stronger
non-Gaussian signatures in the \pdf's.

\rsksubsubsection{Summary}
\label{sec:summary-1}

Klessen (2000) analyzed SPH simulations of driven and decaying, supersonic,
turbulent flows with and without self-gravity, thus extending previous
investigations of mildly supersonic, decaying, {\em non}-self-gravitating
turbulence (Lis \etal\ 1996, 1998) into a regime more relevant molecular
clouds. The flow properties are characterized by using the probability
distribution functions of the density, of the line-of-sight velocity
centroids, and of their increments. Furthermore the dispersion and the
kurtosis of the increment \pdf's are discussed, as well as the spatial
distribution of the velocity increments for the smallest spatial lags.

(1) To asses the influence of variance effects, simple Gaussian
velocity fluctuations are studied. The insufficient sampling of random
Gaussian ensembles leads to distorted \pdf's similar to the observed
ones. For line profiles this has been shown by Dubinski \etal\ (1995).

(2) Decaying, initially highly supersonic turbulence without
self-gravity leads to \pdf's which also exhibit deviations from
Gaussianity. For the trans- and subsonic regime this has been reported
by Lis \etal\ (1996, 1998). However, neglecting gravity and thus not
allowing for the occurrence of collapse motions, these distortions are
not very pronounced and cannot account well for the observational data
(Lis \etal\ 1998, Miesch \etal\ 1998).

(3) When including gravity into the models of decaying initially
supersonic turbulence, the \pdf's get into better agreement with the
observations. During the early dynamical evolution of the system
turbulence carries enough kinetic energy to prevent collapse on all
scales. In this phase the properties of the system are similar to
those of non-gravitating hydrodynamic supersonic turbulence. However, as
turbulent energy decays gravitational collapse sets in. First
localized and on small scales, but as the turbulent support continues
to diminish collapse motions include increasingly larger spatial
scales. The evolution leads to the formation of an embedded cluster of
dense protostellar cores (see also Klessen \& Burkert 2000). As the
collapse scale grows, the $\rho$-, $v$-, and $\Delta v$-\pdf's get
increasingly distorted. In particular, the $\Delta v$-\pdf's for small
spatial lags are strongly peaked and exponential over the entire range
of measured velocities. This is very similar to what is observed in
molecular clouds (for $\rho$-Ophiuchus see Lis \etal\ 1998; for Orion,
Mon R2, L1228, L1551, and HH83 see Miesch \etal\ 1998).

(4) The most realistic model for interstellar turbulence considered
here includes a simple (non-local) driving scheme. It is used to
stabilize the system against collapse on large scales. Again
non-Gaussian \pdf's are observed. Despite global stability, local
collapse is possible and the system again evolves towards the
formation of an embedded cluster of accreting protostellar cores.  As
the adopted driving scheme introduces a bias towards maintaining a
Gaussian velocity distribution, the properties of the \pdf's fall in
between the ones of pure hydrodynamic supersonic turbulence and the
ones observed in systems where self-gravity dominates after sufficient
turbulent decay.  This situation may change when considering more
realistic driving schemes.

(5) A point of caution: The use of $v$- and $\Delta v$-\pdf's to
unambiguously characterize interstellar turbulence and to identify
possible physical driving mechanisms may be limited.  {\em All} models
considered in the current analysis lead to non-Gaussian signatures in
the \pdf's, differences are only gradual. In molecular clouds the
number of physical processes that are expected to give rise to
deviations from Gaussian statistics is large. Simple statistical
sampling effects (Sec.~\ref{sec:init}) and turbulent intermittency
caused by vortex motion (Lis \etal\ 1996, 1998), as well as the effect
self-gravity (Sec.~\ref{sec:decay-with-gravity}) and of shock
interaction in highly supersonic flows (Mac~Low \& Ossenkopf 2000),
{\em all} will lead to non-Gaussian signatures in the observed \pdf's.
Also stellar feedback processes, galactic shear and the presence of
magnetic fields will influence the interstellar medium and create
distortions in the velocity field. This needs to be studied in further
detail.  In addition, the full 3-dimensional spatial and kinematical
information is not accessible in molecular clouds, measured quantities
are always projections along the line-of-sight. As the structure of
molecular clouds is extremely complex, the properties of the \pdf's may
vary considerably with the viewing angle. Attempts to disentangle the
different physical processes influencing interstellar turbulence
therefore should no rely on analyzing velocity \pdf's alone, they
require additional statistical information.

%%%%%%%%%%%%%%%%%%%%%%%%%%%%%%%%%%%%%%%%%%%%%%%%%%%%%%%%%%%%%%%%%%%%%%%%%%%%%%%
% [KHM00] Fourier part of KHM00
%%%%%%%%%%%%%%%%%%%%%%%%%%%%%%%%%%%%%%%%%%%%%%%%%%%%%%%%%%%%%%%%%%%%%%%%%%%%%%%

%\rsksubsection{Fourier analysis of self-gravitating clouds}
\rsksubsection{Fourier Analysis}
\label{subsec:fourier}

In this section we discuss the energy distribution on different
spatial scales during various stages of the dynamical evolution of
supersonically turbulent self-gravitating gaseous systems.
  We perform a Fourier analysis of the energy, computing the
power spectra of kinetic and potential energies from  the
numerical models introduced in Section \ref{sub:beyond} (Table
\ref{tab:models}; for full details see Klessen \etal\ 2000). 
To allow for a direct
comparison, all models are analyzed on a Cartesian grid with $128^3$
cells. For the SPH models this is done using the kernel smoothing
algorithm, while the $256^3$-ZEUS models are simply degraded in
resolution.  For each cell the potential and kinetic energy content is
calculated, and the kinetic energy is further decomposed into its
solenoidal (rotational) and compressional parts. These quantities are
then all transformed into Fourier space, to find the contribution of
different dimensionless wave numbers $k$, or equivalently, to find the
distribution of energy over different spatial scales $\lambda_k =
L/k$.

\begin{figure*}[th]
\unitlength1.0cm
\begin{picture}(16,10.7)
\put(-2.00,-9.50){\epsfbox{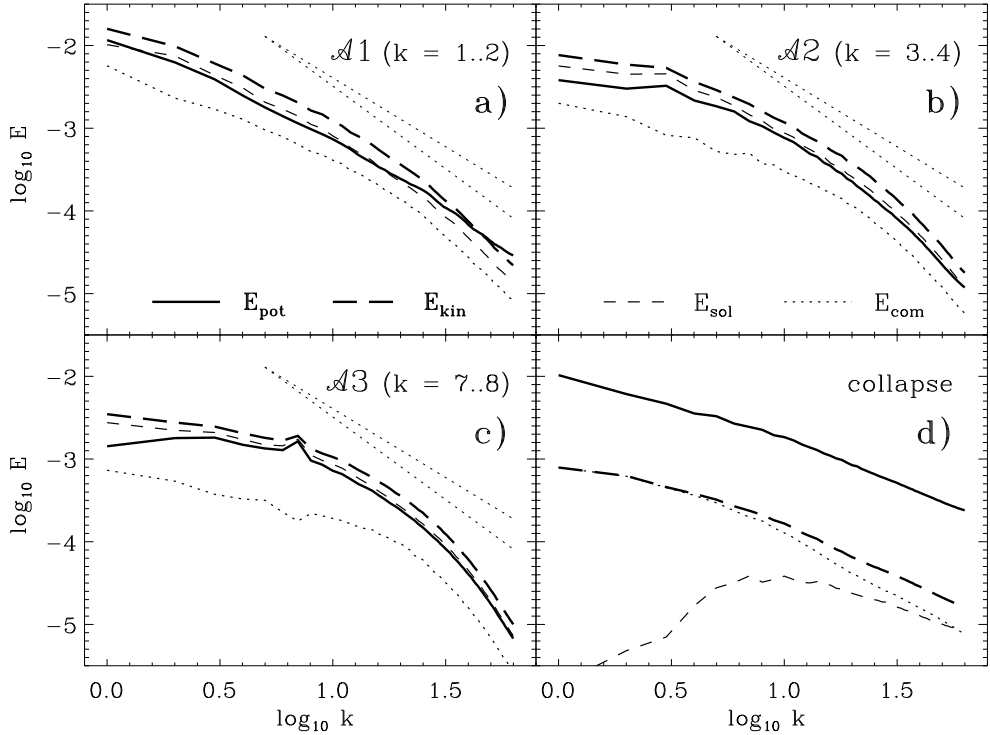}}
\end{picture}
\caption{\label{fig:wave-mode-analysis-1} Energy as function of wave
number $k$ for models with different driving scale: (a) ${\cal A}1$
with $k=1-2$, (b) ${\cal A}2$ with $k=3-4$ and (c) ${\cal A}3$ with
$k=7-8$. The simulations are studied at $t=0.0$, when the hydrodynamic
turbulence is fully developed, immediately after gravity is included.
The plots show potential energy $E_{\rm pot}$ (thick solid lines),
kinetic energy $E_{\rm kin}$ (thick long-dashed lines), its solenoidal
component $E_{\rm sol}$ (short-dashed lines) and its compressional
component $E_{\rm com}$ (dotted lines). The thin dotted lines indicate
the slope $-5/3$ expected from the Kolmogorov (1941) theory and the
slope $-2$ expected for velocity discontinuities associated with
shocks.  For comparison, plot (d) shows a strongly self-gravitating
model that completely lacks turbulent support and therefore contracts
on all scales (data from Klessen \etal\ 1998).  The energy spectra are
computed on a $128^3$ grid onto which the SPH particle distribution
has been assigned using the kernel smoothing procedure. (From Klessen \etal\ 2000) }
\end{figure*}

\rsksubsubsection{Fourier Spectra as Function of Driving Wavelength}
\label{sec:fourier-driving}

The energy spectrum of fully developed turbulence for small-, medium-
and large-scale driving is shown in figure
\ref{fig:wave-mode-analysis-1}. It shows the SPH models (a) ${\cal
A}1$, (b) ${\cal A}2$ and (c) ${\cal A}3$  just at the time $t=0.0$
when gravity is turned on.  In each plot the thick solid lines describe the
potential energy as a function of wave number $k$, and the 
thick long-dashed lines represent the kinetic energy, which can be
decomposed into its solenoidal (rotational) and compressional
components. They are defined via the velocities by $\vec{\nabla}
\cdot \vec{v}_{\rm sol} = 0$ and $\vec{\nabla} \times \vec{v}_{\rm
com} =0$, respectively.

The models ${\cal A}1$ and ${\cal A}2$, which are driven at long and
intermediate wave lengths ($k=1-2$ and $k=3-4$), appear to exhibit an
inertial range below the driving scale, i.e.\ between $0.5 \sil
\log_{10} k \sil 1.5$.  Note that, in real clouds, the dissipation
scale may lie near the upper end of this wave number range as
discussed in \S\ \ref{sub:regions}.  In this interval the energy
distribution  approximately follows a power law very similar to that predicted by
the Kolmogorov (1941) theory ($E_{\rm kin} \propto k^{-5/3}$). This is
understandable given that, in our models, once turbulence is fully
established, the solenoidal component of the kinetic energy always
dominates over the compressible one, $E_{\rm sol} > E_{\rm com}$.  For
a pure shock dominated flow ($E_{\rm com} \gg E_{\rm sol}$) one would
expect a power spectrum with slope $-2$ (Passot \etal\ 1988). To guide
the eye, both slopes are indicated as thin dotted lines in plots (a)
to (c). For model ${\cal A}3$ the smaller number of available modes
between the driving scale $k=7-8$ and the Nyquist frequency does not
allow for an unambiguous identification of a turbulent inertial range.
The permanent energy input necessary to sustain an equilibrium state
of turbulence produces a signature in the energy distribution at the
driving wave length. This is most clearly visible in
figure~\ref{fig:wave-mode-analysis-1}c.
\begin{figure*}[pt]
\unitlength1.0cm
\begin{picture}(16,18.00)
\put(-2.00,-4.50){\epsfbox{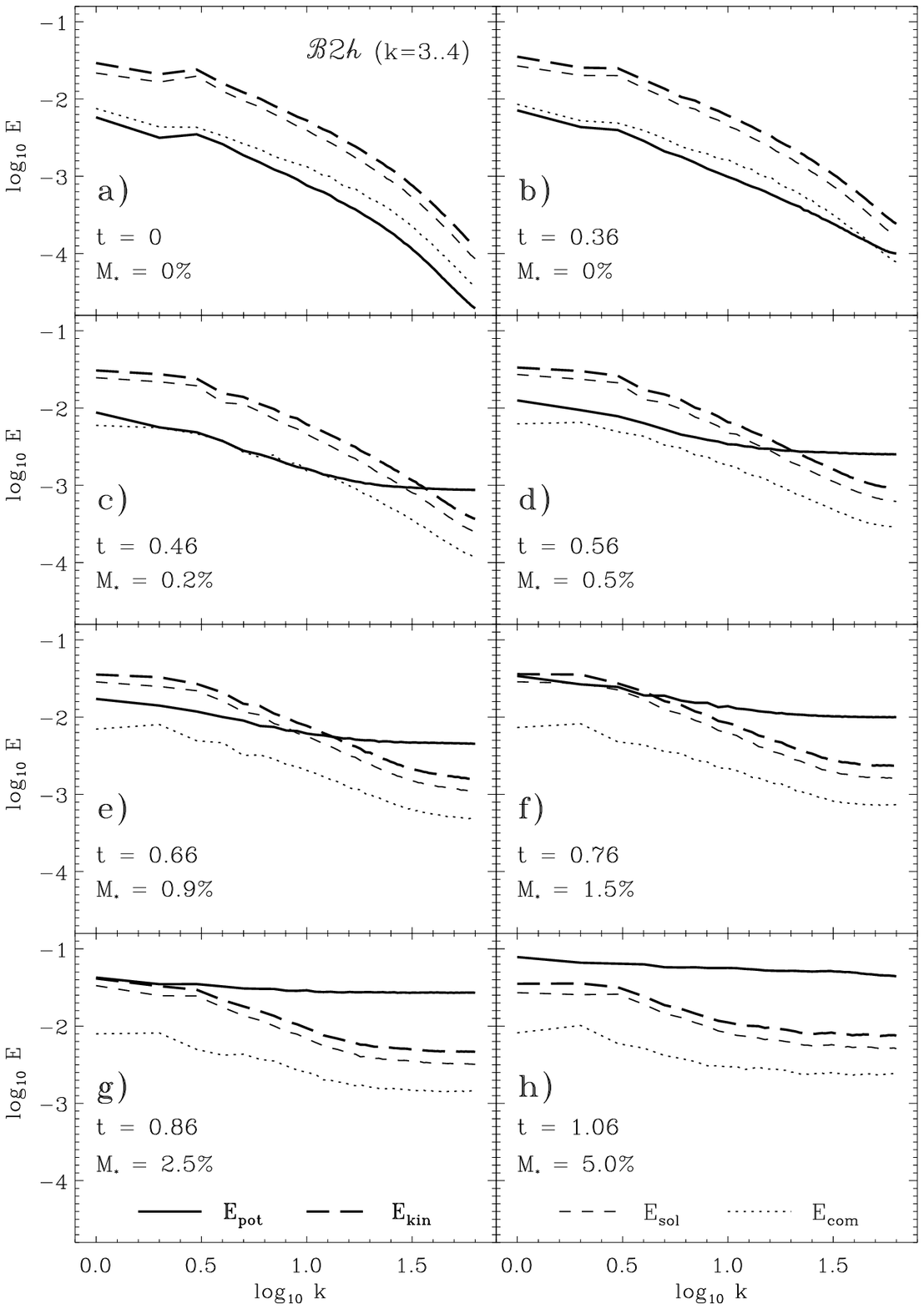}}
\end{picture}
\caption{\label{fig:wave-mode-analysis-2} Fourier analysis of the
high-resolution model ${\cal B}2h$ ($\langle M_{\rm J}\rangle_{\rm
turb} = 3.2$ and $k=3-4$) at different stages of its dynamical
evolution indicated on each plot. Notation and scaling are the same as
in figure \ref{fig:wave-mode-analysis-1}.   (From Klessen \etal\ 2000) }
\end{figure*}

The system is globally stable against gravitational collapse, as
indicated by the fact that for every wave number $k$ the kinetic energy
exceeds the potential energy. For comparison we plot in
figure~\ref{fig:wave-mode-analysis-1}d the energy distribution of a
system without turbulent support. The data are taken from Klessen \etal\ (1998) and stem from an SPH simulation with $500\,000$ particles
containing 220 thermal Jeans masses and no turbulent velocity field,
but otherwise identical physical parameters.  The snapshot is taken at
$t=0.2 \tau_{\rm ff}$ after the start of the simulation. 
This system contracts on {\em all} scales and forms
stars at very high rate within a few free-fall times $\tau_{\rm
ff}$. Contrary to the case of hydrodynamic turbulence, the kinetic
energy distribution is dominated by compressional modes, especially at
small wave numbers. The overall energy budget is determined by the
potential energy $E_{\rm pot}$, which outweighs the kinetic energy
$E_{\rm kin}$ on all spatial scales $k$ by about an order of
magnitude.

\rsksubsubsection{Fourier Spectra During Collapse}
\label{sec:fourier-collapse}

Figure \ref{fig:wave-mode-analysis-2} concentrates on model ${\cal
B}2h$ with $\langle M_{\rm J}\rangle_{\rm turb}=3.2$ and $k=3-4$. It
describes the time evolution of the energy distribution.
Figure~\ref{fig:wave-mode-analysis-2}a shows the state of fully
established turbulence for this model just when gravity is turned on
($t=0.0$).  In the subsequent evolution, local collapse occurs in
shock-generated density enhancements where the potential energy
dominates over the kinetic energy.  This affects the {\em small}
scales first, as seen in the plotted time sequence. As collapse
progresses to higher and higher densities, the scale where the
potential energy dominates rapidly grows. Once the mass fraction in
dense cores has reached about $\sim 3$\%, the potential energy
outweighs the kinetic energy on all scales.  However, this should not
be confused with the signature of global collapse.  The power spectrum
of the potential energy is constant for all $k$. It is the Fourier
transform of a delta function. Local collapse has produced point-like
high-density cores.  The overall filling factor of collapsing clumps
and cores is very low, so most of the volume is dominated by
essentially pure hydrodynamic turbulence. As a consequence, the
velocity field on large scales is not modified much (besides a shift
to higher energies). On small scales, however, the flow is strongly
influenced by the presence of collapsed cores which is noticeable as a
flattening of the power spectra at large wave numbers.  Despite their
small volume filling factor, the cores dominate the overall power
spectrum.  The solenoidal part of the kinetic energy always dominates
over the compressional modes and also the signature of the driving
source in the energy spectrum remains, visible as a `bump' in the
kinetic energy spectrum at $k \approx 8$.

\rsksubsubsection{Summary}
\label{sec:fourier-summary}

In turbulent flows, it is impossible to predict from the start
when and where individual cores form and how they evolve. In all models except
the ones driven below the fluctuation Jeans scale, gravity eventually begins
to dominate over kinetic energy. The Fourier spectra show that this first
occurs on small scales. This indicates the presence of local collapse. As
dense collapsed cores form, the power spectrum of the gravitational energy
becomes essentially flat. The kinetic energy, on the other hand, appears to
follow at intermediate wave numbers a Kolmogorov power spectrum with slope
$-5/3$, less steep than the spectrum expected for pure shock flows. The slope
remains almost unaltered during the course of the evolution, indicating that a
large volume fraction of the system is always well described by pure
hydrodynamic turbulence.  The spatial extent of collapsing regions (where
infall motions dominate over the turbulent flow) is relatively small. This
also explains the fact that the solenoidal component of the flow always
dominates over the compressional part.

%%%%%%%%%%%%%%%%%%%%%%%%%%%%%%%%%%%%%%%%%%%%%%%%%%%%%%%%%%%%%%%%%%%%%%%%%%%%%%%
% [OKH01] Ossenkopf, Klessen, & Heitsch (2001)
%%%%%%%%%%%%%%%%%%%%%%%%%%%%%%%%%%%%%%%%%%%%%%%%%%%%%%%%%%%%%%%%%%%%%%%%%%%%%%%

%\rsksubsection{$\Delta$-Variance of self-gravitating clouds}
\rsksubsection{$\Delta$-Variance}
\label{subsec:delta}

\rsksubsubsection{Introduction}
\label{sec:intro-delta}

In the previous two Sections we have focused on \pdf's and Fourier transforms
to quantify molecular cloud structure -- other statistical measures are
summarized in the reviews by V\'azquez-Semadeni (2000) and Ossenkopf \etal\ (2000).
One further technique especially sensitive to the discrimination of the
relative structural variation on different spatial scales is the
$\Delta$-variance introduced by Stutzki \etal\ (1998). It provides a good
separation of noise and observational artifacts from the real cloud structure
and for isotropic systems its slope is directly related to the spectral index
of the corresponding power spectrum. Bensch \etal\ (2001) and Mac~Low \&
Ossenkopf (2000) have shown that it can be applied in an equivalent way both
to observational data and hydrodynamic and magneto-hydrodynamic turbulence
simulations allowing a direct comparison.

Their investigations, however, neglect the influence of gravitational collapse
on the structure formation so that their conclusions may be limited when
applied to star-forming regions. It is essential to include the effects of
self-gravity for the analysis of star-forming regions. It was the aim of an
investigation by Ossenkopf, Klessen, \& Heitsch (2001) to close this gap and
investigate the interaction between turbulence and self-gravity.  We follow
their line of reasoning and apply the $\Delta$-variance to characterize the
structure in numerical models of driven and decaying self-gravitating
supersonic (magneto-)hydrodynamic turbulence and compare the results to
observed regions of star formation.

% In \S\ref{sec:numerics} we
% introduce the molecular cloud models analyzed here and the numerical
% methods used for their generation. In \S\ref{sec:results} we apply
% the $\Delta$-variance analysis and discuss the time evolution 
% of the structure as local collapse sets in
% and protostellar cores form and accrete mass. Using various models
% and comparing the density and velocity structure we demonstrate how
% power is build up at the different scales during star formation. 
% \S\ref{sec:observations} provides a comparison with observational 
% data including dust emission and molecular lines and discusses the 
% resulting differences.
% We summarize our results in \S\ref{sec:summary}.

\rsksubsubsection{Turbulence Models}
\label{sec:numerics}

The large observed linewidths in molecular clouds imply the presence of
supersonic velocity fields that carry enough energy to counterbalance gravity
on global scales (Section \ref{sub:regions}).  As turbulent energy dissipates
rapidly, i.e.\ roughly on the free-fall time scale (Section \ref{sub:beyond}),
either interstellar turbulence is continuously replenished in order to prevent
or considerably postpone global collapse, or alternatively, molecular clouds
evolve rapidly and never reach dynamical equilibrium between kinetic energy
and self-gravity (Ballesteros-Paredes \etal\ 1999a, Elmegreen 2000b).

We select a set of numerical models mostly from preexisting studies
that spans a large range of the parameter space relevant for 
Galactic molecular clouds. We analyze the time evolution of their
density and velocity structure as gravitational
collapse progresses. Altogether, we include eight models
summarized in Table\ \ref{tab:models-2}.  They differ in the scale on
which turbulent driving occurs, the persistence of this driving, the
inclusion of magnetic fields, and the numerical algorithm used to
solve the hydrodynamic or magnetohydrodynamic equations. The models are
calculated using the particle-based SPH method and the grid-based ZEUS code as
introduced in Section \ref{subsub:numerics} (for full details see Ossenkopf
\etal\ 2001).

In the present analysis we neglect feedback effects from the
produced young stellar objects (like bipolar outflows, stellar winds,
or ionizing radiation from new-born O or B stars).  Thus our results
will hold only for early stages of star-forming systems.

\begin{table*}[t]
\caption{\label{tab:models-2}
Properties of the considered turbulence models together with the resulting time
scales}
\footnotesize
\begin{center}
\begin{tabular}[t]{llcccrr}
\hline\hline
model &
description  &
$k_{\rm d}$$^{a}$ &
numerical method &
further reference$^{b}$ &
$\tau_{5\%}^{c}$ &
$f_{\tau_{\rm ff}}^{d}$\\
\hline
S01 & driven HD turbulence & $1\dots2$ &
SPH & ${\cal B}1h$ in KHM  & 0.6     & 28\%\\
Sd1 & decaying HD turbulence & $1\dots2$ &
SPH & ---                  & 0.6     & 60\%\\
S02 & driven HD turbulence & $7\dots8$ &
SPH & ---                  & $>5.5$  & 0.6\%\\
Sd2 & decaying HD turbulence & $7\dots8$ &
SPH & ---                  & 2.0     & 0.0\% \\
G   & Gaussian density     & --- &
SPH & $\cal I$ in KB       & 1.3     & 9\%\\
H01 & driven HD turbulence & $1\dots2$ & 
ZEUS & ${\cal D}1h$ in KHM & 0.6     & 24\%\\
H02 & driven HD turbulence & $7\dots8$ &
ZEUS & ${\cal D}3h$ in KHM & 4.7     & 0.5\% \\
M01 & driven MHD turbulence & $1\dots2$ & 
ZEUS & ${\cal E}h1i$ in HMK& 1.2     & 6\%  \\
\hline \hline
\end{tabular}
\end{center}
$^{a}${Wavenumber of the original driving}\\ $^{b}${Model names 
in the original papers: KB -- Klessen \& Burkert (2000),
KHM -- Klessen \etal\ (2000), HMK -- Heitsch \etal\ (2001)}\\
$^{c}${Time at which 5\% of the total mass is accreted onto cores
in internal units where $\tau_{\rm ff}=1.5$.}\\
$^{d}${Mass fraction that is accreted onto cores after one global
free-fall time.  --- Adopted from Ossenkopf \etal\ (2001).}\\
\end{table*}

To test the influence of magnetic fields we consider the driven
turbulence model which carries most energy on large scales and
add  magnetic fields to the ZEUS simulations. By comparing
the resulting model M01 with the hydrodynamic model H01 we get
a direct measure for the significance of magnetic fields 
for the interplay of turbulence and self-gravity in  structure
formation. 

The magnetic field in this model is selected to be a major
factor where the ratio between thermal and magnetic pressure $\beta =
p_{\rm th}/p_{\rm mag} = 8 \pi c_{\rm s}^2\rho/B^2 = 0.04$.
With a turbulent Mach number of ${\cal M}_{\rm rms} = 10$ we
find that the turbulent velocity dispersion exceeds the Alfv\'{e}n 
speed by a factor of 1.4 so that the
structure is still essentially determined by supersonic turbulence
and only modified by the magnetic field. The mass in the cloud
still exceeds the critical mass for a  magnetostatically stable
cloud by a factor 2 (Heitsch \etal 2001) so that the field should not
prevent gravitational collapse. Cases of sub-Alfv\'{e}nic
non-self-gravitating turbulence where the whole structure is 
dominated by the magnetic field have been discussed by Ossenkopf
\& Mac~Low (2001).

The models presented here are computed in normalized units. 
Throughout this analisys we give all size values relative to the total 
cube size, all density values relative to the the average density
in the cube, and all velocities relative to the sound speed.
Model G contains 220 thermal Jeans masses, wherease all other models 
have 120 thermal Jeans masses\footnote{We use the 
  spherical definition of the Jeans mass, $M_{\rm J} \equiv
  4/3\,\pi \rho \lambda_{\rm J}^3$, with density $\rho$ and Jeans
  length $\lambda_{\rm J}\equiv \left(\pi{\cal R}T /G
      \rho\right)^{1/2}$, where $G$ and $\cal R$ are the
  gravitational and the gas constant. The mean Jeans mass $\langle
  M_{\rm J} \rangle$ is determined from average density in the
  system $\langle \rho \rangle$. An alternative cubic definition,
  $M_{\rm J} \equiv \rho (2\lambda_{\rm J})^3$, would yield a value
  roughly twice as large.}.
If scaled to mean densities $n({\rm H}_2) = 10^5\,$cm$^{-3}$, 
a value typical for star-forming molecular cloud regions
% (e.g.\ in $\rho$-Ophiuchus, see Motte, Andr{\'e}, \& Neri 1998)
and a temperature of 11.4$\,$K (i.e.\ a sound speed $c_{\rm s} =
0.2\,$km$\,$s$^{-1}$), the mean Jeans mass is one solar mass,
$\langle M_{\rm J} \rangle = 1\,$M$_{\odot}$, and the size of cube G
is $0.34\,$pc whereas all other models have a size of 0.29~pc.
The global free-fall time scale, as defined by $\tau_{\rm ff} =
(3\pi/32G)^{1/2}\,\langle\rho\rangle^{-1/2}$ with $\langle\rho\rangle$
being the average density, is about
$10^5\,$yr. In normalized time units it follows $\tau_{\rm ff}
= 1.5$. The  simulations cover a density range from $n({\rm H}_2) \approx
100\,$cm$^{-3}$ in the lowest density regions to $n({\rm H}_2)
\approx 10^9\,$cm$^{-3}$ where collapsing protostellar cores are
identified and converted into ``sink'' particles in the SPH code.

In this density regime gas cools very efficiently and it is possible
to use an effective polytropic equation-of-state in the simulations
instead of solving the detailed equations of radiative transfer. The
effective polytropic index is typically close to unity, $\gamma_{\rm
eff} \sil 1$, except for densities $10^5\,$cm$^{-3} < n({\rm H}_2) <
10^7\,$cm$^{-3}$, where somewhat smaller values of $\gamma_{\rm eff}$
are expected (Spaans \& Silk 2000). For simplicity, we adopt a value of
$\gamma_{\rm eff}=1$, i.e.\ an isothermal equation of state
for all densities in the simulations. As the choice of
$\gamma_{\rm eff}$ influences the density contrast in shock compressed
gas, this idealisation may introduce some small modifications to the
dynamical behavior compared to real cloud systems (see Scalo \etal\
1998 or Ballesteros-Paredes \etal\ 1999b for further discussion).

\rsksubsubsection{Density Structure}
\label{sec:results}

\rskparagraph{$\Delta$-Variance Analysis}
The $\Delta$-variance analysis was introduced by 
Stutzki \etal\ (1998) as an averaged wavelet transform 
(Zielinsky \& Stutzki 1999)
to measure the amount of structure present at different scales 
in an $E$-dimensional data set. The value of the $\Delta$-variance
at a certain scale is computed by convolving the $E$-dimensional
structure with a normalized spherically symmetric down-up-down 
function of the considered size, and measuring the remaining variance.
For two-dimensional structures, like astrophysical maps,
the filter function is easily visualized as a ``French hat'' wavelet 
with a positive centre surrounded by a negative ring and
equal diameters of the centre and the annulus. The analysis
can be applied in the same way in arbitrary dimensions by extending
the filter function to higher dimensions retaining its radial
symmetry and adapting the value in the negative part to preserve
normalization (Appendix B of Stutzki \etal 1998). 
The resulting $\Delta$-variance as a function of the filter size
measures the amount of structural variation on that scale.

 As the convolution with the filter function corresponds
to a multiplication in Fourier space, we can relate the
$\Delta$-variance to the power spectrum of the data set.
If the structure is characterized by a power law power spectrum
\begin{equation}
P(|\vec{k}|) \propto |\vec{k}|^{-\beta} 
\end{equation}
the slope $\alpha$ of the $\Delta$-variance as a function of 
lag (filter size) is related to the spectral index $\beta$ 
of the power spectrum by $\beta=\alpha+E$ for $0\le \beta < E+4$.
Due to the smooth circular filter function the $\Delta$-variance 
measures the spectral index in a way which is more robust with 
respect to edge and gridding effects than the 
Fourier transform. The $\Delta$-variance provides a clear spatial
separation of different effects influencing observed structures 
like noise or a finite observational resolution.

Mac~Low \& Ossenkopf (2000) have shown that we can translate the
$\Delta$-variance of a three-dimensional isotropic structure into the
$\Delta$-variance of the maps obtained from the projections of this structure
by rescaling with a factor proportional to the lag and an
additional small shift.  This guarantees the preservation of the power
spectral index $\beta$ in projection (Stutzki \etal\ 1998). As we want to
compare the simulations to observational data we will always use the
two-dimensional representation of the $\Delta$-variance results also
when applying the analysis directly to the three-dimensional data
cubes of the simulations. To compute the $\Delta$-variance for our
models the SPH density distribution is assigned onto a Cartesian grid
with $128^3$ cells.  The ZEUS cubes have been analyzed in full
resolution at $256^3$ as well as degraded to $128^3$ for a one-to-one
comparison with the SPH models.  Higher resolution helps to 
extend the dynamic scale range limited by the periodic boundary conditions
at the large scale end and the numerical resolution of the simulations
on the small scale end, but does not change the general behavior
of the resulting $\Delta$-variance.

\begin{figure}[t]
\unitlength1cm
\begin{center}
\begin{picture}(8.6, 5.5)
\put( 8.0,-0.4){\begin{rotate}{90}\epsfysize=8.0cm
    \epsfbox{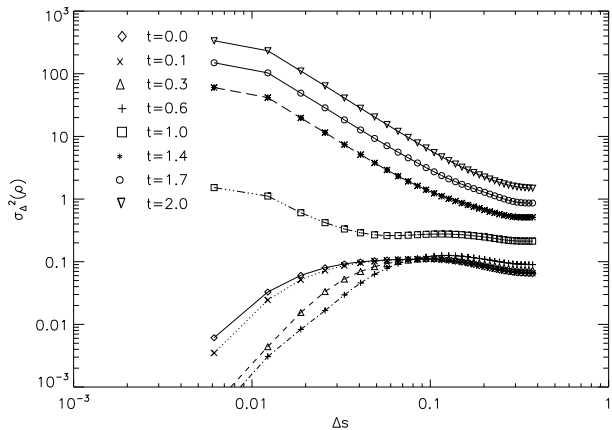}\end{rotate}} 
\end{picture}
\end{center}
\caption{\label{fig:density-Gauss}Time evolution of the strength
of density fluctuations as function of their spatial scale  
measured with the $\Delta$-variance for model G. The density $\rho$
is given in units of the average density in the cube, the lag 
$\Delta s$ in units of the cube size, and the time $t$ in internal
time units where the free-fall time $\tau_{\rm ff}=1.5$. (From Ossenkopf
\etal\ 2001)
} 
\end{figure}

\rskparagraph{Collapse of a Gaussian Density Field}
\label{subsubsec:Gaussian-collapse}

Before we investigate the interplay between supersonic turbulence and
self-gravity, let us consider a system where the density and velocity
structure is dominated by gravity on all scales and at all
times. Model G describes the collapse of Gaussian density fluctuations
with initial power spectrum $P(k) \propto k^{-2}$ and maximum density
contrast $\delta \rho/\rho \approx 50$.  The system is unstable
against gravitational collapse on all scales and forms a cluster
of protostellar cores within about two free-fall time scales (Klessen \&
Burkert 2000, 2001). The velocity structure is coupled to the density
distribution via Poisson's equation and there is no contribution from
interstellar turbulence.

The time evolution of the density structure is illustrated in Figure
\ref{fig:density-Gauss}. Initially, the $\Delta$-variance
$\sigma_{\Delta}^2(n)$ is more or less constant ($\alpha=0$)
on scales $\Delta s \sig 0.02$, in agreement with the initial power 
spectrum $P(k)\propto k^{-2}$. The steepening below  $\Delta s \approx 0.02$
is produced by the finite resolution of the SPH simulations
resulting in the blurring of structures at the smallest scales.

The first changes of the variance $\sigma_{\Delta}^2(n)$ are confined
to small scales.  Initial fluctuations with masses below the
local Jeans limit will quickly smear out by thermal pressure as the
system evolves from purely Gaussian fluctuations into a
hydrodynamically self-consistent state (see Appendix B in Klessen \&
Burkert 2000).  As these fluctuations are by far more numerous than
Jean-unstable contracting ones, the $\Delta$-variance
$\sigma_{\Delta}^2(n)$ begins to decrease on small scales. However, as
the central regions of massive Jeans-unstable fluctuations contract to
sufficiently high densities, $\sigma_{\Delta}^2(n)$ increases again.
This mainly affects the small scales as local collapse modifies the density
structure on time scales of the local free-fall time. At
$t=0.7\tau_{\rm ff}$ the first collapsed core is identified and is soon
followed by others.  Altogether 56 dense protostellar cores build up.
As time advances larger and larger scales exhibit noticeable signs of
contraction.  After about one global free-fall time
collapse starts to involve all spatial modes in the system and
the absolute magnitude of the density fluctuations finally grows on
all scales.  As the small scale structure dominates the density
structure we obtain a negative slope in the $\Delta$-variance
spectrum. In the final step of the simulation roughly 30\,\% of the mass is
accumulated in dense cores and the slope is about --1.7 indicating that a
small but significant contribution of large scale structure is still
present, because an uncorrelated $N$-body system of gravitationally
collapsed points would correspond to a slope of --2 equivalent to a
flat power spectrum $P(k)={\rm const}$.

The flattening at $\Delta s > 0.2$ is due to periodicity. The system
is {\em not} allowed to collapse freely, it is  held up against global
collapse by periodic boundaries which strongly affect the evolution of the
large-scale modes. The graphs of $\Delta$-variance are not extended beyond
effective lags of about 0.4 as the largest filter that we use is half the
cube size and we have to apply an average length reduction factor of $\pi/4$ on
projection to two dimensions.

\begin{figure*}[t]
\unitlength1cm
\begin{picture}(16.6,12.3)
\put(  8.3, 6.2){\begin{rotate}{90}\epsfysize=8.0cm \epsfbox{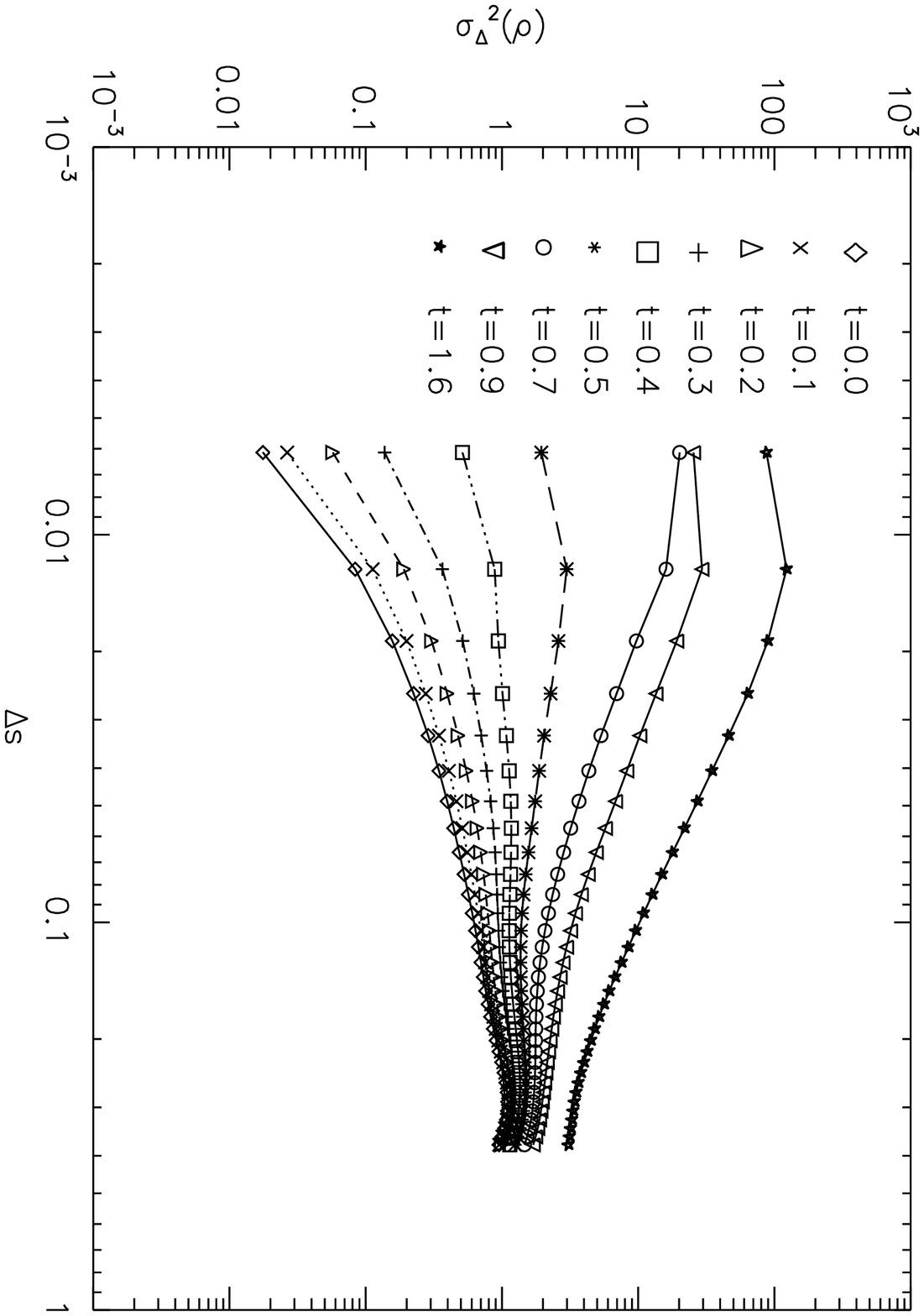}\end{rotate}}
\put( 16.8, 6.2){\begin{rotate}{90}\epsfysize=8.0cm \epsfbox{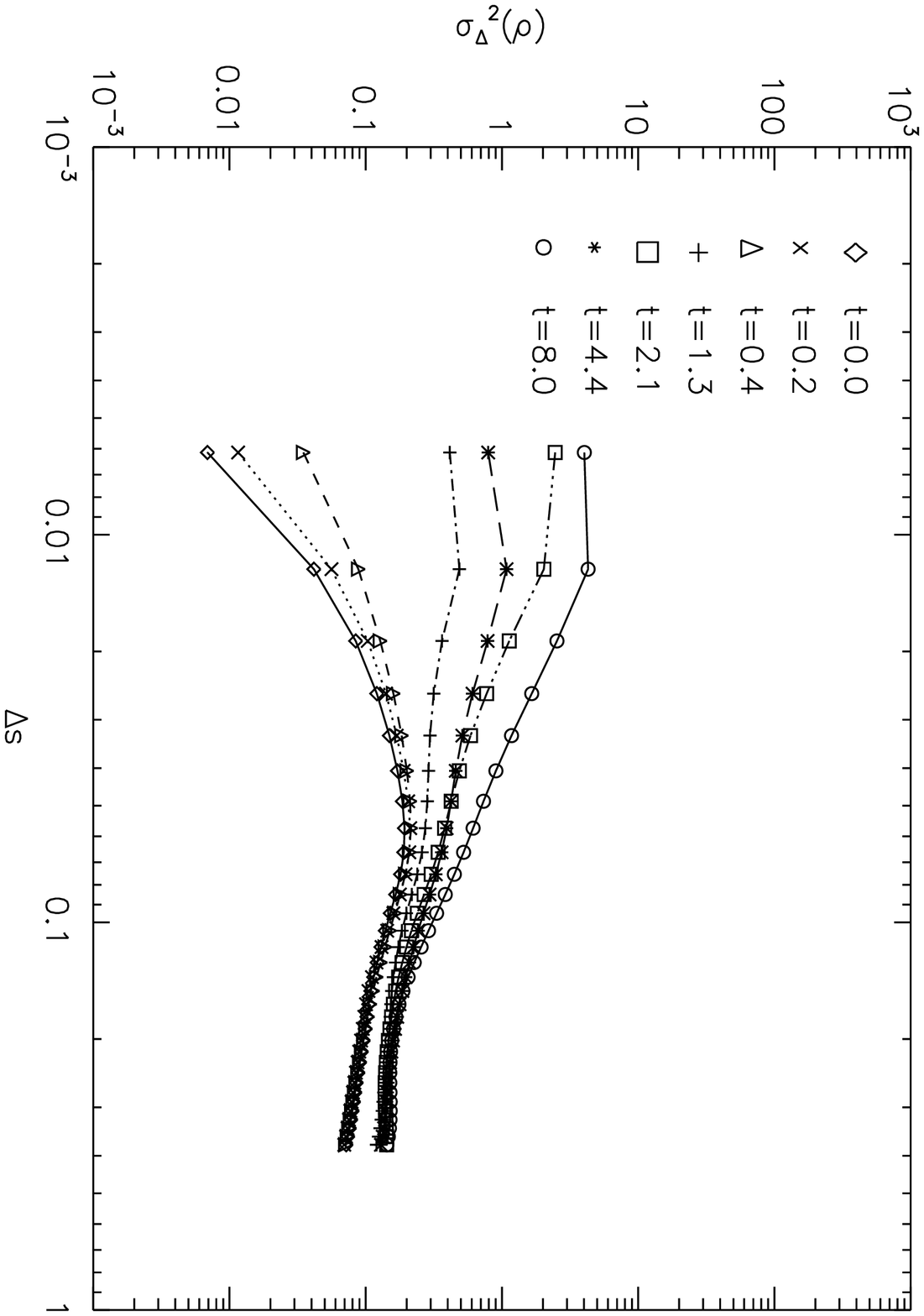}\end{rotate}}
\put(  8.3, 0.0){\begin{rotate}{90}\epsfysize=8.0cm \epsfbox{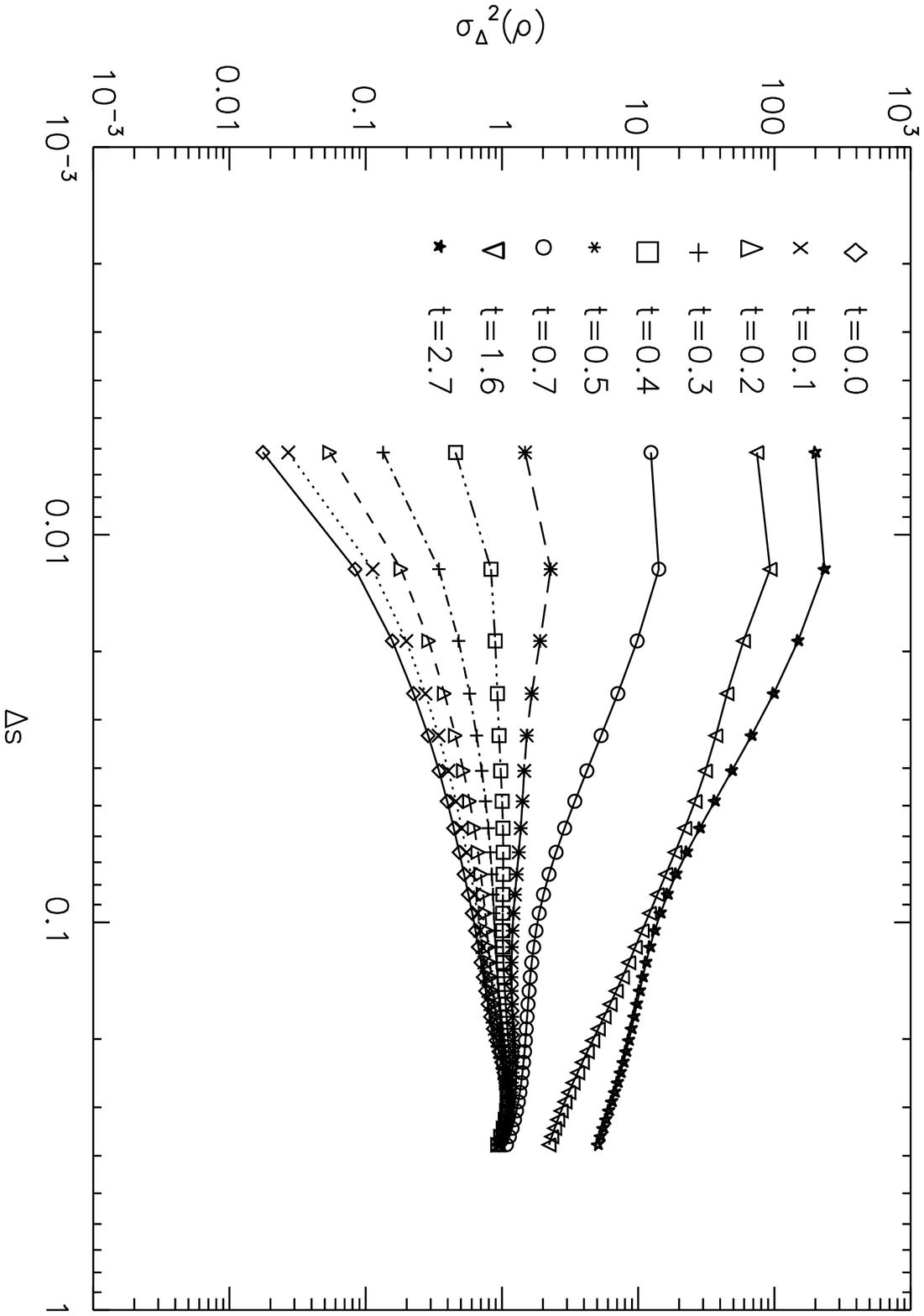}\end{rotate}}
\put( 16.8, 0.0){\begin{rotate}{90}\epsfysize=8.0cm \epsfbox{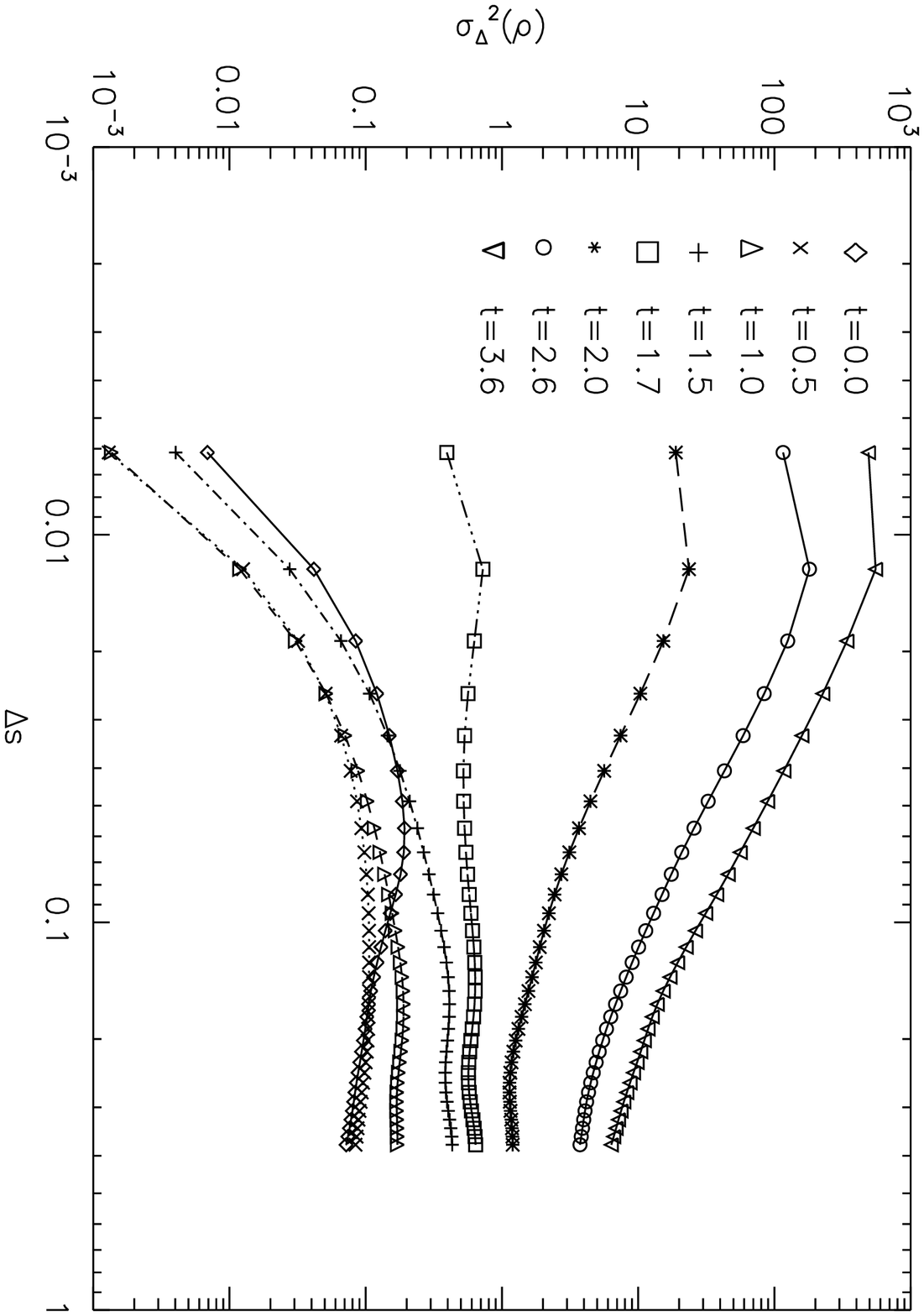}\end{rotate}}
\put(  7.6,11.2){\bf a)}
\put( 16.1,11.2){\bf b)}
\put(  7.6, 5.0){\bf c)}
\put( 16.1, 5.0){\bf d)}
\end{picture}
\vspace{0.2cm}
\caption{\label{fig:density-turb}Time evolution of the size
distribution of density variations as measured with the 
$\Delta$-variance for the four models of (a) S01, (b) S02, (c) Sd1, and (d) Sd2.
The different depicted times are indicated at the
left side of each plot. (From Ossenkopf
\etal\ 2001)}
\end{figure*}

\rskparagraph{Interaction Between Gravity and Turbulence}
\label{subsubsec:HD-turb}
To study the interplay between supersonic turbulence and self-gravity,
we consider four models of interstellar turbulence which probe
very disparate regions of the relevant parameter space. In models S01 and
Sd1 most of the turbulent kinetic energy is carried on large scales,
whereas models S02 and Sd2 involve mainly small-scale turbulent modes. 
In S01 and S02, turbulence is continuously driven such that at any 
moment the overall kinetic energy compensates the global gravitational energy.
In Sd1 and Sd2, the turbulent energy is allowed to decay freely. 

Figure \ref{fig:density-turb} shows $\sigma_{\Delta}^2(n)$ for
all four models as function of time.  In the initial plots one
can clearly see the dominance of the driving scale as discussed
by Mac~Low \& Ossenkopf (2000). The introduction of a velocity field with a
certain scale induces a pronounced peak in the density structure
at a somewhat smaller scale. Thus the curves at $t=0$ show
for the large scale-driven models a power law $\Delta$-variance
from about a third of the cube size down to the numerical
dissipation scale whereas in the small-scale driven model the
driving feature at about 0.07 dominates the structure.

Star formation is a joint feature of all considered models.  Like in the
evolution of the Gaussian density field the gravitational collapse
first modifies only the smallest scales, hardly changing the global
behavior.  As soon as local collapse occurs and the first dense
protostellar cores form and grow in mass by accretion, they represent
the main density fluctuations. Their power is concentrated on small scales 
and the $\Delta$-variance exhibits a negative slope. The structure resulting
from the collapse is very similar in the various models when we
compare evolutionary stages with about the same mass fraction
collapsed onto cores.

In the initially large-scale driven models a turbulent cascade covering
all scales is already present from the beginning shown by the highest 
$\sigma_{\Delta}^2(n)$-values at long  scales. Within this cascade the 
number of small-scale fluctuations is small compared to models S02, Sd2,
and G, and they are typically part of a larger structure so that they
are only weakly dispersed in time. This leads  to a monotonous growth
of the $\Delta$-variance on all scales.
As larger and larger regions become gravitationally unstable the
contraction comprises increasingly larger scales, and finally
$\sigma_{\Delta}^2(n)$ shifts ``upward'' on all scales while
maintaining a fixed slope.  This situation is similar in {\em all}
models with allow for large-scale collapse, i.e.  it is only prevented
in the small scale-driven model S02.

A different temporal behavior is visible in Sd2, the decaying turbulence
which was originally driven at small scales. Like in the collapse of
the Gaussian density field we find in the first steps of the
gravitational evolution a relative reduction of small-scale structure.
This decrease is due to the termination of the initial small-scale
driving resulting in a quick dissipation of the existing fluctuations
by thermal pressure if they are not Jeans supercritical, analogously
to the Gaussian collapse case. In the next steps we notice the
production of structure on larger scales. The smoothing of small-scale
turbulence combined with the onset of self-gravity leads to global
streaming motions which produce density structures correlated on a larger
scale.  Large-scale structures had been initially suppressed 
by the non-local turbulent driving
mechanism. After less than one free-fall time a kind of self-sustaining
inertial cascade with a $\Delta$-variance slope $\alpha$ of about 0.5
is build up like in all other decaying models and in the large scale
driven model. After these initial adjustments the first protostellar
cores form and we find the same dynamical behavior.

The time scale to reach a comparable collapse state and the final 
structure that we reach in the simulations differs
between the models, mainly determined by the strength of the 
turbulent driving. The exponent
$\alpha$ of the $\Delta$-variance in the collapsed state is --1.3 in
the models containing continuous driving, --1.5 in the models where
the turbulence decays during the gravitational collapse, and --1.7 for
the pure collapse of the Gaussian density field. This is
understandable, as decreasing turbulent support leads to enhanced
collapse forming stars in denser clusters. The final deviation from
the uncorrelated field of protostars which has $\alpha=-2$ is thus a
measurable indicator of the turbulent processes in the cloud during
gravitational collapse.  The local collapse produces ``point-like''
high-density cores with small overall filling factor whereas most
of the volume is supported by hydrodynamic turbulence.
In the large-scale driven model 30\,\% of the mass and in the decaying
models 60\,\% of the total mass has turned into cores within the considered 
time interval. In the model
continuously driven at small scales only 3\,\% of the mass is in
stable cores despite a considerably longer simulation time.  The
small-scale driven turbulence leads to the least efficient star
formation in an isolated mode, whereas the other cases result in the
formation of stellar aggregates and clusters (see also Klessen \etal\
2000, Klessen 2001).

Beside the deceleration of collapse the continuation of 
the initial driving has almost no influence on the 
resulting density structure as soon as the first stable 
cores have formed.
It only maintains a constant level of velocity fluctuations
in the main volume of the cloud which is dominated by
low density gas, compared to a homogeneous reduction of these 
fluctuations in the purely decaying case.

\begin{figure}[t]
\unitlength1cm
\begin{picture}(16.0,11.5)
\put(  8.0, 5.5){\begin{rotate}{90}\epsfysize=8.0cm \epsfbox{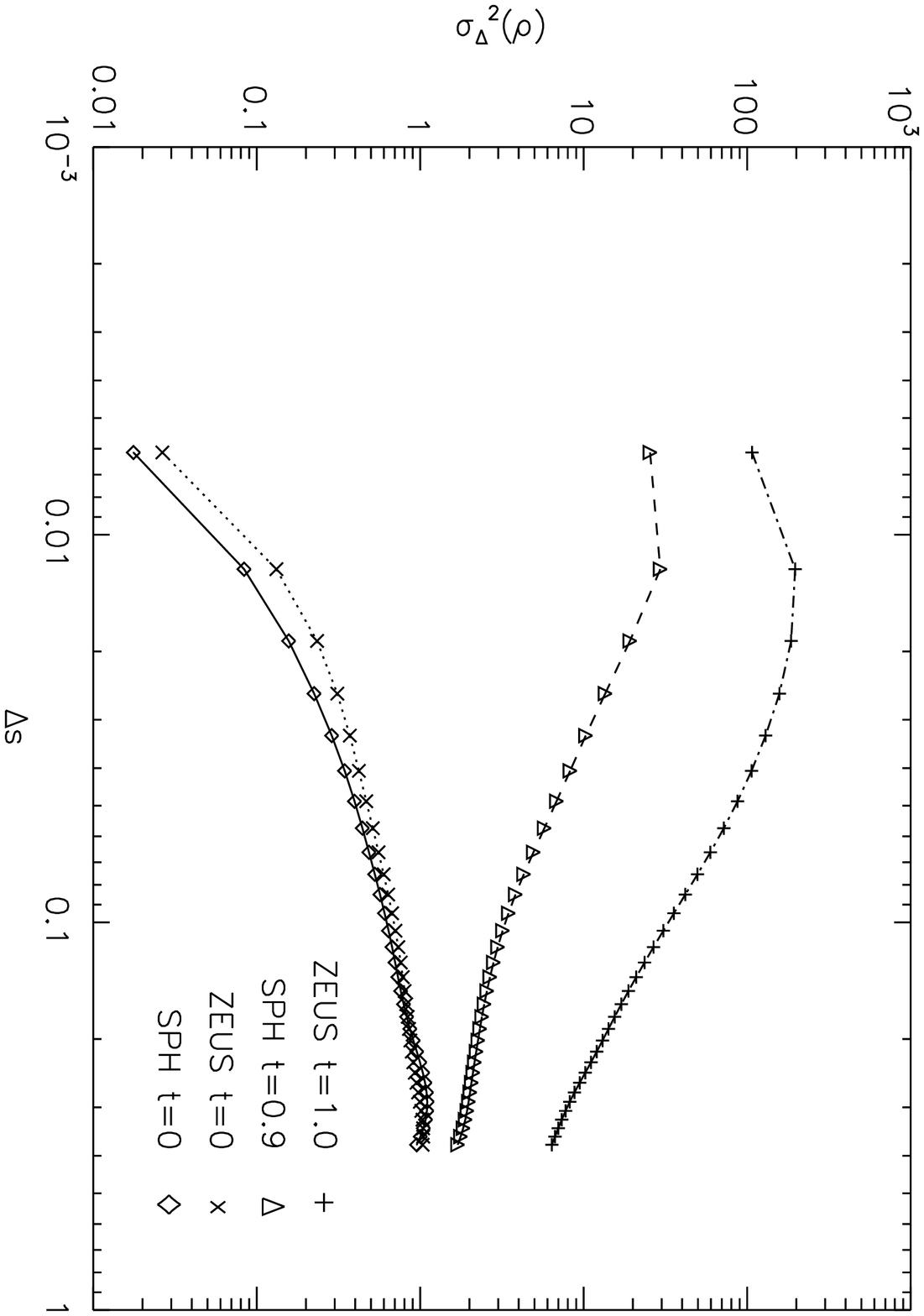}\end{rotate}}
\put(  8.0,-1.0){\begin{rotate}{90}\epsfysize=8.0cm \epsfbox{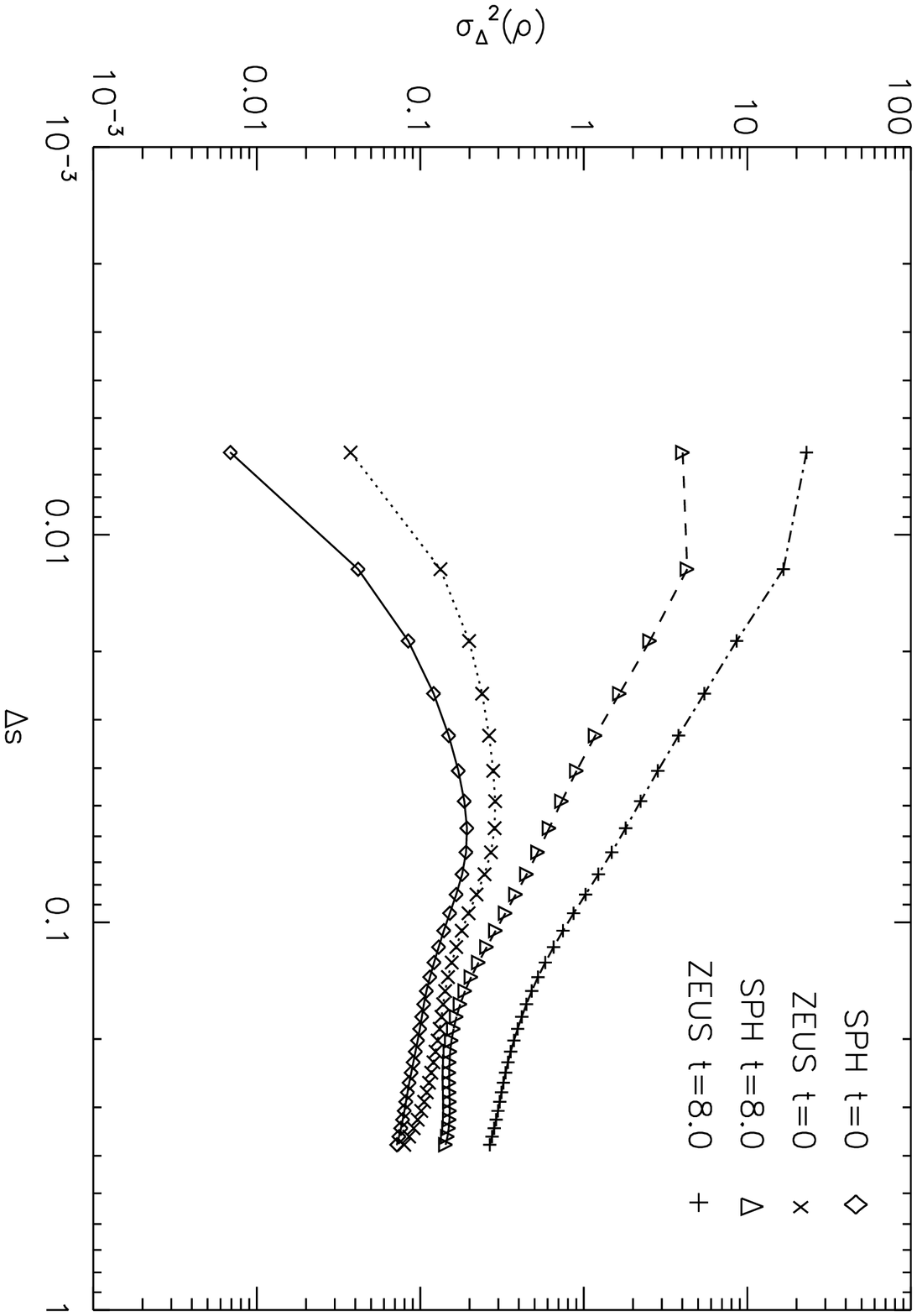}\end{rotate}}
\put( 1.3, 10.4){\bf a)}
\put( 1.3,  3.9){\bf b)}
\end{picture}
\vspace{0.6cm}
\caption{\label{fig:SPH-ZEUS}Comparison of the $\Delta$-variance
measured from the particle-based and the grid-based simulations
at the beginning and at about the same timestep of the gravitationally
collapsed state. Plot (a) shows the large-scale driven model,
plot (b) the small scale-driven situation. (From Ossenkopf
\etal\ 2001)}
\end{figure}

\rskparagraph{Influence of the Numerical Model}
\label{subsubsec:SPH-ZEUS-comparison}

Comparing the results of the particle-based SPH and the grid-based
ZEUS code we can distinguish between numerical artifacts and physical
results, as these two approaches practically bracket the real
dynamical behavior of interstellar turbulence.
In Figure \ref{fig:SPH-ZEUS} we compare the $\Delta$-variance plots
of the density structure obtained for the driven cases using 
either SPH or ZEUS, at the beginning of the gravitational collapse and
in a step where the structure is already dominated by protostellar
cores.

The scaling behavior of the density structure does not differ between
both types of simulations but the absolute magnitude of the density
fluctuations as seen in the total value of the $\Delta$-variance is
somewhat larger for all ZEUS models. In the first steps of the
large-scale driven models both numerical approaches still agree
approximately but during the evolution the scale dependent density
variations become about a factor five higher in the ZEUS model than in
the SPH approach. In the small scale driven models we can notice a
clear difference already at the beginning of the simulations. This is
consistent with the different effective resolution of the methods.
Whereas the SPH code can provide a very good spatial resolution around
the collapsing dense regions the general spatial resolution obtained
with $2\,10^5$ particles is lower than in the ZEUS simulations on a
$256^3$ grid. Thus, the damping of structures at small scales 
due to the finite resolution of the code is slightly stronger in the
SPH simulations than in the corresponding ZEUS models. 
One can for instance see that there is a virtual
reduction of structure below 0.01 which is approximately the radius of
the sink particles in the SPH code. 

As the SPH resolution is explicitly density dependent it is also
reduced on all larger scales in low density regions. This virtually smears
out part of the structure on all scales. Consequently, the $\Delta$-variance
shows lower values on all scales than in the grid-based approach because
it is not biased towards high-density regions like SPH and the observations.
The effect is larger in the small-scale driven models as the same number
of SPH particles has to  represent more shocks than in
large-scale dominated cases further reducing the effective resolution. 
Moreover, the resolution worsens during the collapse evolution
as SPH particles ``vanish'' in the sink particles. Thus, the
ZEUS simulations are preferential due to their higher resolution
if one is interested in the absolute value of the $\Delta$-variance
whereas they provide no essential advantage for the study of the
scaling behavior.

\rskparagraph{Magnetic Fields}
\label{subsubsec:MHD-turb}

\begin{figure}[t]
\unitlength1cm
\begin{center}
\begin{picture}(8.6, 5.5)
\put( 8.0,-0.3){\begin{rotate}{90}\epsfysize=8.0cm \epsfbox{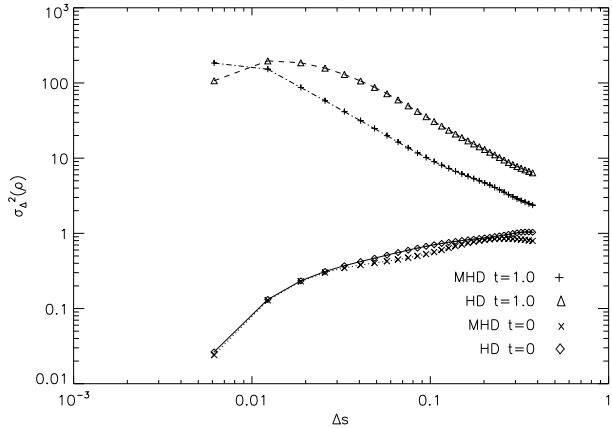}\end{rotate}}
\end{picture}
\end{center}
\caption{\label{fig:density-MHD}Comparison of the $\Delta$-variance
for the large-scale driven model in the hydrodynamic case
or the situation with strong magnetic fields, both computed
with the ZEUS code. (From Ossenkopf \etal\ 2001)}
\end{figure}

As discussed by Mac~Low \& Ossenkopf (2000) magnetic fields hardly change the
general scaling behavior in interstellar turbulence but create
anisotropies in the velocity field and therefore aligned density
structures. Since the $\Delta$-variance cannot measure anisotropies
in the density structure we do not expect to detect the influence
of the magnetic field on the turbulence by the present analysis. 
Figure \ref{fig:density-MHD} shows the $\Delta$-variance
for the initial step and an collapsed stage in a large-scale
driven hydrodynamic model and the equivalent MHD model with a 
strong magnetic field. The initial steps are almost identical but
we find that during collapse the magnetic field effectively helps to transfer 
structure from larger to smaller scales. Thus we confirm the more 
qualitative conclusion of Heitsch \etal\ (2001) that the magnetic
field slightly delays collapse by transferring part of the
turbulent kinetic energy to smaller scales. The general slope
of the $\Delta$-variance is not changed but we obtain somewhat
denser and smaller cores and somewhat less large-scale correlation at
equivalent timesteps.

Computing the $\Delta$-variance for maps projected either in the
direction of the initial magnetic field or perpendicular to it does
not show any significant difference in the density scaling 
behavior as mainly the shape of the collapsed regions is influenced,
towards spiral-shaped structures, which is not measurable with
the isotropic $\Delta$-variance filter.

We have also tested models with a smaller magnetic field where the
magnetic pressure in the order of the thermal pressure or lower.
Here, we find that the field acts like an additional 
contribution to the overall isotropic pressure so that the collapse
is somewhat delayed relative to the hydrodynamic case but the
general structure does not deviate from the hydrodynamic simulations.
Thus there is no need to discuss the weak-field situation here
separately.

\begin{figure}[ht]
\unitlength1cm
\begin{center}
\begin{picture}( 8.6, 5.5)
\put( 8.0,-0.3){\begin{rotate}{90}\epsfysize=8.0cm \epsfbox{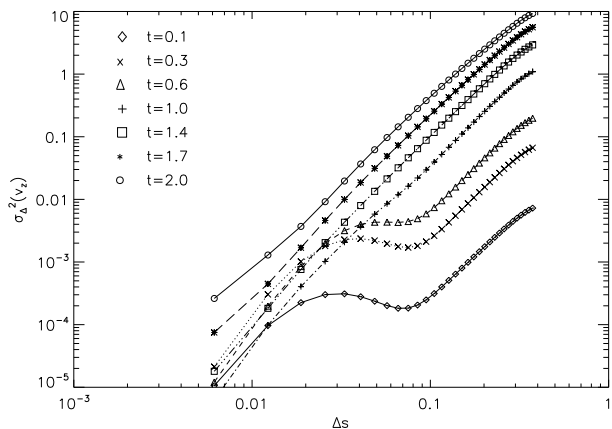}\end{rotate}}
\end{picture}
\end{center}
\caption{\label{fig:velocity-gaussian}Evolution of the $\Delta$-variance of 
the $z$-velocity in the collapse of the Gaussian density distribution
(model G).  The velocity $v_z$ is given here in units of the 
thermal sound speed $c_{\rm s}$. (From Ossenkopf \etal\ 2001)}
\end{figure}

\rsksubsubsection{Velocity Structure}
\label{subsec:velocity}

We can apply the $\Delta$-variance analysis in the same way to the velocity
structure in the simulations (Ossenkopf \& Mac Low 2001).
Figure \ref{fig:velocity-gaussian} shows the evolution of one velocity
component in the collapse of the Gaussian density fluctuations. In the
first steps where we observe a relative reduction of small scale density
fluctuations we find a bimodal velocity distribution with either very
small or very large flows. The surplus of small-scale flows just reflects
the dissipation of the initial small-scale variations by thermal pressure.
When the first stable cores have formed
the picture changes towards that of a typical shock-dominated medium
with a slope $\alpha=2$ (Ossenkopf \& Mac Low 2001) as the result
of supersonic accretion onto dense cores along the emerging filamentary
structure (Klessen \& Burkert 2000, 2001).

\begin{figure}[ht]
\unitlength1cm
\begin{center}
\begin{picture}( 8.6, 5.5)
\put( 8.0,-0.3){\begin{rotate}{90}\epsfysize=8.0cm \epsfbox{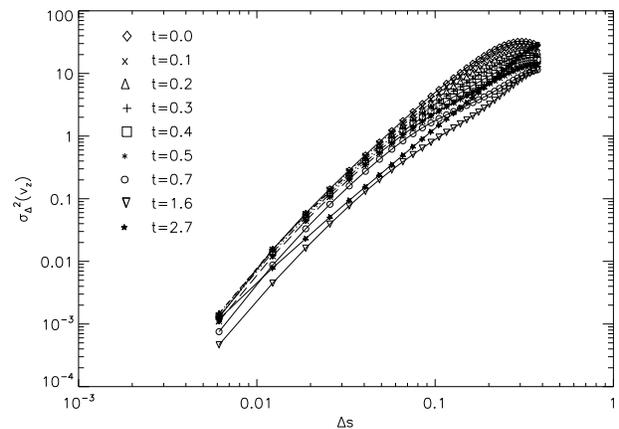}\end{rotate}}
\end{picture}
\end{center}
\caption{\label{fig:velocity-decaying}Evolution of the
$\Delta$-variance of the $z$-velocity component of the decaying
model Sd1. (From Ossenkopf \etal\ 2001)
}
\end{figure}

In all driven models we see no significant changes in the velocity
structure during collapse. This is because the $\Delta$-variance is
not focused towards the dense cores where collapse motions occur, as
these have only a small spatial filling factor. Instead, most of the
volume is occupied by tenuous intercore gas with velocity structure
that is determined by turbulent driving. The $\Delta$-variance
therefore exhibits the power-law behavior of shock dominated gas.

The same hold for decaying turbulence as well. To illustrate that
point, Figure \ref{fig:velocity-decaying} presents the evolution of the
velocity structure for model Sd1, where we drive the turbulence
initially at large scales and switch off the driving during the
gravitational collapse.  The changes in the velocity structure are
only minute. The $\Delta$-variance follows the power law of
shock-dominated flows throughout the entire evolution. Only the total
magnitude of the velocity fluctuations decreases slightly during the
initial decay of turbulence. However, after the onset of collapse,
when the majority of mass is already accumulated in dense cores, the
magnitude of $\sigma_{\Delta}$ increases again. The $\Delta$-variance
becomes dominated by the shock structure arising from the supersonic
accretion flows onto individual cores (similar to the late stages of
model G). The evolution of the small-scale decaying turbulence is not
plotted separately, as we observe the same behavior. Except during the
initial phase of turbulent decay where the velocity structure still
peaks on small scales reflecting the smaller driving wavelength used
to set up the model (see Figure \ref{fig:SPH-ZEUS}d for the density
structure), again after $t\approx 1.5$ when the initial turbulence
is sufficiently decayed away the $\Delta$-variance arrives at the power-law
behavior of shocked gas.

\begin{figure}[ht]
\unitlength1cm
\begin{center}
\begin{picture}( 8.6, 5.5)
\put( 8.0,-0.3){\begin{rotate}{90}\epsfysize=8.0cm \epsfbox{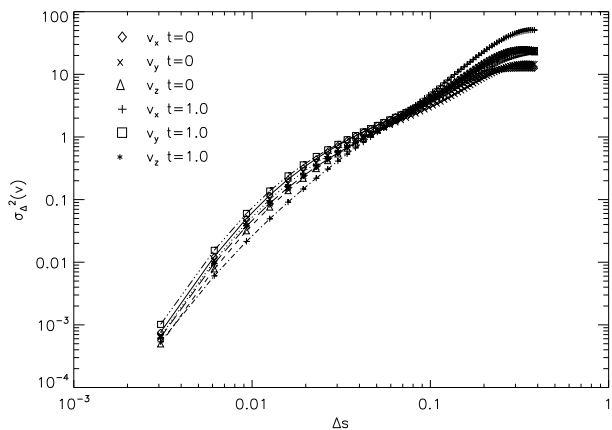}\end{rotate}}
\end{picture}  
\end{center}
\caption{\label{fig:velocity-mhd}$\Delta$-variance of all
three velocity components for model the large-scale driven
MHD model at the initial step and after one free-falling time. (From Ossenkopf \etal\ 2001)
}
\end{figure}

In Figure \ref{fig:velocity-mhd} we show the three velocity components
of the MHD model M01 at the same two timesteps like in the density plot in 
Figure \ref{fig:density-MHD}. In contrast to the findings of
Ossenkopf \& Mac~Low (2002) for sub-Alfv\'{e}nic turbulence, we see no strong 
anisotropy of the velocity field. The velocity structure
along the mean magnetic field ($z$-direction) is very similar to the
perpendicular directions throughout the dynamical
evolution of the system and well within the statistical fluctuations
expected for large-scale turbulence. The velocity structure is still
determined by supersonic turbulence rather than by the magnetic field 
structure because the turbulent rms velocity dispersion exceeds the 
Alfv\'{e}n speed in this model. The local collapse
and the formation of a cluster of collapsed cores tends to make the
influence of the magnetic field on the velocity structure even weaker.
The magnetic field merely decelerates the gravitational
collapse  and changes the geometry of the collapsing regions
as seen in Figure \ref{fig:density-MHD}, but hardly
changes the global velocity structure in this model situation.

Altogether we find that all models that are allowed to evolve freely
or are driven at large scales exhibit a similar velocity scaling
behavior, characteristic of shock-dominated media.  This is the effect of
the undisturbed turbulence evolution and the appearance of
accretion shocks. Both effects lead to remarkably
similar properties of the velocity $\Delta$-variance. Any observed
deviation from this large-range ``Larson'' behavior
indicative of shock-dominated media will hint the presence of
additional physical phenomena and could provide constraints on the
initial conditions and the dynamical state of star-forming regions.

\rsksubsubsection{Comparison with Observations}
\label{sec:observations}

\rskparagraph{Dust Observations}

To compare the simulations of gravitational collapse with
observational data of collapsed regions we have selected the observation
of a star-forming cluster in Serpens by Testi \& Sargent (1998)
as this region represents a state of star formation that may be 
similar to the outcome of our simulations.

\begin{figure}[ht]
\unitlength1cm
\begin{center}

\begin{picture}(8.6, 5.5)
\put( 8.0,-0.3){\begin{rotate}{90}\epsfysize=8.0cm \epsfbox{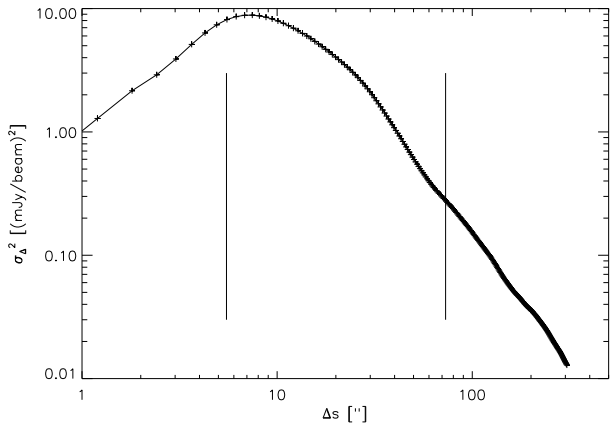}\end{rotate}}
\end{picture}  
\end{center}
\caption{\label{fig:testi}%
  $\Delta$-variance of the dust continuum map in Serpens taken by Testi \&
  Sargent (1998).  The two vertical lines represent the limits of the
  significant range as indicated with the observational data. (From Ossenkopf \etal\ 2001)}
\end{figure}

In Figure \ref{fig:testi} we show the $\Delta$-variance for the 3~mm
dust continuum map of Testi \& Sargent (1998). It reveals an increase of the
relative amount of structure from small scales towards a peak at
7$''$, an intermediate range which can be fitted by a power-law
exponent $\alpha=-1.2$, and a decay with $\alpha=-2$ indicating the
complete lack of large-scale structure at lags above 40$''$. The
behavior at largest and smallest scales can be understood when
looking at the observational base of the map. Testi \& Sargent (1998) give a
resolution for their interferometric observations of $5.5'' \times
4.3''$. Consequently we cannot see any structure below that size. The
fact that our peak falls with 7$''$ somewhat above the 5.5$''$
resolution limit might indicate that the CLEAN beam used in the reduction of
the interferometric data is not exactly Gaussian or slightly wider
than computed. The whole map is taken with an interferometric
mosaicing technique (see for details Testi \& Sargent 2000) without a 
zero-spacing by complementary single dish observations.  Thus the map
cannot contain any structure on scales above the single pointing areas
determined approximately by the size of the primary beam of the OVRO
antennas of 73$''$. This is in agreement with the lack of structure
indicated by the $\Delta$-variance slope of --2 at these scales. 
The two limiting sizes are indicated by vertical lines in
Figure \ref{fig:testi}. Thus we may only discuss the range in between
disregarding other information that is plotted in the interferometric
map but that can eventually not be obtained from the observations.

The steepening of the $\Delta$-variance in the intermediate size
range from $\alpha=-1.2$ to $\alpha=-2$  does qualitatively
agree with the behavior observed in most collapse simulations 
at small scales but does not match any of them quantitatively.
For a detailed comparison the dynamic scale range covered in the simulations
is still insufficient due to the periodic boundary conditions
constraining the large scale behavior. Hence, we can only
conclude that the collapse models show the same general structure
as the dust observations, indicating that they represent a 
realistic scenario but we cannot yet discriminate between different
models using the observational data.

\rskparagraph{Molecular Line Observations}

Bensch \etal\ (2001) provided a detailed $\Delta$-variance analysis 
of the density structure traced by observations in different CO isotopes
for several molecular clouds with different states of
star formation including quiescent clouds like the Polaris Flare
and clouds with violent star formation like Orion A.
They found for all molecular clouds a density structure approximately
characterized by a power law $\Delta$-variance, with an exponent in the
range $0.5 \le \alpha \le 1.3$.  In the best studied cloud one smooth curve
connects scales larger than 10$\,$pc (where turbulence presumably is
driven) with the dissipation scale at 0.05$\,$pc (where ambipolar
diffusion processes become important). The positive slopes 
indicate that the density structure seen in the CO isotopes is dominated by
large-scale modes. This result is consistent with purely supersonic 
turbulence and appears independent of the dynamical
state of the molecular cloud regions studied, i.e. regardless
whether the cloud forms stars or not. 

This is somewhat surprising, since we expect that the density distribution in
star-forming regions is dominated by the collapsing protostellar cores on
small spatial scales.  The $\Delta$-variance spectrum therefore should exhibit
a {\em negative} slope.  The molecular line results are also in obvious
contradiction to the Serpens dust observations discussed above.

The explanation for the difference is hidden in the radiative
transfer problem. A discussion of all major aspects of molecular
line transfer in turbulent media is  provided  by Ossenkopf (2002).
Here, it is sufficient to concentrate on one effect
-- saturation at large optical depths. Molecular lines like the 
lower transitions of ${}^{13}$CO, frequently used to map the
density profile of molecular clouds, become typically optically thick
in the cores of clouds at densities in the order of $10^5$ cm$^{-3}$.
The exact value depends
on the transition, the spatial configuration, temperatures,
and the geometry of the radiation field but one can always assign
a typical density range to the transition from the optically thin
to the optically thick regime. This leads to a saturation of the
line intensities in dense clumps so that the lines do not trace their
internal structure but rather see clump surfaces.
Moreover, the molecules tend to freeze out in dense dark
regions (Kramer \etal\ 1999) amplifying the effect that the line brightness
reflects only part of the column density in dense clumps.

As we do not want to treat the full radiative transfer problem
here, we give only an estimate for the influence of optical
depth effects by including a saturation limit into our
computations. Because the simulations are scale-free and the
typical saturation density varies for different molecules and transitions
there is no particular density value to be used for this limit
so that we have to play with different values.

\begin{figure}[t]
\unitlength1cm
\begin{picture}(8.6,11.3)
\put(  8.0, 5.5){\begin{rotate}{90}\epsfysize=8.0cm \epsfbox{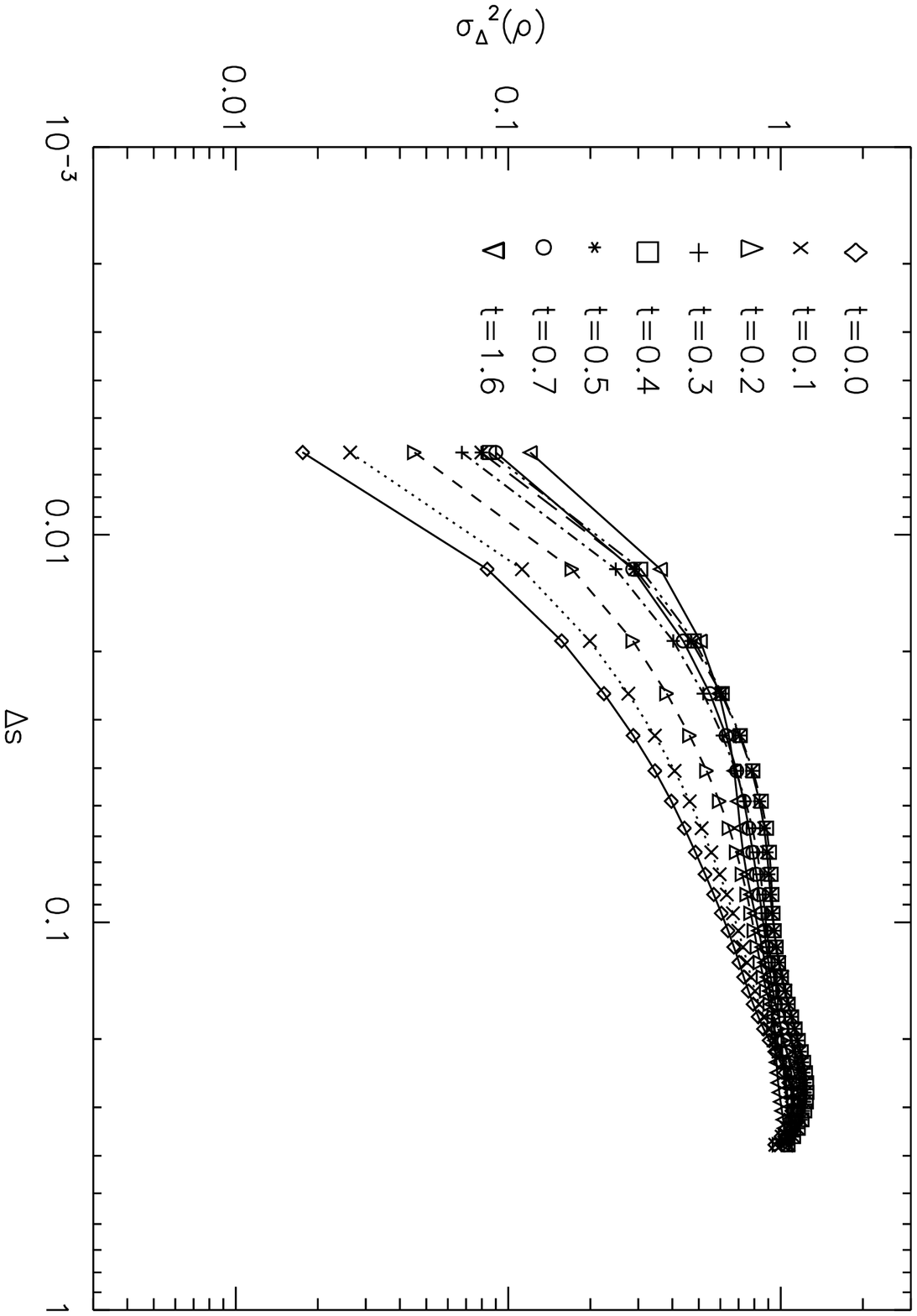}\end{rotate}}
\put(  8.0,-1.0){\begin{rotate}{90}\epsfysize=8.0cm \epsfbox{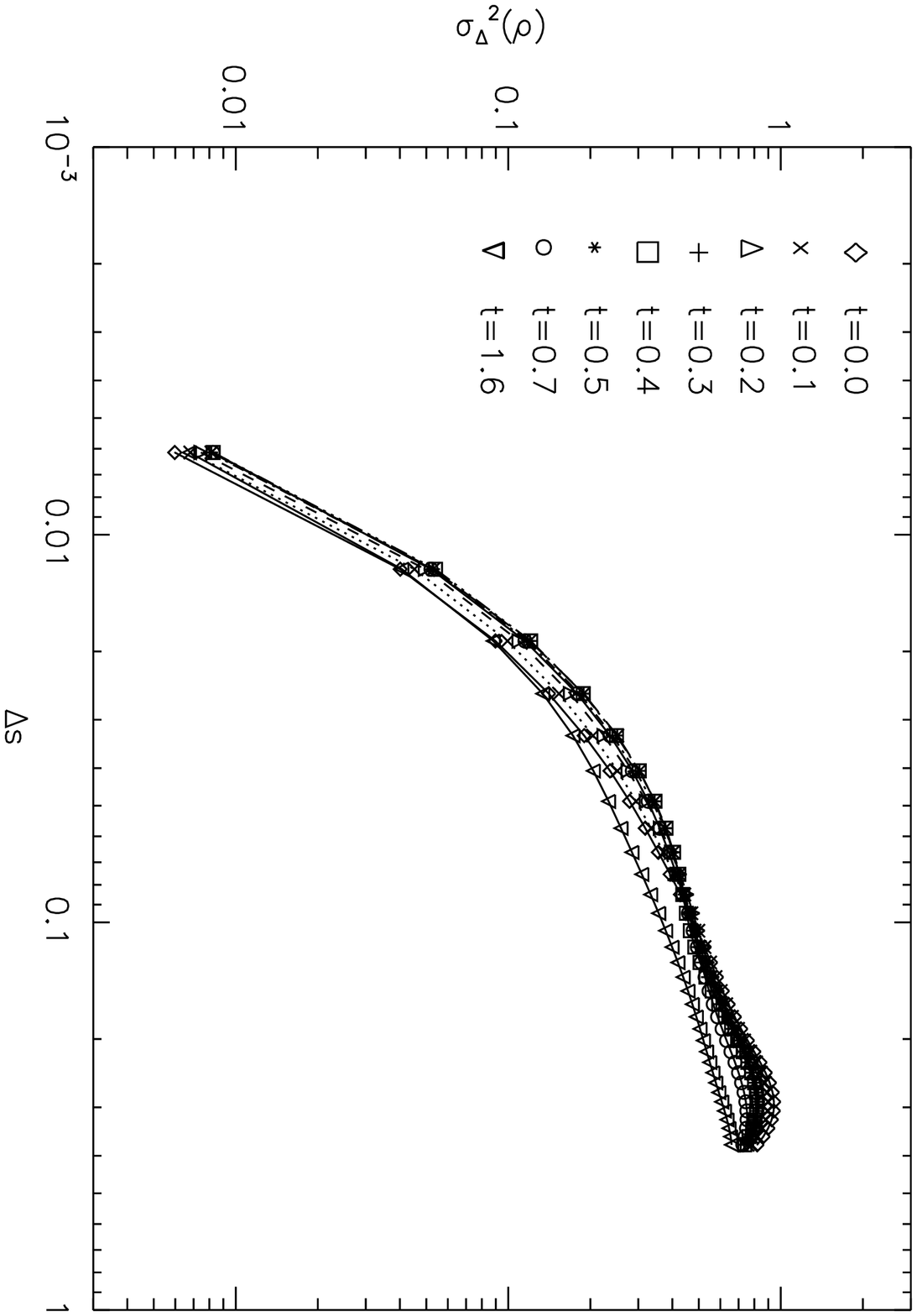}\end{rotate}}
\put( 1.3, 6.4){\bf a)}
\put( 1.3,-0.1){\bf b)}
\end{picture}
\vspace{0.4cm}
\caption{\label{fig:cutted}Time evolution of the $\Delta$-variance for 
the large-scale driven model S01 when the density structure is assumed to
saturate at densities of 240.0 (a) and 24.0 (b). (From Ossenkopf \etal\ 2001)}
\end{figure}

Figure \ref{fig:cutted} illustrates the evolution of the large-scale
driven model shown in Figure \ref{fig:density-turb}a assuming now 
that all densities above a certain threshold are invisible so that 
they are equal to the value of the saturation density. In the upper plot we
have chosen this limit to be the maximum density occurring in the
original turbulent density distribution before gravitational collapse
starts. This is 240 times the average density, i.e.\ a relatively
large value compared to the dynamic range of molecular line observations.
In the lower graph the saturation limit is reduced by a factor
10.

Although only a relatively small fraction of the material appears
at densities above the limit the influence on the $\Delta$-variance is
dramatic. As the collapsing cores produce the relative
enhancement of small scale structure their virtual removal by the saturation
results in an almost constant $\Delta$-variance behavior during the
gravitational collapse. Even in the very conservative upper plot where we
assume that all structures occurring in normal interstellar turbulence
are still optically thin and only the collapsing cores become
optically thick the $\Delta$-variance stays at a positive slope during
the entire evolution.  In the other case where optical depth effects
are assumed to be more important the $\Delta$-variance remains at a
fixed slope of $0.5$ consistent with the molecular line observations
of interstellar clouds.

Hence, optical depth effects can easily prevent the detection of 
gravitational collapse in molecular line observations, since they reproduce
the $\Delta$-variance spectrum of a turbulent molecular cloud even if
collapse has lead already to the formation of protostellar cores.
Even in massively star-forming clouds most molecules used for mapping
trace the diffuse density structure {\em between} Jeans-unstable 
collapsing protostellar cores. This gas is still dominated by interstellar
turbulence. The density contrast in star-forming molecular cloud regions simply
exceeds the density range traceable by molecular transitions.
Protostellar core densities are so high that ${}^{13}$CO at best traces the
outer envelope. Therefore both, star-forming and quiescent molecular 
clouds, exhibit very similar molecular line maps.

It is essential to resolve large density contrasts measuring
the full density structure to study the influence of self-gravity and 
local collapse in star-forming clouds. This can be
achieved using dust continuum emission. Indeed, the 3mm continuum map
of the Serpens cluster by Testi \& Sargent (1998) shows a density
structure that is dominated by small scales as predicted by
our collapse simulations. The drawback
of dust emission observations is the inherent convolution of the
density structure with the unknown temperature profile. Large
dust extinction maps could circumvent this problem but require
long integration times at NIR wavelengths to obtain a dense
sampling with background stars (Lada \etal\ 1999).

\rsksubsubsection{Summary}
\label{sec:summary-2}

Contrary to what is observed for purely hydrodynamic turbulence,
self-gravitating supersonic turbulence yields a density structure that
contains most power on the smallest scales (i.e. in the collapsed
objects) as soon as local collapse has set in. This happens
in all self-gravitating turbulence models regardless of 
the presence or absence of magnetic fields. The $\Delta$-variance 
$\sigma_{\Delta}^2(n)$ exhibits a negative slope
and peaks at small scales as soon as local collapse produces dense
cores. This is in contrast to the case of non-self-gravitating
hydrodynamic turbulence where $\sigma_{\Delta}^2(n)$ has a
positive slope and the maximum at the largest scales. Our results can
therefore be used to differentiate between different
stages of protostellar collapse in star-forming molecular clouds
and to determine scaling properties of the underlying
turbulent velocity field.

The effect of protostellar collapse, however, is not visible in
molecular line maps of star-forming clouds, as all 
molecules trace only a limited dynamic range of
densities. The density contrast in star-forming regions is
much larger. ${}^{12}$CO and ${}^{13}$CO observations, for example, 
trace only the inter-core gas distribution and at best the 
outer parts of individual protostellar cores. Hence, the density
structure seen in these molecules is indistinguishable
for star-forming and non-star-forming regions.

As resolving high density contrasts is the key for
detecting the effect of star formation in the $\Delta$-variance, we
propose observations of dust continuum or of the dust extinction
instead. These techniques do not have the same limitations of the
dynamic range and are
therefore better suited to quantitatively study the full density
evolution during the star-formation process. This is confirmed by a
first comparison of our models with the 3mm dust continuum map 
taken by Testi \& Sargent (1998) in Serpens.

%%%%%%%%%%%%%%%%%%%%%%%%%%%%%%%%%%%%%%%%%%%%%%%%%%%%%%%%%%%%%%%%%%%%%%%%%%%%%%%

%%% Local Variables: 
%%% mode: latex
%%% TeX-master: "habil"
%%% End: 

\rsksection{LOCAL STAR FORMATION}
\label{sec:local}

All present day star formation takes place in molecular clouds (e.g.\
Blitz 1993, Williams, Blitz, \& McKee 2000), so we must understand the
dynamical evolution and fragmentation of molecular clouds to
understand star formation. This section therefore begins with a brief
introduction to molecular clouds properties (\S~\ref{sub:regions}). We
then show how the efficiency and time and length scales of star
formation can depend on the properties of turbulence
(\S~\ref{sub:multiple}), followed by a discussion of the properties of
protostellar cores (\S~\ref{sub:cores}), the immediate progenitors of
individual stars.  We stress the importance of the dynamical
interaction between protostellar cores and their competition for mass
growth in dense, deeply embedded clusters (\S~\ref{sub:clusters}). This
implies strongly time-varying protostellar mass accretion rates
(\S~\ref{sub:accretion}).  Finally, we discuss the consequences of
these probabilistic processes (turbulence and stochastic mass
accretion) for the resulting stellar initial mass function
(\S~\ref{sub:imf}).

\rsksubsection{Molecular Clouds}
\label{sub:regions}

\rsksubsubsection{Composition of Molecular Clouds}
\label{subsubsec:mol-clouds}

Molecular clouds are density enhancements in the interstellar gas dominated by
molecular H$_2$ rather than the atomic H typical of the rest of the ISM,
mainly because they are opaque to the UV radiation that elsewhere dissociates
the molecules.  In the plane of the Milky Way, interstellar gas has been
extensively reprocessed by stars, so the metallicity\footnote{Metallicity is
  usually defined as the fraction of heavy elements relative to hydrogen.  It
  averages over local variations in the abundance of the different elements
  caused by varying chemical enrichment histories.}  $Z \approx Z_{\odot}$, the
solar value, while in other galaxies with lower star formation rates, the
metallicity can be as little as $10^{-3}Z_{\odot}$. This has important
consequences for the radiation transport properties and the optical depth of
the clouds.  The presence of heavier elements such as carbon, nitrogen, and
oxygen determines the heating and cooling processes in molecular clouds (e.g.\ 
Genzel 1991).  Also important, molecules formed from these elements emit the
radiation that traces the extent of molecular clouds.  Radio and submillimeter
telescopes mostly concentrate on the rotational transition lines of carbon,
oxygen and nitrogen molecules (e.g.\ CO, NH$_3$, or H$_2$0). By now, several
hundred different molecules have been identified in the interstellar gas. An
overview of the application of different molecules as tracers for different
physical conditions can be found in the reviews by van Dishoeck \etal\ (1993),
Langer \etal\ (2000), van Dishoeck \& Hogerheijde (2000).

\rsksubsubsection{Density and Velocity Structure of Molecular Clouds}
\label{subsubsec:LSS}

Emission line observations of molecular clouds reveal clumps and filaments on
all scales accessible by present day telescopes. Typical parameters of
different regions in molecular clouds are listed in Table \ref{tab:MC-prop}, 
adopted from Cernicharo (1991). The mass spectrum of clumps in molecular
clouds appears to be well described by a power law, indicating
self-similarity: there is no natural mass or size scale between the lower and
upper limits of the observations.  The largest molecular structures considered
to be single objects are giant molecular clouds (GMCs), which have masses of
$10^5$ to $10^6\,{\rm M}_{\odot}$, and extend over a few tens of parsecs. The
smallest observed structures are protostellar cores with masses of a few solar
masses or less and sizes of $\sil 10^{-2}\,$pc, and less-dense clumps of
similar size.  The volume filling factor of dense clumps, even denser
subclumps and so on, is of the order of 10\% or less.  Star formation always
occurs in the densest regions within a cloud, so only a small fraction of
molecular cloud matter is actually involved in building up stars, while the
bulk of the material remains at lower densities.

\begin{table*}[t]
{\caption{\label{tab:MC-prop}
 Physical properties of interstellar clouds}}
%\begin{tabular}[t]{p{2.5cm}p{3.5cm}p{3.0cm}p{3.0cm}p{3.0cm}}
\begin{tabular}[t]{lllll}
\hline \hline\\[-0.4cm]
& \parbox[t]{3.2cm}{GIANT MOLE\-CULAR CLOUD COMPLEX} & \parbox[t]{3.0cm}{MOLECULAR CLOUD} & \parbox[t]{2.3cm}{STAR-FORMING CLUMP} &
\parbox[t]{2.4cm}{PROTO\-STELLAR CORE$^a$}\vspace{0.2cm} \\
\hline\\[-0.4cm]
%\tableline
Size (pc)                           & $10 - 60$  &$2 - 20$   & $0.1-2$    &$\sil 0.1$\\
Density ($n({\rm H}_2)/{\rm cm}^3$) & $100-500$  &$10^2-10^4$& $10^3-10^5$&$>10^5$ \\
Mass (M$_{\odot}$)                  & $10^4-10^6$&$10^2-10^4$& $10 - 10^3$&$0.1-10$ \\
Line width (km$\,$s$^{-1}$)         & $5-15$     &$1 - 10$   & $0.3-3$    &$0.1-0.7$\\
Temperature (K)                     & $7-15$     &$10-30$    & $10-30$    &$7-15$ \\
Examples                            & \parbox[t]{3.0cm}{W51, W3, M17, \\\mbox{Orion-Monoceros,}\\
Taurus-Auriga-Perseus complex} & \parbox[t]{2.6cm}{L1641, L1630, \\ W33, W3A,\\ B227, L1495,\\
L1529}&& \parbox[t]{2.4cm}{see \\\S~\ref{sub:cores}}\vspace{0.2cm}\\
\hline\hline\\[-0.3cm]
\end{tabular}
{\footnotesize \vspace{0.2cm} $^a$~Protostellar cores in the "prestellar" phase,
  i.e.\ before the formation of the protostar in its interior.
}
\end{table*}

The density structure of molecular clouds is best seen in the column
density of dust, which can be observed via its thermal emission at
millimeter wavelengths in dense regions (e.g.\ Testi \& Sargent 1998,
Motte, Andr\'e, \& Neri 1998), or via its extinction of background
stars in the infrared, if a uniform screen of background stars is
present (Lada \etal\ 1994, Alves, Lada, \& Lada 2001).  The latter
method relies on the near-IR color excess to measure column densities,
which ensures a much greater dynamic range than optical extinction.
This method has been further developed by Cambr\'esy \etal\ (2002) who
use an adaptive grid to extract maximum information from non-uniform
background star fields. It turns out  that the
higher the column density in a region, the higher the variation in
extinction among stars behind that region (Lada \etal\ 1994).  Padoan \& Nordlund (1999)
demonstrated this to be consistent with a super-Alfv\'enic turbulent
flow, while Alves \etal\ (2001) modeled it with a single cylindrical
filament with density $\rho \propto r^{-2}$. Because turbulence forms
many filaments, it is not clear that these two descriptions are
actually contradictory (P. Padoan, 2001, private communication),
although the identification of a single filament would then suggest
that a minimum scale for the turbulence has been identified.

A more general technique is optically thin spectral lines. The best
candidates are $^{13}{\rm CO}$ and C$^{18}$O, though CO freezes out in
the very densest regions with visual extinctions above $A_V \simeq 10$
magnitudes (Alves, Lada, \& Lada, 1999).  Molecular line
observations are therefore only sensitive to gas at relatively low
densities and are limited in dynamic range to at most two decades of
column density.  Nevertheless, the development of sensitive radio
receivers in the 1980's first made it feasible to map an entire
molecular cloud region with high spatial and spectral resolution to
obtain quantitative information about the overall density structure.

\begin{figure*}[tp]
\unitlength1.0cm
\begin{picture}(16,17.9)
%\put(1.0,0.0){\epsfysize=18cm \epsfbox{Falgarone-et-al-1992.eps}}
\put(1.5,0.2){\epsfysize=17.5cm \epsfbox{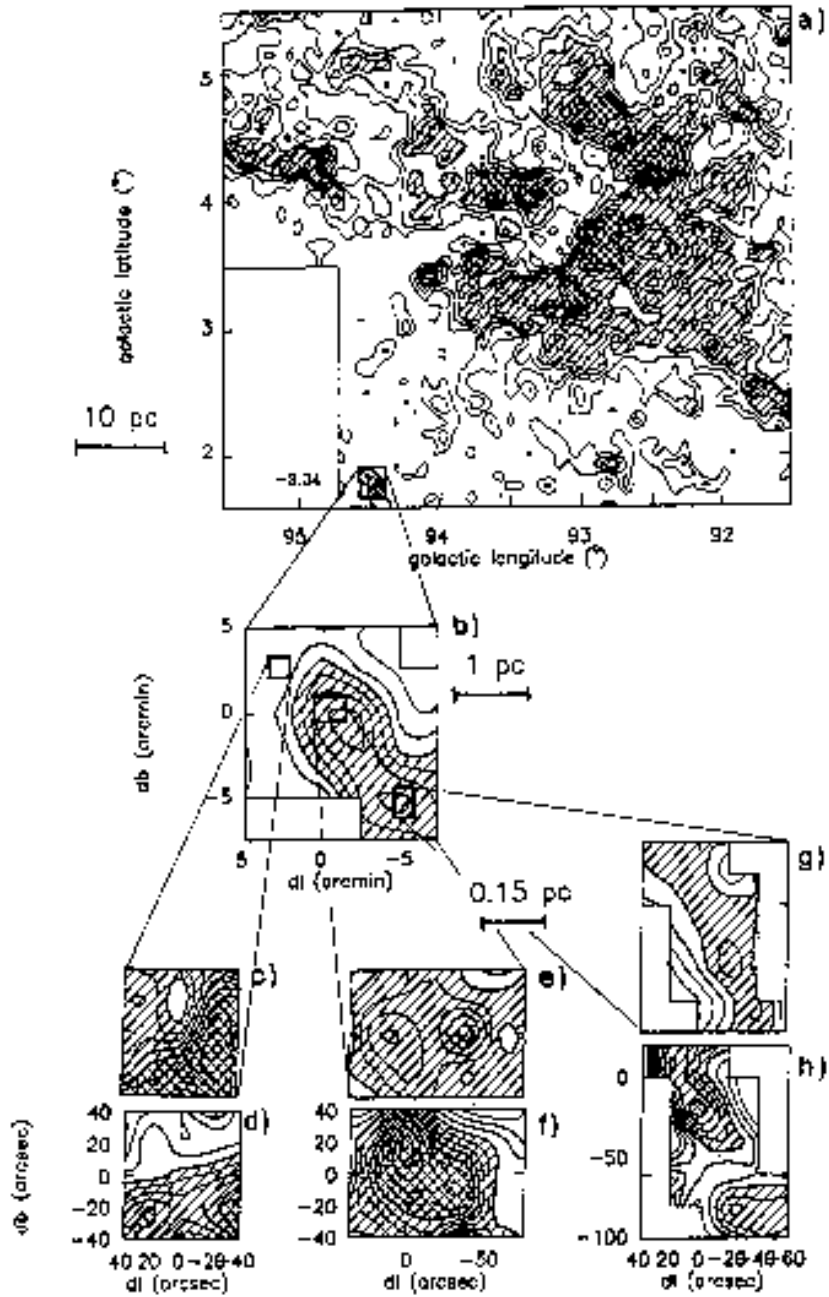}}
\end{picture}
\caption{\label{fig:falg92}
Maps of the molecular gas in the Cygnus OB7 complex. (a) Large scale
map of the $^{13}$CO $({\rm J}=1-0)$ emission. The first level and the
contour spacing are $0.25\,$K. (b) Map of the same transition line of a
sub-region with higher resolution (first contour level and spacing are
$0.3\,$K). Both maps are obtained using the Bordaux telescope. (c)
$^{12}$CO $({\rm J}=1-0)$ and (d) $^{13}$CO $({\rm J}=1-0)$ emission
from the most transparent part of the field. (e) $^{13}$CO $({\rm
J}=1-0)$ and (f) C$^{18}$O $({\rm J}=1-0)$ emission from the most
opaque field. (g) $^{13}$CO $({\rm J}=1-0)$ and (h) C$^{18}$O $({\rm
J}=1-0)$ emission from a filamentary region with medium density. The
indicated linear sizes are given for a distance to Cygnus OB7 of
$750\,$pc. (The figure is from Falgarone \etal\ 1992). 
% e-mail edith@lra.ens.fr 
}
\end{figure*}
The hierarchy of clumps and filaments spans all observable scales (e.g.\ 
Falgarone, Puget, \& Perault 1992, Falgarone \& Phillips 1996, Wiesemeyer
\etal\ 1997) extending down to individual protostars studied with
mm-wavelength interferometry (Ward-Thompson \etal\ 1994, Langer \etal\ 1995,
Gueth \etal\ 1997, Motte \etal\ 1998, Testi \& Sargent 1998,
Ward-Thompson, Motte, \& Andr{\'e} 1999, Bacmann \etal\ 2000, Motte \etal\ 
2001). This is illustrated by Figure \ref{fig:falg92}, which shows $^{13}$CO,
$^{12}$CO and C$^{18}$O maps of a region in the Cygnus OB7 complex at three
levels of successively higher resolution (from Falgarone \etal\ 1992). At each
level, the molecular cloud appears clumpy and highly structured. When observed
with higher resolution, each clump breaks up into a filamentary network of
smaller clumps. Unresolved features exist even at the highest resolution. The
ensemble of clumps identified in this survey covers a mass range from about
$1\,$M$_{\odot}$ up to a few $100\,$M$_{\odot}$ and densities $50\,{\rm
cm}^{-3} < n({\rm H}_2) < 10^4\,{\rm cm}^{-3}$. These values are typical for
all studies of cloud clump structure, with higher densities being reached
primarily in protostellar cores.

The distribution of clump masses is consistent with a power law of the form
\begin{equation}
\label{eqn:clump-spectrum}
\frac{dN}{dm} \propto m^{\alpha}\:,
\end{equation}
with $-1.3 < \alpha < -1.9$ in molecular line studies (Carr 1987, Stutzki \&
G{\"u}sten 1990, Lada, Bally, \& Stark 1991, Williams, de Geus, \& Blitz 1994,
Onishi \etal\ 1996, Kramer 1998, Heithausen \etal\ 1998). Dust continuum
studies, which pick out the densest regions, find steeper values of $-1.9 <
\alpha < -2.5$ (Testi \& Sargent 1998, Motte \etal\ 1998, see also the
discussion in Ossenkopf \etal\ 2001) similar to
the stellar mass spectrum. The power-law mass
spectrum is often interpreted as a manifestation of fractal density structure
(e.g.\ Elmegreen \& Falgarone 1996).  However, the full physical meaning
remains unclear. In most studies molecular cloud clumps are determined either
by a Gaussian decomposition scheme (Stutzki \& G{\"u}sten 1990) or by the
attempt to define (and separate) clumps following density peaks (Williams
\etal\ 1994). There is no one-to-one correspondence between the identified
clumps in either method, however. Furthermore, molecular clouds are only seen
in projection, so one only measures surface density instead of volume
density.  Even when velocity information is taken into account, the real
three-dimensional structure of the cloud remains elusive. In particular, it
can be demonstrated in models of interstellar turbulence that single clumps
identified in a position-position-velocity space tend to separate into
several clumps in real three-dimensional space (Ballesteros-Paredes \& Mac Low
2002).  This projection effect in particular may render clump mass spectra
improper statistical tools for characterizing molecular cloud structure.

Other means to quantify the structural and dynamical properties of molecular
clouds involve correlations and probability distribution functions (PDF's) of
dynamical variables. Two-point correlation functions have been studied by many
authors, including Scalo (1984), Kleiner \& Dickman (1987), Kitamura \etal\ 
(1993), Miesch \& Bally (1994), LaRosa, Shore \& Magnani (1999), and
Ballesteros-Paredes, V{\'a}zquez-Semadeni, \& Goodman (2002), while other
studies have concentrated on analyzing the PDFs of the column density in
observations, both physical and column density in computational models, and of
dynamical observables such as the centroid velocities of molecular lines and
their differences.  The density PDF has been used to characterize numerical
simulations of the interstellar medium by V{\'a}zquez-Semadeni (1994), Padoan,
Nordlund, \& Jones (1997), Passot, \& V{\'a}zquez-Semadeni (1998), Scalo
\etal\ (1998), and Klessen (2000). Velocity PDFs for several star-forming
molecular clouds have been determined by Miesch \& Scalo (1995) and Miesch,
Scalo \& Bally (1999). Lis \etal\~(1996, 1998) analyzed snapshots of a
numerical simulation of mildly supersonic, decaying turbulence (without
self-gravity) by Porter, Pouquet, \& Woodward (1994) and applied the method to
observations of the $\rho$-Ophiuchus cloud. Altogether, the observed PDFs
exhibit strong non-Gaussian features, often being nearly exponential with
possible evidence for power-law tails in the outer parts. Further methods to
quantify molecular cloud structure involve spectral correlation methods
(Rosolowsky \etal\ 1999), principal component analysis (Heyer \& Schloerb
1997), or pseudometric methods used to describe and rank cloud complexity
(Wiseman \& Adams 1994, Adams \& Wiseman 1994).  

A technique especially sensitive to the amount of structure on different
spatial scales is wavelet analysis (e.g.\ Gill \& Henriksen 1990; Langer,
Wilson, \& Anderson 1993). In particular, the $\Delta$-variance, introduced by
Stutzki \etal\ (1998), provides a good separation of noise and observational
artifacts from the real cloud structure. For isotropic systems its slope is
directly related to the spectral index of the corresponding Fourier power
spectrum. It can be applied in an equivalent way both to observational data
and hydrodynamic and MHD turbulence simulations, allowing a direct comparison,
as discussed by Mac~Low \& Ossenkopf (2000), Bensch, Stutzki, \& Ossenkopf
(2001), and Ossenkopf \& Mac Low (2002).  They find that the structure of
low-density gas in molecular clouds is dominated by large-scale modes and,
equivalently, the velocity field by large-scale motions. This means that
molecular cloud turbulence is likely to be driven from the outside, by sources
acting external to the cloud on scales of at least several tens of parsec
(Ossenkopf \& Mac~Low 2002).  The observational findings are different,
however, when focusing on high-density gas in star forming regions. In this
case, the $\Delta$-variance clearly shows that the density structure is
dominated by individual protostellar cores at the smallest resolved scales
(Ossenkopf, Klessen, \& Heitsch 2001). This effect is best seen in dust
emission maps as these are able to trace large density contrasts.
Alternatively, dust extinction maps may also prove to be useful in giving high
resolution and tracing large density contrasts (see e.g.\ Alves \etal\ 2000
for the Bok globule B68; or Padoan, Cambr\'esy, \& Langer 2002 for the Taurus
molecular cloud).  Molecular line observations are only sensitive to gas at
relatively low densities, so they mainly trace the gas between protostellar
cores: no signs of collapse appear in $\Delta$-variance analyses of line maps
(Ossenkopf \etal\ 2001).

\rsksubsubsection{Support of Molecular Clouds}
\label{subsubsec:support}

Molecular clouds are cold (e.g.\ Cernicharo 1991). The kinetic temperature
inferred from molecular line ratios is typically about 10$\,$K for dark,
quiescent clouds and dense cores in GMCs which are shielded from UV radiation
by high column densities of dust, while it can reach $50-100$~K in regions
heated by UV radiation from high-mass stars.  For example, the temperature of
gas and dust behind the Trapezium cluster in Orion is about 50$\,$K.  In cold
regions, the only heat source is cosmic rays, while cooling comes from
emission from dust and the most abundant molecular species.  The thermal
structure of the gas is related to its density pattern and its chemical
abundance, so it is remarkable that over a wide range of gas densities and
metallicities the equilibrium temperature remains almost constant in a small
range around $T\approx 10\,$K (Goldsmith \& Langer 1978, Goldsmith 2001).  The
approximation of isothermality only breaks down when the cloud becomes opaque
to cooling radiation so that heat can no longer be radiated away efficiently,
which occurs when the gas density $n({\rm H}_2) > 10^{10}$cm$^{-3}$.  The
equation of state then moves from isothermal with polytropic exponent $\gamma
=1$ to adiabatic, with $\gamma \sim 7/5$ being appropriate for molecular
hydrogen (see e.g.\ Tohline 1982 and references therein).

Because of their low temperatures, GMCs were traditionally claimed to
be gravitationally bound (Kutner \etal\ 1977; Elmegreen , Lada, \&
Dickinson 1979; Blitz 1993; Williams \etal\ 2000).  Their masses are
orders of magnitude larger than the critical mass for gravitational
stability $M_J$ defined by Equation (\ref{eqn:jeans-mass}), computed from
the average density and temperature.  However, if only thermal
pressure opposed gravitational attraction, they should be collapsing
and very efficiently forming stars on a free-fall timescale, which is
roughly $\tau_{\rm ff} \approx 10^6\,$years, Equation~(\ref{eqn:free-fall-time}). That is not the case. Within molecular
clouds low-mass gas clumps appear highly transient and pressure
confined rather then being bound by self-gravity. This is the case
only for the most massive individual cores. These are the sites where
star formation actually is observed (Williams, Blitz, \& Stark 1995;
Yonekura \etal\ 1997; Kawamura \etal\ 1998; Simon \etal\ 2001).

Molecular cloud lifetimes are estimated to lie between several $10^6$~up to a
few $10^7\,$years (see \S~\ref{sub:clouds} for detailed discussion). With such
short lifetimes, molecular clouds are likely never to reach a state of
dynamical equilibrium (Ballesteros-Paredes \etal\ 1999; Elmegreen 2000).  This is in contrast to the classical picture which
sees molecular clouds as long-lived equilibrium structures (Blitz \& Shu
1980). The overall star formation efficiency on scales of molecular
clouds as a whole is low in our Galaxy, of order of 10\% or smaller.

Except maybe on scales of isolated protostellar cores, the observed line
widths are always wider than implied by the excitation temperature of the
molecules. This is interpreted as the result of bulk motion associated with
turbulence. We argue in this review that it is the interstellar turbulence
that determines the lifetime and fate of molecular clouds, and that regulates
their ability to collapse and form stars.

Magnetic fields have long been discussed as a stabilizing agent in
molecular clouds. However, magnetic fields with average field strength
of 10$\,\mu$G (Verschuur 1995a,b; Troland \etal\ 1996) are not
sufficient to stabilize molecular clouds as a whole. This is in
particular true on scales of individual protostars where magnetic
fields appear too weak to hold gravitational collapse in essentially
all cases observed (see \S~\ref{sub:standardprobs}). Furthermore,
magnetic fields are not capable of preventing turbulent velocity
fields from decaying quickly (see e.g.\ Figure \ref{fig:prlres} and
its discussion). It appears that the role of magnetic fields in stabilizing
giant molecular clouds as a whole is less important than assumed in
the standard theory (\S~\ref{sub:standard}).  The conclusion is, that
molecular cloud turbulence must at least be partially driven by some
energy source to have molecular clouds survive over several dynamical
timescales and to explain the observed low star formation efficiencies
on scales of molecular clouds as a whole\footnote{Note that on scales
  of individual star forming regions the efficiency to convert
  molecular cloud material into stars can be very high and reach
  values up to 50\%.  Only a small fraction of molecular cloud
  material associated with high-density regions is involved in the
  star formation process. The bulk of molecular cloud matter is in
  `inactive' tenous state between individual star forming regions.}.
For a further discussion of possible driving mechanisms for
interstellar turbulence see \S~\ref{sub:efficient}.

\rsksubsubsection{Scaling Relations for Molecular Clouds}
\label{subsubsec:scaling-law}

Observations of molecular clouds exhibit correlations between various
properties, such as clump size, velocity dispersion, density and mass. Larson
(1981) first noted, using data from several different molecular cloud surveys,
that the density $\rho$ and the velocity dispersion $\sigma$ appear to scale
with the cloud size $R$ as
\begin{eqnarray}
\rho &\propto& R^{\alpha}\;\label{eqn:larson-a}\\
\sigma &\propto& R^{\beta}\;,\label{eqn:larson-b}
\end{eqnarray}
with $\alpha$ and $\beta$ being constant scaling exponents. 
Many studies have been done of the scaling properties of molecular clouds. The
most commonly quoted values of the exponents are $\alpha \approx -1.15 \pm
0.15$ and $\beta \approx 0.4 \pm 0.1$ (e.g.\ Dame \etal\ 1986, Myers \&
Goodman 1988, Falgarone \etal\ 1992, Fuller \& Myers 1992, Wood, Myers, \&
Daugherty 1994, Caselli \& Myers 1995).  However, the validity of these
scaling relations is the subject of strong controversy and significantly
discrepant values have been reported by Carr (1987) and Loren (1989), for
example.

The above standard values are often interpreted in terms of the virial
theorem (Larson 1981, Caselli \& Myers 1995).  If one assumes virial
equilibrium, Larson's relations (Equations \ref{eqn:larson-a} and
\ref{eqn:larson-b}) are not independent. For $\alpha = -1$, which
implies constant column density, a value of $\beta = 0.5$ suggests
equipartition between self-gravity and the turbulent velocity
dispersion, that is that the ratio between kinetic and potential
energy is constant with 
%$E_{\rm kin}/|E_{\rm pot}| = \sigma^2R/(2GM)\approx 1/2$.  
$K/|W| = \sigma^2R/(2GM)\approx 1/2$.  
Note, that for any arbitrarily chosen value of the
density scaling exponent $\alpha$, a corresponding value of $\beta$
obeying equipartition can always be found (V{\'a}zquez-Semadeni \&
Gazol 1995).  Equipartition is usually interpreted as indicating
virial equilibrium in a static object.  However, Ballesteros-Paredes
\etal\ (1999b) pointed out that in a dynamic, turbulent environment,
the other terms of the virial equation (McKee \& Zweibel 1992) can
have values as large as or larger than the internal kinetic and
potential energy.  In particular, the changing shape of the cloud will
change its moment of inertia, and turbulent flows will produce large
fluxes of kinetic energy through the surface of the cloud.  As a
result, rough equipartition between internal kinetic and potential
energy does not necessarily imply virial equilibrium.

Furthermore, Kegel (1989) and Scalo (1990) proposed that the density-size
relation may be a mere artifact of the limited dynamic range in the
observations, rather than reflecting a real property of interstellar clouds.
In particular, in the case of molecular line data, the observations are
restricted to column densities large enough that the tracer molecule is
shielded against photodissociating UV radiation. Also, with limited
integration times, most CO surveys tend to select objects of roughly constant
column density, which automatically implies $\rho \propto R$.  Surveys that
use larger integration times and therefore have larger dynamic range seem to
exhibit an increasingly larger scatter in density-size plots, e.g.~as seen in
the data of Falgarone \etal\ (1992).  Results from numerical simulations,
which are free from observational bias, indicate the same trend
(V{\'a}zquez-Semadeni, Ballesteros-Paredes, \& Rodriguez 1997).
Three-dimensional simulations of supersonic turbulence (Mac Low 1999) were
used by Ballesteros-Paredes \& Mac Low (2002) to perform a comparison of 
clumps measured as density enhancements in physical space to clumps measured
column density enhancements in simulated observational space
(position-velocity). They found no relation between density and size in
physical space, but a clear trend of $\rho \propto R$ in the simulated
observations, caused simply by the tendency of clump-finding algorithms to
pick out clumps with column densities close to the local peak values.

There are two other concerns. The proportionality between line
integrated CO intensity and molecular mass surface density has been
reliably established only for extragalactic observations (Dickman,
Snell, \& Schloerb 1986).  This relationship is only valid for scales
larger than a few parsec, at which calibration has been possible.
Also it depends on the assumption of virialization of the gas (e.g.\
Genzel 1991), and local thermodynamic equilibrium.  Padoan \etal\ (2000)
demonstrated that the assumption of local thermodynamic equilibrium
can lead to underestimates of the actual column density in turbulent
molecular clouds by factors of up to 7.  Additionally, for clumps
within molecular clouds, the structures identified in CO often do not
correspond to those derived from higher-density tracers (see e.g.\
Langer \etal\ 1995, Bergin \etal\ 1997, Motte \etal\ 1998 for
observational discussion, and Ballesteros-Paredes \& Mac Low 2002 for
theoretical discussion).  Altogether, the existence of a unique
density-size relation and its astrophysical meaning is not well
established.

The velocity-size relation does not appear to be so prone to
observational artifacts.  However, many measurements of molecular
clouds do not exhibit this correlation (e.g.\ Loren 1989, or Plume
\etal\ 1997).  If it is detected in a molecular cloud, it probably is
a real property of the cloud that may be explained by a number of
physical mechanisms, ranging from the standard (though incomplete)
argument of virial equilibrium to the action of interstellar
turbulence. In supersonic turbulent flows, however, the scaling
relation is a natural consequence of the characteristic energy
spectrum in an ensemble of shocks, even in the complete absence of
self-gravity (Ossenkopf \& Mac Low 2002, Ballesteros-Paredes \&
Mac~Low 2002, Boldyrev, Nordlund, \& Padoan 2002).

\rsksubsection{Star Formation in Mo\-le\-cu\-lar Clouds}
\label{sub:multiple}
%\input{multiple.tex}

%%%
%%% RMP-Module for the "Local star-formation regions"
%%%
%%% checked out by RSK: 16.08.02
%%% checked in  by RSK: 
%%%
%%% \rsksubsection{Multiple star formation}
%%%  / From single stars to stellar
%%%  clusters / Star formation from turbulent molecular cloud fragmentation}
%%% \label{sub:multiple}
%%%
%%%

% We apply our findings to molecular clouds in
% the Milky Way and argue that the control of star formation by
% supersonic turbulence produces the full range of observed star forming
% regions (in \S~\ref{subsub:apply}), ranging from inefficient more or
% less isolated mode of star formation (e.g.\ in Taurus-Aurigae) to the
% build-up of dense rich stellar clusters (e.g.\ the Trapezium cluster
% in Orion).
%\rsksubsubsection{Clustered versus isolated star formation}

%end of M-MML edit 11.9.02

%\rsksubsection{Molecular clouds as sites of star formation}
%\label{subsec:MC-SF}

All giant molecular clouds surveyed within distances less than
3$\,$kpc form stars (Blitz 1993, Williams \etal\ 2000), except
possibly the Maddalena \& Thaddeus (1985) cloud (Lee, Snell, \&
Dickman 1996; Williams \& Blitz 1998).  However this cloud may have
formed just recently.
%Possible formation mechanisms for
%molecular cloud complexes involve spiral arm shocks, shear
%instabilities in the differentially rotating Galactic disk, or
%Rayleigh-Taylor or thermal instabilities on large scales. Furthermore,
%they could form in the converging flows of supershells driven by
%multiple supernovae {\bf\{QUOTE\}}.
%Mooney \& Solomon (1988) argue that at least 25\% of the molecular clouds in the
%inner part of Galaxy are not forming massive stars.
%However, clouds without O or B star formation may still be forming
%stars of lower mass in great abundance, like the cloud in
%Taurus-Auriga. Taking the distance dependent sensitivity limits into
%account, it appears fair to say that all clouds will form stars at
%least at some stage of their evolution (however small their overall
%efficiency may be).
The star formation process in molecular clouds appears to be fast.
Once the collapse of a cloud region sets in, it rapidly forms an
entire cluster of stars within $10^6$ years or less. This is indicated
by the young stars associated with star forming regions,
typically T~Tauri stars with ages less than $10^6$ years (e.g.\ Gomez
\etal\ 1992, Green \& Meyer 1996, Carpenter \etal\ 1997, Hartmann
2001), and by the small age spread in more evolved stellar clusters
(e.g.\ Hillenbrand, Palla \& Stahler 1999, 2001). 

Star-forming molecular cloud regions in our Galaxy can vary enormously
in size and mass. In small, low-density, regions stars form with low
efficiency, more or less in isolation or scattered around in small
groups of up to a few dozen members. Denser and more massive regions
may build up stars in association and clusters of a few hundred
members.  This appears to be the most common mode of star formation in
the solar neighborhood (Adams \& Myers 2001). Examples of star
formation in small groups and associations are found in the
Taurus-Aurigae molecular cloud (e.g.\ Hartmann 2002). Young stellar
groups with a few hundred members form in the Chamaeleon I dark cloud
(e.g.\ Persi \etal\ 2000) or $\rho$-Ophiuchi (Bontemps \etal\
2001). Each of these clouds is at a distance of about 130 to 160$\,$pc
from the Sun.  Many nearby star forming regions have been associated
with a ring-like structure in the Galactic disk called Gould's belt
(P{\"o}ppel 1997), although its reality remains somewhat uncertain.

The formation of dense rich clusters with thousands of stars is rare.
The closest molecular cloud region where this happens 
%lies in the constellation of Orion: it 
is the Orion Nebula Cluster in L1641 (Hillenbrand 1997; Hillenbrand \&
Hartmann 1998), which lies at a distance of $\sim 450\,$pc.  A rich
cluster somewhat further away is associated with the Monoceros R2
cloud (Carpenter \etal\ 1997) at a distance of $\sim 830\,$pc.  The
cluster NGC~3603 is roughly ten times more massive than the Orion
Nebula Cluster.  It lies in the Carina region, at about $7\,$kpc
distance. It contains about a dozen O stars, and is the nearest object
analogous to a starburst knot (Brandl \etal\ 1999, Moffat \etal\
2002). To find star-forming regions that capable of forming hundreds
of O stars one has to look towards giant extragalactic H$_{\rm
II}$-regions, the nearest of which is 30 Doradus in the Large
Magellanic Cloud, a satellite galaxy of our Milky Way at a distance at
55$\,$kpc (for an overview see the book by Chu \etal\ 1999). The giant
star forming region 30 Doradus is thought to contain up to a hundred
thousand young stars, including more than 400 O stars (Hunter \etal\
1995; Walborn \etal\ 1999). Even more massive star forming regions are
associated with tidal knots in interacting galaxies, as observed in
the Antennae (NGC~4038/8, see e.g.\ Zhang, Fall, \& Whitmore 2001) or
as inferred for starburst galaxies at high redshift (Sanders \&
Mirabel 1996).

This sequence demonstrates that the star formation process spans many
orders of magnitude in scale, ranging from isolated single
stars ($M\approx 1$~M$_{\odot}$) to ultra-luminous starburst galaxies
with masses of several $10^{11}$M$_{\odot}$ and star formation rates of
$10^2$--$10^3$~M$_{\odot}$~yr$^{-1}$; for comparison the
present-day rate in the Milky Way is about 1~M$_{\odot}$~yr$^{-1}$.
%We argue in this review article that t
This enormous variety of star
forming regions appears to be controlled by 
%supersonic turbulence, more precisely, by 
the competition between self-gravity and the turbulent
velocity field in interstellar gas. When turbulence dominates, or at least
carries sufficient energy on small scales to prevent collapse, the
star formation process is inefficient and slow, and stars build up in
small groups only. If turbulent gas motions are weak, or dominated by
large-scale modes, stars can form in clusters with locally high
efficiency, since gravity can overwhelm turbulence locally.  The larger
the volume where gravity dominates over turbulent kinetic energy, the
larger and more massive the stellar cluster that will form. In
starburst galaxies, self-gravity may overwhelm kinetic energy on
scales of several kpc.

%start of old multiple...
%Different star formation regions present different distributions of protostars
%and pre-main sequence stars.  In some regions, such as the Taurus molecular
%cloud, stars form isolated from other stars, scattered throughout the cloud
%(Mizuno \etal\~1995).  In other regions, they form in clusters, as in L1630 in
%Orion (Lada 1992), or even more extremely in starburst regions such as
%NGC~3603, the nearest galactic star-forming region resembling a starburst knot
%(e.g.\ Moffat \etal\ 2002), or 30 Doradus in the Large Magellanic Cloud
%(Walborn \etal\ 1999). 

Numerical simulations of self-gravitating, turbulent clouds demonstrate that
the {\em length scale} and {\em strength} of energy injection into the system
determine the structure of the turbulent flow, and therefore the locations at
which stars are most likely to form.  Large-scale driving leads to large
coherent shock structures 
(e.g.~figure~\ref{fig:3D-cubes-1..2+7..8}a). Local collapse occurs
predominantly in filaments and layers of shocked gas and is very efficient in
converting gas into stars (Klessen \etal\ 2000). 
%This leads to what we can identify as a `clustered'
%mode of star formation: stars form in coherent aggregates and clusters. 
Even more so, this applies to regions of molecular gas that have
become decoupled from energy input. As turbulence decays, these
regions begin to contract and form dense clusters or associations of
stars with very high efficiency on about a free-fall time scale
(Klessen \etal\~1998, Klessen \& Burkert 2000, 2001).  The same holds for
insufficient support, i.e.~for regions where energy input is not
strong enough to completely balance gravity. They too will contract to
form dense stellar clusters.

%The `isolated' mode of star formation occurs 
On the other hand, a more isolated mode of star formation occurs in
regions that are supported by driving sources that act
on {\em small} scales, and in an incoherent or stochastic manner. In
this case, individual, shock-induced, density fluctuations form at
random locations and evolve more or less independently of each
other. The resulting stellar population is widely dispersed throughout
the cloud and, since collapsing clumps are frequently destroyed by shock
interaction, the overall star formation rate is low.

\begin{figure*}[th]
\unitlength1.0cm
\begin{picture}(16,11.7)
%\put( 0.0,-6.0){\epsfxsize=14cm \epsfbox{ms50734-fig11.ps}}
\put( 0.0,-7.3){\epsfxsize=14.5cm \epsfbox{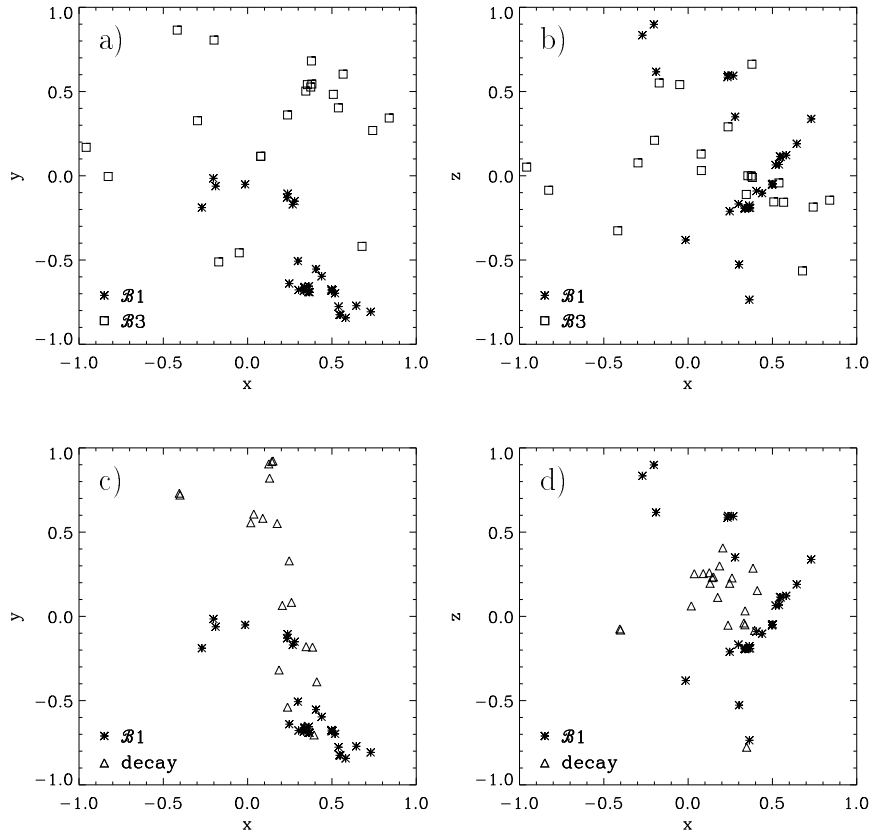}}
\end{picture}
\caption{\label{fig:2D-projection} Comparison of collapsed core
locations between two globally stable models with different driving
wavelength projected into (a) the $xy$-plane and into (b) the
$xz$-plane.  ${\cal B}1$ with $k=1-2$ is driven at large scales, and
${\cal B}3$ with $k=7-8$ is driven at small ones. Plots (c) and (d)
show the core locations for model ${\cal B}1$ now contrasted with a
simulation of decaying turbulence from Klessen~(2000).  The snapshots
are selected such that the mass accumulated in dense cores is $M_*
\sil 20$\%. Note the different times needed for the different models
to reach this point.  For model ${\cal B}1$ data are taken at $t=1.1$,
for ${\cal B}3$ at $t=12.3$. The simulation of freely decaying
turbulence is shown at $t=1.1$. All times are normalized to the global
free-fall time scale of the system. (From Klessen \etal\ 2000.) }
\end{figure*}
These points are illustrated in Figure~\ref{fig:2D-projection},. 
which shows the distribution of collapsed cores in several models with
strong enough turbulence to formally support against collapse.
%projected onto the $xy$- and $xz$-planes.  
Coherent, efficient local collapse occurs in model ${\cal B}1$, where
the turbulence is driven strongly at long wave lengths
%The flow is dominated by large
%coherent shocks, so cores form in aggregates associated with the
%filamentary structure of shock compressed gas 
(compare with Figure~\ref{fig:3D-cubes-1..2+7..8}).
%The overall efficiency of
%converting gas into stars in this `clustered' mode is very high.  The
%upper half of Figure~\ref{fig:2D-projection} compares the model ${\cal
% B}1$ with m
Incoherent, inefficient collapse occurs in model ${\cal B}3$, on the
other hand, where turbulence is driven at small scales.
% and results in incoherent collapse behavior. 
Individual cores form independently of
each other at random locations and random times,
%In this `isolated' mode, cores 
are widely distributed throughout the entire volume, and
exhibit considerable age spread.
%
%In the lower half of Figure~\ref{fig:2D-projection} we contrast the
%large-scale driving model ${\cal B}1$ with a simulation of freely
%decaying turbulence described by Klessen (2000) that has the same
%thermal Jeans mass.  
In the decaying turbulence model, once the kinetic energy level has
decreased sufficiently, all spatial modes of the system contract
gravitationally, including the global ones (Klessen 2000). As in the
case of large-scale shock compression, stars form more or less
coevally in a limited volume with high efficiency. 
%Both insufficient turbulent support and the complete loss of it therefore
%appear to lead to clustered star formation.  The Trapezium cluster in
%Orion may be a good example for the outcome of this mechanism (e.g.\
%Hillenbrand 1997, Hillenbrand \& Hartmann 1998).  All the projections
%shown in figure~\ref{fig:2D-projection} are taken at a stage of the
%dynamical evolution when the mass accumulated in dense cores is $M_*
%\approx 20$\%.  This occurs at very different times, as noted in the
%captions, which directly reflects the varying efficiencies of local
%collapse in these models.

\begin{figure*}[ht]
\unitlength1.0cm
\begin{picture}(16,6.0)
%\put( 0.0,-7.0){\epsfxsize=14cm \epsfbox{ms50734-fig12.ps}}
\put( 0.0,-7.6){\epsfxsize=14.5cm \epsfbox{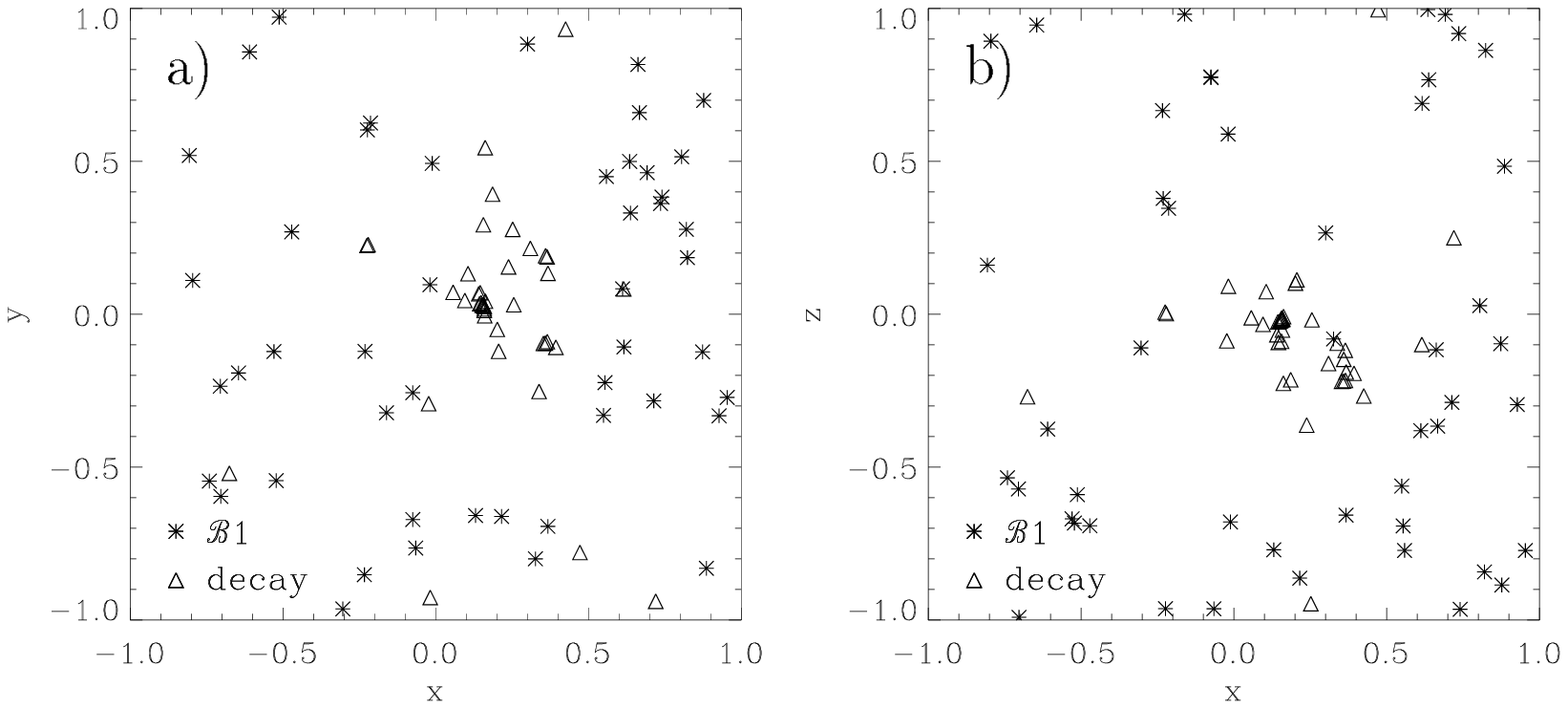}}
\end{picture}
\caption{\label{fig:2D-projection-later-times} Core positions for
  model ${\cal B}1$ ($k=1-2$) and the decay model when the core mass
  fraction is $M_* \approx 65$\%, projected into (a) the $xy$-plane
  and (b) the $xz$-plane (compare with Figure~\ref{fig:2D-projection}c
  \& d).  For ${\cal B}1$ the time is $t=8.7$ and for decay model
  $t=2.1$. Whereas the cluster in ${\cal B}1$ is completely dissolved
  and the stars are widely dispersed throughout the computational
  volume, the cluster in the decay simulation remains bound. (From
  Klessen \etal\ 2000.)}
\end{figure*}

Despite the fact that both turbulence driven on large scales and freely
decaying turbulence lead to star formation in aggregates and clusters,
Figure~\ref{fig:2D-projection-later-times} suggests a possible way to
distinguish between them.
% by taking the long-term evolution of the resulting
%clusters into account. 
Decaying turbulence typically leads to the formation of a bound
stellar cluster, while aggregates associated with large-scale,
coherent, shock fronts often have higher velocity dispersions that
result in their complete dispersal. 
%This is illustrated in
%figure~\ref{fig:2D-projection-later-times}, which compares the core
%distribution in model ${\cal B}1$ and in the decay simulation at $M_*
%\approx 65$\%, when both systems have already undergone considerable
%evolution.  The cores in model ${\cal B}1$ are completely dispersed
%throughout the molecular cloud volume. The cluster that formed during
%the turbulent decay remains bound with a much longer evaporation time
%scale. 
Note, however, that at the late stages of dynamical evolution shown in
Figure~\ref{fig:2D-projection-later-times}, our model becomes less
appropriate, as feedback from newly formed stars is not
included. Ionization and outflows from the stars formed first will
likely retard or even prevent the accretion of the remaining gas onto
protostars, possibly preventing a bound cluster from forming even in
the case of freely decaying turbulence.

The control of star formation by supersonic turbulence gives rise to a
continuous but articulated picture. There may not be physically distinct
modes of star formation, but qualitatively different behaviors do appear
over the range of possible turbulent flows. The apparent
dichotomy between a clustered mode of star formation and an
isolated, as discussed by Lada (1992) for L1630 and  Strom, Strom, \&
Merrill (1993) for L1941, disappears, if a different balance between
turbulent strength and gravity holds at the relevant length scales in
these different clouds.

Turbulent flows tend to have hierarchical structure (e.g.\ She \&
Leveque 1994) which may explain the hierarchical distribution of stars
in star forming regions shown by statistical studies of the
distribution of neighboring stars in young stellar clusters (e.g.\
Larson 1995; Simon 1997; Bate, Clarke, \& McCaughrean 1998, Nakajima
\etal\ 1998; Gladwin \etal\ 1999; Klessen \& Kroupa 2001).
Hierarchical clustering seems to be a common feature of all star
forming regions (e.g.\ Efremov \& Elmegreen 1998). It may be a natural
outcome of turbulent fragmentation.

%{\bf I'm not sure we want to claim the following:}{\em create a
%continuous hierarchy of structure throughout the scales.  The
%clustering of the stars follows a pattern analogous to the
%hierarchical clumping structure in interstellar gas.  The formation of
%rich clusters is associated with a significant peak in a statistical
%fluctuation spectrum.  and the effect of finding OB stars
%preferentially in regions of enhanced stellar density (e.g.\ Testi,
%Palla, \& Natta 1999) is explained as a mere statistical effect.
%Assuming a universal initial stellar mass function (\S~\ref{sub:imf})
%the probability of finding a high-mass star increases linearly with
%the sample size, i.e.~with the total number of stars in the observed
%region.}

\rsksubsection{Properties of Protostellar Cores}% RSK[summer 2002]
\label{sub:cores}
%\input{cores.tex}

%%%
%%% RMP-Module for the "Local star-formation regions"
%%%
%%% checked out by RSK: 25.07.02
%%% checked in  by RSK: 16.08.02
%%%
%%%\rsksubsection{Protostellar core properties}
%%%\label{sub:cores}
%%%

\begin{figure*}[th]
\unitlength1.0cm
%\begin{turn}{180}
\begin{picture}(16.0,11.1)
%\put(17.0,10.5){\begin{rotate}{180}\epsfysize=10.5cm\epsfbox{mk03_fig20.ps}\end{rotate}}
\put(17.0,10.5){\begin{rotate}{180}\epsfysize=10.5cm \epsfbox{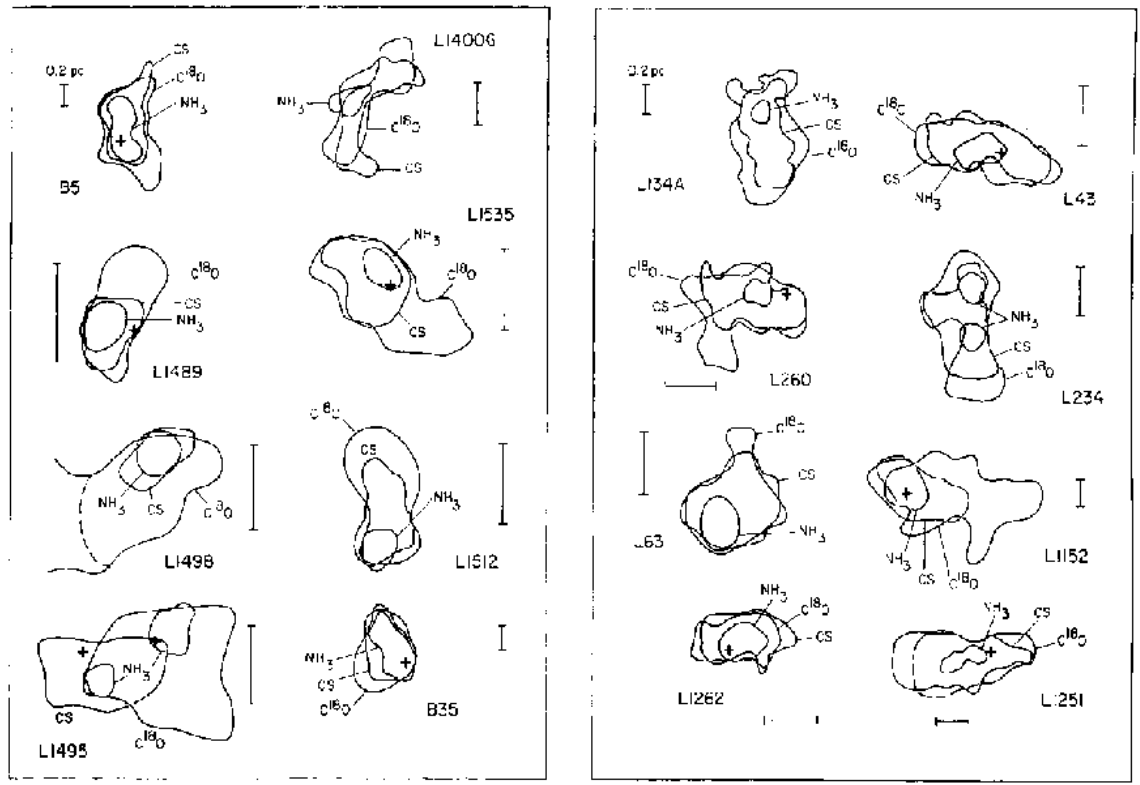}\end{rotate}}
\end{picture}
%\end{turn}
\caption{\label{fig:core-shapes}Intensity contours at half maximum of
    16 dense cores in dark clouds, in the $1.3\,$cm $(J,k) = (1,1)$
    lines of NH$_3$, in the $3.0\,$mm $J=2\rightarrow 1$ line of CS,
    and the $2.7\,$mm $J=1\rightarrow 0$ line of C$^{18}$O. A linear
    scale of $0.2\,$pc is indicated in each individual map and
    associated protostars are specified by a cross. The figure is from
    Myers \etal\ 1991.}
\end{figure*}

Protostellar cores are the direct precursors of stars. The properties of young
stars are thus intimately related to the properties of their parental clumps
and it is therefore important to observationally determine the characteristics
of condensed cores in molecular clouds.  A number of such small, dense
molecular cores have been identified by low angular resolution, molecular line
surveys of nearby dark clouds (e.g.\ Benson \& Myers 1989). These cores are
thought to be the sites of low-mass star formation. About half of them are
associated with low-luminosity IRAS sources and CO outflows, the other half is
designated as `starless' or 'prestellar' (e.g.\ Beichman \etal\ 1986,
Andr{\'e} \etal\ 2000). Those may be in a evolutionary state shortly before forming
stellar objects in their interior, thus they often are referred to as
pre-stellar cores.  One of the most notable properties of the sampled cores
are their very narrow line widths. These are very close to the line widths
expected for thermal broadening alone and, as a result, many of the cores
appear approximately gravitationally virialized (e.g.\ Myers 1983).  They are
thought either to be in the very early stage of gravitational collapse or have
subsonic turbulence supporting the clump. A comparison of the line widths of
cores with embedded protostellar objects (i.e.~with associated IRAS sources)
and the `starless' cores reveals a substantial difference. Typically, cores
with infrared sources exhibit broader lines, which suggests the presence of a
considerable turbulent component not present in `starless' cores. This may be
caused by the central protostar feeding back energy and momentum into its
surrounding envelope. Molecular outflows associated with many of the sources
may be direct indication for this process.  The first submillimeter continuum
maps of dense pre-stellar cores were made by Ward-Thompson \etal\ (1994). By
now many different star forming clouds have been surveyed (e.g.\ 
$\rho$-Ophiuchi: Motte \etal\ 1998, Johnstone \etal\ 2000; NGC
2068/2071: Motte \etal\ 2001; Orion: Johnstone \& Bally 1999, Mitchell \etal\ 
2001).  The authors of this high-resolution studies found that `starless'
cores are larger and less centrally condensed than the cores with embedded
sources. But both groups appear to have similar masses.  Altogether, the
`starless' cores are probable the precursors of class 0 protostellar clumps
and may reflect the very early stages of protostellar collapse: a
gravitationally bound fragment has formed in a molecular cloud which evolves
towards progressively higher degree of central condensation, but has not yet
formed a hydrostatic protostar in the center (i.e.\ a class 0
object). The `starless' cores are observed in the mass range 
from about $0.1\,$M$_{\odot}$ to $10\,$M$_{\odot}$.  Typical line widths are
$0.4\,$km$\!\;$s$^{-1}$ in NH$_3$ and $0.6\,$km$\!\;$s$^{-1}$ in C$^{18}$O for
cores with embedded sources and $0.3\,$km$\!\;$s$^{-1}$ in NH$_3$ and
$0.5\,$km$\!\;$s$^{-1}$ in C$^{18}$O for pre-stellar cores with embedded
sources (e.g.\ Butner, Lada, \& Loren 1995). For comparison, a gas temperature
of 10$\,$K corresponds to a thermal line width of $0.16\,$km$\!\;$s$^{-1}$ for
NH$_3$ and $0.12\,$km$\!\;$s$^{-1}$ for C$^{18}$O. High-resolution maps
suggest that the radial density profiles of the pre-stellar cores on average
follow a $1/r^2$-law but are relatively steep towards their edges and flatten
out near their centers (see e.g.\ Figure \ref{fig:L1689B}). Furthermore, their
2-dimensional shapes deviate quite considerably from spherical symmetry, as
illustrated in Figure \ref{fig:core-shapes}.  The cores are elongated with
ratios between semi-major and semi-minor axis of about 2 -- 3; some even
appear completely irregular.

\begin{figure*}[ht]
\unitlength1.0cm
\begin{picture}(16,10.0)
%\put(0.0,-2.0){\epsfxsize=9.5cm \epsfbox{ms40383-fig12A.ps}}
%\put(8.0,-2.0){\epsfxsize=9.5cm \epsfbox{ms40383-fig12B.ps}}
\put(-0.3,-2.0){\epsfxsize=9.8cm \epsfbox{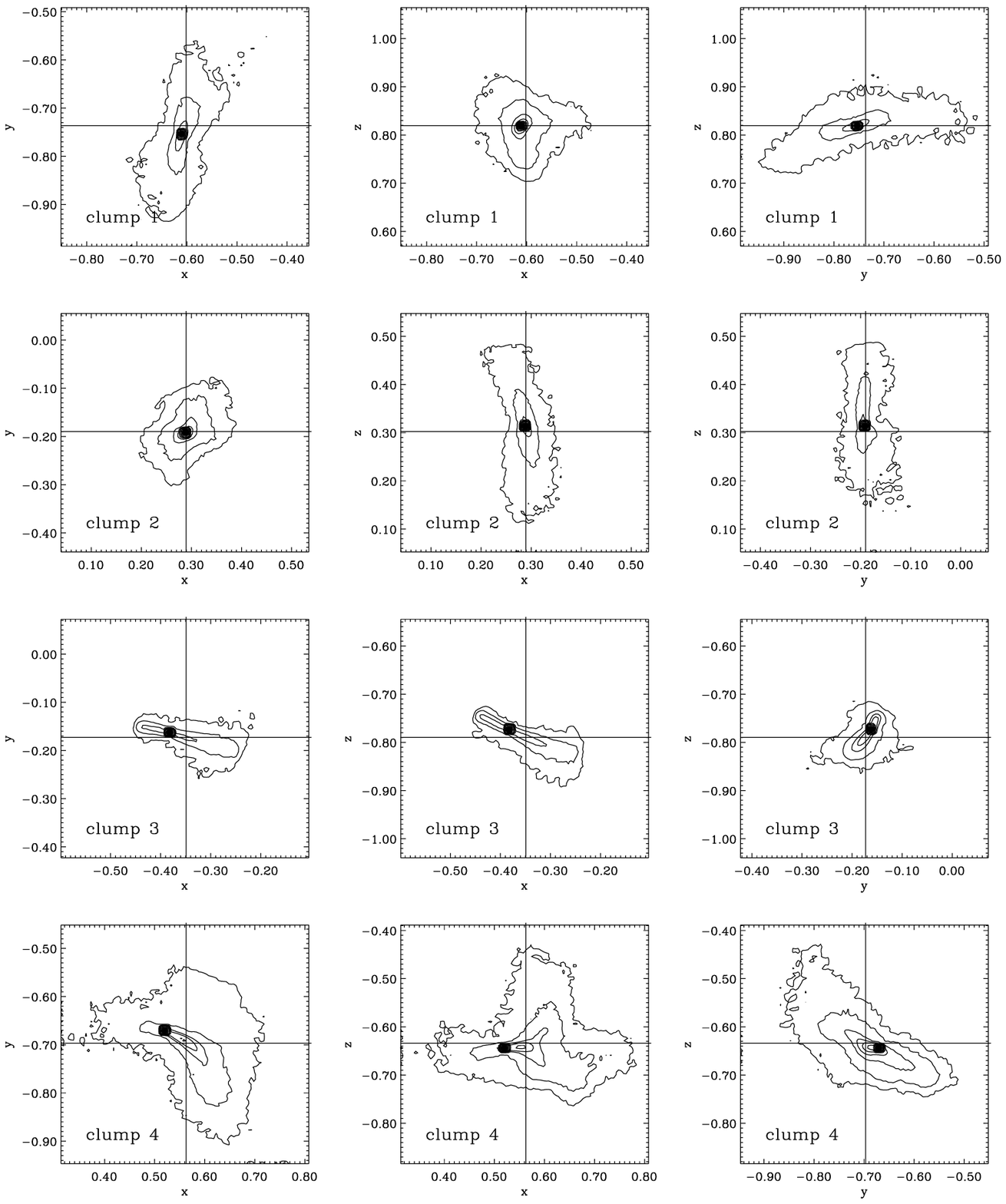}}
\put( 8.0,-2.0){\epsfxsize=9.8cm \epsfbox{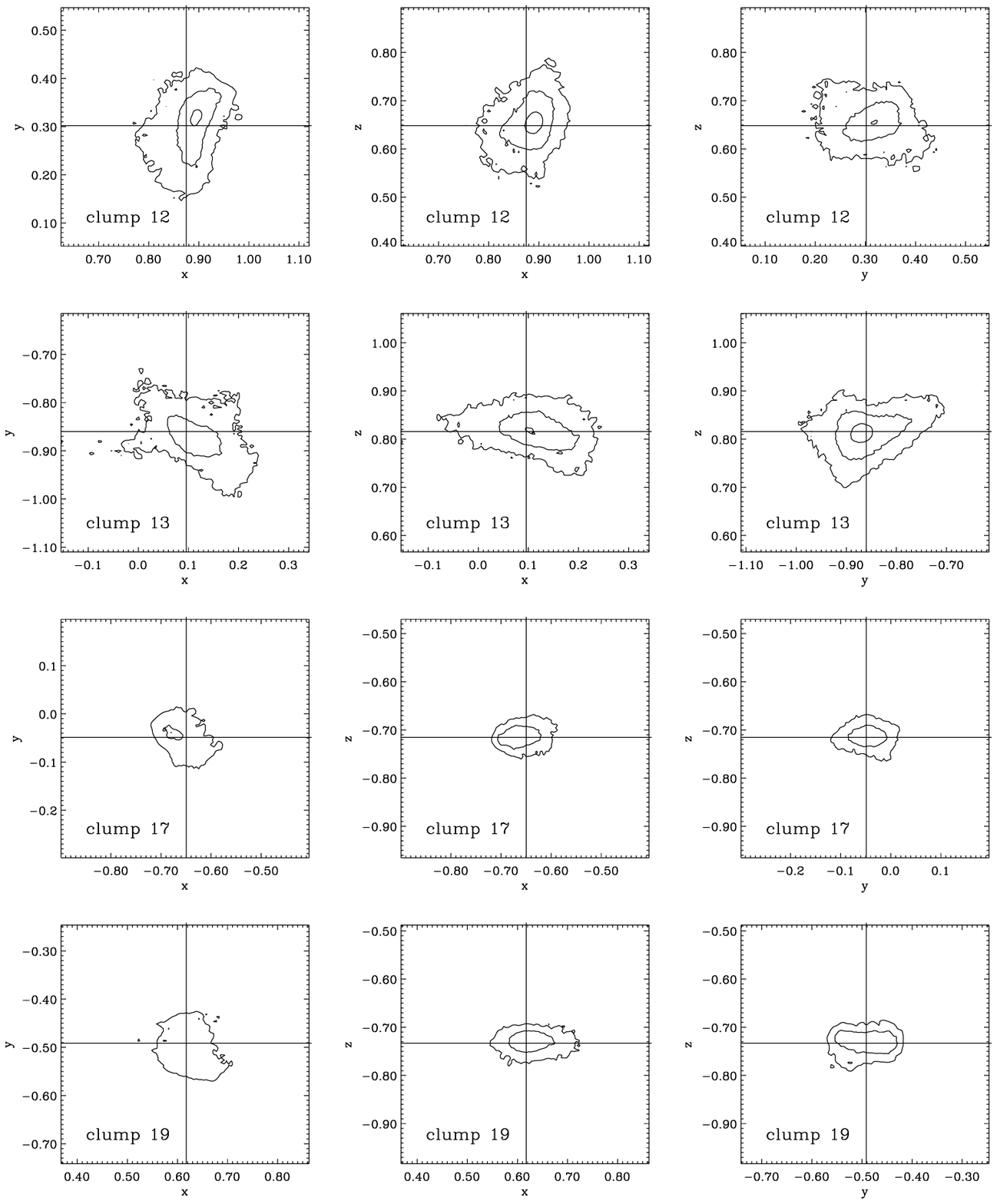}}
\end{picture}
\caption{\label{fig:ind-clumps}Protostellar cores from a model of
  clustered star formation. The left side depicts protostellar cores
  with collapsed central object (indicated by a black dot), the right
  side ``starless'' cores without protostar.  Cores are numbered
  according to their peak density. Surface density contours are spaced
  logarithmically with two contour levels spanning one decade,
  $\log_{10}\Delta \rho = 0.5$. The lowest contour is a factor of
  $10^{0.5}$ above the mean density. (From Klessen \& Burkert 2000.)
  }
\end{figure*}

The observed core properties can be compared with cores identified in
numerical models of interstellar cloud turbulence. Like their
astronomical counterparts, model clumps are generally highly distorted
and triaxial.  Depending on the projection angle, they often appear
extremely elongated, being part of a filamentary structure which may
appear as a chain of connecting, elongated individual clumps. Figure
\ref{fig:ind-clumps} plots a selection of model clumps from Klessen \&
Burkert (2000). Cores that already formed a protostellar object in
their interior are plotted on the left, on the right ``starless'' core
without central protostar are shown. Note the similarity to the
appearance of observed protostellar cores (Figure
\ref{fig:core-shapes}). It is clearly visible that the clumps are very
elongated. The ratios between the semi-major and the semi-minor axis
measured at the second contour level are typically between 2:1 and
4:1. However, there may be significant deviations from simple triaxial
shapes, see e.g.~clump \#4 which is located at the intersection of two
filaments. This clump is distorted by infalling material along the
filaments and appears `y'-shaped when projected into the $xz$-plane.
As a general trend, high density contour levels typically are regular
and smooth, because there the gas is mostly influenced by pressure and
gravitational forces.  On the other hand, the lowest level samples gas
that is strongly influenced by environmental effects. Hence, it
appears patchy and irregular. The location of the condensed core
within a clump is not necessarily identical with the center-of-mass of
the clump, especially in irregularly shaped clumps.

\begin{figure}[ht]
\unitlength01.0cm
\begin{picture}(16,5.8)
%\put( 0.000,0.000){\epsfxsize=8.0cm \epsfbox{ms40383-fig13.ps}}
\put(-0.700,-0.300){\epsfxsize=9.0cm \epsfbox{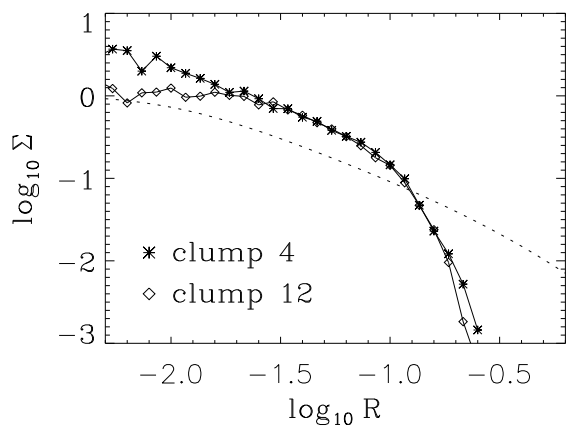}}
\end{picture}
\caption[F13]{\label{fig:radial-density-profile}Radial surface density
  profiles for the $xz$-projection of cores 4 and 12 in Figure
  \ref{fig:ind-clumps}. For the ``starless'' core 12 the density
  profile flattens out at small radii, whereas for core 4 it continues
  as $1/r^2$ all the way towards the center. The
  density profile of a singular isothermal sphere (with $\rho \propto
  1/r^2$) is indicated by the dotted line for comparison. (From Klessen \& Burkert
  2000.)  }
\end{figure}

Typically, the overall density distribution of identified clumps in our
simulations follows a power law and the density increases from the outer
regions inwards to the central part as $\rho(r) \propto 1/r^2$.  For clumps
that contain collapsed cores, this distribution continues all the way to the
central protostellar object. However, for clumps that have not yet formed a
collapsed core in their center, the central density distribution flattens out.
This is a generic property and is illustrated in Figure
\ref{fig:radial-density-profile}, which has to be compared with radial
profiles of observed cores (see Figure \ref{fig:L1689B}). The agreement is
remarkable. It is a natural prediction of turbulent fragmentation that stars
form from cores with initially flat inner density profile, follow by a
power-law decline with moderate slope ($\sim -2$) at intermediate radii, and
which is finally truncated at some maximum radius (i.e.\ may be approximated
by a power-law slope $\ll -2$ beyond that radius).

The surface density profiles of many observed protostellar cores are often
claimed to be matched by fitting quasi-equilibrium Bonnor-Ebert spheres (Ebert
1955, Bonnor 1956). The best example is the Bok globule B68 (Alves \etal\ 
2001).  Ballesteros-Paredes, Klessen, \& V\'azquez-Semadeni (2002) argue that
turbulent fragmentation produces cores that are in $\sim$3/5 of all cases well
fit by Bonnor-Ebert profiles, of which most ($\sim$4/5) again imply stable
equilibrium conditions.  However, none of the cores analyzed by
Ballesteros-Paredes \etal\ (2002) are actually equilibrium configurations, but
instead are dynamically evolving, because supersonic turbulence cannot create
hydrostatic equilibrium structures (V{\'a}zquez-Semadeni \etal\ 2002).  The
method of fitting BE profiles to observed cores to derive their physical
properties therefore appears unreliable (see also Boss \& Hartmann 2002).

The density profile predicted by supersonic turbulence describes the
properties of observed prestellar cores very well. However, it could in
principle also be reproduced by models where molecular gas clumps are confined
by helical magnetic fields (Fiege \& Pudritz 2000a,b). But helical field
structures would tend to unwind, as magnetic fields have the tendency to
``straighten'' themselves out.  Therefore these models require external forces
to continuously exert strong torques.  These forces need to be strong to be
able to twist the field lines, thus they would necessarily induce considerable
gas motions and make it impossible to achieve the static equilibrium
configurations required for the model to work.  The hypothesis of static
helical magnetic fields being responsible for the observed properties of
prestellar cores therefore appears not viable. Other models that have been
proposed to describe the properties of protostellar cores are based
quasi-static equilibrium conditions (e.g.\ with composite polytropic equation
of state, as discussed by Curry \& McKee 2000), or invoke thermal instability
(e.g.\ Yoshii \& Sabano 1980, Gilden 1984b, Graziani \& Black 1987, Burkert \&
Lin 2000), gravitational instability through ambipolar diffusion (e.g.\ Basu
\& Mouschovias 1994, Nakamura, Hanawa, \& Nakano 1995, Indebetouw \& Zweibel
2000, Ciolek \& Basu 2000), or nonlinear Alfv{\'e}n waves (e.g.\ Carlberg \&
Pudritz 1990, Elmegreen 1990, 1997a, 1999b), or rely on clump collisions (e.g.,
Gilden 1984a, Murray \& Lin 1996, Kimura \& Tosa 1996). Altogether models
based on supersonic turbulence as discussed here appear to be the ones most
consistent with observational data.

Besides the direct comparison of projected surface density maps as discussed
in this section, there is ample additional evidence supporting the idea of the
turbulent origin of the structure and kinematics of molecular cloud cores and
clouds as a whole. It comes for example from comparing numerical models of
supersonic turbulence with (1) stellar extinction measurements (Padoan \etal\ 
1997), (2) Zeeman splitting measurements (e.g.\ Padoan \& Nordlund 1999), (3)
polarization maps (e.g.\ Padoan \etal 2001a), (4) Faraday rotation
measurements (e.g.\ Heitsch \etal\ 2001b, Ostriker, Stone, \& Gammie 2001),
(5) determination of the velocity structure of dense cores and their immediate
environment (e.g.\ Padoan \etal\ 2001b), or (6) various other statistical
measures of structure and dynamics of observed clouds as mentioned
in \S~\ref{sub:regions}).

% {\bf\{Just as a reminder of magnetic breaking:
% %
% \begin{itemize}
% \item angular momentum distribution (do it in the second round -- if
%   no time left now). This is what we wanted to include:
%   - binary stars?\\
%   - clusters?\\
%   - turbulent transport?\\
%   - accretion disk transport?\\
%   - magnetic braking?\\
%   - Needs to be investigated further. is this a solved problem?  [see
%   above discussion of standard theory]\\
% \end{itemize}
% %
% \}}

\rsksubsection{Dynamical Interactions in Clusters}% RSK[summer 2002]
\label{sub:clusters}
%\input{clusters.tex}

%%%
%%% RMP-Module for the "Local star-formation regions"
%%%
%%% checked out by RSK: 16.08.02
%%% checked in  by RSK: 
%%%
%%% \rsksubsection{Dynamical interactions in clusters}
%%% \label{sub:clusters}
%%%
%%%

Star forming regions can differ enormously in scale and density as a
consequence of supersonic turbulence (as discussed in
\S~\ref{sub:multiple}). Stars almost never form in isolation, but
instead in groups and clusters.  The number density of protostars and
protostellar cores in rich compact clusters the can be high enough for
mutual dynamical interaction to become important.  This has important
consequences for the mass growth history of individual stars and the
subsequent dynamical evolution of the nascent stellar cluster, because
this introduces a further degree of stochasticity to the star
formation process in addition to the statistical chaos associated with
turbulence and turbulent fragmentation in the first place.

\begin{figure*}[htp]
\begin{center}
\unitlength1cm
\begin{picture}(16.6,5.3)
%\put( 0.00,0.00){\epsfxsize=5.5cm \epsfbox{sink-trajectory-01.eps}}
%\put( 5.20,0.00){\epsfxsize=5.5cm \epsfbox{sink-trajectory-03.eps}}
%\put(10.50,0.00){\epsfxsize=5.5cm \epsfbox{sink-trajectory-02.eps}}
\put( 0.00,-0.30){\epsfxsize=5.7cm \epsfbox{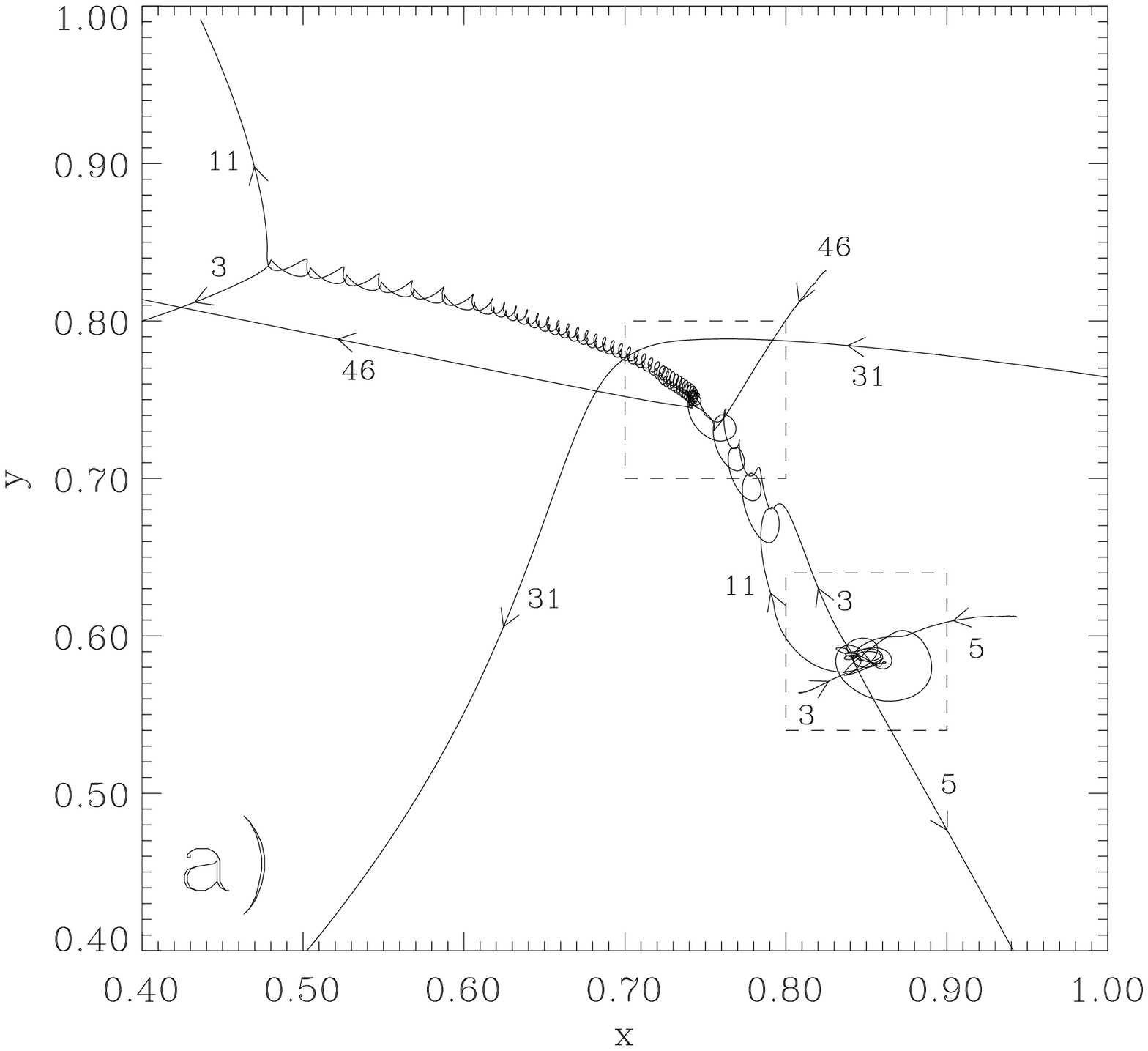}}
\put( 5.40,-0.30){\epsfxsize=5.7cm \epsfbox{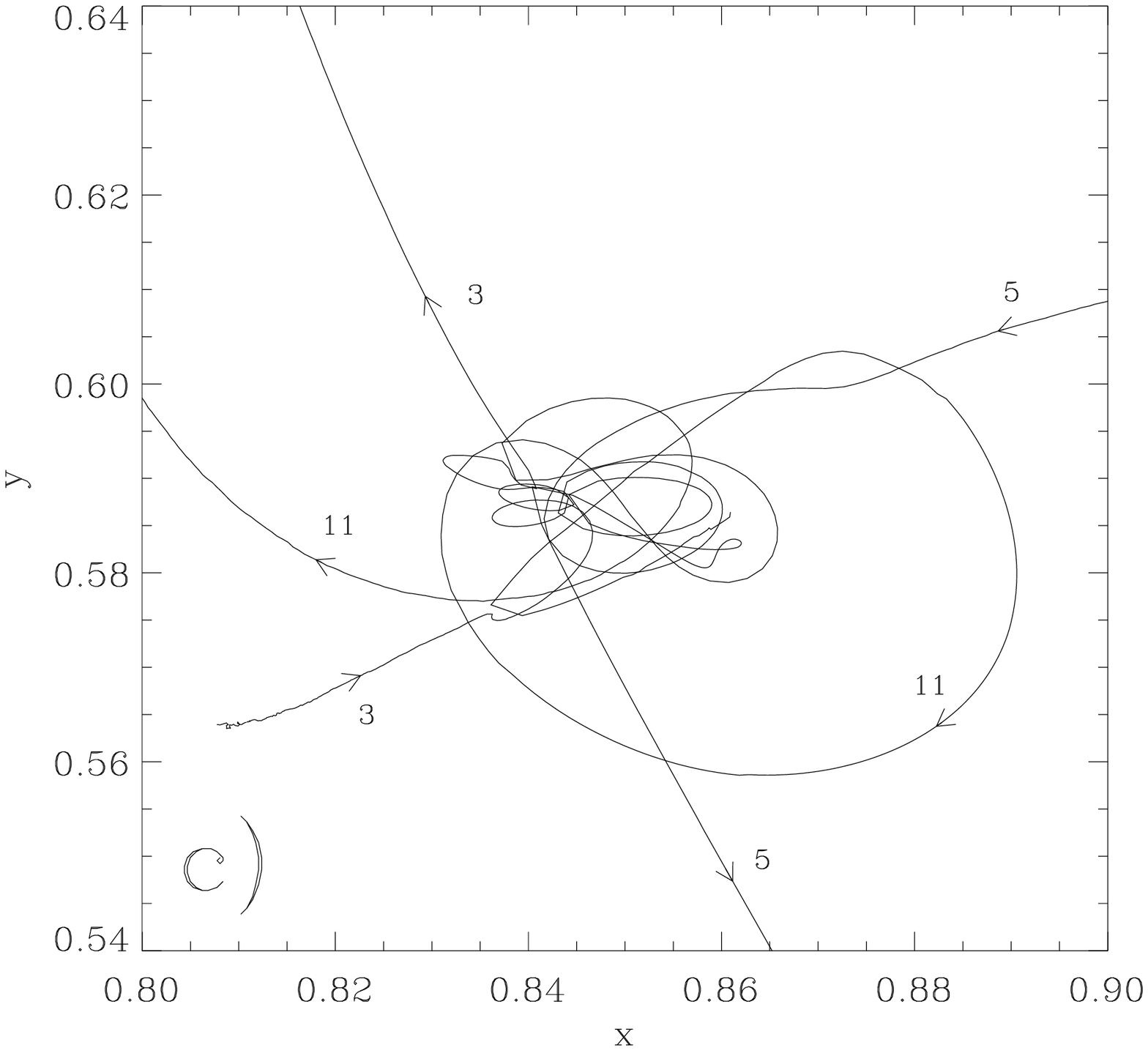}}
\put(10.80,-0.30){\epsfxsize=5.7cm \epsfbox{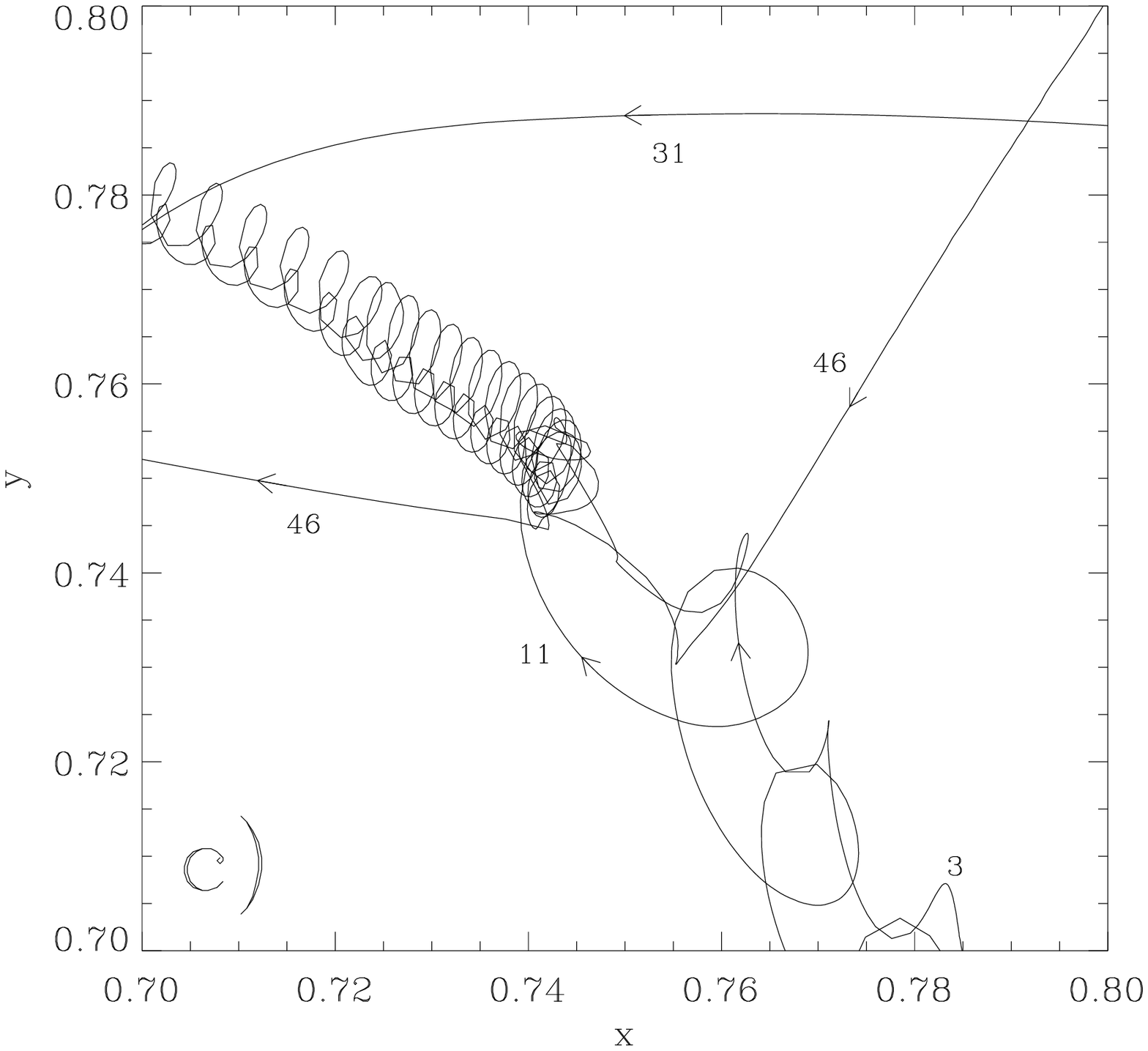}}
\put( 1.05, 0.45){\epsfxsize=0.7cm \epsfbox{white.eps}}
\put( 6.45, 0.45){\epsfxsize=0.7cm \epsfbox{white.eps}}
\put(11.85, 0.45){\epsfxsize=0.7cm \epsfbox{white.eps}}
\put( 1.15, 0.55){\large a)}
\put( 6.55, 0.55){\large b)}
\put(11.95, 0.55){\large c)}
\end{picture}
\end{center}
\caption{\label{fig:trajectory} Example of protostellar
  interactions in an embedded nascent star cluster. Figure (a) shows
  the projected trajectories of five accreting cores in a numerical
  model of star cluster formation by Klessen \& Burkert (2000).  For
  illustration purpose neither the trajectories of other cores in the
  cluster nor the distribution of gas is depicted. We highlight two
  events in the evolutionary sequence, (b) the formation of an
  unstable triple system at the beginning of cluster formation with
  the lowest mass member being expelled from the cluster, and (c)
  binary hardening in a close encounter together with subsequent
  acceleration of the resulting close binary due to another distant
  encounter during late evolution. The corresponding parts of the
  orbital paths are enlarged by a factor of six. Numbers next to the
  trajectory identify the protostellar core. For further detail see
  Klessen \& Burkert (2000).  }
\end{figure*}

When a molecular cloud region of a few hundred solar masses or more
coherently become gravitationally unstable, it will contract and build up
a dense cluster of embedded protostars within one or two free-fall
timescales. While contracting individually to build up a protostar in
their interior, individual protostellar gas clumps still follow the
global flow patterns. They stream towards a common center of
attraction, may undergo further fragmentation or more likely merge
together. The timescales for clump mergers or clump collapse are
comparable. Merged clumps therefore may contain multiple protostars
now compete with each other for further accretion. They are now
embedded in the same limited and rapidly changing reservoir of
contracting gas. As the cores are dragged along with the global gas
flow, quickly a dense cluster of accreting protostellar cores builds
up.  Analogous to dense stellar clusters, the dynamical evolution is
subject to the complex gravitational interaction between the cluster
members, close encounters or even collisions may occur and drastically
alter the orbital parameters of protostars. Triple or higher-order
systems are likely to form. They are generally unstable and
consequently a considerable fraction of protostellar cores becomes
expelled from the parental cloud. The expected complexity of
protostellar dynamics already in the deeply embedded phase of
evolution is illustrated in Figure \ref{fig:trajectory}, which shows
trajectories of five accreting protostars in a calculation of
molecular cloud fragmentation and clustered star formation by Klessen
\& Burkert (2000).

The effects of mutual dynamical interaction of protostellar cores in
the embedded phase of star cluster formation have been investigated by
a variety of authors. Here, we list some basic results.

{\em (a)} Close encounters in nascent star clusters will influence the
accretion disk expected to surround every protostar. These disks may
be tidally truncated or even be disrupted. This influences mass
accretion through the disk, modify the ability to subfragment and form
a binary star, and/or the probability of planet formation (e.g.\ 
Clarke \& Pringle 1991; Murray \& Clarke 1993 ; McDonald \& Clarke
1995; Hall \etal\ 1996; Scally \& Clarke 2001; Kroupa \& Burkert 2001;
Smith \& Bonnell 2001; Bonnell \etal\ 2001c). In particular, Ida,
Larwood, \& Burkert (2000) note that an early stellar encounter may
explain features of our own solar system, namely the high
eccentricities and inclinations observed in the outer part of the
Edgeworth-Kuiper Belt at distances larger than $42\,$AU.

{\em (b)} Stellar systems with more than two members are in general
are unstable. In a triple system, for example, the lowest-mass member
has the highest probability to be expelled. If this happens in the
embedded phase, the protostar leaves a region of high-density gas.
This terminates further mass growth and sets its final mass. Thus, the
dynamical processes have important consequences for the resulting
stellar mass spectrum in dense stellar clusters. This will be
discussed in \S~\ref{sub:imf}.  Ejected objects can travel quite far,
and indeed this has been suggested to account the so called ``run
away'' T-Tauri stars found in X-ray observation in the vicinities of
star-forming molecular clouds (e.g.\ Sterzik \& Durison 1995, 1998,
Smith \etal\ 1997; Klessen \& Burkert 2000; or for observations e.g.\ 
Neuh{\"a}user \etal\ 1995; or Wichmann \etal\ 1997). However, it is
not clear whether the observed extended stellar population is
associated with any currently star forming cloud. These stars may be
as old as 100$\,$Myr and may have formed in clouds that long have been
dispersed by now. Also, many of these stars could not have traveled to
their observed positions if associated with the currently star-forming
cloud unless it were extremely long lived.
 
{\em (c)} Dynamical interaction leads to mass segregation. Star
clusters evolve towards equipartition. For massive stars this means
that they have on average smaller velocities than low-mass stars (in
order keep the kinetic energy $K = 1/2\, m v^2 \approx
$constant). Thus, massive stars ``sink'' towards the cluster center,
while low-mass stars will predominantly populate large cluster radii.
(e.g.\ Kroupa 1995,a,b,c). This holds already for nascent star
clusters in the embedded phase (e.g.\ Bonnell \& Davis 1998). 

{\em (d)} Dynamical interaction and competition for mass accretion
lead to highly time-variable protostellar mass growth rates. This will
be discussed in more detail in \S~\ref{sub:accretion}.

{\em (e)} The radii of stars in the pre-main sequence contraction
phase are several times larger than stellar radii on the main sequence
(for a review on pre-main sequence evolution see, e.g., Palla 2000,
2002).  Stellar collisions are therefore more likely to occur during in
very early evolution of star clusters. During the embedded phase the
encounter probability is further increased by gas drag and dynamical
friction. Collisions in dense protostellar clusters have therefore
been proposed as mechanism to produce massive stars (Bonnell, Bate, \&
Zinnecker 1998; Stahler, Palla, \& Ho 2000). The formation of massive
stars has long been considered a puzzle in theoretical astrophysics,
because one-dimensional calculations predict for stars above $\sim
10\,$M$_{\odot}$ the radiation pressure acting on the infalling dust
grains to be strong enough to halt or even revert further mass
accretion (e.g.\ Yorke \& Kr{\"u}gel 1977; Wolfire \& Cassinelli 1987;
or Palla 2000, 2001). However, detailed two-dimensional calculations
by Yorke \& Sonnhalter (2002) demonstrate that in the more realistic
scenario of mass growth via an accretion disk the radiation barrier
may be overcome. Mass can accrete from the disk onto the star along
the equator while radiation is able to escape along the polar direction.
Massive stars, thus, may form via the same processes as ordinary
low-mass stars. Collisional processes need not to be invoked.

\rsksubsection{Accretion Rates}% RSK[summer 2002]
\label{sub:accretion}

When a gravitationally unstable gas clump collapses to build up the
central star, it follows an observationally well determined sequence.
Prior to the formation of a hydrostatic nucleus, an observed
pre-stellar condensation exhibits a density structure which has a flat
inner part, then decreases outward roughly as $\rho \propto r^{-2}$, and is
truncated at some finite radius (e.g.\ Bacmann \etal\ 2000). Once the
central YSO builds up, the class 0 phase is reached and the density
follows $\rho \propto r^{-2}$ down to the observational resolution
limit. As larger and larger portions of the infalling envelope get
accreted the protostar is identified as a class I object, and when
accretion fades away it enters the T Tauri phase (e.g.\ Andr{\'e},
Ward-Thompson, \& Barsony 2000).  In the main accretion phase, the
energy budget is dominated by the release of gravitational energy in
the accretion process. Hence, protostars exhibit large IR and sub-mm
luminosities and drive powerful outflows.  Both phenomena can be used
to estimate the protostellar mass accretion rate $\dot{M}$; and
observations suggest that $\dot{M}$ varies strongly and declines with
time. Accretion is largest in the class 0 phase and drops
significantly in the subsequent evolution (e.g.\ Andr{\'e} \&
Montmerle 1994, Bontemps \etal\ 1996, Henriksen, Andr{\'e}, \&
Bontemps 1997). The estimated lifetimes are a few $10^4\,$years for
the class 0 and a few $10^5\,$years for the class I phase.

These observational findings favor a dynamical description of the star
formation process (e.g.\ Larson 1969, Penston 1969a, Hunter 1977,
Henriksen \etal\ 1997, Basu 1997), but raise doubts about the
"inside-out" scenario of the collapse of quasi-static isothermal
spheres (Shu 1977) which predicts a constant accretion rate
(\S~\ref{sub:standard}; see \S~\ref{sub:standardprobs} for a critical
discussion).  As the analytical studies, most numerical work of
protostellar core collapse (e.g.\ Foster \& Chevalier 1993, Tomisaka
1996, Ogino, Tomisaka, \& Nakamura 1999, Wuchterl \& Tscharnuter 2002)
concentrates on isolated objects.  However, stars predominantly form
in groups and clusters. Numerical studies that investigate the effect
of the cluster environment on protostellar mass accretion rates are
reported for example by Bonnell \etal\ (1990, 2001a,b), Klessen \&
Burkert (2000, 2001), Klessen \etal\ (2001), Heitsch \etal\ (2001),
Klessen (2001a).

\begin{figure*}[tb]
\unitlength1cm
\begin{picture}(16.0,10.0)
%\put(  1.0,  1.4){\epsfxsize=7.2cm \epsfbox{Klessen-2001-1A.ps}}
%\put(  8.6,  1.4){\epsfxsize=7.2cm \epsfbox{Klessen-2001-1B.ps}}
\put(  1.0,  1.4){\epsfxsize=7.2cm \epsfbox{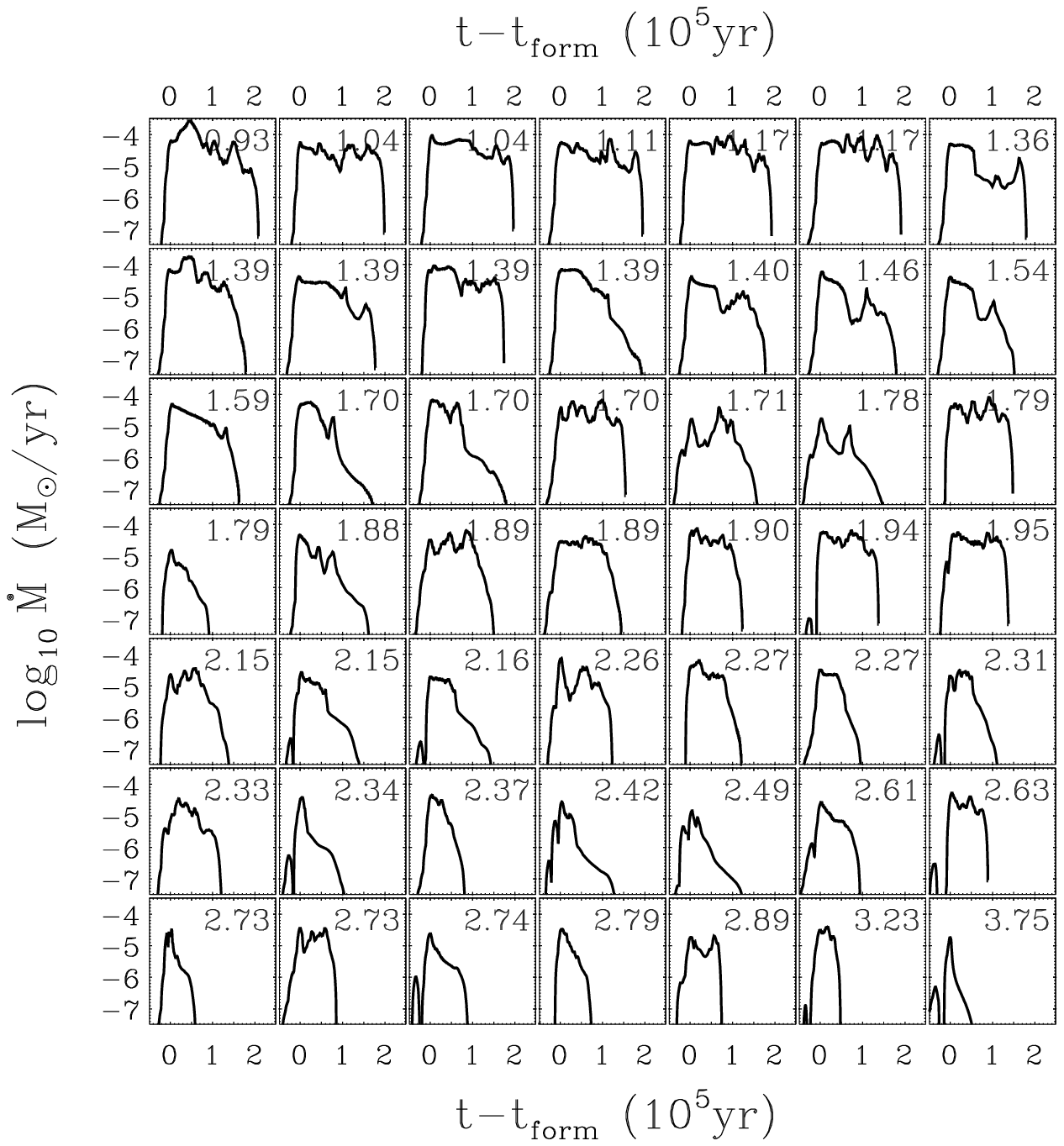}}
\put(  8.6,  1.4){\epsfxsize=7.2cm \epsfbox{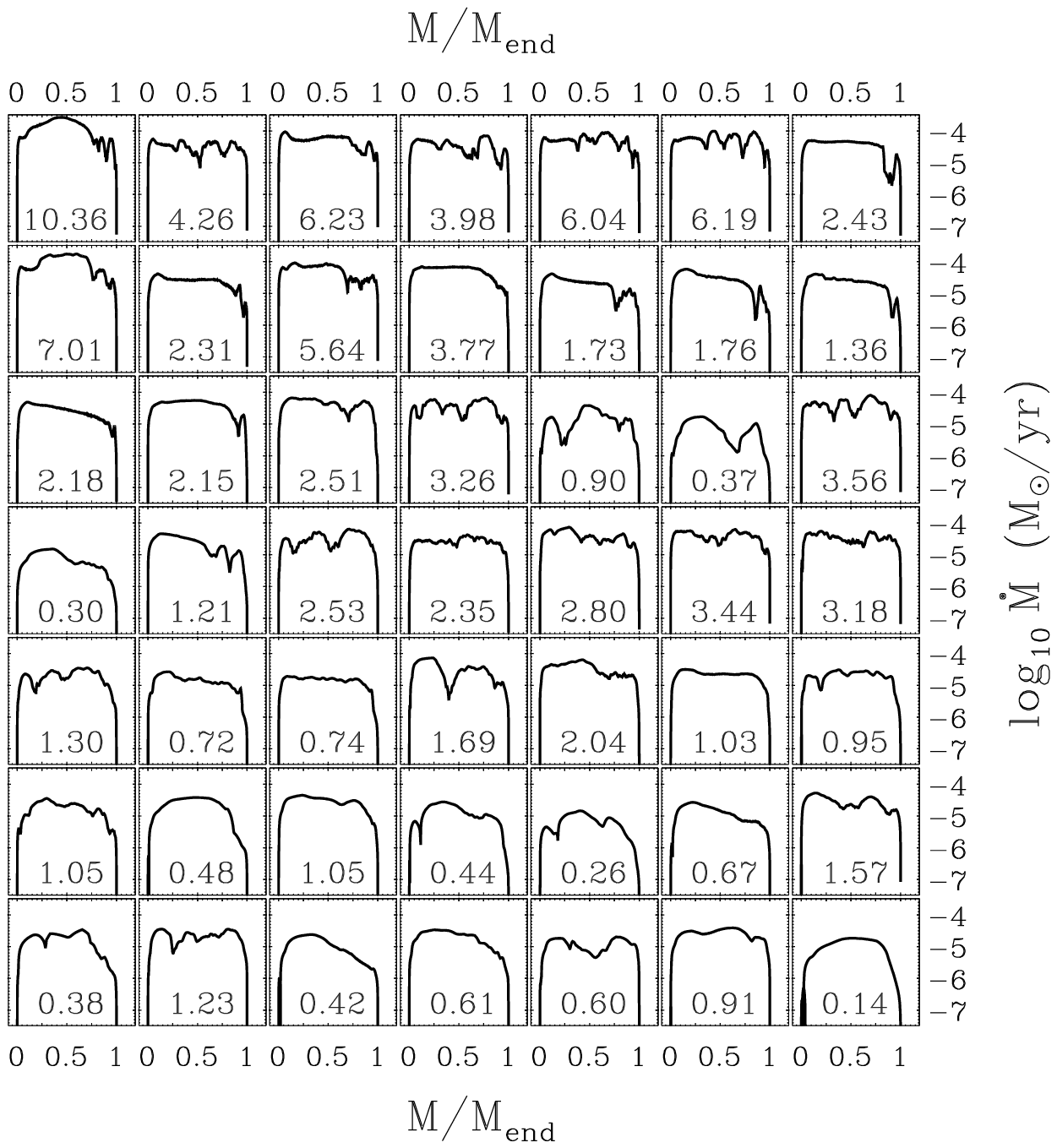}}
\end{picture}
\caption{\label{fig:accretion-rates} Examples of time varying mass accretion rates for
  protostellar cores forming in a dense cluster environment. The left
  panel shows accretion rate $\dot{M}$ versus time after formation
  $t-t_{\rm form}$ for 49 randomly selected protostellar cores in a
  numerical model of molecular cloud fragmentation from Klessen \&
  Burkert (2000).  Formation time $t_{\rm form}$ is defined by the
  first occurrence of a hydrostatic protostellar object deeply embedded
  in the interior of a collapsing gas clump. To link individual
  accretion histories to the overall cluster evolution, $t_{\rm form}$
  is indicated in the upper right corner of each plot and measures the
  elapsed time since the start of the simulation. The free-fall
  timescale of the considered molecular region is $\tau_{\rm ff}
  \approx 10^5\,$years. High-mass stars tend to form early in the
  dynamical evolution and are able to maintain high accretion rates
  throughout the entire simulation. On the contrary, low-mass stars
  tend to form later in the cluster evolution and $\dot{M}$ declines
  strongly after the short initial peak accretion phase.  Altogether,
  the accretion histories of cores (even of those with similar masses)
  differ dramatically from each other due to the stochastic influence
  of the cluster environment, as clumps merge and protostellar cores
  compete for accretion from a common gaseous environment.  The right
  panel plots for the same cores $\dot{M}$ as function of the accreted
  mass $M$ with respect to the final mass $M_{\rm end}$, which is
  indicated in the center of each plot. Note that the mass range spans
  two orders of magnitude.  (From Klessen 2001a.)}
\end{figure*}

Klessen (2001a) considers the dynamical evolution of a molecular cloud
regions with $200\,$M$_{\odot}$ in a volume $\left(0.32\,{\rm
    pc}\right)^3$ where turbulence is assumed to have decayed and left
behind random Gaussian fluctuations in the density structure.  As the
system contracts gravitationally, a cluster of 56 protostellar cores
builds up on a timescale of about two to three free-fall times. These
types of numerical models allow for the following predictions on
protostellar accretion rates in dense clusters:

{\em(a)} Protostellar accretion rates in a dense cluster
  environment are strongly time variable. This is illustrated in
Figure \ref{fig:accretion-rates} for 49 randomly selected cores.

{\em(b)} The typical density profiles of gas clumps that give birth to
protostars exhibit a flat inner core, followed by a density fall-off
$\rho \propto r^{-2}$, and are truncated at some finite radius, which
in the dense centers of clusters often is due to tidal interaction
with neighboring cores (see \S~\ref{sub:cores} and
\S~\ref{sub:clusters}).  As result, a short-lived initial phase of
strong accretion occurs when the flat inner part of the pre-stellar
clump collapses.  This corresponds to the class 0 phase of
protostellar evolution. If these cores were to remain isolated and
unperturbed, the mass growth rate would gradually decline in time as
the outer envelope accretes onto the center. This is the class I
phase.  Once the truncation radius is reached, accretion fades and the
object enters the class II phase. This behavior is expected from
analytical models (e.g.\ Henriksen \etal\ 1997) and agrees with other
numerical studies (e.g.\ Foster \& Chevalier 1993).  However,
collapse does not start from rest for the density fluctuations
considered here, and the accretion rates exceed the theoretically
predicted values even for the most isolated objects in the simulation.

{\em (c)} The mass accretion rates of cores in a dense cluster deviate
strongly from the rates of isolated cores. This is a direct result of
the mutual dynamical interaction and competition between protostellar
cores. While gas clumps collapse to build up protostars, they may
merge as they follow the flow pattern towards the cluster potential
minimum. The timescales for both processes are comparable. The density
and velocity structure of merged gas clumps generally differs
significantly from their progenitor clumps, and the predictions for
isolated cores are no longer valid.  More importantly, these new
larger clumps contain multiple protostars, which subsequently
compete with each other for the accretion from a common gas reservoir.
The most massive protostar in a clump is hereby able to accrete more matter
than its competitors (also Bonnell \etal\ 1997, Klessen \& Burkert
2000, Bonnell \etal\ 2001a,b). Its accretion rate is enhanced through
the clump merger, whereas the accretion rate of low-mass cores
typically decreases.  Temporary accretion peaks in the wake of clump
mergers are visible in abundance in Figure \ref{fig:accretion-rates}.
Furthermore, the small aggregates of cores that build up are
dynamically unstable and low-mass cores may be ejected. As they leave
the high-density environment, accretion terminates and their final
mass is reached.
  
{\em (d)} The most massive protostars begin to form first and continue
to accrete at high rate throughout the entire cluster evolution. As
the most massive gas clumps tend to have the largest density contrast,
they are the first to collapse and constitute the center of the
nascent cluster.  These protostars are fed at high rate and
gain mass very quickly. As their parental clumps merge with others,
more gas is fed into their `sphere of influence'. They are able to
maintain or even increase the accretion rate when competing with
lower-mass objects (e.g.\ core 1 and 8 in Figure
\ref{fig:accretion-rates}).  Low-mass stars, on average, tend to form
somewhat later in the dynamical evolution of the system (as indicated
by the absolute formation times in Figure \ref{fig:accretion-rates};
also Figure 8 in Klessen \& Burkert 2000), and typically have only
short periods of high accretion.

{\em (e)} As high-mass stars are associated with large core masses,
while low-mass stars come from low-mass cores, the stellar population
in clusters is predicted to be mass segregated right from the
beginning. High-mass stars form in the center, lower-mass stars tend
to form towards the cluster outskirts. This is in agreement with
recent observational findings for the cluster NGC330 in the Small
Magellanic Cloud (Sirianni \etal\ 2002). Dynamical effects during the
embedded phase of star cluster evolution will enhance this initial
segregation even further (see \S~\ref{sub:clusters}.c).

\begin{figure*}[th]
\unitlength1cm
\begin{picture}(16.0,10.0)
\put(-1.7,-8.5){\epsfxsize=19.50cm \epsfbox{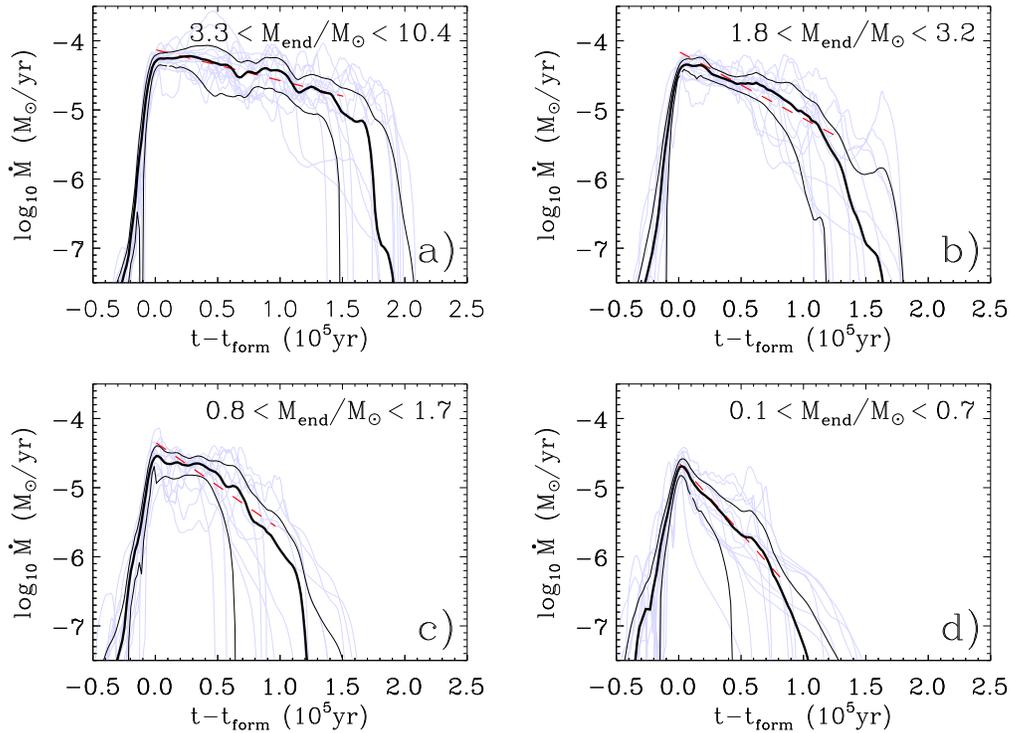}}
\end{picture}
\caption{\label{fig:avg-accretion-rate} Averaged mass accretion rate
  $\langle\dot{M}\rangle$ (thick line) as function of time relative to
  core formation $t-t_{\rm form}$ for four different mass bins
  (ranging from high to low masses as denoted in the top section of
  each plot) overlaid on the contributing individual accretion
  histories. The mean absolute deviation from $\langle\dot{M}\rangle$
  is indicated by thin lines.  An exponential approximation to
  $\langle \dot{M} \rangle$ is indicated by the dashed line. (From
  Klessen 2001a.) }
\end{figure*}
{\em(f)} Individual cores in a cluster environment form and evolve
through a sequence of highly probabilistic events, therefore, their
accretion histories differ even if they accumulate the same final
mass. Accretion rates for protostars of certain mass can only be
determined in a statistical sense. The model predicts that
an exponentially declining rate with a peak value of a few
$10^5\,$M$_{\odot}$yr$^{-1}$, a time constant in the range $0.5$ to $2.5\times
10^5\,$yr, and a cut-off related to gas dispersal from the cluster
offers a reasonable fit to the typical protostellar mass growth in
dense embedded clusters with  uncertainties, however, that remain high
(see Figure \ref{fig:avg-accretion-rate}).

\rsksubsection{Protostellar Evolutionary Tracks}
\label{sub:PMS-tracks}

In this review, we argue that stars are born in interstellar clouds of
molecular hydrogen with the mass growth rates intimately coupled to the
dynamical cloud environment.  Once a prestellar core becomes gravitationally
unstable, it begins to collapse giving birth to a protostar.  While the
structure of molecular clouds is well studied observationally (see \S\ 
\ref{sub:regions}), our knowledge about intrinsic properties of these youngest
stars relies almost entirely on theoretical stellar models. These models give
ages, masses and radii when brightness, distance and effective temperature are
known (e.g.\ Palla 2000, 2002). The so determined ages constitute the only
practical `clock' for tracing the history of star-formation regions and for
studying the evolution of circumstellar disks and planet formation.  They
constitute the basis of our empirical understanding of the evolution of the
young Sun and the origin of solar systems.
 
Until recently (Wuchterl \& Tscharnuter 2003) it was
necessary to {\em assume} a set of initial conditions for the stars at
very young ages (typically at a few $10^5$ years) in order to
calculate the properties at larger ages.  Usually the internal thermal
structure of the star is estimated at a moment when the dynamical
infall motions from the cloud are thought to have faded and the
stellar contraction is sufficiently slow, so that pressure forces
balance gravity.  Then hydrostatic equilibrium is a good
approximation.
%The absolute ages associated with that state are
%obtained by the homological back-extrapolation of the assumed initial
%hydrostatic structure to infinite radius. This leads to typical
%initial ages of $\sim 10^5\,{\rm years}$ which are of the order of the
%free-fall time for a solar mass isothermal equilibrium cloud.  At
%times larger then $10^6\,{\rm years}$ the initial setup of the thermal
%structure is thought to have decayed sufficiently, and the computed
%properties would than agree well with the properties of young stars in
%reality.  
Young star properties are therefore usually calculated without
considering gravitational cloud collapse and protostellar accretion in
detail. See, however, Winkler \& Newman (1980a,b), but also Hartmann,
Cassen \& Kenyon (1997) for a discussion of the possible effects of 
accretion. Altogether, classical PMS calculations typically assume 
fully convective initial conditions as argued for by Hayashi (1961).

However, it can be shown that the assumption of fully convective stellar
structure {\em does not} result from the collapse of isolated, marginally
gravitationally unstable, isothermal, hydrostatic equilibrium, so called
'Bonnor-Ebert' spheres (Wuchterl \& Tscharnuter 2003).  Hence, early stellar
evolution theory has to be reconsidered. This has been done by Wuchterl \&
Klessen (2001), who presented the first calculation of the properties of the
new born star as being a member of a cluster of protostars forming from the
fragmentation of a highly-structured molecular cloud, and followed in detail
the collapse of a one solar mass fragment until it becomes observable in the
visible light. These calculations demonstrated that the newly born star shows
the trace of the fragmentation and collapse process during its main accretion
phase and the early hydrostatic pre-main sequence (PMS) contraction. At an age
of a million years, however, its properties are almost identical for quiet and
turbulent cloud conditions.

\begin{figure*}[th]
\unitlength1.0cm
\begin{picture}(16,11)
%\put(-1.0,-8.0) {\epsfxsize=16cm \epsfbox{paper-fig-subcube.ps}}
\put( 3.00,-0.00){\epsfxsize=10.9cm\epsfbox{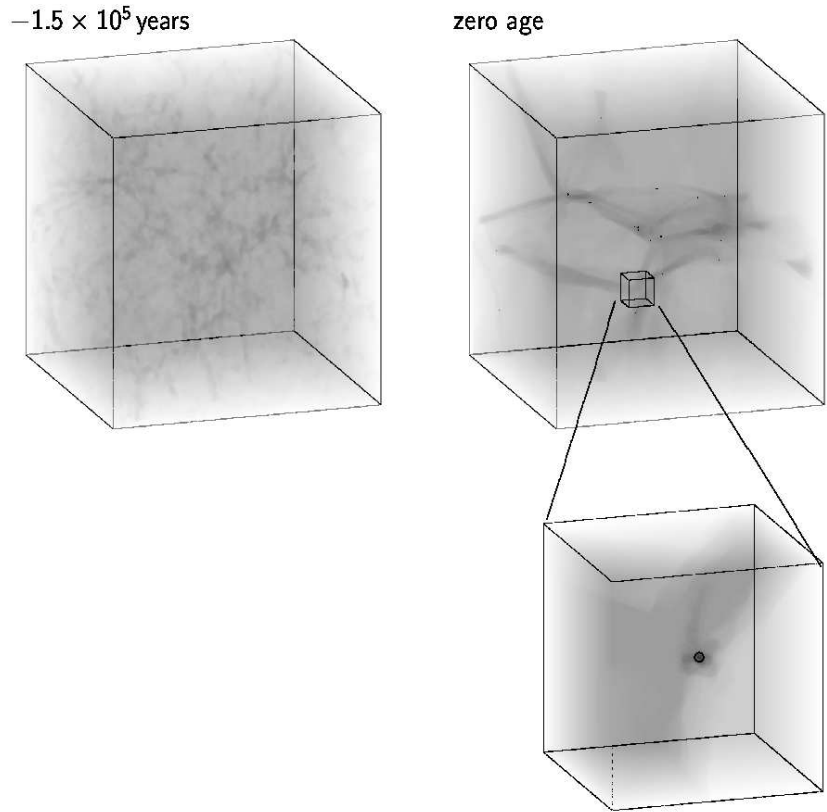}}
\end{picture}
\caption{\label{fig:pms-track-1}%
The 3D density distribution
     of the dynamical molecular cloud fragmentation calculation at two
     different times (model $\cal I$ of KB00, KB01).  As we cannot
     treat the whole cloud, we focus on a sub-volume of mass
     $196\,$M$_{\odot}$ and size $(0.32\,{\rm{pc}})^3$. The left image
     depicts the initial random Gaussian fluctuation field, and the
     central image shows the system when the young proto-Sun reaches
     stellar zero age (i.e.\ when the cloud core for the first time
     becomes optically thick). During this $1.5\times10^5\,$year period
     local collapse occurs and a cluster of deeply embedded and
     heavily accreting pre- and protostellar condensations begins to
     build up (i.e.\ objects without and with central hydrostatic
     core).  The region where the proto-solar condensation forms is
     shown enlarged on the right-hand side. The volume considered in
     the 1D-RHD simulation is indicated by the circle.  --- Adopted from Wuchterl \& Klessen (2001). }
\end{figure*}

\rsksubsubsection{Dynamical PMS Calculations}
\label{subsub:calc}
Wuchterl \& Klessen (2001) considered a molecular cloud region where
turbulence is decayed and has left behind density fluctuations characterized
by a Gaussian random field which follows a power spectrum $P(k) \propto 1/k^2$
(as discussed in Klessen \& Burkert 2000). To describe molecular cloud
fragmentation, they solved the equations of hydrodynamics using a particle
based method (SPH --- smoothed particle hydrodynamics, see Benz 1990, or
Monaghan 1992) in combination with the special-purpose hardware device GRAPE
(GRavity Pipe, see Sugimoto \etal\ 1990, Ebisuzaki \etal\ 1993), focusing on
a sub-region of the cloud with mass $196\,$M$_{\odot}$ and size
$(0.32\,{\rm{pc}})^3$ and adopting periodic boundary conditions (Klessen
1997).  With a mean density of $n({\rm H_2}) = 10^5\,$cm$^{-3}$ and
temperature $T=10\,$K, the simulated volume contained 222 thermal Jeans
masses.  To be able to continue the calculation beyond the formation of the
first collapsing object, compact cores had been replaced by `sink' particles
(Bate \etal\ 1995) once they exceed a density of
$n({\rm{H}}_2)=10^9\,{\rm{cm}}^{-3}$, where we keep track of the mass
accretion, and the linear and angular momenta. The `sink' particle size
defined the volume of a detailed collapse calculation.  The system is
gravitationally unstable and begins to form a cluster of 56 protostellar cores
(as illustration see Figure 1), corresponding to the `clustered' mode of star
formation (Sections \ref{subsub:global} and \ref{sub:multiple}).

Besides that specific choice of the initial molecular cloud conditions, the
only free parameters that remain in this dynamical star formation model are
introduced by the time-dependent convection-model, needed to describe stellar
structure in a `realistic' way. These parameters are determined as usual in
stellar structure theory by demanding agreement between the model-solution and
the actual solar convection zone as measured by helioseismology.

Aiming to describe the birth and the first million years of the Sun, Wuchterl
\& Klessen (2001) selected from the 3D cloud simulation that protostellar core
with final mass closest to $1\,$M$_{\odot}$ and use its mass accretion history
(see Section \ref{sub:accretion}) to determine the mass flow into a spherical
control volume centered on the star. For the stellar mass range considered
here feedback effects are not strong enough to halt or delay accretion into
this protostellar `feeding zone'.  Thus, the core accretion rates are good
estimates for the actual stellar accretion rates. Deviations may be expected
only if the protostellar cores form a binary star, where the infalling mass
must be distributed between two stars, or if very high-angular momentum
material is accreted, where a certain mass fraction may end up in a
circumbinary disk and not accrete onto a star at all. For single stars matter
accreting onto a protostellar disk may be temporarily `stored' in the disk
before getting transported onto the star.  The flow {\em within} the control
volume is calculated by solving the equations of radiation hydrodynamics (RHD)
in the grey Eddington approximation and with spherical symmetry (Castor 1972,
Mihalas \& Mihalas 1984), in their integral form (Winkler \& Norman 1986).  At
any time it is required that the total mass in the volume agrees with the 3D
fragmentation calculation.  Convective energy transfer and mixing is treated
by using a time-dependent convection scheme (Wuchterl \& Feuchtinger 1998)
derived from the model of Kuhfu{\ss} (1987) and including detailed equations
of state and opacities.  Deuterium burning processes are computed with with
standard reaction rates (Caughlan \& Fowler 1988) including convective mixing.
The 1D-RHD calculation covers a spherical volume of radius $R = 160\, {\rm AU}
= 2.46\times10^{15}\,{\rm cm}$ and contains a mass $0.028\, {\rm M_\odot}$ at
$t=0$. The calculation started $1.5\times10^5\,$ years before the moment of
stellar zero age, with a mass $2\times10^{-5}\,{\rm M_\odot}$.  The final mass
of the star is $0.971\, {\rm M_\odot}$.

\begin{figure*}[ht]
\begin{center}
\unitlength1.0cm
\begin{picture}(16,9)
\put(1.0,-0.4) {\epsfxsize=13cm \epsfbox{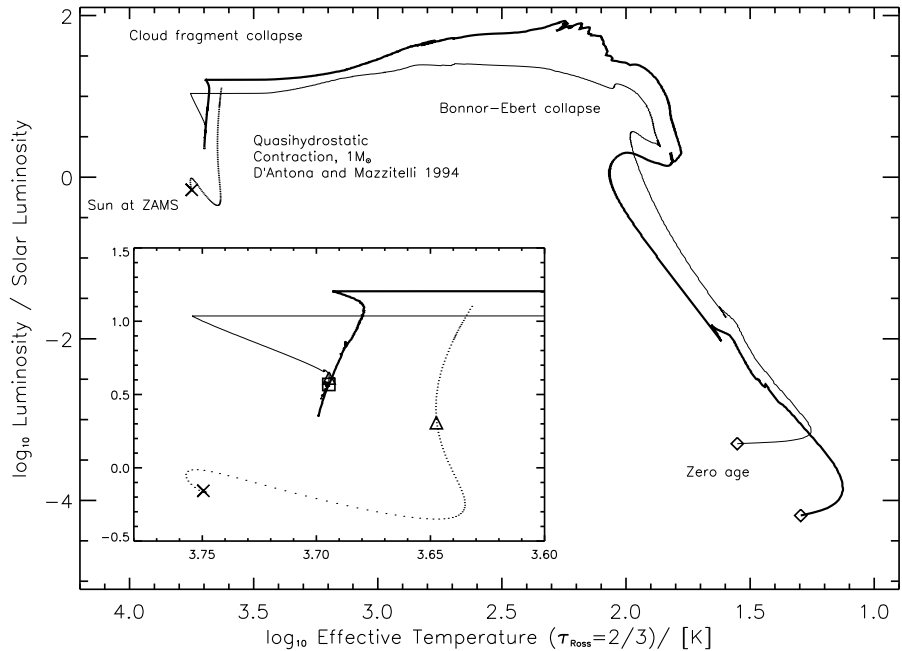}}
\end{picture}
\end{center}
\caption{\label{fig:pms-track-2}%
  Early stellar evolution in the Hertzsprung-Russell diagram.  Three
  evolutionary effective-temperature-luminosity relations (tracks)
  relevant to the young Sun are compared. The dotted line is a
  classical stellar structure, hydrostatic-equilibrium PMS track for
  $1\,{\rm M_\odot} $, for an initially fully convective gas sphere
  (`MLT Alexander' model of D'Antona \& Mazzitelli 1994). The two
  other lines are obtained by describing the formation of the star as
  a result of the collapse of an interstellar cloud.  The {\em thin
    line} is for a cloud fragment in initial equilibrium (a so called
  `Bonnor-Ebert' sphere of a solar mass, see Wuchterl \& Tscharnuter
  2003 for details). The (thick line) is for a cloud fragment that
  results from the dynamical fragmentation of a molecular cloud (KB00,
  KB01).  The two diamonds, in the lower right, indicate zero age for
  the two collapse-results.  Triangles (Bonnor-Ebert from Wuchterl \&
  Tscharnuter 2003), squares (cluster model from Wuchterl \& Klessen
  2001) and crosses (D'Antona \& Mazzitelli 1994) along the respective
  evolutionary tracks mark ages of 1, 10, 100, $\rm 350\, kyr$, 0.5
  and $\rm 1\,Myr$.  The cross at the end of the hydrostatic track
  denotes the moment when energy generation by nuclear reactions in
  the stellar interior, for the first time in stellar live {\em fully}
  compensates the energy losses due to radiation from the stellar
  photosphere, i.e.\ the zero age main sequence (ZAMS). Corresponding
  age-marks for 0.1, 0.35, 0.5 and $\rm 1\,Myr$ are connected by
  dashed lines in the insert. --- Adopted from Wuchterl \& Klessen
  (2001). }
\end{figure*}

\rsksubsubsection{Formation of a $1\,$M$_{\odot}$-Star}

The fragment we have chosen is highlighted in the 3D cloud structure in Figure
\ref{fig:pms-track-1} at the moment when it becomes optically thick and departs from
isothermality, as determined by the 1D solution. This defines the stellar zero
age (Wuchterl \& Tscharnuter 2003).  The mass accretion rates obtained for the
selected fragment are strongly time varying and peak around $1.5\times 10^{-5}
\,$M$_{\odot}\,$yr$^{-1}$ (Section \ref{sub:accretion}).

Before reaching zero age, the temperature is close to the initial
cloud value of 10$\,$K and densities are still low enough so that the
heat produced by the collapse is easily radiated away from the
transparent cloud. Once the envelope becomes optically thick, the
temperature increases rapidly as the accretion luminosity rises. We
determine the effective temperature at the radius where the optical
depth $\tau_{\rm Ross} = 2/3$ (see Baschek, Scholz, \& Wehrse 1991 for
a careful discussion).  Luminosity and temperature for the
non-isothermal phase obtained from the 1D-RHD calculation are shown in
Figure \ref{fig:pms-track-2}.  The zero age is marked by a diamond,
close to the beginning of the thick line in the lower right of Figure
\ref{fig:pms-track-2}. For comparison the results of a
Bonnor-Ebert-collapse and a classical hydrostatic stellar evolution
calculation (D'Antona \& Mazzitelli 1994) are shown as well. The
equations used in the latter study correspond to the hydrostatic limit
of the current dynamical model, all physical parameters (opacities,
etc.) are identical or match closely (see Wuchterl \& Tscharnuter
2003).

The non-isothermal phase can be divided into three parts: (1) There is
a first luminosity increase up to $20\,{\rm L_\odot}$ with the
temperature staying below about 100$\,$K.  The central density of the
fragment rises until a hydrostatic core forms and the accretion flow
onto that core is accelerated 
% until statt as
until quasi-steady state is established. (2) The subsequent main
accretion phase leads to an increase in temperature to 2000$\,$K while
the luminosity shows a broad maximum at $\sim 100\,$L$_{\odot}$.
Compared to the isolated `Bonnor-Ebert case' the violent accretion in
the cluster-environment produces a considerably higher luminosity, and
the oscillations around maximum reflect the variable rates at which
mass is supplied to the accreting protostar as it travels through the
dynamical environment of the proto-cluster. (3) Once accretion fades,
the star approaches
% fade statt fade away (sterben?)
its final mass. The stellar photosphere becomes visible and the
luminosity decreases at roughly constant temperature. This is the
classical pre-main sequence (PMS)
%RSK: Ist das nicht genauer...
contraction phase, shown as a blow up in Figure \ref{fig:pms-track-2}.
The luminosity decreases at almost constant temperature and the
evolutionary tracks are nearly vertical being approximately parallel
to the classical `hydrostatic' track.

\rsksubsubsection{Implications}

The dynamical model allows us to address the question of whether a
trace of the initial fragmentation and collapse process can be found
once the young star arrives at its final mass and becomes optically
visible.  Indeed, the star that forms in a dynamical cloud environment
is brighter when it reaches the PMS phase compared to the isolated
Bonnor-Ebert case. This is due to the higher accretion in the
dynamically evolving cluster environment.

As the mass accretion rates of evolving protostars in dense clusters
are influenced by mutual stochastic interactions and differ
significantly from isolated ones, the positions of stars in the main
accretion phase in the HR diagram are not functions of mass and age
alone, but also depend on the statistical properties of the
protostellar environment.  This affects attempts to infer age and mass
at this very early phase using bolometric temperatures and
luminosities of protostellar cores (see e.g.\ Myers \& Ladd 1993,
Myers \etal\ 1998). It is only possible 
%  in an approximated way -- 
as the statistical average over many different theoretical accretion
histories for different cluster environments or for an observational
sample of protostars with similar cloud conditions.

As the accretion flow fades away, however, the evolutionary tracks of
protostars {\em converge}, and the memory of environmental and initial
conditions is largely lost in the sense that one (final) mass 
corresponds to one track. For given mass and elemental composition
the stellar properties then depend on age only. For our one solar mass 
stars this happens at
$0.95\times10^6\,$years where the effective temperatures become equal
and remain within $20\,$K until the end of the 1D-RHD calculation at
$2\times10^6\,$years.  However, substantial {\em differences} remain
compared to the classical hydrostatic calculations.  The temperature
obtained from collapse models is consistently higher by about 500$\,$K
compared to classical hydrostatic computations at corresponding
luminosities.%
\footnote{ To indicate the consequences for stellar mass
determinations, we point out that during the second million years the
temperature of our one solar mass star corresponds to stars with
$2\,$M$_{\odot}$ on the classical hydrostatic tracks. The differences
in temperature and luminosity equivalently imply corrections for the
inferred ages: the `classical' luminosity at $10^6\,$years is
$6.2\,$L$_{\odot}$, while the corresponding `collapse' values are
smaller, $3.8\,$L$_{\odot}$ and $4.2\,$L$_{\odot}$, respectively. At
$2\times10^6\,$years, the new calculations give luminosities of twice
the solar value. The `classical' age for equivalent luminosities is
$0.8\times10^6\,$years.}

This deviation is the result of a {\em qualitatively} different stellar
structure (Wuchterl \& Tscharnuter 2003).  Most notable, the solar mass stars
resulting from collapse are {\em not} fully convective as is assumed in the
hydrostatic calculations, instead they do have a radiative core of similar
relative size as the present Sun. Convection is confined to a shell in the
outer third of the stellar radius.  The star builds up from material with
increasing entropy as it passes through the accretion shock.  Stellar
structure along the new dynamical evolutionary tracks can be viewed as
homologous to the present Sun rather than to a fully convective structure.
Consequently, the proto-Sun does not evolve along the classical Hayashi track
for a solar mass star, but roughly parallel to that.  It has a higher
effective temperature corresponding to the smaller radius of a partially
radiative object of the same luminosity as a fully convective one.

As the dynamical PMS tracks converge for the two most extreme assumptions
about the stellar environment (dense stellar clusters vs.\ isolated stars) we
predict that a solar mass star at an age of $10^6\,$years will have a
luminosity of $4\pm0.4\,$L$_{\odot}$ and an effective temperature of
$4950\pm20\,$K. The uncertainties reflect the fading traces of the adopted two
highly disparate initial and environmental cloud conditions.  For identical
assumptions made about convection theory and stellar opacities (D'Antona \&
Mazzitelli 1994), the classical values are $2.0\,{\rm L_\odot}$ and $4440\,$K,
respectively.

These prediction are highly controversial and rely on the correctness of the
following assumptions: (1) The structure of young stars can be calculated in
spherical symmetry, (2) the prescription of convection used is sufficiently
accurate outside the regime where it has been tested, i.e.\ the present-day
Sun, and (3) the radiative transfer treatment and especially the present
sources of stellar opacity are sufficiently complete. However, it needs to be
pointed out that {\em all} those assumptions are also made for the classical
calculations that we have used for reference (e.g.\ D'Antona \& Mazzitelli
1994, or Palla 2000, 2002).

\rsksubsection{Initial Mass Function}% RSK[summer 2002]
\label{sub:imf}

The  distribution of stellar masses at birth, described by the initial
mass function (IMF), is a necessary ingredient for the understanding
of many astrophysical phenomena, but no analytic derivation of the
observed IMF has yet stood the test of time. In fact, it appears likely
that a fully deterministic formula for the IMF does not exist.
Rather, any viable theory must take into account the probabilistic
nature of the turbulent process of star formation, which is inevitably
highly stochastic and indeterministic.  We here give a brief overview
of the observational constraints on the IMF, followed by a review of
models for it.

\rsksubsubsection{The Observed IMF} 
\label{subsubsec:IMF-observed}
Hydrogen-burning stars can only exist in a finite mass range
\begin{equation}
\label{eqn:mass-range}
0.08 \sil m \sil 100\:,
\end{equation}
where the dimensionless mass $m\equiv M/(1\mbox{ M}_{\odot})$ is
normalized to solar masses.  Objects with masses less than about
0.08~M$_{\odot}$ do not have central temperatures hot enough for
hydrogen fusion to occur. If they are larger than about ten times the
mass of Jupiter, M$_J = 0.001$~M$_{\odot}$, they are called brown
dwarfs, or more generally substellar objects (e.g.\ Burrows \etal\
1993, Laughlin \& Bodenheimer 1993; or for a review Burrows \etal\
2001). Stars with masses greater than about 100~M$_{\odot}$, on the
other hand, blow themselves apart by radiation pressure (e.g.\
Phillips 1994).  

It is complicated and laborious to estimate the IMF in our Galaxy empirically.
The first such determination from the solar neighborhood (Salpeter 1955)
showed that the number $\xi(m)dm$ of stars with masses in the range $m$ to $m +
dm$ can be approximated by a power-law relation
\begin{equation}
  \label{eqn:salpeter}
  \xi(m)dm \propto m^{-\alpha}dm\;,
\end{equation}
with the index $\alpha \approx 2.35$ for stars in the mass range $0.4
\le m \le 10$. However, approximation of the IMF with a single
power-law is too simple.  Miller \& Scalo (1979) introduced a
log-normal functional form, again to describe the IMF for Galactic
field stars in the vicinity of the Sun,
\begin{eqnarray}
  \label{eqn:miller-scalo}
  \lefteqn{\log_{10}\xi(\log_{10}\,\!m) =}&&\nonumber\\ 
&&\!\!\!\!A -
  \frac{1}{2(\log_{10}\sigma)^2} \left[ \log_{10}\left(\frac{m}{m_0}\right)\right]^2\!\!.
\end{eqnarray}
% \begin{equation}
%   \label{eqn:miller-scalo}
%   \log_{10}\xi(\log_{10}\,\!m) = A -
%   \frac{1}{2(\log_{10}\sigma)^2} \left[ \log_{10}\left(\frac{m}{m_0}\right)\right]^2\;.
% \end{equation}
This analysis has been repeated and improved by Kroupa, Tout, \&
Gilmore (1990), who derive values 
\begin{eqnarray}
  \label{eqn:IMF-kroupa}
   m_0 &=& 0.23 \nonumber \\ 
   \sigma &=& 0.42\\
   A&=&0.1\nonumber  
\end{eqnarray}

\begin{figure}[ht]
\begin{center}
\unitlength1.0cm
\begin{picture}(8,9.6)
%\put( 0.00,0.0){\includegraphics[width=0.4\textwidth]{Kroupa-2002-4a.ps}}
\put( 0.00,-0.6){\includegraphics[width=0.46\textwidth]{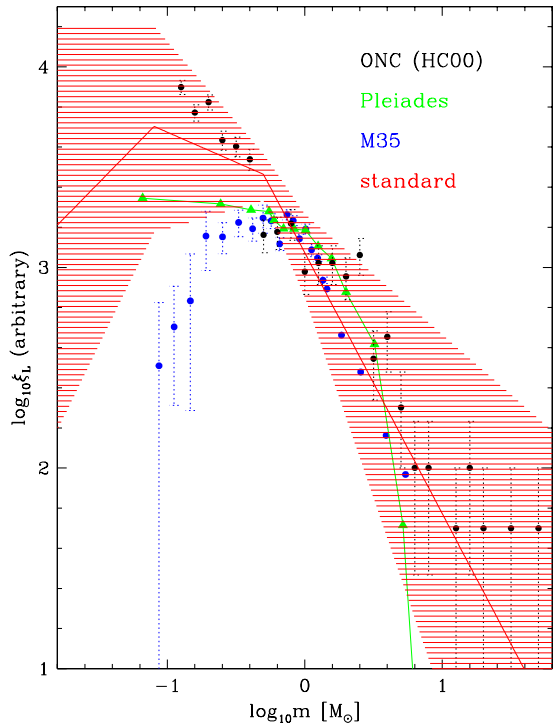}}
\end{picture}
\end{center}
\caption{\label{fig:Kroupa-2002-4a} The measured stellar mass
  function $\xi$ as function of logarithmic mass $\log_{10} m$ in the
  Orion nebular cluster (solid black circles), the Pleiades (green
  triangles), and the cluster M35 (blue circles). None of the mass
  functions is corrected for unresolved multiple stellar systems. The
  average initial stellar mass function derived from Galactic field
  stars in the solar neighborhood is shown as red line with the
  associated uncertainty range indicated by the hatched area. (From
  Kroupa 2002.)}
\end{figure}
The IMF can also be estimated, probably more directly, by studying
individual young star clusters. Typical examples are given in
Figure~\ref{fig:Kroupa-2002-4a} (taken from Kroupa 2002), which
plots the mass function derived from star counts in the Trapezium
Cluster in Orion (Hillenbrand \& Carpenter 2000), in the Pleiades
(Hambly \etal\ 1999) and in the cluster M35 (Barrado~y Navascu\'es
\etal\ 2001). 

The most popular approach to approximating the IMF empirically is to
use a multiple-component power-law of the form (\ref{eqn:salpeter})
with the following parameters (Scalo 1998, Kroupa 2002):
\begin{equation}
\label{eqn:3power-law}
\xi(m) =  \left \{
\begin{array}{ll} \!0.26\,m^{-0.3}& {\rm for } \; 0.01\le m<0.08,\nonumber\\
\!0.035\,m^{-1.3}& {\rm for } \; 0.08\le m<0.5, \\
\!0.019\,m^{-2.3}& {\rm for } \; 0.5\le m<\infty\,.\nonumber\\
\end{array}\right.
\end{equation} 

This representation of the IMF is statistically corrected for
unresolved binaries and multiple stellar systems.  Binary and higher
multiple stars can only be identified as such if their angular
separation exceeds the angular resolution of the telescope used to
survey the sky, or if they are close enough for radial velocity
measurements to detect them as spectroscopic binaries.  Stars in the
middle are falsely counted as single stars.  Neglecting this effect
overestimates the masses of stars, as well as reducing inferred
stellar densities.  These mass overestimates influence the derived
stellar mass distribution, underestimating the number of low-mass
stars.  The IMF may steepen further towards high stellar masses and a
fourth component could be defined with $\xi(m)=0.019\,m^{-2.7}$ for
$m>1.0$ thus arriving at the IMF proposed by Kroupa, Tout, \& Gilmore
(1993).  In equation~(\ref{eqn:3power-law}), the exponents for masses
$m < 0.5$ are very uncertain due to the difficulty of detecting and
determining the masses of very young low-mass stars.  The exponent for
$0.08\le m<0.5$ could vary between $-0.7$ and $-1.8$, and the value in
the substellar regime is even less certain.

 \begin{figure}[ht]
\begin{center}
 \unitlength1.0cm
 \begin{picture}(8,7.5)
% \put( 0.00,-1.0){\includegraphics[width=0.4\textwidth]{Kroupa-2002-5a.ps}}
 \put( 0.20,-2.6){\includegraphics[width=0.45\textwidth]{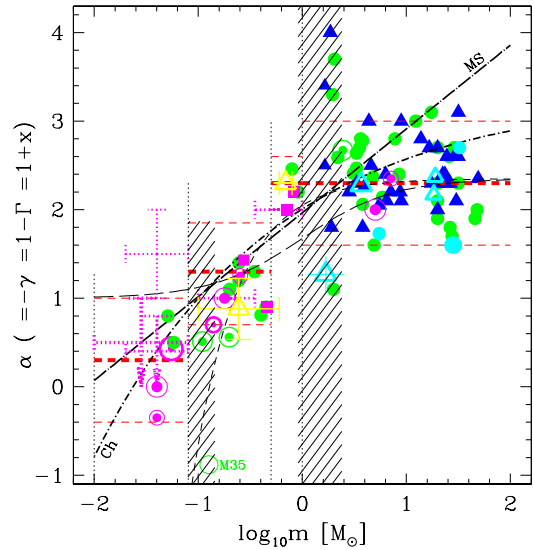}}
 \end{picture}
\end{center}
 \caption{\label{fig:Kroupa-2002-5a} A plot of power-law exponents
   determined for various stellar clusters in the mass range $-2 <
   \log_{10} m < 2$, to illustrate the observed scatter.  The solid
   green and blue dots and solid triangles are from measurements of
   Galactic and Large Magellanic Cloud clusters and OB associations,
   respectively. Globular cluster data are indicated by open yellow
   triangles. None of these measurements is corrected for unresolved
   binaries.  The mean values of the exponent $\alpha$ derived in the
   solar neighborhood, equation \ref{eqn:3power-law}, and the
   associated uncertainties are indicated by horizontal red dashed
   lines. Note that for low stellar masses the values of $\alpha$
   determined from observations in young stellar clusters lie
   systematically lower due to the inability to resolve close binaries
   and multiple stellar systems. Black lines indicate alternative
   functional forms for the IMF, e.g.\ $MS$ gives the Miller-Scalo (1979)
   IMF or $Ch$ the one suggested by Chabrier (2001, 2002).  For a more
   detailed discussion see Kroupa (2002), where the figure is adopted
   from.  }
 \end{figure}
There are some indications that the slope of the mass spectrum
obtained from field stars may be slightly shallower than the one
obtained from observing stellar clusters (Scalo 1998).  The reason for
this difference is unknown, and is somehow surprising given the fact
most field stars may come from dissolved clusters (Adams \& Myer
2001). It is possible that the field star IMF is inaccurate because of
incorrect assumptions about past star formation rates and age
dependences for the stellar scale height. Both issues are either known
or irrelevant for the IMF derived from cluster surveys. On the other
hand, the cluster surveys could have failed to include low-mass stars
due to extinction or crowding.
%
%
% 20??]  suggest that field stars in the solar
% neighborhood come primarily from small groups, which do not appear to
% generate massive stars, perhaps because they never achieve high enough
% stellar densities for collisions to occur, and might have been thought
% to have a steeper IMF.)
% %
% It is possible that the field star IMF is inaccurate because of
% incorrect assumptions about past star formation rates and age
% dependences for the stellar scale height. Both issues are either known
% or irrelevant for the IMF derived from cluster surveys. On the other
% hand, the cluster surveys could have failed to include low-mass stars
% due to extinction or crowding, 
% %MM: corrected to have high-mass stars forming from stellar
% %collisions.  Correct?
% or high-mass stars formed later by stellar collisions between lower
% mass stars (Bonnell, Bate, \& Zinnecker 1998).  However, this trend is
% not well-established because of the different observational methods
% applied and the uncertainties involved.
%
It has also been claimed that the IMF may vary between different
stellar clusters (Scalo 1998), as the measured exponent $\alpha$ in
each mass interval exhibits considerable scatter when comparing
different star forming regions. This is illustrated in Figure
\ref{fig:Kroupa-2002-5a}, which is again taken from Kroupa
(2002). This scatter, however, may be entirely due to effects related to the
dynamical evolution of stellar clusters (Kroupa 2001). 

Despite these differences in detail, all IMF determinations share the
same basic features, and it appears reasonable to say that the basic
shape of the IMF is a universal property common to all star forming
regions in the present-day Galaxy, perhaps with some intrinsic
scatter. There still may be some dependency on the metallicity of the
star forming gas, but changes in the IMF do not seem to be gross even
in that case. There is no compelling evidence for qualitatively
different behavior such as truncation at the low or high-mass end.
%MM: deleted the following:
%stellar mass spectrum cannot be large
%otherwise it would have been observed. Altogether the current evidence
%on this issue is not fully conclusive yet.

\rsksubsubsection{Models of the IMF}
\label{subsubsec:IMF-models}

Existing models to explain the IMF can be divided into five major
groups.  In the first group feedback from the stars themselves
determines their masses. Silk (1995) suggests that stellar masses are
limited by the feedback from both ionization and protostellar
outflows. Nakano, Hasegawa, \& Norman (1995) describe a model in which
stellar masses are sometimes limited by the mass scales of the
formative medium and sometimes by stellar feedback. The most detailed
model in this category stems from Adams \& Fatuzzo (1996) and provides
a transformation between initial conditions in molecular clouds and
final stellar masses. They apply the central limit theorem to the
hypothesis that many independent physical variables contribute to the
stellar masses to derive a log-normal IMF regulated by protostellar
feedback. However, for the overwhelming majority of stars (with masses
$M\sil 5\,M_{\odot}$) protostellar feedback (i.e.\ winds, radiation
and outflows) are unlikely to be strong enough to halt mass accretion,
as shown by detailed protostellar collapse calculations (e.g.\
Wuchterl \& Klessen 2001, Wuchterl \& Tscharnuter 2002).

In the second group of models, initial and environmental conditions
determine the IMF. In this picture, the structural properties of
molecular clouds determine the mass distribution of Jeans-unstable gas
clumps, and the clump properties determine the mass of the stars that
form within. If one assumes a fixed star formation efficiency for
individual clumps, there is a one-to-one correspondence between the
molecular cloud structure and the final IMF.  The idea that
fragmentation of clouds leads directly to the IMF dates back to Hoyle
(1953) and later Larson (1973). More recently, this concept has been
extended to include the observed fractal and hierarchical structure of
molecular clouds Larson (1992, 1995). Indeed random sampling from a
fractal cloud seems to be able to reproduce the basic features of the
observed IMF (Elmegreen \& Mathieu 1983, Elmegreen 1997, 1999,
2000a,c, 2002). A related approach is to see the IMF as a domain
packing problem (Richtler 1994).

The hypothesis that stellar masses are determined by clump masses in
molecular clouds is supported by observations of the dust continuum
emission of protostellar condensations in the Serpens, $\rho$ Ophiuchi,
and Orion star forming regions (Testi \& Sargent 1998; Motte \etal\
1998, 2001; Johnstone \etal\ 2000, 2001). These protostellar cores are
thought to be in a phase immediately before they build up a star in
their interior. Their mass distribution resembles the stellar IMF
reasonably well, suggesting a close correspondence between
protostellar clump masses and stellar masses, leaving little room for
stellar feedback processes, competitive accretion or collisions to act
to determine the stellar mass spectrum. 

A third group of models relies on competitive coagulation or accretion
processes to determine the IMF.  This has a long tradition and dates
back to investigations by Oort (1954) and Field \& Saslaw (1965), but
the interest in this concept continues to the present day (e.g.\ Silk
\& Takahashi 1979; Lejeune \& Bastien 1986; Price \& Podsiadlowski
1995; Murray \& Lin 1996; Bonnell \etal\ 2001a,b; Durisen, Sterzik, \&
Pickett 2001).  Stellar collisions require very high stellar
densities, however, for which observational evidence and theoretical
mechanisms remain scarce.

Fourth, there are models that connect the supersonic turbulent
motions in molecular clouds to the IMF. In particular there are a
series of attempts to find an analytical relation between the stellar
mass spectrum and statistical properties of interstellar turbulence
(e.g.\ Larson 1981, Fleck 1982, Hunter \& Fleck 1982, Elmegreen 1993,
Padoan 1995, Padoan 1995, Padoan \etal\ 1997, Myers 2000, Padoan \&
Nordlund 2002).  However, properties such as the probability
distribution of density in supersonic turbulence in the absence of
gravity have never successfully been shown to have a definite
relationship to the final results of gravitational collapse (Padoan
\etal\ 1997).  Even the more sophisticated attempt by Padoan \&
Nordlund (2002) neglects to take into account that it is likely that
not single shock compressions but multiple compressions and
rarefactions that determine the density structure of supersonic
turbulence (Passot \& V\'azquez-Semadeni 1998).  Furthermore, such
models neglect the effects of competitive accretion in dense cluster
environments (\S~\ref{sub:accretion}), which may be important for
determining the upper end of the IMF.  This makes the attempt to
derive stellar mass spectra from the statistical properties difficult.

Finally there is a more statistical approach.  Larson (1973) and
Zinnecker (1984, 1990) argued that whenever a large set of parameters
is involved in determining the masses of stars, invoking the central
limit theorem of statistics naturally leads to a log-normal stellar
mass spectrum (Adams \& Fatuzzo [1996] made similar arguments).

Regardless of the detailed physical processes involved, the common
theme in all of these models is the probabilistic nature of star
formation.  It appears impossible to predict the formation of specific
individual objects.  Only the fate of an ensemble of stars can be
described ab initio.  The implication is that the star formation
process can only be understood within the framework of a probabilistic
theory.

%\rsksubsubsection{The IMF as result of turbulent fragmentation}
%\label{subsubsec:IMF-clt}
%\mbox{~}\\[2cm]
%{\bf\{Mordecai: Maybe the following is too long. Please try to shorten
%  it. Also THIS subsection may be used in part in the Section ``A new theory
%  of star formation'' (\S~\ref{sub:new})....\}}
%\mbox{~}

 \begin{figure*}[th]
 \unitlength1.0cm
 \begin{picture}(16.0,13.3)
% \put( 2.3,-0.0){\epsfxsize=11.7cm\epsfbox{figure-massspectra-02.ps}}
 \put( 1.0,-1.0){\epsfxsize=15.7cm\epsfbox{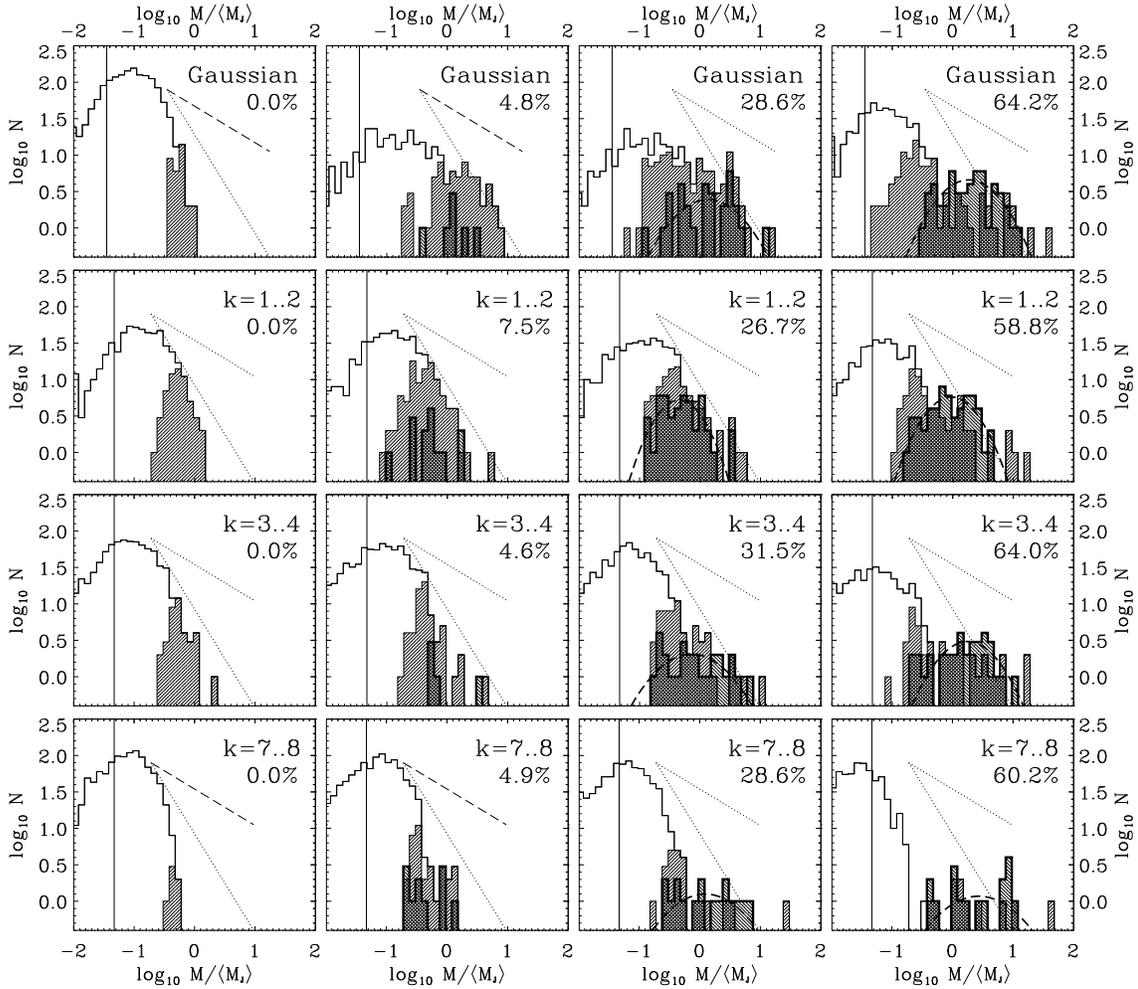}}
 \end{picture}
 \caption{\label{fig:massspectra} Mass spectra of dense collapsed
   cores (hatched thick-lined histograms), of gas clumps (thin lines),
   and of the subset of Jeans unstable clumps (thin lines, hatched
   distribution) for four different models.  The decaying model
   started with Gaussian density perturbations and no turbulence,
   while the other three models were nominally supported by turbulence
   driven at long, intermediate or short scales as indicated by the
   driving wavenumbers $k$.  Masses are binned logarithmically and
   normalized to the average Jeans mass $\langle m_{\rm
   J}\rangle$. The left column gives the initial state of the system
   when the turbulent flow has reached equilibrium but gravity has not
   yet been turned on, the second column shows the mass spectra when
   $m_{\rm *} \approx 5$\% of the mass is accreted onto dense cores,
   the third column shows $m_{\rm *} \approx 30$\%, and the last one
   $m_{\rm *} \approx 60$\%. For comparison with power-law spectra
   ($dN/dm \propto m^{\nu}$), a slope $\alpha = -1.5$ typical for the
   observed clump mass distribution, and the Salpeter slope
   $\alpha=-2.33$ for the IMF, are indicated by the dotted lines.  The
   vertical line shows the resolution limit of the numerical model. In
   columns 3 and 4, the long dashed curve shows the best log-normal
   fit. (From Klessen 2001b.) } \end{figure*}
\rsksubsubsection{Mass Spectra from Turbulent Fragmentation}
\label{subsubsec:mass-spectra}
To illustrate some of the issues discussed above, we examine the mass
spectra of gas clumps and collapsed cores from models of
self-gravitating, isothermal, supersonic turbulence driven with
different wavelengths (Klessen 2001b).  In the absence of magnetic
fields and more accurate equations of state, these models can only be
illustrative, not definitive, but nevertheless they offer insight into
the processes acting to form the IMF.  Figure~\ref{fig:massspectra}
plots for four different models the mass distribution of gas clumps,
of the subset of gravitationally unstable clumps, and of collapsed
cores, at four different evolutionary phases.  In the initial phase,
before local collapse begins to occur, the clump mass spectrum is not
well described by a single power law.  During subsequent evolution, as
clumps merge and grow bigger, the mass spectrum extends towards larger
masses, approaching a power law with slope $\alpha \approx -1.5$.
Local collapse sets in and results in the formation of dense cores
most quickly in the freely collapsing model.  The influence of gravity
on the clump mass distribution weakens when turbulence dominates over
gravitational contraction on the global scale, as in the other three
models. The more the turbulent energy dominates over gravity, the more
the spectrum resembles the initial case of pure hydrodynamic
turbulence. This suggests that the clump mass spectrum in molecular
clouds will be shallower in regions where gravity dominates over
turbulent energy. This may explain the observed range of slopes for
the clump mass spectrum in different molecular cloud regions
(\S~\ref{subsubsec:LSS}).

Like the distribution of Jeans-unstable clumps, the mass spectrum of
dense protostellar cores resembles a log-normal in the model without
turbulent support and in the one with long-wavelength turbulent
driving, with a peak at roughly the average thermal Jeans mass
$\langle m_{\rm J}\rangle$ of the system. These models also predict
initial mass segregation (Section \ref{sub:accretion}.{\em e}).
However, models supported at shorter wavelength have mass spectra much
flatter than observed, suggesting that clump merging and competitive
accretion are important factors leading to a log-normal mass spectrum.
The protostellar clusters discussed here only contain between 50 and
100 cores. This allows for comparison with the IMF only around the
characteristic mass scale, typically about 1~M$_{\odot}$, since the
numbers are too small to study the very low- and high-mass end of the
distribution. Focusing on low-mass star formation, however, Bate,
Bonnell, \& Bromm (2002) demonstrate that brown dwarfs are a natural
and frequent outcome of turbulent fragmentation. In this model, brown
dwarfs form when dense molecular gas fragments into unstable multiple
systems that eject their smallest members from the dense gas before
they have been able to accrete to stellar masses.  Numerical models
with sufficient dynamic range to treat the full range of stellar
masses (Equation \ref{eqn:mass-range}) remain yet to be done.

\rsksection{GALACTIC SCALE STAR FORMATION}
\label{sec:galactic}

How do the mechanisms that control local star formation determine the
global rate and distribution of star formation in galaxies?  In this
section we begin by examining what determines the efficiency of star
formation in \S~\ref{sub:efficient}.  We argue that the balance
between the density of available gas and its turbulent velocity
determines where star formation will occur, and how strongly.  Even if
the turbulent velocity in a region is relatively high, if the density
in that region is also high, the region may still not be supported
against gravitational collapse and prompt star formation.  

Therefore, any mechanism that increases the local density without
simultaneously increasing the turbulent velocity sufficiently can lead
to star formation, via molecular cloud formation, as we discuss in
\S~\ref{sub:clouds}.  Most mechanisms that increase local density
appear to be external to the star formation process, however.
Accretion during initial galaxy formation, interactions and collisions
between galaxies, spiral gravitational instabilities of galactic
disks, and bar formation are major examples. In this review we cannot
do justice to the vast literature on galactic dynamics and
interactions that determine the density distribution in galaxies.  We
do, however, examine what physical mechanisms control the velocity
dispersion in \S~\ref{sub:driving}.  

Finally, in \S~\ref{sub:applications} we briefly speculate on how
turbulent control of star formation may help explain objects with very
different star formation properties, including low surface brightness
galaxies, normal galactic disks, globular clusters, galactic nuclei,
and primordial dwarf galaxies.

\rsksubsection{When is Star Formation Efficient?}
\label{sub:efficient}
%\input{efficient.tex}
%%%
%%%
%%%
%%%\rsksubsection{When is star formation efficient?}
%%%\label{sub:efficient}
%%%

\rsksubsubsection{Overview}
Observers have documented a surprisingly strong connection between the
star formation rate and the local velocity dispersion, column density
and rotational velocity of disk galaxies (Kennicutt 1998a, Martin \&
Kennicutt 2001).  A global Schmidt (1959) law relating star formation
rate surface density to gas surface density as
\begin{equation}
\Sigma_{\rm SFR} = A \Sigma_{\rm gas}^N,
\end{equation}
where a value of $N = 1.4 \pm 0.05$ can be derived from the
observations (Kennicutt 1989, 1998b).  A threshold to star formation
is also found (Kennicutt 1989, Martin \& Kennicutt 2001) in most
galaxies, which also appears related to the gas surface density.  The
Schmidt law can be interpreted as reflecting star formation on a
free-fall timescale, so that (following Wong \& Blitz 2001 for
example) the star formation rate per unit volume of gas with density
$\rho$ is
\begin{equation}
\rho_{\rm SFR} =\epsilon_{\rm SFR} \frac{\rho}{\tau_{\rm ff}} = 
\epsilon_{\rm SFR} \frac{\rho}{(G\rho)^{-1/2}} \propto \rho^{1.5},
\end{equation}
where $\epsilon_{\rm SFR}$ is an efficiency factor observed to be
substantially less than unity.  

The connection between magnetically controlled small-scale star
formation and large-scale star formation is not clear in the
standard theory. Shu, Adams, \& Lizano (1987) did indeed suggest that
OB associations were formed by freely collapsing gas that had
overwhelmed the local magnetic field, but that still implied that the
star formation rate was controlled by the details of the magnetic
field structure, which in turn is presumably controlled by the
galactic dynamo.  If turbulence, as represented by the velocity
dispersion, controls the star formation rate, though, the connection
appears clearer.  The same physical mechanism controls star formation
at all scales.  Regions that are globally supported by turbulence
still engage in inefficient star formation, but the overall star
formation rate will be determined by the frequency of regions of
efficient star formation.

The big open question in this area remains the importance of radiative
cooling to efficient star formation, either on its own or induced by
turbulent compression.  Is cooling, and indeed molecule formation,
necessary for gravitational collapse to begin, or is it rather a
result of already occurring collapse in gravitationally unstable gas?
Certainly there are situations where cooling will make the difference
between gravitational stability and instability, but are those just
marginal cases or the primary driver for star formation in galaxies?

\begin{figure*}[tp]
\begin{center}
\includegraphics[width=0.6\textwidth,angle=270]{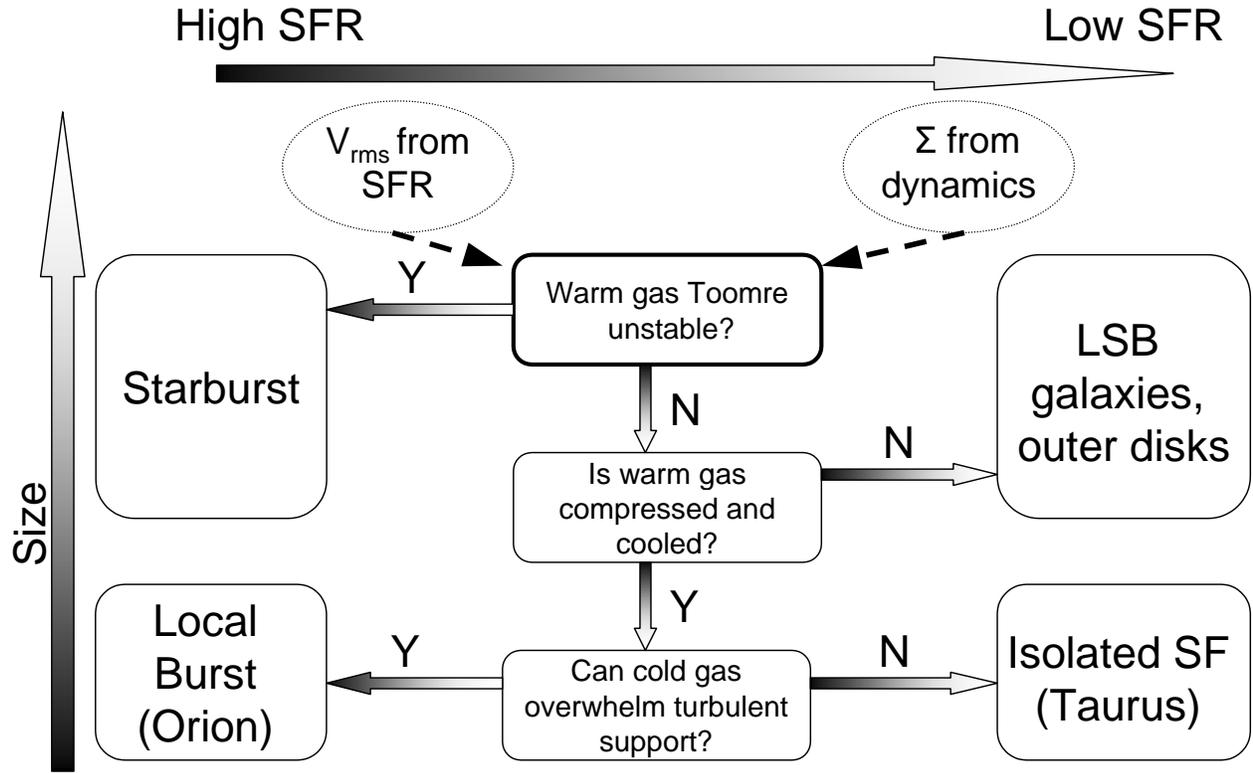}
\end{center}
\caption[Star formation in galaxies]{Illustration of physics
determining star formation in galaxies.}
\label{fig:sf-flowchart}
\end{figure*}

In Figure~\ref{fig:sf-flowchart} we outline a unified picture that
depends on turbulence and cooling to control the star formation rate.  
After describing the different elements of this picture, we will
discuss the steps that we think will be needed to move from this
cartoon to a quantitative theory of the star formation rate.

Probably the factor that determines the star formation rate more than
any other is whether the gas is sufficiently dense to be
gravitationally unstable without additional cooling.  Galactic
dynamics and interactions with other galaxies and the surrounding
intergalactic gas determine the average gas densities in different
regions of galaxies.  The gravitational instability criterion here
includes both turbulent motions and galactic shear, as well as
magnetic fields.  If gravitational instability sets in at the large
scale, collapse will continue so long as sufficient cooling mechanisms
exist to prevent the temperature of the gas from rising (effective
adiabatic index $\gamma_{\rm eff} \leq 1$).  In this situation, molecular
clouds can form in less than $10^5$ yr, as the gas passes through
densities of $10^4$~cm$^{-3}$ or higher, as an incidental effect of
the collapse.  A starburst results, with stars forming efficiently in
compact clusters.  The size of the gravitationally unstable region
determines the size of the starburst.

If turbulent support, rather than thermal support, prevents the gas from
immediately collapsing, compression-induced cooling can become important.
Supersonic turbulence compresses the gas strongly, and most cooling mechanisms
depend on the gas density, usually non-linearly.  Galactic dynamics will again
determine local average density; this mechanism will be more efficient in
regions of higher average density.  If the cooled regions reach high densities
(again, of order $10^4$~cm$^{-3}$), molecule formation will occur quickly,
whether or not gravitational instability sets in (see \S~\ref{sub:clouds}). If
the gas cools in compressed regions, they can become gravitationally unstable
even if they were not before.  Molecule formation will, of course, be more
efficient in regions that collapse gravitationally, but it can occur elsewhere
as well.  The large-scale star formation efficiency will already be much
reduced in this case, as much of the gas will not be compressed sufficiently
to cool.

If the turbulence in the cooled regions does support the gas against general
gravitational collapse, isolated, low-rate star formation can still occur in
regions further compressed by the turbulence.  This may describe regions of
low-mass star formation like the Taurus clouds.  On the other hand, if the
cooled gas does begin to collapse gravitationally, locally efficient star
formation can occur.  The size of the gravitationally unstable region then
really determines whether a group, OB association, or bound cluster eventually
forms.  Star formation in regions like Orion may result from this branch.

%\rsksubsection
%starbursts by accumulation of material (see Lehnert \& Heckman)
\rsksubsubsection{Gravitational Instabilities in Galactic Disks}
\label{subsub:disk-grav}
Now let us consider the conditions under which gravitational instability will
set in.  On galactic scales, the Jeans instability criterion for gravitational
instability must be modified to include the additional support offered by the
shear coming from differential rotation, as well as the effects of magnetic
fields.  The gravitational potential of the stars can also contribute to
gravitational instability on large scales.  Which factor determines the onset
of gravitational instability remains unknown.  Five that have been proposed
are the temperature of the cold phase, the surface density, the local shear,
the presence of magnetic fields, and the velocity dispersion, in different
combinations.

We can heuristically derive the Toomre (1964) criterion for stability
of a rotating, thin disk with uniform velocity dispersion $\sigma$ and
surface density $\Sigma$ using time scale arguments (Schaye 2002).
First consider the Jeans criterion for instability in a thin disk,
which requires that the time scale for collapse of a perturbation of
size $\lambda$
\begin{equation}
t_{\rm coll} = \sqrt{\lambda / G \Sigma}
\end{equation}
be shorter than the time required for the gas to respond to the
collapse, the sound crossing time
\begin{equation}
t_{\rm sc} = \lambda / c_{\rm s}.
\end{equation}
This implies that gravitational stability requires perturbations
with size
\begin{equation} \label{eqn:press}
\lambda < c_{\rm s}^2 / G \Sigma.
\end{equation}
Similarly, in a disk rotating differentially, a perturbation will
rotate around itself, generating centrifugal motions that can also
support against gravitational collapse.  This will be effective if the
collapse time scale $t_{\rm coll}$ exceeds the rotational period
$t_{\rm rot} = 2\pi/\kappa$, where $\kappa$ is the epicyclic
frequency, so that stable perturbations have
\begin{equation} \label{eqn:rot}
\lambda > 4 \pi^2 G \Sigma \kappa^2.
\end{equation}
A regime of gravitational instability occurs if there are
wavelengths that lie between the regimes of pressure and rotational
support, with 
\begin{equation}
\frac{c_{\rm s}^2}{G\Sigma} < \lambda < \frac{4 \pi^2 G\Sigma}{\kappa^2}.
\end{equation}
This will occur if 
\begin{equation} \label{eqn:Toomre}
c_{\rm s}\kappa / 2\pi G \Sigma < 1,
\end{equation}
which is the Toomre criterion for gravitational instability to within
a factor of two.  The full criterion from a linear analysis of
the equations of motion of gas in a shearing disk gives a factor of
$\pi$ in the denominator (Safronov 1960, Goldreich \& Lynden-Bell
1965), while a kinetic theory approach appropriate for a collisionless
stellar system gives a factor of 3.36 (Toomre 1964).

Kennicutt (1989) and Martin \& Kennicutt (2001) have demonstrated that
the Toomre criterion generally can explain the location of the edge of
the star-forming disk in galaxies, although they must introduce a
correction factor $\alpha = 0.69\pm 0.2$ into the left-hand-side of
equation~(\ref{eqn:Toomre}). Schaye (2002) notes that this factor
should be corrected to $\alpha = 0.53$ to correct for the use of both
the velocity dispersion rather than sound velocity, and the exact
Toomre criterion for a stellar rather than a gas disk.

The Toomre criterion given in Equation (\ref{eqn:Toomre}) was derived
for a pure gas disk with uniform temperature and velocity dispersion,
and no magnetic field.  Relaxation of each of these assumptions
modifies the criterion, and indeed each has been argued to be the
controlling factor in determining star formation thresholds by
different authors.

Stars in a gas disk will respond as a collisionless fluid to density
perturbations large compared to their mean separation.  Jog \& Solomon
(1984a) computed the Toomre instability in a disk composed of gas
and stars, and found it to always be more unstable than either
component considered individually.  Both components contribute to the
growth of density perturbations, allowing gravitational collapse to
occur more easily.   Taking into account both
gas (subscript $g$) and stars (subscript $r$), instability occurs when
\begin{equation}
2\pi G k \left(\frac{\Sigma_r}{\kappa^2 + k^2 c_{\rm s} r^2} +
\frac{\Sigma_g}{\kappa^2 + k^2 c_{\rm s} g^2} \right) > 1,
\end{equation}
where $k = 2\pi/\lambda$ is the wavenumber of the perturbation
considered.  Taking into account the effects of the stars always makes
a disk more unstable than it is due to its gas
content alone.  Jog \& Solomon (1984a) and Romeo (1992) extended this model
to include the effect of the finite thickness of the disk.  Elmegreen
(1995) was able with some effort to derive an effective Toomre
parameter that includes the effects of both stars and gas, but that
can only be analytically computed in the thin disk limit.  To compute
it, independent measures of the velocity dispersion of the stars and
of the gas are, of course, needed. Jog (1996) numerically computed the
effective stability parameter for a wide range of values of stellar
and gas disk parameters. The contribution of the stellar disk may
alone be sufficient to explain the correction factors found by
Kennicutt (1989) and Martin \& Kennicutt (2001).

Magnetic fields offer direct support against collapse through their
magnetic pressure and tension.  However, Chandrasekhar (1954) and
Lynden-Bell (1966) were the first to note that they can also have the
less expected effect of destabilization of a rotating system.  The
magnetic field in this case acts to brake the shear that would
otherwise prevent collapse, redistributing angular momentum and
allowing collapse to occur down field lines.  Elmegreen (1987)
performed a linear analysis of the growth rate of gravitational
instability in a rotating, magnetized disk, which was extended by Fan
\& Lou (1997) to follow the excitation of the different modes.

Kim \& Ostriker (2001) were able to identify the regimes in which the
magnetic field acts to either stabilize against collapse or promote
it.  When shear is strong, as it is in the parts of galactic disks
with flat rotation curves, and the field is moderate or weak, with
plasma $\beta \leq 1$, swing amplification stabilized by magnetic
pressure dominates.  Sufficiently unstable disks, with Toomre $Q \leq
$1.0--1.1 (depending on field strength), collapse due to nonlinear
secondary instabilities despite magnetic stabilization.  On the other
hand, if shear is weak, and fields are stronger ($\beta > 1$),
magnetic tension forces act against epicyclic motions, reducing their
stabilizing effect, and producing magneto-Jeans instabilities along
the field lines.

Numerical models by Kim \& Ostriker (2001) showed these mechanisms in
operation, generating large regions of gravitational collapse,
although in the outer parts of disks, the collapse rate from swing
amplification is so slow that additional effects such as spiral arm
amplification may be important to drive the formation of observed
regions of star formation.  Kim \& Ostriker (2002) show that the
introduction of spiral arms indeed produce feathers similar to those
observed, with masses comparable to the largest star forming regions.
% with masses of [{\bf check!}] $10^8 \mbox{ M}_{\odot}$.  These regions are
%larger than even the largest giant molecular clouds, but may be
%comparable in size to the super star clouds identified by Efremov
%(199?) and Efremov \& Elmegreen (199?).  This mechanism also appears
%capable of generating the spurs and feathers seen on spiral arms, as
%described by Kim \& Ostriker (200?).  
These results suggest that the presence of magnetic fields may
actually enhance the star formation rate in disks.

The temperature and the velocity of the coldest gas in a multi-phase
interstellar medium at any point in the disk may be the determining
factor for gravitational instability, rather than some average
temperature.  Schaye (2002) suggests that the sharp rise in
temperature associated with the lack of molecular gas causes the sharp
drop in the star formation rate at the edges of disk galaxies.  He
derives the disk surface density required to allow molecule formation
in the presence of the intergalactic ultraviolet background field and
suggests that this is consistent with the observed threshold column
densities.  However, Martin \& Kennicutt (2001) show a wide variation
in the atomic gas fraction at the critical radius (see their
Figure~9a), calling this idea into question.

The balance between gravitation and local shear is argued by Hunter,
Elmegreen, \& Baker (1998) to be a better criterion than the Toomre
(1964) criterion, which balances gravitation against Coriolis
forces. Effectively this substitutes the Oort A constant (Binney \&
Tremaine 1997) for the epicyclic frequency $\kappa$ in
Equation\ (\ref{eqn:Toomre}).  The difference is small (of order 10\%)
in galaxies with flat rotation curves, but can lower the critical
density substantially in galaxies with rising rotation curves, such as
dwarf galaxies.

\rsksubsubsection{Thermal Instability}
\label{subsub:thermal}
Thermal instability has been the organizing principle behind the most
influential models of the ISM (Pikel'ner 1968; Field, Goldsmith, \&
Habing 1969; McKee \& Ostriker 1977; Wolfire \etal\ 1995). Under the
assumption of approximate pressure and thermal equilibrium, thermal
instability can explain the widely varying densities observed in the
ISM. It can not explain the order of magnitude higher pressures
observed in molecular clouds, though, so it was thought that most
molecular clouds must be confined by their own self-gravity. Turbulent
pressure fluctuations in a medium with effective adiabatic index less
than unity (that is, one that cools when compressed, like the ISM) can
provide an alternative explanation for both pressure and density
fluctuations. Although thermal instability exists, it does not
necessarily act as the primary structuring agent, nor, therefore, as
the determining factor for the star formation rate.

Thermal instability occurs when small perturbations from thermal
equilibrium grow.  The dependence on density $\rho$ and temperature
$T$ of the heat-loss function ${\cal L} = \Lambda - \Gamma$, the sum
of energy losses minus gains per gram per second, determines whether
instability occurs.  Parker (1953) derived the isochoric instability
condition, while Field (1965) pointed out that cooling inevitably
causes density changes, either due to dynamical flows if the region is
not isobaric, or due to pressure changes if it is.  He then derived
the isobaric instability condition.  The alternative of dynamical
compression in a region large enough to be unable to maintain isobaric
conditions has received renewed attention as described below.

The isobaric instability condition derived by Field (1965) is
\begin{equation} \label{eq:isobaric}
\left(\frac{\partial {\cal L}}{\partial T}\right)_P =
\left(\frac{\partial {\cal L}}{\partial T}\right)_{\rho} - \frac{\rho_0}{T_0}
\left(\frac{\partial {\cal L}}{\partial \rho}\right)_T < 0,
\end{equation}
where $\rho_0$ and $T_0$ are the equilibrium values.  Optically thin
radiative cooling in the interstellar medium gives a cooling function
that can be expressed as a piecewise power law $\Lambda \propto \rho^2
T^{\beta_i}$, where $\beta_i$ gives the value for a temperature range
$T_{i-1} < T < T_i$, while photoelectric heating is independent of
temperature.  Isobaric instability occurs when $\beta_i < 1$, while
isochoric instability only occurs with $\beta_i < 0$
(e.g. Field 1965).

In interstellar gas cooling with equilibrium ionization, there are two
temperature ranges subject to thermal instability.  In the standard
picture of the three-phase interstellar medium governed by thermal
instability (McKee \& Ostriker 1977), the higher of these, with
temperatures $10^{4.5}$~K$<T<10^7$~K (Raymond, Cox, \& Smith 1977),
separates hot gas from the warm ionized medium.  The lower range of
$10^{1.7}$~K$<T<10^{3.7}$~K (Figure 3a of Wolfire \etal\ 1995)
separates the warm neutral medium from the cold neutral medium.
Cooling of gas out of ionization equilibrium has been studied in a
series of papers by Spaans (1996, Spaans \& Norman 1997, Spaans \& Van
Dishoeck 1997, Spaans \& Carollo 1998) as described by Spaans \& Silk
(2000).  The effective adiabatic index depends quite strongly on the
details of the local chemical, dynamical and radiation environment, in
addition to the pressure and temperature of the gas.  Although regions
of thermal instability occur, the pressures and temperatures may
depend strongly on the details of the radiative transfer in a
turbulent medium, the local chemical abundances, and other factors.

When thermal instability occurs, it can drive strong motions that
dynamically compress the gas nonlinearly.  Thereafter, neither the
isobaric nor the isochoric instability conditions hold, and the
structure of the gas is determined by the combination of dynamics and
thermodynamics (Meerson 1996, Burkert \& Lin 2000, Lynden-Bell \&
Tout 2001, Kritsuk \& Norman 2002). 

V\'azquez-Semadeni, Gazol, \& Scalo (2000) examined the behavior of
thermal instability in the presence of driven turbulence, magnetic
fields, and Coriolis forces and concluded that the structuring effect
of the turbulence overwhelmed that of thermal instability in a
realistic environment.  Gazol \etal\ (2001) found that about half of
the gas in such a turbulent environment will actually have
temperatures falling in the thermally unstable region, and emphasize
that a bimodal temperature distribution may simply be a reflection of
the gas cooling function, not a signature of a discontinuous phase
transition.  Mac Low \etal\ (2002) examined supernova-driven
turbulence and found a broad distribution of pressures, which were
more important than thermal instability in producing a broad range of
densities in the interstellar gas.

The discovery of substantial amounts of gas out of thermal equilibrium
by Heiles (1999) has provided observational support for a picture in
which turbulent flows rather than thermal instability dominates
structure formation prior to gravitational collapse.  Heiles (1999)
measured the temperature of gas along lines of sight through the warm
and cold neutral medium by comparing absorption and emission profiles
of the H{\sc i} 21 cm fine structure line.  He found that nearly half
of the warm neutral clouds measured showed temperatures that are
unstable according to the application of the isobaric instability
condition, Equation\ (\ref{eq:isobaric}), to the Wolfire \etal\ (1995)
equilibrium ionization phase diagram.

%{\bf add thoughts on Schaye 2002 here}

Although the heating and cooling of the gas clearly plays an important
role in the star formation process, the presence or absence of an
isobaric instability may be less important than the effective
adiabatic index, or similar measures of the behavior of the gas on
compression, in determining its ultimate ability to form stars.
However, Schaye (2002) has made the argument that it is exactly the
ability of the gas to cool above some critical column density that
determines the edge of the star-forming region in disks.

%\rsksubsection{Galactic dynamics determines surface density}
%\label{sub:density}
%\input{density.tex}

\rsksubsection{Formation and Lifetime of Molecular Clouds}
\label{sub:clouds}
%\input{clouds.tex}
%% revisions still to be done (9 Aug 01): 
%%  - Ron Allen observations supporting quick formation as well as
%%  agglomeration of cold unobservable H2
%%  - Gravitational contraction forming H2 incidentally
%%
How do molecular clouds form?  Any explanation must account for the
low star-formation efficiencies observed in nearby molecular clouds,
as well as the broad linewidths observed at scales larger than about
0.1$\,$pc in such clouds.  At the same time, the efficient star
formation seen in regions of massive star formation must still be
permitted.  
%This topic has recently been reviewed by ???, so we will
%focus on the implications of shock compression.

Molecular gas forms on dust grains at a rate calculated by Hollenbach,
Werner, \& Salpeter (1971) to be 
\begin{equation}
t_{\rm form} = (1.5 \times 10^9 \mbox{ yr}) \left(\frac{n}{1\,{\rm cm}^{-3}}\right)^{-1},
\end{equation}
where $n$ is the number density of gas particles.  Experimental work
by Piranello \etal\ (1997a, 1997b, 1999) on molecular hydrogen
formation on graphite and olivine suggests that rates may be strongly
temperature dependent and that the Hollenbach \etal\ (1971) result
may be a lower limit to the formation time. However, the same group
reports that molecule formation is rather more efficient on amorphous
ices (Manic\'o \etal\ 2001) such as would be expected on grain
surfaces deep within dark clouds, so that the rates computed by
Hollenbach \etal\ (1971) may be reached after all.  Further
experimental investigation of molecule formation appears necessary.

The linear density dependence of the formation rate implies that
molecular gas either forms very slowly, over tens of millions of
years at the average densities of order 10$^2\,$cm$^{-3}$ in molecular
regions, or else it forms at very high densities, $\sig
10^4\,$cm$^{-3}$ in a few hundred thousand years.  The latter idea  becomes increasingly attractive.

When molecular clouds were first discovered, they were thought to have
lifetimes of over 100$\,$Myr (e.g.\ Scoville \& Hersh 1979) because of
their apparent predominance in the inner galaxy.  These estimates were
shown to depend upon too high a conversion factor between CO and H$_2$
masses by Blitz \& Shu (1980).  They revised the estimated lifetime
down to roughly 30$\,$Myr based on the association with spiral arms,
apparent ages of associated stars, and overall star formation rate in
the Galaxy.  

Chemical equilibrium models of dense cores in molecular clouds (as
reviewed, for example, by Irvine, Goldsmith, \& Hjalmarson 1986)
showed disagreements with observed abundances in a number of
molecules. These cores would take as much as 10$\,$Myr to reach
equilibrium, which could still occur in the standard model. However,
Prasad, Heere, \& Tarafdar (1991) demonstrated that the abundances of
the different species agreed much better with the results at times of
less than 1$\,$Myr from time-dependent models of the chemical
evolution of collapsing cores.  Bergin \etal\ (1997) came to a similar
conclusion from a careful study of several giant cores in comparison
to an extensive chemical model network, while Saito \etal\ (2002)
studied deuterium fractionation, also finding short lifetimes.

Ballesteros-Paredes, Hartmann, \& V\'azquez-Semadeni (1999) have
argued for a lifetime of less than 10$\,$Myr for molecular clouds as a whole.
They base their argument on the notable lack of a population of $5-20\,$Myr old
stars in molecular clouds.  Stars in the clouds typically have ages under
$3-5\,$Myr, judging from their position on pre-main-sequence evolutionary tracks
in a Hertzsprung-Russell diagram (D'Antona \& Mazzitelli 1994; Swenson \etal\ 
1994; with discrepancies resolved by Stauffer, Hartmann, \& Barrado y
Navascues 1995).  Older weak-line T Tauri stars identified by X-ray surveys
with {\em Einstein} (Walter \etal\ 1988) and {\em ROSAT} (Neuh\"auser \etal\ 
1995) are dispersed over a region as much as 70$\,$pc away from molecular gas,
suggesting that they were not formed in the currently observed gas (Feigelson
1996).  Leisawitz, Bash, \& Thaddeus (1989), Fukui \etal\ (1999) and Elmegreen
(2000) have made similar arguments based on the observation that only stellar
clusters with ages under about 10$\,$Myr are associated with substantial amounts
of molecular gas in the Milky Way and the LMC.

\begin{figure*}[th]
\begin{center}
\includegraphics[width=0.7\textwidth,angle=270]{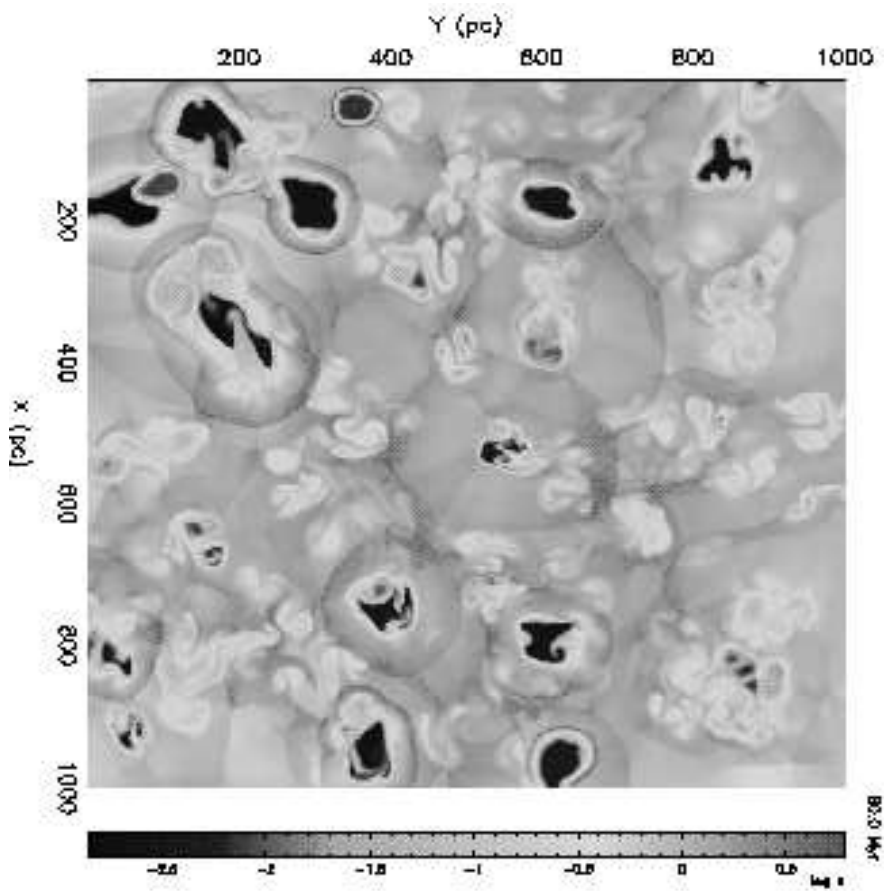}
%Figure 1a from Mac Low, Balsara, Avillez, & Kim (2001)
\end{center}
\caption{\label{fig:mbak-1a}
Log of number density from a three-dimensional SN-driven model of the
ISM with resolution of 1.25~pc, including radiative
cooling and the gravitational field of the stellar disk, as described
by Mac Low \etal\ (2001).  High-density, shock-confined regions are
naturally produced by intersecting SN-shocks from field SNe.
 }
\end{figure*}
For these short lifetimes to be plausible, either molecule formation
must proceed quickly, and therefore at high densities, or observed
molecular clouds must be formed from preexisting molecular gas, as
suggested by Pringle, Allen, \& Lubow (2001).  A plausible place for
fast formation of H$_2$ at high density is the shock compressed layers
naturally produced in a SN-driven ISM, as shown in
Figure\ \ref{fig:mbak-1a} from Mac Low \etal\ (2001)\footnote{Similar
morphologies have been seen in many other global simulations of the
ISM, including Rosen, Bregman, \& Norman (1993), Rosen \& Bregman
(1995), Rosen, Bregman, \& Kelson (1996), Korpi \etal\ (1999), Avillez
(2000), and Wada \& Norman (1999, 2001). Mac Low (2000) reviews such
simulations.}.  They showed that pressures in the ISM are broadly
distributed, with peak pressures in cool gas ($T \sim 10^3$ K) as much
as an order of magnitude above the average because of shock
compressions.  This gas is swept up from ionized $10^4\,$K gas, so
between cooling and compression its density has already been raised up
to two orders of magnitude from $n\approx 1\,$cm$^{-3}$ to $n\approx
100\,$cm$^{-3}$.  These simulations did not include a correct cooling
curve below $10^4\,$K, so further cooling could not occur even if
physically appropriate, but it would be expected.

We can understand this compression quantitatively.  The sound speed in
the warm gas is $(8.1\,$km$\,$s$^{-1})(T/10^4\mbox{ K})^{1/2}$, taking
into account the mean mass per particle $\mu = 2.11 \times 10^{-24}\,$g
for gas 90\% H and 10\% He by number.  The typical velocity dispersion
for this gas is $10-12\,$km$\,$s$^{-1}$ (e.g.\ Dickey \& Lockman 1990,
Dickey, Hanson, \& Helou 1990), so that shocks with Mach numbers
${\cal M} =$ 2--3 are moderately frequent.  Temperatures in these
shocks reach values $T \le 10^5\,$K, which is close to the peak of
the interstellar cooling curve (e.g.\ Dalgarno \& McCray 1972;
Raymond, Cox, \& Smith 1976), so the gas cools quickly back to
$10^4\,$K.  The density behind an isothermal shock is $\rho_1 = {\cal
M}^2 \rho_0$, where $\rho_0$ is the pre-shock density, so order of
magnitude density enhancements occur easily.  The optically-thin
radiative cooling rate $\Lambda(T)$ drops off at $10^4\,$K as H~atoms
no longer radiate efficiently (Dalgarno \& McCray 1972; Spaans \&
Norman 1997), but the radiative cooling $L \sim n^2 \Lambda(T)$.  The
quadratic sensitivity to density means that density enhancements
strongly enhance cooling.  Hennebelle \& P\'erault (1999) show that
such shock compressions can trigger the isobaric thermal instability
(Field, Goldsmith, \& Habing 1969; Wolfire \etal\ 1995), reducing
temperatures to of order 100$\,$K or less.  Heiles (2000) observes a
broad range of temperatures for neutral hydrogen from below 100$\,$K to a
few thousand$\,$K.  The reduction in temperature by two orders of
magnitude from $10^4\,$K to 100$\,$K raises the density correspondingly,
for a total of as much as three orders of magnitude of compression.
Gas that started at densities somewhat higher than average of say
10$\,$cm$^{-3}$ can be compressed to densities of $10^4\,$cm$^{-3}$,
enough to reduce H$_2$ formation times to a few hundred thousand
years.

\begin{figure}[th]
\begin{center}
\includegraphics[width=0.3\textwidth,angle=270]{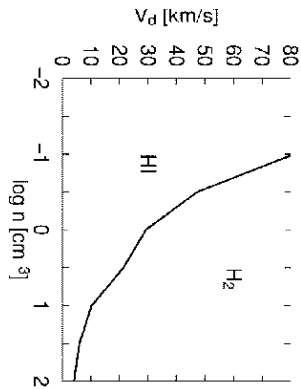}
%Figure 10 from Koyama & Inutsuka 2000
\end{center}
\caption{ \label{fig:ko-in-10}
Shock velocities $V_d$ and pre-shock number densities $n$ at which
the cold post-shock layer is more than 8\% molecular, taken from
one-dimensional simulations by Koyama \& Inutsuka (2000) that include
H$_2$ formation and dissociation, and realistic heating and cooling
functions from Wolfire \etal\ (1995).  
}
\end{figure}
Koyama \& Inutsuka (2000) have demonstrated numerically that
shock-confined layers do indeed quickly develop high enough densities
to form H$_2$ in under a million years, using one-dimensional
computations including heating and cooling rates from Wolfire \etal\
(1995) and H$_2$ formation and dissociation.  In Figure\ \ref{fig:ko-in-10}
we show the parameter space in which they find H$_2$ formation is
efficient.  Hartmann, Ballesteros-Paredes, and Bergin (2001) make a
more general argument for rapid H$_2$ formation, based in part on
lower-resolution, two-dimensional simulations described by Passot,
V\'azquez-Semadeni, \& Pouquet (1995) that could not fully resolve
realistic densities like those of Koyama \& Inutsuka (2000), but do
include larger-scale flows showing that the initial conditions for the
one-dimensional models are not unreasonable.  Hartmann \etal\ (2001)
further argue that the self-shielding against the background UV field
also required for H$_2$ formation will become important at
approximately the same column densities required to become
gravitationally unstable.

As was already noted by Ballesteros-Paredes \etal\ (1999b), shock-confined
layers were shown numerically to be unstable by Hunter \etal\ (1986) in the
context of colliding spherical density enhancements, and by Stevens, Blondin,
\& Pollack (1992) in the context of colliding stellar winds.  Vishniac (1994)
demonstrated analytically that isothermal, shock-confined layers are subject
to a nonlinear thin shell instability (NTSI).  The physical mechanism can be
seen by considering a shocked layer perturbed sinusoidally.  The ram pressure
on either side of the layer acts parallel to the incoming flow, and thus at an
angle to the surface of the perturbed layer.  Momentum is deposited in the
layer with a component parallel to the surface, which drives material towards
extrema in the layer, causing the perturbation to grow.  A careful numerical
study by Blondin \& Mark (1996) in two dimensions demonstrated that the NTSI
saturates in a thick layer of transsonic turbulence when the flows become
sufficiently chaotic that the surface no longer rests at a substantial angle
to the normal of the incoming flow.

Thermal instability will act in conjunction with shock confinement
(see \S~\ref{subsub:thermal}).
% .
% Isochoric thermal instability occurs in the ISM whenever 
% \begin{equation}
% \left(\frac{\partial \Lambda(T)}{\partial T}\right)_{\rho} < 0
% \end{equation}
% (Field 1965), where the rate of energy loss in unit volume is $L =
% \rho^2 \Lambda(T)$. Regions of instability exist in temperature ranges
% of $10^2$ to $10^3\,$K (Wolfire \etal\ 1995), and in the range $10^4$
% to $10^6\,$K (Dalgarno \& McCray 1972, Raymond, Cox, \& Smith 1975).
Burkert \& Lin (2000) computed the nonlinear development of the
thermal instability, demonstrating that shock waves form during the
dynamical collapse of nonlinear regions.  Hennebelle \& P\'erault
(1999) demonstrated that shock compression can trigger thermal
instability in otherwise stable regions in the diffuse ISM, even in
the presence of magnetic fields (Hennebelle \& P\'erault 2000), so
that compressions much greater than the isothermal factor of ${\cal
M}^2$ can occur.

\begin{figure}[ht]
\vspace*{0.2cm}
\begin{center}
\includegraphics[width=0.45\textwidth,angle=0]{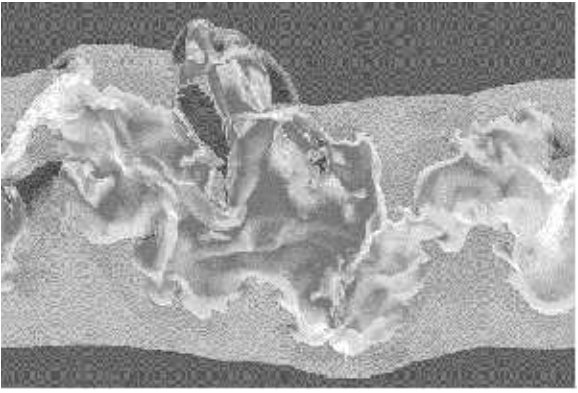}
% Figure 2 from Walder \& Folini 2000 (in color)
\end{center}
\vspace*{-0.1cm}
\caption{\label{fig:wal-fol-2}
Instability of cooled layer confined by strong radiative shocks (from
above and below, Mach number $M\sim 20$), computed in two dimensions
with an adaptive mesh refinement technique by Walder \& Folini (2000).
Darkest regions have densities of 14~cm$^{-3}$, while white represents
a density of $2 \times 10^4$ cm$^{-3}$.}
\end{figure}
These cold, dense layers are themselves subject to dynamical instabilities, as
has  been shown in two-dimensional computations by Koyama \& Inutsuka
(2002).  The instabilities they found are caused by some combination of
thermal instability and mechanisms very similar to the NTSI studied by
Vishniac (1994) for the isothermal case. Figure\ \ref{fig:wal-fol-2} shows another
example of these instabilities from a numerical study by Walder \& Folini
(2000).  These dynamical instabilities can drive strongly supersonic motions
in the cold, dense layer.  If that layer is dense enough for molecule
formation to proceed quickly, those molecules will show strongly supersonic
linewidths on all but the very smallest scales, as seen in the models of
Koyama \& Inutsuka (2002), in agreement with the observations of molecular
clouds.  It remains to be shown whether this scenario can quantitatively
explain the full ensemble of molecular clouds observed in the solar
neighborhood, or elsewhere in our own and external galaxies.

\rsksubsection{Driving Mechanisms}
\label{sub:driving}
%\input{driving.tex}
%%%
%%%
%%%
%%%\rsksubsection{Driving mechanisms determine velocity dispersion}
%%%\label{sub:driving}
%%%

Both support against gravity and maintenance of observed motions
appear to depend on continued driving of the turbulence, which has
kinetic energy density $e = (1/2) \rho v_{\rm rms}^2$.  Mac Low
(1999, 2002) estimates that the dissipation rate for isothermal, supersonic
turbulence is 
\begin{eqnarray} \label{eqn:dissip}
\dot{e} &\!\!\!\!\simeq\!\!\!\!& -(1/2)\rho v_{\rm rms}^3/L_d  \\
&\!\!\!\!=\!\!\!\!& -(3 \times 10^{-27} \,\mbox{erg}\,\mbox{cm}^{-3}\,\mbox{s}^{-1})\nonumber
\\
&\!\!\!\!\times\!\!\!\!& \left(\frac{n}{1\,\mbox{cm}^{-3}}\right)
\left(\frac{v_{\rm rms}}{10\,\mbox{km}\,\mbox{s}^{-1}}\right)^3 
\left(\frac{L_d}{100 \,\mbox{pc}}\right)^{-1},\nonumber
\end{eqnarray}
% \begin{equation} \label{eqn:dissip}
% \dot{e} \simeq -(1/2)\rho v_{\rm rms}^3/L_d 
% = -(3 \times 10^{-27} \,\mbox{erg}\,\mbox{cm}^{-3}\,\mbox{s}^{-1}) 
% \left(\frac{n}{1\,\mbox{cm}^{-3}}\right)
% \left(\frac{v_{\rm rms}}{10\,\mbox{km}\,\mbox{s}^{-1}}\right)^3 
% \left(\frac{L_d}{100 \,\mbox{pc}}\right)^{-1},
% \end{equation}
where $L_d$ is the driving scale, which we have somewhat arbitrarily
taken to be 100~pc (though it could well be smaller), and we have
assumed a mean mass per particle $\mu= 2.11\times 10^{-24}$~g.  The
dissipation time for turbulent kinetic energy
\begin{eqnarray} \label{eqn:disstime}
\tau_d &\!\!\!\!=\!\!\!\!& e / \dot{e} \simeq L/v_{\rm rms} =\\
&\!\!\!\!=\!\!\!\!& (9.8\, \mbox{Myr})
\left(\frac{L_d}{100\,\mbox{pc}}\right)
\left(\frac{v_{\rm rms}}{10\,\mbox{km}\,\mbox{s}^{-1}}\right)^{-1},\nonumber 
\end{eqnarray}
% \begin{equation} \label{eqn:disstime}
% \tau_d = e / \dot{e} \simeq L/v_{\rm rms} = (9.8\, \mbox{Myr})
% \left(\frac{L_d}{100\,\mbox{pc}}\right)
% \left(\frac{v_{\rm rms}}{10\,\mbox{km}\,\mbox{s}^{-1}}\right)^{-1},
% \end{equation}
which is just the crossing time for the turbulent flow across the
driving scale (Elmegreen 2000b).  What then is the energy source for
this driving? We here review the energy input rates for a number of
possible mechanisms.

\rsksubsubsection{Magnetorotational Instabilities}

One energy source for interstellar turbulence that has long been
considered is shear from galactic rotation (Fleck 1981).  However, the
question of how to couple from the large scales of galactic rotation
to smaller scales remained open.  Work by Sellwood \& Balbus (1999)
has shown that the magnetorotational instability (Balbus \& Hawley
1991, 1998) could couple the large-scale motions to small scales
efficiently.  The instability generates Maxwell stresses (a positive
correlation between radial $B_R$ and azimuthal $B_{\Phi}$ magnetic
field components) that transfer energy from shear into turbulent
motions at a rate $\dot{e} = T_{R\Phi} \Omega$ (Sellwood \& Balbus
1999).  Numerical models suggest that the Maxwell stress tensor
$T_{R\Phi} \simeq 0.6 B^2/(8\pi)$ (Hawley, Gammie \& Balbus 1995).
For the Milky Way, the value of the rotation rate recommended by the
IAU is $\Omega = (220 \mbox{ Myr})^{-1} = 1.4 \times 10^{-16} \mbox{ rad
s}^{-1}$, though this may be as much as 15\% below the true value
(Olling \& Merrifield 1998, 2000).  The magnetorotational instability
may thus contribute energy at a rate
% \begin{equation}
% \dot{e} = (3 \times 10^{-29}\,\mbox{erg}\,\mbox{cm}^{-3}\,\mbox{s}^{-1})
% \left(\frac{B}{3 \mu\mbox{G}}\right)^2 \left(\frac{\Omega}{(220\,\mbox{Myr})^{-1}}\right). 
% \end{equation}
 \begin{eqnarray}
\!\!\!\!\lefteqn{\dot{e} = (3 \times 10^{-29}\,\mbox{erg}\,\mbox{cm}^{-3}\,\mbox{s}^{-1})}\nonumber\\
&\!\!\!\!\!\!\!\!\!\!\!\!\!&\times \left(\frac{B}{3 \mu\mbox{G}}\right)^2 \left(\frac{\Omega}{(220\,\mbox{Myr})^{-1}}\right). 
 \end{eqnarray}
For parameters appropriate to the H{\sc i} disk of a sample small
galaxy, NGC~1058, including $\rho = 10^{-24}\,$g$\,$cm$^{-3}$, Sellwood \&
Balbus (1999) find that the magnetic field required to produce the
observed velocity dispersion of 6~km~s$^{-1}$ is roughly 3 $\mu$G, a
reasonable value for such a galaxy.  Similar arguments would hold for
the outer disk of the Milky Way.  This instability may provide a base
value for the velocity dispersion below which no galaxy will fall.  If
that is sufficient to prevent collapse, little or no star formation
will occur, producing something like a low surface brightness galaxy
with large amounts of H{\sc i} and few stars. This may also apply to
the outer disk of our own Galaxy.

\rsksubsubsection{Gravitational Instabilities}

Motions coming from gravitational collapse have often been suggested as a
local driving mechanism in molecular clouds, but fail due to the quick decay
of the turbulence (\S~\ref{subsub:global}).  If the turbulence decays in less
than a free-fall time, as suggested by Equation\ (\ref{eqn:decay}), then it
cannot delay collapse for substantially longer than a free-fall time.

On the galactic scale, spiral structure can drive turbulence in gas
disks.  Roberts (1969) first demonstrated that shocks would form in
gas flowing through spiral arms formed by gravitational instabilities
in the stellar disk (Lin \& Shu 1964, Lin, Yuan, \& Shu 1969).  These
shocks were studied in thin disks by Tubbs (1980) and Soukoup \& Yuan
(1981), who found few vertical motions.  It has been
realized that in a more realistic thick disk, the spiral shock will
take on some properties of a hydraulic bore, with gas passing through
a sudden vertical jump at the position of the shock (Martos \& Cox
1998, G\'omez \& Cox 2002).  Behind the shock, downward flows of as
much as 20~km~s$^{-1}$ appear (G\'omez \& Cox 2002).  Some portion of
this flow will contribute to interstellar turbulence.  However, the
observed presence of interstellar turbulence in irregular galaxies
without spiral arms, as well as in the outer regions of spiral
galaxies beyond the regions where the arms extend suggest that this
cannot be the only mechanism driving turbulence.  A more quantitative
estimate of the energy density contributed by spiral arm driving
has not yet been done.

The interaction between rotational shear and gravitation can, at least
briefly, drive turbulence in a galactic disk, even in the absence of
spiral arms. This process has been numerically modeled at high
resolution (sub-parsec zones) in two dimensions in a series of papers
by Wada \& Norman (1999, 2001), Wada, Spaans, \& Kim (2000), and Wada,
Meurer \& Norman (2002).  However, these models all share two
limitations: they do not include the dominant stellar component, and
gravitational collapse cannot occur beneath the grid scale. The
computed filaments of dense gas are thus artificially supported, and
would actually continue to collapse to form stars, rather than driving
turbulence in dense disks, see S\'anchez-Salcedo (2001) for a detailed
critique.  In very low density disks, where even the dense filaments
remained Toomre stable, this mechanism might operate, however.

Wada \etal\ (2002) estimated the energy input from this mechanism
following the lead of Sellwood \& Balbus (1999), but substituting
Newton stresses (Lynden-Bell \& Kalnajs 1972) for Maxwell stresses.
The Newton stresses will only add energy if a positive correlation
between radial and azimuthal gravitational forces exists, however,
which is not demonstrated by Wada \etal\ (2002).  Nevertheless, they
estimate the order of magnitude of the energy input from Newton
stresses as
% \begin{eqnarray}
% \dot{e} & \simeq & G (\Sigma_g/H) \lambda^2 \Omega \nonumber \\
%         & \simeq & (4 \times 10^{-29} \mbox{ erg cm$^{-3}$ s}^{-1})
% \left(\frac{\Sigma_g}{10 \mbox{ M$_{\odot}$ pc}^{-2}} \right)^2
% \left(\frac{H}{100 \mbox{ pc}} \right)^{-2}
% \left(\frac{\lambda}{100 \mbox{ pc}} \right)^{2}
% \left(\frac{\Omega}{(220 \mbox{ Myr})^{-1}}\right), 
% \end{eqnarray}
\begin{eqnarray}
\dot{e} &\!\!\! \simeq \!\!\!& G (\Sigma_g/H) \lambda^2 \Omega  \\
        &\!\!\! \simeq \!\!\!& (4 \times 10^{-29} \mbox{ erg cm$^{-3}$
        s}^{-1})\nonumber \\
&\!\!\!\!\!\!\!\!\!\!\!&\times
\left(\frac{\Sigma_g}{10 \mbox{ M$_{\odot}$ pc}^{-2}} \right)^2
\left(\frac{H}{100 \mbox{ pc}} \right)^{-2}\nonumber \\
&\!\!\!\!\!\!\!\!\!\!\!&\times
\left(\frac{\lambda}{100 \mbox{ pc}} \right)^{2}
\left(\frac{\Omega}{(220 \mbox{ Myr})^{-1}}\right),\;\;\;\;\;\;\;\;\;\;\;\;\nonumber
\end{eqnarray}
where $G$ is the gravitational constant, $\Sigma_g$ the density of
gas, $H$, the scale height of the gas, $\lambda$ a length scale of
turbulence, and $\Omega$ the angular velocity of the disk.  Values
chosen are appropriate for the Milky Way.  This is two orders of
magnitude below the value required to maintain interstellar
turbulence, see Equation~(\ref{eqn:dissip}).

\rsksubsubsection{Protostellar Outflows}

Protostellar jets and outflows are a popular suspect for the energy
source of the observed turbulence.  We can estimate their average
energy input rate, following McKee (1989), by assuming that some
fraction $f_{\rm w}$ of the mass accreted onto a star during its formation
is expelled in a wind traveling at roughly the escape velocity.  Shu
\etal\ (1988) argue that $f_{\rm w} \approx 0.4$, and that most of the mass
is ejected from close to the stellar surface, where the escape
velocity 
\begin{eqnarray}
\lefteqn{v_{\rm esc} = \left(\frac{2GM}{R}\right)^{1/2} = (200\,{\rm km}\,{\rm
  s}^{-1}) }\nonumber\\
&&\mbox{\hspace{1.0cm}}\times\left(\frac{M}{1\,{\rm M}_{\odot}}\right)^{1/2} \left(\frac{R}{10\,{\rm R}_{\odot}}\right)^{-1/2},\nonumber\\[-0.3cm]
\end{eqnarray}
% \begin{equation}
% v_{\rm esc} = \left(\frac{2GM}{R}\right)^{1/2} = (200\,{\rm km}\,{\rm
%   s}^{-1}) 
% \left(\frac{M}{1\,{\rm M}_{\odot}}\right)^{1/2} \left(\frac{R}{10\,{\rm R}_{\odot}}\right)^{-1/2},
% \end{equation}
where the scaling is appropriate for a solar-type protostar with
radius $R = 10\,\mbox{R}_{\odot}$. 
Observations of neutral atomic winds from protostars suggest outflow
velocities of roughly this value (Lizano \etal\ 1988, Giovanardi et
al.\ 2000).

The total energy input from protostellar winds will substantially
exceed the amount that can be transferred to the turbulence due to
radiative cooling at the wind termination shock.  We represent the
fraction of energy lost there by $\eta_{\rm w}$.  A reasonable upper limit
to the energy loss is offered by assuming fully effective radiation
and momentum conservation, so that 
 \begin{equation}
\eta_{\rm w} < \frac{v_{\rm rms}}{v_{\rm w}} = 0.05 \left(\frac{v_{\rm
       rms}}{10\,\mbox{km}\,{\rm s}^{-1}}\right) \!\left(\frac{200\,\mbox{km}\,{\rm s}^{-1}}{v_{\rm w}}\!\right)\mbox{\vspace{0.0cm}}
 \end{equation}
%  \begin{eqnarray}
% \lefteqn{\eta_{\rm w} < \frac{v_{\rm rms}}{v_{\rm w}} = 0.05 \left(\frac{v_{\rm
%        rms}}{10\,\mbox{km}\,{\rm s}^{-1}}\right) \!\left(\frac{200\,\mbox{km}\,{\rm s}^{-1}}{v_{\rm w}}\!\right)}&&\nonumber\\[0.1cm]
%  \end{eqnarray}
where $v_{\rm rms}$ is the rms velocity of the turbulence, and we have
assumed that the flow is coupled to the turbulence at typical
velocities for the diffuse ISM.  If we assumed that most of the energy
went into driving dense gas, the efficiency would be lower, as typical
rms velocities for CO outflows are 1--2~km~s$^{-1}$. The energy
injection rate 
\begin{eqnarray}
\dot{e}&\!\!=\!\!& \frac12 f_{\rm w} \eta_{\rm w} \frac{\dot{\Sigma}_*}{H} v_{\rm w}^2  \\
       &\!\!\!\simeq\!\!&  (2 \times 10^{-28} \,\mbox{erg}\,\mbox{cm}^{-3}\,\mbox{s}^{-1})
       \left(\frac{H}{200\,\mbox{pc}}\right)^{-1} \nonumber \\ 
        &\!\!\!\!\!\!\!\!& \times\left(\frac{f_{\rm w}}{0.4}\right) \times
     \left(\frac{v_{\rm w}}{200\,\mbox{km}\,\mbox{s}^{-1}}\right) 
       \left(\frac{v_{\rm rms}}{10\, \mbox{km}\,{\rm s}^{-1}}\right) \nonumber \\
       &\!\!\!\!\!\!\!\!& \times\left(\frac{\dot{\Sigma}_*}{4.5 \times 10^{-9}\, \mbox{M$_{\odot}\,$pc$^{-2}\,$yr$^{-1}$}}\right),\nonumber
\end{eqnarray}
where $\dot{\Sigma}_*$ is the surface density of star formation, and
$H$ is the scale height of the star-forming disk.  The scaling value
used for $\dot{\Sigma}_*$ is the solar neighborhood value (McKee 1989).

Although protostellar jets and winds are indeed quite energetic, they
deposit most of their energy into low density gas (Henning 1989), as
is shown by the observation of multi-parsec long jets extending
completely out of molecular clouds (Bally \& Devine 1994).
Furthermore, observed motions of molecular gas show increasing power
on scales all the way up to and perhaps beyond the largest scale of
molecular cloud complexes (Ossenkopf \& Mac Low 2002).  It is hard to
see how such large scales could be driven by protostars embedded in
the clouds.

\rsksubsubsection{Massive Stars}
In active star-forming galaxies, however, massive stars appear likely
to dominate the driving.  They do so through ionizing radiation and
stellar winds from O~stars, and clustered and field supernova
explosions, predominantly from B~stars no longer associated with their
parent gas.  The supernovae appear likely to dominate, as we now show.

\rskparagraph{Stellar Winds}
First, we consider stellar winds.  The total energy input from a
line-driven stellar wind over the main-sequence lifetime of an early
O~star can equal the energy from its supernova explosion, and the
Wolf-Rayet wind can be even more powerful.  However, the mass-loss
rate from stellar winds drops as roughly the sixth power of the star's
luminosity if we take into account that stellar luminosity varies as
the fourth power of stellar mass (Vink, de Koter \& Lamers 2000), and
the powerful Wolf-Rayet winds (Nugis \& Lamers 2000) last only $10^5$
years or so, so only the very most massive stars contribute
substantial energy from stellar winds.  The energy from supernova
explosions, on the other hand, remains nearly constant down to the
least massive star that can explode.  As there are far more lower-mass
stars than massive stars, with a Salpeter IMF giving a power-law in
mass of $\alpha = -2.35$ (Equation\ \ref{eqn:salpeter}), supernova
explosions will inevitably dominate over stellar winds after the first
few million years of the lifetime of an OB association.
% , until the
% lifetime of the least massive star to explode of around 40--50 Myr.

\rskparagraph{{\rm H}{\sc ii} Region Expansion}
Next, we consider ionizing radiation from OB stars.  The total
amount of energy contained in ionizing radiation is vast.  Abbott
(1982) estimates the total luminosity of ionizing radiation in the
disk to be 
\begin{equation}
\dot{e} = 1.5 \times 10^{-24} \mbox{ erg s$^{-1}$ cm}^{-3}.
\end{equation}
However, only a very small fraction of this total energy goes to driving
interstellar motions.

Ionizing radiation primarily contributes to interstellar turbulence by
ionizing H{\sc ii} regions, heating them to 7000--10,000~K, and raising
their pressures above that of surrounding neutral gas, so that they
expand supersonically.  Matzner (2002) computes the momentum input
from the expansion of an individual H{\sc ii} region into a
surrounding molecular cloud, as a function of the cloud mass and the ionizing
luminosity of the central OB association.  By integrating over the
H{\sc ii} region luminosity function derived by McKee \& Williams
(1997), he finds that the average momentum input from a Galactic
region is 
\begin{eqnarray}
\!\!\!\!\!\!\!\!\!\!\!\!\!\!\!\!\!\!\!\!\!\!\!\!\!\!\!\!\!\!\!\!\!\!\lefteqn{\langle \delta p \rangle \simeq (260 \,\mbox{km}\,\mbox{s}^{-1})\!
\left(\frac{N_{\rm H}}{1.5 \times
    10^{22}\,\mbox{cm}^{-2}}\right)^{-3/14}}\mbox{\hspace{2.90cm}}\nonumber \\
&&\!\!\!\!\!\!\!\!\!\!\!\!\!\!\!\!\!\!\!\!\!\!\!\!\!\!\!\!\!\!\!\times\left(\frac{M_{cl}}{10^6\,\mbox{M}_{\odot}}\right)^{1/14} 
\langle M_* \rangle.
\end{eqnarray}
% \begin{equation}
% \langle \delta p \rangle \simeq (260 \,\mbox{km}\,\mbox{s}^{-1})
% \left(\frac{N_{\rm H}}{1.5 \times 10^{22}\,\mbox{cm}^{-2}}\right)^{-3/14}
% \left(\frac{M_{cl}}{10^6\,\mbox{M}_{\odot}}\right)^{1/14} 
% \langle M_* \rangle.
% \end{equation}
The column density $N_{\rm H}$ is scaled to the mean value for Galactic
molecular clouds (Solomon \etal\ 1987), which varies little as cloud
mass $M_{cl}$ varies.  The mean stellar mass per cluster in the Galaxy
$\langle M_* \rangle = 440 \,\mbox{M}_{\odot}$ (Matzner 2002).

The number of OB associations contributing substantial amounts of
energy can be drawn from the McKee \& Williams (1997) cluster
luminosity function
\begin{equation}
{\cal N} (> S_{49}) = 6.1 \left(\frac{108}{S_{49}} - 1 \right),
\end{equation}
where ${\cal N}$ is the number of associations with ionizing photon
luminosity exceeding $S_{49} = S/(10^{49}$~s$^{-1})$.  The
luminosity function is rather flat below $S_{49} = 2.4$, the
luminosity of a single star of 120~M$_{\odot}$, which was the highest
mass star considered, so taking its value at $S_{49} = 1$ is about
right, giving ${\cal N}(> 1) = 650$ clusters.

To derive an energy input rate per unit volume $\dot{e}$ from the mean
momentum input per cluster $\langle \delta p \rangle$, we need to
estimate the average velocity of momentum input $v_{i}$, the time over
which it occurs $t_{i}$, and the volume $V$ under consideration.
Typically expansion will not occur supersonically with respect to the
interior, so $v_{i} < c_{{\rm s},i}$, where $c_{{\rm s},i} \simeq 10$~km~s$^{-1}$
is the sound speed of the ionized gas.  McKee \& Williams (1997) argue
that clusters typically last for about five generations of massive
star formation, where each generation lasts $\langle t_* \rangle =
3.7$~Myr. The scale height for massive clusters is $H_c \sim 100$~pc
(e.g.\ Bronfman \etal\ 2000), and the radius of the star-forming disk
is roughly $R_{sf} \sim 15$~kpc, so the relevant volume $V = 2 \pi
R_{sf}^2 H_c$.  The energy input rate from H{\sc ii} regions is then
\begin{eqnarray}
\dot{e}&\!\!=\!\!& \frac{\langle \delta p \rangle {\cal N}(>1) v_{i}}{V
t_{i}} \\
       &\!\!=\!\!& (3 \times 10^{-30} \mbox{ erg s$^{-1}$ cm}^{-3})\nonumber \\
&&\!\!\!\!\!\!\times\!\left(\frac{N_{\rm H}}{1.5\times 10^{22} \mbox{ cm}^{-2}}\right)^{-3/14}\!\!
\left(\frac{M_{cl}}{10^6 \mbox{ M}_{\odot}}\right)^{1/14}\nonumber \\
&&\!\!\!\!\!\!\times\!\left(\frac{\langle M_* \rangle}{440 \mbox{ M}_{\odot}}\right)\!
\left(\frac{{\cal N}(>1)}{650}\right) \!
         \left(\frac{v_{i}}{10 \mbox{ km s}^{-1}}\right)\nonumber \\
&&\!\!\!\!\!\!\times\!\left(\frac{H_c}{100 \mbox{ pc}}\right)^{-1}\!\!
\left(\frac{R_{sf}}{15 \mbox{ kpc}}\right)^{-2}\!\!
\left(\frac{t_{i}}{18.5 \mbox{ Myr}}\right)^{-1}\!\!\!\!\!, \nonumber
\end{eqnarray}
% \begin{eqnarray}
% \dot{e}& =& \frac{\langle \delta p \rangle {\cal N}(>1) v_{i}}{V
% t_{i}} \nonumber \\
%        & = & (3 \times 10^{-30} \mbox{ erg s$^{-1}$ cm}^{-3})
% \left(\frac{N_{\rm H}}{1.5\times 10^{22} \mbox{ cm}^{-2}}\right)^{-3/14}
% \left(\frac{M_{cl}}{10^6 \mbox{ M}_{\odot}}\right)^{1/14}
% \left(\frac{\langle M_* \rangle}{440 \mbox{ M}_{\odot}}\right)
% \left(\frac{{\cal N}(>1)}{650}\right) \times \nonumber \\
%        & \times  &  \left(\frac{v_{i}}{10 \mbox{ km s}^{-1}}\right)
% \left(\frac{H_c}{100 \mbox{ pc}}\right)^{-1}
% \left(\frac{R_{sf}}{15 \mbox{ kpc}}\right)^{-2}
% \left(\frac{t_{i}}{18.5 \mbox{ Myr}}\right)^{-1}, 
% \end{eqnarray}
where all the scalings are appropriate for the Milky Way as discussed
above.  Nearly all of the energy in ionizing radiation goes towards
maintaining the ionization degree of the diffuse  medium, and hardly
any towards driving turbulence.  Flows of ionized gas may be important
very close to young clusters and may terminate star formation locally
(\S~\ref{subsub:termination}), but do not appear to contribute
significantly on a global scale.

\rskparagraph{Supernovae}
The largest contribution from massive stars to interstellar turbulence
comes from supernova explosions.  To estimate their energy input rate,
we begin by finding the supernova rate in the Galaxy $\sigma_{SN}$.
Cappellaro \etal\ (1999) estimate the total supernova rate in
supernova units to be $0.72 \pm 0.21$ SNu for galaxies of type S0a-b
and $1.21 \pm 0.37$~SNu for galaxies of type Sbc-d, where 1 SNu = 1 SN
(100~yr)$^{-1} (10^{10} L_B/\mbox{L}_{\odot})^{-1}$, and $L_B$ is the
blue luminosity of the galaxy.  Taking the Milky Way as lying between
Sb and Sbc, we estimate $\sigma_{SN} = 1$~SNu.  Using a Galactic
luminosity of $L_B = 2 \times 10^{10} \mbox{ L}_{\odot}$, we find a
supernova rate of (50~yr)$^{-1}$, which agrees well with the estimate
in equation~(A4) of McKee (1989).  If we use the same scale height
$H_c$ and star-forming radius $R_{sf}$ as above, we can compute the
energy input rate from supernova explosions with energy $E_{SN} =
10^{51}$~erg to be
\begin{eqnarray}
\dot{e} &\!\!=\!\!&\frac{\sigma_{SN} \eta_{SN} E_{SN}}{\pi R_{sf}^2 H_c}
       \\ 
       &\!\!=\!\!& (3 \times 10^{-26} \mbox{ erg s$^{-1}$ cm}^{-3})\nonumber \\
&&\times\left(\frac{\eta_{SN}}{0.1} \right)
\left(\frac{\sigma_{SN}}{1 \mbox{ SNu}} \right) 
\left(\frac{H_c}{100 \mbox{ pc}} \right)^{-1}\nonumber \\
&&\times\left(\frac{R_{sf}}{15 \mbox{ kpc}} \right)^{-2}
\left(\frac{E_{SN}}{10^{51} \mbox{ erg}} \right).\nonumber 
\end{eqnarray}
The efficiency of energy transfer from supernova blast waves to the
interstellar gas $\eta_{SN}$ depends on the strength of radiative
cooling in the initial shock, which will be much stronger in the
absence of a surrounding superbubble (e.g.\ Heiles 1990).  Substantial
amounts of energy can escape in the vertical direction in superbubbles
as well, however.  Norman \& Ferrara (1996) make an analytic estimate
of the effectiveness of driving by SN remnants and superbubbles. The
scaling factor $\eta_{SN} \simeq 0.1$ used here was derived by
Thornton \etal\ (1998) from one-dimensional numerical simulations of
SNe expanding in a uniform ISM, or can alternatively be drawn from
momentum conservation arguments comparing a typical expansion velocity
of 100$\,$km$\,$s$^{-1}$ to typical interstellar turbulence velocity of
10$\,$km$\,$s$^{-1}$.  Detailed multi-dimensional modeling of the
interactions of multiple SN remnants (e.g.\ Avillez 2000) will be
required to better determine it.

Supernova driving appears to be powerful enough to maintain the
turbulence even with the dissipation rates estimated in
equation~(\ref{eqn:dissip}).  It provides a large-scale
self-regulation mechanism for star formation in disks with sufficient
gas density to collapse despite the velocity dispersion produced by
the magnetorotational instability.  As star formation increases in
such galaxies, the number of OB stars increases, ultimately increasing
the supernova rate and thus the velocity dispersion, which restrains
further star formation.

Supernova driving not only determines the velocity dispersion, but may
actually form molecular clouds by sweeping gas up in a turbulent
flow. Clouds that are turbulently supported will experience
inefficient, low-rate star formation, while clouds that are too
massive to be supported will collapse (e.g. Kim \& Ostriker 2001),
undergoing efficient star formation to form OB associations or even
starburst knots.

%Argument:

%velocity dispersions supersonic
%multiple driving mechanisms available, but energetics favors SNe
%continuous spectrum suggests that molecular clouds not separately
%driven

%Turbulent support appears more important than thermal support in
%determining whether and where gravitational instability occurs in
%galaxies. 

%turbulent support in explosion-driven turbulence (beyond N\&F)

\rsksubsection{Applications}
\label{sub:applications}
%\input{applications.tex}
%%%
%%%
%%%
%%%\rsksubsection{Applications}
%%%\label{sub:applications}

Different types of objects with different star formation properties
can be qualitatively explained by the combination of density
determined by galactic dynamics and turbulence driven by different
mechanisms.  We here present illustrative scenarios for different
objects, moving from low to high star formation efficiency.

\rsksubsubsection{Low Surface Brightness Galaxies}

Low surface brightness galaxies have large fractions of their baryonic
mass in gas, whether they have masses typical of massive (Schombert
\etal\ 1992, McGaugh \& de Blok 1997) or dwarf galaxies
(Schombert, McGaugh, \& Eder 2001).  Nevertheless, their star
formation rates lie well below typical values for high surface
brightness galaxies (van der Hulst \etal\ 1993; McGaugh \& de Blok
1997). Their rotation curves have been derived from both H{\sc i}
measurements (van der Hulst \etal\ 1993, de Blok, McGaugh, \& van der
Hulst, 1996), and higher resolution H$\alpha$ measurements (Swaters,
Madore, \& Trewhalla, 2000; McGaugh, Rubin, \& de Blok, 2001, Matthews
\& Gallagher 2002) which may sometimes disagree with the H{\sc
i} in the innermost regions (Swaters \etal\ 2000), but are in
generally good agreement (McGaugh \etal\ 2001).  They have lower gas
and stellar surface densities than high surface brightness galaxies
(van der Hulst \etal\ 1987; de Blok \& McGaugh 1996).

The question of whether their disks have surface densities lying below the
Kennicutt (1989) threshold for star formation has been studied using rotation
curves derived from H{\sc i} measurements for both massive (van der Hulst
\etal\ 1993) and dwarf (van Zee \etal\ 1997) galaxies.  In the case of massive
galaxies, surface densities beneath the Kennicutt (1989) threshold do indeed
appear to explain the lack of star formation (van der Hulst 1993).  The
moderate levels of turbulence required to maintain the observed velocity
dispersions may be produced by magnetorotational instabilities (Sellwood \&
Balbus 1999).  Other explanations for the lack of star formation, such as an
inability to form molecular hydrogen (Gerritsen \& de Blok 1999) or to cool it
(Mihos, Spaans, \& McGaugh 1999), were derived from numerical models that did
not include magnetic effects, and thus had no source of support other than
thermal pressure to counteract gravitational collapse and star formation. If
magnetorotational instability is the dominant support mechanism, then star
formation will not be suppressed in the center where the rotational shear
drops.  This is in fact where star formation is found in low surface
brightness galaxies.

In the case of dwarf galaxies (Hunter 1997), the situation appears to be
slightly more complex. Van Zee \etal\ (1997) demonstrate that the surface
density in a sample of low surface brightness dwarf galaxies falls
systematically below the Kennicutt threshold, with star formation observed in
regions that approach the threshold, while van Zee, Skillman, \& Salzer (1998)
show that blue compact dwarf galaxies have surface densities exceeding the
threshold in their centers.  Hunter, Elmegreen, \& Baker (1998), on the other
hand, argue that a criterion based on local shear correlates better with the
observations, especially in galaxies with rising rotation curves.  Another
factor that may be contributing to the star formation histories of dwarf
galaxies is that starbursts in the smaller ones (under $10^8$~M$_{\odot}$) can
actually push all the gas well out into the halo, from where it will take some
hundreds of millions of years to collect back in the center (Mac Low \&
Ferrara 1999).  This scenario may be consistent with observations in some
galaxies, as summarized by Simpson \& Gottesman (2000).

\rsksubsubsection{Galactic Disks}

In normal galactic disks, where SNe appear to dominate the driving of
the turbulence, most regions will have a star formation rate just
sufficient to produce turbulence that can balance the local surface
density in a self-regulating fashion.  However, as spiral arms or
other dynamical features increase the local density, this balance
fails, leading to higher local star formation rates.  Because the
increase in star formation rate as turbulence is overwhelmed is
continuous, the enhanced star formation in spiral arms and similar
structures does not globally approach starburst rates except when the
densities are greatly enhanced.  Locally, however, even relatively
small regions can reach starburst-like star formation efficiencies if
they exceed the local threshold for turbulent support and begin to
collapse freely.  A classic example of this is the massive star
formation region NGC~3603, which locally resembles a starburst knot,
even though the Milky Way globally does not have a large star
formation rate.  Also the Trapezium cluster in Orion is thought to be
formed with efficiency of $\sil 50$\% (Hillenbrand \& Hartmann 1998).
%
%Even Orion has a local star formation
%efficiency greater than 50\% (ref).

\rsksubsubsection{Globular Clusters}

Globular clusters may simply be the upper end of range of normal
cluster formation.  Whitmore (2000) reviews evidence showing that
young clusters have a power-law distribution reaching up to globular
cluster mass ranges.  The luminosity function for old globular
clusters is log normal, which Fall \& Zhang (2001) attribute to the
evaporation of the smaller clusters by two-body relaxation, and the
destruction of largest clusters by dynamical interactions with the
background galaxy.  They suggest that the power-law distribution of
young clusters is related to the power-law distribution of molecular
cloud masses found by Harris \& Pudritz (1995).  However, numerical
models of gravitational collapse tend to produce mass distributions
that appear more log-normal, and are not closely related in shape to
the underlying mass distributions of density peaks (Klessen 2001,
Klessen \etal\ 2000).  It remains unknown whether cluster masses are
determined by the same processes as the masses of individual
collapsing objects, but the simulations do not include any physics
that would limit them to one scale and not the other.  Further
investigation of this question will be interesting.

\rsksubsubsection{Galactic Nuclei}
In galaxies with low star formation rates, the galactic nucleus is often the
only region with substantial star formation occurring. As rotation curves
approach solid body in the centers of galaxies, magnetorotational
instabilities will die away, leaving less turbulent support and perhaps
greater opportunity for star formation. In more massive galaxies, gas is often
funneled towards the center by bars and other disk instabilities, again
increasing the local density sufficiently to overwhelm local turbulence and
drive star formation.

Hunter \etal\ (1998) and Schaye (2002) note that central regions of galaxies
appear to have normal star formation despite having surface densities that
appear to be stable according to the Toomre criterion.  This could be due to
reduced turbulence in these regions reducing the surface density required.
This has classically been difficult to measure because H~{\sc i} observations
with sufficient velocity resolution to measure typical turbulent linewidths of
6--12~km~s$^{-1}$ have generally had rather low spatial resolution, with just
a few beams across the galaxy.  Most calculations of the critical surface
density just assume a constant value of the turbulent velocity dispersion,
which may well be incorrect (Wong \& Blitz 2002).

As an alternative, or perhaps additional explanation, Kim \& Ostriker
(2001) point out that the magneto-Jeans instability acts strongly in
the centers of galaxies.  The magnetic tension from strong magnetic
fields can reduce or eliminate the stabilizing effects from Coriolis
forces in these low shear regions, effectively reducing the problem to
a two-dimensional Jeans stability problem along the field lines.

\rsksubsubsection{Primordial Dwarfs}

In the complete absence of metals, cooling becomes much more
difficult.  Thermal pressure supports gas that accumulates in dark
matter haloes until the local Jeans mass is exceeded.  The first
objects that can collapse are the ones that can cool from H$_2$
formation through gas phase reactions.  Abel, Bryan, \& Norman (2000,
2002) and Bromm, Coppi, \& Larson (1999) have computed models of the
collapse of these first objects.  Abel \etal\ (2000, 2002) used
realistic cosmological initial conditions, and found that inevitably a
single star formed at the highest density peak before substantial
collapse had occurred elsewhere in the galaxy.  Bromm \etal\ (1999)
used a flat-top density perturbation that was able to fragment in many
places simultaneously, due to its artificial symmetry.

Work by Li, Klessen, \& Mac Low (2003) suggests that the
lack of fragmentation seen by Abel \etal\ (2000, 2002) may be due to
the relatively stiff equation of state of metal-free gas.  Li \etal\
found that fragmentation of gravitationally collapsing gas is strongly
influenced by the polytropic index $\gamma$ of the gas, with
fragmentation continuously decreasing from $\gamma \sim 0.2$ to
$\gamma \sim 1.3$.  The limited cooling available to primordial gas
even with significant molecular fraction may raise its polytropic
index sufficiently to suppress fragmentation.  Abel and coworkers argue that
the resulting stars are likely to have masses exceeding
100$\,$M$_{\odot}$, leading to prompt supernova explosions with
accompanying metal pollution and radiative dissociation of H$_2$.

\rsksubsubsection{Starburst Galaxies}

Star burst galaxies convert gas into stars at such enormous rates,
that the timescale to exhaust the available material becomes short
compared to the age of the universe (see the review by Sanders \&
Mirabel 1996). Starbursts are therefore short-lived phenomena
typically lasting for a few tens of Myr, however, may occur several
times during the lifespan of a galaxy.  The star formation rates in
starburst galaxies can be as high as 1000$\,$M$_{\odot}\,$yr$^{-1}$
(Kennicutt 1998) which is three orders of magnitude above the current
rate of the Milky Way.  Starbursting galaxies are rare in the local
universe, but rapidly increase in frequency at larger lookback times
suggesting that the starburst phenomenon was a dominant phase of early
galaxy evolution at high redshifts. The strongest star bursts are
observed towards galactic nuclei or in circumnuclear regions. However,
in interacting galaxies, star formation is also triggered far away
from the nucleus in the overlapping regions or in spiral arms or
sometimes even in tidal tales. In such an interaction a significant
number of super-star clusters are formed, which may be the progenitors
of present-day globular clusters. The Antennae galaxy, the product of
a major merger of to spiral galaxies (NGC~4038 and 4039), is a famous
example where star formation is most intense in the overlap region
between the two galaxies (Whitmore \& Schweizer 1995). Merging events
seem always associated with the most massive and most luminous
starburst galaxies, the `ultraluminous IR galaxies' -- ULIRG's --
(Sanders \& Mirabel 1996).  However, the starburst phenomenon can also
be triggered in a more gentle, minor merger. Such an event disturbs
but does not disrupt the primary galaxy. It will recover from the
interaction without dramatic changes in its overall morphology. In
particular, this may apply to the lower-mass `luminous blue compact
galaxies' which often show very little or no sign of interaction
(e.g.\ van Zee, Salzer, \& Skillman 2001). Alternative triggers of the
starburst phenomenon that have have been suggested for these galaxies
including bar instabilities in the galactic disk (Shlosman, Begelman,
\& Frank 1990), or also the compressional effects of multiple
supernovae and winds from massive stars (e.g.\ Heckman, Armus, \&
Miley 1990) which then would lead to very localized burst of star
formation. Regardless of the specific nature of the triggering
mechanism, the relevant property is a fast and efficient flow of gas
into a concentrated region on timescales short enough to beat stellar
feedback processes. This can only be provided by gravitational torques
(Combes 2001). We conclude that starburst galaxies are extreme
examples of a continuum of star formation phenomena, with gravity
overwhelming any resistive effects of turbulent gas motions on kpc
scales.

%Starburst galaxies are extremely luminous objects with an abundance of
%young massive stars and a disturbed optical morphology. Famous
%examples such as the Antennae, involve major mergers, but the
%starburst phenomenon can also be triggered in a more gentle, minor
%merger. Such an event disturbs but does not disrupt the primary galaxy
%(i.e. the primary will recover from the interaction without a drastic
%jump along the Hubble sequence). In such an interaction a significant
%number of super-star clusters (SSCs) are formed, which may be the
%progenitors of present-day globular clusters. The ubiquity of globular
%cluster systems among a wide range of galaxy types suggests that these
%interactions play a significant role in the evolution of `normal'
%galaxies. Therefore to understand the formation and evolution of
%galaxies, it is essential to understand such processes. Here we
%present a HST WFPC2 broad-band study of three starburst galaxies with
%different merger histories.

\rsksection{CONCLUSIONS}
\label{sec:conclusions}
% \section{CONCLUSIONS}
% \label{sec:conclusions}
%
\rsksubsection{Summary}
\label{sub:summary}

The formation of stars represents the triumph of gravity over a
succession of opponents.  These include thermal pressure, turbulent
flows, magnetic flux, and angular momentum.  
For several decades, magnetic fields were thought to dominate the
resistance against gravity, with star formation occurring
quasistatically.  A growing body of observational evidence suggests
that when star formation actually occurs it does so quickly and
dynamically, with a rate controlled by driven supersonic turbulence. 
Such turbulence is required to explain the broad linewidths observed
in star-forming clouds, as magnetic fields cannot explain them.  The
varying balance between turbulence and gravity then provides a natural
explanation for the widely varying star formation rates seen both
locally and globally.  Scattered, inefficient star formation is a
signpost of turbulent support, while clustered, efficient star
formation occur in regions lacking support.  In this picture, gravity
has already won in all or nearly all observed dense protostellar
cores: dynamical collapse seems to explain their observed properties
better than the alternatives.  The  mass distribution of stars then depends
at least partly on the density and velocity structure resulting
from the turbulence, perhaps explaining the apparent local variations
of the stellar initial mass function (IMF)  despite its broad universality.

In Section \ref{sec:paradigm} we summarize and critically discuss the
physical phenomena that regulate stellar birth. We begin with a
historical overview of the classical dynamical theory of star
formation (\S\ \ref{sub:classical}), which already included turbulent
flows, but only in the microturbulent approximation, treating them as
an addition to the thermal pressure.  We then turn to the development
of the standard theory of star formation (\S\ \ref{sub:standard})
which was motivated by growing understanding of the importance of the
interstellar magnetic field in the 1960's and 1970's.

The standard theory relies on ion-neutral drift, also known as ambipolar
diffusion, to solve the magnetic flux problem for protostellar cores,
which were thought to be initially magnetohydrostatically
supported. At the same time magnetic tension resulted in braking of
rotating protostellar cores, thus solving the angular momentum problem
as well. The timescale for ambipolar diffusion to remove enough
magnetic flux from the cores for gravitational collapse to set in can
exceed the free-fall time by as much as an order of magnitude,
suggesting that magnetic support could also explain low observed
star formation rates.  Finally, magnetic fields were also
invoked to explain observed supersonic motions.

However, both observational and theoretical results have
begun to cast doubt on the standard theory. In \S\
\ref{sub:standardprobs} we summarize theoretical limitations
of the standard isothermal sphere model that forms the basis for many
of the practical applications of the standard theory.  We discuss
several observational findings that put the fundamental assumptions of
that theory into question. The observed magnetic field strengths in
molecular cloud cores appear too weak to support against gravitational
collapse. At the same time, the infall motions measured around star
forming cores extend too broadly, while the central density profiles
of cores are flatter than expected for isothermal spheres.
Furthermore, the chemically derived ages of cloud cores are
comparable to the free-fall time instead of the much longer ambipolar
diffusion timescale. Observations of young stellar objects also appear
discordant.  Accretion rates appear to decrease rather than remaining
constant, far more embedded objects have been detected in cloud cores
than predicted, and the spread of stellar ages in young clusters does
not approach the ambipolar diffusion time.

New theoretical and numerical studies of turbulence that point beyond
the standard theory while looking back to the classical dynamical
theory for inspiration have now emerged (\S\ \ref{sub:beyond}).
Numerical studies demonstrated that supersonic turbulence decays
rapidly, in roughly a crossing time of the region under consideration,
regardless of magnetic field strength.  Under molecular cloud
conditions, it decays in less than a free-fall time.  This implies
that the turbulence in star-forming clouds needs to be continuously
driven in order to maintain the observed motions. 
%(\S\ \ref{subsub:motions}).  
Driven turbulence has long been thought
capable of supporting gas against gravitational collapse.
%(\S~\ref{subsub:self-grav}).  
Numerical models 
%described in \S~\ref{subsub:numerics} 
were used to test this, showing that turbulence indeed can offer
global support, while at the same time leading to local collapse
on small scales. In strongly compressible turbulence, 
gravitational collapse  occurs localized in the density enhancements
produced by shocks.
% (\S\ \ref{subsub:local}).  
The rate of local collapse depends strongly on
the strength and driving scale of the turbulence. This gives a natural
explanation for widely varying star formation rates.  Magnetic fields
not strong enough to provide static support make a quantitative but
not a qualitative difference, reducing the collapse rate somewhat, but
not preventing local collapse.
%(\S\ \ref{subsub:MHD}).  
They may still act to transfer angular momentum so long as they are
coupled to the gas, however.

We outline the shape of the new theory in \S\ \ref{sub:new}.  Rather
than relying on quasistatic evolution of magnetostatically supported
objects, it suggests that supersonic turbulent support controls star
formation.  
%mm: added 
Inefficient, isolated star formation is a hallmark of turbulent
support, while efficient, clustered star formation occurs in its
absence.  When stars form, they do so dynamically, collapsing on the
local free-fall time.  The initial conditions of clusters appear
largely determined by the properties of the turbulent gas, as is the
rate of mass accretion onto these objects.
%mm
The balance between turbulent support and local density then
determines the star formation rate.  Turbulent support is provided by
some combination of supernovae and galactic rotation, along with
possible contributions from other processes.  Local density is
determined by galactic dynamics and interactions, along with the
balance between heating and cooling in a region.  The initial mass
function is at least partly determined by the initial distribution of
density resulting from turbulent flows, although a contribution from
stellar feedback and interactions with nearby stars cannot be ruled
out. The initial conditions for stellar clusters in this theory come
from the turbulent flow from which they formed.
%{\bf Mordecai: THE MORE I THINK OF IT, THE MORE I LIKE THE IDEA OF
%EXTENDING THIS PARAGRAPH TO CONTAIN THE BASIC IDEAS OF THE NEW PARADIGM.}

We explore the implications of the control of star formation by
supersonic turbulence at the scale of individual stars and stellar
clusters in \S\ \ref{sec:local}.  We begin by looking more closely at
the structure of turbulent molecular clouds (\S\ \ref{sub:regions}),
noting that some well known descriptions like Larson's (1981) laws may
be natural consequences of a turbulent flow observed in projection.
%In particular, Larson's mass-radius relation likely an observational
%artifact, so most molecular clouds are probably not in virial
%equilibrium, although their largest cores do appear to be just over
%the verge of collapse and so not far from equilibrium mass.
Observations indicate that interstellar  turbulence is driven on
large scales, quite likely on 
scales substantially larger than the clouds themselves (see also \S\
\ref{sub:driving}). 
%(\S\ \ref{subsubsec:LSS}, and see also \S\ \ref{sub:driving}).  
We examine how turbulent fragmentation determines the star forming
properties of molecular clouds (\S\ \ref{sub:multiple}), and then turn
to discuss protostellar cores (\S\ \ref{sub:cores}) and stellar
clusters (\S~\ref{sub:clusters}) in particular.  Strongly time-varying
protostellar mass growth rates may result as a natural consequence of
competitive accretion in nascent embedded clusters
%competitive accretion in turbulent star forming regions 
(\S\ \ref{sub:accretion}).  
Turbulent models predict protostellar mass distributions that appear
roughly consistent with the observed stellar mass spectrum (\S\
\ref{sub:imf}), although more work needs to be done to arrive at a
full understanding of the origin of stellar masses.

The same balance between turbulence and gravity that seems to
determine the efficiency of star formation in molecular clouds may
also work at galactic scales, as we discuss in
\S~\ref{sec:galactic}. We begin by describing the effects of
differential rotation and thermal instability competing and
cooperating with turbulence to determine the overall star formation
efficiency in \S~\ref{sub:efficient}.
%It remains unclear which factor dominates, although it appears on
%scales of galactic disks that
%turbulent support likely does, at least in star-forming regions of
%disks.  
The transient nature of molecular clouds suggests that they form from
gas compressed by large-scale turbulent flows in galactic disks. This
very same flow may also drive the turbulent motions observed within
the clouds, and furthermore, may also be responsible for their
destruction on a short timescale
%perhaps they are again destroyed by the same flow 
(\S\ \ref{sub:clouds}).  We then examine the physical mechanisms that
could drive the interstellar turbulence, focusing our discussion on
the energy available from each mechanism in \S\ \ref{sub:driving}.  In
star-forming regions of disks, supernovae appear to overwhelm all
other possibilities.  In outer disks and low surface brightness
galaxies, on the other hand, the situation is not so clear:
magnetorotational or gravitational instabilities look most likely to
drive the observed flows.  Finally, in
\S\ \ref{sub:applications}, we give examples of how this picture may
apply to different types of objects, including low surface brightness,
normal, and starburst galaxies, as well as galactic nuclei and
globular clusters.

\rsksubsection{Future Research Problems}
\label{sub:future}

Although the outline of a new theory of star formation has emerged, it
is by no means complete.  The ultimate goal of a predictive,
quantitative theory of the star formation rate and initial mass
function remains elusive.  It may be that the problem is intrinsically
so complex, like terrestrial climate, that no single solution exists,
but only a series of temporary, quasi-steady states.  Certainly our
understanding of the details of the star formation process can be
improved, though.  Ultimately, coupled models capturing different
scales are likely necessary to capture the interaction of the
turbulent cascade with the varying thermodynamics, chemistry, and
opacity of gas at different densities.  We can identify several major
questions that capture the outstanding problems. As we merely want to
summarize these open issues in star formation, we refrain from giving
an in-depth discussion and the associated references, which
may largely be found in the body of the review.

{\em How can we describe turbulence driven by astrophysical
processes?}  It is not uniform at the driving scale, both because of
its magnetization, and because of the non-uniformity of explosions and
other drivers.  The scalefree nature of the turbulent cascade is
further perturbed by the drastic changes in the equation of state that
occur as densities increase, leading to stronger radiative cooling and
the reduction of heating by the exclusion of cosmic rays.  At small
scales, diffusion and dissipation mechanisms determine the structure.
Although ambipolar diffusion probably limits the production of
small-scale magnetic field structures, there is increasing theoretical
support for additional density and velocity structure at scales below
the ambipolar diffusion cutoff, whose interaction with self-gravity
needs to be investigated.  Ultimately, the dense regions produced by
this imperfect cascade play one of the major roles in determining the
initial mass function of stars.

{\em What determines the masses of  individual stars?}  The size of the
initial reservoir of collapsing gas, determined by turbulent flow,
must be one element.  Subsequent accretion from the turbulent gas,
perhaps in competition with other stars, or even by collisions between
either protostellar cores or stars, could also be important, but must
be shown to occur, in particular in a magnetized medium.  The
properties of protostellar objects depend on the time history of this
accretion. Feedback from the newly formed star itself, or from its
neighbors, in the form of radiation pressure, ionizing radiation, or
stellar winds and jets, may yet prove to be another bounding term on
stellar mass.

{\em At what scales does the conservation of angular momentum and
magnetic flux fail?}  That they must fail is clear from the vast
discrepancy between galactic and stellar values.  Protostellar jets
almost certainly form when magnetic fields redistribute angular
momentum away from accreting gas. This demonstrates that the
conservation of flux and angular momentum must be coupled at least at
small scales.  However, the observational hint that molecular cloud
cores may be lacking substantial magnetic flux from the galactic value
suggests that magnetic flux may already be lost at rather large
scales.  Sweeping gas along field lines during the formation of
molecular clouds, magnetic reconnection processes, and ambipolar
diffusion in combination with turbulent transport offer possible
solutions that require further investigation. These processes allow
for the necessary compression of gas to higher densities while at the
same time increase the mass-to-flux ratio.  However, on scales of
individual stellar systems, the observed high fraction of binary stars
suggests that magnetic braking cannot be completely efficient at
draining angular momentum from collapsing protostars, and indicates
that ambipolar diffusion may limit the effectiveness of braking.

{\em What determines the initial conditions of stellar clusters?} The
spatial distribution of stars of different masses in a stellar cluster
or association, the initial velocity dispersion of its stars, and its
binary distribution probably all depend largely on the properties of
the turbulent flow from which it formed.  How much depends on the
details of the turbulence, and how much depends on the properties of
dynamically collapsing gas must still be determined.  The influence of
magnetic fields on these properties remains an almost unexplored
field, although their ability to redistribute angular momentum
suggests that they must play at least some role.

{\em What controls the distribution and metallicity of gas in
star-forming galaxies?}  At the largest scale, gas follows the
potential of a galaxy just as do all its other constituents.  The
dissipative nature of gas can allow it to quickly shed angular
momentum in disturbed potentials and fall to the centers of galaxies,
triggering starbursts.  Even in normal galaxies, gravitational
instability may determine the location of the largest concentrations
of gas available for star formation.  How important is turbulence in
determining the location and properties of molecular clouds formed
from that gas?  Are the molecular clouds destroyed again by the same
turbulent flow that created them, or do they decouple from the flow,
only to be destroyed by star formation within them?  How slowly do
turbulent flows mix chemical inhomogeneities, and can the scatter of
metallicities apparent in stars of apparently equal age be explained
by the process?

{\em Where and how fast do stars form in galaxies?}  The existence of
the empirical Schmidt Law relating gas column density to star
formation rate, probably with a threshold at low column density, still
needs to be definitively explained.  Can the threshold be caused by
a universal minimum level of turbulence, or by a minimum column
density below which it is difficult for gas to cool?  In either case,
examination of low-metallicity and dwarf galaxies may well provide
examples of objects sufficiently different from massive disk galaxies
in both cooling and rotation to demonstrate one or the other of these
possibilities.

{\em What determines the star formation efficiency of galaxies?}  The
relative importance of turbulence, rotation, gravitational
instability, and thermal instability remains unresolved.  At this
scale, turbulence can only play an instrumental role, transmitting the
influence of whatever drives it to the interstellar gas.  One
possibility is that galaxies are essentially self-regulated, with
Type~II supernovae from recent star formation determining the level of
turbulence, and thus the ongoing star formation rate.  Another
possibility is that a thermal or rotational bottleneck to star
formation exists, and that galaxies actually form stars just as fast
as they are able, more or less regardless of the strength of the
turbulence in most reasonable regimes.   Finding observational and
theoretical means to distinguish these scenarios represents the great
challenge of understanding the large-scale behavior of star formation
in galaxies.

%%% References
\rsksection*{BIBLIOGRAPHY}
\addcontentsline{toc}{chapter}{BIBLIOGRAPHY}
\label{sec:bib}
%\begin{thebibliography}{}
{
\setlength{\parskip}{0.05cm}
\setlength{\itemsep}{0.05cm}
\footnotesize
\begin{list}{}{\itemindent-0.6cm \leftmargin0.6cm}

\newcommand{\rskbibitem}{\item}

%Journal definitions
\def\araa{{\em Ann.\ Rev.\ Astron.\ Astrophys.}}
\def\aas{{\em Astron.\ Astrophys.\ Suppl.\ Ser.}}
\def\aj{{\em Astron.\ J.}}
\def\anap{{\em Ann.\ Astrophys.}}   %{\em Annales d'Astrophysique}
\def\an{{\em Astron. Nach.}}
\def\apj{{\em Astrophys.\ J.}}
\def\apjl{{\em Astrophys.\ J.\ (Letters)}}
\def\apjs{{\em Astrophys.\ J.\ Suppl.\ Ser.}}
\def\aap{{\em Astron.\ Astrophys.}}
\def\apss{{\em Astrophys.\ Space Science}}
\def\baas{{\em Bull.\ Amer.\ Astron.\ Soc.}}
\def\bain{{\em Bull.\ Astron.\ Inst.\ Netherlands}}
\def\fcp{{\em Fund.\ Cosm.\ Phys.}}
\def\jcam{{\em J.\ Comput.\ Appl.\ Math.}}
\def\jcp{{\em J.\ Comput.\ Phys.}}
\def\jfm{{\em J.\ Fluid Mech.}}
\def\jt{{\em J.\ Turbulence}}
\def\mnras{{\em Mon.\ Not.\ R.\ Astron.\ Soc.}}
\def\nat{{\em Nature}}
\def\pd{{\em Physica D}}
\def\pla{{\em Phys.\ Lett. A}}
\def\ppl{{\em Phys.\ Plasmas}}
\def\pta{{\em Phil.\ Trans.\ A.}}
\def\prp{{\em Phys.\ Report}}
\def\ptp{{\em Prog.\ Theo.\ Phys.}}
\def\prd{{\em Phys.\ Rev.\ D}}
\def\pre{{\em Phys.\ Rev.\ E}}
\def\prl{{\em Phys.\ Rev.\ Lett.}}
\def\prsa{{\em Proc.\ R.\ Soc.\ London A}}
\def\pasj{{\em Pub.\ Astron.\ Soc.\ Japan}}
\def\pasp{{\em Pub.\ Astron.\ Soc.\ Pacific}}
\def\pfl{{\em Phys.\ Fluids}}
\def\rmp{{\em Rev.\ Mod.\ Phys.}}
\def\zp{{\em Z.\ Phys.}}
\def\za{{\em Z.\ Astrophys.}}

%%% References
%\begin{references}
%\begin{thebibliography}{}

\rskbibitem{} Abbott, D. C., 1982, \apj, {\bf 263}, 723 

\rskbibitem{} Adams, F.\ C., and M.\ Fatuzzo, 1996, \apj, {\bf 464}, 256

\rskbibitem{} Adams, F.\ C., and P.\ C.\ Myers, 2001, \apj, {\bf 553}, 744

\rskbibitem{} Adams, F.\ C., and J.\ J.\ Wiseman, 1994, \apj, {\bf 435}, 693

\rskbibitem{} Aikawa, Y., N.\ Ohashi, S.\ Inutsuka, E.\ Herbst, and S.\ Takakuwa,   2001,  \apj,  {\bf 552}, 639 

\rskbibitem{} Alves, J.\ F., C.\ J.\ Lada, and E.\ A.\ Lada,  1999,  \apj,   {\bf
515}, 265 

\rskbibitem{} Alves, J.\ F., C.\ J.\ Lada, and E.\ A.\ Lada, 2001,
\nat, {\bf 409}, 159

\rskbibitem{} Andr{\'e}, P.,\ and T.\ Montmerle,  1994,  \apj,  {\bf 420}, 837 

\rskbibitem{} Andr{\'e}, P., D.\ Ward-Thompson, and M.\ Barsony, 2000, in
{\em Protostars and Planets IV}, edited by V.\ Mannings, A.\ P.\ Boss,
and S.\ S.\ Russell (University of Arizona Press, Tucson), p.\ 59

\rskbibitem{} Andr{\'e}, P., D.\ Ward-Thompson, and F.\ Motte,  1996,
\aap,  {\bf  314}, 625  

\rskbibitem{} Avillez, M.\ A.,  2000,  \mnras,  {\bf
    315}, 479 

\rskbibitem{} Arons, J., and C.\ E.\ Max, 1975, \apjl, {\bf 196}, L77

\rskbibitem{} Bacmann, A., P.\ Andr{\'e}, J.\ -.L\ Puget, A.\ Abergel,
S.\  Bontemps, and D.\ Ward-Thompson,  2000,  \aap,  {\bf 361}, 555  

\rskbibitem{} Balbus, S.\ A., and J.\ F.\ Hawley, 1991, \apj, {\bf 376}, 214 

\rskbibitem{} Balbus, S.\ A., and J.\ F.\ Hawley, 1998, \rmp, {\bf 70}, 1

\rskbibitem{} Baldry, I.\ K.\ et al.\  2002, \apj, {\bf 569}, 582 

\rskbibitem{} Balk, A.\ M., 2001, \pla, {\bf 279}, 370

\rskbibitem{} Balkovsky, E., and G.\ Falkovich, 1998, \pre, {\bf 57}, R1231 

\rskbibitem{} Balkovsky, E., G.\ Falkovich, I.\ Kolokolov, V.\ Lebedev, 1997,
\prl, {\bf 78}, 1452  

\rskbibitem{} Ballesteros-Paredes, J.,
  and M.-M.\ Mac Low, 2002, \apj, {\bf  570},  734

\rskbibitem{} Ballesteros-Paredes, J., L.\ Hartmann, and E.\ V{\'
a}zquez-Semadeni, 1999a, \apj, {\bf 527}, 285 

\rskbibitem{} Ballesteros-Paredes, J., R.\ S.\ Klessen, and E.\ V{\'
a}zquez-Semadeni, 2003, \apj, in press  (astro-ph/0301546)

\rskbibitem{} Ballesteros-Paredes, J., E.\ V{\'a}zquez-Semadeni, and A.\ A.\
Goodman, 2002, \apj {\bf 571}, 334 

\rskbibitem{} Ballesteros-Paredes, J., E.\ V{\'a}zquez-Semadeni, and J.\ Scalo,
1999b,  \apj,  {\bf 515}, 286  

\rskbibitem{} Bally, J., and D.\ Devine, 1994, \apj, {\bf 428}, L65 

\rskbibitem{} Bally, J., D.\ Devine, and B.\ Reipurth, 1996, \apj, {\bf 473}, L49

\rskbibitem{} Balsara, D. S., 1996, \apj, {\bf 465}, 775

\rskbibitem{} Balsara, D., and A.\ Pouquet, 1999, \ppl, {\bf 6}, 89 

\rskbibitem{} Balsara, D. S., D.\ Ward-Thompson, and R.\ M.\ Crutcher, 2001, \mnras, {\bf 327}, 715

\rskbibitem{} Balsara, D. S., R.\ M.\ Crutcher, and A.\ Pouquet, 2001, \apj, {\bf 557}, 451

\rskbibitem{} Barrado y Navascu{\' e}s, D., J.\ R.\ Stauffer, J.\ Bouvier, and  E.\ L.\ Mart{\' i}n, 2001, \apj, {\bf 546}, 1006
  
\rskbibitem{} Baschek, B., M.\ Scholz, and R.\ Wehrse, 1991, \aap, {\bf 246}, 374

\rskbibitem{} Bastien, P., J.\ Arcoragi, W.\ Benz, I.\ A.\ Bonnell, and H.\  Martel,  1991,  \apj,  {\bf 378}, 255 

\rskbibitem{} Basu, S., 1997, \apj, {\bf 485}, 240

\rskbibitem{} Basu, S.\ and T.\ C.\ Mouschovias,  1994,  \apj,  {\bf 432}, 720 

\rskbibitem{} Basu, S.\ and T.\ C.\ Mouschovias,  1995a,  \apj,  {\bf 452}, 386 

\rskbibitem{} Basu, S.\ and T.\ C.\ Mouschovias,  1995b,  \apj,  {\bf 453}, 271 

\rskbibitem{} Batchelor, G.\ K., 1949, {Austral.\ J.\ Sci.\
  Res.}, {\em 2}, 437
 
\rskbibitem{} Bate, M.\ R., 2000, \mnras, {\bf 314}, 33 

\rskbibitem{} Bate, M.\ R., and A.\ Burkert,  1997,  \mnras,  {\bf 288}, 1060 

\rskbibitem{} Bate, M.\ R., C.\ J.\ Clarke, and M.\ J.\ McCaughrean, 1998, \mnras, {\bf 297}, 1163

\rskbibitem{} Bate, M.\ R., I.\ A.\ Bonnell, and V.\ Bromm, 2002, \mnras, {\bf 332}, L65

\rskbibitem{} Bate, M.\ R., I.\ A.\ Bonnell, and N.\ M.\ Price, 1995, \mnras, {\bf 277}, 362

\rskbibitem{} Baureis, P., R.\ Ebert, and F.\ Schmitz,  1989,  \aap,  {\bf 225},  405 

\rskbibitem{} Beckwith, S.\ V.\ W., 1999, in {\em NATO ASIC Proc.\ 540: The Origin of Stars  and Planetary Systems},  edited by C.\ J.\ Lada and N.\ D.\ Kylafis (Kluwer Academic Publishers), p.\ 579

\rskbibitem{} Beichman, C.\ A., P.\ C.\ Myers, J.\ P.\ Emerson, S.\ Harris,  R.\ Mathieu, P.\ J.\ Benson, and R.\ E.\ Jennings,  1986,  \apj,  {\bf 307}, 337 

\rskbibitem{} Bensch F., J.\ Stutzki,  and V.\ Ossenkopf, 2001, \aap {\bf
336}, 636

\rskbibitem{} Benson, P.\ J., and P.\ C.\ Myers, 1989, \apjs, {\bf 71}, 89

\rskbibitem{} Benz, W., 1990, in {\em The Numerical Modelling of Nonlinear
Stellar Pulsations}, edited by J. R. Buchler (Kluwer, Dordrecht), 269

\rskbibitem{} Bergin, E.\ A., and W.\ D.\ Langer,  1997,  \apj,  {\bf 486}, 316 

\rskbibitem{} Bergin, E.\ A., P.\ F.\ Goldsmith, R.\ L.\ Snell, and W.\  D.\
Langer, 1997a, \apj, {\bf 428}, 285 

\rskbibitem{} Bergin, E.\ A., H.\ Ungerechts, P.\ F.\ Goldsmith,  R.\ L.\ Snell,
W.\ M.\ Irvine, and F.\ P.\ Schloerb, 1997b, \apj,   {\bf 482}, 267  

\rskbibitem{} Bertoldi, F., and C.\ C.\ McKee, 1996, in {\em {Amazing   Light: A Volume Dedicated to C.\ H.\ Townes on his 80th Birthday}},   edited by R.\ Chiao (Springer, New York), p.\ 41

\rskbibitem{} Bertout, C.,  1989,  \araa,  {\bf 27}, 351 

\rskbibitem{} Bertschinger, E., 1998, \araa, {\bf 36}, 599 

\rskbibitem{} Binney, J., and S.\ Tremaine, 1987, {\em Galactic Dynamics}
(Princeton University Press, Princeton)

\rskbibitem{} Biskamp, D., W.\ C.\ M{\" u}ller,
  1999, \prl, {\bf 83}, 2195  

\rskbibitem{} Biskamp, D., and W.-C.\ M\"uller,
  2000, \ppl, {\bf 7}, 4889 

\rskbibitem{} Black, D.\ C., and P.\ Bodenheimer, 1976, \apj, {\bf 206}, 138

\rskbibitem{} Blitz, L., 1993,  in {\em Protostars and Planets III}, edited by E.\ H.\ Levy and J.\ I.\ Lunine (University of Arizona Press, Tucson), p.\ 125

\rskbibitem{} Blitz, L.,\ and F.\ H.\ Shu, 1980, \apj, {\bf 238}, 148

\rskbibitem{} Bodenheimer, P.,  1995,  \araa,  {\bf 33}, 199 

\rskbibitem{} Bodenheimer, P., and A.\ Sweigart, 1969, \apj, {\bf 152}, 515

\rskbibitem{} Bodenheimer, P., and W.\ Tscharnuter,  1979,  \aap,  {\bf 74},  288 

\rskbibitem{} Bodenheimer, P., A.\ Burkert, R.\ I.\ Klein, and A.\ P.\
Boss,   2000,  in {\em Protostars and Planets IV},  edited by V.\
Mannings, A.\ P.\ Boss, and S.\ S.\ Russell (University of Arizona
Press, Tucson), p.\ 327  

\rskbibitem{} Boldyrev, S., 2002, \apj, {\bf 569}, 841

\rskbibitem{} Boldyrev, S., \AA.\ Nordlund, and P.\ Padoan, 2002a, \apj, {\bf 573}, 678 

\rskbibitem{} Boldyrev, S., \AA.\ Nordlund, and P.\ Padoan, 2002b, \prl, submitted (astro-ph/0203452)

\rskbibitem{} Bonazzola, S., E.\ Falgarone, J.\ Heyvaerts, M.\  Perault, and J.\ L.\ Puget, 1987, \aap, {\bf 172}, 293

\rskbibitem{} Bonazzola, S., M.\ Perault, J.\ L.\ Puget, J.\ Heyvaerts,  E.\ Falgarone, J.\ F.\ Panis, 1992, \jfm, {\bf 245}, 1

\rskbibitem{} Bonnell, I.\ A., and M.\ R.\ Bate, 2002, \mnras, in press

\rskbibitem{} Bonnell, I.\ A., and M.\ B.\ Davies, 1998, \mnras, {\bf 295}, 691.

\rskbibitem{} Bonnell, I.\ A., M.\ R.\ Bate, and H.\ Zinnecker, 1998, \mnras, {\bf 298}, 93

\rskbibitem{} Bonnell, I.\ A., M.\ R.\ Bate, C.\ J.\ Clarke, and J.\ E.\  Pringle,  1997,  \mnras,  {\bf 285}, 201 

\rskbibitem{} Bonnell, I.\ A., M.\ R.\ Bate, C.\ J.\ Clarke, and J.\ E.\ Pringle, 2001a, \mnras, {\bf 323}, 785

\rskbibitem{} Bonnell, I.\ A., C.\ J.\ Clarke, M.\ R.\ Bate, and J.\ E.\ Pringle, 2001b, \mnras, {\bf 324}, 573 

\rskbibitem{} Bonnell, I.\ A., K.\ W.\ Smith, M.\ B.\ Davies, and K.\ Horne, 2001c, \mnras, {\bf 322}, 859

\rskbibitem{} Bonnor, W.\ B., 1956, \mnras, {\bf 116}, 351

\rskbibitem{} Bontemps, S., P.\ Andr{\'e}, S.\ Terebey, and S.\ Cabrit, 1996, \aap, {\bf 311}, 858

\rskbibitem{} Bontemps, S.\ and 21 colleagues, 2001, \aap, {\bf 372}, 173
  
\rskbibitem{} Boratav, O., A.\ Eden, A., and A.\ Erzan (eds.), 1997, {\em
    Turbulence Modeling and Vortex Dynamics} (Springer Verlag, Heidelberg)

\rskbibitem{} Boss, A.\ P.,  1980a,  \apj,  {\bf 237}, 563 

\rskbibitem{} Boss, A.\ P.,  1980b,  \apj,  {\bf 237}, 866 

\rskbibitem{} Boss, A.\ P.,  1996,  \apj,  {\bf 468}, 231

\rskbibitem{} Boss, A. P., 2000, \apjl, {\bf 545}, 61

\rskbibitem{} Boss, A. P., 2002, \apj, {\bf 568}, 743

\rskbibitem{} Boss, A.\ P.\ and P.\ Bodenheimer,  1979,  \apj,  {\bf 234}, 289 

\rskbibitem{} Boss, A.\ P., and L.\ Hartmann, 2002, \apj, {\bf 562}, 842

\rskbibitem{} Boss, A.\ P., and E.\ A.\ Myhill, 1995, \apj, {\bf 451}, 218

\rskbibitem{} Boss, A.\ P., R.\ T.\ Fisher, R.\ I.\ Klein, and C.\ F.\ McKee,
2000,  \apj,  {\bf 528}, 325  

\rskbibitem{} Bourke, T.\ L., P.\ C.\ Myers, G.\ Robinson, and A.\ R.\ Hyland,
2001,  \apj,  {\bf 554}, 916  

\rskbibitem{} Brandl, B., W.\ Brandner, F.\ Eisenhauer, A.\ F.\ J.\ Moffat, F.\
Palla, and H.\ Zinnecker, 1999, \aap, {\bf 352}, L69 

\rskbibitem{} Bronfman, L., S. Casassus, J. May, and L.-\AA. Nyman, 2000,
\aap, {\bf 358}, 521

\rskbibitem{} Bronstein, I.\ N., and K.\ A.\ Semendjajew, 1979, {\em Taschenbuch
    der Mathematik} (Teubner Verlagsgesellschaft, Leibzig)

\rskbibitem{} Burkert, A., M.\ R.\ Bate, and P.\ Bodenheimer,  1997,  \mnras,
{\bf 289}, 497  

\rskbibitem{} Burkert, A., and P.\ Bodenheimer,  1993,  \mnras,  {\bf 264},  798 

\rskbibitem{} Burkert, A., and P.\ Bodenheimer,  1996,  \mnras,  {\bf 280},  1190 

\rskbibitem{} Burkert, A.\ and D.\ N.\ C.\ Lin,  2000,  \apj,  {\bf 537}, 270 

\rskbibitem{} Burkert, A., M.\ R.\ Bate, and P.\ Bodenheimer,  1997, \mnras,   {\bf 289}, 497 

\rskbibitem{} Burrows, A., W.\ B.\ Hubbard, J.\ I.\ Lunine, and J.\ Liebert, 2001, {\em Rev.\ Mod.\ Phys.}, {\bf 73}, 719

\rskbibitem{} Burrows, A., W.\ B.\ Hubbard, D.\ Saumon, and J.\ I.\ Lunine, 1993, \apj, {\bf 406}, 158 

\rskbibitem{} Butner, H.\ M., E.\ A.\ Lada, and R.\ B.\ Loren, 1995, \apj, {\bf 448}, 207 

\rskbibitem{} Cabrit, S., and C.\ Bertout,  1992,  \aap,  {\bf 261}, 274 
  
\rskbibitem{} Cao, N., S.\ Chen, and Z.-S.\ She, 1996, \prl, {\bf 76}, 3711 

\rskbibitem{} Cambr\'esy, L., C. A. Beichman, T. H. Jarrett, and R. M. Cutri,
2002, \aj, {\bf 123}, 2559

\rskbibitem{} Cappellaro, E., R. Evans, and M. Turatto, 1999, \aap, {\bf
351}, 459

\rskbibitem{} Carlberg, R.\ G., and R.\ E.\ Pudritz, 1990, \apj, {\bf 247}, 353

\rskbibitem{} Carpenter, J.\ M., M.\ R.\ Meyer, C.\ Dougados, S.\ E.\   Strom, and L.\ A.\ Hillenbrand, 1997, \aj, {\bf 114}, 198

\rskbibitem{} Carr, J.\ S.,  1987, \apj, {\bf 323}, 170

\rskbibitem{} Caselli, P.\ and P.\ C.\ Myers, 1995, \apj, {\bf 446}, 665
  
\rskbibitem{} Castiglione, P., A.\ Mazzino, P.\ Muratore-Ginanneschi,
  and A.\ Vulpiani, 1999, \pd, {\bf 134}, 75

\rskbibitem{} Castor, J.\ I., 1972, \apj, {\bf 178}, 779
  
\rskbibitem{} Caughlan, G.\ R., and W.\ A.\ Fowler, 1988, {\em Atomic Data and
    Nuclear Data Tables}, {\bf 40}, 283

\rskbibitem{} Cernicharo, J., 1991, in {\em The {Physics} {of}   {Star}{Formation} and {Early} {Stellar} {Evolution} }, edited by C.\   J.\ Lada and N.\ Kylafis (Kluwer Academic Publishers, Dordrecht),   p.\ 287 

\rskbibitem{} Chabrier, G., 2001, \apj, {\bf 554}, 1274

\rskbibitem{} Chabrier, G., 2002, \apj, {\bf 567}, 304

\rskbibitem{} Chandrasekhar, S., 1949, \apj, {\bf 110}, 329

\rskbibitem{} Chandrasekhar, S., 1951a, \prsa, {\bf 210}, 18

\rskbibitem{} Chandrasekhar, S., 1951b, \prsa, {\bf 210}, 26 

\rskbibitem{} Chandrasekhar, S., 1953, \apj, {\bf 118}, 116

\rskbibitem{} Chandrasekhar, S., 1954, \apj, {\bf 119}, 7

\rskbibitem{} Chandrasekhar, S., and E.\ Fermi, 1953a, \apj, {\bf 118}, 113

\rskbibitem{} Chandrasekhar, S., and E.\ Fermi, 1953b, \apj, {\bf 118}, 116

\rskbibitem{} Chapman, S., H.\ Pongracic, M.\ Disney, A.\ Nelson, J.\
Turner,  and A.\ Whitworth,  1992,  \nat,  {\bf 359}, 207  

\rskbibitem{} Chen, H., P.\ C.\ Myers, E.\ F.\ Ladd, and D.\ O.\ S.\
Wood,  1995,   \apj,  {\bf 445}, 377  

\rskbibitem{} Chertkov, M., I.\ Kolokolov, and M.\ Vergassola, 1997, \pre, {\bf
    56}, 5483
  
\rskbibitem{} Cho, J., and A.\ Lazarian, 2002, in {Acoustic emission
    and scattering by turbulent flows}, edited by M.\ Rast (Springer
  Verlag, Heidelberg) (astro-ph/0301462)

\rskbibitem{} Cho, J., A.\ Lazarian, and E.\ T.\ Vishniac, 2002, \apj,
{\bf 566}, L49

\rskbibitem{} Chu, Y.-H., N.\ B.\ Suntzeff, J.\ E.\ Hesser, and D.\ A.\
Bohlender, 1999, {\em New Views of the Magellanic Clouds} (Proceedings
of IAU Symposium 190, ASP Conference Series, Vol.\ 190) 

\rskbibitem{} Ciolek, G.\ E., and S.\ Basu, 2000, \apj, {\bf 529}, 925

\rskbibitem{} Ciolek, G.\ E., and S.\ Basu, 2001, \apj, {\bf 547}, 272

\rskbibitem{} Ciolek, G.\ E.\ and T.\ C.\ Mouschovias,  1993,  \apj,
{\bf 418},  774  

\rskbibitem{} Ciolek, G.\ E.\ and T.\ C.\ Mouschovias,  1994,  \apj,
{\bf 425},  142  

\rskbibitem{} Ciolek, G.\ E.\ and T.\ C.\ Mouschovias,  1995,  \apj,
{\bf 454},  194  

\rskbibitem{} Ciolek, G.\ E.\ and T.\ C.\ Mouschovias,  1996,  \apj,
{\bf 468},  749  

\rskbibitem{} Ciolek, G.\ E.\ and T.\ C.\ Mouschovias,  1998,  \apj,
{\bf 504},  280  

\rskbibitem{} Clarke, C.\ J., and  J.\ E.\ Pringle, 1991, \mnras, {\bf 249}, 584

\rskbibitem{} Clarke, C.\ J., I.\ A.\ Bonnell, and L.\ A.\ Hillenbrand,
2000, in {\em Protostars and Planets IV}, edited by V.\ Mannings,
A.\ P.\ Boss, and  S.\ S.\ Russell (University of Arizona Press,
Tucson), p.\ 151  

\rskbibitem{} Clarke, D. 1994, National Center for Supercomputing
  Applications Technical Report

\rskbibitem{} Clemens, D.\ P., 1985, \apj, {\bf 295}, 422

\rskbibitem{} Coleman, G., J.\ Kim, and R.\ D.\ Moser,
  1995, \jfm, {\bf 305}, 159 

\rskbibitem{} Combes, F., 2001, in {\em The Central Kiloparsec of
Starbursts and AGN}, edited by J.\ H.\ Knapen, J.\ E.\ Beckman, I.\
Shlosman, and T.\ J.\ Mahoney (ASP Conference Series 249), p.\ 475.

\rskbibitem{} Cook, T.\ L., and F.\ H.\ Harlow,  1978,  \apj,  {\bf 225}, 1005 

\rskbibitem{} Crutcher, R.\ M.,  1999,  \apj,  {\bf 520}, 706 

\rskbibitem{} Crutcher, R.\ M.\ and T.\ H.\ Troland,  2000,  \apj,  {\bf
537},  L139  

\rskbibitem{} Curry, C.\ L., 2002, \apj, {\bf 576}, 849

\rskbibitem{} Curry C.\ L., and C.\ F.\ McKee, 2000, \apj,  {\bf 528}, 734 

\rskbibitem{} Dalgarno, A.\ and R.\ A.\ McCray,  1972,  \araa,  {\bf 10}, 375 

\rskbibitem{} Dame, T.\ M., B.\ G.\ Elmegreen,  R.\ S.\ Cohen, and P.\
Thaddeus, 1986, \apj, {\bf 305}, 892 

\rskbibitem{} D'Antona, F., and I. Mazzitelli, 1994, \apjs, {\bf 90}, 467

\rskbibitem{} Davies, R.\ D.\ and W.\ L.\ H.\ Shuter,  1963,  \mnras,  {\bf 126},  369 

\rskbibitem{} de Blok, W. J. G., and S. S. McGaugh, 1996, \apjl, {\bf 469}, L89

\rskbibitem{} Dehnen, W., and J.\ J.\ Binney, 1998, \mnras, {\bf 294}, 429

\rskbibitem{} Deiss, B.\ M., A.\ Just, and  W.\ H.\ Kegel, 1990, \aap,
{\bf 240}, 123  

\rskbibitem{} Desch, S.\ J.\ and T.\ C.\ Mouschovias,  2001,  \apj,  {\bf
550},  314  

\rskbibitem{} de\ Vega, H.\ J., and N.\ S{\'a}nchez, 2000, {\em Phys.\
Lett.\ B}, {\bf 490}, 180 

\rskbibitem{} de\ Vega, H.\ J., and N.\ S{\'a}nchez, 2001a, (astro-ph/0101568)

\rskbibitem{} de\ Vega, H.\ J., and N.\ S{\'a}nchez, 2001a, (astro-ph/0101568)

\rskbibitem{} de\ Vega, H.\ J., N.\ S{\'a}nchez, and F.\ Combes, 1996a,
\nat, {\bf 383}, 56 

\rskbibitem{} de\ Vega, H.\ J., N.\ S{\'a}nchez, and F.\ Combes, 1996b,
\prd, {\bf 54}, 6008 

\rskbibitem{} Dewar, R.\ L., 1970, \pfl, {\bf 13}, 2710

\rskbibitem{} Diaz-Miller, R.~I., J.~Franco, and S.~N.~Shore, 1998, \apj,
{\bf 501}, 192.

\rskbibitem{} Dickey, J. M., M. M. Hanson, and G. Helou, 1990, \apj, {\bf
352}, 522

\rskbibitem{} Dickey, J. M., and F. J. Lockman, 1990, \araa, {\bf 28}, 215

\rskbibitem{} Dickman, R.\ L., R.\ L.\ Snell, and F.\ P.\ Schloerb, 1986,
\apj, {\bf 309}, 326  

\rskbibitem{} Domolevo, K., and L.\
  Sainsaulieu, 1997,   \jcp, {\bf 133}, 256

\rskbibitem{} Draine, B.\ T., 1980, \apj, {\bf 241}, 1021
  
\rskbibitem{} Dubinski, J., R.\ Narayan, and T.\ G.\ Phillips, 1995, \apj, {\bf
    448}, 226

\rskbibitem{} Duquennoy, A., and M.\ Mayor,  1991,  \aap,  {\bf 248}, 485 

\rskbibitem{} Durisen, R.\ H., M.\ F.\ Sterzik, and B.\ K.\ Pickett,
2001, \aap< {\bf 371}, 952 

\rskbibitem{} Ebert, R., 1955, \za, {\bf 36}, 222

\rskbibitem{} Ebert, R., 1957, \za,  {\bf 42}, 263 

\rskbibitem{} Ebert, R., S. von Hoerner, and St. Temesv\'ary, 1960, in
Die Entstehung von Sternen durch Kondensation diffuser Materie,
authored by G. R. Burbidge, F. D. Kahn, R. Ebert, S. von Hoerner, and
St. Temesv\'ary (Springer Verlag, Berlin), 184

\rskbibitem{} Ebisuzaki T., Makino J., Fukushige
  T., Taiji M., Sugimoto D., Ito T., Okumura S.K. 1993, \pasj 45, 269

\rskbibitem{} Efremov, Y.\ N., and B.\ G.\ Elmegreen,  1998,  \mnras,
{\bf 299},  588  

\rskbibitem{} Elmegreen, B.\ G., 1979, \apj, {\bf 232}, 729

\rskbibitem{} Elmegreen, B.\ G., 1985, \apj, {\bf 299}, 196

\rskbibitem{} Elmegreen, B.\ G., 1991,  in {\em NATO ASIC Proc.\ 342: The
Physics of Star Formation and Early Stellar Evolution},  edited by C.\
J.\ Lada and N.\ D.\ Kylafis (Kluwer Academic Publishers), p.\ 35

\rskbibitem{} Elmegreen, B.\ G., 1990, \apj, {\bf 361}, L77 

\rskbibitem{} Elmegreen, B.\ G., 1993, \apj, {\bf 419}, L29

\rskbibitem{} Elmegreen, B.\ G., 1995, \mnras, {\bf 275}, 944

\rskbibitem{} Elmegreen, B.\ G., 1997a, \apj, {\bf 480}, 674. 

\rskbibitem{} Elmegreen, B.\ G., 1997b, \apj, {\bf 486}, 944 

\rskbibitem{} Elmegreen, B.\ G., 1999a, \apj, {\bf 515}, 323

\rskbibitem{} Elmegreen, B.\ G., 1999b, \apj, {\bf 527}, 266

\rskbibitem{} Elmegreen, B.\ G., 2000a, \mnras, {\bf 311}, L5

\rskbibitem{} Elmegreen, B.\ G., 2000b, \apj, {\bf 530}, 277  

\rskbibitem{} Elmegreen, B.\ G., 2000c, \apj, {\bf 539}, 342

\rskbibitem{} Elmegreen, B.\ G., 2002a, \apj, {\bf 564}, 773

\rskbibitem{} Elmegreen, B.\ G., 2002b, \apj, {\bf 577}, in press (astro-ph/02027114) 

\rskbibitem{} Elmegreen, B.\ G., and E.\ Falgarone, 1996, \apj, {\bf 471}, 816

\rskbibitem{} Elmegreen, B.\ G., and R.\ D.\ Mathieu, 1983, \mnras, {\bf 203}, 305

\rskbibitem{} Elmegreen, B.\ G., and F.\ Combes,  1992,  \aap,  {\bf 259}, 232 

\rskbibitem{} Elmegreen, B.\ G., C.\ J.\ Lada, and D.\ F.\ Dickinson, 1979 \apj,
{\bf 230}, 415 

\rskbibitem{} Elmegreen, B.\ G., Y.\ Efremov, R.\ E.\ Pudritz, and H.\ Zinnecker,
2000, in {\em Protostars and Planets IV},  edited by V.\ Mannings, A.\ P.\
Boss, and S.\ S.\ Russell (University of Arizona Press, Tucson), p.\ 179 

\rskbibitem{} Falgarone, E., and T.\ G.\ Phillips, 1996,  {\em \apj}, {\bf 472},
191  

\rskbibitem{} Falgarone, E., J.\ L.\ {Puget}, and M.\ {Perault}, 1992, \aap, {\bf 257}, 715 
  
\rskbibitem{} Falgarone, E., D.\ C.\ Lis, T.\ G.\  Phillips, A.\ Pouquet, D.\ H.\
  Porter,  and P.\ R.\ Woodward,  1994, \apj, {\bf 436},728

\rskbibitem{} Falgarone, E.,  J.-F.\ Panis, A.\ Heithausen, M.\ P\'{e}rault,
  J.\ Stutzki, J.-L.\ Puget, and F.\ Bensch, 1998, \aap, {\bf 331}, 669 

\rskbibitem{} Falkovich, G., and V.\ Lebedev, 1997, \prl, {\bf 79}, 4159

\rskbibitem{} Feigelson, E. D., 1996, \apj, {\bf 468}, 306

\rskbibitem{} Fiedler, R.\ A.\ and T.\ C.\ Mouschovias,  1992,  \apj,  {\bf 391},
199  

\rskbibitem{} Fiedler, R.\ A.\ and T.\ C.\ Mouschovias,  1993,  \apj,  {\bf 415},
680  

\rskbibitem{} Fiege, J.\ D., and R.\ E.\ Pudritz, 1999, in {\em New
                  Perspectives on the Interstellar Medium}, edited by
                  A.\ R.\ Taylor, T.\ L.\ Landecker, and G.\ Joncas
                  (ASP Conference Series 168), p.\ 248

\rskbibitem{} Fiege, J.\ D., and R.\ E.\ Pudritz, 2000a, \mnras, {\bf 311}, 85

\rskbibitem{} Fiege, J.\ D., and R.\ E.\ Pudritz, 2000b, \mnras, {\bf 311}, 105 

\rskbibitem{} Field, G.\ B., 1965, \apj, {\bf 142}, 531

\rskbibitem{} Field, G.\ B., 1978, in {\em Protostars and Planets}, edited by T.\ Gehrels (University of Arizona Press, Tucson), p.\ 243 

\rskbibitem{} Field, G.\ B., D.\ W.\ Goldsmith, and H.\ J.\ Habing, 1969,
\apj, {\bf 155}, L49 

\rskbibitem{} Fischer, D.\ A., and G.\ W.\ Marcy,  1992,  \apj,  {\bf 396},  178 

\rskbibitem{} Fleck, R.\ C.,  1981, \apj, {\bf 246}, L151

\rskbibitem{} Fleck, R.\ C., 1982, \mnras, {\bf 201}, 551

\rskbibitem{} Forbes, D.,  1996,  \aj,  {\bf 112}, 1073 

\rskbibitem{} Franco, J., and A.\ Carraminana (eds.), 1999, {\em Interstellar
Turbulence} (Cambridge University Press)

\rskbibitem{} Foster, P.\ N., and R.\ A.\ Chevalier, 1993, \apj, {\bf 416}, 303

\rskbibitem{} Franco, J., S.~N.~Shore, and G.~Tenorio-Tagle, 1994, \apj,
{\bf 436}, 795

\rskbibitem{} Fricke, K.\ J., C.\ Moellenhoff, and W.\ Tscharnuter,  1976,   \aap,  {\bf 47}, 407 

\rskbibitem{} Frisch, U., 1995, {\em Turbulence -- The Legacy of A.\ N.\
Kolmogorov} (Cambridge University Press)

\rskbibitem{} Fukui, Y.\ {\em et al.}, 1999, \pasj, {\bf 51}, 745 

\rskbibitem{} Fuller, G.\ A., and P.\ C.\ Myers, 1992, \apj, {\bf 384}, 523

\rskbibitem{} Galli, D., F.\ H.\ Shu, G.\ Laughlin, and S.\ Lizano,  2001,  \apj,   {\bf 551}, 367 

\rskbibitem{} Galli, D., S.\ Lizano, Z.\ Y.\ Li, F.\ C.\ Adams, and F.\ H.\ Shu,   1999,  \apj,  {\bf 521}, 630 

\rskbibitem{} Galli, D.\ and F.\ H.\ Shu,  1993a,  \apj,  {\bf 417}, 220 

\rskbibitem{} Galli, D.\ and F.\ H.\ Shu,  1993b,  \apj,  {\bf 417}, 243 

\rskbibitem{} Gammie, C.\ F., and E.\ C.\ Ostriker, 1996, \apj, {\bf 466}, 814

\rskbibitem{} Gazol, A., E.\ V\'azquez-Semadeni, F.\ J.\ S\'anchez-Salcedo, and
J.\ Scalo, 2001, \apj, {\bf 557}, L121 

\rskbibitem{} Genzel, R., 1991, in {\em { The {Physics} {of}   {Star}{Formation}
and {Early} {Stellar} {Evolution} }}, edited by   C.\ J.\ Lada and N.\ D.\
Kylafis (Kluwer Academic Publishers, Dordrecht),   p.\ 155 

\rskbibitem{} Gilden, D.\ L, 1984a, \apj, {\bf 279}, 335

\rskbibitem{} Gilden, D.\ L, 1984b, \apj, {\bf 283}, 679

\rskbibitem{} Gill, A.\ G., and R.\ N.\ Henriksen,  1990, \apj, {\bf 365}, L27

\rskbibitem{} Giovanardi, C., L. F. Rodr\'iguez, S. Lizano, and J. Cant\'o,
2000, \apj, {\bf 538}, 728

\rskbibitem{} Gladwin, P.\ P., S.\ Kitsionas, H.\ M.\ J.\ Boffin, and A.\
P.\ Whitworth, 1999, \mnras, {\bf 302}, 305 

\rskbibitem{} Goldreich, P.\ and S.\ Sridhar, 1995, \apj, {\bf 438}, 763

\rskbibitem{} Goldreich, P.\ and S.\ Sridhar, 1997, \apj, {\bf 485}, 680

\rskbibitem{} Goldsmith, P., 2001, \apj, {\bf 557}, 736

\rskbibitem{} Goldsmith, P.\ F., and W.\ D.\ Langer, 1978, \apj,  {\bf 222}, 881 

\rskbibitem{} G\'omez, G. C., and D. P. Cox, 2002, \apj {\bf 580}, 235

\rskbibitem{} G\'omez, M., B.\ F.\ Jones, L.\ Hartmann, S.\ J.\ Kenyon, J.\
R.\ Stauffer, R.\ Hewett, and I.\ N.\ Reid, 1992, \aj, {\bf 104}, 762 

\rskbibitem{} Gomez, T., Politano, H., Pouquet, A.,  
Larchev{\^ e}que, M., 2001, \pfl, {\bf 13}, 2065 

\rskbibitem{} Graziani, F., and D.\ C.\ Black, 1981, \apj, {\bf 251}, 337

\rskbibitem{} Greene, T.\ P., and M.\ R.\ Meyer, 1995, \apj, {\bf 450}, 233 

\rskbibitem{} Gueth, F., S.\ Guilloteau, A.\ Dutrey, R.\ {Bachiller}, 1997, {\em \aap}, {\bf 323}, 943

\rskbibitem{} Hall, S.\ M., C.\ J.\ Clarke,  and J.\ E.\ Pringle, 1996, \mnras, {\bf 278}, 303

\rskbibitem{} Hambly, N.\ C., S.\ T.\ Hodgkin, M.\ R.\ Cossburn, and R.\ F.\ Jameson, 1999, \mnras, {\bf 303}, 835  

\rskbibitem{} Hanawa, T., and K.\ Nakayama, 1997, \apj, {\bf 484}, 238

\rskbibitem{} Hartigan, P., S.\ Edwards, and L.\ Ghandour,  1995,  \apj,  {\bf  452}, 736 

\rskbibitem{} Hartmann, L.,  1998,  Accretion processes in star formation, Cambridge astrophysics series, Vol.\ 32, (Cambridge University Press) 

\rskbibitem{} Hartmann, L., 2001, \aj, {\bf 121}, 1030

\rskbibitem{} Hartmann, L., 2002, \apj, submitted (astro-ph/0207216)

\rskbibitem{} Hartmann, L., J.\ Ballesteros-Paredes, and E.\ A.\ Bergin, 2001,
\apj, {\bf 562}, 852  

\rskbibitem{} Hartmann, L., P.\ Cassen, and S.\ J.\ Kenyon, 1997, \apj,
{\bf 475}, 770

\rskbibitem{} Hawley, J. F., and J. M. Stone, 1995, {\em Computer
Phys. Comm.}, {\bf 89}, 127 

\rskbibitem{} Hawley, J. F., C. F. Gammie, and S. A. Balbus, 1996, /apj,
{\bf 464}, 690

\rskbibitem{} Hayashi, C., 1961,  \pasj, {\bf 13}, 450 

\rskbibitem{} Hayashi, C., 1966, \araa, {\bf 4}, 171

\rskbibitem{} Heckman, T.~M., L.~Armus, and G.~K.~Miley, 1990, \apjs,
{\bf 74}, 833

\rskbibitem{} Heiles, C., 1990, \apj, {\bf 354}, 483

\rskbibitem{} Heiles, C., A.\ A.\ Goodman, C.\ F.\ McKee, and E.\ G.\ Zweibel,
1993,  in {\em Protostars and Planets III},  edited by E.\ H.\ Levy and J.\
I.\ Lunine (University of Arizona Press, Tucson), p.\ 279  

\rskbibitem{} Heisenberg, W., 1948a, \zp, {\bf 124}, 628

\rskbibitem{} Heisenberg, W., 1948b, \prsa, {\bf 195}, 402

\rskbibitem{} Heithausen, A., F.\ Bensch, J.\ Stutzki, E.\ Falgarone, and
J.\ F.\ Panis, 1998, \aap, {\bf 331}, L68 

\rskbibitem{} Heitsch, F., M.\ Mac Low, and R.\ S.\
  Klessen,  2001a,  \apj,   {\bf 547}, 280  

\rskbibitem{} Heitsch, F., E.\ G.\ Zweibel, M.-M.\ Mac Low, P.\ Li, and
M.\ L.\ Norman, 2001b, \apj, {\bf 561}, 800

\rskbibitem{} Hendriksen, R.\ N., 1989, \mnras, {\bf 240}, 917

\rskbibitem{} Hendriksen, R.\ N., P.\ Andr{\'e}, and S.\ Bontemps, 1997, \aap, {\bf 323}, 549

\rskbibitem{} Hennebelle, P.\ and M.\ P{\'e}rault,  1999,  \aap,  {\bf 351},  309 

\rskbibitem{} Hennebelle, P.\ and M.\ P{\'e}rault,  2000,  \aap,  {\bf
359},  1124  

\rskbibitem{} Henning, T., 1989, \an, {\bf 310}, 363

\rskbibitem{} Henning, T., and R.\ Launhardt,  1998,  \aap,  {\bf 338}, 223 

\rskbibitem{} Henriksen, R., P.\ Andr{\'e}, and S.\ Bontemps, 1997, \aap,   {\bf 323}, 549

\rskbibitem{} Heyer, M.\ H., and F.\ P.\ Schloerb, 1997, \apj, {\bf 475}, 173 

\rskbibitem{} Hillenbrand, L.\ A., 1997, \aj, {\bf 113}, 1733

\rskbibitem{} Hillenbrand, L.\ A., and J.\ M.\ Carpenter, 2000, \apj,
{\bf 540}, 236 

\rskbibitem{} Hillenbrand, L.\ A., and L.\ W.\ Hartmann, 1998, \apj, {\bf 492}, 540

\rskbibitem{} Hiltner, W.\ A., 1949, \apj, {\bf 109}, 471

\rskbibitem{} Hiltner, W.\ A., 1951, \apj, {\bf 114}, 241
  
\rskbibitem{} Hockney, R.\ W., and J.\ W.\ Eastwood, 1988, {\em Computer
    Simulation using Particles} (IOP Publishing Ltd., Bristol and
  Philadelphia)

\rskbibitem{} Hodapp, K.\ and J.\ Deane,  1993,  \apjs,  {\bf 88}, 119 

\rskbibitem{} Hogerheijde, M.\ R., E.\ F.\ van Dishoeck, G.\ A.\ Blake, and  H.\ J.\ van Langevelde,  1998,  \apj,  {\bf 502}, 315 

\rskbibitem{} Hollenbach, D.\ J., M.\ W.\ Werner, and E.\ E.\ Salpeter, 1971, \apj, {\bf 163}, 165

\rskbibitem{} Hoyle, F.,  1953,  \apj,  {\bf 118}, 513 

\rskbibitem{} Huang, P.\ G., G.\ N.\ Coleman, and P.\
  Bradshaw, 1995, \jfm, {\bf 305},  185

\rskbibitem{} Hubbard, W.\ B.,  A.\ Burrows, and J.\ I.\ Lunine, 2002, \araa, {\bf 40}, 103 

\rskbibitem{} Hunter, C.,  1977,  \apj,  {\bf 218}, 834 

\rskbibitem{} Hunter, C.,  1986,  \mnras,  {\bf 223}, 391 

\rskbibitem{} Hunter, D. A., 1997, \pasp, {\bf 109}, 937

\rskbibitem{} Hunter, D. A., B. G. Elmegreen, and A. L. Baker, 1998, \apj, {\bf
493}, 595

\rskbibitem{} Hunter, D.\ A., E.\ J.\ Shaya, P.\ Scowen, J.\ J.\ Hester, E.\ J.\
Groth, R.\ Lynds, and E.\ J.\ O'Neil,  1995, \apj, {\bf 444}, 758 

\rskbibitem{} Hunter, J.\ H., and R.\ C.\ Fleck, 1982, \apj, {\bf 256}, 505

\rskbibitem{} Ida, S., J.\ Larwood, and A.\ Burkert, 2000, \apj, {\bf 528}, 351

\rskbibitem{} Inutsuka, S., and S.\ M.\ Miyama,  1992,  \apj,  {\bf 388}, 392 

\rskbibitem{} Inutsuka, S., and S.\ M.\ Miyama,  1997,  \apj,  {\bf 480}, 681 

\rskbibitem{} Indebetouw, R., and E.\ G.\ Zweibel, 2000, \apj, {\bf 532}, 361

\rskbibitem{} Irvine, W. M., P. F. Goldsmith, and A. Hjalmarson, 1986, in {\em
Interstellar Processes}, edited by D. J. Hollenbach and H. A. Thronson,
Jr. (Reidel, Dordrecht), 561
 
\rskbibitem{} Isichenko, M.\ B., 1992, \rmp, {\bf 64}, 961

\rskbibitem{} Jahyesh,  and Z.\ Warhaft, 1991, \prl, {\bf 67}, 3503 

\rskbibitem{} Jayawardhana, R., L.\ Hartmann, and N.\ Calvet,  2001,  \apj,   {\bf 548}, 310 

\rskbibitem{} Jeans, J.\ H., 1902, \pta, {\bf 199}, 1

\rskbibitem{} Johnstone, D.\ and J.\ Bally, 1999, \apj, {\bf 510}, L49

\rskbibitem{} Johnstone, D., C.\ D.\ Wilson, G.\ Moriarty-Schieven, G.\ Joncas, G.\ Smith, E.\ Gregersen, and M.\ Fich, 2000, \apj, {\bf 545}, 327

\rskbibitem{} Johnstone, D., M.\ Fich, G.\ F.\ Mitchell, and G.\ Moriarty-Schieven, 2001, \apj, {\bf 559}, 307

\rskbibitem{} Jones, C.\ E., S.\ Basu, and J.\ Dubinski,  2001,  \apj,  {\bf
551}, 387  

\rskbibitem{} A. Kawamura, A., T. Onishi,  Y. Yonekura, K. Dobashi, A. Mizuno,
H. Ogawa, Y. Fukui, 1998, \apjs, {\bf 117}, 387

\rskbibitem{} Kegel, W.\ H., 1989, \aap, {\bf 225}, 517

\rskbibitem{} Kennicutt, R.\ C., Jr., 1998a, \araa, {\bf 36}, 189

\rskbibitem{} Kennicutt, R.\ C., Jr., 1998b, \apj, {\bf 498}, 541

\rskbibitem{} Keto, E.\ R., J.\ C.\ Lattanzio, and J.\ J.\ Monaghan,
1991,   \apj,  {\bf 383}, 639  

\rskbibitem{} Kida, S., and Y.\ Murakami, 1989, {\em Fluid Dyn.\ Res.}, {\bf 4},
  347

\rskbibitem{} Kim, W.-T., and E.\ C.\ Ostriker, 2001, \apj, {\bf 559}, 70

\rskbibitem{} Kim, W.-T., and E.\ C.\ Ostriker, 2002, \apj, {\bf 570}, 132

\rskbibitem{} Kimura, T., and M.\ Tosa, 1996, \aap, {\bf 308}, 979

\rskbibitem{} Kippenhahn, R., and A.\ Weigert, 1990, {\em Stellar
Structure and Evolution} (Springer Verlag, Berlin, Heidelberg)

\rskbibitem{} Kitamura, Y., K.\ Sunada, M.\ Hayashi,T.\ Hasegawa, 1993,
\apj, {\bf 413}, 221

\rskbibitem{} Klapp, J., L.\ D.\ G.\ Sigalotti, and F.\ de Felice, 1993, \aap, {\bf 273}, 175

\rskbibitem{} Klein, R.\ I., 1999, \jcam, {\bf 109}, 123 

\rskbibitem{} Klein, R.\ I., C.\ F.\ McKee, and P.\ Colella, 1994, \apj,
{\bf 420}, 213

\rskbibitem{} Kleiner, S.\ C., and R.\ L.\ Dickman, 1987, \apj, 312, 837

\rskbibitem{} Klessen, R. S., 1997, \mnras, {\bf 292}, 11

\rskbibitem{} Klessen, R. S., 2000, \apj, {\bf 535}, 869

\rskbibitem{} Klessen, R.\ S., 2001a, \apj, {\bf 550}, L77

\rskbibitem{} Klessen, R.\ S., 2001b, \apj, {\bf 556}, 837

\rskbibitem{} Klessen, R.\ S., and A.\ Burkert,  2000,  \apjs,  {\bf 128}, 287 

\rskbibitem{} Klessen, R.\ S., and A.\ Burkert,  2001,  \apj,  {\bf 549}, 386 

\rskbibitem{} Klessen, R.\ S., and P.\ Kroupa, 2001, \aap, {\bf 372}, 105 

\rskbibitem{} Klessen, R.\ S., A.\ Burkert, and M.\ R.\ Bate, 1998, \apj, {\bf 501}, L205 

\rskbibitem{} Klessen, R.\ S., F.\ Heitsch, and M.-M.\
  Mac Low,  2000, \apj,   {\bf 535}, 887  

\rskbibitem{} Kolmogorov, A.\ N., 1941a, {\em Dokl.\ Akad.\ Nauk SSSR}, {\bf 30}, 301 --- translated and reprinted 1991 in \prsa, {\bf 343}, 9

\rskbibitem{} Kolmogorov, A.\ N., 1941b, {\em Dokl.\ Akad.\ Nauk SSSR},
{\bf 31}, 538 

\rskbibitem{} K{\"o}nigl, A., and R. E. Pudritz, in {\em Protostars and Planets
IV}, edited by V. Manning, A. P. Boss, and S. S. Russell (U. of
Arizona Press, Tucson), 759

\rskbibitem{} Korpi, M.\ J., A.\ Brandenburg, A.\ Shukorov, I.\ Tuominen, and \AA.\ Nordlund, 1999, \apjl, {\bf 514}, L99

\rskbibitem{} Koyama, H., and S.\ Inutsuka,  2000, \apj, {\bf 532}, 980

\rskbibitem{} Koyama, H., and S.\ Inutsuka,  2002, \apjl, {\bf 564}, L97

\rskbibitem{} Kramer, C., J.\ {Stutzki}, R.\ {Rohrig},  U.\ {Corneliussen}, 1998, \aap, {\bf 329}, 249 

\rskbibitem{} Kramer C., J.\ Alves, C.\ J.\ Lada, E.\ A.\ Lada, A.\ 
Sievers, H.\ Ungerechts, and C.\ M.\ Walmsley, 1999, \aap, {\bf 342},
257

\rskbibitem{} Krebs, J., and W.\ Hillebrandt, 1983, \aap, {\bf 128}, 411

\rskbibitem{} Kroupa, P., 1995a, \mnras, {\bf 277}, 1491

\rskbibitem{} Kroupa, P., 1995b, \mnras, {\bf 277}, 1507

\rskbibitem{} Kroupa, P., 1995c, \mnras, {\bf 277}, 1522

\rskbibitem{} Kroupa, P., 2001, \mnras, {\bf 322}, 231

\rskbibitem{} Kroupa, P., 2002, {\em Science}  {\bf 295}, 82 

\rskbibitem{} Kroupa, P., and A.\ Burkert, 2001, \apj, {\bf 555}, 945

\rskbibitem{} Kroupa, P., C.\ A.\ Tout, G.\ Gilmore, 1990, \mnras, {\bf 244}, 76 

\rskbibitem{} Kroupa, P., C.\ A.\ Tout, G.\ Gilmore, 1993,  \mnras, {\bf 262}, 545 

\rskbibitem{} Kuhfu{\ss}, R., 1987,  PhD-Thesis, TU M\"unchen

\rskbibitem{} Kulsrud, R.\ M., and W.\ P.\ Pearce, 1969, \apj, {\bf 156}, 445

\rskbibitem{} Kutner, M.\ L., K.\ D.\ Tucker, G.\ Chin, and P.\ Thaddeus, 1977,
\apj, {\bf 215}, 521

\rskbibitem{} Leisawitz, D., F. N. Bash, and P. Thaddeus, 1989, \apjs, {\bf 70},
731 

\rskbibitem{} L\'eorat, J., T.\ Passot, and A.\ Pouquet, 1990, \mnras, {\bf 243},
293 

\rskbibitem{} LaRosa, T.\ N., S.\ N.\ Shore, and L.\ Magnani, 1999, \apj,
{\bf 512}, 761

\rskbibitem{} Lada, C.\ J., J.\ Alves, and E.\ A.\ Lada, 1996, \aj,
{\bf 111}, 1964

\rskbibitem{} Lada, C.\ J., E.\ A.\ Lada, D.\ P.\ Clemens, and J.\ 
Bally, 1994, \apj, {\bf 429}, 694

\rskbibitem{} Lada, E.\ A., 1992, \apj, {\bf 393}, L25
  
\rskbibitem{} Lada, C.\ J., J.\ Alves, and E.\ A.\ Lada,
  1999, \apj, {\bf 515}, 265
  
\rskbibitem{} Lada, E.\ A., J.\ Bally, and A.\ A.\ {Stark}, 1991, \apj, {\bf
    368}, 432
  
\rskbibitem{} Lamballais, E., M.\ Lesieur, and O.\ M{\'e}tais, 1997, \pre,
  {\bf 56}, 761

\rskbibitem{} Langer, W., R.\ Wilson, and C.\ Anderson, 1993, \apj, {\bf
408}, L25  

\rskbibitem{} Langer, W.\ D., T.\ Velusamy, T.\ B.\ H.\ Kuiper, W.\
Levin,   E.\ Olsen,V.\ Migenes, 1995, {\em \apj}, {\bf 453}, 293 

\rskbibitem{} Langer, W.\ D., E.\ F.\ van Dishoeck, E.\ A.\ Bergin, G.\
A.\ Blake, A.\ G.\ G.\ M.\ Tielens, T.\ Velusamy, and D.\ C.\ B.\
Whittet, 2000, in {\em Protostars and Planets IV}, edited by V.\
Mannings, A.\ P.\ Boss, and S.\ S.\ Russell (University of Arizona
Press, Tucson), p.\ 29 

\rskbibitem{} Lanzetta, K.\ M., A.\ Yahil, and A.\ Fern\'andez-Soto,
1996, Nature, {\bf 381}, 759 

\rskbibitem{} Lanzetta, K.\ M.,  N.\ Yahata, S.\ Pascarelle, H.\ Chen,
and A.\ Fern{\'a}ndez-Soto, 2002,  \apj, {\bf 570}, 492  

\rskbibitem{} LaRosa, T.\ N., S.\ N.\ Shore, and L.\ Magnani,1999, \apj, {\bf
    512}, 761

\rskbibitem{} Larson, R.\ B.,  1969, \mnras,  {\bf 145}, 271 

\rskbibitem{} Larson, R.\ B.,  1972, \mnras,  {\bf 156}, 437

\rskbibitem{} Larson, R.\ B.,  1973, \mnras,  {\bf 161}, 133

\rskbibitem{} Larson, R.\ B.,  1981, \mnras,  {\bf 194}, 809

\rskbibitem{} Larson, R.\ B.,  1992, \mnras,  {\bf 256}, 641

\rskbibitem{} Larson, R.\ B.,  1995, \mnras,  {\bf 272}, 213

\rskbibitem{} Laughlin, G., and P.\ Bodenheimer, 1993, \apj, {\bf 403}, 303

\rskbibitem{} Lechner, R., J.\ Sesterhenn, and R.\ Friedrich, 2001,
  \jt, {\bf 2}, 001

\rskbibitem{} Lee, C.\ W., P.\ C.\ Myers, and M.\ Tafalla,  1999,  \apj,
{\bf  526}, 788  

\rskbibitem{} Lee, C.\ W., P.\ C.\ Myers, and M.\ Tafalla,  2001,  \apjs,
{\bf  136}, 703  

\rskbibitem{} Lee, Y., R. L. Snell, \& R. L. Dickman, 1996, \apj, {\bf
472}, 275

\rskbibitem{} Leinert, C., T.\ Henry, A.\ Glindemann, and D.\ W.\
McCarthy,   1997,  \aap,  {\bf 325}, 159  

\rskbibitem{} Leisawitz, D., F.\ N.\ Bash, and P.\ Thaddeus, 1989, \apjs,
{\bf 70}, 731

\rskbibitem{} Lejeune, C., and P.\ Bastien, 1986, \apj, {\bf 309}, 167

\rskbibitem{} Lesieur, M., 1997, {\em Turbulence in Fluids}, 3rd ed.\
(Kluwer, Dordrecht), p.\ 245 

\rskbibitem{} Li, P., M.\ L.\ Norman, F.\ Heitsch, and M.-M.\ Mac Low,
2000, \baas, {\bf 197}, 05.02 

\rskbibitem{} Li, Y., R.\ S.\ Klessen, and M.-M. Mac Low, 2003, in {\em
Extragalactic Globular Cluster Systems}, edited by M. Kissler-Patig
(Springer, Heidelberg), in press (astro-ph/0210479)

\rskbibitem{} Li, Z., and F.\ H.\ Shu,  1996,  \apj,  {\bf 472}, 211 

\rskbibitem{} Li, Z., and F.\ H.\ Shu,  1997,  \apj,  {\bf 475}, 237 

\rskbibitem{} Lillo, F.\ and R.\ N.\ Mantegna, 2000, \pre,
  {\bf 61}, R4675 

\rskbibitem{} Lilly, S.\ J., O.\ Le Fevre, F.\ Hammer, and D.\ Crampton,
1996, \apjl, {\bf 460}, L1  

\rskbibitem{} Lin, C. C., and F. H. Shu, 1964, \apj, {\bf 140}, 646

\rskbibitem{} Lin, C. C., C. Yuan, and F. H. Shu, 1969, \apj, {\bf 155}, 721

\rskbibitem{} Lis, D.\ C., J.\ Keene, T.\ G.\ Phillips, and J.\ Pety,
1998, \apj, {\bf 504}, 889   

\rskbibitem{} Lis, D.\ C., J.\ Pety, T.\ G.\ Phillips, and E.\ Falgarone,
1996, \apj, {\bf 463}, 623  

\rskbibitem{} Lissauer, J.\ J.,  1993,  \araa,  {\bf 31}, 129 

\rskbibitem{} Lithwick, Y., and P.\ Goldreich, 2001, \apj, {\bf 562}, 279

\rskbibitem{} Lizano, S., and F.\ H.\ Shu,  1989,  \apj,  {\bf 342}, 834 

\rskbibitem{} Lizano, S., C. Heiles, L. F. Rodr\'iguez, B.-C. Koo,
F. H. Shu, T. Hasegawa, S. Hayashi, I. F. Mirabel, 1988, \apj, {\bf
328}, 763

\rskbibitem{} Lombardi, M., and G.\ Bertin, 2001, \aap, submitted
(astro-ph/0106336) 

\rskbibitem{} Lombardi, J.\ C.,  A.\ Sills, F.\ A.\
  Rasio, and S.\ L.\ Shapiro, 1999, \jcp,  {\bf 152}, 687

\rskbibitem{} Loren, R.\ B., 1989, \apj, {\bf 338}, 902

\rskbibitem{} Lynden-Bell, D., and A. J. Kalnajs, 1972, \mnras, {\bf 157}, 1

\rskbibitem{} M\"uller, W.-C., and D.\ Biskamp, 2000, \prl, {\bf 84}, 475

\rskbibitem{} Machiels, L.,  and M.\ O.\ Deville, 1998, \jcp, {\bf 145}, 256 

\rskbibitem{} Mac Low, M.-M., 1999, \apj, {\bf 524}, 169

\rskbibitem{} Mac Low, M.-M., 2000, in  {\em Stars, Gas and Dust in Galaxies:
Exploring the Links}, edited by D.\ Alloin, K.\ Olson, and G.\ Galaz, ASP
Conf.\ Ser.\ No.\ 221 (ASP, San Francisco) 55 

\rskbibitem{} Mac Low, M.-M., 2002, in {\em Simulations of
magnetohydrodynamic turbulence in astrophysics}, edited by T. Passot
and E. Falgarone (Springer, Heidelberg) in press (astro-ph/0201157)

\rskbibitem{} Mac Low, M.-M., D.\ S.\ Balsara, M.\ A.\ de Avillez, and
J.\ Kim, 2001, \apj, submitted (astro-ph/0106509) 

\rskbibitem{} Mac Low, M.-M., and A. Ferrara, 1999, \apj, {\bf 513}, 142

\rskbibitem{} Mac Low, M.-M., R.\ S.\ Klessen, A.\
  Burkert, and M.\ D.\ Smith, 1998, \prl, {\bf 80}, 2754 

\rskbibitem{} Mac Low, M.-M.,, C.\ F.\ McKee, R.\ I.\ Klein, J.\ M.\
Stone, and M.\ L.\ Norman, 1994, \apj, {\bf 433}, 757

\rskbibitem{} Mac Low M.-M.\ and V.\ Ossenkopf, 2000, \aap {\bf 353}, 339

\rskbibitem{} Madau, P., H.\ C.\ Ferguson, M.\ E.\ Dickinson, M.\
Giavalisco, C.\ C.\ Steidel, and A.\ Fruchter, 1996, \mnras, {\bf
283}, 1388 

\rskbibitem{} Maddalena, R.\ J., and P.\ Thaddeus, 1985, \apj, {\bf 294}, 231

\rskbibitem{} Magnani, L., L.\ Blitz, L.,  and L.\ Mundy, 1985, \apj,
{\bf 295}, 402 

\rskbibitem{} Manic\'o, G.,  G.\ Ragun{\'\i}, V.\ Pirronello, J.\ E.\
Roser, and G.\ Vidali, 2001, \apjl, {\bf 548}, L253 

\rskbibitem{} Maron, J., and P.\ Goldreich, 2001,
  \apj, {\bf 554}, 1175 

\rskbibitem{} Martin, C.\ L., and R.\ C.\ Kennicutt, Jr., 2001, \apj,
{\bf 555}, 301 

\rskbibitem{} Martos, M. A., and D. P. Cox, 1998, \apj, {\bf 509}, 703

\rskbibitem{} Masunaga, H., S.\ M.\ Miyama, and S.\ Inutsuka,  1998,  \apj,  {\bf
495}, 346 

\rskbibitem{} Masunaga, H., and S.\ Inutsuka, 2000a, \apj, {\bf 531}, 350 

\rskbibitem{} Masunaga, H., and S.\ Inutsuka, 2000b, \apj, {\bf 536}, 406

\rskbibitem{} Mathieu, R.\ D., A.\ M.\ Ghez, E.\ L.\ N.\ Jensen, and M.\
Simon,  2000,  in {\em Protostars and Planets IV},  edited by V.\
Mannings, A.\ P.\ Boss, and S.\ S.\ Russell (University of Arizona
Press, Tucson),  p.\ 703  

\rskbibitem{} Matthews, L. D., and J. S. Gallagher, III, 2002, \apjs,
{\bf 141}, 429

\rskbibitem{} Matzner, C. D., 2002, \apj, {\bf 566}, 302

\rskbibitem{} McDonald, J.\ M., and C.\ J.\ Clarke, 1995, \mnras, {\bf 275}, 671

\rskbibitem{} McGaugh, S. S., and W. J. G. de Blok, 1997, \apj, {\bf 481}, 689

\rskbibitem{} McGaugh, S. S., V. C. Rubin, and W. J. G. de Blok, 2001,
\apj, {\bf 122}, 2381

\rskbibitem{} McKee, C.\ F., 1989, \apj, {\bf 345}, 782

\rskbibitem{} McKee,  C.\ F., 1999, in {\em NATO ASIC Proc.\ 540: The
Origin of Stars  and Planetary Systems},  edited by C.\ J.\ Lada and
N.\ D.\ Kylafis (Kluwer Academic Publishers), p.\ 29 

\rskbibitem{} McKee, C.\ F., and J.\ P.\ Ostriker, 1977, \apj, {\bf 218},
148

\rskbibitem{} McKee, C. F., and J. P. Williams, 1997, \apj, {\bf 476}, 144

\rskbibitem{} McKee, C.\ F., and E.\ G.\ Zweibel, 1992, \apj, {\bf 399},
551

\rskbibitem{} McKee, C.\ F., and E.\ G.\ Zweibel, 1995, \apj, {\bf 440},
686

\rskbibitem{} McKee, C.\ F., E.\ G.\ Zweibel, A.\ A.\ Goodman, and C.\ Heiles,
1993,  in {\em Protostars and Planets III}, edited by E.\ H.\ Levy and J.\ I.\
Lunine (University of Arizona Press, Tucson), p.\ 327  

\rskbibitem{} Mestel, L., and L.\ Spitzer, Jr., 1956, \mnras, {\bf 116}, 503

\rskbibitem{} Metzler, R., and J.\ Klafter, 2000,
  \prp, {\bf 339}, 1

\rskbibitem{} Miesch, M.S., and J.\ M.\ Bally, 1994, \apj, {\bf 429}, 645 

\rskbibitem{} Miesch, M.S., and J.\ M.\ Scalo,  1995, \apj, {\bf 450}, L27 

\rskbibitem{} Miesch, M.S., J.\ M.\ Scalo, and J.\ Bally, 1999, \apj, {\bf 524},
895

\rskbibitem{} Mihalas, D., and B.\ W.\ Mihalas, 1984, {\em Foundations of
    Radiation Hydrodynamics} (Oxford University Press, New York)

\rskbibitem{} Mihos, J. C., M. Spaans, and S. S. McGaugh, 1999, \apj,
{\bf 515}, 89

\rskbibitem{} Miller, G.\ E., and J.\ M.\ Scalo, 1979, \apjs, {\bf 41}, 513

\rskbibitem{} Mitchell, G.\ F., D.\ Johnstone, G.\ Moriarty-Schieven, M.\
Fich, and N.\ F.\ H.\ Tothill, 2001, \apj, {\bf 556}, 215

\rskbibitem{} Mizuno, A., T.\ Onishi, Y.\ Yonekura, T.\ Nagahama, H.\
Ogawa, and Y.\ Fukui, 1995, \apj, {\bf 445}, L161

\rskbibitem{} Moffat, A. F. J., M. F. Corcoran, I. R. Stevens, G. Skalkowski,
S. V. Marchenko, A. M\"ucke, A. Ptak, B. S. Koribalski, L. Brenneman,
R. Mushotzky, J. M. Pittard, A. M. T. Pollock, and W. Brandner, 2002, \apj,
573, 191 

\rskbibitem{} Monaghan, J. J., 1992, \araa, {\bf 30}, 543

\rskbibitem{} Mooney, T.\ J., and P.\ M.\ Solomon, 1988, \apj, {\bf 334}, L51 

\rskbibitem{} Morton, S.\ A., T.\ C.\ Mouschovias, and G.\ E.\ Ciolek,  1994,
\apj,  {\bf 421}, 561  

\rskbibitem{} Moser, R.\ D., J.\ Kim, and N.\ N.\ Mansour, 1999,
  \pfl, {\bf 11}, 943

\rskbibitem{} Motte, F., and P.\ Andr{\' e},  2001,  \aap,  {\bf 365}, 440 

\rskbibitem{} Motte, F., P.\ Andr{\'e}, and R.\ Neri,  1998,  \aap,  {\bf 336},
150  

\rskbibitem{} Motte, F., P.\ Andr{\' e}, D.\ Ward-Thompson, and S.\
Bontemps, 2001, \aap, {\bf 372}, L41

\rskbibitem{} Mouschovias, T.\ C., 1991a,  \apj,  {\bf 373}, 169 

\rskbibitem{} Mouschovias, T.\ Ch., 1991b, in {\em The Physics of Star
Formation and Early Stellar Evolution}, edited by C.\ J.\ Lada and N.\ D.\
Kylafis (Kluwer, Dordrecht), p.\ 61 

\rskbibitem{} Mouschovias, T.\ Ch., 1991c, in {\em Physics of Star
Formation and Early Stellar Evolution}, edited by C.\ J.\ Lada and N.\
D\.\ Kylafis (Kluwer, Dordrecht), p.\ 449 

\rskbibitem{} Mouschovias, T.\ C., and S.\ A.\ Morton, 1991, \apj, {\bf
371}, 296

\rskbibitem{} Mouschovias, T.\ C., and S.\ A.\ Morton,  1992a,  \apj,  {\bf 390},  144 

\rskbibitem{} Mouschovias, T.\ C., and S.\ A.\ Morton,  1992b,  \apj,  {\bf 390},  166 

\rskbibitem{} Mouschovias, T.\ C., and E.\ V.\ Paleologou, 1979, \apj, {\bf 230},
204

\rskbibitem{} Mouschovias, T.\ C., and E.\ V.\ Paleologou, 1980, \apj, {\bf 237},
877

\rskbibitem{} Mouschovias, T.\ Ch., and L.\ Spitzer, Jr., 1976, \apj, {\bf 210}, 326 

\rskbibitem{} M{\" u}ller, W.~C., and D.\
  Biskamp, 2000, \prl, {\bf 84}, 475  

\rskbibitem{} Murray, S.\ D., and C.\ J.\ Clarke, 1993, \mnras, {\bf 265}, 169

\rskbibitem{} Murray, S.\ D., and D.\ N.\ C.\ Lin,  1996,  \apj,  {\bf 467},  728 

\rskbibitem{} Myers, P.\ C., 1983, \apj, {\bf 270}, 105

\rskbibitem{} Myers, P.\ C., 2000, \apj, {\bf 530}, L119

\rskbibitem{} Myers, P.\ C., and A.\ A.\ Goodman, 1988 \apj, {\bf 326}, L27

\rskbibitem{} Myers, P.\ C., and V.\ K.\ Khersonsky,  1995,  \apj,  {\bf 442}, 186 

\rskbibitem{} Myers, P.\ C., and E.\ E.\ Ladd, 1993,  \apj, {\bf 413}, L47

\rskbibitem{} Myers, P.\ C., N.\ J.\ Evans, and N.\ Ohashi, 2000, in {\em
Protostars and Planets IV}, edited by V.\ Mannings, A.\ P.\ Boss, and
S.\ S.\ Russell (University of Arizona Press, Tucson), p.\ 217 

\rskbibitem{} Myers, P.\ C., F.\ C.\ Adams, H.\ Chen, and E.\ Schaff,
1998,  \apj,  {\bf 492}, 703  

\rskbibitem{} Myers, P.\ C., G.\ A.\ Fuller,  A.\ A.\ Goodman, and P.\
J.\ Benson,   1991, \apj, {\bf 376}, 561 

\rskbibitem{} Myers, P.\ C., D.\ Mardones, M.\ Tafalla, J.\ P.\ Williams,
and  D.\ J.\ Wilner,  1996,  \apj,  {\bf 465}, L133  

\rskbibitem{} Nakajima, Y., K.\ Tachihara, T.\ Hanawa, and M.\ Nakano,
1998, \apj, {\bf 497}, 721 

\rskbibitem{} Nakamura, F., T.\ Hanawa, and T.\ Nakano, 1995, \apj, {\bf 444}, 770

\rskbibitem{} Nakano, T., 1976, \pasj, {\bf 28}, 355

\rskbibitem{} Nakano, T., 1979, \pasj, {\bf 31}, 697

\rskbibitem{} Nakano, T., 1982, \pasj, {\bf 34}, 337 

\rskbibitem{} Nakano, T., 1983, \pasj, {\bf 35}, 209 

\rskbibitem{} Nakano, T., 1998, \apj, {\bf 494}, 587

\rskbibitem{} Nakano, T., and T.\ Nakamura, 1978, \pasj, {\bf 30}, 681

\rskbibitem{} Nakano, T., T.\ Hasegawa, and C.\ Norman, 1995, \apss, {\bf
224}, 523 

\rskbibitem{} Nakazawa, K., C.\ Hayashi, and M.\ Takahara, 1976, \ptp,
{\bf 56}, 515 

\rskbibitem{} Narlikar, J.\ V., and T.\ Padmanabhan, 2001, \araa, {\bf 39}, 211

\rskbibitem{} Neuh{\"a}user, R.,  M.\ F.\ Sterzik, G.\ Torres, E.\ L.\ Mart\'in,
1995, \aap, {\bf 299}, L13 

\rskbibitem{} Ng, C.\ S., and A.\ Bhattacharjee, 1996, \apj, {\bf 465}, 845

\rskbibitem{} Norman, C. A., and A. Ferrara, 1996, \apj, {\bf 467}, 280

\rskbibitem{} Norman, C. A., and J.\ Silk, 1980, \apj,  {\bf 239}, 968 

\rskbibitem{} Norman, M.\ L., J.\ R.\ Wilson, and R.\ T.\ Barton,  1980,
\apj,  {\bf 239}, 968  

\rskbibitem{} Nugis, T., and H. J. G. L. M. Lamers, 2000, \aap, {\bf
360}, 227

\rskbibitem{} Obukhov, A.\ M., 1941, {\em Dokl.\ Akad.\ Nauk
    SSSR}, {\bf 32}, 22 

\rskbibitem{} Ogino, S., K.\ Tomisaka, and F.\ Nakamura,  1999,  \pasj,
{\bf  51}, 637  

\rskbibitem{} Olling, R. P., and M. R. Merrifield, 1998, \mnras, {\bf 297}, 943

\rskbibitem{} Olling, R. P., and M. R. Merrifield, 2000, \mnras, {\bf 311}, 361

\rskbibitem{} Olmi, L., and L. Testi, 2002, \aap, {\bf 392}, 1053

\rskbibitem{} Onishi, T., A.\ {Mizuno}, A.{Kawamura}, H.\ {Ogawa}, Y.\
{Fukui},   1996, \apj, {\bf 465}, 815 

\rskbibitem{} Oort, J.\ H., 1954, \bain, {\bf 12}, 177

\rskbibitem{} Oort, J. H., \& L. Spitzer, Jr., 1955, \apj, {\bf 121}, 6

\rskbibitem{} Ossenkopf, V.\ 2002, \aap, 
{\bf 391}, 295 

\rskbibitem{} Ossenkopf V., and M.-M.\ Mac\ Low, 2002, \aap,  {\bf 390}, 307 

\rskbibitem{} Ossenkopf V., R.\ S.\ Klessen, and F.\ Heitsch, 2001, \aap,
{\bf 379}, 1005  

\rskbibitem{} Ossenkopf V., Bensch F.,
  Stutzki J. 2000, in {\em The Chaotic Universe}, edited by V.\ G.\ Gurzadyan
  and R.\ Ruffini (World Sci.), p. 394

\rskbibitem{} Ossia, S., and M.\ Lesieur, \jt, {\bf
    2}, 013

\rskbibitem{} Osterbrock, D.\ E., 1961, \apj, {\bf 134}, 270

\rskbibitem{} Ostriker, E.\ C., C.\ F.\ Gammie, and J.\ M.\ Stone, 1999, \apj, {\bf 513}, 259

\rskbibitem{} Ostriker, E.\ C., J.\ M.\ Stone, and C.\ F.\ Gammie, 2001, \apj,
{\bf 546}, 980

\rskbibitem{} Padoan, P., 1995, \mnras, {\bf 277}, 377

\rskbibitem{} Padoan, P., and \AA.\ Nordlund, 1999, \apj, {\bf 526}, 279 

\rskbibitem{} Padoan, P., and \AA.\ Nordlund, 2002, \apj, {\bf 576}, 870 

\rskbibitem{} Padoan, P., L.\ Cambr\'esy, and W.\ Langer, 2002, \apj submitted
(astro-ph/0208217)  

\rskbibitem{} Padoan, P., A.\ A.\ Goodman, B.\ T.\ Draine, M.\ Juvela, \AA.\ 
Nordlund, and {\"O}.\ E.\ R{\" o}gnvaldsson, 2001a, \apj, {\bf 559}, 1005

\rskbibitem{} Padoan, P., M.\ Juvela, J. Bally, and \AA.\ Nordlund, 2000,
\apj, {\bf 529}, 259

\rskbibitem{} Padoan, P., M.\ Juvela, A.\ A.\ Goodman, and {\AA}.\ Nordlund,
2001b, {\bf 553}, 227 

\rskbibitem{} Padoan, P., \AA.\ Nordlund, and B.\ J.\ T.\ Jones, 1997, \mnras,
{\bf 288}, 145  

\rskbibitem{} Padoan, 
P., E.\ Zweibel, and    {\AA}.\ Nordlund, 2000, \apj, {\bf 540}, 332 

\rskbibitem{} Padoan, P., \AA\ Nordlund, \"O.\ E.\ R\"ognvaldsson, and A.\
Goodman, 2002, \apj, submitted (astro-ph/0011229) 

\rskbibitem{} Palla, F., 2000, in {\em The Origin of Stars and Planetary Systems}
edited by C.\ J.\ Lada and N.\ D.\ Kylafis (Kluwer Academic Publisher,
Dordrecht), p.\ 375 

\rskbibitem{} Palla, F., 2002, in {\em Physics on Star Formation in Galaxies,
Saas-Fee Advanced Course 29}, edited by A.\ Maeder and G.\ Meynet
(Springer-Verlag, Heiderberg), p.\ 9 

\rskbibitem{} Palla, F., and S.\ W.\ Stahler, 1999, \apj, {\bf 525}, 77

\rskbibitem{} Palla, F., and S.\ W.\ Stahler, 2000, \apj, {\bf 540}, 255

\rskbibitem{} Palmer, P., and B.\ Zuckerman, 1967, \apj, {\bf 148}, 727

\rskbibitem{} Passot, T., and E.\ V{\'a}zquez-Semadeni,  1998, \pre, 58, 4501 

\rskbibitem{} Passot, T., A.\ Pouquet, and P.\ R.\ Woodward,  1988, \aap, {\bf 197}, 392

\rskbibitem{} Passot, T., E.\ V{\'a}zquez-Semadeni, and A.\ Pouquet, 1995, \apj, {\bf 455}, 536

\rskbibitem{} Penston, M.\ V., 1969a, \mnras, {\bf 144}, 425

\rskbibitem{} Penston, M.\ V., 1969b,  \mnras,  {\bf 145}, 457 

\rskbibitem{} Persi, P., and 20 colleagues, 2000, \aap, {\bf 357}, 219

\rskbibitem{} Phillips, A.\ C., 1994,  {\em The Physics of Stars} (Wiley, Chichester, New York) 

\rskbibitem{} Pikel'ner, S.\ B.\ 1968, {\em Sov.\ Astron.}, {\bf 11}, 737.

\rskbibitem{} Pirronello, V.,  O.\ Biham, C.\ Liu, L.\ Shen, and G.\ Vidali, 1997a, \apjl, {\bf 483}, L131 

\rskbibitem{} Pirronello, V., C.\ Liu, J.\ E.\ Roser, and G.\ Vidali, 1999, \aap, {\bf 344}, 681 

\rskbibitem{} Pirronello, V., C.\ Liu, L.\ Shen, and G.\ Vidali, 1997b, \apjl, {\bf 475}, L69 

\rskbibitem{} Plume, R., D.\ T.\ Jaffe, N.\ J.\ Evans, I., J.\
Mart\'{\i}n-Pintado, and J.\ G\'omez-Gonz\'alez, 1997, \apj, {\bf 476}, 730  

\rskbibitem{} Porter, D.\ H., and P.\ R.\ Woodward, 1992, \apjs, {\bf 93}, 309

\rskbibitem{} Porter D.\ H., and P.\ R.\ Woodward, 2000, \apj, {\bf
  127}, 159
  
\rskbibitem{} Porter, D.\ H., A.\ Pouquet, and P.\ R.\ Woodward,
  1992, \prl, {\bf 68}, 3156
  
\rskbibitem{} Porter, D.\ H., A.\ Pouquet, and P.\ R.\ Woodward,
  1994, \pfl, {\bf 6}, 2133

\rskbibitem{} Porter, D.\ H., A.\ Pouquet, I.\ V.\ 
  Sytine, and P.\ R.\ Woodward, 1999, {\em Phys.\ A}, {\bf 263}, 263
  
\rskbibitem{} Potter, D., 1977, {\em Computational Physics}
  (Wiley, New York)

\rskbibitem{} Prasad, S. S., K. R. Heere, and S. P. Tarafdar, 1991, {\bf 373},
123 

\rskbibitem{} Pratap, P., J.\ E.\ Dickens, R.\ L.\ Snell, M.\ P.\ Miralles,  E.\
A.\ Bergin, W.\ M.\ Irvine, and F.\ P.\ Schloerb,  1997,  \apj,  {\bf 486},
862 

\rskbibitem{} Price, N.\ M., and P.\ {Podsiadlowski}, 1995, \mnras, {\bf
273}, 1041 

\rskbibitem{} Pringle, J. E., R. J. Allen, and S. H. Lubow, 2001, \mnras,
{\bf 327}, 663

\rskbibitem{} Prosser, C.\ F., J.\ R.\ Stauffer, L.\ Hartmann, D.\ R.\ Soderblom,  B.\ F.\ Jones, M.\ W.\ Werner, and M.\ J.\ McCaughrean,  1994,  \apj,  {\bf 421},  517 

\rskbibitem{} P{\" o}ppel, W.\ G.\ L., \fcp, {\bf 18}, 1

\rskbibitem{} Rana, N.\ C., 1991, \araa, {\bf 29}, 129

\rskbibitem{} Rand, R. J., and S. R. Kulkarni, 1989, \apj, {\bf 343}, 760

\rskbibitem{} Rand, R. J., and A. G. Lyne, 1994, \mnras, {\bf 268}, 497

\rskbibitem{} Raymond, J.\ C., D.\ P.\ Cox, and B.\ W.\ Smith,  1976,
\apj,  {\bf  204}, 290  

\rskbibitem{} Reipurth, B., and C.\ Clarke, 2001, \aj, {\bf 122}, 432

\rskbibitem{} Richardson, L.\ F., 1922, {\em Weather Prediction by
  Numerical Process} (Cambridge University Press, Cambridge)

\rskbibitem{} Richer, J.\ S., D.\ S.\ Shepherd, S.\ Cabrit, R.\
Bachiller, and  E.\ Churchwell, 2000, in {\em Protostars and Planets
IV}, edited by V.\ Mannings, A.\ P.\ Boss, and S.\ S.\ Russell
(University of Arizona Press, Tucson), p.\ 867 

\rskbibitem{} Richtler, T., 1994, \aap, {\bf 287}, 517

\rskbibitem{} Roberts, W. W., 1969, \apj, {\bf 158}, 123

\rskbibitem{} Rodriguez-Gaspar, J.~A., G.~Tenorio-Tagle, and J.~Franco,
1995, \apj, {\bf 451}, 210

\rskbibitem{} Romeo, A.\ B., 1992, \mnras, {\bf 256}, 307

\rskbibitem{} Rosen, A.\ and J.\ N.\ Bregman,  1995,  \apj,  {\bf 440}, 634 

\rskbibitem{} Rosen, A., J.\ N.\ Bregman, and D.\ D.\ Kelson,  1996,
\apj,  {\bf  470}, 839  

\rskbibitem{} Rosen, A., J.\ N.\ Bregman, and M.\ L.\ Norman,  1993,
\apj,  {\bf  413}, 137  

\rskbibitem{} Rosolowsky, E.\ W., A.\ A.\ Goodman, D.\ J.\ Wilner, and
J.\ P.\ Williams, 1999, \apj, {\bf 524}, 887 

\rskbibitem{} Rozyczka, M., W.\ M.\ Tscharnuter, K.\ -H.\ Winkler, and H.\ W.\  Yorke,  1980,  \aap,  {\bf 83}, 118 

\rskbibitem{} Ruden, S.\ P., 1999, in {\em NATO ASIC Proc.\ 540: The Origin of Stars  and Planetary Systems},  edited by C.\ J.\ Lada and N.\ D.\ Kylafis (Kluwer Academic Publishers), p.\ 643 

\rskbibitem{} Ryden, B.\ S., 1996, \apj, {\bf 471}, 822

\rskbibitem{} Saito, S., Y. Aikawa, E. Herbst, M. Ohishi, T. Hirota, S. Yamamoto,
and N. Kaifu, 2002, \apj, {\bf 569}, 836 

\rskbibitem{} S\'anchez-Salcedo, F.\ J., 2001, \apj, {\bf 563}, 867

\rskbibitem{} Safier, P.\ N., C.\ F.\ McKee, and S.\ W.\ Stahler, 1997, \apj, {\bf 485}, 660

\rskbibitem{} Safronov, V.\ S.,  1960, \anap,  {\bf 23},  979 

\rskbibitem{} Salpeter, E.\ E., 1955, \apj, {\bf 121}, 161

\rskbibitem{} Sanders, D.\ B., and I.\ F.\ Mirabel, 1996, \araa, {\bf 34}, 749

\rskbibitem{} Saraceno, P., P.\ Andr{\'e}, C.\ Ceccarelli, M.\ Griffin,
and  S.\ Molinari,  1996,  \aap,  {\bf 309}, 827  

\rskbibitem{} Sarma, A.\ P., T.\ H.\ Troland, D.\ A.\ Roberts, and R.\
M.\ Crutcher,   2000,  \apj,  {\bf 533}, 271  

\rskbibitem{} Scally, A., and C.\ Clarke, 2001, \mnras, {\bf 325}, 449

\rskbibitem{} Scalo, J.\ M., 1984, \apj, {\bf 277}, 556

\rskbibitem{} Scalo, J.\ M., 1986, {\em Fund.\ Cos.\ Phys.}, {\bf 11}, 1 

\rskbibitem{} Scalo, J.\ M., 1990, in {\em {Physical   Processes in Fragmentation and Star Formation}} edited by R.\   Capuzzo-Dolcetta and C.\ Chiosi (Kluwer Academic Publishers,   Dordrecht), P.\ 151

\rskbibitem{} Scalo, J.\ M., 1998, in {\em The Stellar Initial Mass Function (38th Herstmonceux Conference)}, edited by G.\ Gilmore, I.\ Parry, and S.\ Ryan (ASP Conf.\ Ser.\ Vol.\ 142), p.\ 201

\rskbibitem{} Scalo, J.\ M., E.\ {\'a}zquez-Semadeni, D.\ Chappell, T.\
Passot,  1998, \apj, {\bf 504}, 835   

\rskbibitem{} Schombert, J. M., G. D. Bothun, S. E. Schneider,
and S. S. McGaugh, 1992, \apj, {\bf 103}, 1107

\rskbibitem{} Schombert, J. M., S. S. McGaugh, and J. A. Eder, 2001, \apj,
{\bf 121}, 2420

\rskbibitem{} Schmitz, F.,  1983,  \aap,  {\bf 120}, 234 

\rskbibitem{} Schmitz, F.,  1984,  \aap,  {\bf 131}, 309 

\rskbibitem{} Schmitz, F.,  1986,  \aap,  {\bf 169}, 171 

\rskbibitem{} Schmitz, F.,  1987,  \aap,  {\bf 179}, 167 

\rskbibitem{} Schmitz, F.,  1988,  \aap,  {\bf 200}, 127 

\rskbibitem{} Schmitz, F., and R.\ Ebert,  1986,  \aap,  {\bf 154}, 214 

\rskbibitem{} Schmitz, F., and R.\ Ebert,  1987,  \aap,  {\bf 181}, 41 

\rskbibitem{} Scoville, N.\ Z.\ and K.\ Hersh, 1979, \apj, {\bf 229}, 578

\rskbibitem{} Sellwood, J.\ A., and S.\ A.\ Balbus, 1999, \apj, {\bf 511}, 660

\rskbibitem{} She, Z.-S., 1991, {\em Fluid Dyn.\ Res.}, {\bf 8}, 143

\rskbibitem{} She, Z.-S., and E.\ Leveque, 1994, \prl, {\bf 72}, 336

\rskbibitem{} She, Z.-S., E.\ Jackson, and  S.\ A.\ Orszag,  1991, \prsa, {\bf
    434}, 101 

\rskbibitem{} Shlosman, I., M.~C.~Begelman, and J.~Frank, 1990, \nat,
{\bf 345}, 679

\rskbibitem{} Shu, F.\ H., 1977, \apj, {\bf 214}, 488

\rskbibitem{} Shu, F.\ H., 1991, in {\em Physics of Star Formation and
Early Stellar Evolution}, edited by C.\ J.\ Lada and N.\ D.\ Kylafis
(Kluwer, Dordrecht), p.\ 365

\rskbibitem{} Shu, F.\ H., and Z.\ Li,  1997,  \apj,  {\bf 475}, 251 

\rskbibitem{} Shu, F. H., S. Lizano, S. P. Ruden, and J. Najita, 1998,
\apjl, {\bf 328}, L19

\rskbibitem{} Shu, F.\ H., J.\ Najita, D.\ Galli, E.\ Ostriker, and S.\
Lizano,  1993,  in {\em Protostars and Planets III},  edited by E.\
H.\ Levy and J.\ I.\ Lunine (University of Arizona Press, Tucson), p.\
3  

\rskbibitem{} Shu, F.\ H., F.\ C.\ Adams, and S.\ Lizano,  1987,  \araa,
{\bf  25}, 23  

\rskbibitem{} Shu, F.\ H., A.\ Allen, H.\ Shang, E.\ C.\ Ostriker, and
Z.\ Li,   1999,  in {\em NATO ASIC Proc.\ 540: The Origin of Stars and
Planetary Systems}, edited by C.\ J.\ Lada and N.\ D.\ Kylafis (Kluwer
Academic Publishers), p.\ 193  

\rskbibitem{} Shu, F.\ H., G.\ Laughlin, S.\ Lizano, and D.\ Galli,
2000,  \apj,  {\bf 535}, 190  

\rskbibitem{} Silk, J., 1995, \apj, {\bf 438}, L41

\rskbibitem{} Silk, J., and T.\ Takahashi, 1979, \apj, {\bf 229}, 242

\rskbibitem{} Silk, J., and Y.\ Suto,  1988,  \apj,  {\bf 335}, 295 

\rskbibitem{} Simon, M., 1997, \apj, {\bf 482}, L81

\rskbibitem{} Simon, R., J. M. Jackson, D. P. Clemens, T. M. Bania, M. H. Heyer,
2001, \apj, {\bf 551}, 747

\rskbibitem{} Simpson, C. E., and S. T. Gottesman, 2000, \aj, {\bf 120}, 2975

\rskbibitem{} Sirianni, M., A.~Nota, G.~De Marchi, C.~Leitherer, and
M.~Clampin, 2002, \apj, {\bf 579}, 275

\rskbibitem{} Smith, K.\ W., I.\ A.\ Bonnell, and M.\ R.\ Bate, 1997, \mnras,
{\bf 288}, 1041 

\rskbibitem{} Smith, K.\ W.\ and I.\ A.\ Bonnell, 2001, \mnras, {\bf 322}, L1

\rskbibitem{} Smith, M.\ D., and M.-M.\ Mac Low,  1997, \aap, {\bf 326}, 801

\rskbibitem{} Smith, M.~D., M.-M.\ Mac Low, and F.\
  Heitsch, 2000, \aap, {\bf 362}, 333 

\rskbibitem{} Snell, R. L., R. B. Loren, and R. L. Plambeck, 1980, \apjl,
{\bf 239}, L17

\rskbibitem{} Solomon, P. M., A. R. Rivolo, J. Barrett, and A. Yahil,
1987, \apj, {\bf 319}, 730

\rskbibitem{} Soukup, J. E., and C. Yuan, \apj, {\bf 246}, 376

\rskbibitem{} Spaans, M., 1996,  \aap, {\bf 307}, 271

\rskbibitem{} Spaans, M., and C.\ M.\ Carollo, 1998, \apj, {\bf 502},  640

\rskbibitem{} Spaans, M., and C.\ A.\ Norman, 1997, \apj, {\bf 488},  27

\rskbibitem{} Spaans, M., and J., Silk, 2000, \apj, {\bf 538}, 115

\rskbibitem{} Spaans, M., and E.\ F.\ van Dishoeck, 1997, \apj,  {\bf 323}, 953

\rskbibitem{} Spitzer, L., Jr., 1968, {\em Diffuse Matter in Space}, (Wiley
Interscience, New York)

\rskbibitem{} Stahler, S.\ W., 1988, \apj, {\bf 332}, 804

\rskbibitem{} Stahler, S.\ W., F.\ Palla, and P.\ T.\ P.\ Ho, 2000, in {\em
Protostars and Planets IV}, edited by V.\ Mannings, A.\ P.\ Boss, \& S.\ S.\
Russell (University of Arizona Press, Tucson), p.\ 327  

\rskbibitem{} Stauffer, J. R., L. W., Hartmann, and D. Barrado y Navascues, 1995,
\apj, {\bf 454}, 910

\rskbibitem{} Steinmetz, M., 1996, \mnras, {\bf 278}, 1005

\rskbibitem{} Sterzik, M.\ F., and R.\ H.\ Durisen, 1995, \aap, {\bf 304} L9.

\rskbibitem{} Sterzik, M.\ F., and R.\ H.\ Durisen, 1998, \aap, {\bf 339}, 95

\rskbibitem{} Stone, J. M., and M. L. Norman, 1992a, \apjs, {\bf 80}, 753

\rskbibitem{} Stone, J. M., and M. L. Norman, 1992b, \apjs, {\bf 80}, 791

\rskbibitem{} Stone, J.\ M., E.\ C.\ Ostriker, and C.\ F.\ Gammie, 1998, \apj, {\bf 508}, L99 

\rskbibitem{} Strittmatter, P.\ A., 1966, \mnras, {\bf 132}, 359

\rskbibitem{} Strom, K.\ M., S.\ E.\ Strom,  and K.\ M.\ Merrill, 1993, \apj, {\bf 412}, 233 

\rskbibitem{} Stutzki, J., and R.\ G{\"u}sten, 1990, \apj, {\bf 356}, 513
  
\rskbibitem{} Stutzki J., F.\ Bensch, A.\ Heithausen,
  V.\ Ossenkopf, and M.\ Zielinsky, 1998, \aap {\bf 336}, 697

\rskbibitem{}  Sugimoto, D., Y.\ Chikada, J.\ Makino, T.\ Ito, T.\
  Ebisuzaki, and M.\ Umemura, 1990, \nat, {\bf 345}, 33 

\rskbibitem{} Suto, Y., and J.\ Silk, 1988, \apj, {\bf 326}, 527

\rskbibitem{} Swaters, R. A., B. F. Madore, \& M. Trewhella, 2000, \apjl,
{\bf 531}, L107

\rskbibitem{} Swenson, F. J., J. Faulkner, F. J. Rogers, and C. A. Iglesias,
1994, \apj, {\bf 425}, 286

\rskbibitem{} Sytine, I.~V., D.\ H.\ Porter, P.\ R.\
  Woodward, S.\ W.\ Hodson,  and K.-H.\ Winkler, 2000, \jcp, {\bf 158}, 225 

\rskbibitem{} Tafalla, M., D.\ Mardones, P.\ C.\ Myers, P.\ Caselli, R.\ Bachiller,  and P.\ J.\ Benson,  1998,  \apj,  {\bf 504}, 900 

\rskbibitem{} Tatematsu, K., and 15 colleagues,  1993,  \apj,  {\bf 404}, 643 

\rskbibitem{} Taylor, G.\ I., 1921, {\em Proc. London Math.\
    Soc.}, {\bf 20}, 126

\rskbibitem{} Testi, L., and A.\ I.\ Sargent, 1998,
  \apjl, {\bf 508}, L91 

\rskbibitem{} Testi, L., and A.\ I.\ Sargent, 2000,
  in {\em Imaging at Radio through Sub-millimeter wavelengths}, edited by J.\
  Magnum and S.\ Radford (ASP Conf.\ Series 217), p.\ 283

\rskbibitem{} Testi, L., F.\ Palla, and A.\ Natta, 1999, \aap, {\bf 342}, 515

\rskbibitem{} Terebey, S., F. H. Shu, and P. Cassen, 1984, \apj, {\bf 286}, 529

\rskbibitem{} Thornton, K., M. Gaudlitz, H.-Th. Janka, and M. Steinmetz,
1998, \apj, {\bf 500}, 95

\rskbibitem{} Tohline, J.\ E.,  1980,  \apj,  {\bf 235}, 866 

\rskbibitem{} Tohline, J.\ E.,  1982,  {\em Fund.\ Cosmic Phys.},  {\bf  8}, 1 

\rskbibitem{} Tomisaka, K.,  1991,  \apj,  {\bf 376}, 190 

\rskbibitem{} Tomisaka, K.,  1995,  \apj,  {\bf 438}, 226 

\rskbibitem{} Tomisaka, K.,  1996a,  \pasj,  {\bf 48}, 701 

\rskbibitem{} Tomisaka, K.,  1996b,  \pasj,  {\bf 48}, L97 

\rskbibitem{} Tomisaka, K., S.\ Ikeuchi, and T.\ Nakamura,  1988a,  \apj,
{\bf  326}, 208  

\rskbibitem{} Tomisaka, K., S.\ Ikeuchi, and T.\ Nakamura,  1988b,  \apj,
{\bf  335}, 239  

\rskbibitem{} Tomisaka, K., S.\ Ikeuchi, and T.\ Nakamura,  1989a,  \apj,
{\bf  341}, 220  

\rskbibitem{} Tomisaka, K., S.\ Ikeuchi, and T.\ Nakamura,  1989b,  \apj,
{\bf  346}, 1061  

\rskbibitem{} Tomisaka, K., S.\ Ikeuchi, and T.\ Nakamura,  1990,  \apj,
{\bf  362}, 202  

\rskbibitem{} Toomre, A., 1964, \apj, {\bf 139}, 1217

\rskbibitem{} Troland, T.\ H., R.\ M.\ Crutcher, A.\ A.\ Goodman, C.\
Heiles, I.\ Kazes, P.\ C.\ Myers, 1996, \apj, {\bf 471}, 302

\rskbibitem{} Troland, T. H., and C. Heiles, 1986, \apj, {\bf 301}, 339

\rskbibitem{} Truelove, J.\ K., R.\ I.\ Klein, C.\ F.\ McKee, J.\ H.\  Holliman, L.\ H.\ Howell, and J.\ A.\ Greenough,  1997,  \apj,  {\bf 489},  L179 

\rskbibitem{} Truelove, J.\ K., R.\ I.\ Klein, C.\ F.\ McKee, J.\ H.\  Holliman, L.\ H.\ Howell, J.\ A.\ Greenough, and D.\ T.\ Woods,  1998,   \apj,  {\bf 495}, 821 

\rskbibitem{} Tsai, J.\ C., and J.\ J.\ L.\ Hsu, 1995, \apj, {\bf 448}, 774

\rskbibitem{} Tscharnuter, W.,  1975,  \aap,  {\bf 39}, 207 

\rskbibitem{} Tsuribe, T.\ and S.\ Inutsuka,  1999a,  \apj,  {\bf 523}, L155 

\rskbibitem{} Tsuribe, T.\ and S.\ Inutsuka,  1999b,  \apj,  {\bf 526}, 307 

\rskbibitem{} Tubbs, A. D., 1980, \apj, {\bf 239}, 882

\rskbibitem{} Turner, J.\ A., S.\ J.\ Chapman, A.\ S.\ Bhattal, M.\ J.\
Disney, H.\ Pongracic, and A.\ P.\ Whitworth,  1995,  \mnras,  {\bf
277},  705  

\rskbibitem{} Umemura, M., T.\ Fukushige, J.\ Makino, T.\ Ebisuzaki, D.\ Sugimoto,
  E.\ L.\ Turner, and  A.\ Loeb,  1993, \pasj,  {\bf 45}, 311

\rskbibitem{} Vainshtein, S.\ I., 1997, \pre, {\bf 56}, 6787 

\rskbibitem{} van der Hulst, J. M., E. D. Skillman, T. R. Smith,
G. D. Bothun, S. S. McGaugh, and W. J. G. de Blok, 1993, \apj, {\bf 106},
548

\rskbibitem{} van der Hulst, J. M., E. D. Skillman, R. C. Kennicutt, \&
G. D. Bothun, 1987, \aap, {\bf 177}, 63

\rskbibitem{} van der Marel, R.~P., and M.\ Franx,  1993, \apj, {\bf 407}, 525 

\rskbibitem{} van Dishoeck, E.\ F.\ and G.\ A.\ Blake,  1998,  \araa,  {\bf 36},  317 

\rskbibitem{} van Dishoeck, E.\ F., and M.\ R.\ Hogerheijde, 2000, in {\em The Origin of Stars and Planetary Systems} edited by C.\ J.\ Lada and N.\ D.\ Kylafis (Kluwer Academic Publisher, Dordrecht), p.\ 97

\rskbibitem{} van Dishoeck, E.\ F., G.\ A.\ Blake, B.\ T.\ Draine, and J.\ I.\ Lunine, 1993, in {\em Protostars and Planets III}, edited by  E.\ H.\ Levy and J.\ I.\ Lunine (University of Arizona Press, Tucson), p.\ 163

\rskbibitem{} van Leer, B., 1977, \jcp, {\bf 23}, 276

\rskbibitem{} van Zee, L., M. P. Haynes, J. J. Salzer, and A. H. Broeils, 1997,
\aj, {\bf 113}, 1618

\rskbibitem{} van Zee, L., J.~J.~Salzer, and E.~D.~Skillman, 2001, \aj,
{\bf 122}, 121

\rskbibitem{} van Zee, L., E.\ D.\ Skillman, and J.\ J.\ Salzer, 1998, \aj,
{\bf 116}, 1186

\rskbibitem{} V{\'a}zquez-Semadeni, E.\ 1994, \apj, {\bf 423}, 681 
  
\rskbibitem{} V\'azquez-Semadeni, E.,
  2000, in {\em The Chaotic Universe}, edited by V.\ G.\ Gurzadyan and R.\ 
  Ruffini (World Sci.), p.\ 379

\rskbibitem{} V{\'a}zquez-Semadeni, E., and A.\ Gazol, 1995, \aap, {\bf 303}, 204

\rskbibitem{} {V{\'a}zquez-Semadeni},
  E., and J.\ Scalo, 1992, \prl, {\bf 68}, 2921   

\rskbibitem{} V{\'a}zquez-Semadeni, E., J.\ Ballesteros-Paredes, and L.\
F.\ Rodr\'iguez,  1997, \apj, {\bf 474}, 292 

\rskbibitem{} V{\'a}zquez-Semadeni, E., J.\ Ballesteros-Paredes, and R.\
S.\ Klessen, 2003, \apj, submitted

\rskbibitem{} V{\'a}zquez-Semadeni, E.,
  T.\ Passot, and A.\ Pouquet, 1995, \apj, {\bf 441}, 702  

\rskbibitem{} V{\'a}zquez-Semadeni, E., T.\ Passot, and A.\ Pouquet, 1996 \apj,
  {\bf 473}, 881 

\rskbibitem{} V{\'a}zquez-Semadeni, E., M.\ Shadmehri, J.\ Ballesteros-Paredes, 2002, \apj,
submitted (astro-ph/0208245) 

\rskbibitem{} V{\'a}zquez-Semadeni, E.,
  E.\ C.\ Ostriker, T.\ Passot, C.\ F.\ Gammie, and J.\ M.\ Stone,  2000, in
  {\em Protostars and Planets IV},  edited by V.\ Mannings,  A.\ P.\ Boss, and
  S.\ S.\ Russell (University of Arizona Press, Tucson), p.\ 3 

\rskbibitem{} Verschuur, G.\ L., 1995a, \apj, {\bf 451}, 624

\rskbibitem{} Verschuur, G.\ L., 1995b, \apj, {\bf 451}, 645

\rskbibitem{} Vincent, A., and M.\ Meneguzzi, 1991, \jfm, {\bf 225}, 1  

\rskbibitem{} Vink, J. S., A. de Koter, and H. J. G. L. M. Lamers, 2000,
\aap, {\bf 362}, 295

\rskbibitem{} Vio, R., G.\ Fasano, M.\ Lazzarin, and O.\ Lessi, 1994, \aap, {\bf
    289}, 640

\rskbibitem{} von Weizs\"acker, C.\ F., 1943, \za, {\bf 22}, 319

\rskbibitem{} von Weizs\"acker, C.\ F., 1951, \apj, {\bf 114}, 165

\rskbibitem{} Wada, K., G. Meurer, and C.\ A.\ Norman,  2002,  \apj,
{\bf 577}, 197  

\rskbibitem{} Wada, K., and C.\ A.\ Norman,  1999,  \apjl,  {\bf 516}, L13 

\rskbibitem{} Wada, K., and C.\ A.\ Norman,  2001,  \apj,  {\bf 547}, 172 

\rskbibitem{} Wada, K., M. Spaans, and S. Kim, 2000, \apj, {\bf 540}, 797

\rskbibitem{} Walborn, N.\ R., R.\ H.\ Barb{\'a}, W.\ Brandner, M.\ ;.\
Rubio, E.\ K.\ Grebel, and R.\ G.\ Probst, 1999, \aj, {\bf 117}, 225 

\rskbibitem{} Walder, R.\ and D.\ Folini, 2000, \apss, {\bf 274}, 343

\rskbibitem{} Walker, T.\ P., G.\ Steigman, H.-S.\ Kang, D.\ M.\ Schramm, and K.\ A.\ Olive, 1991, \apj, {\bf 376}, 51

\rskbibitem{} Walter, F. M., A. Brown, R. D. Mathieu, P. C. Myers, and
F. V. Vrba, 1988, \aj, {\bf 96}, 297

\rskbibitem{} Ward-Thompson, D., F.\ Motte, and P.\ Andr{\'e},  1999,
\mnras,  {\bf  305}, 143  

\rskbibitem{} Ward-Thompson, D., J.\ M.\ Kirk, R.\ M.\ Crutcher, J.\ S.\
Greaves,  W.\ S.\ Holland, and P.\ Andr{\'e},  2000,  \apj,  {\bf
537}, L135  

\rskbibitem{} Ward-Thompson, D., P.\ F.\ Scott, R.\ E.\ Hills, and P.\
Andr{\'e},   1994,  \mnras,  {\bf 268}, 276  

\rskbibitem{} Wardle, M., 1990, \mnras, {\bf 246}, 98

\rskbibitem{} Whitmore, B.\ C., and F.\ Schweizer, 1995, \aj, {\bf 109}, 960

\rskbibitem{} Whitmore, B.\ C., 2000, in {\em Space Telescope Symposium
Series, No.\ 14}, edited by M.\ Livio (Cambridge U.\ Press, Cambridge)
in press (astro-ph/0012546) 

\rskbibitem{} Whitworth, A. P., 1979, \mnras, {\bf 186}, 59

\rskbibitem{} Whitworth, A.\ P., and D.\ Summers, 1985, \mnras, {\bf 214}, 1

\rskbibitem{} Whitworth, A.\ P., and D.\ Ward-Thompson, 2001, \apj,  {\bf
547}, 317  

\rskbibitem{} Whitworth, A.\ P., A.\ S.\ Bhattal, N.\ Francis, and S.\
J.\ Watkins, 1996, \mnras, {\bf 283}, 1061 

\rskbibitem{} Whitworth, A.\ P., S.\ J.\ Chapman, A.\ S.\ Bhattal, M.\
J.\  Disney, H.\ Pongracic, and J.\ A.\ Turner,  1995,  \mnras,  {\bf
277}, 727  

\rskbibitem{} Wichmann, R., J.\ Krautter, E.\ Covino, J.\ M.\ Alcala, R.\
Neuh{\"a}user, J.\ H.\ M.\ M.\ Schmitt, 1997, \aap, {\bf 320}, 185 

\rskbibitem{} Wiesemeyer, H., R.\ Guesten, J.\ E.\ Wink, H.\ W.\ Yorke,
1997,  {\em \aap}, {\bf 320}, 287 

\rskbibitem{} Wilden, B.\ S.,  B.\ F.\ Jones, D.\ N.\ C.\
  Lin, and D.\ R.\ Soderblom,  2002, \aj,  {\bf 124}, 2799  

\rskbibitem{} Williams, J.\ P., and L.\ Blitz, 1995, \apj, {\bf
451}, 252

\rskbibitem{} Williams, J.\ P., L.\ Blitz, and C.\ F.\ 
  McKee, 2000, in {\em Protostars and Planets IV}, edited by V.\ Mannings, A.\ 
  P.\ Boss, and S.\ S.\ Russell (University of Arizona Press, Tucson), p.\ 97

\rskbibitem{} Williams, J.\ P., L.\ Blitz, and A. A. Stark, 1995, \apj, {\bf
494}, 657

\rskbibitem{} Williams, J.\ P., E.\ J.\ de Geus, L.\ Blitz, 1994, \apj,
{\bf 428}, 693  

\rskbibitem{} Williams, J.\ P., P.\ C.\ Myers, D.\ J.\ Wilner, and J.\ di
Francesco,  1999,  \apj,  {\bf 513}, L61  

\rskbibitem{} Wilner, D.\ J., P.\ C.\ Myers, D.\ Mardones, and M.\
Tafalla,  2000,   \apj,  {\bf 544}, L69  

\rskbibitem{} Winkler, K.-H.\ A., and  M.\ J.\  Newman, 1980a,
  \apj, {\bf 236}, 201  
\rskbibitem{} Winkler, K.-H.\ A., and  M.\ J.\  Newman, 1980b,
  \apj, {\bf 238}, 311  
 
\rskbibitem{} Winkler, K.-H.\ A., and M.\ L.\ Norman, 1986, in {\em Astrophysical
    Radiation Hy\-dro\-dynamics}, editors K.-H.\ A.\ Winkler and M.\ L.\ 
  Norman (Reidel Publishing, Dordrecht)

\rskbibitem{} Wiseman, J.\ J., and F.\ C.\ Adams, 1994, \apj, {\bf 435}, 708

\rskbibitem{} Wolfire, M.\ G, and J.\ P.\ Cassinelli, 1987, \apj, {\bf 319}, 850

\rskbibitem{} Wolfire, M.\ G., D.\ Hollenbach, C.\ F.\ McKee, A.\ G.\ G.\ M.\ Tielens,  and E.\ L.\ O.\ Bakes, 1995, \apj, {\bf 443}, 152

\rskbibitem{} Wong, T., and L.~Blitz, 2002, \apj, {\bf 569}, 157

\rskbibitem{} Wood, D.\ O.\ S., P.\ C.\ Myers, and  D.\ A.\ Daugherty, 1994, \apjs, {\bf 95}, 457

\rskbibitem{} Wuchterl, G., and M.\ U.\ Feuchtinger,  1998,
\aap,   {\bf 340}, 419
 
\rskbibitem{} Wuchterl, G., and R.\ S.\ Klessen, 2001, \apj, {\bf 560},  L185

\rskbibitem{} Wuchterl, G., and W.\ Tscharnuter, 2003, \aap, {\bf
  398}, 1081

\rskbibitem{} Yonekura, Y., K. Dobashi, A. Mizuno, H. Ogawa, Y. Fukui, 1997,
\apjs, {\bf 110}, 21

\rskbibitem{} Yorke, H.\ W., and E.\ Kr{\"u}gel, 1977, \aap, {\bf 54}, 183

\rskbibitem{} Yorke, H.\ W., and C.\ Sonnhalter, 2002, \apj, {\bf 569}, 846

\rskbibitem{} Yorke, H. W., Tenorio-Tagle, G., Bodenheimer, P., and
M. R\'o\.zyczka, 1989, \aap, {\bf 216}, 207  

\rskbibitem{} Yoshii, Y, and Y.\ Sabano, 1980, \pasj, {\bf 32}, 229

\rskbibitem{} Zhang, Q., S.\ M.\ Fall,  and B.\ C.\ Whitmore, 2001, \apj,
{\bf 561}, 727 

\rskbibitem{} Zhou, S., N.\ J.\ Evans, C.\ Koempe, and C.\ M.\ Walmsley,  1993,   \apj,  {\bf 404}, 232 

\rskbibitem{}
Zielinsky M., and J.\ Stutzki, 1999, \aap, {\bf 347}, 633

\rskbibitem{} Zinnecker, H., 1984, \mnras, {\bf 210}, 43

\rskbibitem{} Zinnecker, H., 1990, in  {\em {Physical   Processes in Fragmentation and Star Formation}}, edited by R.\   Capuzzo-Dolcetta and C.\ Chiosi (Kluwer Academic Publisher,   Dordrecht), p.\ 201

\rskbibitem{} Zuckerman, B., 2001, \araa, {\bf 39}, 459

\rskbibitem{} Zuckerman, B., and P.\ Palmer, 1974, \araa, {\bf 12}, 279

\rskbibitem{} Zweibel, E.\ G., and A.\ Brandenburg, 1997, \apj, {\bf 478},
563

\rskbibitem{} Zweibel, E.\ G., and K.\ Josafatsson, 1983, \apj, {\bf 270}, 511

%%%%%%%%%%%%%%%%%%%%%%%%%%%%%%%%%%%%%%%%%%%%%%%%%%%%%%%%%%%

%\end{thebibliography}
%\end{references}

%%% Local Variables: 
%%% mode: latex
%%% TeX-master: "habil"
%%% End: 

\end{list}
}
%\end{thebibliography}

%%%%%%%%%%%%%%%%%%%%%%%%%%%%%%%%%%%%%%%%%%%%%%%%%%%%%%%%%%%%%%%%%%%%%%
\backmatter
\thispagestyle{empty}
{\onecolumn
\rsksection*{Thanks...}
\addcontentsline{toc}{chapter}{THANKS...}

... to all those people whose help and friendship made this work
possible!

Special thanks to Mordecai-Mark Mac~Low, and to Javier Ballesteros-Paredes, Peter 
Bodenheimer, Andreas Burkert, Fabian Heitsch, Pavel Kroupa, Douglas N.\ C.\ Lin,
Enrique V\'azquez-Semadeni, and Hans  Zinnecker. They have accompanied my
scientific career with  long-term collaborations, discussions, and
exchange of ideas and results.

I acknowledges support by the Emmy Noether Program of the Deutsche
Forschungsgemeinschaft (DFG: KL1358/1) and funding by the NASA Astrophysics
Theory Program through the Center for Star Formation Studies at NASA's Ames
Research Center, UC Berkeley, and UC Santa Cruz.  Preparation of this work
made furthermore extensive use of the NASA Astrophysical Data System Abstract
Service.

{\vfill %\begin{center}
\begin{minipage}[t]{11.0cm}
\em \Large Beyond all others, I thank Sibylle, Johanna, Jonathan and my mother
for love and faith and everything.\end{minipage}%\end{center}
}
\vspace*{2cm}
}

\newpage 
{\thispagestyle{empty}
\onecolumn
\vspace*{5.0cm}
\begin{center}
\includegraphics{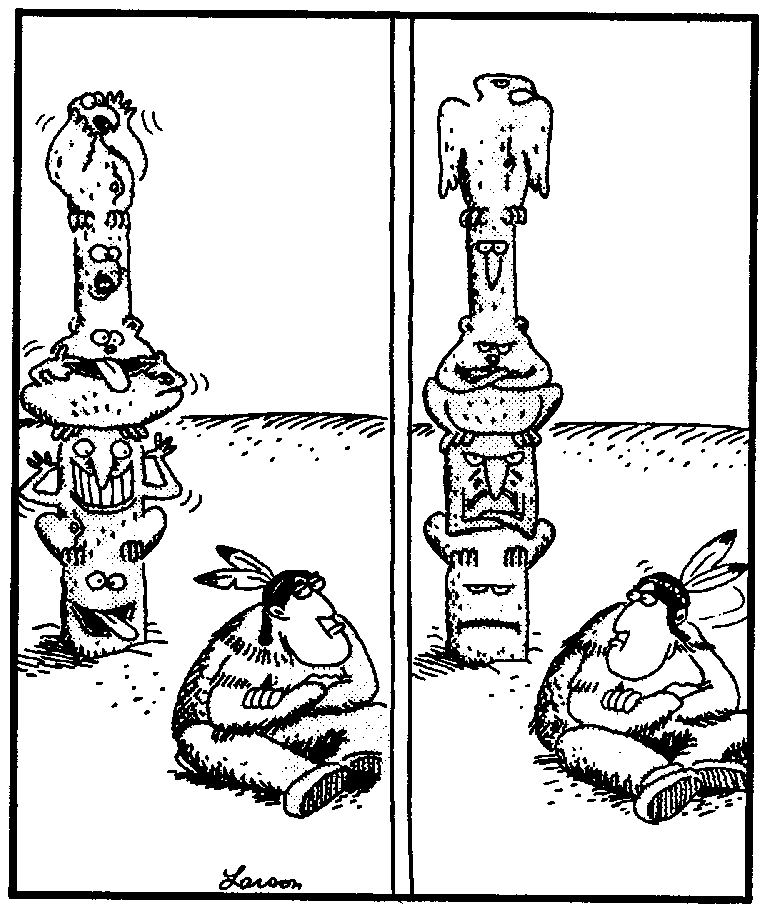}
\end{center}
}

\end{document}